\documentclass[final,3p,times]{elsarticle}

\usepackage{graphicx}
\usepackage{amssymb}
\usepackage{amsmath}
\usepackage{booktabs}
\usepackage{bm}
\usepackage{color}
\usepackage{multirow}
\usepackage{arydshln}
\usepackage{rotating}


\def\qq{\langle\bar qq\rangle}
\def\GGa{\langle GG\rangle}

\def\qGqa{\langle\bar qGq\rangle}

\def\f(s){\left[(\alpha+\beta)m_c^2-\alpha\beta s\right]}
\def\FF(s){\left[(\alpha+\beta)m_c^2-\alpha\beta s\right]}
\def\HH(s){\left[m_c^2-\alpha(1-\alpha) s\right]}

\usepackage[colorlinks, citecolor=blue,anchorcolor=red,menucolor=red, linkcolor=red,filecolor=red,runcolor=red,urlcolor=blue,frenchlinks=red]{hyperref}

\journal{Physics Reports}

\begin{document}

\begin{frontmatter}



\title{The hidden-charm pentaquark and tetraquark states}


\author[PKU,BEIHANG]{Hua-Xing Chen\footnotemark[1]}
\ead{hxchen@buaa.edu.cn}

\author[USASK]{Wei Chen\footnotemark[1]}
\ead{wec053@mail.usask.ca}

\author[LZU,IMP]{Xiang Liu\corref{cor1}}
\cortext[cor1]{Corresponding author} \ead{xiangliu@lzu.edu.cn}

\author[PKU,PKU1,PKU2]{Shi-Lin Zhu\corref{cor2}}
\cortext[cor2]{Corresponding author} \ead{zhusl@pku.edu.cn}

\address[PKU]{School of Physics and State Key Laboratory of Nuclear Physics and Technology, Peking University, Beijing 100871, China}
\address[BEIHANG]{School of Physics and Nuclear Energy Engineering, Beihang University, Beijing 100191, China}
\address[USASK]{Department of Physics and Engineering Physics, University of Saskatchewan, Saskatoon, Saskatchewan S7N 5E2, Canada}
\address[LZU]{School of Physical Science and Technology, Lanzhou University, Lanzhou 730000, China}
\address[IMP]{Research Center for Hadron and CSR Physics, Lanzhou University and Institute of Modern Physics of CAS, Lanzhou 730000, China}
\address[PKU1]{Collaborative Innovation Center of Quantum Matter, Beijing 100871, China}
\address[PKU2]{Center of High Energy Physics, Peking University, Beijing 100871, China}
\footnotetext[1]{These authors equally contribute to this work.}

\begin{abstract}

In the past decade many charmonium-like states were observed
experimentally. Especially those charged charmonium-like $Z_c$
states and bottomonium-like $Z_b$ states can not be accommodated
within the naive quark model. These charged $Z_c$ states are good
candidates of either the hidden-charm tetraquark states or molecules
composed of a pair of charmed mesons. Recently, the LHCb
Collaboration discovered two hidden-charm pentaquark states, which
are also beyond the quark model. In this work, we review the current
experimental progress and investigate various theoretical
interpretations of these candidates of the multiquark states. We
list the puzzles and theoretical challenges of these models when
confronted with the experimental data. We also discuss possible
future measurements which may distinguish the theoretical schemes on
the underlying structures of the hidden-charm multiquark states.

\end{abstract}


\end{frontmatter}


\tableofcontents

%-------------Section 1-----------------------------------
%
%=====================================================================================
%=====================================================================================
\section{Introduction}
\label{sect:1}

\subsection{Quark model and the multiquark states}
\label{Sect.1.1}
%=====================================================================================
%=====================================================================================
%

Quantum Chromodynamics (QCD) is the underlying theory of strong
interaction. According to QCD, quarks and anti-quarks are in the
fundamental representation of the non-Abelian SU(3) color gauge
group while gluons belong to the adjoint representation. The QCD
Lagrangian reads
\begin{eqnarray}\label{QCD}
\mathfrak{L} &=& \bar \psi_i \Big ( i \gamma^\mu (D_\mu)_{ij} - m
\delta_{ij} \Big ) \psi_j - {1 \over 4} G_{\mu\nu}^a G^{\mu\nu}_a \,
,
\end{eqnarray}
where the covariant derivative is defined as
\begin{eqnarray}
(D_\mu)_{ij} &=& \partial_\mu \delta_{ij} - i g A_\mu^a T^a_{ij} \, .
\end{eqnarray}
In Eq. (\ref{QCD}), $\psi_i(x)$ is the quark field, and $A_\mu^a$ is
the gluon field, both of which carry the color charge. $\gamma_\mu$
is the Dirac matrix and $T^a_{ij}={\lambda^a_{ij} / 2}$ is the
generator of the SU(3) gauge group.

QCD has three important properties: asymptotic freedom, confinement,
approximate chiral symmetry and its spontaneous breaking. Quarks and
gluons are confined within the mesons and baryons. Their color
interactions increase as the involved energy scale decreases. At the
hadronic scale, QCD is highly non-perturbative due to the
complicated infrared behavior of the non-Abelian SU(3) gauge group.

At present it is still impossible for us to derive the hadron
spectrum analytically from the QCD Lagrangian. Lattice QCD was
invented to solve QCD numerically through simulations on the
lattice, which has proven very powerful in the calculation of the
hadron spectrum and hadronic matrix elements. Besides lattice QCD,
many phenomenological models with some kind of QCD spirit were
proposed. Among them, the quark model may be the most successful one,
which categorizes hadrons into two families: mesons and baryons. The
former are made of one quark and one antiquark, and the latter are
made of three quarks.

With the experimental progress in the past decade, dozens of
charmonium-like $XYZ$ states have been reported~\cite{pdg}. They provide
good opportunities to identify tetraquark states, which are made of
two quarks and two antiquarks. Moreover, the LHCb Collaboration
recently observed two hidden-charm pentaquark resonances,
$P_c(4380)$ and $P_c(4450)$, in the $J/\psi p$ invariant mass
spectrum~\cite{Aaij:2015tga}. They are good candidates of pentaquark
states, which are made of four quarks and one antiquark.
We shall briefly review the experimental progress on these charmonium-like
states and hidden-charm pentaquark resonances in Sec.~\ref{sect:2}.

To study these tetraquark and pentaquark candidates, the traditional
quark model as well as its updated version seems to be incapable any
more. Various theoretical frameworks were proposed to interpret
these new multiquark systems, such as the one-boson-exchange (OBE)
model, the one-pion-exchange (OPE) model, the chiral unitary model,
the QCD sum rule, the chiral quark model, the diquark-antidiquark
model etc. We shall briefly introduce these models and review their
applications on the hidden-charm pentaquark resonances in
Sec.~\ref{sect:3}, and review the applications of more theoretical
frameworks on the charmonium-like states in Sec.~\ref{sect:4}. An
outlook and a brief summary will be given in Sec.~\ref{sect:5}.

\subsubsection{Quark model} \label{Sect.1.1.1}
%=====================================================================================
%=====================================================================================
%

According to the traditional quark model, a meson is composed of a
pair of quark and antiquark and a baryon is composed of three
quarks. Both mesons and baryons are color singlets. The quark in the
quark model is sometimes denoted as the constituent quark, which is
different from the current quark in the QCD Lagrangian. For example,
the constituent up/down quark mass is about one third of the nucleon mass
or one half of the $\rho$ meson mass. At the energy scale around 2 GeV,
the up/down current quark mass is around several MeV.

Within the quark model, each quark carries the energy
$\sqrt{m^2+{\bf p}^2}$, where $m$ is the constituent quark mass and $\bf p$ denotes
its momentum. In the non-relativistic limit, the energy term is
expanded as the sum of the mass and kinetic energy. The inter-quark
interactions include the linear confinement force and the one gluon
exchange force. There also exist various hyperfine interactions such
as the spin-spin interaction, the color-magnetic interaction, the
spin-orbit interaction, and the tensor force etc. Up to now, nearly
all the mesons and baryons can be classified within such a simple
quark model picture.

The discovery of $J/\psi$ \cite{Aubert:1974js,Augustin:1974xw} in
1974 inspired theorists to propose potential models
\cite{Eichten:1974af,Appelquist:1974yr}. The mass spectrum of the
charmonium family was obtained by solving the Schr\"odinger
equation. The hadron spectroscopy was reconsidered in the framework
of quark model
\cite{DeRujula:1975qlm,Sakharov:1980ph,Federman:1981nk,Stanley:1980zm,Stanley:1980fe,Koniuk:1979vy,Richard:1983mu,Richard:1983tc,Ono:1982ft,Basdevant:1985ux,Godfrey:1985xj}.
In the following, we take the well-known Godfrey-Isgur (GI) quark
model as an example and introduce it briefly \cite{Godfrey:1985xj}.

In the GI model~\cite{Godfrey:1985xj}, the interaction between the
quark and antiquark is described by the Hamiltonian
\begin{eqnarray}
\mathcal{H}=\sqrt{m_1^2+ \textbf{p}^2} + \sqrt{m_2^2+ \textbf{p}^2} + V_{\mathrm{eff}}\left(\textbf{p},\textbf{r}\right) \, ,
\end{eqnarray}
where the subscripts 1 and 2 denote the quark and the antiquark,
respectively. $V_{\mathrm{eff}}(\textbf{p},\textbf{r})$ is the
effective potential of the $q\bar{q}$ system, and contains the
short-distance one-gluon-exchange interaction and the long-distance
linear confining interaction. The latter was at first employed by
the Cornell group and later confirmed by the lattice QCD
simulations.

$V_{\mathrm{eff}}(\textbf{p},\textbf{r})$ is obtained from the
on-shell $q\bar{q}$ scattering amplitudes in the center-of-mass (CM)
frame \cite{Godfrey:1985xj}. In the non-relativistic limit,
$V_{\mathrm{eff}}(\textbf{p},\textbf{r})$ is transformed into the
standard non-relativistic potential $V_{\mathrm{eff}}(r)$:
\begin{eqnarray}
V_{\mathrm{eff}}(r)=H^{\mathrm{conf}} + H^{\mathrm{hyp}} +
H^{\mathrm{SO}} \, .
\end{eqnarray}
The first term $H^{\mathrm{conf}}$ includes the spin-independent
linear confinement and Coulomb-type interactions
\begin{eqnarray}
H^{\mathrm{conf}} = - \Bigg[ {3\over4} c + {3\over4} br -
\frac{\alpha_s(r)}{r} \Bigg] \bm{F}_1\cdot\bm{F}_2 \, ,
\end{eqnarray}
the second term $H^{\mathrm{hyp}}$ is the color-hyperfine
interaction
\begin{eqnarray}
H^{\mathrm{hyp}}&=&-\frac{\alpha_s(r)}{m_1m_2}\Bigg[\frac{8\pi}{3}\bm{S}_1\cdot\bm{S}_2\delta^3
(\bm r)+\frac{1}{r^3}\Big(\frac{3\bm{S}_1\cdot\bm r \bm{S}_2\cdot\bm
r}{r^2}-\bm{S}_1\cdot\bm{S}_2\Big)\Bigg] \bm{F}_1\cdot\bm{F}_2 \, ,
\end{eqnarray}
and the third term $H^{\mathrm{SO}}$ is the spin-orbit interaction
\begin{eqnarray}
H^{\mathrm{SO}}=H^{\mathrm{SO(cm)}}+H^{\mathrm{SO(tp)}} \, ,
\end{eqnarray}
where $H^{\mathrm{SO(cm)}}$ is the color-magnetic term and
$H^{\mathrm{SO(tp)}}$ is the Thomas-precession term, i.e.,
\begin{eqnarray}
H^{\mathrm{SO(cm)}} &=&
-\frac{\alpha_s(r)}{r^3}\left(\frac{1}{m_1}+\frac{1}{m_2}\right)\left(\frac{\bm{S}_1}{m_1}+\frac{\bm{S}_2}{m_2}\right)\cdot
\bm{L}~~\bm{F}_1\cdot\bm{F}_2,
\\
H^{\mathrm{SO(tp)}} &=& \frac{-1}{2r}\frac{\partial
H^{\mathrm{conf}}}{\partial
r}\Bigg(\frac{\bm{S}_1}{m^2_1}+\frac{\bm{S}_2}{m^2_2}\Bigg)\cdot
\bm{L}.
\end{eqnarray}
In the above expressions, $\bm{S}_1/\bm{S}_2$ denotes the spin of
the quark/antiquark and $\bm{L}$ is the orbital momentum between the
quark and the antiquark. $\bm{F}$ is related to the Gell-Mann
matrix, $\bm{F}_1=\bm{\lambda}_1/2$ and
$\bm{F}_2=-\bm{\lambda}^*_2/2$. Especially, we have
$\langle\bm{F}_1\cdot\bm{F}_2\rangle=-4/3$ for the mesons.

The relativistic effects were also taken into account in the GI
model. More details of the GI model can be found in Appendices of
Ref. \cite{Godfrey:1985xj}. The GI quark model was very successful
in the description of the spectrum and static properties of the
mesons and baryons.

\subsubsection{Exotic states and multiquark states} \label{Sect.1.1.2}
%=====================================================================================
%=====================================================================================
%

According to the quark model, the parity for a
meson is $P=(-)^{L+1}$ and
the $C$-parity for a neutral meson $C=(-)^{L+S}$, where $L$ and $S$ are
the orbital and spin angular momentum, respectively. The allowed
$J^{PC}$ reads: $0^{-+}$, $0^{++}$, $1^{--}$, $1^{+-}$, $1^{++}$,
$\cdots$. In contrast, a conventional $q\bar q$ meson in the quark
model can not carry the following quantum numbers: $0^{--}$,
$0^{+-}$, $1^{-+}$, $2^{+-}$, $\cdots$. States with these $J^{PC}$
quantum numbers are beyond the naive quark model, which are
sometimes denoted as exotic or non-conventional states.
Different from the meson case, the $qqq$ baryon in the quark model
can exhaust all the $J^{P}$ quantum numbers, i.e., $J^P={1\over
2}^{\pm}, {3\over 2}^{\pm}, {5\over 2}^{\pm}, \cdots$.

However, the constituent quark model can not be derived rigorously
from QCD. The quark model spectrum is not necessarily the same as
the QCD hadron spectrum. QCD may allow a much richer hadron spectrum.

In fact, at the birth of the quark model
\cite{GellMann:1964nj,Zweig:1981pd}, Gell-Mann and Zweig proposed
not only the existence of the $q\bar q$ mesons and $qqq$ baryons but
also the possible existence of the $q\bar qq\bar q$ tetraquarks and
$qqqq\bar q$ pentaquarks. The concept of the multiquarks was
proposed even before the advent of quantum chromodynamics (QCD)!

In Ref.~\cite{GellMann:1964nj}, M.~Gell-Mann wrote: ``{\it Baryons
can now be constructed from quarks by using the combinations
$(qqq)$, $(qqqq\bar q)$, etc., while mesons are made out of $(q \bar
q)$, $(qq \bar q \bar q)$, etc.}''

In Ref.~\cite{Zweig:1981pd}, G.~Zweig also wrote: ``{\it In general,
we would expect that baryons are built not only from the product of
these aces, $AAA$, but also from $\bar A AAAA$, $\bar A \bar A
AAAAA$, etc., where $\bar A$ denotes an anti-ace. Similarly, mesons
could be formed from $\bar A A$, $\bar A \bar A AA$, etc.}''

The multiquarks can be further classified into tetraquarks ($qq\bar
q \bar q$), pentaquarks ($qqqq\bar q$), dibaryon $(qqqqqq)$ and
baryonium $(qqq\bar q\bar q\bar q)$ etc. Jaffe studied the
tetraquark states within the framework of the MIT bag model in
1976~\cite{Jaffe:1976ig,Jaffe:1976ih}. This subject was later
studied by Chan and Hogaasen~\cite{Chan:1977st}, and many other
theorists. The tetraquark states containing heavy quarks were
investigated by Chao in 1979~\cite{Chao:1979tg,Chao:1979mm}. The
pentaquarks ($qqqq\bar q$) composed of light quarks were
investigated by Hogaasen and Sorba~\cite{Hogaasen:1978jw} in 1978
and Strotmann in 1979~\cite{Strottman:1979qu}. The name
``pentaquark'' was first proposed by Lipkin in
1987~\cite{Lipkin:1987sk}. Two groups studied possible pentaquarks
containing one charm quark in 1987
\cite{Gignoux:1987cn,Lipkin:1987sk}.

In 2003, the LEPS Collaboration announced the observation of the
$\Theta$ pentaquark which is composed of $uudd\bar s$
\cite{Nakano:2003qx}. However, this state was not confirmed by the
subsequent more advanced experiments \cite{Hicks:2012zz}.

Jaffe also discussed the H-dibaryon, where six light quarks $uuddss$
are confined within one MIT bag \cite{Jaffe:1976yi}. In nature there
exists the deuteron which is also composed of six light quarks (see
discussions in Refs.~\cite{Matveev:1977xt,Hogaasen:1979qa}). The
difference between the dibaryon and deuteron lies in their color
configurations. Within the deuteron, there are two quark clusters,
both of which are color singlets. For the dibaryon, one expects six
quarks within one cluster.

In QCD, the gluons not only mediate the strong interaction between
quarks but also interact among themselves since they carry color
charges. Two or more gluons may form the color singlet, which is
called the glueball. One or more gluons may interact with a pair of
quark and antiquark to form the hybrid meson. The hybrid mesons or
tetraquark states or the glueballs can carry all the so-called
exotic $J^{PC}$ quantum numbers in the quark model. Strictly
speaking, there does not exist any exotic quantum number from the
viewpoint of QCD. One can construct color-singlet local operators
to verify that these quantum numbers are allowed in QCD. We shall
illustrate this point in the following sections. Throughout this
review, either the word ``exotic'' or ``non-conventional'' should be
understood within the context of the quark model.

\subsubsection{Comparison of QED and QCD and their spectrum} \label{Sect.1.1.3}
%=====================================================================================
%=====================================================================================
%

Quantum Electrodynamics (QED) is very different from QCD. The gauge
group of QED is U(1). The photon mediates the electromagnetic
interactions between charges. However, the photon is neutral and
does not carry charge. There does not exist the photon
self-interaction. We do not have the analogue of the glueball and
hybrid meson in QED. Instead there are free electrons and photons
while all quarks and gluons are confined within the hadrons.

Except the above big difference, it's intriguing to notice the
similarity between QED and QCD. In the following, we compare the
well-known bound states in QED and possible hadrons in QCD. In QED
we have the bound states composed of $e^+e^-$, $\mu^+\mu^-$,
$\mu^+e^-$. In QCD we have the light mesons composed of $q\bar q$,
$s\bar s$, $s\bar q$, where $q=u, d$ is the up/down quark, and $s$
is the strange quark. For the hydrogen atom in QED, the electron
circles around the proton. For the heavy-flavored meson/baryon in
QCD, the light quarks circle around the heavy charm or bottom quark.

In QED there exist the bound states composed of $e^+e^-e^+e^-$ and
$e^+e^-\mu^+\mu^-$ \cite{Hylleraas:1947zza,Cassidy:2007}. In QCD
some of the scalar mesons below 1 GeV may have the flavor
configurations $q\bar q q \bar q$ and $q\bar q s\bar s$. In QED we
have the hydrogen molecule where two electrons are shared by the two
protons and the valence bond binds this system tightly. In QCD we
may expect the $q\bar Q q \bar Q$ and $\bar q Q q \bar Q$ tetraquark
states within one MIT bag, where the two light quarks are shared by
the two heavy quarks.

In QED there exist many molecules which are loosely bound by the van
der Vaals force. The van der Vaals force is nothing but the residual
electromagnetic force arising from the two-photon exchange process
in QED. In QCD we have the deuteron which is the hadronic molecular
state bound by the meson exchange force. At the quark-gluon level,
the meson exchange force is the residual strong interaction force
arising from the gluon and quark exchange process. In QCD we may
also expect other loosely bound deuteron-like molecular states
composed of two heavy flavored hadrons.

\subsection{General status of hadron spectroscopy}
\label{Sect.1.2}
%=====================================================================================
%=====================================================================================
%

Although most of the observed hadrons can be classified as the
ordinary $q\bar q$ mesons and $qqq$ baryons, there have been huge
theoretical and experimental efforts to search for the candidates of
the exotic hadrons. These exotic states encode important information
of QCD. For example, the identification of the glueballs and hybrid
mesons will establish the direct evidence of the dynamical role of
the gluons in the low energy sector.

The exotic $J^{PC}$ quantum numbers provide a convenient handle in
the search of the nonconventional states. If a resonance decays into
the final state with $J^{PC}=1^{-+}$, the parent resonance is a good
candidate of the hybrid meson.

On the other hand, the exotic flavor quantum number is also a
valuable asset in the experimental search of the exotic states. If
the resonance carries the isospin $I=2$ or a meson has an isospin
$I=3/2$, it may be a multiquark candidate. For the $\Theta$
resonance, its baryon number is $1$ and strangeness $S=+1$. Hence,
it must be a candidate of pentaquarks.

Some hadrons do not have exotic $J^{PC}$ or flavor quantum numbers.
But they may have exotic color or flavor or spatial configurations.
These states can be searched for through the overpopulation of the
quark model spectrum. Sometimes the deviation from the quark model
predictions of their masses, decay widths, various reactions,
production and decay behaviors may also provide insightful clues in
the search of the exotic states.

Let's take the $J^{PC}=0^{++}$ scalar isoscalar mesons as an
example. Below 2 GeV, we have $\sigma$, $f_0(980)$, $f_0(1370)$,
$f_0(1500)$, $f_0(1710)$, $f_0(1790)$, $f_0(1810)$~\cite{pdg}.
Within the quark model, there are only four scalar isoscalar mesons
within this mass range even if we consider the radial excitations.
Clearly there is serious overpopulation of the scalar spectrum. The
quark content of some of the above states can not be $q\bar q$.
Overpopulation of the spectrum provides another useful window in the
experimental search of the non-conventional states.

Let's move on to the nine scalar mesons below 1 GeV, which play a
fundamental role in the spontaneous breaking of the chiral symmetry
in QCD. The scalar meson carries one orbital excitation. Hence, its
mass is expected to be several hundred MeV higher than the $\rho$
meson mass in the quark model. Either lattice QCD simulation or
other theoretical approaches indicates the $L=1$ $q\bar q$ state
lies around 1.2 GeV. Within the quark model, the $f_0(980)$ meson
with the quark content $s\bar s$ should be $200\sim 300$ MeV heavier
than the $a_0(980)$ with the quark content $q\bar q$. However, they
are almost degenerate in reality. The unusual low mass of the scalar
nonet and the abnormal mass ordering of the $f_0(980)$ and
$a_0(980)$ are two puzzles in the quark model. In contrast, both
puzzles can be solved very naturally if the scalar mesons belong to
the tetraquark nonet \cite{Jaffe:1976ih,Chen:2007xr}. The chiral
unitary approach for the interaction of pseudoscalar mesons may also
give rise to these
structures~\cite{Oller:1997ti,Kaiser:1998fi,Pelaez:2015qba}.

Since 2003, many charmonium-like states have been observed through
$B$ meson decays, the initial state radiation (ISR), double
charmonium production, two photon fusion, and excited charmonium or
bottomonium decays. Some of them do not fit into the quark model
spectrum easily and are proposed as the candidates of the
hidden-charm exotic mesons, including the di-meson molecular states,
tetraquarks, hybrid charmonium states and conventional charmonium
states distorted by the coupled-channel effects, etc. Molecular
states are loosely bound states composed of a pair of heavy mesons.
They are probably bound by the long-range color-singlet pion
exchange. Tetraquarks are bound states of two quarks and two
antiquarks, which are bound by the colored force between quarks and
antiquarks. There are many states within the same tetraquark
multiplet. Some members are charged or even carry strangeness.
Hybrid charmonia are bound states composed of a charm
quark-antiquark pair and one excited gluon.

In this review, we focus on the recent experimental and theoretical
progress on the hidden-charm multiquark systems such as hybrid
charmonia, hidden-charm tetraquarks, hidden-charm pentaquarks, and
hadronic molecules composed of a pair of heavy-flavored hadrons.
Interested readers may also consult reviews in Refs.
\cite{Jaffe:2004ph,Valcarce:2005em,Swanson:2006st,Eichten:2007qx,Zhu:2007wz,Klempt:2007cp,Voloshin:2007dx,Godfrey:2008nc,Klempt:2009pi,Nielsen:2009uh,Brambilla:2010cs,Druzhinin:2011qd,Li:2012me,Liu:2013waa,Chen:2013wva,Chen:2014fza,Brambilla:2014jmp,Esposito:2014rxa,Olsen:2014qna,Briceno:2015rlt,Pelaez:2015qba}.

%=====================================================================================
%=====================================================================================
%

%-------------Section 3-----------------------------------
%
%=====================================================================================
%=====================================================================================
\section{Experimental progress on the hidden-charm multiquark states}
\label{sect:2}

\subsection{The Charmonium-like $XYZ$ states}
\label{Sect.2.0}
%=====================================================================================
%=====================================================================================
%

With the experimental progress, the family of the charmonium-like
states has become more and more abundant. To date, dozens of
charmonium-like states have been observed by several major particle
physics experimental collaborations such as CLEO-c, BaBar, Belle,
BESIII, CDF, D$\O$, LHCb, CMS and so on. Since 2003, these
collaborations have been continuing to surprise us with novel
discoveries, which have inspired theorists' extensive interests in
exploring the underlying mechanism behind those exotic phenomena. As
one of the most important issues in hadron physics, the study of the
charmonium-like states provides us a good chance to deepen our
understanding of the complicated non-perturbative behavior of QCD in
the low energy regime. Especially, investigations of the underlying
structures of the charmonium-like states may help us to understand
the mechanism of the confinement and chiral symmetry breaking
better.

All the above major particle physics experimental collaborations
have contributed to the observations of the charmonium-like states.
Before reviewing the experimental status of the charmonium-like
states, we would like to introduce the collaborations briefly:
\begin{enumerate}
\item {\bf CLEO-c}: CLEO-c was the experiment at the Cornell
Electron Storage Ring (CESR) located at Wilson Laboratory of Cornell
University.  As the upgrade of CLEO, the CLEO-c experiment ran at
lower energies and carried out the study of charmonia and charmed
mesons {due to the competition from two $B$ factories BaBar and
Belle}. The CLEO-c experiment confirmed the observation of the
$Y(4260)$ \cite{He:2006kg}. Although CLEO-c finished data-collecting
on 3 March 2008, the accumulated CLEO-c data was applied to confirm
the observation of the charged charmonium-like structure $Z_c(3900)$
\cite{Xiao:2013iha}.

\item{\bf BaBar}: As one of the two $B$ factories, the BaBar
experiment was designed to study CP violation in the $B$ meson
system. Its detector was located at SLAC National Accelerator
Laboratory, which ceased operation on 7 April 2008. However, its
data analysis is still ongoing. Due to the development of the
Initial State Radiation (ISR) technique, the BaBar experiment also
focused on the study of charmonia and charmonium-like states. In the
past decade, BaBar has played a crucial role in the discoveries of
many charmonium-like states. For example, BaBar first observed the
famous $Y(4260)$ in the $e^+e^-\to J/\psi\pi^+\pi^-$ process
\cite{Aubert:2005rm}.

\item{\bf Belle}: As the other $B$ factory, the Belle experiment
was located at the High Energy Accelerator Research Organization
(KEK), which was also set up to study CP violation in the $B$ meson
system. As a byproduct, the Belle Collaboration discovered many
charmonium-like states. For example, Belle reported the observation
of the $X(3872)$ in 2003 \cite{Choi:2003ue}, which is the first
member in the family of the charmonium-like states. Although Belle
finished its data-taking on 30 June 2010, its data analysis is going
on.

\item{\bf BESIII}: BESIII is the experiment at Beijing
Electron-Positron Collider II (BEPC II), located at Institute of
High Energy Physics (IHEP). Since its center of mass energy can go
up to 4.6 GeV, the BESIII experiment has become an ideal platform to
explore the charmonium-like states. In 2013, BESIII announced the
observation of the charged charmonium-like structure $Z_c(3900)$
\cite{Ablikim:2013mio}.

\item{\bf CDF and \bf D{\O}}: CDF and D{\O} were the two
particle experiments located at the Tevatron at Fermilab. They
discovered the top quark in 1995 \cite{Abe:1995hr,Abachi:1994td}.
The CDF and {D\O} experiments both confirmed the $X(3872)$
\cite{Acosta:2003zx,Abazov:2004kp}. The CDF Collaboration also
reported the charmonium-like state $Y(4140)$ \cite{Aaltonen:2009tz}.

\item{\bf LHCb}: As one of seven particle physics experiments
at the Large Hadron Collider (LHC) at CERN, the LHCb experiment
focuses on $B$-physics. LHCb also studies the productions of the
charmonium-like states through the direct $pp$ collisions and the
$B$ meson decays. For example, the LHCb Collaboration measured the
spin-parity quantum number of the $X(3872)$ \cite{Aaij:2013zoa}.

\item{\bf CMS}: CMS is another important experiment at LHC
at CERN. The CMS and ATLAS collaborations discovered the Higgs Boson
in July 2012 \cite{Chatrchyan:2012xdj,Aad:2012tfa}. In recent years,
CMS also contributed to the search of the charmonium-like states
such as the $X(3872)$ \cite{Chatrchyan:2013cld} and the $Y(4140)$
\cite{Chatrchyan:2013dma}.

\end{enumerate}

\begin{figure}[hbtp]
\begin{center}
\includegraphics[width=10cm]{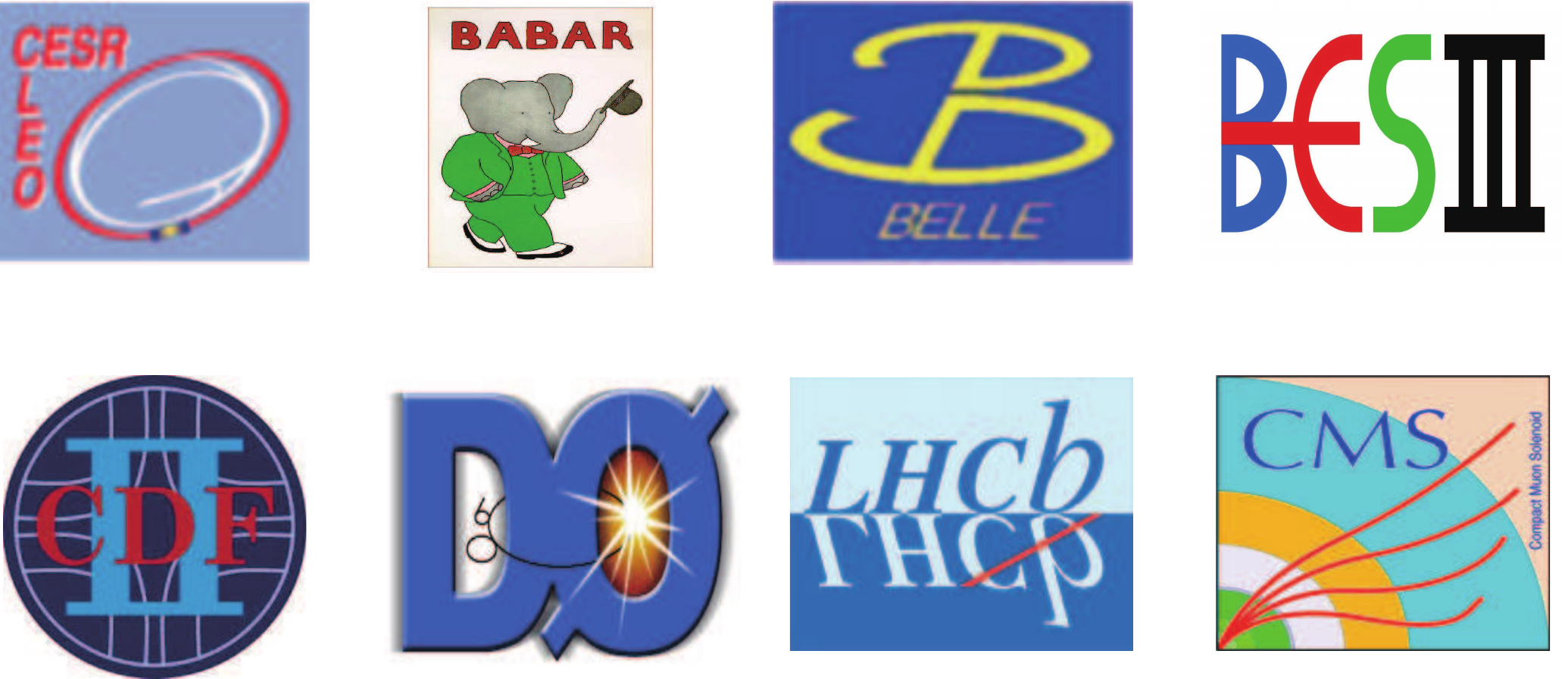}
\caption{(Color online) The logos of the experimental collaborations which
contributed to the observation of the charmonium-like states. }
\label{Fig:2.0.logo}
\end{center}
\end{figure}

In Fig. \ref{Fig:2.0.logo}, we collect Logos of these experimental
collaborations which have contributed to the observations of the
charmonium-like states.

\begin{figure}[hbtp]
\begin{center}
\includegraphics[width=15cm]{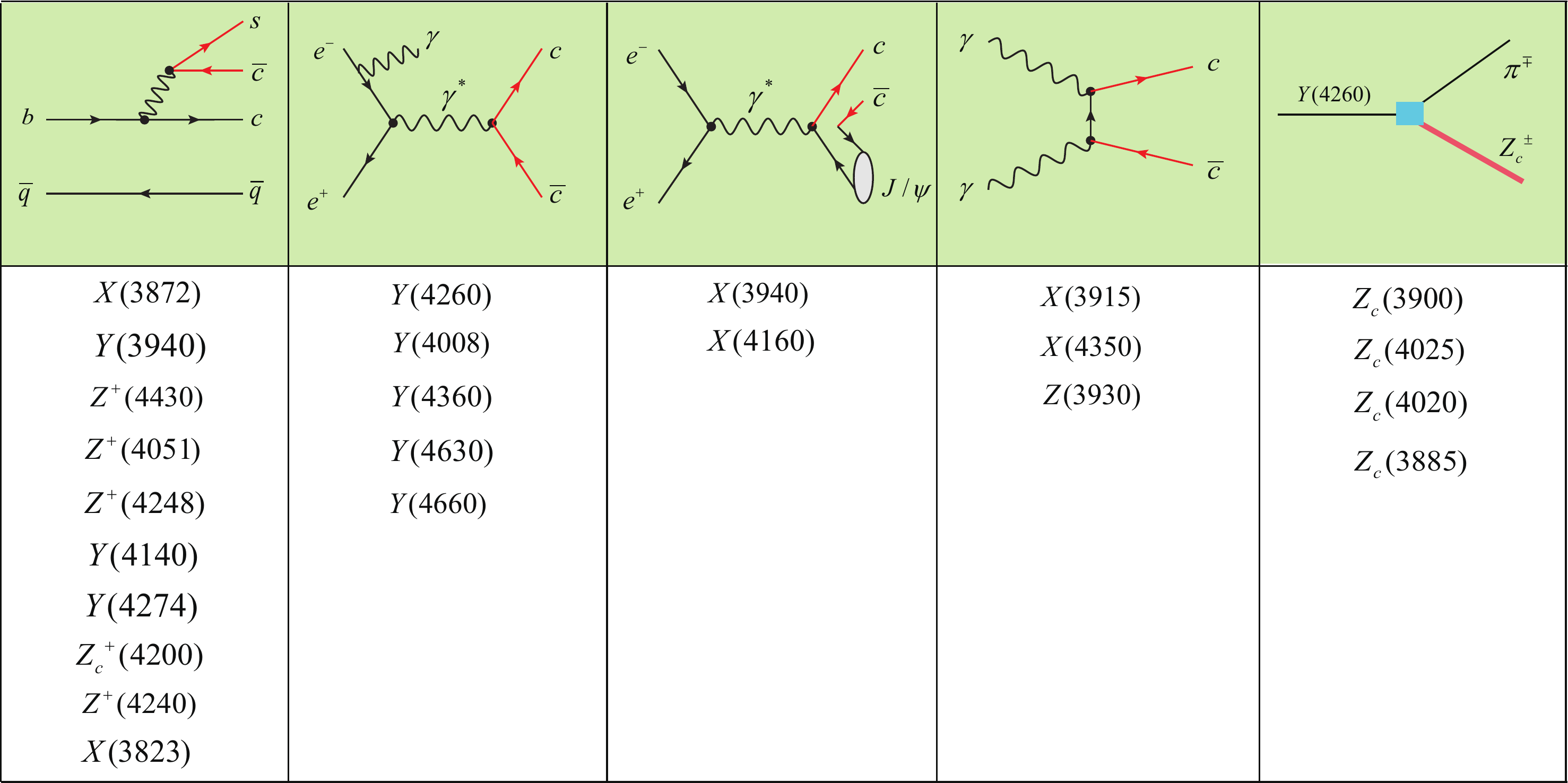}
\caption{(Color online) Five groups of the charmonium-like states corresponding to
five production mechanisms. } \label{Fig:2.0.production}
\end{center}
\end{figure}

According to the different production mechanisms, all the observed
charmonium-like states can be categorized into five groups as shown
in Fig. \ref{Fig:2.0.production}. The states collected in the first,
second, third, and fourth columns are produced via the $B$ meson
decays, initial state radiation technique (ISR) in the $e^+e^-$
annihilation, the double charmonium production processes, and two
photon fusion processes, respectively. The
$Z_c(3900)/Z_c(4025)/Z_c(3885)/Z_c(4200)$ listed in the fifth column
are produced from the hadronic decays of the $Y(4260)$.

%
%=====================================================================================
%=====================================================================================
\subsubsection{$XYZ$ states produced through $B$ meson decays}
\label{Sect:2.1}
%=====================================================================================
%=====================================================================================
%

\paragraph{$X(3872)$}
\label{Sect:2.1.1}

\begin{figure}[hbtp]
\begin{center}
\includegraphics[width=15cm]{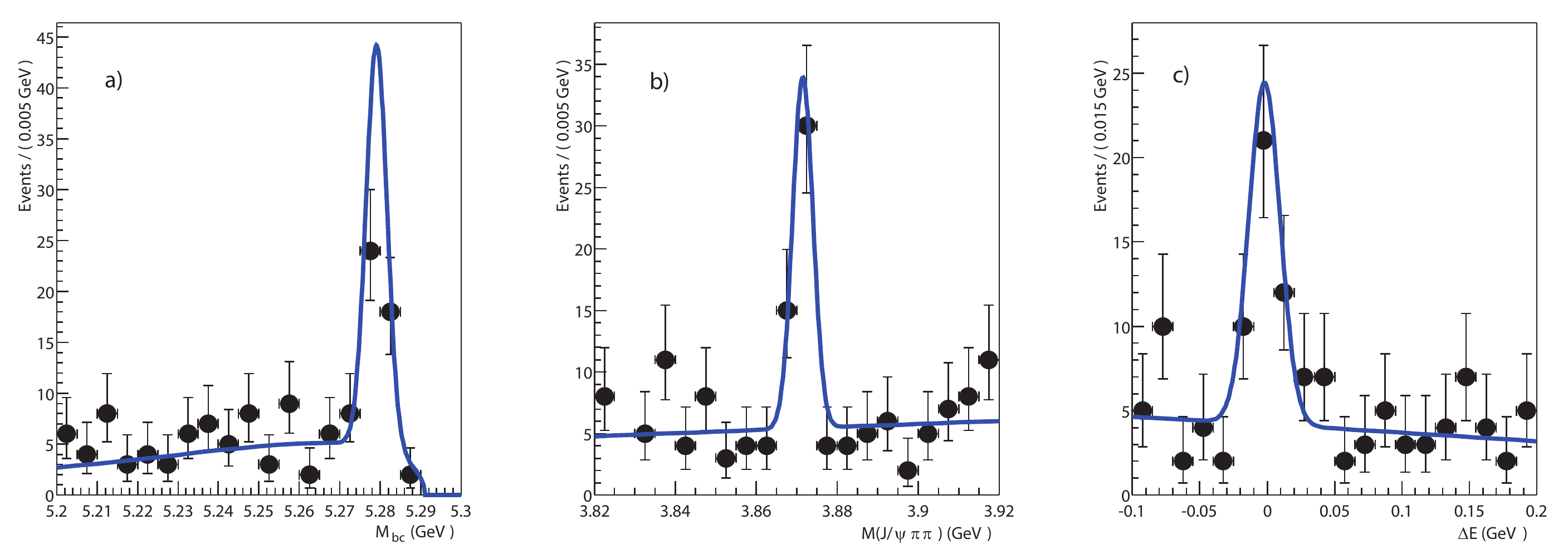}
\caption{(Color online) The beam-energy constrained mass $M_{bc} = \sqrt{(E^{\rm
CM}_{\rm beam})^2 - (p^{\rm CM}_B)^2}$ (left), the $\pi^+ \pi^-
J/\psi$ invariant mass (middle), and the energy difference $\Delta E
= E^{\rm CM}_B - E^{\rm CM}_{\rm beam}$ (right) for the $X(3872) \to
\pi^+ \pi^- J/\psi$ signal region, from Belle~\cite{Choi:2003ue}.}
\label{Fig:2.1.X3872Belle}
\end{center}
\end{figure}

The $X(3872)$ resonance was first observed by the Belle
Collaboration in 2003~\cite{Choi:2003ue}. Since its discovery, its
existence was confirmed by many subsequent experiments
\cite{Abe:2005ix,Gokhroo:2006bt,Adachi:2008te,Adachi:2008sua,Bhardwaj:2011dj,
Choi:2011fc,Acosta:2003zx,Abulencia:2005zc,Abulencia:2006ma,Aaltonen:2009vj,
Abazov:2004kp,Aubert:2004ns,Aubert:2005zh,Aubert:2006aj,Aubert:2007rva,Aubert:2008gu,
Aubert:2008ae,delAmoSanchez:2010jr,Aaij:2011sn,Aaij:2013zoa,Aaij:2014ala,Chatrchyan:2013cld,Ablikim:2013dyn}
as shown in Fig. \ref{Fig:2.1.X3872Belle}. Despite the huge
experimental efforts, we still do not fully understand its nature.

\begin{figure}[hbtp]
\begin{center}
\includegraphics[width=10cm]{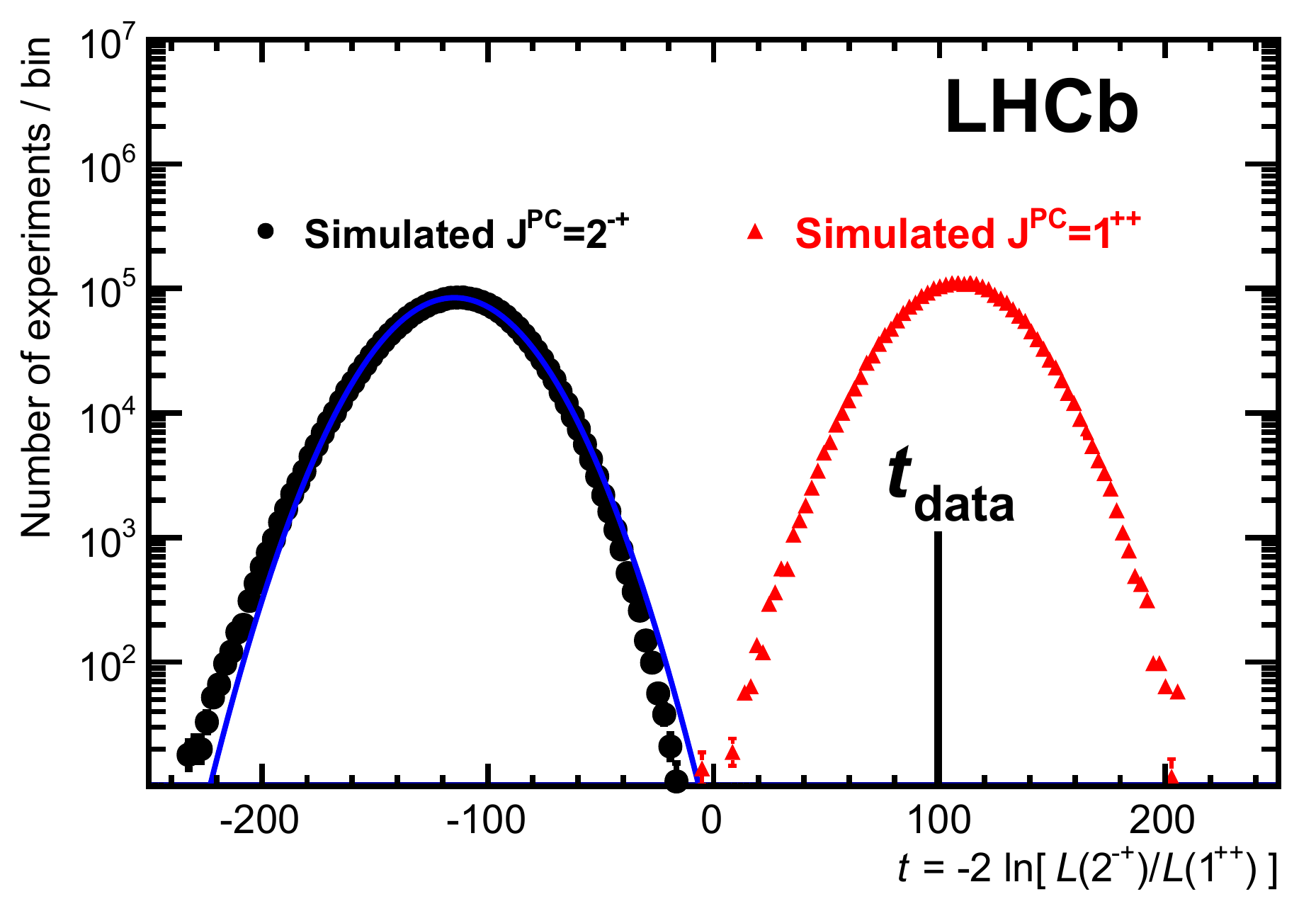}
\caption{(Color online) Distribution of the test statistic $t$ for the simulated
experiments with $J^{PC} = 2^{-+}$ and $1^{++}$, from
LHCb~\cite{Aaij:2013zoa}, with $t_{\rm data}$ the value of the test
statistic for the data.} \label{Fig:2.1.X3872LHCb}
\end{center}
\end{figure}

The mass and width of the $X(3872)$ from different experiments are
summarized in Table~\ref{Table:2.1.X3872}. A fit to these parameters
yields an average mass $(3871.69 \pm 0.17)$ MeV~\cite{pdg} and a
width $<1.2$ MeV at 90\% C.L.~\cite{Choi:2011fc}. Its mass is
extremely close to the $D^0 \bar D^{*0}$ mass threshold, $(3871.81
\pm 0.09)$ MeV. We also collect its productions and decay modes in
Table~\ref{Table:2.1.X3872}. The $X(3872)$ was mostly observed in the
$B$ meson decay process $B^{\pm,0} \to K^{\pm,0}_{(S)} X(3872)$ with
the $X(3872)$ decaying into $\pi^+ \pi^- J/\psi$. The $X(3872)$ was also
produced in $p \bar p$ annihilations, $pp$ collisions, and $e^+ e^-$
annihilations (possibly through the $Y(4260)$, see Sec.~\ref{Sect:2.2.1})
and decays into $D^{*0} \bar D^0$, $D^0 \bar D^0 \pi^0$, $\gamma
J/\psi$, $\gamma \psi(3686)$, and $\omega J/\psi$ with $\omega$
decaying into $\pi^+ \pi^- \pi^0$. Its quantum numbers have been
studied by Belle, BaBar and CDF, and determined to be
$I^G J^{PC} = 0^+ 1^{++}$ by the recent LHCb
experiment~\cite{Aaij:2013zoa}, as shown in
Fig.~\ref{Fig:2.1.X3872LHCb}.

\renewcommand{\arraystretch}{1.4}
\begin{table}[hbtp]
\caption{The resonance parameters of the $X(3872)$ and its observed
productions and decay channels. Here the $X(3872)$ is abbreviated as
$X$. \label{Table:2.1.X3872}}
\begin{center}
\begin{tabular}{c|cccc} \toprule[1pt]
%  \multicolumn{3}{c}{Y(4260)}\\ \midrule[1pt]
  Experiment                           & Mass [MeV]                                  & Width [MeV]                               & Productions and Decay Modes & $J^{PC}$ \\\midrule[1pt]
  Belle~\cite{Choi:2003ue}           & $3872 \pm 0.6 \pm 0.5$                      & $<2.3$                                    & $B \to K X (\to \pi^+ \pi^- J/\psi)$ \\
  Belle~\cite{Abe:2005ix}            & --                                          & --                                        & $B \to K X (\to \gamma J/\psi \, , \, \omega J/\psi \to \pi^+ \pi^- \pi^0 J/\psi)$  & $C=+1$ \\
  Belle~\cite{Gokhroo:2006bt}        & $3875.4 \pm 0.7 {^{+0.4}_{-1.7}} \pm 0.9$   & --                                        & $B \to K X (\to D^0 \bar D^0 \pi^0)$  &  $1^{++}/2^{++}$ \\
  Belle~\cite{Adachi:2008te}         & $3871.46 \pm 0.37 \pm 0.07$                 & --                                        & $B \to K X (\to \pi^+ \pi^- J/\psi)$ \\
  Belle~\cite{Adachi:2008sua}        & $3872.9 {^{+0.6}_{-0.4}} {^{+0.4}_{-0.5}}$  & $3.9 {^{+2.8}_{-1.4}} {^{+0.2}_{-1.1}}$   & $B \to K X (\to D^{*0} \bar D^0)$ \\
  Belle~\cite{Bhardwaj:2011dj}       & --                                          & --                                        & $B \to K X (\to \gamma J/\psi)$ \\
  Belle~\cite{Choi:2011fc}           & $3871.84 \pm 0.27 \pm 0.19$                 & $<1.2$                                    & $B \to K X (\to \pi^+ \pi^- J/\psi)$ \\
  CDF~\cite{Acosta:2003zx}           & $3871.3 \pm 0.7 \pm 0.4$                    & --                                        & $p \bar p \to {\rm anything} + X (\to \pi^+ \pi^- J/\psi)$ \\
  CDF~\cite{Abulencia:2005zc}        & --                                          & --                                        & $p \bar p \to {\rm anything} + X (\to \pi^+ \pi^- J/\psi)$  & $C=+1$ \\
  CDF~\cite{Abulencia:2006ma}        & --                                          & --                                        & $p \bar p \to {\rm anything} + X (\to \pi^+ \pi^- J/\psi)$  &  $1^{++}/2^{-+}$ \\
  CDF~\cite{Aaltonen:2009vj}         & $3871.61 \pm 0.16 \pm 0.19$                 & --                                        & $p \bar p \to {\rm anything} + X (\to \pi^+ \pi^- J/\psi)$ \\
  D\O\,~\cite{Abazov:2004kp}              & $3871.8 \pm 3.1 \pm 3.0$                    & --                                        & $p \bar p \to {\rm anything} + X (\to \pi^+ \pi^- J/\psi)$ \\
%  BaBar-1~\cite{Aubert:2004fc}        & --                                          & --                                        & not seen in $B \to K X (\to \eta J/\psi)$ \\
  BaBar~\cite{Aubert:2004ns}         & $3873.4 \pm 1.4$                            & --                                        & $B^- \to K^- X (\to \pi^+ \pi^- J/\psi)$ \\
%  BaBar-3~\cite{Aubert:2005eg}        & --                                          & --                                        & $B \to K X (\to \gamma J/\psi \, , \, \to \gamma \psi(3686))$ \\
  BaBar~\cite{Aubert:2005zh}         & $3871.3 \pm 0.6 \pm 0.1$                    & $<4.1$                                    & $B^- \to K^- X (\to \pi^+ \pi^- J/\psi)$ \\
                                       & $3868.6 \pm 1.2 \pm 0.2$                    & --                                        & $B^0 \to K^0 X (\to \pi^+ \pi^- J/\psi)$ \\
%  BaBar-3~\cite{Aubert:2005vi}        & --                                          & --                                        & $B \to K X (\to \gamma J/\psi \, , \, \to \gamma \psi(3686))$ \\
  BaBar~\cite{Aubert:2006aj}         & --                                          & --                                        & $B \to K X (\to \gamma J/\psi)$  & $C=+1$ \\
  BaBar~\cite{Aubert:2007rva}        & $3875.1 {^{+0.7}_{-0.5}} \pm 0.5$           & $3.0 {^{+1.9}_{-1.4}} \pm 0.9$            & $B \to K X (\to \bar D^{*0} D^0)$ \\
  BaBar~\cite{Aubert:2008gu}         & $3871.4 \pm 0.6 \pm 0.1$                    & $<3.3$                                    & $B^+ \to K^+ X (\to \pi^+ \pi^- J/\psi)$ \\
                                       & $3868.7 \pm 1.5 \pm 0.4$                    & --                                        & $B^0 \to K^0 X (\to \pi^+ \pi^- J/\psi)$ \\
  BaBar~\cite{Aubert:2008ae}         & --                                          & --                                        & $B \to K X (\to \gamma J/\psi \, , \, \to \gamma \psi(3686))$ \\
  BaBar~\cite{delAmoSanchez:2010jr}  & $3873.0 {^{+1.8}_{-1.6}} \pm 1.3$           & --                                        & $B \to K X (\to \omega J/\psi \to \pi^+ \pi^- \pi^0 J/\psi)$  &  $2^{-}$\\
  LHCb~\cite{Aaij:2011sn}            & $3871.95 \pm 0.48 \pm 0.12$                 & --                                        & $p p \to {\rm anything} + X (\to \pi^+ \pi^- J/\psi)$ \\
  LHCb~\cite{Aaij:2013zoa}           & --                                          & --                                        & $p p \to {\rm anything} + X (\to \pi^+ \pi^- J/\psi)$ & $1^{++}$ \\
  LHCb~\cite{Aaij:2014ala}           & --                                          & --                                        & $p p \to {\rm anything} + X (\to \gamma J/\psi \, , \, \to \gamma \psi(3686))$  \\
  CMS~\cite{Chatrchyan:2013cld}        & --                                          & --                                        & $p p \to {\rm anything} + X (\to \pi^+ \pi^- J/\psi)$ \\
  BESIII~\cite{Ablikim:2013dyn}        & $3871.9 \pm 0.7 \pm 0.2$                    & $<2.4$                                    & $e^+ e^- [\to Y(4260)] \to \gamma X (\to \pi^+ \pi^- J/\psi)$ \\
\bottomrule[1pt]
\end{tabular}
\end{center}
\end{table}

\renewcommand{\arraystretch}{1.4}
\begin{table}[hbtp]
\caption{Some product branching fractions of the $X(3872)$
resonance. Here, the $X(3872)$ is abbreviated as $X$.
\label{Table:2.1.X3872BR}}
\begin{center}
\begin{tabular}{c|cccc} \toprule[1pt]
%  \multicolumn{3}{c}{Y(4260)}\\ \midrule[1pt]
  Experiment                           & Product Branching Fractions \\\midrule[1pt]
  Belle~\cite{Choi:2003ue}           & ${{\cal B}(B^+ \to K^+ X) \times {\cal B}(X \to \pi^+ \pi^- J/\psi) \over {\cal B}(B^+ \to K^+ \psi(3686)) \times {\cal B}(\psi(3686) \to \pi^+ \pi^- J/\psi)} = 0.063 \pm 0.012 \pm 0.007$ \\
  Belle~\cite{Abe:2005ix}            & ${{\cal B}(B \to K X) \times {\cal B}(X \to \gamma J/\psi)} = (1.8 \pm 0.6 \pm 0.1) \times 10^{-6}$ \\
  Belle~\cite{Gokhroo:2006bt}        & ${\cal B}(B \to K D^0 \bar D^0 \pi^0) = (1.27 \pm 0.31 {^{+0.22}_{-0.39}}) \times 10^{-4}$, near $X$ threshold \\
  Belle~\cite{Adachi:2008sua}        & ${{\cal B}(B \to K X) \times {\cal B}(X \to D^{*0} \bar D^0)} = (0.80 \pm 0.20 \pm 0.10) \times 10^{-4}$ \\
  Belle~\cite{Bhardwaj:2011dj}       & ${{\cal B}(B^+ \to K^+ X) \times {\cal B}(X \to \gamma J/\psi)} = (1.78 {^{+0.48}_{-0.44}} \pm 0.12) \times 10^{-6}$ \\
  Belle~\cite{Choi:2011fc}           & ${{\cal B}(B^+ \to K^+ X) \times {\cal B}(X \to \pi^+ \pi^- J/\psi)} = (8.61 \pm 0.82 \pm 0.52) \times 10^{-6}$ \\
  BaBar~\cite{Aubert:2004ns}         & ${{\cal B}(B^- \to K^- X) \times {\cal B}(X \to \pi^+ \pi^- J/\psi)} = (1.28 \pm 0.41) \times 10^{-5}$ \\
  BaBar~\cite{Aubert:2005zh}         & ${{\cal B}(B^0 \to K^0 X) \times {\cal B}(X \to \pi^+ \pi^- J/\psi)} = (5.1 \pm 2.8 \pm 0.7) \times 10^{-6}$ \\
                                       & ${{\cal B}(B^- \to K^- X) \times {\cal B}(X \to \pi^+ \pi^- J/\psi)} = (10.1 \pm 2.5 \pm 1.0) \times 10^{-6}$ \\
  BaBar~\cite{Aubert:2006aj}         & ${{\cal B}(B^+ \to K^+ X) \times {\cal B}(X \to \gamma J/\psi)} = (3.3 \pm 1.0 \pm 0.3) \times 10^{-6}$ \\
  BaBar~\cite{Aubert:2008gu}         & ${{\cal B}(B^+ \to K^+ X) \times {\cal B}(X \to \pi^+ \pi^- J/\psi)} = (8.4 \pm 1.5 \pm 0.7) \times 10^{-6}$ \\
                                       & ${{\cal B}(B^0 \to K^0 X) \times {\cal B}(X \to \pi^+ \pi^- J/\psi)} = (3.5 \pm 1.9 \pm 0.4) \times 10^{-6}$ \\
  BaBar~\cite{Aubert:2008ae}         & ${{\cal B}(B^\pm \to K^\pm X) \times {\cal B}(X \to \gamma J/\psi)} = (2.8 \pm 0.8 \pm 0.1) \times 10^{-6}$ \\
                                       & ${{\cal B}(B^\pm \to K^\pm X) \times {\cal B}(X \to \gamma \psi(3686))} = (9.5 \pm 2.7 \pm 0.6) \times 10^{-6}$ \\
  BaBar~\cite{delAmoSanchez:2010jr}  & ${{\cal B}(B^+ \to K^+ X) \times {\cal B}(X \to \omega J/\psi)} = (0.6 \pm 0.2 \pm 0.1) \times 10^{-5}$ \\
                                       & ${{\cal B}(B^0 \to K^0 X) \times {\cal B}(X \to \omega J/\psi)} = (0.6 \pm 0.3 \pm 0.1) \times 10^{-5}$ \\
  LHCb~\cite{Aaij:2011sn}            & ${\sigma(pp \to {\rm anything} + X) \times {\cal B}(X \to \pi^+ \pi^- J/\psi)} = (5.4 \pm 1.3 \pm 0.8) $ nb, for $5 < p_T < 20$ GeV \\
  CMS~\cite{Chatrchyan:2013cld}        & ${\sigma(pp \to {\rm anything} + X) \times {\cal B}(X \to \pi^+ \pi^- J/\psi)} = (1.06 \pm 0.11 \pm 0.15) $ nb, for $10 < p_T < 30$ GeV \\
  BESIII~\cite{Ablikim:2013dyn}        & ${\sigma^B(e^+ e^- \to \gamma X) \times {\cal B}(X \to \pi^+ \pi^- J/\psi)} = (0.33 \pm 0.12 \pm 0.02) $ nb, at $\sqrt{s} = 4.260$ GeV \\
\bottomrule[1pt]
\end{tabular}
\end{center}
\end{table}

Besides the resonance parameters listed in
Table~\ref{Table:2.1.X3872}, these experiments provided many
branching fractions. They are also useful experimental information,
and we summarize some of them in Table~\ref{Table:2.1.X3872BR}.
Particularly, we list the following isospin-violating branching
fractions
\begin{eqnarray}
{\Gamma(X \to \pi^+ \pi^- \pi^0 J/\psi) \over \Gamma(X \to \pi^+
\pi^- J/\psi)} &=& 1.0 \pm 0.4 \pm 0.3 \, ,\label{Eq.2.1.3872iso1}
\\ {\Gamma(X \to \omega J/\psi) \over \Gamma(X \to \pi^+ \pi^- J/\psi)} &=&
\left\{\begin{array}{cc} 0.7 \pm 0.3 &
\mbox{for}~~B^+~~\mbox{events}
\\ 1.7 \pm 1.3 & \mbox{for}~~B^0~~\mbox{events}
\end{array}\right. \, ,\label{Eq.2.1.3872iso2}
\end{eqnarray}
which were observed by Belle~\cite{Abe:2005ix} and
BaBar~\cite{delAmoSanchez:2010jr}, respectively.

The difference between $B^0 \to K^0 X$ and $B^\pm \to K^\pm X$ also
attracted much experimental interest. The Belle Collaboration
measured the ratio of branching fractions ${\Gamma(B^0 \to K^0 X) /
\Gamma(B^+ \to K^+ X)}$ to be $(0.82 \pm 0.22 \pm
0.05)$~\cite{Adachi:2008te}, $(1.26 \pm 0.65 \pm
0.06)$~\cite{Adachi:2008sua} and $(0.50 \pm 0.14 \pm
0.04)$~\cite{Choi:2011fc}, and the mass difference between the
$X(3872)$ states produced in $B^+$ and $B^0$ decay to be $\delta M =
M_{(B^+ \to K^+X)} - M_{(B^0 \to K^0 X)} = (0.18 \pm 0.89 \pm 0.26)$
MeV~\cite{Adachi:2008te} and $(-0.69 \pm 0.97 \pm 0.19)$
MeV~\cite{Choi:2011fc}, while the BaBar Collaboration measured this
ratio to be $(0.50 \pm 0.30 \pm 0.05)$~\cite{Aubert:2005zh}, and $(0.41
\pm 0.24 \pm 0.05)$~\cite{Aubert:2008gu}, and this mass difference to
be $(2.7 \pm 1.3 \pm 0.2)$ MeV~\cite{Aubert:2005zh} and $(2.7 \pm
1.6 \pm 0.4)$ MeV~\cite{Aubert:2008gu}, where we have assumed
experimental results from $B^+$ and $B^-$ are the same.

Besides the above observations, the $X(3872)$ resonance was not seen
(all values are given at 90\% confidence level (C.L.).):
\begin{enumerate}

\item in the $\gamma \chi_{c1}$ decay mode in Belle~\cite{Choi:2003ue},
and the upper limit was measured to be
\begin{eqnarray}
{\Gamma(X(3872) \to \gamma \chi_{c1}) \over \Gamma(X(3872) \to \pi^+
\pi^- J/\psi)} < 0.89  \, .
\end{eqnarray}

\item in the $\eta J/\psi$ decay mode in BaBar~\cite{Aubert:2004fc},
and the upper limit was measured to be
\begin{eqnarray}
{\cal B}(B^\pm \to K^\pm X(3872)) \times {\cal B}(X(3872) \to \eta
J/\psi) < 7.7 \times 10^{-6}  \, .
\end{eqnarray}

\item in the $\gamma \psi(3686)$ decay mode in Belle~\cite{Bhardwaj:2011dj},
and the upper limit was measured to be
\begin{eqnarray}
{\Gamma(X(3872) \to \gamma \psi(3686)) \over \Gamma(X(3872) \to
\gamma J/\psi)} < 2.1  \, .
\end{eqnarray}
However, both the BaBar and LHCb experiments observed this decay
mode, and this branching fraction was measured to be $(3.4 \pm
1.4)$~\cite{Aubert:2008ae} and $(2.46 \pm 0.64 \pm
0.29)$~\cite{Aaij:2014ala}, respectively.

\end{enumerate}

Another useful upper limit is
\begin{eqnarray}
{\cal B}(B^\pm \to K^\pm X(3872)) < 3.2 \times 10^{-4} \, ,
\end{eqnarray}
which was given by the BaBar Collaboration at 90\%
C.L.~\cite{Aubert:2005vi}.

\paragraph{$Y(3940)$}
\label{Sect:2.1.2}

\begin{figure}[hbtp]
\begin{center}
\includegraphics[width=11cm]{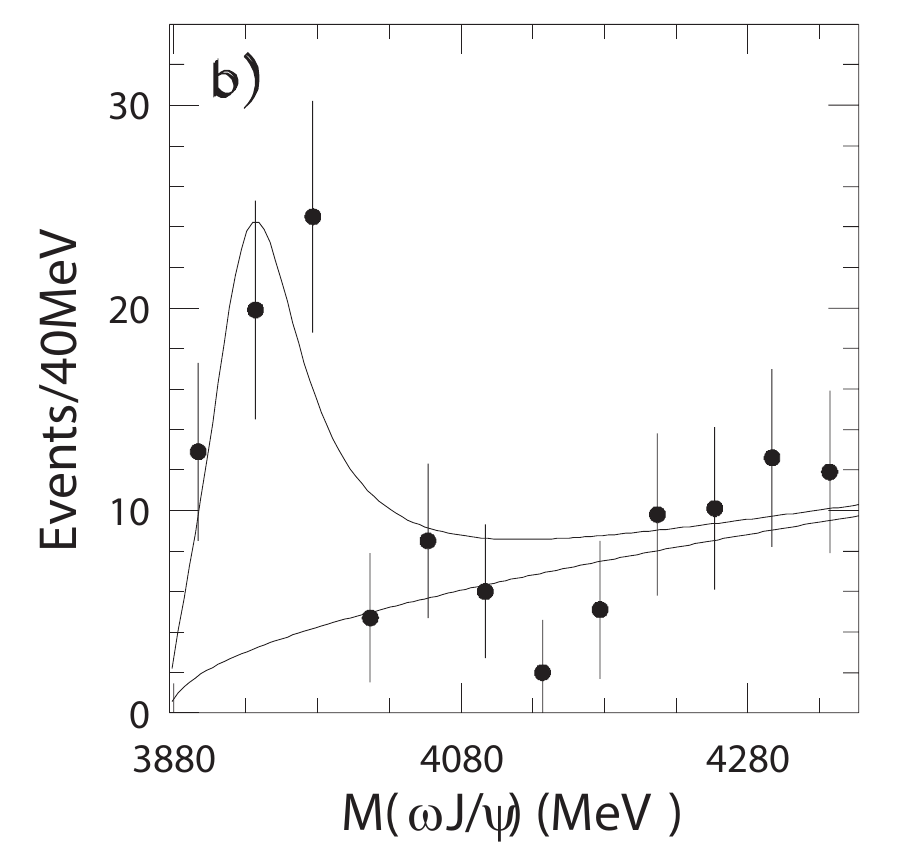}
\caption{The $\omega J/\psi$ invariant mass spectrum of $B \to K
\omega J/\psi$ from Belle \cite{Abe:2004zs}.}
\label{Fig:2.1.Y3940Belle}
\end{center}
\end{figure}

The charmonium-like state $Y(3940)$ was firstly reported by the
Belle Collaboration in the $\omega J/\psi$ invariant mass
distribution in the exclusive $B \to K \omega J/\psi$ decay in
2004~\cite{Abe:2004zs}, as shown in Fig.~\ref{Fig:2.1.Y3940Belle}.
Its statistical significance was estimated to be greater than
$8\sigma$, and its mass and width were measured to be $M
=(3943\pm11\pm13)$ MeV and $\Gamma = (87\pm22\pm26)$ MeV,
respectively. This mass value is very close to that of the $X(3940)$
resonance, which was observed in the double charmonium production
and will be discussed in Sec.~\ref{Sect:2.3.1}. The Belle Collaboration
also measured the product of branching fractions, i.e.,
\begin{eqnarray}
\mathcal{B}(B \to K Y(3940)) \times\mathcal{B}(Y(3940) \to \omega
J/\psi) = (7.1 \pm 1.3 \pm 3.1) \times 10^{-5} \, .
\end{eqnarray}
Later, the BaBar Collaboration confirmed this observation in the
same process with a lower mass. In 2007,  its mass and width were
measured to be $M =(3914.6{^{+3.8}_{-3.4}}\pm2.0)$ MeV and $\Gamma
=(34{^{+12}_{-8}}\pm5)$ MeV, respectively~\cite{Aubert:2007vj}. In
the subsequent experiment in 2010~\cite{delAmoSanchez:2010jr} its
mass and width were measured to be $M
=(3919.1{^{+3.8}_{-3.5}}\pm2.0)$ MeV and $\Gamma
=(31{^{+10}_{-8}}\pm5)$ MeV, respectively.

\begin{figure}[hbtp]
\begin{center}
\includegraphics[width=8cm]{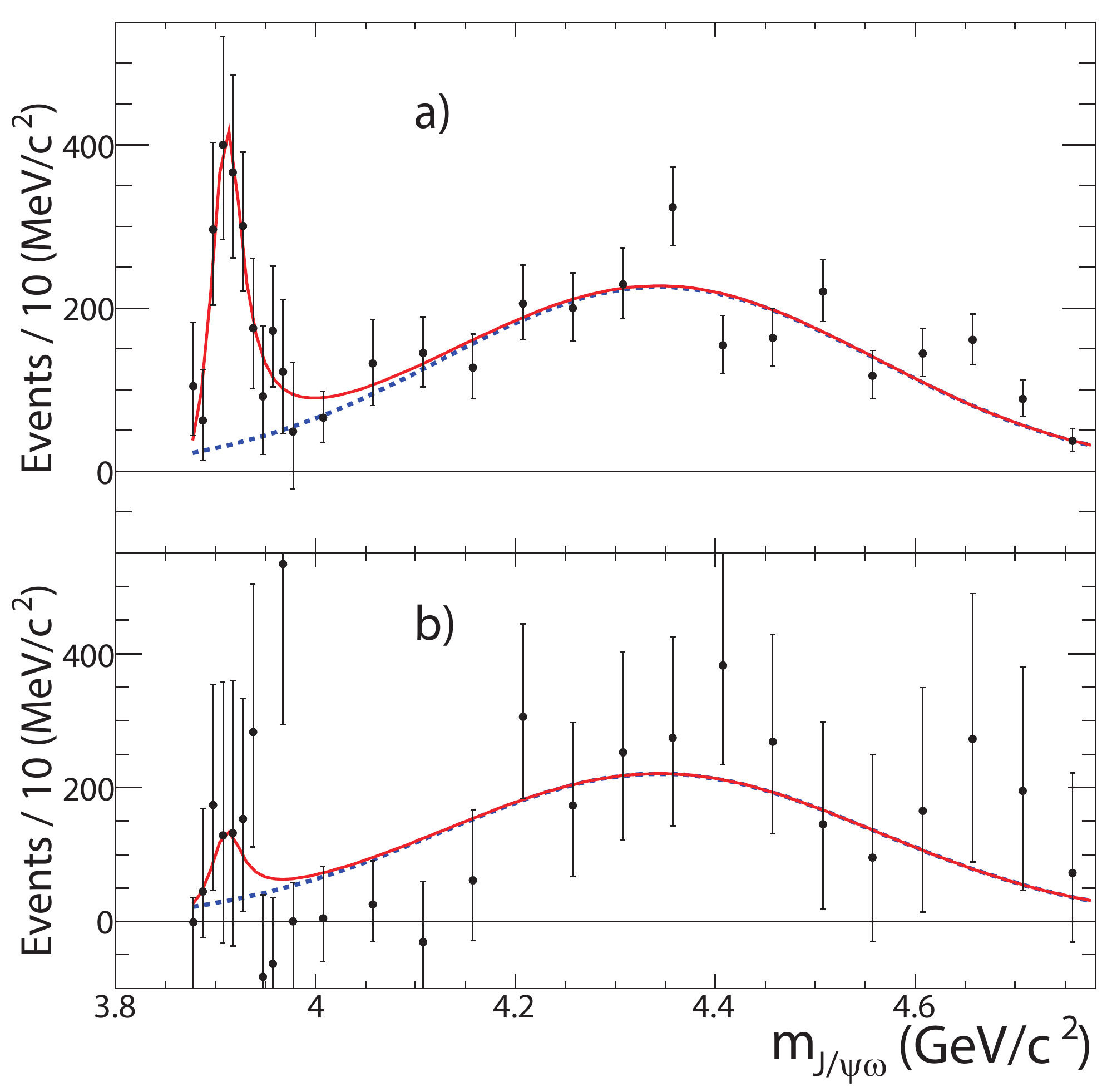}
\caption{(Color online) The $\omega J/\psi$ invariant mass spectrum of $B \to K
\omega J/\psi$ from BaBar~\cite{Aubert:2007vj}. Here, $B^+$ and
$B^0$ decays are shown in the upper and lower panels, respectively.
} \label{Fig:2.1.Y3940BaBar}
\end{center}
\end{figure}

Since these two BaBar experiments studied both the $B^0 \to K^0
\omega J/\psi$ and $B^+ \to K^+ \omega J/\psi$ processes, it allows
them to evaluate the two parameters, $R_Y$ and $R_{NR}$, defined as
the ratios between the number of $B^0$ and $B^+$ events
for the $Y(3940)$ signal and for the nonresonant contribution,
respectively.
The results are shown in Fig.~\ref{Fig:2.1.Y3940BaBar}. The former
experiment~\cite{Aubert:2007vj} obtained
\begin{eqnarray}
\mathcal{B}( B^+ \to Y(3940) K^+ ) \times\mathcal{B}( Y(3940) \to
J/\psi \omega ) &=& ( 4.9 {^{+1.0}_{-0.9}} \pm 0.5 ) \times 10^{-5}
\, ,
\\ \nonumber \mathcal{B}( B^0 \to Y(3940) K^0 )  \times\mathcal{B}( Y(3940) \to J/\psi \omega ) &=& ( 1.3 {^{+1.3}_{-1.1}} \pm 0.2 ) \times 10^{-5} \, ,
\end{eqnarray}
and $R_Y = 0.27{^{+0.28}_{-0.23}}{^{+0.04}_{-0.01}}$ and $R_{NR} =
0.97{^{+0.23}_{-0.22}}{^{+0.03}_{-0.02}}$. $R_Y$ is three standard
deviations below the isospin expectation, but agrees with that of
$X(3872)$~\cite{Aubert:2008gu}, while $R_{NR}$ agrees with the
isospin expectation. The latter
experiment~\cite{delAmoSanchez:2010jr} got
\begin{eqnarray}
\mathcal{B}( B^+ \to Y(3940) K^+ ) \times\mathcal{B}( Y(3940) \to
J/\psi \omega ) &=& ( 3.0 {^{+0.7}_{-0.6}} {^{+0.5}_{-0.3}} ) \times
10^{-5} \, ,
\\ \nonumber \mathcal{B}( B^0 \to Y(3940) K^0 )  \times\mathcal{B}( Y(3940) \to J/\psi \omega ) &=& ( 2.1 \pm 0.9 \pm 0.3 ) \times 10^{-5} \, ,
\end{eqnarray}
and $R_Y = 0.7{^{+0.4}_{-0.3}} \pm 0.1$ and $R_{NY} =
0.7\pm0.1\pm0.1$, which are both consistent
with the previous results~\cite{Aubert:2007vj}, and at the same time
(almost) agree with the isospin expectation.

To date, the only observed decay mode of the $Y(3940)$ is $\omega
J/\psi$. The Belle Collaboration searched for its open charm decay
mode $D^{*0} \bar D^0$ and set an upper limit~\cite{Adachi:2008sua}
\begin{eqnarray}
\mathcal{B}( B \to Y(3940) K ) \times \mathcal{B}( Y(3940) \to
D^{*0} \bar D^0 ) &<& 0.67 \times 10^{-4} \, .
\end{eqnarray}
at 90\% C.L.. They further used this value, together with the results
from Refs.~\cite{Abe:2004zs,Aubert:2007vj,Abe:2007jna}, to indicate
that the $X(3940)$ and $Y(3940)$ are different states.

Besides the $Y(3940)$ and $X(3940)$ states, two other resonances
$X(3915)$ and $Z(3930)$ were observed in the $\gamma\gamma$ fusion
and will be discussed in Sec.~\ref{Sec:2.4}. These four neutral
states were all discovered in the $3.90$-$3.95$ GeV mass region, and
need to be carefully studied and classified. We note that the two
charmonium-like states, $Y(3940)$ and $X(3915)$, were identified
as the same state $\chi_{c0}(2P)$ in Particle Data Group
(PDG)~\cite{pdg}. In this review we will discuss them in Sec.
\ref{Sec:2.4.2}.

\paragraph{$Y(4140)$ and $Y(4274)$}
\label{Sect:2.1.3}

\begin{figure}[hbtp]
\begin{center}
\includegraphics[width=11cm]{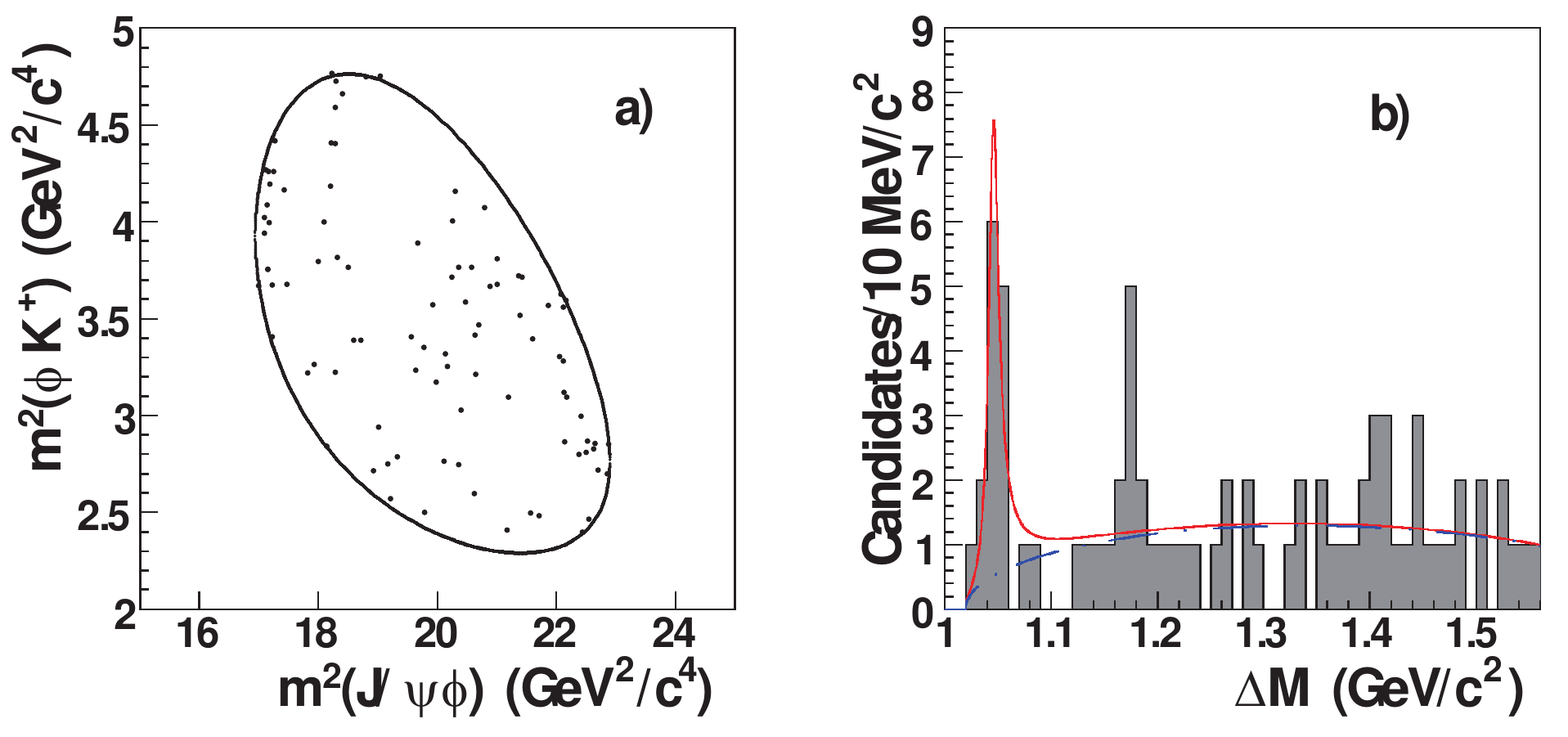}
\caption{(Color online) The mass difference, $\Delta M$, between $\mu^+\mu^-K^+K^-$
and $\mu^+\mu^-$, in the $B^+$ mass window from
CDF \cite{Aaltonen:2009tz}.}
\label{Fig:2.1.Y4140CDF}
\end{center}
\end{figure}

In 2009, the $Y(4140)$ state was first announced by the CDF
Collaboration~\cite{Aaltonen:2009tz}, where they reported evidence
for a narrow structure near the $J/\psi\phi$ threshold in the
exclusive $B \to K J/\psi \phi$ decay in $\bar p p$ collisions
at $\sqrt s = 1.96$ TeV, with an integrated luminosity of 2.7
fb$^{-1}$ and a statistical significance of $3.8\sigma$. The results
are shown in Fig.~\ref{Fig:2.1.Y4140CDF}. The
mass and width of the $Y(4140)$ were measured to be $ M = (4143.0 \pm 2.9
\pm 1.2)$ MeV and $\Gamma = (11.7{^{+8.3}_{-5.0}} \pm 3.7)$ MeV,
respectively.

However, this structure was not seen in the following Belle
experiment, which measured the $J/\psi\phi$ invariant mass in the
process $\gamma \gamma \to \phi J/\psi$ with the $J/\psi$ decaying
into lepton pairs and $\phi$ decaying into $K^+K^-$
pairs~\cite{Shen:2009vs}. They gave the following upper limits on
the branching fraction of $Y(4140) \to \phi J/\psi$:
\begin{eqnarray}
\Gamma(Y(4140)\to {\gamma\gamma}) \times{\cal B}(Y(4140) \to \phi
J/\psi ) < \left\{\begin{array}{cc}
 41~{\rm eV} \, ,& {\rm for} \,  J^P = 0^+ \, ,
\\
 6.0~{\rm eV} \, ,& {\rm for} \,  J^P = 2^+ \, ,
\end{array}\right.
\end{eqnarray}
at 90\% C.L.. Instead, the Belle Collaboration observed another
resonance, the $X(4350)$, which will be discussed in Sec. \ref{Sec:2.4.3}.

Later in 2011, the CDF Collaboration reported a further study based
on the increased $B^+ \to J/\psi \phi K^+$ sample, and confirmed the
$Y(4140)$ structure, with an integrated luminosity of 6.0 fb$^{-1}$
and a significance greater than $5\sigma$~\cite{Aaltonen:2011at}.
%Their results are shown in the lower panel of Fig.~\ref{Fig:2.1.Y4140CDF}.
The mass and width of the $Y(4140)$ were
measured slightly more precisely, to be $M = (4143.4
{^{+2.9}_{-3.0}} \pm 0.6)$ MeV and $\Gamma = (15.3{^{+10.4}_{-6.1}}
\pm 2.5)$ MeV. They also extracted the relative branching fraction
$\mathcal{B}_{rel}$ to be
\begin{eqnarray}
\mathcal{B}_{rel} = {{\cal B}(B^+ \to Y(4140) K^+) \times {\cal
B}(Y(4140) \to J/\psi \phi) \over {\cal B}(B^+ \to J/\psi \phi
K^+)} = 0.149 \pm 0.039 \pm 0.024 \, . \label{Eq2.1.Y4140CDF}
\end{eqnarray}
Besides the $Y(4140)$, the CDF Collaboration reported another
structure, named as $Y(4274)$, in the $J/\psi\phi$ invariant mass
spectrum, with a significance of $3.1\sigma$~\cite{Aaltonen:2011at}.
Its mass and width were measured to be $M = (4274.4{^{+8.4}_{-6.7}}
\pm 1.9)$ MeV and $\Gamma = (32.3{^{+21.9}_{-15.3}} \pm 7.6)$ MeV.

\begin{figure}[hbtp]
\begin{center}
\includegraphics[width=8cm]{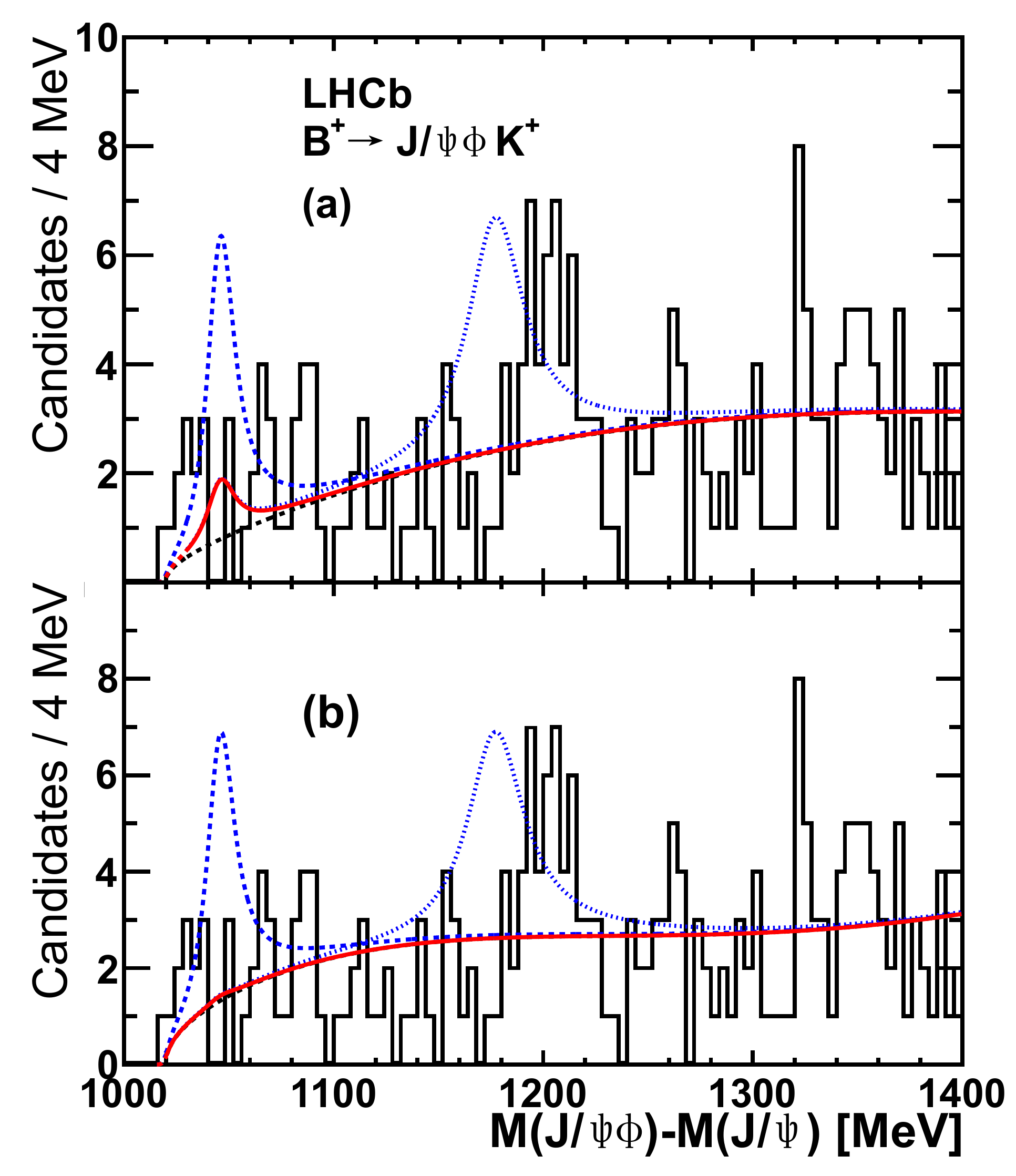}
\caption{(Color online) Distribution of the mass difference $M(J/\psi \phi) -
M(J/\psi)$ for the $B^+ \to J/\psi \omega K^+$ from
LHCb~\cite{Aaij:2012pz}.} \label{Fig:2.1.Y4140BelleLHCb}
\end{center}
\end{figure}

One year later, the LHCb Collaboration searched for the $Y(4140)$
decaying to $J/\psi\phi$ using $B \to K J/\psi \phi$ decay events
with an integrated luminosity of 0.37 fb$^{-1}$ in $pp$ collisions
and at significantly larger energy $\sqrt s = 7$
TeV~\cite{Aaij:2012pz}. They did not observe these two states, and
set two upper limits on the relative branching fractions:
\begin{eqnarray}
\mathcal{B}_{rel} = {{\cal B}(B^+ \to Y(4140) K^+) \times {\cal
B}(Y(4140) \to J/\psi \phi) \over {\cal B}(B^+ \to J/\psi \phi
K^+)} < 0.07 \, ,\label{Eq2.1.Y4140LHCb}
\end{eqnarray}
at 90\% C.L., and
\begin{eqnarray}
{{\cal B}(B^+ \to Y(4274) K^+) \times {\cal B}(Y(4274) \to J/\psi
\phi) \over {\cal B}(B^+ \to J/\psi \phi K^+)} < 0.08 \, ,
\end{eqnarray}
at 90\% C.L..

\begin{figure}[hbtp]
\begin{center}
\includegraphics[width=15cm]{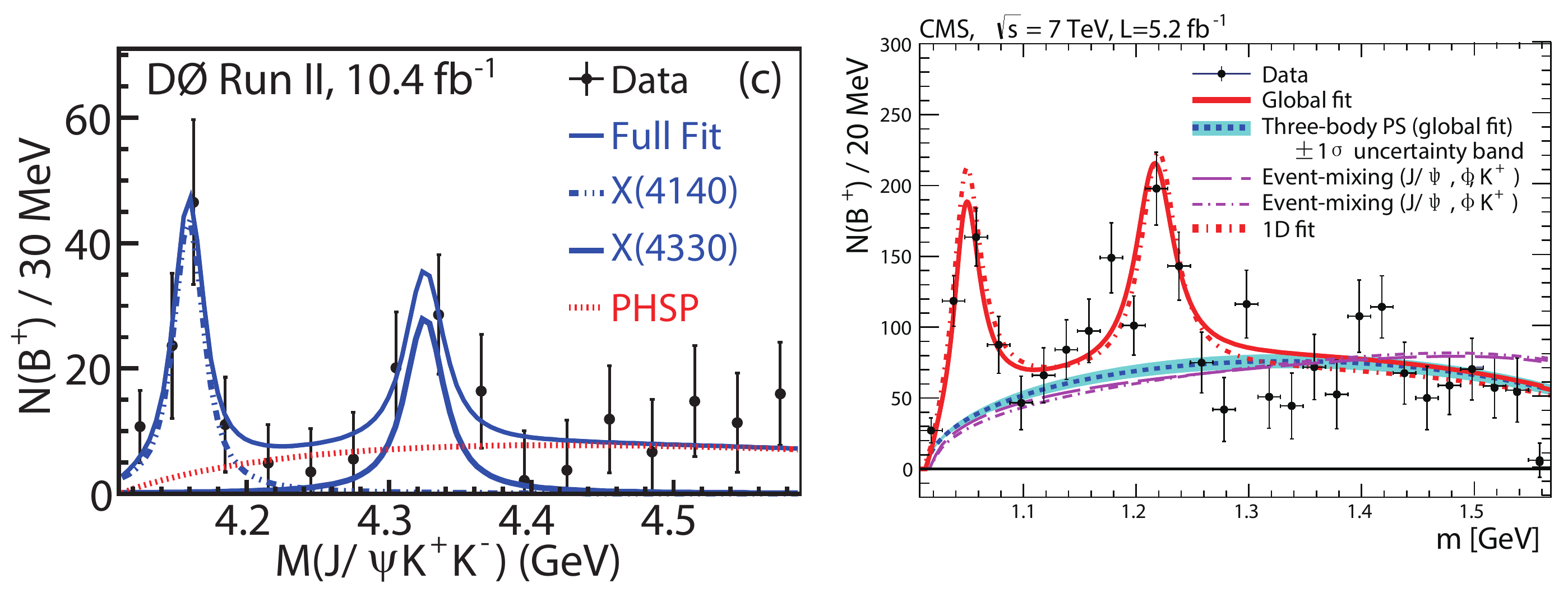}
\caption{(Color online) Left: The $J/\psi K^+ K^-$ invariant mass spectrum of $B^+
\to J/\psi \phi K^+$ from D\O\,~\cite{Abazov:2013xda}. Right: The
number of $B^+ \to J/\psi \phi K^+$ candidates as a function of
$\Delta m = m(\mu^+\mu^-K^+K^-) - m(\mu^+\mu^-)$ from
CMS~\cite{Chatrchyan:2013dma}.} \label{Fig:2.1.Y4140D0CMS}
\end{center}
\end{figure}

The situation changed in 2013. Both the D\O\, and CMS collaborations
confirmed the observation of the
$Y(4140)$~\cite{Abazov:2013xda,Chatrchyan:2013dma}. The D\O\, experiment investigated the decay process $B^+ \to J/\psi \phi K^+$ produced in $\bar p p$ collisions
at $\sqrt s = 1.96$ TeV, as shown in the left panel of
Fig.~\ref{Fig:2.1.Y4140D0CMS}. Their results supported the existence
of the $Y(4140)$, with an integrated luminosity of 10.4 fb$^{-1}$ and a
statistical significance of $3.1\sigma$. The relative branching
fraction was extracted to be
\begin{eqnarray}
{{\cal B}(B^+ \to Y(4140) K^+) \over {\cal B}(B^+ \to J/\psi \phi
K^+)} = (21 \pm 8 \pm 4) \% \, , {\rm for} \, M(J/\psi \phi) < 4.59
\, {\rm GeV} \, .
\end{eqnarray}
Their data also indicated the possible existence of a structure
around 4300 MeV, but they did not obtain a stable fit with an
unconstrained width.

The CMS experiment studied the $J/\psi\phi$ mass spectrum in the
same decay process $B^\pm \to J/\psi \phi K^\pm$ produced in $pp$
collisions at $\sqrt s = 7$ TeV, and confirmed the existence of the
$Y(4140)$ resonance, with an integrated luminosity of 5.2 fb$^{-1}$
and a significance greater than $5\sigma$~\cite{Chatrchyan:2013dma}.
Their results are shown in the right panel of
Fig.~\ref{Fig:2.1.Y4140D0CMS}. The relative branching fraction
$\mathcal{B}_{rel}$ was extracted to be about 0.10 with a
statistical uncertainty of about 30\%, which is consistent with
Eq.~(\ref{Eq2.1.Y4140CDF}), the value measured by CDF, and
Eq.~(\ref{Eq2.1.Y4140LHCb}), the upper limit given by LHCb. The CMS
Collaboration also confirmed the existence of the second structure,
$Y(4274)$.

We summarize the information of the $Y(4140)$ and $Y(4274)$
resonance parameters from different experiments in
Table~\ref{Table:2.1.Y4140}. Since these two resonances were both
observed in the $J/\psi\phi$ decay mode, their $C$-parity and
$G$-parity should be even.

\renewcommand{\arraystretch}{1.6}
\begin{table}[hbtp]
\caption{The resonance parameters of the $Y(4140)$ and $Y(4274)$. Here, all results are in units of MeV.
\label{Table:2.1.Y4140}}
\begin{center}
\begin{tabular}{c|cc} \toprule[1pt]
%  \multicolumn{3}{c}{Y(4260)}\\ \midrule[1pt]
  Experiment                                & $Y(4140)$                     & $Y(4274)$  \\\midrule[1pt]
  CDF~\cite{Aaltonen:2009tz}        &$M=4143.0 \pm 2.9 \pm 1.2$, $\Gamma=11.7{^{+8.3}_{-5.0}} \pm 3.7$               & --    \\
  CDF~\cite{Aaltonen:2011at}        &$M=4143.4 {^{+2.9}_{-3.0}} \pm 0.6$, $\Gamma=15.3{^{+10.4}_{-6.1}} \pm 2.5$               & $M=4274.4{^{+8.4}_{-6.7}} \pm 1.9$, $\Gamma=32.3{^{+21.9}_{-15.3}} \pm 7.6$   \\
  D\O\,~\cite{Abazov:2013xda}        &$M=4159.0 \pm 4.3 \pm 6.6$, $\Gamma=19.9 \pm 12.6 {^{+1.0}_{-8.0}}$               & --   \\
  CMS~\cite{Chatrchyan:2013dma}        &$M=4148.0 \pm 2.4 \pm 6.3$, $\Gamma=28{^{+15}_{-11}} \pm 19$               & $M=4313.8 \pm 5.3 \pm 7.3$, $\Gamma=38{^{+30}_{-15}} \pm 16$  \\
 % \multicolumn{3}{c}{Y(4008)}\\ \midrule[1pt]
 % Belle \cite{Yuan:2007sj}          &$4008\pm40^{+114}_{-28}$            &$226\pm44\pm87$\\
 % Belle \cite{Liu:2013dau}          &$3890.8\pm40.5\pm11.5$              &$254.5\pm39.5\pm13.6$\\
\bottomrule[1pt]
\end{tabular}
\end{center}
\end{table}

\paragraph{$Z^+(4430)$}
\label{Sect:2.1.4}

\begin{figure}[hbtp]
\begin{center}
\includegraphics[width=13cm]{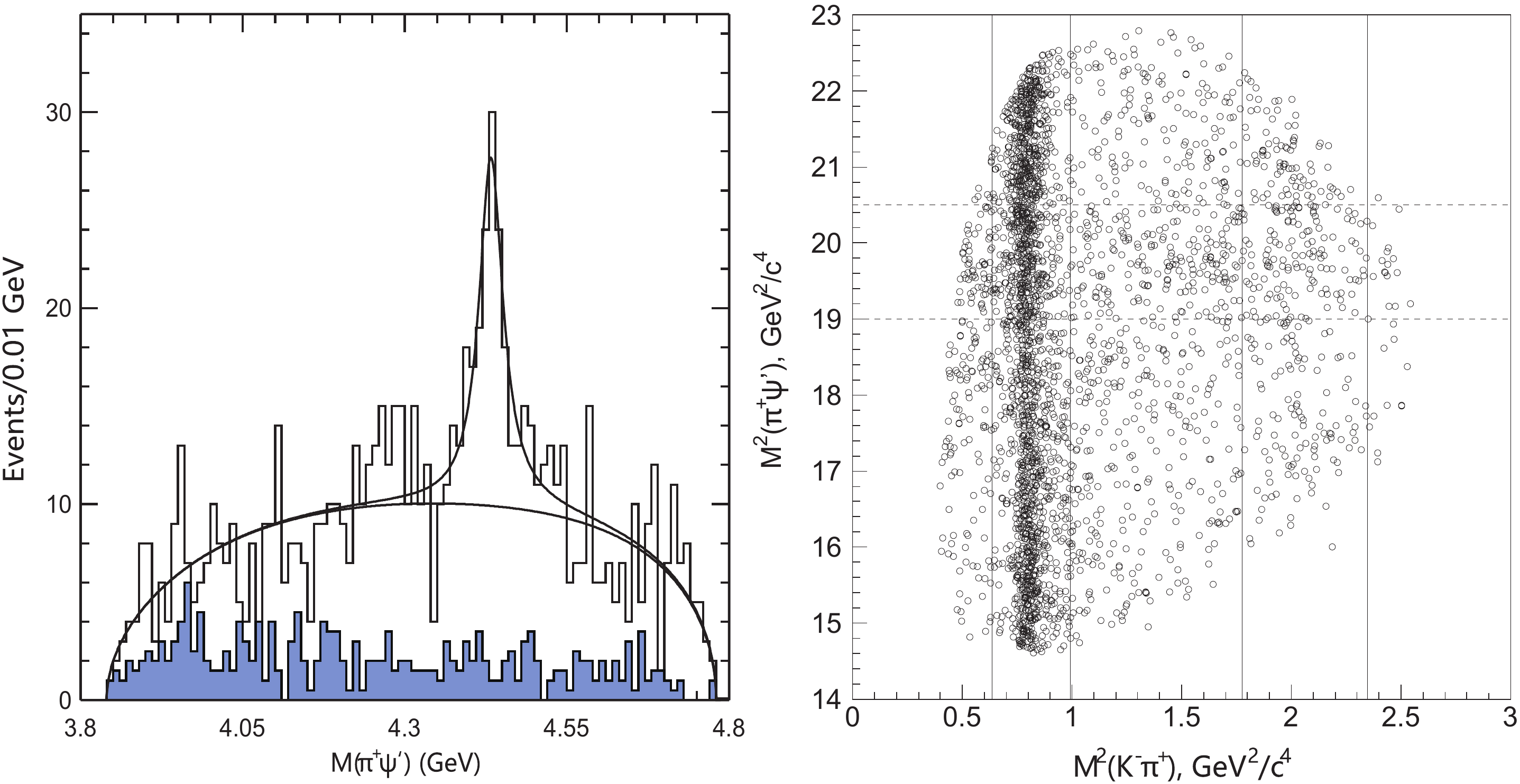}
\caption{(Color online) The $\pi^+ \psi(3686)$ invariant mass spectrum of $\bar B^0 \to
K^- \pi^+ \psi(3686)$ (left) from Belle~\cite{Choi:2007wga}, and its
Dalitz plot (right) from Belle~\cite{Mizuk:2009da}.}
\label{Fig:2.1.Z4430Belle1}
\end{center}
\end{figure}

The charged charmonium-like state $Z^+(4430)$ was first observed by
the Belle Collaboration in the $\pi^\pm \psi(3686)$ invariant mass
distribution in $B \to K \pi^\pm \psi(3686)$ decays in
2007~\cite{Choi:2007wga}, with a statistical significance of
$6.5\sigma$. The results are shown in the left panel of
Fig.~\ref{Fig:2.1.Z4430Belle1}. Its mass and width were measured to
be $ M = (4433 \pm 4 \pm 2)$ MeV and $\Gamma = (45{^{+18}_{-13}}
{^{+30}_{-13}})$ MeV, respectively. They also extracted the product
of branching fractions as
\begin{eqnarray}
{{\cal B}(\bar B^0 \to K^- Z^+(4430)) \times {\cal B}(Z^+(4430)
\to \pi^+ \psi(3686))} = (4.1 \pm 1.0 \pm 1.4) \times 10^{-5} \, .
\end{eqnarray}

\begin{figure}[hbtp]
\begin{center}
\includegraphics[width=14cm]{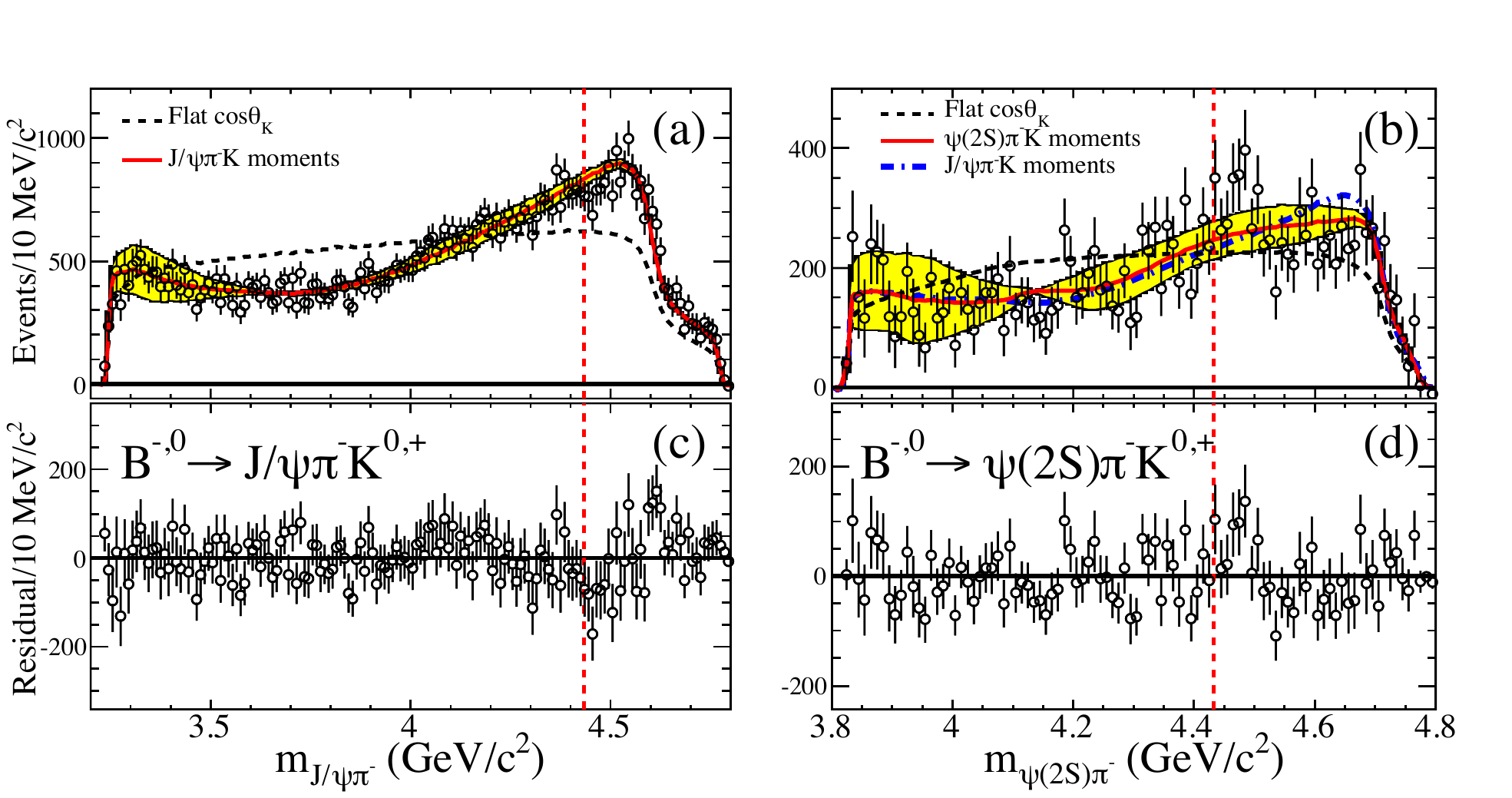}
\caption{(Color online) The $\pi^- J/\psi$ and $\pi^- \psi(3686)$ mass
distributions from BaBar~\cite{Aubert:2008aa} for the combined decay
modes (a) $B^{0,-} \to K^{+,0} \pi^- J/\psi$ and (b) $B^{0,-} \to
K^{+,0} \pi^- \psi(3686)$. (c) and (d) are the residuals (data-solid
curve) for (a) and (b), respectively.} \label{Fig:2.1.Z4430BaBar}
\end{center}
\end{figure}

However, this signal was not seen in the following BaBar experiment,
which analyzed both $\pi^- J/\psi$ and $\pi^- \psi(3686)$ invariant
masses in $B^{0,-} \to K^{+,0} \pi^- J/\psi$ and $B^{0,-} \to
K^{+,0} \pi^- \psi(3686)$ decays~\cite{Aubert:2008aa}, respectively.
The results are shown in Fig.~\ref{Fig:2.1.Z4430BaBar}.  They found
no evidence for $Z^-(4430)$ in the $\pi^- J/\psi$ mass distributions
in $B^{0,-} \to K^{+,0} \pi^- J/\psi$ decays, and no clear evidence
for $Z^-(4430)$ in the $\pi^- \psi(3686)$ mass distributions in
$B^{0,-} \to K^{+,0} \pi^- \psi(3686)$  decays. More precisely, they
fitted the $\pi^- \psi(3686)$ invariant mass distribution and obtained
a 2.7 standard deviation signal. This signal has a fitted width
consistent with the value obtained by Belle~\cite{Choi:2007wga}, but
its central mass value was 43 MeV higher than that reported by
Belle~\cite{Choi:2007wga}, with a difference of $+4.7\sigma$. They
set two upper limits on the branching fractions
\begin{eqnarray}
{{\cal B}(B^0 \to K^+ Z^-(4430)) \times {\cal B}(Z^-(4430)
\to \pi^- \psi(3686))} &<& 3.1 \times 10^{-5} \, , \\
\nonumber {{\cal B}(B^- \to \bar K^0 Z^-(4430)) \times {\cal
B}(Z^-(4430) \to \pi^- \psi(3686))} &<& 4.7 \times 10^{-5} \, ,
\end{eqnarray}
at 95\% C.L..

In the following years, the Belle Collaboration continued their
studies on the
$Z^+(4430)$~\cite{Mizuk:2009da,Chilikin:2013tch,Chilikin:2014bkk}.
We summarize the information of the resonance parameters from all
these Belle experiments in Table~\ref{Table:2.1.Z4430}.

In Ref.~\cite{Mizuk:2009da} the Belle Collaboration performed a
Dalitz plot analysis of $B \to K \pi^+ \psi(3686)$, and observed a
signal for $Z^+(4430) \to \pi^+ \psi(3686)$ with a significance of
$6.4\sigma$. The results are shown in the right panel of
Fig.~\ref{Fig:2.1.Z4430Belle1}. They measured the product of
branching fractions, i.e.,
\begin{eqnarray}
{{\cal B}(\bar B^0 \to K^- Z^+(4430)) \times {\cal B}(Z^+(4430)
\to \pi^+ \psi(3686))} = (3.2 {^{+1.8}_{-0.9}} {^{+5.3}_{-1.6}})
\times 10^{-5} \, .
\end{eqnarray}
They also determined the branching fraction
\begin{eqnarray}
{{\cal B}(B^0 \to K^{*0}(892) \psi(3686))} = (5.52
{^{+0.35}_{-0.32}} {^{+0.53}_{-0.58}}) \times 10^{-4} \, ,
\end{eqnarray}
and the fraction of the $K^*(892)$ meson that was longitudinally
polarized to be $f_L = (44.8 {^{+4.0}_{-2.7}} {^{+4.0}_{-5.3}}) \%$.

In Ref.~\cite{Chilikin:2013tch} the Belle Collaboration performed a
full amplitude analysis of $B^0 \to K^+ \pi^- \psi(3686)$ decays to
determine the spin and parity of the $Z^+(4430)$. Their results show
that the $Z^+(4430)$ has quantum numbers $J^P = 1^+$, this hypothesis
being favored over the $0^-$, $1^-$, $2^-$ and $2^+$ hypotheses at the
levels of $3.4\sigma$, $3.7\sigma$, $4.7\sigma$ and $5.1\sigma$,
respectively. They also calculated the following branching fractions
\begin{eqnarray}
\nonumber {{\cal B}(B^0 \to K^+ \pi^- \psi(3686))} &=& (5.80 \pm 0.39) \times
10^{-4} \, ,
\\ {{\cal B}(B^0 \to K^{*0}(892) \psi(3686))} &=& (5.55 {^{+0.22}_{-0.23}} {^{+0.41}_{-0.84}}) \times 10^{-4} \, ,
\\ \nonumber {{\cal B}(B^0 \to K^+ Z^-(4430)) \times {\cal B}(Z^-(4430) \to \pi^- \psi(3686))} &=& (6.0 {^{+1.7}_{-2.0}} {^{+2.5}_{-1.4}}) \times 10^{-5} \, ,
\end{eqnarray}
together with $f_L = (45.5 {^{+3.1}_{-2.9}} {^{+1.4}_{-4.9}}) \%$.

\begin{figure}[hbtp]
\begin{center}
\includegraphics[width=13cm]{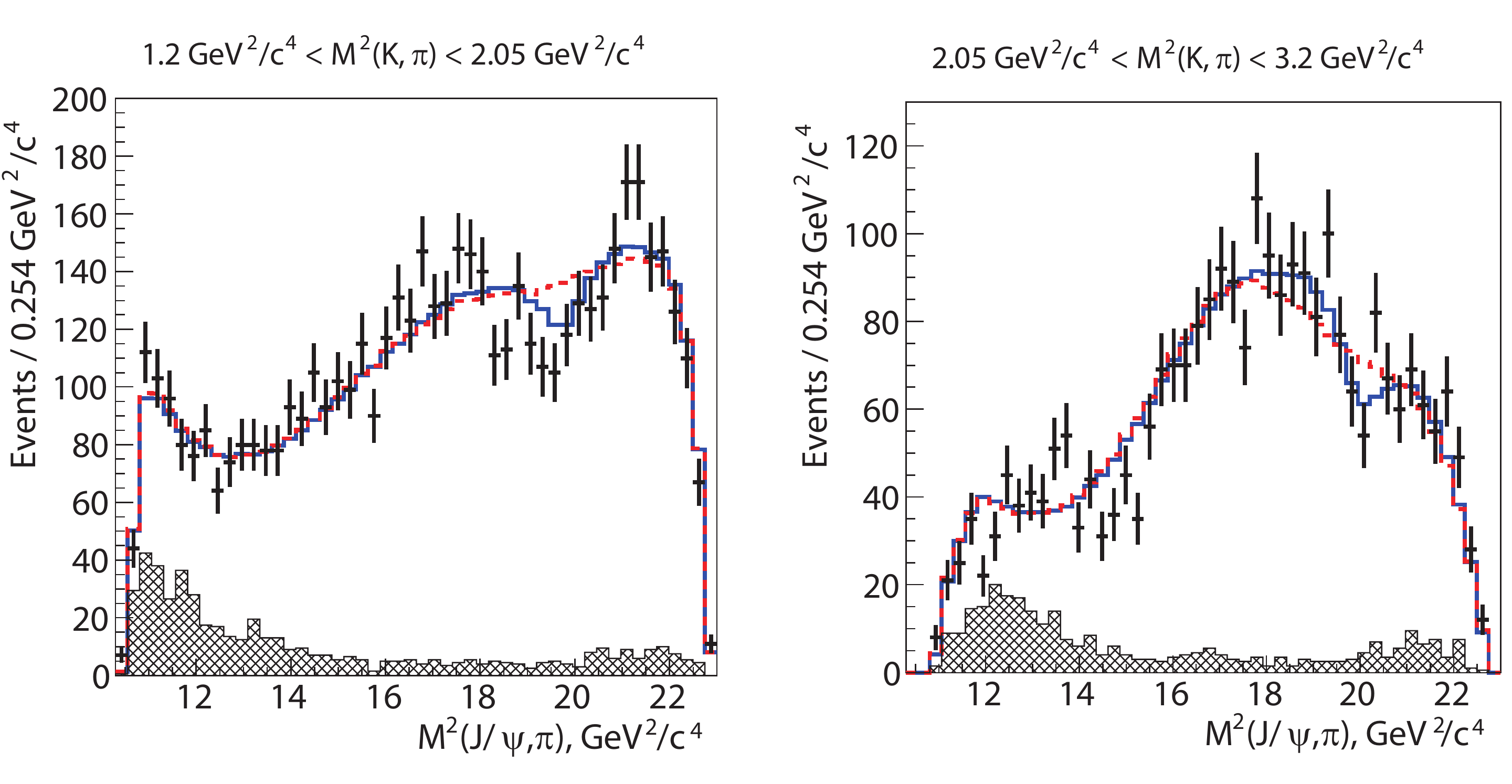}
\caption{(Color online) The fit results with (solid line) and without (dashed line)
the $Z^+(4430)$ from Belle~\cite{Chilikin:2014bkk}, where the
$Z_c(4200)$ contribution has been included.}
\label{Fig:2.1.Z4430Belle2}
\end{center}
\end{figure}

In Ref.~\cite{Chilikin:2014bkk} the Belle Collaboration found
evidence for $Z^+(4430) \to \pi^+ J/\psi$ in $\bar B^0 \to K^- \pi^+
J/\psi$ decays, as shown in Fig.~\ref{Fig:2.1.Z4430Belle2}. They
also measured its branching fraction
\begin{eqnarray}
{{\cal B}(\bar B^0 \to K^- Z^+(4430)) \times {\cal B}(Z^+(4430)
\to \pi^+ J/\psi)} = (5.4 {^{+4.0}_{-1.0}} {^{+1.1}_{-0.9}}) \times
10^{-6} \, .
\end{eqnarray}

\begin{figure}[hbtp]
\begin{center}
\includegraphics[width=12cm]{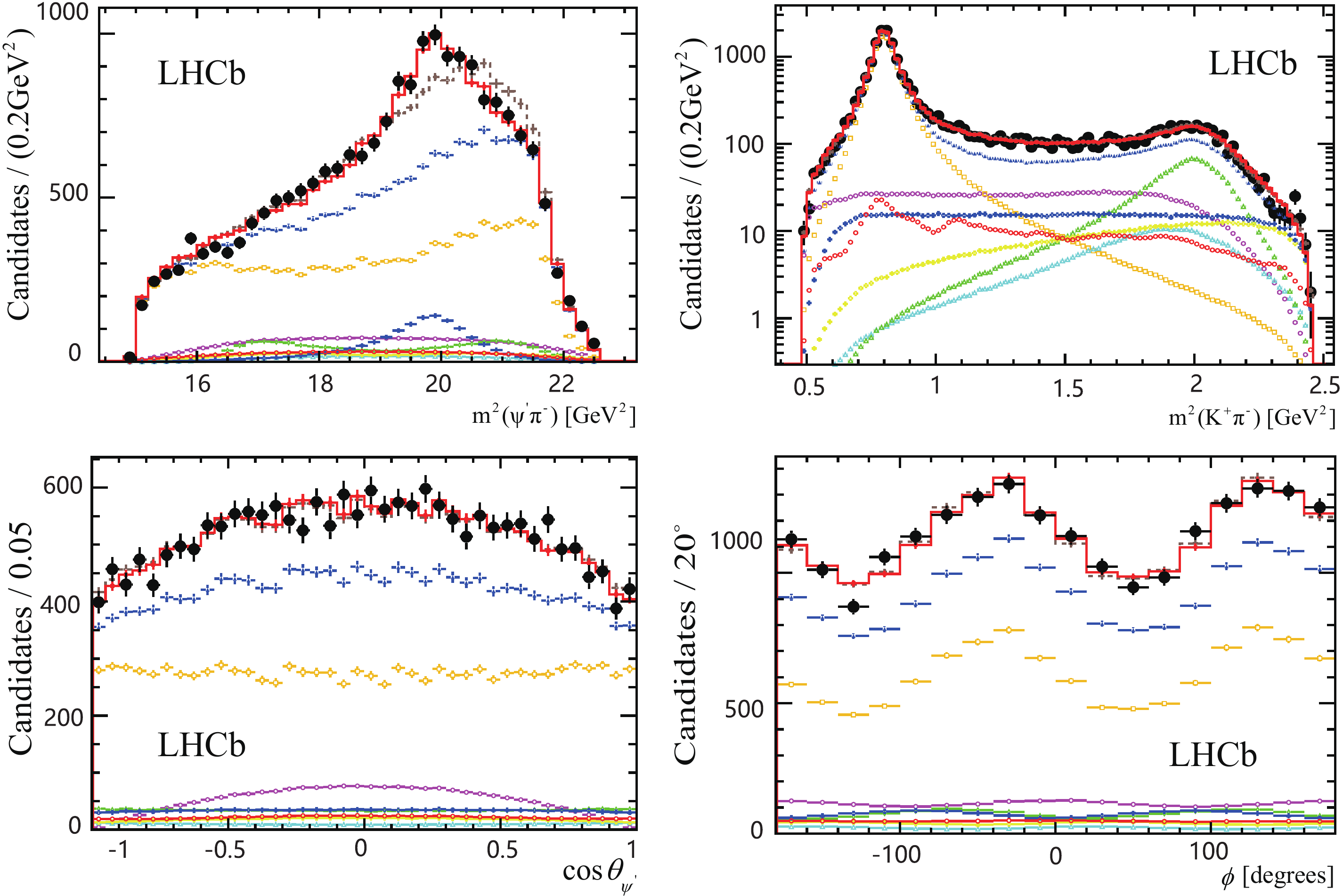}
\caption{(Color online) The four-dimensional fit to study the $B^0 \to K^+ \pi^-
\psi(3686)$ decays, performed by LHCb~\cite{Aaij:2014jqa}. The red
solid and brown dashed histograms represent the total amplitude with
and without the $Z^+(4430)$, respectively.}
\label{Fig:2.1.Z4430LHCb}
\end{center}
\end{figure}

Besides these Belle experiments, the LHCb Collaboration also
confirmed the existence of the $Z^+(4430)$, by performing a
four-dimensional fit in the analysis of the $\pi^- \psi(3686)$
invariant mass distribution in the $B^0 \to K^+ \pi^- \psi(3686)$
decay~\cite{Aaij:2014jqa}, as shown in Fig.~\ref{Fig:2.1.Z4430LHCb}. They measured its mass and width, as
listed also in Table~\ref{Table:2.1.Z4430}, and established its
spin-parity to be $J^P = 1^+$, both with very high significance.
Moreover, they ruled out the $0^-$, $1^-$, $2^+$ and $2^-$
hypotheses for its spin-parity by at least $9.7\sigma$,
$15.8\sigma$, $16.1\sigma$ and $14.6\sigma$, respectively. Its
amplitude fraction was determined to be $f_{Z^+(4430)} = (5.9 \pm
0.9 {^{+1.5}_{-3.3}}) \%$, whose definition is $f_R \equiv { \int
S_R(\Phi)d\Phi / \int S(\Phi)d\Phi}$, for the component $R$, where
in $S_R(\Phi)$ all except the $R$ amplitude terms are set to zero.

\renewcommand{\arraystretch}{1.6}
\begin{table}[hbtp]
\caption{The resonance parameters for the $Z^+(4430)$ and the observed
decay channels. \label{Table:2.1.Z4430}}
\begin{center}
\begin{tabular}{c|ccc} \toprule[1pt]
%  \multicolumn{3}{c}{Y(4260)}\\ \midrule[1pt]
  Experiment                                & Mass [MeV]                    & Width [MeV] & Decay Mode\\\midrule[1pt]
  Belle$^1$~\cite{Choi:2007wga}        & $4433 \pm 4 \pm 2$               &  $45{^{+18}_{-13}} {^{+30}_{-13}}$   &   $Z^+(4430) \to \pi^+ \psi(3686)$ \\
  Belle$^2$~\cite{Mizuk:2009da}        & $4443 {^{+15}_{-12}} {^{+19}_{-13}}$             & $107 {^{+86}_{-43}} {^{+74}_{-56}}$  &   $Z^+(4430) \to \pi^+ \psi(3686)$ \\
  Belle$^3$~\cite{Chilikin:2013tch}    & $4485 \pm22 {^{+28}_{-11}}$                    & $200 {^{+41}_{-46}} {^{+26}_{-35}}$  &   $Z^-(4430) \to \pi^- \psi(3686)$ \\
  Belle$^4$~\cite{Chilikin:2014bkk}    & --            & --  &   evidence for $Z^+(4430) \to \pi^+ J/\psi$\\
  LHCb~\cite{Aaij:2014jqa}    & $4475 \pm 7 {^{+15}_{-25}}$            & $172 \pm13 {^{+37}_{-34}}$  &   $Z^-(4430) \to \pi^- \psi(3686)$\\
 % \multicolumn{3}{c}{Y(4008)}\\ \midrule[1pt]
 % Belle \cite{Yuan:2007sj}          &$4008\pm40^{+114}_{-28}$            &$226\pm44\pm87$\\
 % Belle \cite{Liu:2013dau}          &$3890.8\pm40.5\pm11.5$              &$254.5\pm39.5\pm13.6$\\
\bottomrule[1pt]
\end{tabular}
\end{center}
\end{table}

\paragraph{$Z^+(4051)$ and $Z^+(4248)$}
\label{Sect:2.1.5}

\begin{figure}[hbtp]
\begin{center}
\includegraphics[width=13cm]{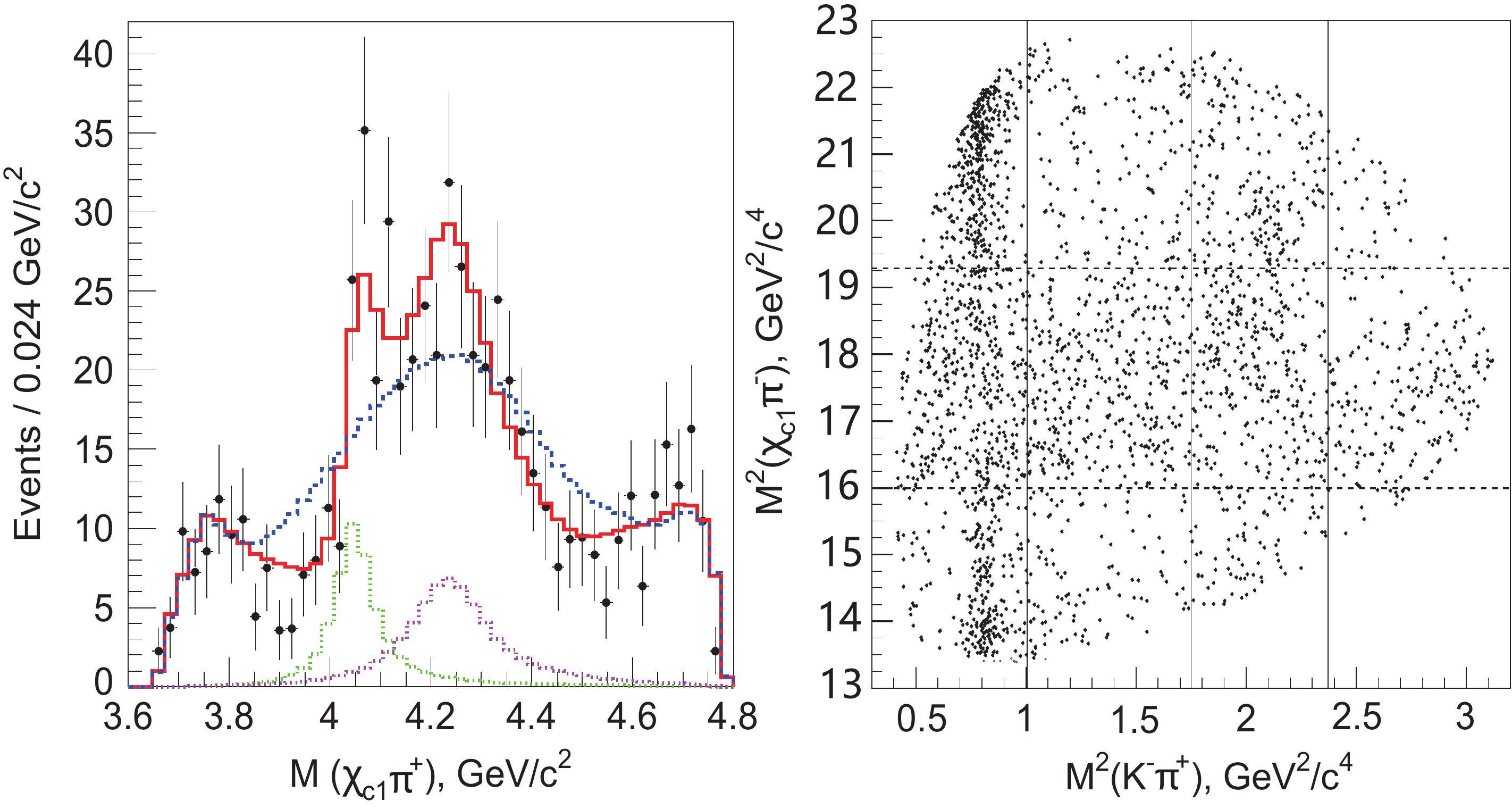}
\caption{(Color online) The $\pi^+ \chi_{c1}$ invariant mass spectrum of $\bar B^0
\to K^- \pi^+ \chi_{c1}$ (left), and its Dalitz plot (right) from
Belle~\cite{Mizuk:2008me}.} \label{Fig:2.1.Z4050Belle}
\end{center}
\end{figure}

The two charged charmonium-like states, $Z^+(4051)$ and $Z^+(4248)$,
were first observed by the Belle Collaboration in
2008~\cite{Mizuk:2008me}, one year after they reported the
observation of the $Z^+(4430)$~\cite{Choi:2007wga}. They studied the
$\pi^+ \chi_{c1}$ invariant mass distribution in the exclusive $\bar B^0
\to K^- \pi^+ \chi_{c1}$ decay, as shown in
Fig.~\ref{Fig:2.1.Z4050Belle}. After performing a Dalitz plot
analysis, they reported two resonance-like structures, $Z^+(4051)$
and $Z^+(4248)$. The masses and widths of the $Z^+(4051)$ and
$Z^+(4248)$ were determined to be $ M = (4051 \pm 14 {^{+20}_{-41}})$ MeV,
$\Gamma = (82{^{+21}_{-17}} {^{+47}_{-22}})$ MeV and $ M = (4248
{^{+44}_{-29}} {^{+180}_{-35}})$ MeV, $\Gamma = (177
{^{+54}_{-39}} {^{+316}_{-61}})$ MeV, respectively. They also provided
the following branching fractions
\begin{eqnarray}
{{\cal B}(\bar B^0 \to K^- Z^+(4051)) \times {\cal B}(Z^+(4051)
\to \pi^+ J/\psi)} &=& (3.0 {^{+1.5}_{-0.8}} {^{+3.7}_{-1.6}})
\times 10^{-5} \, ,\label{Eq.2.1.Z4051h1}
\\ {{\cal B}(\bar B^0 \to K^- Z^+(4248)) \times {\cal B}(Z^+(4248) \to \pi^+ J/\psi)} &=& (4.0 {^{+2.3}_{-0.9}} {^{+19.7}_{-0.5}}) \times 10^{-5}\label{Eq.2.1.Z4051h2} \, .
\end{eqnarray}

\begin{figure}[hbtp]
\begin{center}
\includegraphics[width=10cm]{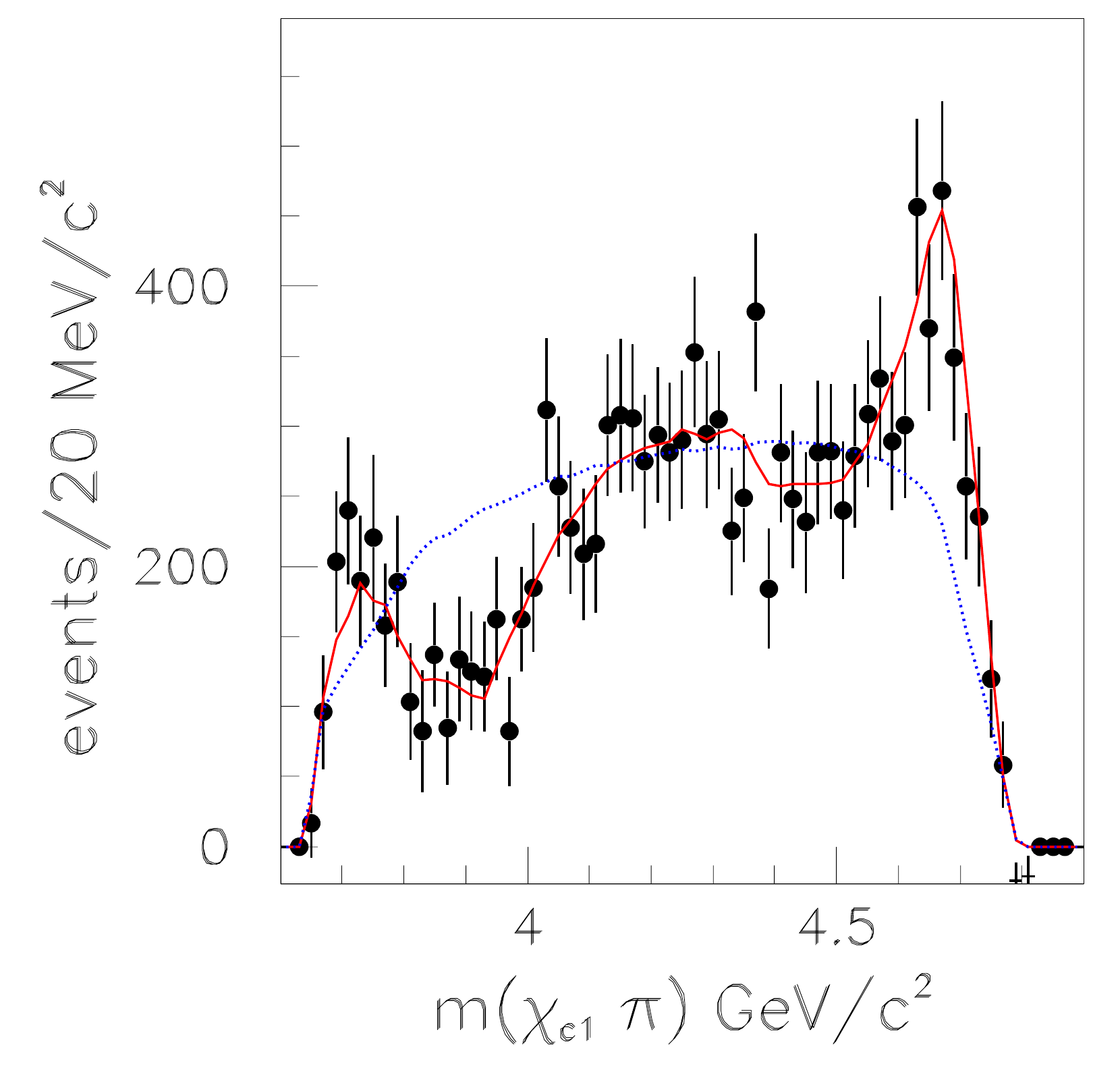}
\caption{(Color online) The $\pi^+ \chi_{c1}$ invariant mass spectrum of $\bar B
\to K \pi^+ \chi_{c1}$ from BaBar~\cite{Lees:2011ik}, with
background subtracted and efficiency corrected.}
\label{Fig:2.1.Z4050BaBar}
\end{center}
\end{figure}

Neither of these two observations was seen in the following BaBar
experiment, which studied the $\pi^+ \chi_{c1}$ invariant mass in
$\bar B^0 \to K^- \pi^+ \chi_{c1}$ and $B^+ \to K^0_S \pi^+
\chi_{c1}$ decays~\cite{Lees:2011ik}. The results are shown in
Fig.~\ref{Fig:2.1.Z4050BaBar}.  They found no significant resonant
structure in both the $\pi \chi_{c1}$ and $\pi J/\psi$ mass
distributions in $B \to K \pi \chi_{c1}$ decays. They set two upper
limits on the branching fractions at 90\% C.L.
\begin{eqnarray}
{{\cal B}(\bar B^0 \to K^- Z^+(4051)) \times {\cal B}(Z^+(4051)
\to \pi^+ J/\psi)} &<& 1.8 \times 10^{-5} \, ,
\\
{{\cal B}(\bar B^0 \to K^- Z^+(4248)) \times {\cal B}(Z^+(4248)
\to \pi^+ J/\psi)} &<& 4.0 \times 10^{-5} \, ,
\end{eqnarray}
%and together
%\begin{eqnarray}
%{{\cal B}(\bar B^0 \to K^- Z^+) \times {\cal B}(Z^+ \to \pi^+ J/\psi)} < 4.7 \times 10^{-5} \, .
%\end{eqnarray}
which are consistent with the Belle results listed in Eqs.~(\ref{Eq.2.1.Z4051h1}) and (\ref{Eq.2.1.Z4051h2}).

\begin{figure}[hbtp]
\begin{center}
\includegraphics[width=10cm]{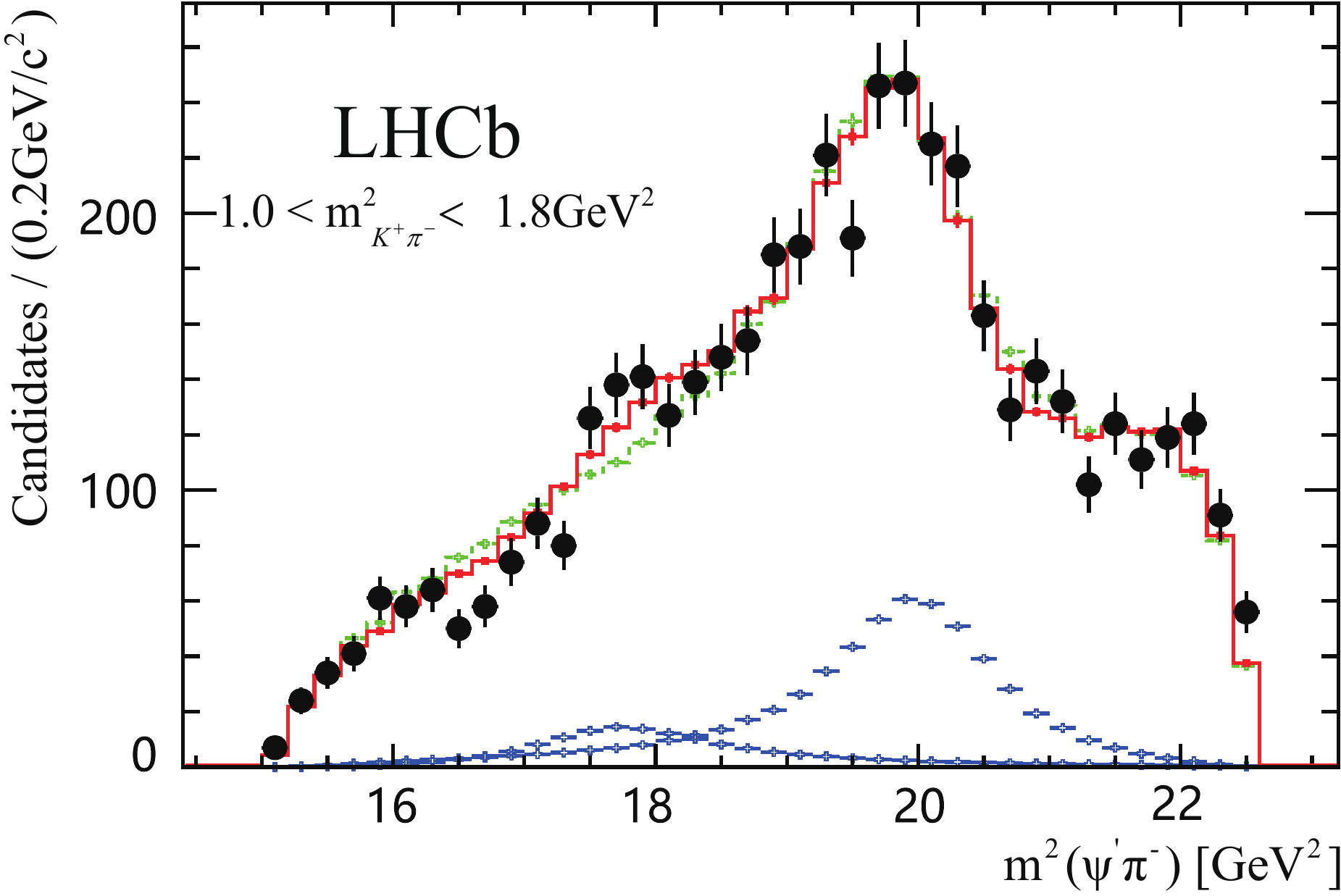}
\caption{(Color online) The square of the $\pi^- \psi(3686)$ invariant mass
spectrum of $B^0 \to K^+ \pi^- \psi(3686)$ from
LHCb~\cite{Aaij:2014jqa}. The solid-line red and dashed-line green
histograms represent the fit with two $Z^+$'s ($Z^+(4430)$ and
$Z^+(4240)$) and the fit with only one $Z^+$ ($Z^+(4430)$),
respectively.} \label{Fig:2.1.Z4240LHCb}
\end{center}
\end{figure}

\paragraph{$Z_c^+(4200)$ and $Z^+(4240)$}
\label{Sect:2.1.6}

Besides finding the evidence of $Z^+(4430) \to \pi^+ J/\psi$ in the
$\bar B^0 \to K^- \pi^+ J/\psi$ decay, Belle observed a new charged
charmonium-like structure $Z_c^+(4200)$~\cite{Chilikin:2014bkk},
which decays into $J/\psi\pi^+$ with a significance of 6.2$\sigma$.
The measured mass and width of the $Z_c^+(4200)$ are
$(4196^{+31+17}_{-29-13})$ MeV and $(370^{+70+70}_{-70-132})$ MeV,
respectively. It is obvious that the $Z_c^+(4200)$ is a very broad
structure~\cite{Chilikin:2014bkk}. The spin-parity quantum number of
the $Z_c^+(4200)$ favors $J^P=1^+$ according to the analysis of
Belle~\cite{Chilikin:2014bkk}. Additionally, the branching fraction
relevant to the $Z_c^+(4200)$~\cite{Chilikin:2014bkk} was measured,
i.e.,
\begin{eqnarray}
{\mathcal{B}}(\bar{B}^0\to Z_c^+(4200)K^-)\times
{\mathcal{B}}(Z_c^+(4200)\to
J/\psi\pi^+)=(2.2^{+0.7+1.1}_{-0.5-0.6})\times 10^{-5}.
\end{eqnarray}

In Ref.~\cite{Aaij:2014jqa}, LHCb not only confirmed the existence
of the $Z^+(4430)$, but also found a new structure the $Z^+(4240)$ in the
$\pi^- \psi(3686)$ invariant mass distribution of $B^0 \to K^+ \pi^-
\psi(3686)$ decay with a statistical significance of $6\sigma$, as
shown in Fig.~\ref{Fig:2.1.Z4240LHCb}. Its mass and width were
measured to be $ M = (4239 \pm 18 {^{+45}_{-10}})$ MeV and $\Gamma =
(220 \pm 47 {^{+108}_{-74}})$ MeV, respectively. Its spin-parity
quantum number $J^P = 0^-$ was preferred over $1^-$, $2^-$ and $2^+$
by $8\sigma$. But $J^P = 0^-$ was preferred over $1^+$ only by
$1\sigma$. In other words, the $J^P=1^+$ assignment is not fully
excluded for the $Z^+(4240)$. In addition, its amplitude fraction was
determined to be $f_{Z^+(4240)} = (1.6 \pm 0.5 {^{+1.9}_{-0.4}})
\%$.

To date, three charged charmonium-like structures around 4.2 GeV
were observed in $B$ meson decays, the $Z^+(4248)$ which is discussed
in Sec. \ref{Sect:2.1.5}, and the $Z_c^+(4200)$ and $Z^+(4240)$. We note
that the $Z^+(4248)$ and $Z^+(4240)$ are denoted as $X(4250)^\pm$ and
$X(4240)^\pm$ in PDG~\cite{pdg}, respectively.

We need to mention an opinion from Belle. In
Ref.~\cite{Chilikin:2014bkk}, Belle indicated that the resonance
parameters of the $Z^+(4240)$ reported by LHCb were close to those of
the $Z_c^+(4200)$ while $J^P=1^+$ is not excluded for the $Z^+(4240)$. Thus,
the $Z_c^+(4200)$ and $Z^+(4240)$ may be the same state.

In order to further clarify the above three charged charmonium-like
structures, more precise experimental studies are needed, especially
the measurement of their spin-parity quantum numbers.

\paragraph{$X(3823)$}
\label{Sect:2.1.7}

In 1994, the E705 Collaboration reported a $2.8\sigma$ structure at
$3.836$ GeV in the $J/\psi\pi^+\pi^-$ channel
\cite{Antoniazzi:1993jz}. If this structure was a resonance, the
$^3D_2(2^{--})$ assignment was favored by the experimental data.
However, there were only $58\pm21$ events in the E705 data \cite{Antoniazzi:1993jz}.

\begin{figure}[hbtp]
\begin{center}
\includegraphics[width=10cm]{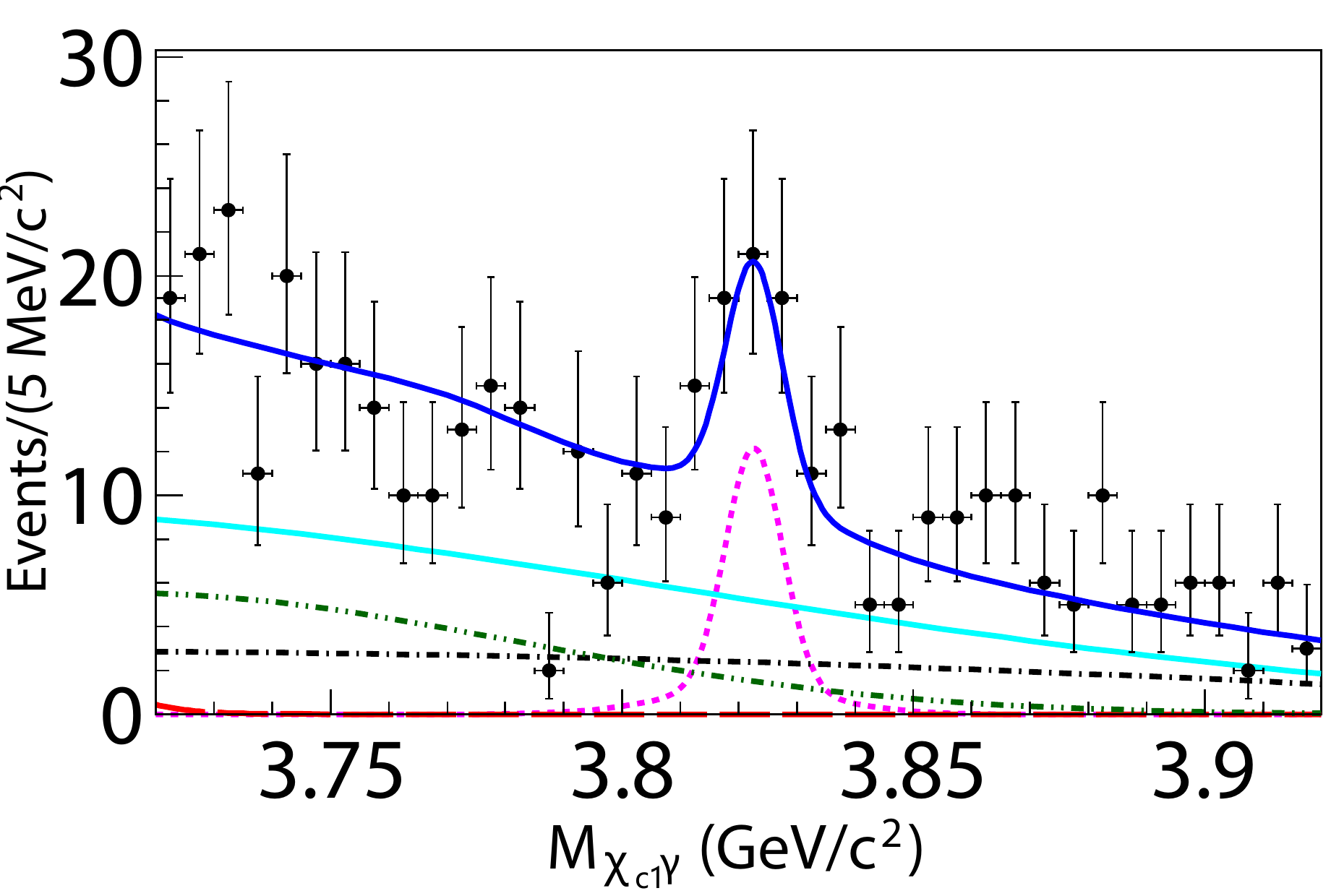}
\caption{(Color online) The invariant mass distribution $M_{\chi_{c1}\gamma}$ in
the $B\to\chi_{c1}\gamma K$ process for $M_{\rm bc} > 5.27 $
GeV$/c^2$, from Belle~\cite{Bhardwaj:2013rmw}.}
\label{Fig:2.1.X3823Belle}
\end{center}
\end{figure}

In 2013, the Belle Collaboration observed a new narrow resonance
decaying to $\chi_{c1}\gamma$ in the $B\to\chi_{c1}\gamma K$ process
with a statistical significance of $3.8\sigma$
\cite{Bhardwaj:2013rmw}. This state has a mass of
$(3823.1\pm1.8\pm0.7)$ MeV. The invariant mass of the
$\chi_{c1}\gamma$ distribution is shown in Fig.
\ref{Fig:2.1.X3823Belle}. Belle measured the branching fraction
product $\mathcal{B}(B^{\pm} \to X(3823) K^{\pm})\times
\mathcal{B}(X(3823) \to \chi_{c1}\gamma)$ $=$ $(9.7 \pm 2.8 \pm
1.1)\times 10^{-6}$. They found no evidence for $X(3823)\to
\chi_{c2}\gamma$ decay and set an upper limit of the ratio
$R_B\equiv \frac{\mathcal{B}(X(3823)\to
\chi_{c2}\gamma)}{\mathcal{B}(X(3823)\to \chi_{c1}\gamma)} <$ 0.41
at 90\% C.L. They suggested this new resonance $X(3823)$ as the $1
^3D_2$ charmonium state with $J^{PC}=2^{--}$. The mass and radiative
decay behavior agree with the theoretical predictions for the $1
^3D_2$ state
\cite{Eichten:1978tg,Eichten:1979ms,Buchmuller:1980su,Godfrey:1985xj,Ebert:2002pp,Eichten:2004uh}.

\begin{figure}[hbtp]
\begin{center}
\includegraphics[width=10cm]{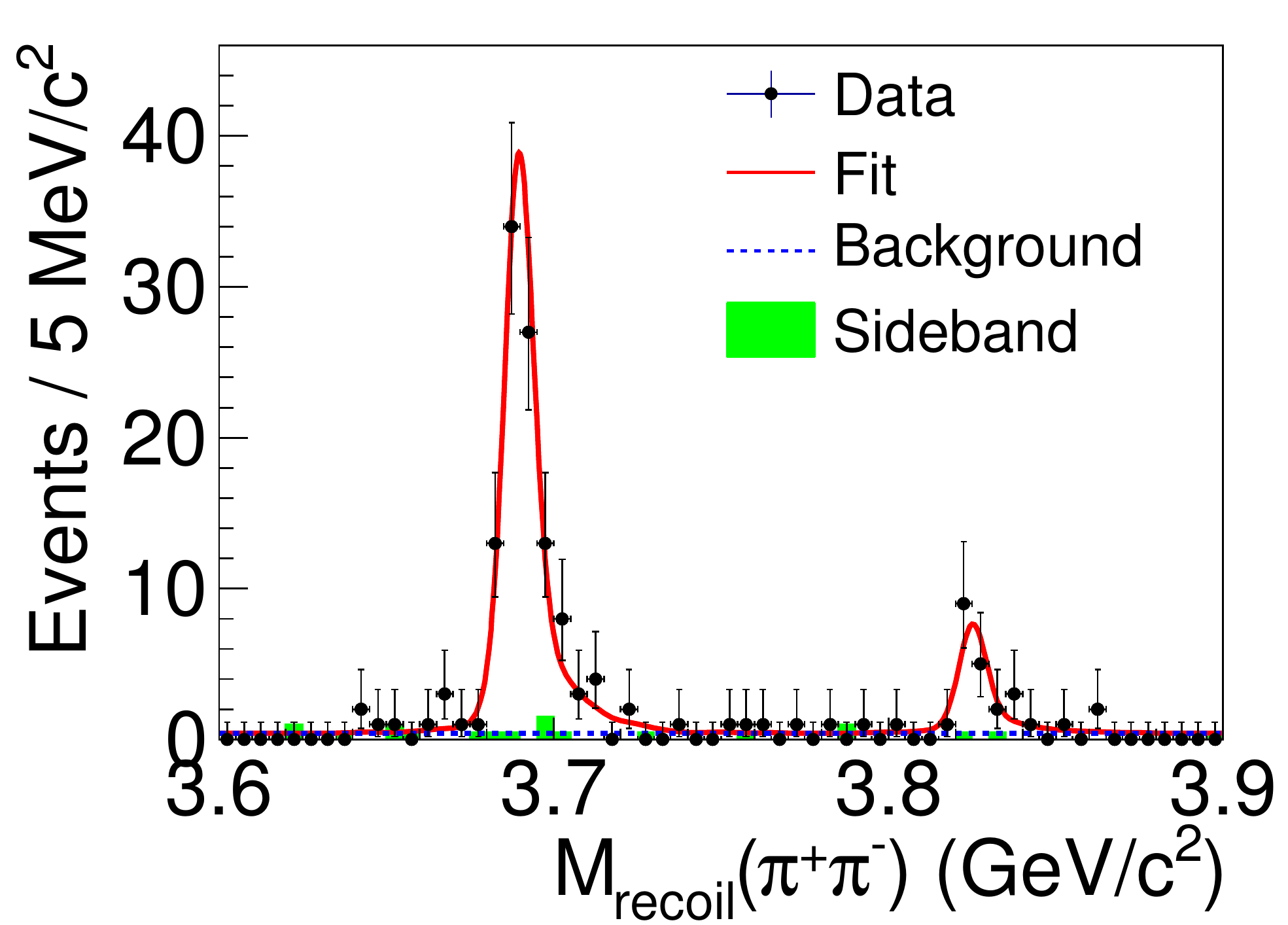}
\caption{(Color online) Simultaneous fit to the $M_{\rm recoil}(\pi^+\pi^-)$
distribution of the $\gamma\chi_{c1}$ events, from
BESIII~\cite{Ablikim:2015dlj}.} \label{Fig:2.1.X3823BESIII}
\end{center}
\end{figure}

Recently, the BESIII Collaboration confirmed the $X(3823)$ resonance
in the process of $e^+e^-\to \pi^+\pi^-X(3823) \to
\pi^+\pi^-\gamma\chi_{c1}$ at $6.2\sigma$, with a mass $(3821.7\pm
1.3\pm 0.7)$ MeV and width less than 16 MeV at the 90\% C.L.
\cite{Ablikim:2015dlj}. This observation is consistent with the
measurement by Belle \cite{Bhardwaj:2013rmw}. The simultaneous fit
to the $M_{\rm recoil}(\pi^+\pi^-)$ distribution of
$\gamma\chi_{c1}$ is shown in Fig. \ref{Fig:2.1.X3823BESIII}. BESIII
also provided the production cross sections of $\sigma^{B}(e^+e^-\to
\pi^+\pi^-X(3823))\cdot\mathcal{B}_1(X(3823)\to \gamma\chi_{c1})$ at
$\sqrt{s}=4.230$, 4.260, 4.360, 4.420, and 4.600~GeV, as shown in
Table \ref{Table:2.1.X3823}.

\renewcommand{\arraystretch}{1.4}
\begin{table}[hbtp]
\caption{ The production cross sections of
$\sigma^{B}(e^+e^-\to \pi^+\pi^-X(3823))\cdot\mathcal{B}_1(X(3823)\to
\gamma\chi_{c1})$ ($\sigma^B_{X}\cdot \mathcal{B}_1$) and
$\mathcal{B}_2(X(3823)\to \gamma\chi_{c2})$ ($\sigma^B_{X}\cdot
\mathcal{B}_2$), and the Born cross section $\sigma^B(e^+e^-\to
\pi^+\pi^-\psi^\prime)$ ($\sigma^B_{\psi^\prime}$) at different
energies from BESIII \cite{Ablikim:2015dlj}.
The relative ratio $\mathcal{R}_{\psi^\prime}$ corresponds to $\frac{\sigma^B[e^+e^-\to\pi^+\pi^- X(3823)]\mathcal{B}(X(3823)\to\gamma\chi_{c1})}
{\sigma^B[e^+e^-\to\pi^+\pi^-\psi^\prime]\mathcal{B}(\psi^\prime\to\gamma\chi_{c1})}$. \label{Table:2.1.X3823}}
\begin{center}
\begin{tabular}{cccccc} \toprule[1pt]
  $\sqrt{s}$~(GeV) & $\sigma^B_X\cdot\mathcal{B}_1$~(pb) & $\sigma^B_X\cdot\mathcal{B}_2$~(pb) & $\sigma^B_{\psi^\prime}$ (pb) & $\mathcal{R}_{\psi^\prime}$ \\
  \hline
  4.230 & $0.12^{+0.24}_{-0.12}\pm 0.02$ $(<0.64)$ & - & $34.1\pm8.1\pm4.7$ &- \\
  4.260 & $0.23^{+0.38}_{-0.24}\pm 0.04$ $(<0.98)$ & - & $25.9\pm8.1\pm3.6$ & - \\
  4.360 & $1.10^{+0.64}_{-0.47}\pm 0.15$ $(<2.27)$ & $(<1.92)$ & $58.6\pm14.2\pm8.1$ & $0.20^{+0.13}_{-0.10} $ \\
  4.420 & $1.23^{+0.59}_{-0.46}\pm 0.17$ $(<2.19)$ & $(<0.54)$ & $33.4\pm7.8\pm4.6$ & $0.39^{+0.21}_{-0.17}$ \\
  4.600 & $0.47^{+0.44}_{-0.27}\pm 0.07$ $(<1.32)$ & - & $10.4^{+6.4}_{-4.7}\pm1.5$ & - \\
\bottomrule[1pt]
\end{tabular}
\end{center}
\end{table}

%
%=====================================================================================
%=====================================================================================
\subsubsection{$Y$ states produced through the $e^+e^-$ annihilation}
\label{Sect.2.2}
%=====================================================================================
%=====================================================================================
%

So far, five charmonium-like states, the $Y(4260)$, $Y(4008)$,
$Y(4360)$, $Y(4660)$ and $Y(4630)$, have been reported in the
$e^+e^-$ annihilation processes, which are due to the development of
the initial-state radiation (ISR) technique. As the first $Y$ state
in the charmonium-like state family, the $Y(4260)$ has attracted
great attention from both experimentalists and theorists. To some
extent, the $X(3872)$ and $Y(4260)$ are the two superstars among all
the observed charmonium-like states. Those charmonium-like states
observed in the $e^+e^-$ annihilation follow the same naming
convention as that of the $Y(4260)$.

Before giving the experimental details of these $Y$ states, we
briefly list their discovery modes:
\begin{eqnarray*}
Y(4260)\, {\mathrm{and}}\, Y(4008)&:& \quad e^+e^-\to \gamma_{\mathrm{ISR}}\pi^+\pi^- J/\psi,\\
Y(4360)\, {\mathrm{and}}\, Y(4660)&:& \quad e^+e^-\to \gamma_{\mathrm{ISR}}\pi^+\pi^- \psi(3686),\\
Y(4630)&:& \quad e^+e^-\to
\gamma_{\mathrm{ISR}}\Lambda_c\bar{\Lambda}_c.
\end{eqnarray*}

\begin{figure}[hbtp]
\begin{center}
\includegraphics[width=10cm]{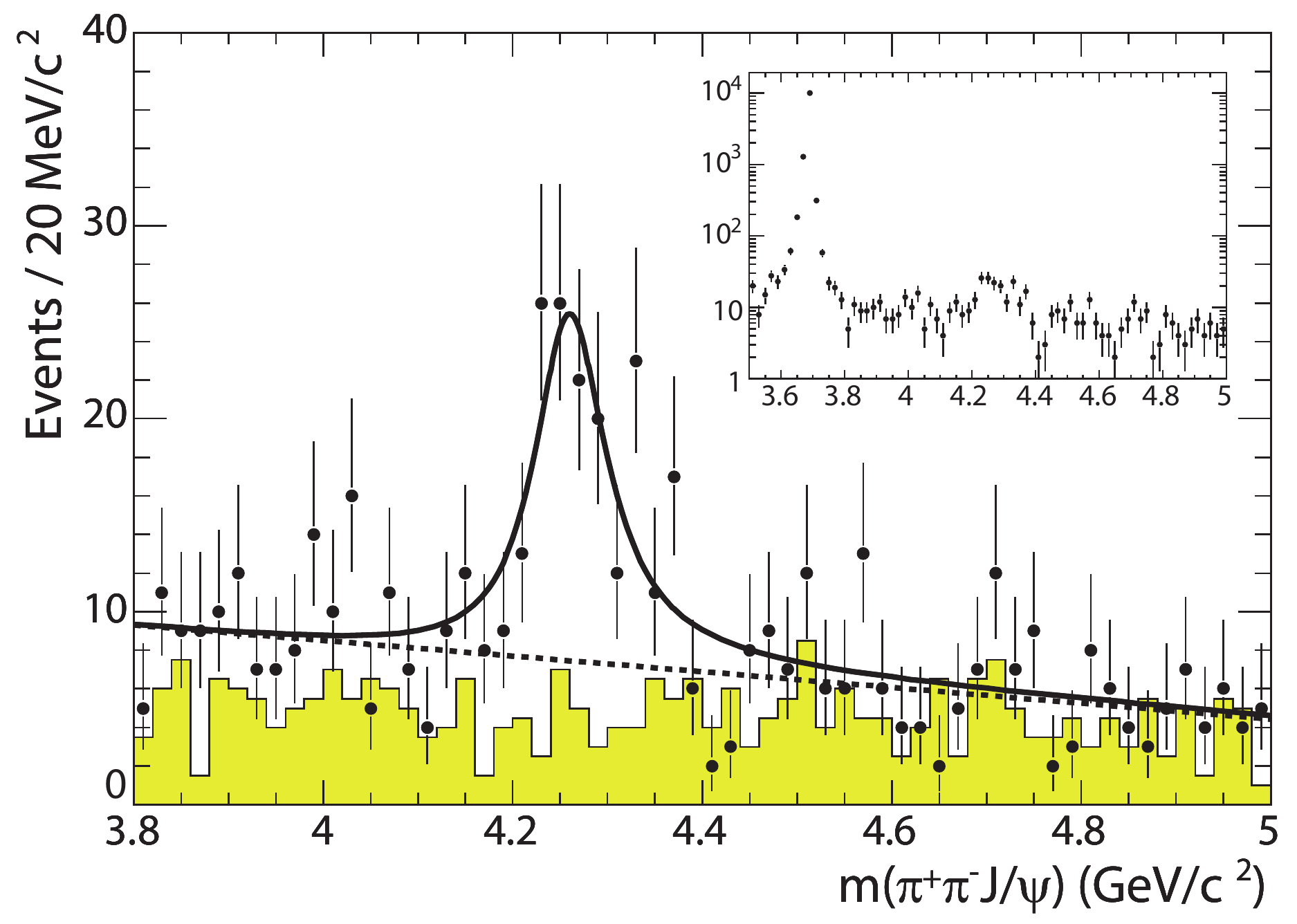}
\caption{(Color online) The $\pi^+\pi^-J/\psi$ invariant mass spectrum of
$e^+e^-\to \gamma_{\mathrm{ISR}}\pi^+\pi^- J/\psi$ at
$\sqrt{s}=3.8-5.0$ GeV$/c^2$ and the $Y(4260)$ structure
(black solid curve), from BaBar \cite{Aubert:2005rm}.} \label{Fig:2.2.Y4260}
\end{center}
\end{figure}

\paragraph{$Y(4260)$ and $Y(4008)$}
\label{Sect:2.2.1} The observation of the $Y(4260)$ was first
announced by the BaBar Collaboration in 2005~\cite{Aubert:2005rm},
where they used the ISR technique to study the process $e^+e^-
\rightarrow \gamma_{\rm ISR} \pi^+\pi^- J/\psi$ at $\sqrt s=10.58$
GeV, as shown in Fig.~\ref{Fig:2.2.Y4260}. Later, the $Y(4260)$ was
confirmed by both the CLEO~\cite{He:2006kg} and
Belle~\cite{Yuan:2007sj} collaborations in the same process. Besides
the above experimental measurement of the $Y(4260)$, CLEO analyzed
the data at the CESR $e^+ e^-$ collisions at $\sqrt s= 3.97-4.26$
GeV~\cite{Coan:2006rv}. They confirmed the $Y(4260)\to \pi^+\pi^-
J/\psi$ decay channel at $11\sigma$ significance and observed a new
decay mode $Y(4260)\to \pi^0\pi^0 J/\psi$ at $5.1\sigma$
significance. CLEO also found the evidence of $Y(4260)\to K^+K^-
J/\psi$~\cite{Coan:2006rv}. The Belle Collaboration set the
following upper limits at 90\% C.L.~\cite{Yuan:2007bt,Shen:2014gdm}:
\begin{eqnarray}
\nonumber \Gamma(Y \to e^+ e^-) \times {\cal B}(Y \to K^+K^- J/\psi) &<&
1.2~{\rm eV}~\mbox{\cite{Yuan:2007bt}} \, ,
\\ \Gamma(Y \to e^+ e^-) \times {\cal B}(Y \to K^+K^- J/\psi) &<& 1.7~{\rm eV}~\mbox{\cite{Shen:2014gdm}} \, ,
\\ \nonumber \Gamma(Y \to e^+ e^-) \times {\cal B}(Y \to K_S^0K_S^0 J/\psi) &<& 0.85~{\rm eV}~\mbox{\cite{Shen:2014gdm}} \, .
\end{eqnarray}
The BaBar Collaboration searched for the signal of the $Y(4260)$ via $B^-\to
J/\psi \pi^+\pi^- K^-$ \cite{Aubert:2005zh}, and set an upper limit
${{\cal B}}(B^-\to Y(4260)K^-,Y(4260)\to J/\psi \pi^+
\pi^-)<2.9\times10^{-5}$ \cite{Aubert:2005zh}. Thus, there does not
exist direct evidence for the $Y(4260)$ in the $B$ meson decay at
present.

A fit to the $Y(4260)$ resonance yielded a mass $(4251 \pm 9)$ MeV
and a decay width $(120 \pm 12)$ MeV~\cite{pdg}. We also summarize
its resonance parameters from different experiments in
Table~\ref{Table:2.2.Y4260}. Since the $Y(4260)$ was directly
produced from the $e^+e^-$ annihilation, its quantum number is
$J^{PC} = 1^{--}$.

\begin{figure}[hbtp]
\begin{center}
\includegraphics[width=15cm]{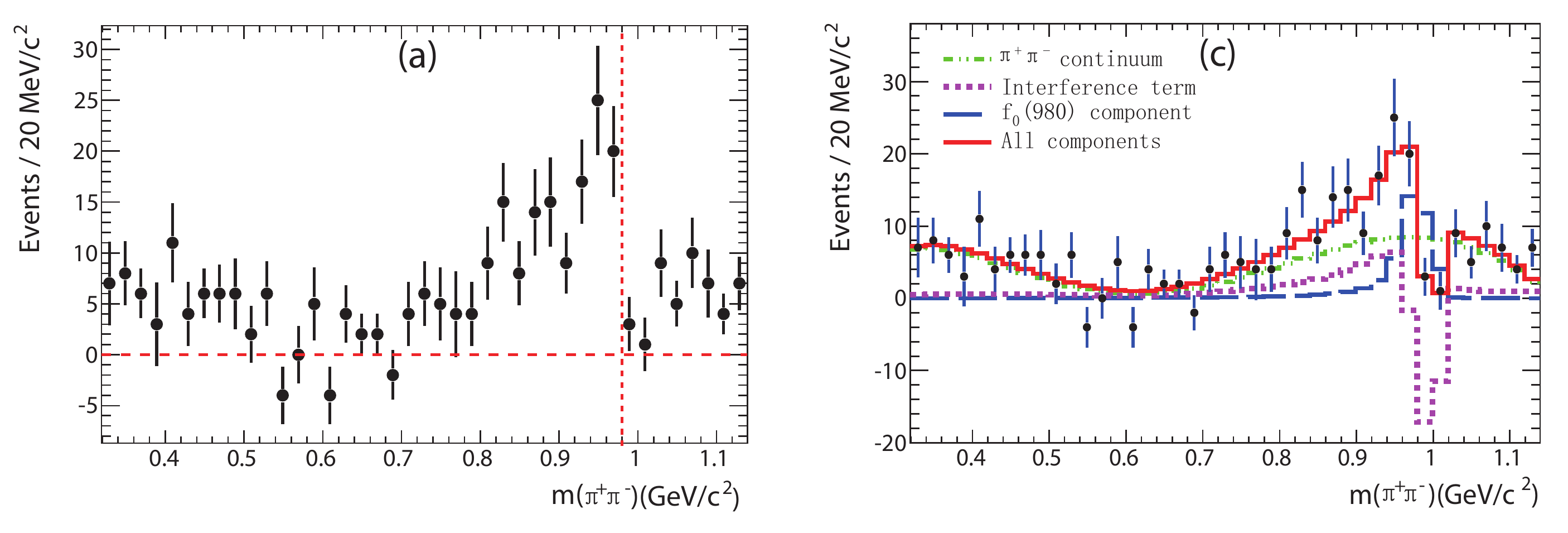}
\caption{(Color online) The dipion invariant mass spectrum (left) of $e^+e^-\to
\gamma_{\mathrm{ISR}}\pi^+\pi^- J/\psi$ and the corresponding
experimental analysis (right), from BaBar \cite{Lees:2012cn}.}
\label{Fig:2.2.Y4260-dipion}
\end{center}
\end{figure}

The observed decay modes of the $Y(4260)$ include $J/\psi
\pi^+\pi^-$~\cite{Aubert:2005rm,He:2006kg,Yuan:2007sj}, $J/\psi
\pi^0\pi^0$~\cite{Coan:2006rv}, and possibly $J/\psi K^+ K^-$
\cite{Coan:2006rv}. Additionally, the dipion mass distribution of
$Y(4260)\to \pi^+\pi^- J/\psi$ was measured
\cite{Aubert:2005rm,He:2006kg,Yuan:2007sj,Lees:2012cn}, and there
exists an enhancement around 980 MeV, which may arise from the
scalar meson $f_0(980)$. Especially, BaBar further carried out the
analysis of the dipion mass distribution of $Y(4260)\to \pi^+\pi^-
J/\psi$ (see Fig. \ref{Fig:2.2.Y4260-dipion}) and obtained the
branching ratio ${\cal B}(Y(4260)\to J/\psi f_0(980), f_0(980)\to
\pi^+\pi^-)/{\cal B}(Y(4260)\to J/\psi\pi^+\pi^-)=(0.17\pm0.13)$
\cite{Lees:2012cn}. Thus, the intermediate $f_0(980)$ contribution
to $Y(4260)\to \pi^+\pi^- J/\psi$ is not
dominant~\cite{Lees:2012cn}. Besides $Y(4260) \rightarrow J/\psi
f_0(980) \rightarrow J/\psi \pi^+\pi^-$, the $Y(4260) \rightarrow
Z_c(3900)^\pm \pi^\mp \rightarrow J/\psi \pi^+\pi^-$ mode was also
seen~\cite{Ablikim:2013mio,Liu:2013dau}, where the charged
charmonium-like structure $Z_c(3900)$ was observed in the
corresponding $J/\psi\pi^\pm$ invariant mass spectrum, which we will
discuss in detail later.

The product branching fractions of the $Y(4260)$ in the $e^+ e^-$
annihilations were measured by several experiments
\cite{Aubert:2005rm,He:2006kg,Lees:2012cn}:
\begin{eqnarray}
\nonumber \Gamma(Y(4260)\to e^+ e^-) \times {\cal B}(Y(4260) \to \pi^+ \pi^-
J/\psi) &=& (5.5 \pm 1.0 {^{+0.8}_{-0.7}}) \,{\rm
eV}~\mbox{\cite{Aubert:2005rm}} \, ,
\\ \Gamma(Y(4260) \to e^+ e^-) \times {\cal B}(Y(4260) \to \pi^+ \pi^- J/\psi) &=& (8.9 {^{+3.9}_{-3.1} \pm 1.8 })\, {\rm eV}~\mbox{\cite{He:2006kg}} \, ,
\\ \nonumber \Gamma(Y(4260) \to e^+ e^-) \times {\cal B}(Y(4260) \to \pi^+ \pi^- J/\psi) &=& (9.2 \pm 0.8 \pm 0.7)\, {\rm eV}~\mbox{\cite{Lees:2012cn}} \, .
\end{eqnarray}
Its open charm decay modes $D \bar D$, $D^* \bar D$, $D^* \bar D^*$,
$D_s^+ D_s^-$, $D_s^{*+} D_s^-$, $D_s^{*+} D_s^{*-}$, $D^0 D^{*-}
\pi^+$, $D^* \bar D \pi$, $D^* \bar D^* \pi$ were not seen \cite{Aubert:2006mi,Abe:2006fj,Pakhlova:2007fq,Pakhlova:2009jv,CroninHennessy:2008yi,Aubert:2009aq,delAmoSanchez:2010aa}
with the following upper limits:
\begin{eqnarray}
\nonumber {\cal B}(Y(4260) \to D \bar D) / {\cal B}(Y(4260) \to \pi^+ \pi^- J/\psi) &<& 7.6~\mbox{\cite{Aubert:2006mi}} \, ,
\\ {\cal B}(Y(4260) \to D_s^+ D_s^-) / {\cal B}(Y(4260) \to \pi^+ \pi^- J/\psi) &<& 0.7~\mbox{\cite{delAmoSanchez:2010aa}} \, ,
\\ \nonumber {\cal B}(Y(4260) \to D_s^{*+} D_s^-) / {\cal B}(Y(4260)\to \pi^+ \pi^- J/\psi) &<& 44~\mbox{\cite{delAmoSanchez:2010aa}} \, ,
\\ \nonumber {\cal B}(Y(4260) \to D_s^{*+} D_s^{*-}) / {\cal B}(Y(4260) \to \pi^+ \pi^- J/\psi) &<& 30~\mbox{\cite{delAmoSanchez:2010aa}} \, ,
\end{eqnarray}
at 95\% C.L., and
\begin{eqnarray}
\nonumber {\cal B}(Y(4260) \to D^0 D^{*-} \pi^+) / {\cal B}(Y(4260) \to \pi^+ \pi^- J/\psi) &<& 9~\mbox{\cite{Pakhlova:2009jv}} \, ,
\\ \nonumber \sigma(Y(4260) \to D \bar D) / \sigma(Y(4260) \to \pi^+ \pi^- J/\psi) &<& 4.0~\mbox{\cite{CroninHennessy:2008yi}} \, ,
\\ \nonumber \sigma(Y(4260) \to D^* \bar D) / \sigma(Y(4260) \to \pi^+ \pi^- J/\psi) &<& 45~\mbox{\cite{CroninHennessy:2008yi}} \, ,
\\ \nonumber \sigma(Y(4260) \to D^* \bar D^*) / \sigma(Y(4260) \to \pi^+ \pi^- J/\psi) &<& 11~\mbox{\cite{CroninHennessy:2008yi}} \, ,
\\ \nonumber \sigma(Y(4260) \to D_s^+ D_s^-) / \sigma(Y(4260) \to \pi^+ \pi^- J/\psi) &<& 1.3~\mbox{\cite{CroninHennessy:2008yi}} \, ,
\\ \sigma(Y(4260) \to D_s^{*+} D_s^-) / \sigma(Y(4260) \to \pi^+ \pi^- J/\psi) &<& 0.8~\mbox{\cite{CroninHennessy:2008yi}} \, ,
\\ \nonumber \sigma(Y(4260) \to D_s^{*+} D_s^{*-}) / \sigma(Y(4260) \to \pi^+ \pi^- J/\psi) &<& 9.5~\mbox{\cite{CroninHennessy:2008yi}} \, ,
\\ \nonumber \sigma(Y(4260) \to D^* \bar D \pi) / \sigma(Y(4260) \to \pi^+ \pi^- J/\psi) &<& 15~\mbox{\cite{CroninHennessy:2008yi}} \, ,
\\ \nonumber \sigma(Y(4260) \to D^* \bar D^* \pi) / \sigma(Y(4260) \to \pi^+ \pi^- J/\psi) &<& 8.2~\mbox{\cite{CroninHennessy:2008yi}} \, ,
\\ \nonumber {\cal B}(Y(4260) \to D^* \bar D) / {\cal B}(Y(4260) \to \pi^+ \pi^- J/\psi) &<& 34~\mbox{\cite{Aubert:2009aq}} \, ,
\\ \nonumber {\cal B}(Y(4260) \to D^* \bar D^*) / {\cal B}(Y(4260)\to \pi^+ \pi^- J/\psi) &<& 40~\mbox{\cite{Aubert:2009aq}} \, ,
\end{eqnarray}
at 90\% C.L.. Its hidden charm decay modes $J/\psi \eta$, $J/\psi
K_s^0 K_s^0$ \cite{Liu:2008hja,Wang:2012bgc,Shen:2014gdm}, and
charmless decay modes $\pi^+ \pi^- \phi$, $K^+ K^- \pi^0$, $K^0_s
K^\pm \pi^\mp$, $p \bar
p$~\cite{Aubert:2006bu,Aubert:2007ym,Aubert:2005cb} were not seen in
experiments, neither, and the following upper limits were given:
\begin{eqnarray}
\nonumber \Gamma(Y(4260) \to e^+ e^-) \times {\cal B}(Y(4260) \to \eta J/\psi) &<& 14.2 \,{\rm eV}~\mbox{\cite{Wang:2012bgc}} \, ,
\\ \nonumber \Gamma(Y(4260) \to e^+ e^-) \times {\cal B}(Y(4260) \to \pi^+ \pi^- \phi) &<& 0.4\, {\rm eV}~\mbox{\cite{Aubert:2006bu}} \, ,
\\ \Gamma(Y(4260) \to e^+ e^-) \times {\cal B}(Y(4260) \to K^+ K^- \pi^0) &<& 0.6 \,{\rm eV}~\mbox{\cite{Aubert:2007ym}} \, ,
\\ \nonumber \Gamma(Y(4260) \to e^+ e^-) \times {\cal B}(Y(4260) \to K^0_S K^\pm \pi^\mp) &<& 0.5\, {\rm eV}~\mbox{\cite{Aubert:2007ym}} \, ,
\\ \nonumber {\cal B}(Y(4260)\to p \bar p) / {\cal B}(Y(4260)\to \pi^+ \pi^- \phi) &<& 13\%~\mbox{\cite{Aubert:2005cb}} \, ,
\end{eqnarray}
at 90\% C.L.. A puzzling phenomenon of the $Y(4260)$ is that it
is absent in the $R$ value scan
\cite{Burmester:1976mn,Brandelik:1978ei,Siegrist:1981zp,Bai:1999pk,CroninHennessy:2008yi,Ablikim:2009ad},
which is the challenge to the traditional vector charmonium
interpretation of the $Y(4260)$. We note that the
$R$ value denotes the ratio of the rate of hadron production 
to that for muon pairs in the annihilation of $e^+e^-$ pair, 
i.e., $R=\sigma(e^+e^-\to hadrons)/\sigma(e^+e^-\to \mu^+\mu^-)$.

In Ref.~\cite{Ablikim:2013dyn}, the BESIII Collaboration studied the
$e^+e^-\to \gamma X(3872)$ process and measured the product of the
cross section $\sigma(e^+e^-\to \gamma X(3872))$ and the branching
ratio ${\cal B}(X(3872)\to \pi^+\pi^- J/\psi)$ at center-of-mass
energies 4.009, 4.229, 4.26 and 4.360 GeV, which hinted the
existence of the $Y(4260)\to \gamma X(3872)$ radiative decay. More
experimental information is needed to confirm the $Y(4260)\to \gamma
X(3872)$ mode. The experimental observation of the new decay mode of
the $Y(4260)$ is important to reveal its underlying structure.

\renewcommand{\arraystretch}{1.6}
\begin{table}[htb]
\caption{The resonance parameters for the $Y(4260)$ and the observed
decay channels.   \label{Table:2.2.Y4260}}
\begin{center}
\begin{tabular}{c|ccc} \toprule[1pt]
%  \multicolumn{3}{c}{Y(4260)}\\ \midrule[1pt]
  Experiment                                & Mass (MeV)                        & Width (MeV)  & Decay Mode\\\midrule[1pt]
  BaBar \cite{Aubert:2005rm}        &$4259\pm8^{+2}_{-6}$               &$88\pm23^{+6}_{-4}$   &   $J/\psi \pi^+\pi^-$ \\
  CLEO \cite{He:2006kg}             &$4284^{+17}_{-16}\pm4$             &$73^{+39}_{-25}\pm5$  &   $J/\psi \pi^+\pi^-$ \\
  Belle \cite{Abe:2006hf}           &$4295\pm10^{+10}_{-3}$             &$133\pm26^{+13}_{-6}$  &   $J/\psi \pi^+\pi^-$\\
  Belle \cite{Yuan:2007sj}          &$4247\pm12^{+17}_{-32}$            &$108\pm19\pm10$  &   $J/\psi \pi^+\pi^-$\\
  BaBar \cite{Aubert:2008aj}        &$4252\pm6^{+2}_{-3}$               &$105\pm18^{+4}_{-6}$  &   $J/\psi \pi^+\pi^-$\\
  BaBar \cite{Lees:2012cn}          &$4244\pm5\pm4$                     &$114^{+16}_{-15}\pm7$ &   $J/\psi f_0(980) (\to \pi^+\pi^-)$, \\
  Belle \cite{Liu:2013dau}          &$4258.6\pm8.3\pm12.1$              &$134.1\pm16.4\pm5.5$ &   $\pi^\mp Z_c(3900)^\pm (\to J/\psi \pi^\pm)$ \\     %                               \midrule[1pt]
 % \multicolumn{3}{c}{Y(4008)}\\ \midrule[1pt]
 % Belle \cite{Yuan:2007sj}          &$4008\pm40^{+114}_{-28}$            &$226\pm44\pm87$\\
 % Belle \cite{Liu:2013dau}          &$3890.8\pm40.5\pm11.5$              &$254.5\pm39.5\pm13.6$\\
\bottomrule[1pt]
\end{tabular}
\end{center}
\end{table}

\begin{figure}[hbtp]
\begin{center}
\includegraphics[width=16cm]{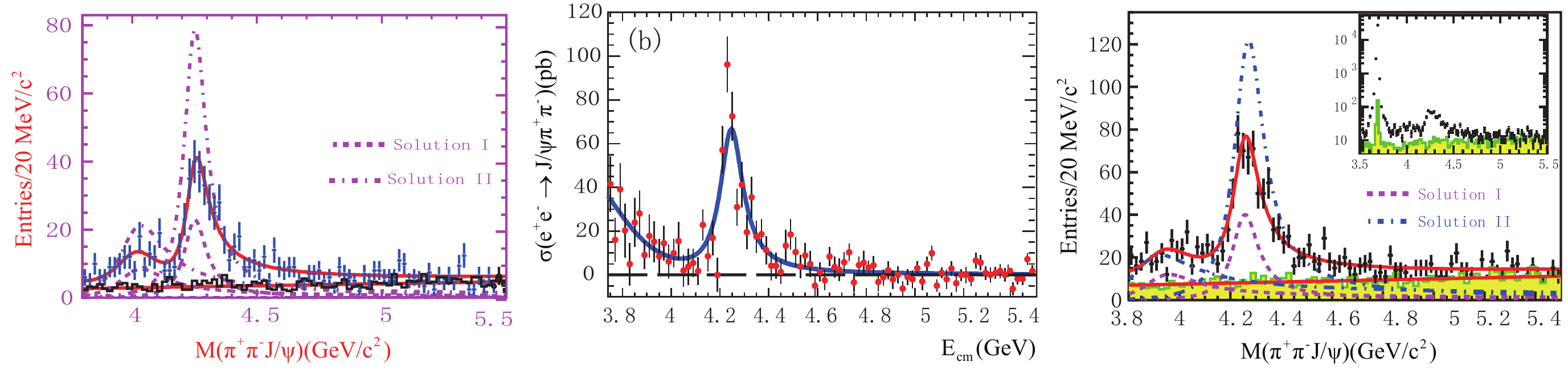}
\caption{(Color online) The $\pi^+\pi^-J/\psi$ invariant mass spectrum of
$e^+e^-\to \gamma_{\mathrm{ISR}}\pi^+\pi^- J/\psi$ from Belle
\cite{Yuan:2007sj} (left), BaBar  \cite{Lees:2012cn} (middle) and
Belle \cite{Liu:2013dau} (right).} \label{Fig:2.2.Y4008}
\end{center}
\end{figure}

Besides the $Y(4260)$, the Belle Collaboration indicated that there may
exist another very broad structure $Y(4008)$ around 4.05 GeV in the
measured $\pi^+\pi^-J/\psi$ mass spectrum \cite{Yuan:2007sj}. By
adopting two interfering Breit-Wigner formalism to fit the
experimental data, a broad structure with the mass
$(4008\pm40^{+114}_{-28})$ MeV and width $(226\pm44\pm87)$ MeV was
extracted, which was named as $Y(4008)$. However, BaBar did not found
the $Y(4008)$ signal in the same process $e^+e^-\to \pi^+\pi^- J/\psi$
\cite{Lees:2012cn}. Later, based on the new measurement of $e^+e^-\to
\pi^+\pi^- J/\psi$ with a 967 fb$^{-1}$ data sample, Belle confirmed
that there exists an event cluster around 4.08 GeV
\cite{Liu:2013dau}. The inconsistency between BaBar  \cite{Lees:2012cn}
and Belle \cite{Yuan:2007sj,Liu:2013dau}
results of the $Y(4008)$ should be clarified in future experiments. In
Fig. \ref{Fig:2.2.Y4008}, we list the Belle and BaBar experimental
data of the $\pi^+\pi^- J/\psi$ mass spectrum of  $e^+e^-\to
\pi^+\pi^- J/\psi$ for comparison.

\paragraph{$Y(4360)$ and $Y(4660)$}
\label{Sect:2.2.2}

After the observation of the $Y(4260)$ in $e^+e^- \rightarrow
\gamma_{\rm ISR} \pi^+\pi^- J/\psi$, the BaBar Collaboration
analyzed a similar process $e^+e^- \rightarrow \gamma_{\rm ISR}
\pi^+\pi^- \psi(3686)$, where a resonant structure $Y(4360)$ was
observed~\cite{Aubert:2007zz}. Later, the Belle Collaboration
confirmed the existence of the $Y(4360)$ in the same process.

\begin{figure}[hbtp]
\begin{center}
\includegraphics[width=13cm]{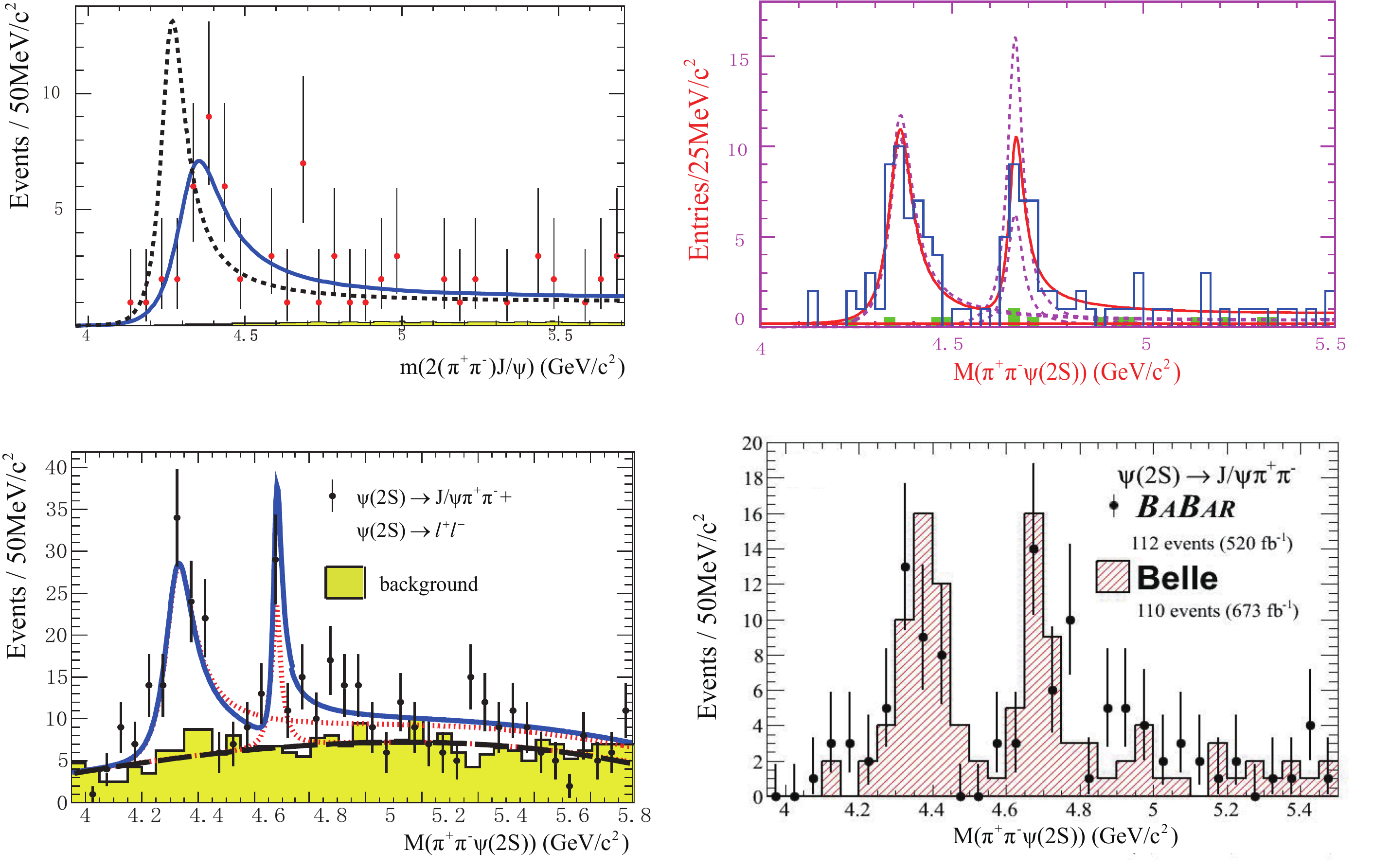}
\caption{(Color online) The measured $\pi^+\pi^-\psi(2S)$ invariant mass spectrum
of $e^+e^-\to \gamma_{\mathrm{ISR}}\pi^+\pi^- \psi(2S)$ from BaBar
and Belle. Here, the upper left and lower left panels are taken from BaBar
\cite{Aubert:2007zz,Lees:2012pv}, and the upper right panel is taken from Belle \cite{Wang:2007ea}.
The lower right panel is a comparison of the Belle (dots with
errors) and BaBar (hatched histogram) data from BaBar \cite{Lees:2012pv}.} \label{Fig:2.2.Y4360}
\end{center}
\end{figure}

Besides the $Y(4360)$, Belle further indicated that there was
another enhancement structure $Y(4660)$~\cite{Wang:2007ea}
associated with the $Y(4360)$ in the $\pi^+\pi^- \psi(3686)$
invariant mass spectrum of $e^+e^- \rightarrow \gamma_{\rm ISR}
\pi^+\pi^- \psi(3686)$. Belle's observation of the
$Y(4660)$~\cite{Wang:2007ea} was not confirmed by BaBar
\cite{Aubert:2007zz}, which resulted in a long-term debate whether
there is a $Y(4660)$ structure in $e^+e^- \rightarrow \gamma_{\rm
ISR} \pi^+\pi^- \psi(3686)$. Finally in 2012, BaBar confirmed the
existence of the $Y(4660)$ with new data on the $e^+e^- \rightarrow
\gamma_{\rm ISR} \pi^+\pi^- \psi(3686)$ process~\cite{Lees:2012pv}.
In Fig. \ref{Fig:2.2.Y4360}, the $\pi^+\pi^- \psi(3686)$ invariant
mass spectrum of $e^+e^- \rightarrow \gamma_{\rm ISR} \pi^+\pi^-
\psi(3686)$ is presented.

A fit to the $Y(4360)$ structure yielded a mass $(4354 \pm 10)$ MeV
and a width $(78 \pm 16)$ MeV while a fit to the $Y(4660)$ yielded a
mass $(4665 \pm 10)$ MeV and a width $(53 \pm 16)$ MeV, where these
resonance parameters of the $Y(4360)$ and $Y(4660)$ are the averaged
values listed in PDG \cite{pdg}. We summarize their resonance
parameters from different experiments in
Table~\ref{Table:2.2.Y4360}. Since they are produced from the
$e^+e^-$ annihilation, the quantum numbers of the $Y(4360)$ and
$Y(4660)$ are $J^{PC} = 1^{--}$.

\renewcommand{\arraystretch}{1.6}
\begin{table}[htb]
\caption{The resonance parameters for the $Y(4360)$ and $Y(4660)$ and
the observed decay channels. The results shown in the last row were
obtained by Liu, Qin and Yuan by performing a combined fit to BaBar
and Belle data of $e^+e^-\to \psi(3686)\pi^+\pi^-$. Here, the mass
and width are in units of MeV. \label{Table:2.2.Y4360}}
\begin{center}
\begin{tabular}{c|ccc} \toprule[1pt]
%  \multicolumn{3}{c}{Y(4260)}\\ \midrule[1pt]
  Experiment                                & $Y(4360)$                      & $Y(4660)$ & Decay Mode\\\midrule[1pt]
  BaBar \cite{Aubert:2007zz}        &$M=4324\pm24$, $\Gamma=172\pm33$               & --   &   $\psi(2S) \pi^+\pi^-$ \\
  Belle \cite{Wang:2007ea}           &$M=4361\pm9\pm9$, $\Gamma=74\pm15\pm10$             &$M=4664\pm11\pm5$, $\Gamma=48\pm15\pm3$  &   $\psi(2S) \pi^+\pi^-$\\
  BaBar \cite{Lees:2012pv}          &$M=4340\pm16\pm9$, $\Gamma=94\pm32\pm13$                     &$M=4669\pm21\pm3$, $\Gamma=104\pm48\pm10$  &   $\psi(2S) \pi^+\pi^-$\\
   LQY \cite{Liu:2008hja}          &$M=4355^{+9}_{-10}\pm9$, $\Gamma=103^{+17}_{-15}\pm11$            &$M=4661^{+9}_{-8}\pm6$, $\Gamma=42^{+17}_{-12}\pm6$  &   $\psi(2S) \pi^+\pi^-$\\
 % \multicolumn{3}{c}{Y(4008)}\\ \midrule[1pt]
 % Belle \cite{Yuan:2007sj}          &$4008\pm40^{+114}_{-28}$            &$226\pm44\pm87$\\
 % Belle \cite{Liu:2013dau}          &$3890.8\pm40.5\pm11.5$              &$254.5\pm39.5\pm13.6$\\
\bottomrule[1pt]
\end{tabular}
\end{center}
\end{table}

For the $Y(4360)$ and $Y(4660)$, only the $\psi(3686) \pi^+\pi^-$
decay mode was observed
\cite{Aubert:2007zz,Wang:2007ea,Lees:2012pv}. Their open charm decay
modes such as $Y(4360)/Y(4660) \to D^0 D^{*-} \pi^+$ are still
missing \cite{Pakhlova:2009jv}. Belle \cite{Wang:2012bgc} also
indicated that there was no evidence for the $Y(4360)$ and $Y(4660)$
in the $J/\psi \eta$ final state from the $e^+e^-$ annihilations.

\paragraph{$Y(4630)$}
\label{Sect:2.2.3}

\begin{figure}[hbtp]
\begin{center}
\includegraphics[width=7.3cm]{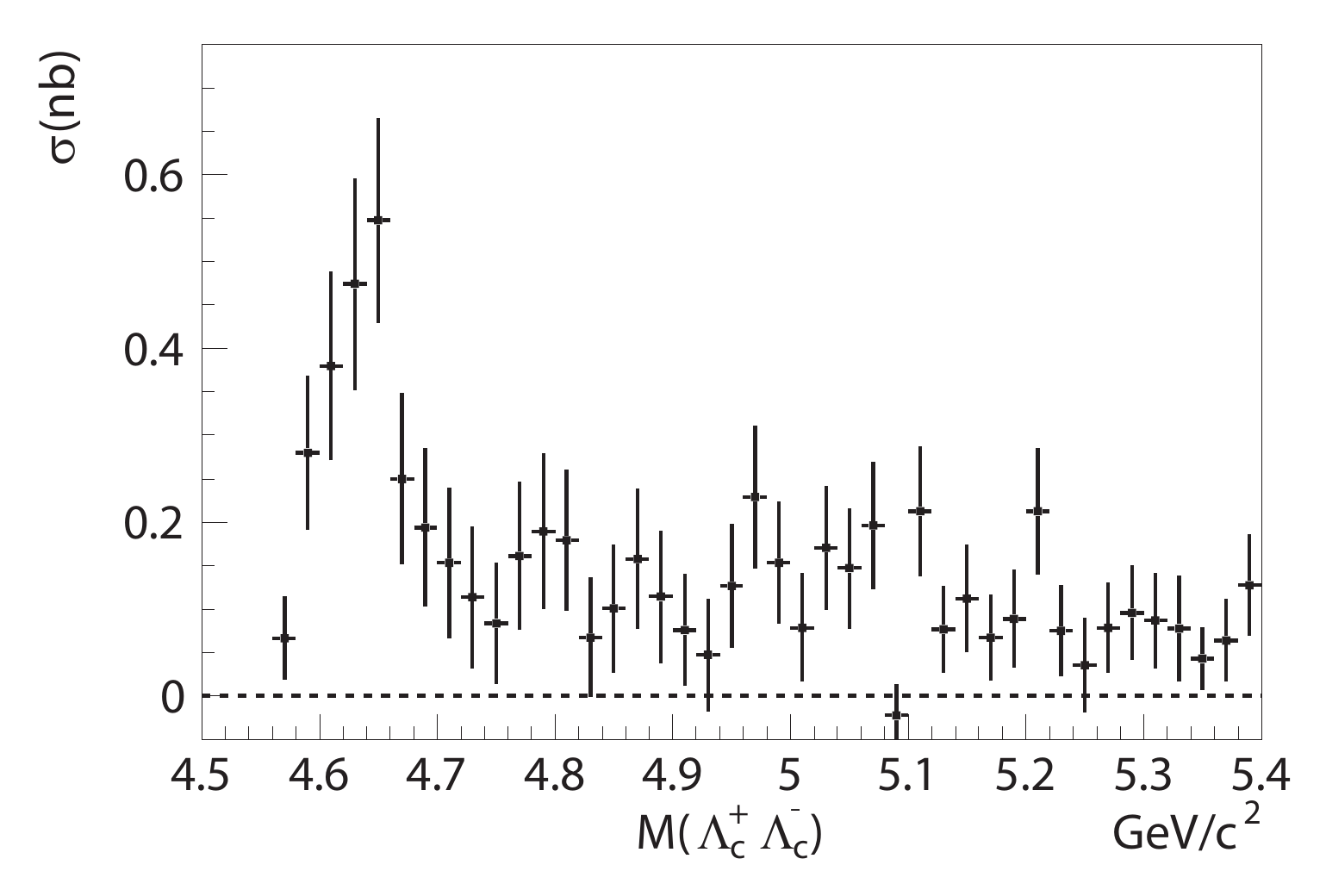}
\caption{The cross section of the $e^+e^- \to \Lambda_c \bar
\Lambda_c$ process~\cite{Pakhlova:2008vn}, where an enhancement
structure $Y(4630)$ appears in the $\Lambda_c \bar \Lambda_c$
invariant mass spectrum.} \label{Fig:2.2.Y4630}
\end{center}
\end{figure}

In 2008, the Belle Collaboration measured the exclusive $e^+e^- \to
\Lambda_c \bar \Lambda_c$ cross section, and observed an enhancement
$Y(4630)$ with a significance of $8.2\sigma$, which was close to the
$\Lambda_c \bar \Lambda_c$ threshold~\cite{Pakhlova:2008vn}, as
shown in Fig.~\ref{Fig:2.2.Y4630}. Its mass and width were
determined to be $M = (4634{^{+8}_{-7}}{^{+5}_{-8}})$ MeV and
$\Gamma = (92{^{+40}_{-24}}{^{+10}_{-21}})$ MeV, respectively, which
are consistent within errors with the mass and width of the
$Y(4660)$ resonance~\cite{Pakhlova:2008vn}. Further experiments are
needed to determine whether the $Y(4630)$ and $Y(4660)$ are the same
structure.

%
%=====================================================================================
%=====================================================================================
\subsubsection{$X$ states produced through double charmonium production}
\label{Sect:2.3}
%=====================================================================================
%=====================================================================================
%

Two charmonium-like states $X(3940)$ and $X(4160)$ were observed
through the double charmonium production. If one compares the
Feynman diagrams shown in the second and third columns of Fig.
\ref{Fig:2.0.production}, one would notice the similarity between the
double charm production and the $e^+e^-$ annihilation. However,
there still exists some difference, i.e., the final states of the double
charmonium production process include a charmonium-like state plus a
$J/\psi$. In contrast, those $Y$ states discussed in
Sec.~\ref{Sect.2.2} were directly produced via the $e^+e^-$
annihilation. The two $X$ states, $X(3940)$ and $X(4160)$, were produced from the following
processes:
\begin{eqnarray*}
X(3940)&:& \quad e^+e^-\to J/\psi \bar{D} D^*,\\
X(4160)&:& \quad e^+e^-\to J/\psi {D}^{*+} D^{*-}.
\end{eqnarray*}

\paragraph{$X(3940)$}
\label{Sect:2.3.1}

\begin{figure}[hbtp]
\begin{center}
\includegraphics[width=15cm]{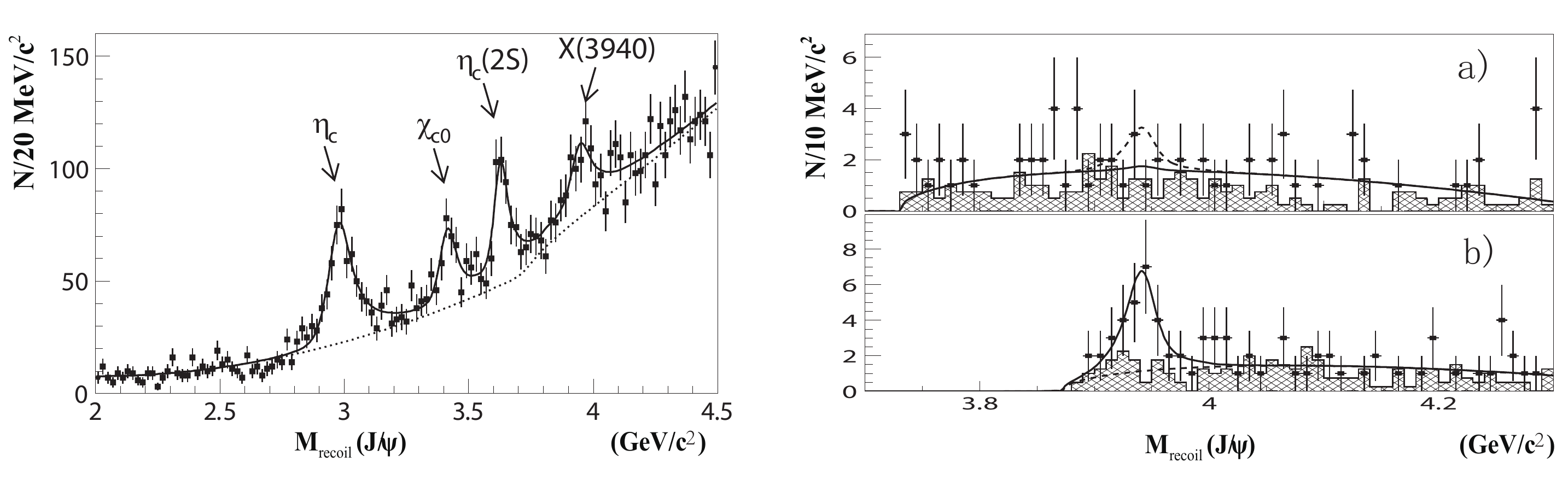}
\caption{The distribution of masses recoiling against the
reconstructed $J/\psi$ in inclusive $e^+e^- \to J/\psi X$ events (left)
. (a) and (b) are $M_{\mathrm{recoil}}(J/\psi)$
distributions for events tagged and constrained as $e^+e^-\to J/\psi
D\bar{D}$ and $e^+e^-\to J/\psi D^*\bar{D}$,
respectively. Both taken from Belle \cite{Abe:2007jna}.} \label{Fig:2.3.X3940}
\end{center}
\end{figure}

The charmonium-like state $X(3940)$ was first observed by the Belle
Collaboration in the process $e^+ e^- \to J/\psi X(3940)$ with a
significance of $5.0\sigma$~\cite{Abe:2007jna}, as shown in
Fig.~\ref{Fig:2.3.X3940}. Besides the $X(3940)$, three conventional
charmonia $\eta_c$, $\chi_{c0}$, and $\eta_c(2S)$ were seen very
clearly \cite{Abe:2007jna}. The mass of the $X(3940)$ was measured
to be $(3943\pm6\pm6)$ MeV.

By performing a fit to the $M_{\mathrm{recoil}}(J/\psi)$ distributions
for events tagged and constrained as $e^+e^-\to J/\psi D\bar{D}$ and
$e^+e^-\to J/\psi D^*\bar{D}$ \cite{Abe:2007jna} (see Fig.~\ref{Fig:2.3.X3940}),
Belle found that there was no evidence of the
$X(3940)\to D\bar{D}$ decay mode, while the $X(3940)\to D^*\bar{D}$
mode was observed. From this analysis, the width of the $X(3940)$ was
determined to be $\Gamma=(15.4\pm10.1)$ MeV, and the upper limit of
the branching ratio of the $X(3940)\to D\bar{D}$ was set, i.e.,
\begin{eqnarray}
{\cal B}(X(3940)\to D\bar{D})<41\% \, ,
\end{eqnarray}
at 90\% C.L.. In addition, Belle also set the
upper limit for the hidden-charm decay $X(3940)\to J/\psi\omega$,
i.e.,
\begin{eqnarray}
{\cal B}(X(3940)\to J/\psi\omega)<26\% \, ,
\end{eqnarray}
at 90\% C.L..

Later, Belle confirmed the $X(3940)$ in the process $e^+e^-\to J/\psi
D^*\bar{D}$ with a significance of $5.7\sigma$~\cite{Abe:2007sya},
where the resonance parameters of the $X(3940)$ were measured to be
$M=(3942^{+7}_{-6}\pm6)$ MeV and $\Gamma=(37^{+26}_{-15}\pm8)$
MeV~\cite{Abe:2007sya}.

At present, the only observed decay mode of the $X(3940)$ was $D^* \bar
D$. All the other decay modes like $D \bar D$ and $J/\psi \omega$
were not seen~\cite{Abe:2007jna}. If $e^+e^-\to J/\psi X(3940)$ is
dominated by $e^+e^-\to \gamma^*\to J/\psi X(3940)$, the $C$ parity
of the $X(3940)$ should be even, i.e., $C=+$.

\paragraph{$X(4160)$}
\label{Sect:2.3.2}

\begin{figure}[hbtp]
\begin{center}
\includegraphics[width=13cm]{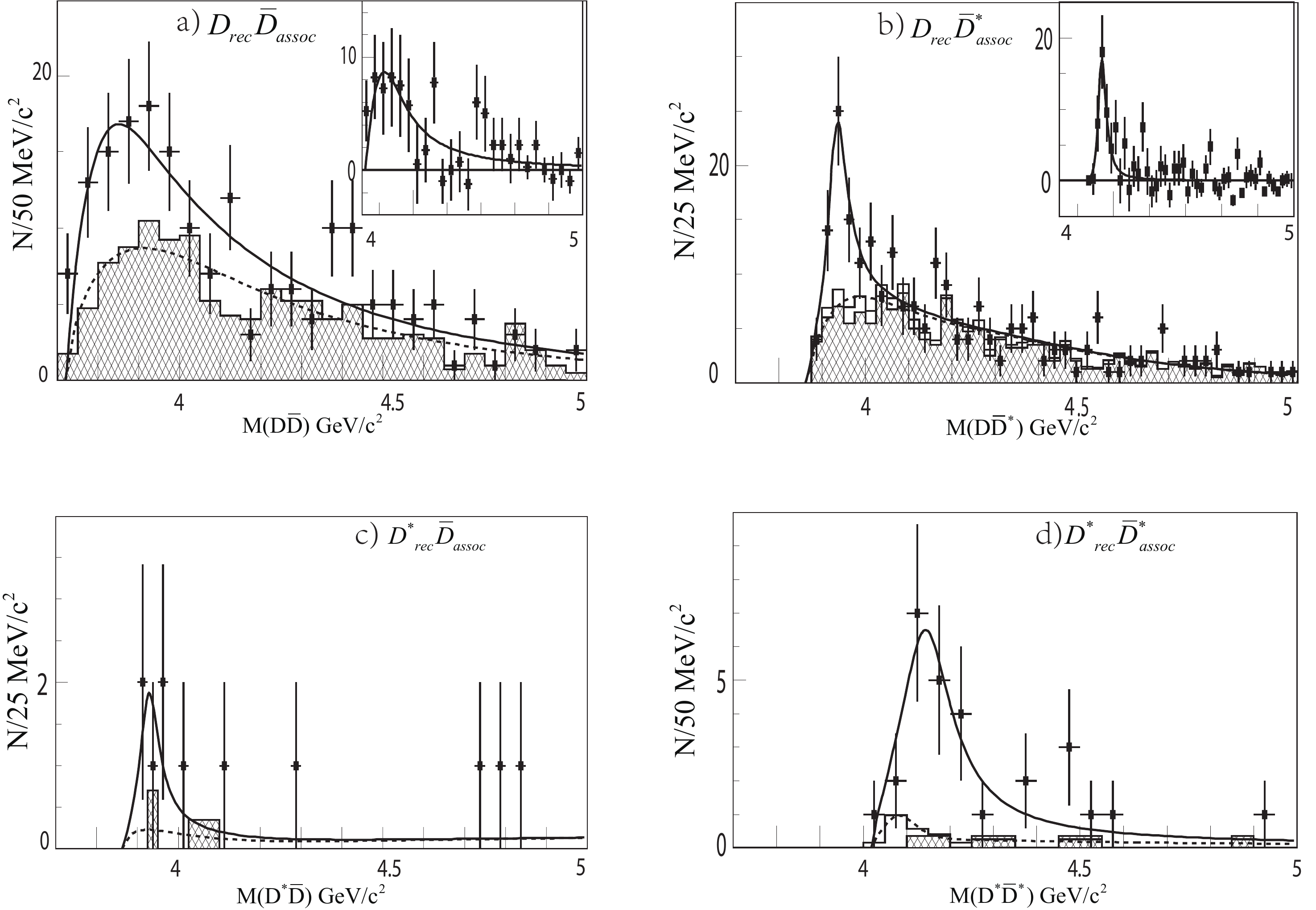}
\caption{The $M(D^{(*)}_{\rm rec} \bar D^{(*)}_{\rm assoc})$ spectra
for events tagged and constrained as (a) $e^+ e^- \to J/\psi D\bar
D$, (b) $e^+ e^- \to J/\psi D\bar D^{*} $, (c) $e^+ e^- \to J/\psi
D^{*} \bar D$, and (d) $e^+ e^- \to J/\psi D^{*} \bar
D^{*}$, from Belle~\cite{Abe:2007sya}.} \label{Fig:2.3.X4160}
\end{center}
\end{figure}

Using a data sample with an integrated luminosity of 693 fb$^{-1}$
near the $\Upsilon(4S)$ resonance, the Belle Collaboration also
analyzed the $e^+ e^- \to J/\psi D^{*+} D^{*-}$ process and found a
new charmonium-like state $X(4160)$ with a significance of
$5.1\sigma$~\cite{Abe:2007sya}. The $D^{*+} D^{*-}$ invariant mass
spectrum is shown in Fig.~\ref{Fig:2.3.X4160}. The mass and width of
the $X(4160)$ were measured to be $M = (4156{^{+25}_{-20}}\pm15)$
MeV and $\Gamma = (139{^{+111}_{-61}}\pm21)$ MeV~\cite{Abe:2007sya},
respectively. The $C$ parity of the $X(4160)$ is also even.

Besides the above observations, the analysis of
Belle~\cite{Abe:2007sya} showed that there may exist a broad
structure in the $M_{D\bar{D}}$ distribution (see Fig.
\ref{Fig:2.3.X4160} (a)), which can not be described by the
non-resonant $e^+e^-\to J/\psi D\bar{D}$ events. This structure was
quite puzzling. Unfortunately the present data sample was not large
enough to analyze this possible resonant structure~\cite{Abe:2007sya}.

Although both the $X(3940)$ and $X(4160)$ have a statistical
significance larger than $5\sigma$, they were only observed by the
Belle experiments, and still need to be confirmed by other
experiments.

%================================================================================
%================================================================================
\subsubsection{The $XYZ$ states from $\gamma\gamma$ fusion processes}\label{Sec:2.4}
%================================================================================
%================================================================================
The $\gamma\gamma$ fusion process $\gamma\gamma\to X$ produces
$C$-even charmonium states in $B$ factories. According to the
Landau-Yang selection rule \cite{Landau:1948kw,Yang:1950rg}, two
photons do not couple to any $J=1$ state. Therefore, $\gamma\gamma$
fusion process can only produce charmonium-like states with quantum
numbers $I^GJ^{PC}=0^+0^{++}$ and $0^+2^{++}$. To date, three new
charmonium-like states were reported in the $\gamma\gamma$ fusion
processes. They are the $Z(3930)$ in the $\gamma\gamma\to D\bar D$
process \cite{Uehara:2005qd}, the $X(3915)$ in the $\gamma\gamma\to
\omega J/\psi$ process \cite{Uehara:2009tx}, and the $X(4350)$ in
the $\gamma\gamma\to \phi J/\psi$ process \cite{Shen:2009vs}.
\begin{eqnarray*}
Z(3930)&:& \quad \gamma\gamma \to D\bar{D},\\
X(3915)&:& \quad \gamma\gamma\to J/\psi \omega,\\
X(4350)&:& \quad \gamma\gamma\to J/\psi \phi.
\end{eqnarray*}
The possible quantum numbers of the $Z(3930), X(3915)$ and $X(4350)$
states are either $I^GJ^{PC}=0^+0^{++}$ or $0^+2^{++}$.

%================================================================================
\paragraph{$Z(3930)$}\label{Sec:2.4.1}
%================================================================================

\begin{figure}[hbtp]
\begin{center}
\includegraphics[width=14cm]{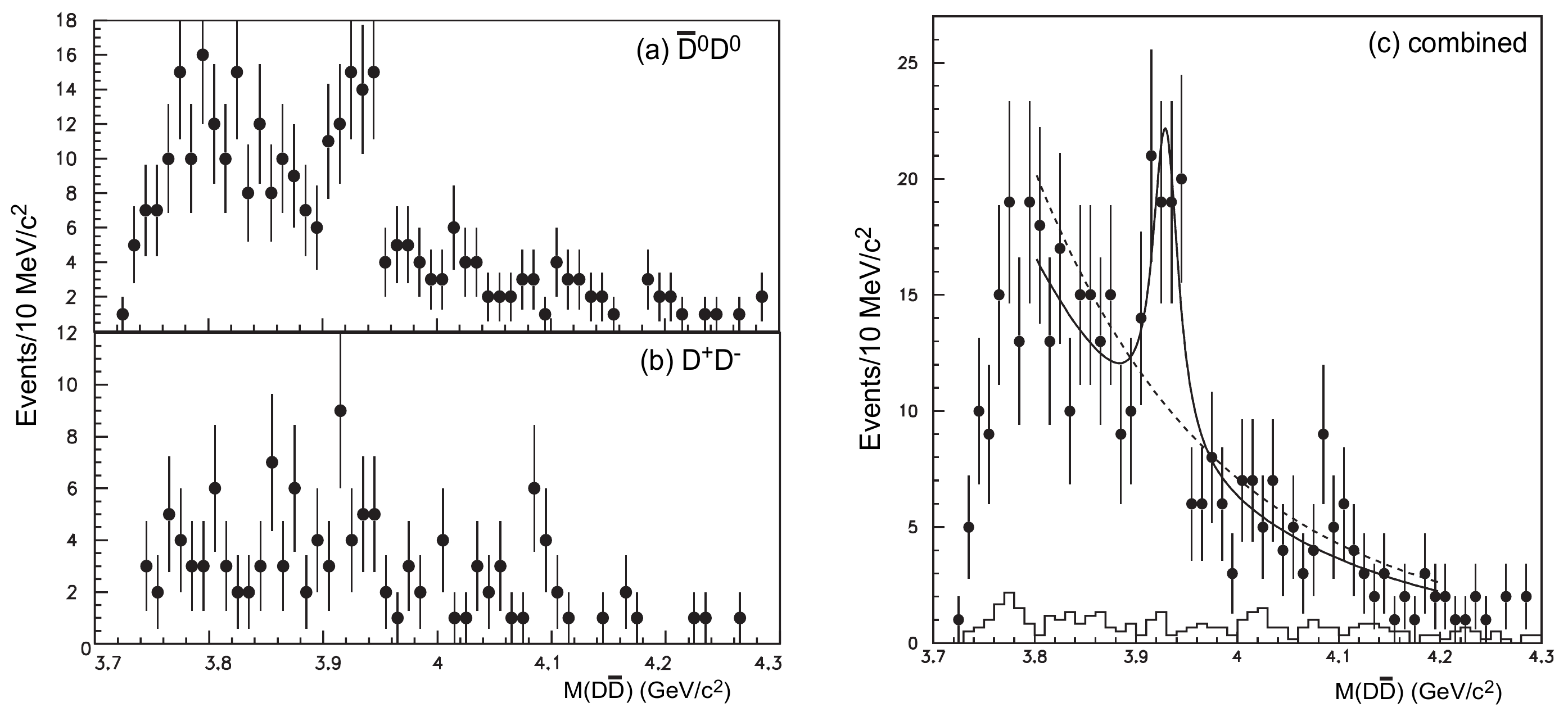}
\caption{The $D\bar D$ invariant mass distribution from
Belle\cite{Uehara:2005qd}, in which (a) is the $D^0\bar D^0$, (b)
the $D^+D^-$ mode and (c) the combined distribution.}
\label{Fig:2.4.Z3930}
\end{center}
\end{figure}

The $Z(3930)$ is one of the four charmonium-like states around 3940
MeV reported by the Belle Collaboration in 2005
\cite{Uehara:2005qd}. The others are the $Y(3940)$, $X(3940)$, and
$X(3915)$, which are discussed in Sec.~\ref{Sect:2.1.2},
\ref{Sect:2.3.1} and \ref{Sec:2.4.2}, respectively. The $Z(3930)$
was observed in the $D\bar D$ invariant mass spectrum of the process
of $\gamma\gamma\to D\bar D$ (see Fig. \ref{Fig:2.4.Z3930}), with
the mass $M=(3929\pm 5\pm 2)$ MeV, width $\Gamma=(29\pm 10\pm 2)$
MeV, and its two-photon decay width times the branching fraction
$\Gamma({Z(3930)\to\gamma\gamma})\times{\cal B}(Z(3930)\to
D\bar{D})=(0.18 \pm 0.05 \pm 0.03)$ keV (for $J=2$)
\cite{Uehara:2005qd}.

\begin{figure}[hbtp]
\begin{center}
\includegraphics[width=14cm]{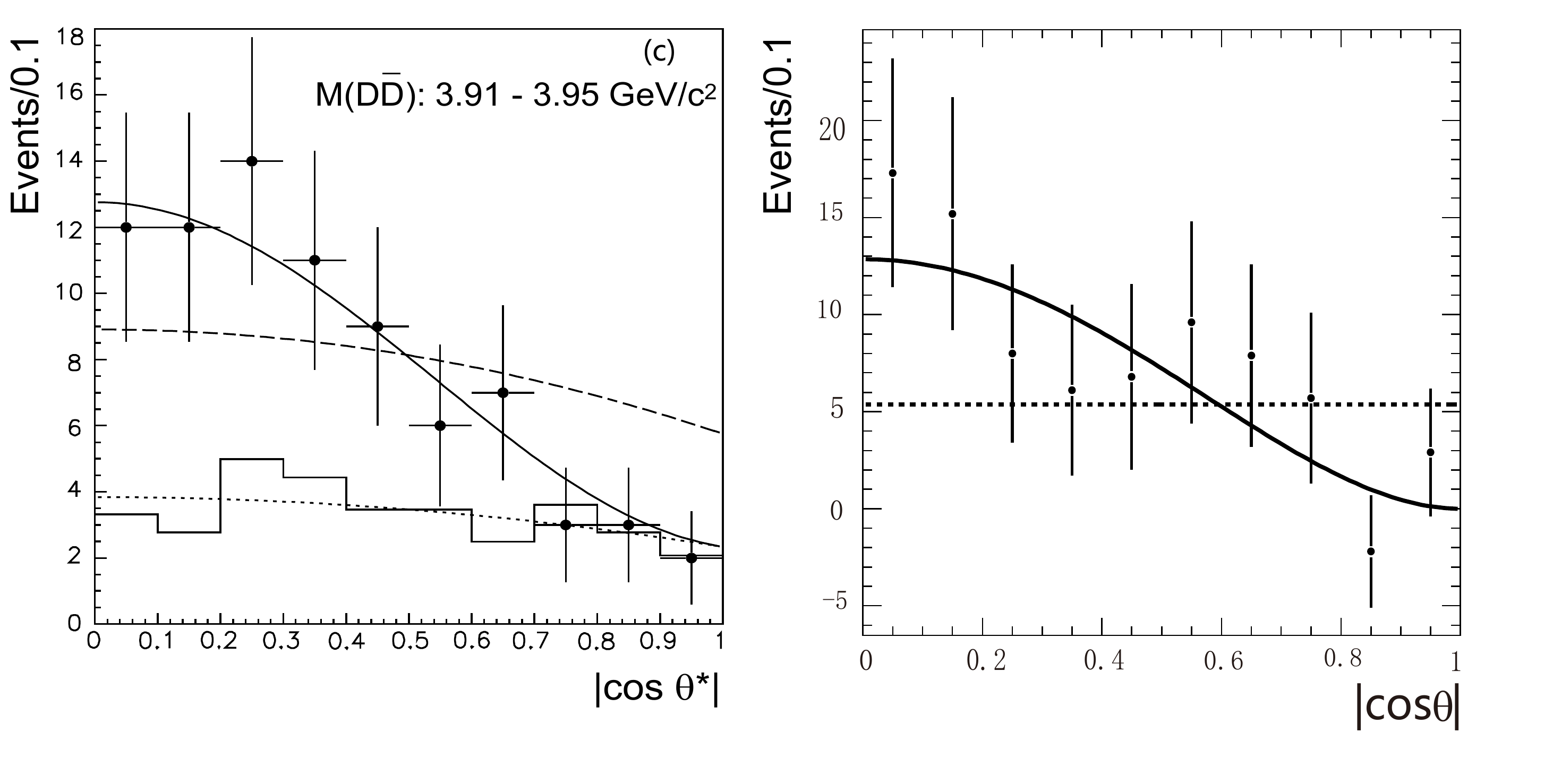}
\caption{The $|\cos\theta^*|$ and $|\cos\theta|$ distributions from
Belle \cite{Uehara:2005qd} and BaBar \cite{Aubert:2010ab},
respectively.} \label{Fig:2.4.3930cos}
\end{center}
\end{figure}

In Ref. \cite{Uehara:2005qd}, Belle also measured the $\cos\theta^*$
distributions in the $3.91<M(D\bar{D})<3.95$ GeV region, where
$\theta^*$ denotes the angle of a $D$ meson relative to the beam
axis in the $\gamma\gamma$ center of mass frame \cite{Uehara:2005qd}
(see the left panel in Fig. \ref{Fig:2.4.3930cos}). The experimental
analysis indicated that the $J=2$ assignment was favored
significantly. In addition, Belle measured the ratio of the
branching fractions for the $D^{0}\bar{D}^{0}$ and $D^+D^-$ modes to
be ${\cal B}(Z(3930) \to D^+D^-)/{\cal B}(Z(3930) \to D^0\bar{D}^0)=
(0.74 \pm 0.43 \pm 0.16)$, suggesting isospin invariance as expected
for the conventional $c\bar c$ states. As indicated in Ref.
\cite{Uehara:2005qd}, the measured mass, decay width, decay angular
distribution and $\Gamma(Z(3930)\to\gamma\gamma)\times{\cal
B}(Z(3930)\to D\bar{D})$ of the $Z(3930)$ state supported the
$Z(3930)$ as the candidate of the missing $2^3P_2$ charmonium state
$\chi^{\prime}_{c2}(2P)$, which was predicted in Refs.
\cite{Uehara:2005qd,Godfrey:1985xj,Eichten:2004uh,Munz:1996hb}.

The BaBar Collaboration confirmed the $Z(3930)$ state in the $D\bar
D$ invariant mass distribution of $\gamma\gamma\to D\bar D$ process,
with the mass $M=(3926.7 \pm 2.7 \pm 1.1)$ MeV and width
$\Gamma=(21.3 \pm 6.8 \pm 3.6)$ MeV, respectively
\cite{Aubert:2010ab}. They identified the $Z(3930)$ as a tensor
state with $J^{PC}=2^{++}$ as shown in the right panel of Fig.
\ref{Fig:2.4.3930cos}.

The open charm decay mode $Z(3930)\to D\bar D^*$ is also expected
since the  $Z(3930)$ lies above the $D\bar D^*$ threshold, which will test the
assignment of the $Z(3930)$ as the $\chi_{c2}^\prime (2P)$ charmonium state. In PDG
\cite{pdg}, the $Z(3930)$ state was assigned as the
radially excited charmonium $\chi^{\prime}_{c2}(2P)$ with
$J^{PC}=2^{++}$.

%%%%%%%%%%%%%%%%%%%%%%%%%%%%%%%%%%%%%%%%%%%%%%%%%%%%%

%%%%%%%%%%%%%%%%%%%%%%%%%%%%%%%%%%%%%%%%%%%%%%%%%%%%%

%================================================================================
\paragraph{$X(3915)$}\label{Sec:2.4.2}
%================================================================================

\begin{figure}[hbtp]
\begin{center}
\includegraphics[width=15cm]{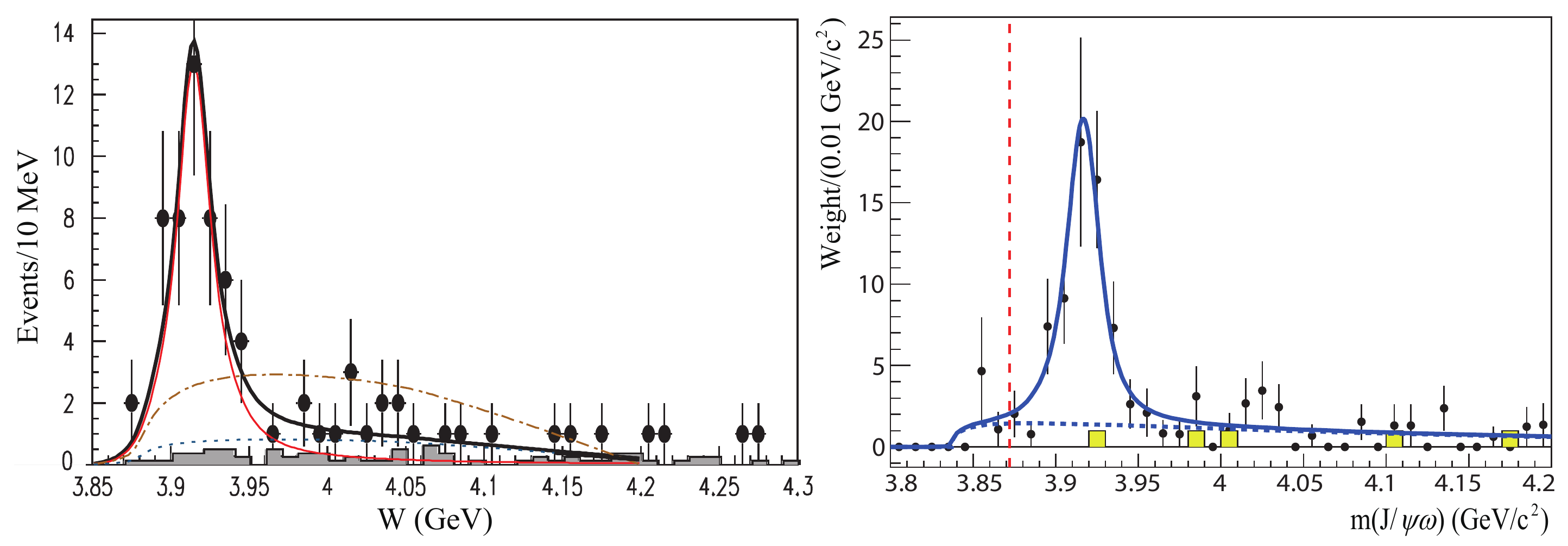}
\caption{(Color online) The $X(3915)\to\omega J/\psi$ signals in the $\gamma\gamma\to
\omega J/\psi$ process from Belle \cite{Uehara:2009tx} (left) and
BaBar \cite{Lees:2012xs} (right).} \label{Fig:2.4.X3915}
\end{center}
\end{figure}

The $X(3915)$ ($\chi_{c0}^\prime(2P)$ in PDG \cite{pdg}) state was
first reported by the Belle Collaboration in $\gamma\gamma\to \omega
J/\psi$ process \cite{Uehara:2009tx} (see the left panel in Fig.
\ref{Fig:2.4.X3915}). The measured mass and decay width were
$M=(3915\pm 3\pm 2)$ MeV and $\Gamma=(17\pm 10\pm 3)$ MeV,
respectively. Since the quantum numbers of both the $J/\psi$ and
$\omega$ are $I^GJ^{PC}=0^-1^{--}$, the $X(3915)$ carries positive
$C$-parity and $G$-parity. As discussed above for the $\gamma\gamma$
fusion process, the possible spin-parity for the $X(3915)$ is
$J^P=0^+$ or $2^+$. For these two assignments, Belle gave the
products of the two-photon decay width and the branching fraction to
$\omega J/\psi$ as \cite{Uehara:2009tx}
\begin{eqnarray}
\Gamma(X(3915)\to {\gamma \gamma}) \times{\cal B}(X(3915) \to
\omega J/\psi) = \left\{ \begin{array}{ll}
(61 \pm 17 \pm 8)~ {\rm eV} & \mbox{for $J^P=0^+$}\, , \\
(18 \pm 5 \pm 2)~ {\rm eV} & \mbox{for $J^P=2^+$\, , helicity-2\, .}
\end{array} \right. \label{Eq:2.4.X3915}
\end{eqnarray}
%%%%%%%%%%%%%%%%%%%%%%%%%%%%%%%%%%%%%%%%%%%%%%%%%%%%%

%%%%%%%%%%%%%%%%%%%%%%%%%%%%%%%%%%%%%%%%%%%%%%%%%%%%%

The BaBar Collaboration confirmed the existence of the $X(3915)$
decaying into $\omega J/\psi$ in $\gamma\gamma\to \omega J/\psi$
process (see the right panel in Fig. \ref{Fig:2.4.X3915}), with the
mass $(3919.4 \pm 2.2 \pm 1.6)$ MeV and width $(13 \pm 6 \pm 3)$ MeV
\cite{Lees:2012xs}. Their analysis favored the $J^{P}=0^{+}$
assignment. BaBar also measured $\Gamma(X(3915)\to {\gamma
\gamma})\times {\cal B}(X(3915) \to \omega J/\psi)=(52\pm 10\pm 3)$
eV, which was consistent with Belle's measurement for the $J^P=0^+$
assignment \cite{Uehara:2009tx}.

According to the BaBar's measurement, PDG assigned the $X(3915)$ as
the charmonium radial excitation $\chi_{c0}^\prime(2P)$ \cite{pdg}.
The open-charm decay mode $X(3915)\to D^{(*)}\bar D^{(*)}$ was
expected for the conventional $\chi_{c0}^\prime(2P)$ state. However,
there were no signs of the $D^{(*)}\bar D^{(*)}$ peak around 3915
MeV in both Belle \cite{Brodzicka:2007aa} and BaBar's
\cite{Aubert:2007rva} analyses of the $B\to K D^{(*)}\bar D^{(*)}$
process. This puzzling phenomenon should be clarified in future
experiments.

Belle also studied the radiative decay of $\Upsilon(2S)$ to search
for the $X(3915)$. They found no significant signals and set an
upper limit ${\cal B}(\Upsilon(2S) \to \gamma X(3915))\times {\cal
B}(X(3925)\to \omega J/\psi)<2.8\times 10^{-6}$ at $90\%$ C.L.
\cite{Shen:2010iu,Wang:2011qm}.

%================================================================================
\paragraph{$X(4350)$}\label{Sec:2.4.3}
%================================================================================

\begin{figure}[hbtp]
\begin{center}
\includegraphics[width=9cm]{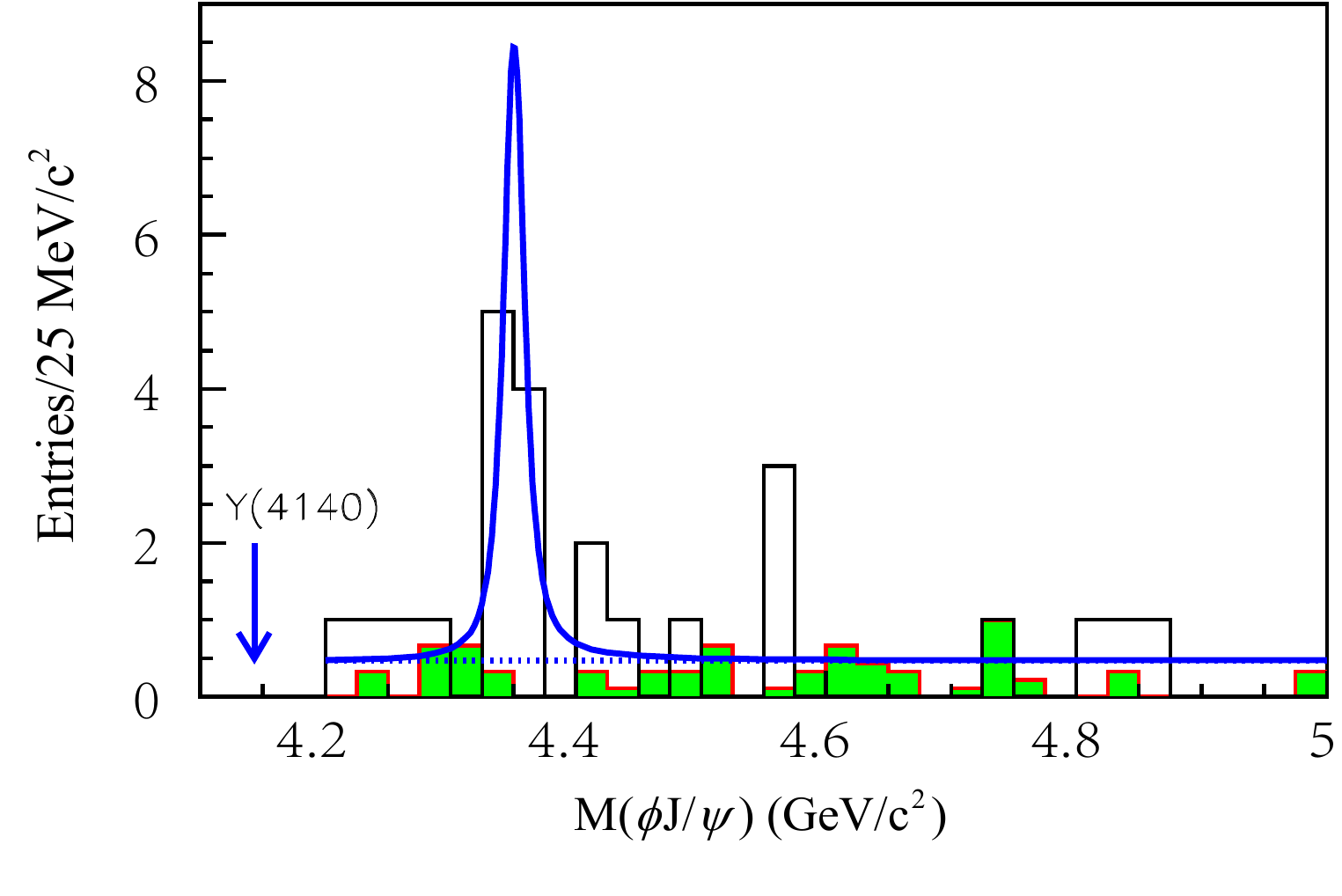}
\caption{(Color online) The $\omega J/\psi$ invariant mass distribution of
$\gamma\gamma\to \omega J/\psi$ from Belle \cite{Shen:2009vs}.}
 \label{Fig:2.4.X4350}
\end{center}
\end{figure}

In Sec. \ref{Sect:2.1.3}, we introduced the discovery of the
$Y(4140)$ by the CDF Collaboration \cite{Aaltonen:2009tz}. The Belle
Collaboration studied the $\gamma\gamma\to \phi J/\psi$ process to
search for the $Y(4140)$ state \cite{Shen:2009vs}. However, they did
not see the $Y(4140)$ signal in the $\gamma\gamma$ fusion process.
Unexpectedly, they observed another new narrow structure around 4.35
GeV in the $\phi J/\psi$ invariant mass distribution (see Fig.
\ref{Fig:2.4.X4350}), which was named as $X(4350)$. The mass and
width of this charmonium-like state was measured to be
$(4350.6^{+4.6}_{-5.1}\pm0.7)$ MeV and $(13^{+18}_{-9}\pm4)$ MeV,
respectively. Similar to the $Y(4140)$ and $X(3915)$, the possible
quantum number of the $X(4350)$ is either $I^GJ^{PC}=0^+0^{++}$ or
$0^+2^{++}$. Belle also measured the product of the two-photon decay
width of the $X(4350)$ and its branching fraction to $\phi J/\psi$
\cite{Shen:2009vs}, i.e.,
\begin{eqnarray}
\Gamma(X(4350)\to {\gamma \gamma}) \times{\cal B}(X(4350) \to \phi
J/\psi) = \left\{ \begin{array}{ll}
(6.7^{+3.2}_{-2.4}\pm1.1)~ {\rm eV} & \mbox{for $J^P=0^+$}\, , \\
(1.5^{+0.7}_{-20.6}\pm0.3)~ {\rm eV} & \mbox{for $J^P=2^+$\, .}
\end{array} \right. \label{Eq:2.4.X4350}
\end{eqnarray}

Later, Belle also tried to search for the $X(4350)$ via the radiative
decay of $\Upsilon(2S)$, and only obtained the upper limit ${\cal
B}(\Upsilon(2S) \to \gamma X(4350)) \times{\cal B}(X(4350)\to \phi
J/\psi)<1.3\times 10^{-6}$ at $90\%$ C.L. \cite{Shen:2010iu,Wang:2011qm} .

%================================================================================
%================================================================================
\subsubsection{Charged charmonium-like $Z_c$ states}\label{Sec:2.5}
%================================================================================
%================================================================================
Until now, there have accumulated abundant experimental observations
of the charged charmonium-like states. In Sec. \ref{Sect:2.1}, we have
introduced the charged charmonium-like states observed in the $B$
meson decays. In this section, we focus on the charged
charmonium-like states from the hadronic decays of the $Y(4260)$ and
higher charmonia. We collect the productions and decay modes of the
four charged charmonium-like states $Z_c(3900)$, $Z_c(3885)$,
$Z_c(4025)$ and $Z_c(4020)$ below,
\begin{eqnarray*}
e^+e^-&&\to \left\{\begin{array}{l} \left.
\begin{array}{l}
Z_c(3900)\pi^\mp \to J/\psi\pi^\pm\pi^\mp \, , \\
Z_{c}(4025)\pi^\mp \to (D^{*} \bar{D}^{*})^{\pm} \pi^\mp \, , \\
Z_c(4020)\pi^\mp\to h_c\pi^\pm \pi^\mp \, , \\
Z_c(3885)\pi^+\to (D\bar{D}^*)^-\pi^+ \, . \\
\end{array}\right.
\end{array}\right.
\end{eqnarray*}

%======================================================================================================================
\paragraph{$Z_c(3900)$ and $Z_c(3885)$}\label{Sec:2.5.1}
%======================================================================================================================

Since 2013, the BESIII Collaboration announced several charged
charmonium-like states $Z_c(3900)$ \cite{Ablikim:2013mio},
$Z_c(3885)$ \cite{Ablikim:2013xfr}, $Z_c(4020)$
\cite{Ablikim:2013wzq} and $Z_c(4025)$ \cite{Ablikim:2013emm} at
$\sqrt{s}=4.26$ GeV. All these structures were observed in the
process $Y(4260)\to \pi^-Z_c^+$ (as shown in Table
\ref{Table:2.5.Zc}). The $Z_c(3900)$ state was also observed by
Belle \cite{Liu:2013dau} and confirmed later using data of CLEO-c
\cite{Xiao:2013iha}. Before we discuss these charged charmonium-like
states, we collect the experimental information of these states in
Table \ref{Table:2.5.Zc}, including their masses, widths, production
processes from different experiments.
%%%%%%%%%%%%%%%%%%%%%%%%%%%%%%%%%%%%%%%%%%%%%%%%%%%%%%%%%%%%%%%%%%%%%%%%%%%%%%%%%%%%%%%%%%%%%%%%%%%%%%%%%%%%%%%%%%%%%%%%%%%%%%%%%%%%%%%%%%%%%
\renewcommand{\arraystretch}{1.6}
\begin{table}[htb]
\caption{Experimental information of the charged charmonium-like
states $Z_c(3900)$, $Z_c(3885)$, $Z_c(4020)$ and $Z_c(4025)$,
including their masses, widths and production processes.
\label{Table:2.5.Zc}}
\begin{center}
\begin{tabular}{ccccc} \toprule[1pt]
State             & $M$ (MeV)   & $\Gamma$~(MeV)  & Process~(decay mode) & Experiment \\
\midrule[1pt]
${Z_c(3900)}$     & $3899.0\pm3.6\pm4.9$ &  $46\pm10\pm20$    &  $e^+e^-\to Y(4260) \to \pi^- + (J/\psi\, \pi^+)$   & BESIII \cite{Ablikim:2013mio} \\
                  & $3894.5\pm6.6\pm4.5$ &  $63\pm24\pm26$    &  $e^+e^-\to Y(4260) \to \pi^- + (J/\psi\, \pi^+)$   & Belle  \cite{Liu:2013dau} \\
                  & $3886\pm4\pm2$       &  $37\pm4\pm8$      &  $e^+e^-\to \psi(4160) \to \pi^- + (J/\psi\, \pi^+)$& Xiao \textit{et al.} \cite{Xiao:2013iha} \\
${Z_c(3885)}$     & $3882.2\pm1.1\pm1.5$ &  $26.5\pm1.7\pm2.1$&  $e^+e^-\to Y(4260) \to \pi^- + (D\bar{D}^{*})^+$   & BESIII \cite{Ablikim:2013xfr,Ablikim:2015swa} \\
${Z_c(4020)}$     & $4022.9\pm0.8\pm2.7$ &  $7.9\pm2.7\pm2.6$ &  $e^+e^-\to Y(4260) \to \pi^- + (h_c\, \pi^+)$      & BESIII \cite{Ablikim:2013wzq} \\
${Z_c(4025)}$     & $4026.3\pm2.6\pm3.7$ &  $24.8\pm5.6\pm7.7$&  $e^+e^-\to Y(4260) \to \pi^- + (D^*\bar{D}^{*})^+$ & BESIII \cite{Ablikim:2013emm} \\
\bottomrule[1pt]
\end{tabular}
\end{center}

\end{table}
%%%%%%%%%%%%%%%%%%%%%%%%%%%%%%%%%%%%%%%%%%%%%%%%%%%%%%%%%%%%%%%%%%%%%%%%%%%%%%%%%%%%%%%%%%%%%%%%%%%%%%%%%%%%%%%%%%%%%%%%%%%%%%%%%%%%%%%%%%%%%

\begin{figure}[htpb]
\begin{center}
\scalebox{0.55}{\includegraphics{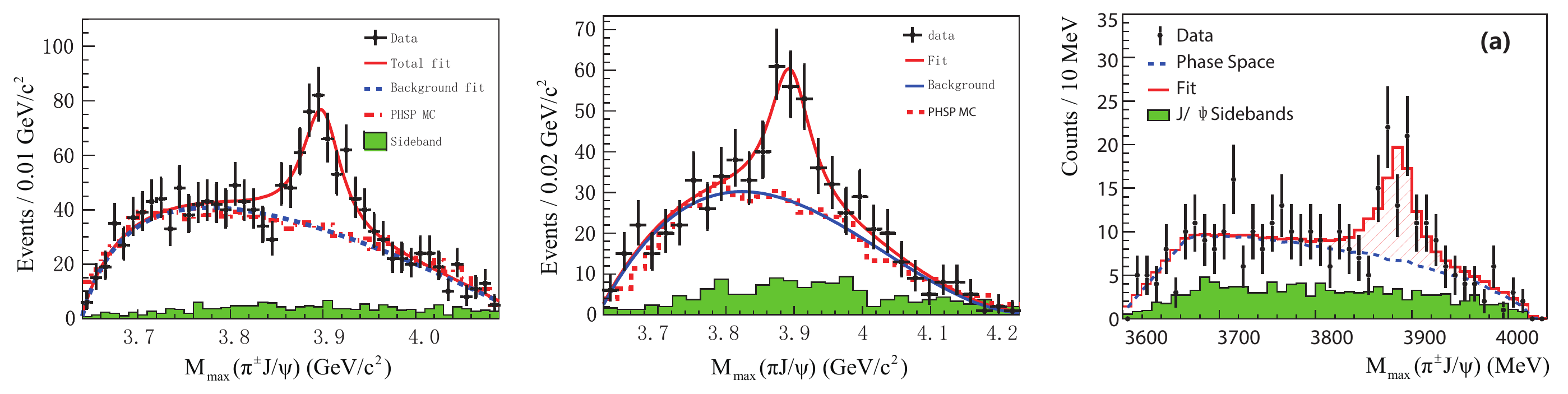}}
\end{center}
\caption{(Color online) The $Z_c(3900)$ enhancements in the $J/\psi \pi^\pm$ mass
spectrum from BESIII \cite{Ablikim:2013mio} (left), Belle
\cite{Liu:2013dau} (middle) and Xiao \textit{et al.} \cite{Xiao:2013iha} (right)
respectively.}
\label{Fig:2.5.Zc3900Psi}
\end{figure}

The $Z_c(3900)$ was observed in the $J/\psi \pi^\pm$ invariant mass
distribution of the $e^+e^-\to J/\psi \pi^+\pi^-$ process by the
BESIII Collaboration \cite{Ablikim:2013mio}. As shown in Fig.
\ref{Fig:2.5.Zc3900Psi}, the mass peak of the $Z_c(3900)$ in this
channel lies about 23 MeV above the open-charm threshold $D^+\bar
D^{*0}$ (or ${D^{*+}\bar D^{0}}$). In the same production channel,
Belle also reported the $Z_c(3900)$ structure \cite{Liu:2013dau}.
Meanwhile, Xiao \textit{et al.} analyzed the decay $\psi(4160) \to
J/\psi \pi^+\pi^-$ and observed the charged $Z_c(3900)$
\cite{Xiao:2013iha}. As shown in Table \ref{Table:2.5.Zc}, the mass
and decay width of the charged $Z_c(3900)$ from different
experiments are consistent with each other. From these three
experiments, the quantum number of the $Z_c(3900)$ was argued to be
$I^GJ^P=1^+1^+$ assuming the orbital angular momentum between the
$J/\psi$ and $\pi$ is zero.
%%%%%%%%%%%%%%%%%%%%%%%%%%%%%%%%%%%%%%%%%%%%%%%%%%%%%%%%%%%%%%%%%%%%%%%%%%%%%%%%%%%%%%%%%%%%%%%%%%%%%%%%%%%%%%%%%%%%%%%%%%%%%%%%%%%%%%%%%%%%%
%%%%%%%%%%%%%%%%%%%%%%%%%%%%%%%%%%%%%%%%%%%%%%%%%%%%%%%%%%%%%%%%%%%%%%%%%%%%%%%%%%%%%%%%%%%%%%%%%%%%%%%%%%%%%%%%%%%%%%%%%%%%%%%%%%%%%%%%%%%%%

\begin{figure}[htbp]
\begin{center}
\scalebox{0.55}{\includegraphics{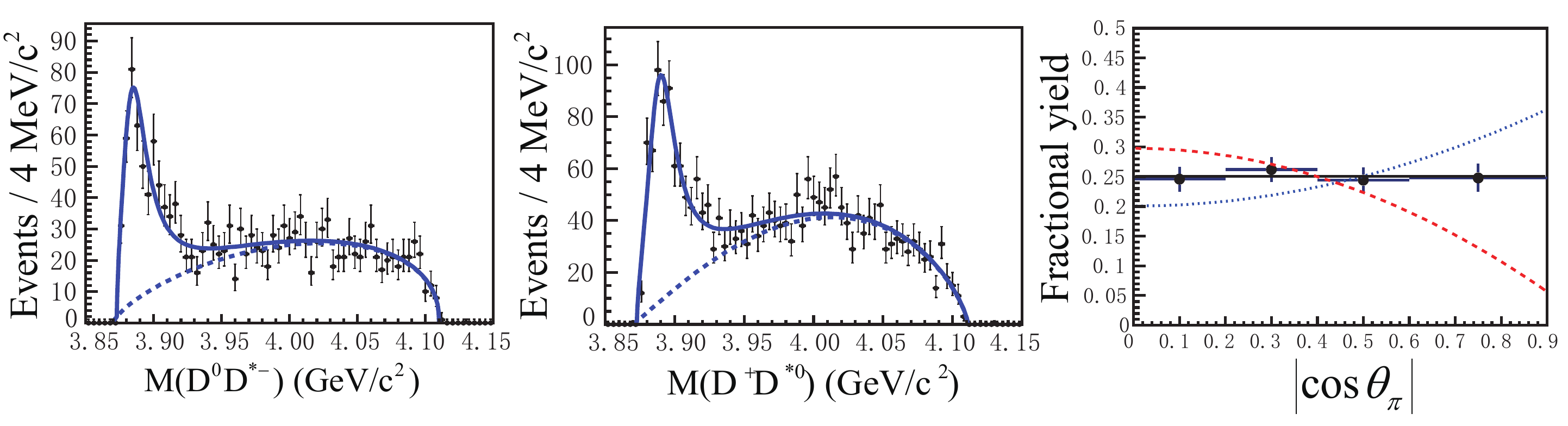}}
\end{center}
\caption{(Color online) The $Z_c(3900)$ structure in the $(D\bar D^*)^\pm$ final
states from BESIII \cite{Ablikim:2013xfr}.}
 \label{Fig:2.5.Zc3900DD}
\end{figure}

BESIII also studied the open-charm decay $e^+e^-\to Y(4260) \to
(D\bar{D}^*)^\pm\pi^\mp$ \cite{Ablikim:2013xfr} and found a charged
structure in the $(D\bar D^*)^\pm$ mass spectrum, as shown in Fig.
\ref{Fig:2.5.Zc3900DD}. BESIII named this structure as the
$Z_c(3885)$ since its measured mass (see Table \ref{Table:2.5.Zc})
was slightly lower than that of the $Z_c(3900)$ measured in the
$J/\psi\pi$ channel by BESIII \cite{Ablikim:2013mio} and Belle
\cite{Liu:2013dau}. However, the measured mass and width of the
$Z_c(3885)$ \cite{Ablikim:2013xfr} were consistent with those of the
$Z_c(3900)$ state obtained by Xiao \textit{et al.}
\cite{Xiao:2013iha}.
{Later in Ref.~\cite{Ablikim:2015swa}, BESIII studied the same process
and improved the statistical significance of the $Z_c(3885)$ signal
to be greater than 10$\sigma$.}
If we consider the $Z_c(3900)$ and $Z_c(3885)$
as the same state, the $Z_c(3900)/Z_c(3885)$ was observed in both
the hidden-charm $J/\psi\pi$ and open-charm $D\bar D^*$ decay
channels.

BESIII also performed the analysis on the angular distribution of
the $\pi Z_c(3885)$ system \cite{Ablikim:2013xfr}. Their results are shown in Fig.
\ref{Fig:2.5.Zc3900DD}, and their data supported the $J^P=1^+$ assignment
and ruled out the $J^P=0^-, 1^-$ possibilities. With the same
spin-parity and similar mass and width, the $Z_c(3900)$ and
$Z_c(3885)$ were probably the same state. Under this assumption, the
ratio of the partial decay width of these two decay modes was
measured as \cite{Ablikim:2013xfr}
\begin{eqnarray}
\frac{\Gamma(Z_c(3885)\to D\bar D^*)}{\Gamma(Z_c(3900)\to
J/\psi\pi)}=(6.2\pm1.1\pm2.7)\, .
\end{eqnarray}
In other words, $D\bar D^*$ was the dominant decay mode of the
$Z_c(3900)/Z_c(3885)$. However, this ratio is still much
smaller than those of the established conventional charmonium states
above the open-charm threshold, such as the $\psi(3770)$ and
$\psi(4040)$:
\begin{eqnarray}
\frac{\Gamma (\psi(3770)\to D\bar{D})}{\Gamma(\psi(3770)\to \pi\pi
J/\psi)}&=&(482\pm 84)~\mbox{\cite{pdg}}\, ,
\\ \nonumber \frac{\Gamma
(\psi(4040)\to D^{(*)}\bar{D}^{(*)})}{\Gamma(\psi(4040)\to \eta
J/\psi)}&=&(192\pm 27)~\mbox{\cite{Ablikim:2012ht}}\, .
\end{eqnarray}
%%%%%%%%%%%%%%%%%%%%%%%%%%%%%%%%%%%%%%%%%%%%%%%%%%%%%%%%%%%%%%%%%%%%%%%%%%%%%%%%%%%%%%%%%%%%%%%%%%%%%%%%%%%%%%%%%%%%%%%%%%%%%%%%%%%%%%%%%%%%%
%%%%%%%%%%%%%%%%%%%%%%%%%%%%%%%%%%%%%%%%%%%%%%%%%%%%%%%%%%%%%%%%%%%%%%%%%%%%%%%%%%%%%%%%%%%%%%%%%%%%%%%%%%%%%%%%%%%%%%%%%%%%%%%%%%%%%%%%%%%%%

\begin{figure}[htbp]
\begin{center}
\scalebox{0.5}{\includegraphics{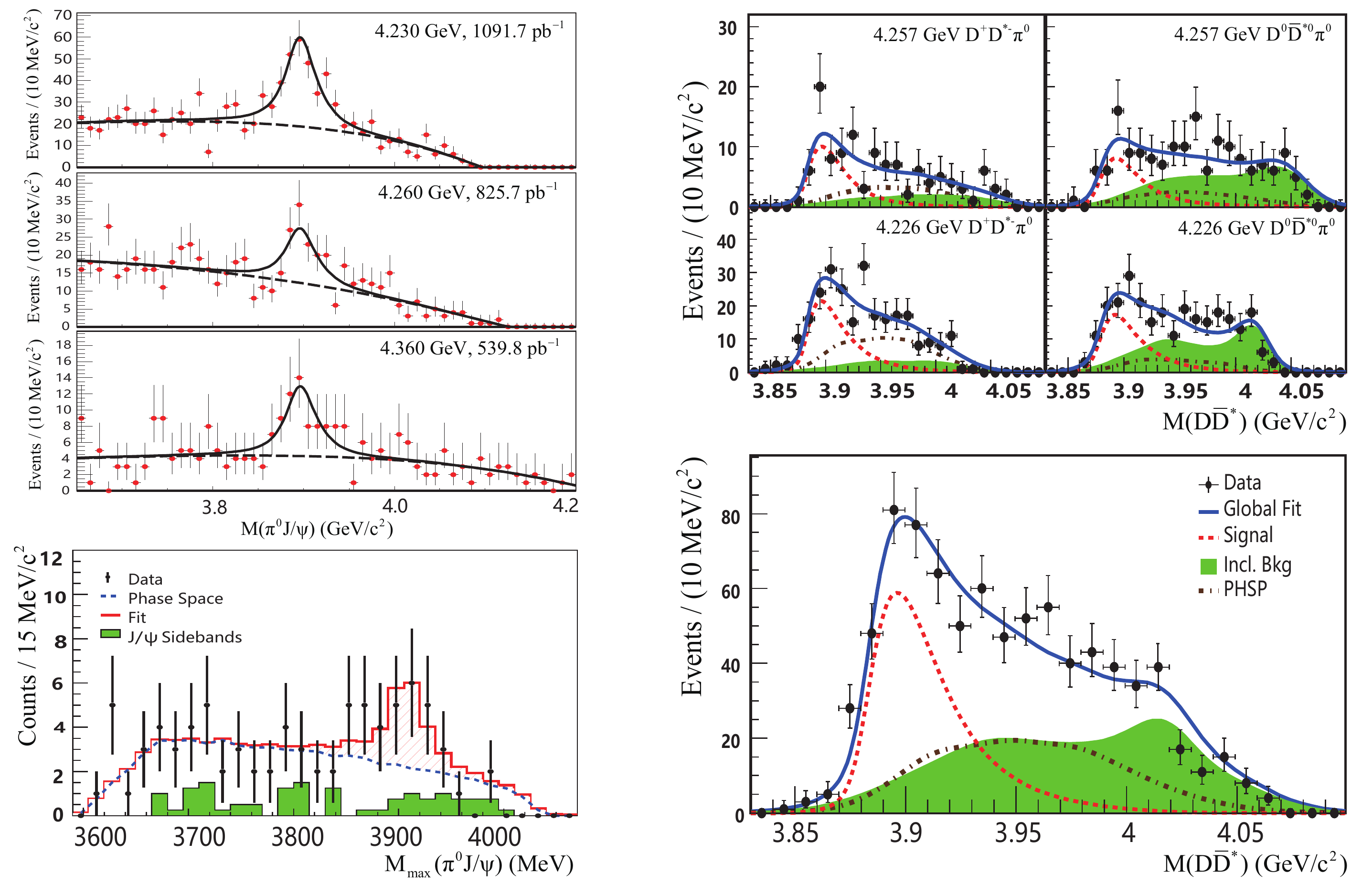}}
\end{center}
\caption{(Color online) The neutral $Z_c(3900)^0$ state in $\pi^0J/\psi$ mass
spectrum from BESIII \cite{Ablikim:2015tbp} and
Xiao \textit{et al.} \cite{Xiao:2013iha} and $(D\bar D^*)^0$ final states from BESIII
\cite{Ablikim:2015gda}.}
 \label{Fig:2.5.Zc03900}
\end{figure}

The neutral partners of the charged $Z_c(3900)^\pm$ and
$Z_c(3885)^\pm$ states were also reported. In Ref.
\cite{Xiao:2013iha}, Xiao \textit{et al.} provided evidence of the
neutral state $Z_c(3900)^0$ decaying into $\pi^0J/\psi$ at a
$3.5\sigma$ significance level, as shown in Fig.
\ref{Fig:2.5.Zc03900}. The mass and decay width of the $Z_c(3900)^0$
were obtained as $(3904\pm9\pm5)$ MeV and $37$ MeV
\cite{Xiao:2013iha}, respectively. Recently, the $Z_c(3900)^0$ was
discovered by BESIII in the $e^+e^-\to \pi^0Z_c(3900)^0 \to \pi^0
\pi^0 J/\psi$ process with a significance of $10.4\sigma$
\cite{Ablikim:2015tbp} (see Fig. \ref{Fig:2.5.Zc03900}). The
measured mass $(3894.8 \pm 2.3 \pm 3.2)$ MeV and width $(29.6 \pm
8.2 \pm 8.2)$ MeV were consistent with the results obtained in Ref.
\cite{Xiao:2013iha}. BESIII also reported a neutral state
$Z_c(3885)^0$ in the $e^+e^-\to (D\bar D^*)^0\pi^0$ process with the
mass $M=(3885.7^{+4.3}_{-5.7} \pm 8.4)$ MeV and width
$\Gamma=(35^{+11}_{-12} \pm 15)$ MeV \cite{Ablikim:2015gda}.
%%%%%%%%%%%%%%%%%%%%%%%%%%%%%%%%%%%%%%%%%%%%%%%%%%%%%%%%%%%%%%%%%%%%%%%%%%%%%%%%%%%%%%%%%%%%%%%%%%%%%%%%%%%%%%%%%%%%%%%%%%%%%%%%%%%%%%%%%%%%%
%%%%%%%%%%%%%%%%%%%%%%%%%%%%%%%%%%%%%%%%%%%%%%%%%%%%%%%%%%%%%%%%%%%%%%%%%%%%%%%%%%%%%%%%%%%%%%%%%%%%%%%%%%%%%%%%%%%%%%%%%%%%%%%%%%%%%%%%%%%%%

Besides, a search for the $Z_c(3900)^\pm$ state in the exclusive
production process by virtual photons was performed in the channel
$Z_c(3900)^\pm\to J/\psi\pi^\pm$ at COMPASS \cite{Adolph:2014hba}.
There was no signal of the exclusive photoproduction of the
$Z_c(3900)^\pm$ state and its decay into $J/\psi\pi^\pm$. The upper
limit of the ratio ${\cal B}(Z_c(3900)^\pm\to J/\psi \pi^{\pm}
)\times \sigma_{\gamma N \to Z_c(3900)^\pm N} /\sigma_{ \gamma N \to
J/\psi N}$ was determined to be $3.7\times10^{-3}$, which suggested
that the hidden-charm decay mode $Z_c(3900)^\pm\to J/\psi \pi^{\pm}$
was not its dominant decay mode.

BESIII studied the $Z_c(3900)^\pm \to\omega\pi^\pm$ decay and found
no significant $Z_c(3900)^\pm$ signals \cite{Ablikim:2015cag}. In
Ref. \cite{Ablikim:2015xfo}, the isospin violating decay $Y(4260)\to
J/\psi\eta\pi^0$ was also studied by BESIII. No signal was observed
and the upper limit of the branching fraction ratio ${\cal
B}(Z_c^0\to J/\psi\eta)/{\cal B}(Z_c^0\to J/\psi\pi^0)$ was
measured to be 0.15 at $\sqrt{s}=4.226$ GeV and 0.65 at
$\sqrt{s}=4.257$ GeV.

%======================================================================================================================
\paragraph{$Z_c(4025)$ and $Z_c(4020)$}\label{Sec:2.5.2}
%======================================================================================================================

\begin{figure}[htbp]
\begin{center}
\scalebox{0.5}{\includegraphics{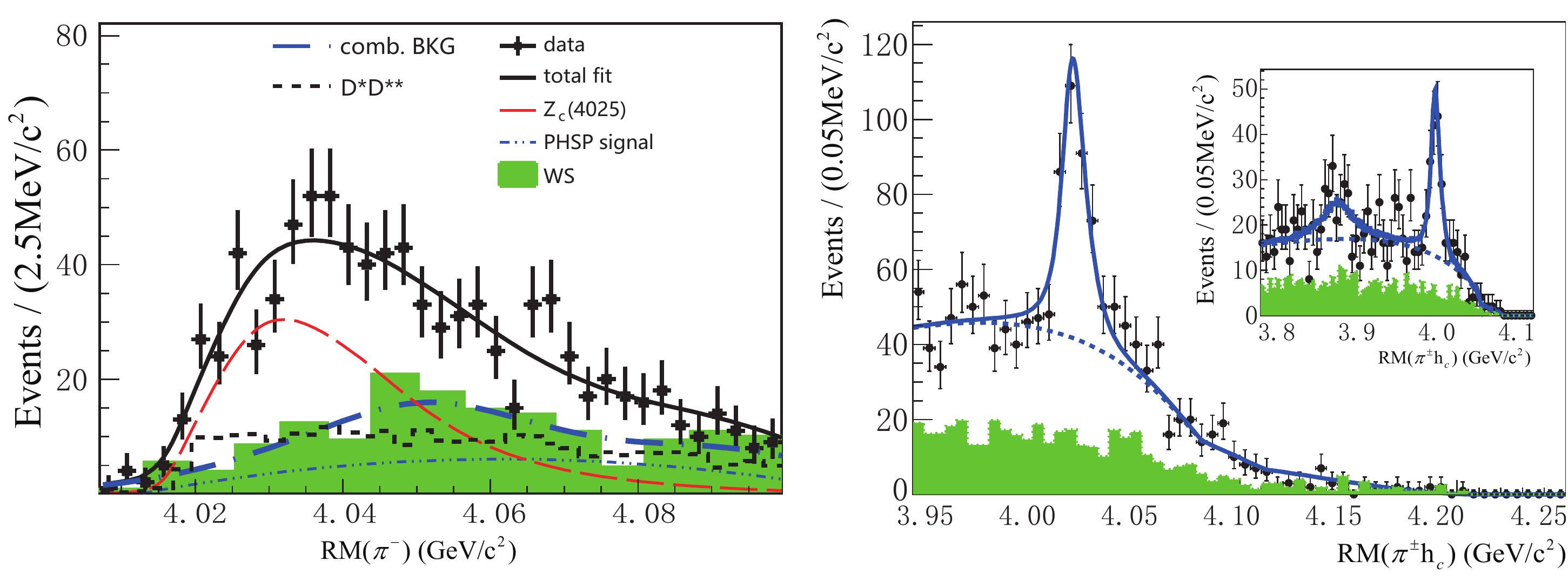}}
\end{center}
\caption{(Color online) The mass peaks of the $Z_c(4025)$ \cite{Ablikim:2013emm}
and $Z_c(4020)$ \cite{Ablikim:2013wzq} resonances in
$(D^*\bar{D}^{*})^\pm$ (left) and $\pi^\pm h_c$ (right) invariant mass
distributions, respectively. The inset in the right panel shows the result of
distributions including  both the $Z_c(4020)$ and $Z_c(3900)$
decaying into $\pi^+ h_c$.}
\label{Fig:2.5.Zc4020}
\end{figure}

The $Z_c(4025)$ state was first observed in the
$(D^*\bar{D}^{*})^\pm $ mass spectrum in the $e^+e^-\to Y(4260) \to
(D^*\bar{D}^{*})^\pm\pi^\mp$ process by BESIII
\cite{Ablikim:2013emm}. Almost at the same time, BESIII reported
another charged charmonium-like structure $Z_c(4020)$ in the
$\pi^\pm h_c$ invariant mass distribution in the process of
$e^+e^-\to Y(4260) \to \pi^- \pi^+ h_c$ \cite{Ablikim:2013wzq}. The
masses and widths of the $Z_c(4025)$ and $Z_c(4020)$ resonances are
collected in Table \ref{Table:2.5.Zc}. As shown in Fig.
\ref{Fig:2.5.Zc4020}, the mass of the $Z_c(4025)$ state is very
close to that of the $Z_c(4020)$ while the $Z_c(4025)$ is much
broader than the $Z_c(4020)$. In general, the resonance parameters
of the $Z_c(4020)$ agree with those of the $Z_c(4025)$ state within
$1.5\sigma$ \cite{Ablikim:2013wzq}. If the $Z_c(4025)$ and
$Z_c(4020)$ are the same state, its quantum number are probably
$I^GJ^P=1^+1^+$ \cite{He:2013nwa}.
%%%%%%%%%%%%%%%%%%%%%%%%%%%%%%%%%%%%%%%%%%%%%%%%%%%%%%%%%%%%%%%%%%%%%%%%%%%%%%%%%%%%%%%%%%%%%%%%%%%%%%%%%%%%%%%%%%%%%%%%%%%%%%%%%%%%%%%%%%%%%
%%%%%%%%%%%%%%%%%%%%%%%%%%%%%%%%%%%%%%%%%%%%%%%%%%%%%%%%%%%%%%%%%%%%%%%%%%%%%%%%%%%%%%%%%%%%%%%%%%%%%%%%%%%%%%%%%%%%%%%%%%%%%%%%%%%%%%%%%%%%%

The $Z_c(3900)^\pm\to h_c\pi^\pm$ process was also included in the
fit, which is shown as the inset in the right panel in Fig. \ref{Fig:2.5.Zc4020}
\cite{Ablikim:2013wzq}. There was a weak signal of the $Z_c(3900)$
with a statistical significance of $2.1\sigma$ in this situation. The
production cross section of $Z_c(3900)^\pm\to \pi^\pm h_c$ was
found to be smaller than 11 pb at $\sqrt{s}=4.26$ GeV at $90\%$
C.L. \cite{Ablikim:2013wzq}. This upper limit was lower than the
production cross section of $Z_c(3900)^\pm\to \pi^\pm J/\psi$
obtained in Ref. \cite{Ablikim:2013mio}, which is about $14$ pb.

\begin{figure}[htbp]
\begin{center}
\scalebox{0.5}{\includegraphics{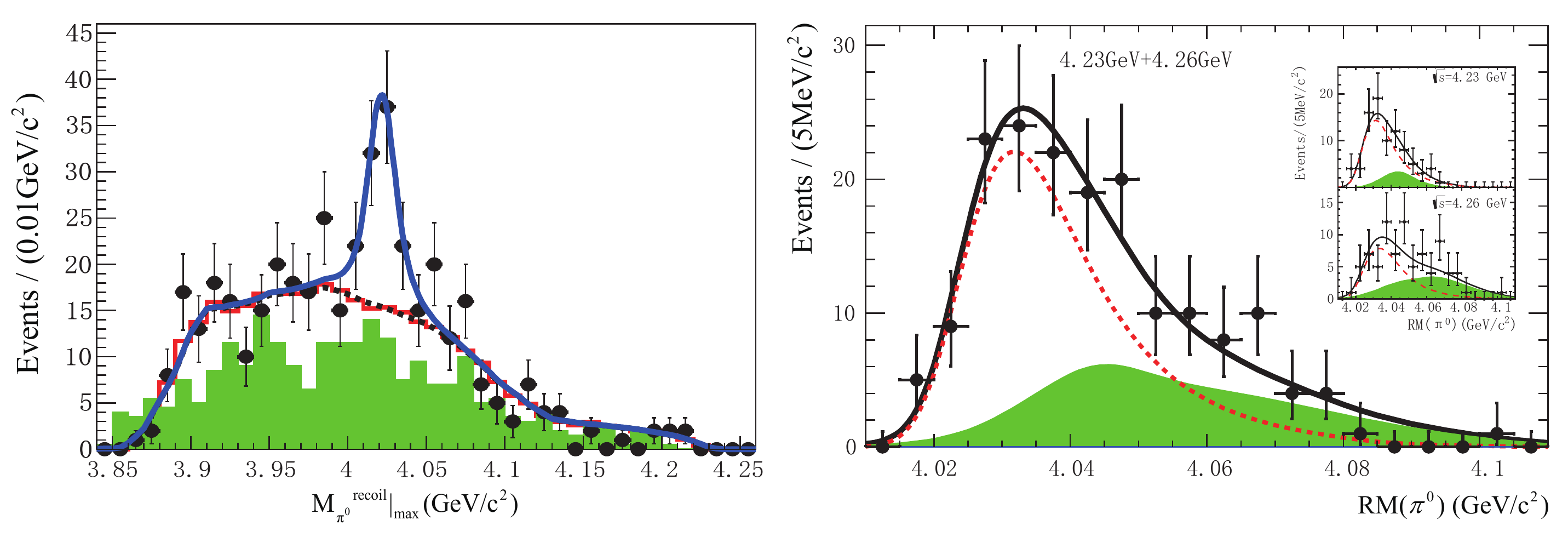}}
\end{center}
\caption{(Color online) The neutral states $Z_c(4020)^0$ and $Z_c(4025)^0$ in the
$\pi^0 h_c$ \cite{Ablikim:2014dxl} and $(D^*\bar D^*)^0$
\cite{Ablikim:2015vvn} invariant mass distributions.
For the $(D^*\bar D^*)^0$ distribution, the main fit is for the
combination of all data while the inset plots are for the two
collision energy.}
 \label{Fig:2.5.Zc04020}
\end{figure}

Recently, the neutral partners of the $Z_c(4025)$ and $Z_c(4020)$
states were also observed by BESIII. In Ref. \cite{Ablikim:2014dxl},
a neutral state $Z_c(4020)^0$ was reported in the $e^+e^-\to \pi^0
\pi^0 h_c$ process. Its production cross section was half of that in
the $e^+e^-\to \pi^+ \pi^- h_c$ process within less than $2\sigma$.
The mass of the $Z_c(4020)^0$ was $(4023.9 \pm 2.2 \pm 3.8)$ MeV,
which was consistent with that of the charged $Z_c^\pm(4020)$ state.
Later, BESIII observed another neutral state $Z_c(4025)^0$ in the
$(D^*\bar D^*)^0$ invariant mass distribution of $e^+e^-\to
\pi^0(D^*\bar D^*)^0$ process at $\sqrt{s}=4.23$ GeV and 4.26 GeV
\cite{Ablikim:2015vvn}. The measured mass and decay width were
$(4025.5^{+2.0}_{-4.7}\pm3.1)$ MeV and $(23.0\pm6.0\pm1.0)$ MeV,
respectively. The production cross section $\sigma(e^{+}e^{-} \to
Z_c(4025)^0\pi^0 \to\pi^0(D^*\bar D^*)^0)$ was measured to be
$(43.4\pm8.0 \pm 5.4)\, \rm{pb}$ at $\sqrt{s}=4.26$ GeV. Thus, the
ratio $\frac{\sigma(e^+e^- \to Z_c(4025)^0\pi^0 \to\pi^0(D^*\bar
D^*)^0)}{\sigma(e^+e^- \to Z_c^+(4025)\pi^- \to\pi^-(D^*\bar
D^*)^+)}\sim 1$ at $\sqrt{s}=4.26$ GeV \cite{Ablikim:2015vvn}. This
result is consistent with the expectation of isospin symmetry. The
mass peaks of the neutral states $Z_c(4020)^0$ and $Z_c(4025)^0$ in
the $\pi^0 h_c$ and $(D^*\bar D^*)^0$ invariant mass distributions
are shown in Fig. \ref{Fig:2.5.Zc04020}.
%%%%%%%%%%%%%%%%%%%%%%%%%%%%%%%%%%%%%%%%%%%%%%%%%%%%%%%%%%%%%%%%%%%%%%%%%%%%%%%%%%%%%%%%%%%%%%%%%%%%%%%%%%%%%%%%%%%%%%%%%%%%%%%%%%%%%%%%%%%%%
%%%%%%%%%%%%%%%%%%%%%%%%%%%%%%%%%%%%%%%%%%%%%%%%%%%%%%%%%%%%%%%%%%%%%%%%%%%%%%%%%%%%%%%%%%%%%%%%%%%%%%%%%%%%%%%%%%%%%%%%%%%%%%%%%%%%%%%%%%%%%

%================================================================================
\subsection{Charged bottomonium-like states $Z_b(10610)$ and $Z_b(10650)$}\label{Sec:2.7}
\begin{figure}[htpb]
\begin{center}
\scalebox{0.5}{\includegraphics{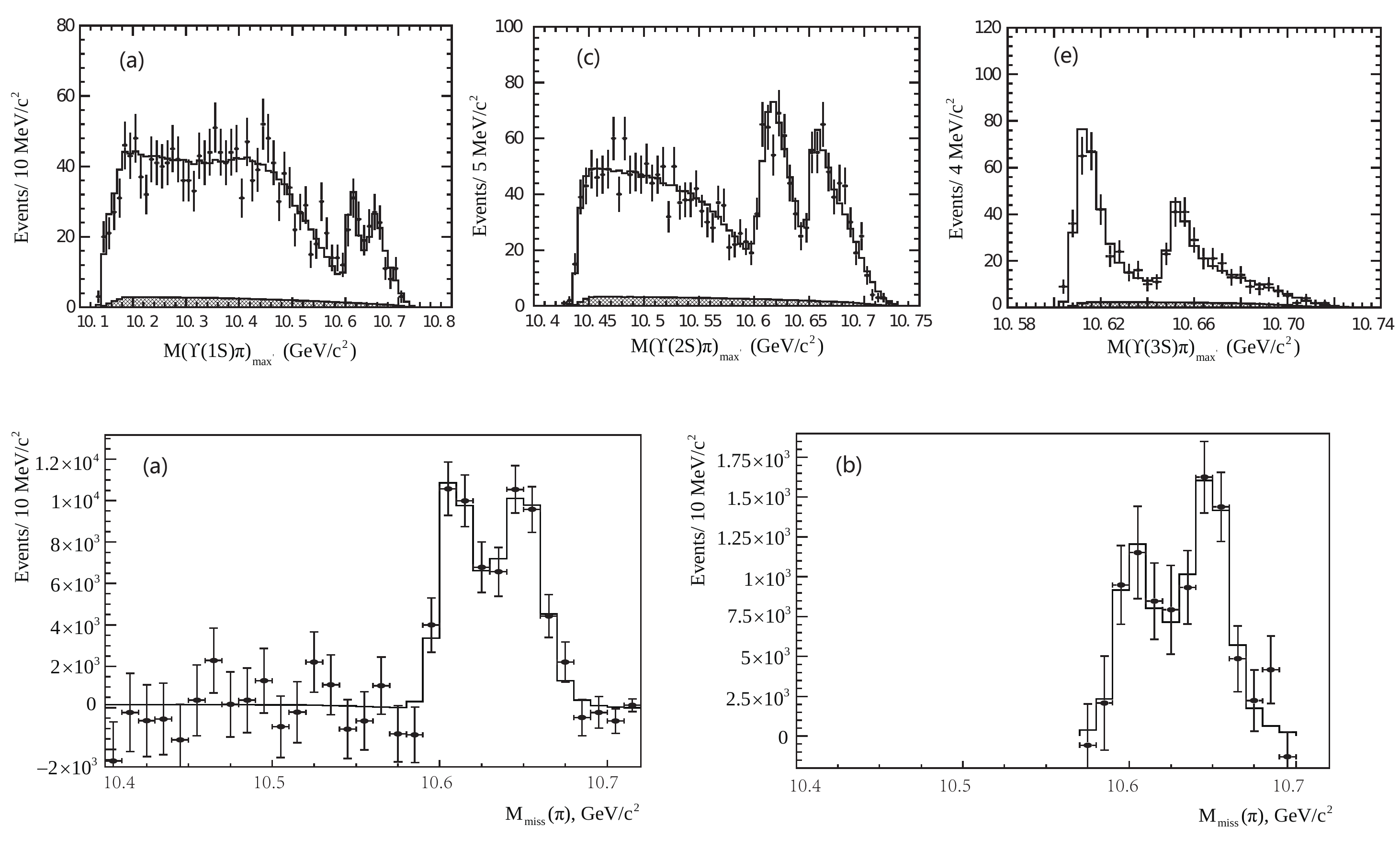}}
\end{center}
\caption{Signals for the $Z_b(10610)$ and $Z_b(10650)$ structures in
$\Upsilon(1S)\pi$, $\Upsilon(2S)\pi$, $\Upsilon(3S)\pi$,
$h_b(1P)\pi$ and (e) $h_b(2P)\pi$ from Belle
\cite{Belle:2011aa}.}
\label{Fig:2.7.ZbU}
\end{figure}

%================================================================================
In 2011, the Belle Collaboration reported two narrow structures in
the invariant mass distributions of the $\pi^\pm\Upsilon(nS)$ $(n=1,
2, 3)$ and $\pi^\pm h_b(mP)$ $(m=1, 2)$ final states in the
processes $\Upsilon(5S)\to \Upsilon(nS)\pi^\pm\pi^\mp$ $(n=1, 2, 3)$
and $\Upsilon(5S)\to h_b(mP)\pi^\pm\pi^\mp$ $(m=1, 2)$
\cite{Belle:2011aa} (see Fig. \ref{Fig:2.7.ZbU}). We collect the
observed channels, masses and widths of these two bottomonium-like
states in Table \ref{Table:2.7.Zb}. The averaged masses and widths
over the five final states are $M_1=(10607.2 \pm 2.0)$ MeV,
$\Gamma_1=(18.4 \pm 2.4)$ MeV for $Z_b(10610)$ and $M_2=(10652.2 \pm
1.5)$ MeV, $\Gamma_2=(11.5 \pm 2.2)$ MeV for $Z_b(10650)$, which are
also listed in the Table. The observed $Z_b(10610)$ and $Z_b(10650)$
states lie slightly above the ${B\bar B^*}$ and ${B^*\bar B^*}$
thresholds, respectively.

Belle also performed the charged pion angular distribution analysis
in Ref. \cite{Collaboration:2011gja}, which favored the $J^P=1^+$
spin-parity assignment for both the $Z_b(10610)$ and $Z_b(10650)$.
Since the initial state $\Upsilon(5S)$ has $I^G=1^-$, the isospin and
$G$-parity of the $Z_b$ states should be $I^G=1^+$ due to the pion
emission \cite{Collaboration:2011gja}. The recent amplitude analysis
of the three-body $\Upsilon(nS)\pi^+\pi^-$ final states confirmed
the $I^GJ^P=1^+1^+$ assignment for both the $Z_b(10610)$ and
$Z_b(10650)$ states \cite{Garmash:2014dhx}.

%%%%%%%%%%%%%%%%%%%%%%%%%%%%%%%%%%%%%%%%%%%%%%%%%%%%%%%%%%%%%%%%%%%%%%%%%%%%%%%%%%%%%%%%%%%%%%%%%%%%%%%%%%%%%%%%%%%%%%%%%%%%%%%%%%%%%%%%%%%%%
\renewcommand{\arraystretch}{1.6}
\begin{table}[htb]
\caption{The resonance parameters for the $Z_b(10610)$ and $Z_b(10650)$
from the $\pi^\pm\Upsilon(nS)$ $(n=1, 2, 3)$ and $\pi^\pm h_b(mP)$
$(m=1, 2)$ decay channels \cite{Belle:2011aa}.   \label{Table:2.7.Zb}}
\begin{center}
\begin{tabular}{c|cc|cc} \toprule[1pt]
                        & \multicolumn{2}{c}{$Z_b(10610)$}                                         &    \multicolumn{2}{c}{$Z_b(10650)$}        \\
  Channels              & Mass (MeV)            & Width (MeV)                   & Mass (MeV)          & Width (MeV) \\
\midrule[1pt]
$\pi^\pm\Upsilon(1S)$   & $10611 \pm 4 \pm 3$      &$22.3 \pm 7.7^{+3.0}_{-4.0}$   & $10657 \pm 6 \pm 3$     & $16.3 \pm 9.8^{+6.0}_{-2.0}$\\
$\pi^\pm\Upsilon(2S)$   & $10609 \pm 2 \pm 3$      &$24.2 \pm 3.1^{+2.0}_{-3.0}$   & $10651 \pm 2 \pm 3$     & $13.3 \pm 3.3^{+4.0}_{-3.0}$\\
$\pi^\pm\Upsilon(3S)$   & $10608 \pm 2 \pm 3$      &$17.6 \pm 3.0 \pm 3.0$         & $10652 \pm 1 \pm 2$     & $8.4  \pm 2.0 \pm 2.0$      \\
$\pi^\pm h_b(1P)$       & $10605 \pm 2^{+3}_{-1}$  &$11.4^{+4.5+2.1}_{-3.9-1.2}$   & $10654 \pm 3^{+1}_{-2}$ & $20.9^{+5.4+2.1}_{-4.7-5.7}$\\
$\pi^\pm h_b(2P)$       & $10599^{+6+5}_{-3-4}$    &$13^{+10+9}_{-8-7}$            & $10651^{+2+3}_{-3-2}$   & $19 \pm 7^{+11}_{-7}$ \\
Averaged                & $10607.2 \pm 2.0$        &$18.4 \pm 2.4$                 & $10652.2 \pm 1.5$       & $11.5 \pm 2.2$ \\
\bottomrule[1pt]
\end{tabular}
\end{center}
\end{table}
%%%%%%%%%%%%%%%%%%%%%%%%%%%%%%%%%%%%%%%%%%%%%%%%%%%%%%%%%%%%%%%%%%%%%%%%%%%%%%%%%%%%%%%%%%%%%%%%%%%%%%%%%%%%%%%%%%%%%%%%%%%%%%%%%%%%%%%%%%%%%
%%%%%%%%%%%%%%%%%%%%%%%%%%%%%%%%%%%%%%%%%%%%%%%%%%%%%%%%%%%%%%%%%%%%%%%%%%%%%%%%%%%%%%%%%%%%%%%%%%%%%%%%%%%%%%%%%%%%%%%%%%%%%%%%%%%%%%%%%%%%%%%%%%%%%%%%%%%%%%%%%%%%%%%%%%%%%%%%%%%%%%%%%%%%%%%%%%%%%%%%%%%%%%%%%%%%%%%%%%%%%%%%%%%%%%%%%%%%%%

\begin{figure}[htbp]
\begin{center}
\scalebox{0.5}{\includegraphics{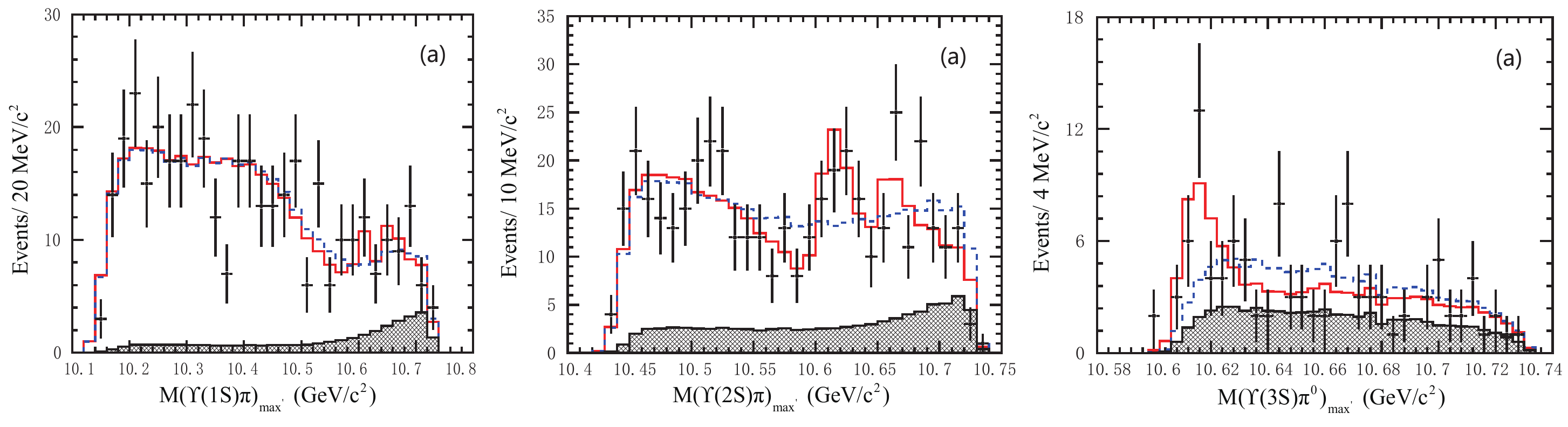}}
\end{center}
\caption{(Color online) The neutral partner the $Z_b(10610)^0$ in the $\Upsilon(nS)\pi^0\,(n=1,2,3)$ final states from Belle \cite{Krokovny:2013mgx}.}
\label{Fig:2.7.Zb0U}
\end{figure}

The neutral $Z_b^0(10610)$ state was also observed soon in a Dalitz
analysis of $\Upsilon(10860)\to\Upsilon(nS)\pi^0\pi^0, (n=1,2, 3)$
decays by the Belle Collaboration \cite{Adachi:2012im,Krokovny:2013mgx}.
The observed mass of the $Z_b(10610)^0$ was $(10609\pm 4\pm 4)$ MeV,
which was consistent with that of the charged $Z_b(10610)^\pm$ state.
There was no significant signal for the $Z_b(10610)^0$ in the $\Upsilon(1S)\pi^0$
final states, as shown in Fig. \ref{Fig:2.7.Zb0U}. However, the present data are
insufficient to observe the neutral partner of the $Z_b(10650)$ in the
$\Upsilon(nS)\pi^0\pi^0$ $(n=1,2,3)$ channels
\cite{Adachi:2012im,Krokovny:2013mgx}.
%%%%%%%%%%%%%%%%%%%%%%%%%%%%%%%%%%%%%%%%%%%%%%%%%%%%%%%%%%%%%%%%%%%%%%%%%%%%%%%%%%%%%%%%%%%%%%%%%%%%%%%%%%%%%%%%%%%%%%%%%%%%%%%%%%%%%%%%%%%%%
%%%%%%%%%%%%%%%%%%%%%%%%%%%%%%%%%%%%%%%%%%%%%%%%%%%%%%%%%%%%%%%%%%%%%%%%%%%%%%%%%%%%%%%%%%%%%%%%%%%%%%%%%%%%%%%%%%%%%%%%%%%%%%%%%%%%%%%%%%%%%

In Ref. \cite{Adachi:2012cx}, Belle observed the open-bottom decay
modes of the $Z_b(10610)$ and $Z_b(10650)$ states via
$\Upsilon(10860)\to Z_b(10610)^\pm\pi^\mp\to [B\bar{B}^*+{\rm
c.c.}]^{\pm}\pi^\mp$ and $\Upsilon(10860)\to
Z_b(10650)^\pm\pi^\mp\to [B^*\bar{B}^*]^\pm\pi^\mp$ processes. Meanwhile, they also studied the
$\Upsilon(10860)\to \Upsilon(nS)\pi^+\pi^-\,(n=1, 2, 3)$ and
$\Upsilon(10860)\to h_b(mP)\pi^+\pi^-\,(m=1, 2)$ decays. They measured
the ratios of the branching fractions \cite{Adachi:2012cx}:

\begin{eqnarray}
\frac{{\cal{B}}(Z_b(10610)\to B\bar B^*)}{\sum_{n=1, 2, 3}{{\cal{B}}(Z_b(10610)\to\Upsilon(nS)\pi)}+
                          \sum_{m=1, 2}{Z_b(10610)\to h_b(mP)\pi}} =
6.2\pm0.7\pm1.3^{+0.0}_{-1.8}\, ,
\end{eqnarray}
and
\begin{equation}
\frac{{\cal{B}}(Z_b(10650)\to B^*\bar B^*)}{\sum_{n=1, 2,
3}{{\cal{B}}(Z_b(10650)\to\Upsilon(nS)\pi)}+
                          \sum_{m=1, 2}{Z_b(10650)\to h_b(mP)\pi}} =
2.8\pm0.4\pm0.6^{+0.0}_{-0.4}\, , \label{eq:zbhfrac}
\end{equation}
which indicated that the open-bottom decays $Z_b(10610)\to B\bar B^*$
and $Z_b(10610)\to B^*\bar B^*$ were the dominant decay modes for the
$Z_b(10610)$ and $Z_b(10650)$ respectively.

%%%%%%%%%%%%%%%%%%%%%%%%%%%%%%%%%%%%%%%%%%%%%%%%%%%%%%%%%%%%%%%%%%%%%%%%%%%%%%%%%%%%%%%%%%%%%%%%%%%%%%%%%%%%%%%%%%%%%%%%%%%%%%%%%%%%%%%%%%%%%
%\begin{figure}[htbp]
%\begin{center}
%\scalebox{0.45}{\includegraphics{figs/27Zb12BB.pdf}}
%\end{center}
%\caption{The $Z_b(10610)$ and $Z_b(10650)$ mass peaks in (a)
%$\Upsilon(10860)\to B\bar B^*\pi$ and (b) $\Upsilon(10860)\to B\bar
%B^*\pi$, respectively \cite{Adachi:2012cx}.}
%\label{Fig:2.7.ZbBB}
%\end{figure}
%%%%%%%%%%%%%%%%%%%%%%%%%%%%%%%%%%%%%%%%%%%%%%%%%%%%%%%%%%%%%%%%%%%%%%%%%%%%%%%%%%%%%%%%%%%%%%%%%%%%%%%%%%%%%%%%%%%%%%%%%%%%%%%%%%%%%%%%%%%%%

Recently, the Belle Collaboration studied the $\Upsilon(11020)\to
h_b(nP)\pi^+\pi^-\,(n=1, 2)$ transitions and found evidence for the
$\Upsilon(11020)$ decays into the charged bottomonium-like
$Z_b(10610)$ and/or $Z_b(10650)$ states \cite{Abdesselam:2015zza}.

There were some other efforts to search for the new bottomonium-like
states. In Ref. \cite{He:2014sqj}, Belle studied the $X_b\to
\omega\Upsilon(1S)$ process and found no significant
bottomonium-like signal decaying into $\omega\Upsilon(1S)$ with a
mass between $10.55$ GeV and $10.65$ GeV. The ATLAS Collaboration
tried to search for the bottomonium-like states in the $X_b\to
\Upsilon(1S)\pi^+\pi^-$ channel near $\sqrt{s}=8$ TeV at LHC
\cite{Aad:2014ama}. However, they found no evidence for any new
structure in the mass ranges $10.05-10.31$ GeV and $10.40-11.00$
GeV.

%
%=====================================================================================
%=====================================================================================

\subsection{The hidden-charm pentaquark states observed by LHCb}
\label{Sect.2.6}
%=====================================================================================
%=====================================================================================
%

Recently, the LHCb Collaboration reported the observation of two
exotic structures, denoted as the $P_c(4380)^+$ and $P_c(4450)^+$,
in the $J/\psi p$ invariant mass spectrum of the $\Lambda_b^0 \to
J/\psi K^- p$ decay~\cite{Aaij:2015tga}. They used the data of $pp$
collisions corresponding to 1 fb$^{-1}$ of integrated luminosity at
7 TeV, and 2 fb$^{-1}$ at 8 TeV. The significance of the lower mass
and higher mass states is $9\sigma$ and $12\sigma$, respectively.
Both resonances decay into the $J/\psi p$ final states. They must
have minimal quark contents $c\bar cuud$, and thus are good
candidates of exotic hidden-charm pentaquarks. Further experimental
research should be pursued to confirm these pentaquark states.

\begin{figure}[hbtp]
\begin{center}
\includegraphics[width=13cm]{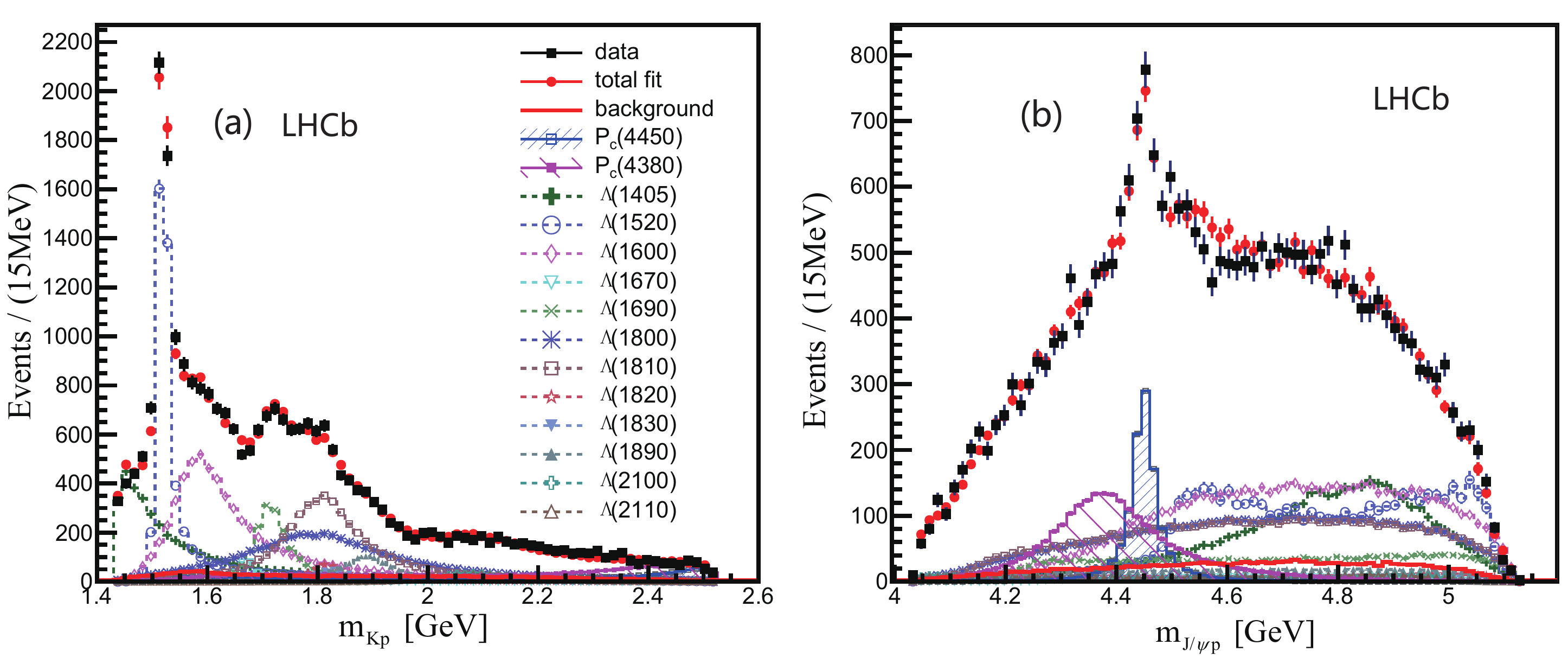}
\caption{(Color online) The $K^- p$ (left) and $J/\psi p$ (right) invariant mass
spectrum of $\Lambda_b^0 \to J/\psi K^- p$ from LHCb
\cite{Aaij:2015tga}, where the background has been subtracted. Fit
projections for the reduced $\Lambda^*$ model with two $P^+_c$
states are also shown.} \label{Fig:2.6.Pc}
\end{center}
\end{figure}

The LHCb Collaboration used an amplitude analysis of the three-body
final-state, and extracted the masses and widths of the
$P_c(4380)^+$ and $P_c(4450)^+$ to be
\begin{eqnarray}
\nonumber M_{P_c(4380)^+} &=& (4380 \pm 8 \pm 29)\, {\rm MeV} \, ,
\\ \Gamma_{P_c(4380)^+} &=& (205 \pm 18 \pm 86) \,{\rm MeV} \, ,
\\ \nonumber M_{P_c(4450)^+} &=& (4449.8 \pm 1.7 \pm 2.5)\, {\rm MeV} \, ,
\\ \nonumber \Gamma_{P_c(4450)^+} &=& (39 \pm 5 \pm 19) \,{\rm MeV} \, .
\end{eqnarray}
The $P_c(4380)^+$ and $P_c(4450)^+$ states preferred the $J^P$
assignments $(3/2^-, 5/2^+)$, but LHCb also said that ``Other combinations are
less likely''~\cite{Aaij:2015tga}, i.e., the $-2\ln{\cal L}$ values
were only 1 unit better than those of the parity reversed
combination $(3/2^+, 5/2^-)$, and $2.3^2$ units better than those of
the $(5/2^+, 3/2^-)$ assignment. All the other combinations from
$1/2^\pm$ through $7/2^\pm$ were tested and ruled out.

\begin{figure}[hbtp]
\begin{center}
\includegraphics[width=15cm]{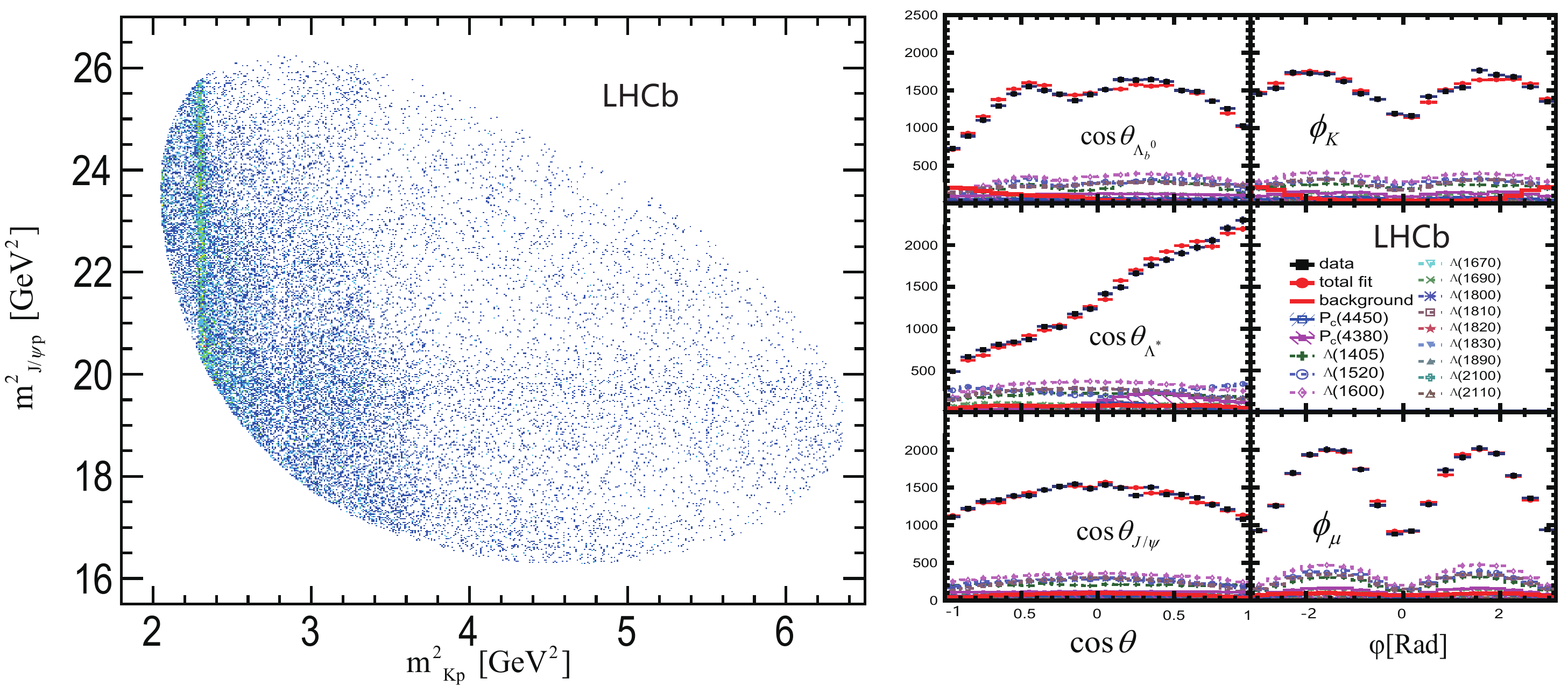}
\caption{(Color online) Left: the Dalitz plot using the $K^-p$ and $J/\psi p$
invariant masses-squared from LHCb \cite{Aaij:2015tga}. Right:
various decay angular distributions for the best fit with the two
$P^+_c$ states.} \label{Fig:2.6.PcDLZ}
\end{center}
\end{figure}

Their best fit projections with these two $P^+_c$ states are shown
in Fig.~\ref{Fig:2.6.Pc}, where the fractions of the total sample
due to the $P_c(4380)^+$ and $P_c(4450)^+$ are $(8.4 \pm 0.7 \pm
4.2)\%$ and $(4.1 \pm 0.5 \pm 1.1)\%$, respectively. For comparison,
the fractions of the $\Lambda(1405)$ and $\Lambda(1520)$ are $(15
\pm 1 \pm 6)\%$ and $(19 \pm 1\pm 4)\%$, respectively. The Dalitz
plot and the decay angular distributions for this best fit are shown
in Fig.~\ref{Fig:2.6.PcDLZ}.

The LHCb Collaboration reported the branching fraction of the
$\Lambda^0_b \to J/\psi K^- p$ decay recently~\cite{Aaij:2015fea}:
\begin{eqnarray}
{\cal B}(\Lambda_0^b \to J/\psi K^- p) = (3.04 \pm 0.04 \pm 0.06
\pm 0.33 {^{+0.43}_{-0.27}}) \times 10^{-4} \, .
\end{eqnarray}
Hence, the product branching ratios of the $P_c(4380)^+$ and
$P_c(4450)^+$ were determined to be
\begin{eqnarray}
{\cal B}(\Lambda_0^b \to K^- P_c(4380)^+) \times {\cal
B}(P_c(4380)^+ \to J/\psi p) &=& (2.56 \pm 0.22 \pm 1.28
{^{+0.46}_{-0.36}}) \times 10^{-5} \, ,
\\ {\cal B}(\Lambda_0^b \to K^- P_c(4450)^+) \times {\cal B}(P_c(4450)^+ \to J/\psi p) &=& (1.25 \pm 0.15 \pm 0.33 {^{+0.22}_{-0.18}}) \times 10^{-5} \, .
\end{eqnarray}

%%-------------Section 3-----------------------------------
\section{Theoretical interpretations of the hidden-charm pentaquark states}
\label{sect:3}

At the birth of the quark model (QM), the multiquark states with
configurations like $(qq\bar{q}\bar{q})$ and $(qqqq\bar{q})$ were
proposed together with the conventional mesons and baryons
\cite{GellMann:1964nj,Zweig:1964jf}. In general, the possible hadron
configurations include
\begin{eqnarray*}
\mathrm{Hadron}\left\{
  \begin{array}{cccccc}
  \mathrm{Conventional\,\, QM\,\, states}&\left\{
  \begin{array}{cccccc}
    \mathrm{Meson}: & q\bar{q},\quad Q\bar{q},\quad Q\bar{Q}\\
    \mathrm{Baryon}: & qqq,\quad Qqq,\quad QQq,\quad...\\
    \end{array}
\right. \\
    \mathrm{Exotic\,\, states}&\left\{
  \begin{array}{cccccc}
  \mathrm{Molecular\,\,state}\\
   \mathrm{Hybrid \,\,meson}\\
    \mathrm{Glueball}\\
     \mathrm{Tetraquark}\\
      \mathrm{Pentaquark}\\
    ...
  \end{array}
\right.
  &
  \end{array}
\right. ,
\end{eqnarray*}
where $q$ and $Q$ denote the light ($u,d,s$) and heavy ($c,b$)
quarks, respectively. Among the above configurations, the molecular
states, hybrid mesons, glueballs, tetraquarks, and pentaquarks etc,
are exotic states within the context of the quark model as
emphasized in the introduction. Exploration of these exotic states
has been one of the central topics of hadron physics in the past
several decades.

\subsection{The molecular scheme}
\label{sect:3.1}

Among these exotic states, the molecular states, which are composed of two
color-singlet hadrons, have received
extensive attention. The deuteron is a very typical example of the molecular
state, which is a loosely bound state of a proton and a neutron with a
binding energy around $2.225$ MeV only. Sometimes the molecular
states are also denoted as the multiquark states in literatures.
However, they are loosely bound by the color-singlet meson exchange
force with the binding energy around several MeV to several tens MeV
only. In contrast, the ``genuine" multiquark states are confined
within one box via the color force. In some phenomenological
models, their building blocks are colored objects such as
quarks, diquarks and triquarks etc.

Forty years ago, Voloshin and Okun investigated the interaction of a
pair of charmed mesons and the possible molecular states
\cite{Voloshin:1976ap}. de Rujula, Georgi, and Glashow studied the
possibility of the $\psi(4040)$ as a $D^*\bar{D}^*$ molecular charmonium
\cite{DeRujula:1976qd}. With the quark-pion interaction model,
T\"ornqvist calculated the possible $D\bar{D}^*$ and $D^*\bar{D}^*$
molecular states in Refs. \cite{Tornqvist:1993vu,Tornqvist:1993ng}.
Later, Dubynskiy and Voloshin also indicated the existence of a
possible resonance around the $D\bar{D}^*$ threshold
\cite{Voloshin:2006pz,Dubynskiy:2006sg}.

Lipkin discussed the molecular anticharmed strange pentaquark
$P_{\bar{c}s}$ composed of a nucleon and ${\bar D}_s$ meson with a
simplified version of the color-spin hyperfine interaction model
\cite{Lipkin:1987sk}. In the light quark sector, Weinstein and Isgur
proposed the $f_0(980)$ and $a_0(980)$ as the $K\bar{K}$ molecular
states in Refs.
\cite{Weinstein:1982gc,Weinstein:1983gd,Weinstein:1990gu}. However,
their partner states $\sigma$ and $\kappa$ within the nonet can not
be explained within this molecular scheme.
The advent of the chiral unitary approach has brought new light into 
this issue and all those states are generated from a coupled channels 
unitary approach based on the interaction provided by the chiral 
Lagrangians~\cite{Oller:1998hw,Pelaez:2015qba}.

Unfortunately, the early-stage theoretical efforts on the
hidden-charm molecular states were not supported by the subsequent
experimental progress. Before 2003, all the experimental observed
charmonium states could be accommodated within the quark model
easily. In fact, there was no need to introduce the molecular
picture into the charmonium spectroscopy at that time.

The story of the molecular states changed dramatically in 2003,
which is the renaissance year of the hadron spectroscopy. Since
2003, more and more new light hadron states and charmonium-like
states were reported experimentally. Some of them lie close to the
threshold of two mesons and are considered as good candidates of the
molecular states. In the following, we list several typical
examples:

\begin{itemize}
\item In 2003, the $X(1835)$ was observed by the BESII Collaboration
in the $p\bar{p}$ mass spectrum of the $J/\psi\to \gamma p\bar{p}$
radiative decay \cite{Bai:2003sw}, with mass
$M=(1859{^{+3}_{-10}}{^{+5}_{-25}})$ MeV and width $\Gamma<30$ MeV. Since this
enhancement structure is close to the $p\bar{p}$ threshold,
the $X(1835)$ was suggested to be a baryonium state in Refs.
\cite{Datta:2003iy,Gao:2003ka,Liu:2004er}.

\item In 2003, the charmonium-like state $X(3872)$ was announced
by the Belle Collaboration \cite{Choi:2003ue} (see Sec.
\ref{Sect:2.1.1} for its experimental information). The $X(3872)$
sits on the $D\bar{D}^*$ threshold, which inspired the $D\bar{D}^*$
molecular explanation \cite{Swanson:2003tb}. Over the past 13 years,
there have been heated discussions on this issue. In
Sec.~\ref{Sect:4.5}, we will introduce the current status of the
$X(3872)$ in detail.

\item BaBar observed a narrow state $D_{s0}(2317)$
in the $D_s^+\pi^0$ invariant mass spectrum from the $e^+e^-$
annihilation \cite{Aubert:2003fg}. The observation of the $D_{s0}(2317)$
also stimulated discussions of the $D{K}$ molecular state
\cite{Barnes:2003dj}.

\item The $Y(3930)$ and $Y(4140)$ were reported by the Belle
\cite{Abe:2004zs} and CDF~\cite{Aaltonen:2009tz} collaborations (see
Sec. \ref{Sect:2.1.2} and Sec. \ref{Sect:2.1.3} for more details), respectively.
Due to their similarity and proximity to the two-meson thresholds,
the $Y(3930)$ and $Y(4140)$ were proposed as the $D\bar{D}^*$
and $D_s\bar{D}_s^*$ molecular states in Ref. \cite{Liu:2009ei}, respectively.

\item{As the first observed charged charmonium-like state, the $Z^+(4430)$
was once suggested to be the $D_1(D_1^\prime) \bar D^*$ molecule in Ref.
\cite{Meng:2007fu} and was reexamined in Refs. \cite{Liu:2007bf,Liu:2008xz}}
by a dynamical calculation. Later, more charged states $Z_b(10610)$,
$Z_b(10650)$ and $Z_c(3900)$ were reported. Lying very close to the
$B\bar{B}^*$ and $B^*\bar{B}^*$ thresholds, the $Z_b(10610)$ and
$Z_b(10650)$ were proposed as the $B\bar{B}^*$ and $B^*\bar{B}^*$
molecular states in Ref. \cite{Sun:2011uh}, respectively. Their hidden-charm
partners were also predicted, which can be related to the $Z_c(3900)$ \cite{Sun:2012zzd}.

\end{itemize}

\subsubsection{The deuteron as a hadronic molecule}

To date, the deuteron is the only well-established hadronic
molecular state. As a loosely bound state composed of a proton and a
neutron, the deuteron is the only bound state of the $NN$ system
with $J^P=1^+$ and a binding energy $E=2.225$ MeV. The deuteron is a
very typical molecular system, where the internal motion of nucleons
is governed by the non-relativistic nuclear force. By solving the
Schr\"odinger equation, one can get useful information of the
deuteron.

Yukawa first proposed that the nucleon-nucleon interaction is mediated
through the exchange of the $\pi$ meson, which contributes to the
long-range part of the nuclear force. The effective $\pi NN$
interaction Lagrangian reads
\begin{eqnarray}
\mathcal{L}=g_{NN\pi} \bar{\psi}i\gamma_5\boldsymbol{\tau} \psi\cdot
\boldsymbol{\pi},\label{Eq.3.1.nnpi}
\end{eqnarray}
with $\psi=(p,n)$,  $\boldsymbol{\pi}=(\pi_1,\pi_2,\pi_3)$ in the
isospin space, and %, and $\pi^\pm=(\pi_1\mp i \pi_2)/\sqrt{2}$ and $\pi_3=\pi^0$.
$g_{NN\pi}$ the coupling constant. With Eq.
(\ref{Eq.3.1.nnpi}), the non-relativistic nucleon-nucleon potential via
$\pi$ meson exchange can be obtained as
\begin{eqnarray}
V_{\pi}=\frac{g_{NN\pi}^2}{4\pi}\frac{m_{\pi}^2}{12m_N^2}(\bm{\tau_1}\cdot\bm{\tau_2})\left\{
  \bm{\sigma_1}\cdot\bm{\sigma_2}
  +\left[\frac{3(\bm{\sigma_1}\cdot\bm{r})(\bm{\sigma_2}\cdot\bm{r})}{r^2}
  -\bm{\sigma_1}\cdot\bm{\sigma_2}\right]
  \left[1+\frac{3}{m_{\pi}r}+\frac{3}{m_{\pi}^2r^2}\right]\right\}\frac{e^{-m_{\pi}r}}{r}\, ,
\end{eqnarray}
where $m_N$ and $m_\pi$ denote the masses of nucleon and $\pi$
meson, respectively.

There exists strong attraction between two nucleons in the medium
range. Such attraction can be reproduced well through the scalar
meson $\sigma$ exchange with a mass around 600 MeV, which mimics the
correlated two-pion exchange in the modern version of the nuclear
force based on the chiral perturbation theory~\cite{Oset:2000gn}. The short-range
nuclear force is strongly repulsive. The repulsion is described by
the exchange of the vector mesons $\rho$ and $\omega$, which play
the same role as the multiple pion exchange and the
low-energy-constant contributions in the chiral perturbation theory.
The meson exchange model is the basis of the Nijimegen potential and
Bonn potential.

The deuteron is a very shallow bound state with a large spatial
distribution. Its radial wave function extends to 2 fm. In fact, a
small binding energy and a large radius are key features of the
hadronic molecular states. The long-range attraction through the
pion exchange, the S-wave and D-wave channel coupling, the tensor
force and short-range repulsion work together to form the extremely
loosely bound deuteron. This is an important lesson we learn from
the deuteron, which should guide us in the exploration of the
hidden-charm hadronic molecules. 

An interesting test to distinguish the deuteron as a proton neutron 
bound state from a more elementary structure was provided by 
Ref.~\cite{Weinberg:1965zz}. This compositeness condition has been applied 
to claim other molecular states~\cite{Baru:2003qq} and generalized to 
coupled channels in Refs.~\cite{Hyodo:2013nka,Gamermann:2009uq} and to 
higher partial waves in Ref.~\cite{Aceti:2012dd}.

\subsubsection{The meson exchange model}

\begin{table}[htb]
\caption{Some new hadron states which are close to the two-hadron
thresholds. \label{Table.3.1.threshold}}
\begin{center}
\begin{tabular}{cccccccc}
\toprule[1pt]
Observation & Threshold &Observation & Threshold\\
\midrule[1pt]
$X(1860)$ \cite{Bai:2003sw}&$p\bar{p}$&$D_{s0}(2317)$ \cite{Aubert:2003fg}&$DK$\\
$D_{s1}(2460)$ \cite{Besson:2003cp}&$D^*K$&$X(3872)$ \cite{Choi:2003ue}&$D^*D$\\
$Y(3940)$ \cite{Abe:2004zs}&$D^*D^*$&$Y(4140)$ \cite{Aaltonen:2009tz}&$D_s^*D_s^*$\\
$Y(4274)$ \cite{Aaltonen:2011at}&$D_{s0}(2317)D$&$Y(4630)$ \cite{Pakhlova:2008vn}&$\Lambda_c\Lambda_c$\\
$Z^+(4430)$&$D_1D^*/D_1^\prime D^*$&$Z^{+}(4250)$ \cite{Mizuk:2008me}&$D_1D/D_0D^*$\\
$\Lambda_c(2940)$ \cite{Aubert:2006sp}&$D^*N$&$\Sigma_c(2800)$ \cite{Mizuk:2004yu}&$DN$\\
\bottomrule[1pt]
\end{tabular}
\end{center}
\end{table}

Since 2003, many enhancement structures near the two-hadron
thresholds have been reported as shown in Table \ref{Table.3.1.threshold}.
The proximity of their masses to the thresholds inspired molecular
explanations of these structures
\cite{Liu:2004er,He:2006is,Liu:2007ez,Liu:2007bf,Liu:2007fe,Tornqvist:2003na,Swanson:2004pp,Liu:2008fh,
Close:2009ag,Close:2010wq,Lee:2009hy,Xu:2010fc,
Liu:2008du,Liu:2008xz,
Liu:2008tn,Liu:2009ei,Hu:2010fg,Shen:2010ky,He:2010zq,Liu:2010hf,Liu:2008mi,Liu:2008qb,
Liu:2009wb,Ding:2007ar,Ding:2008mp,Ding:2008gr,Ding:2009zq,Lee:2011rka,Chen:2011cta}.
With the refinement of the meson exchange model, many subtle aspects
of this framework were investigated, such as the S-D wave mixing
effect \cite{Tornqvist:1993vu,Swanson:2003tb}, coupled-channel
effect \cite{Li:2012bt,Li:2012cs,Li:2012ss}, and recoil correction
\cite{Zhao:2014gqa,Zhao:2015mga}.

In the deduction of the effective potential of the molecular system,
one first derives the relativistic scattering amplitude at the tree
level
\begin{eqnarray}
\langle f|S|i \rangle = \delta_{fi}+i\langle
f|T|i\rangle=\delta_{fi}+i(2\pi)^4\delta^4(p_f-p_i)\mathcal{M}_{fi}\, ,
\end{eqnarray}
where $T$ is the interaction part of the $S$ matrix and
$\mathcal{M}$ denotes the invariant matrix element. After applying
the Bonn approximation to the Lippmann-Schwinger equation, the $S$
matrix reads
\begin{eqnarray}
\langle f|S|i\rangle =\delta_{fi}-i2\pi\delta(E_f-E_i)V_{fi}\, ,
\end{eqnarray}
where $V_{fi}$ is the effective potential in the momentum space.
Considering the different normalization conventions adopted for the
scattering amplitude $\mathcal{M}_{fi}$ and the $T$-matrix $T_{fi}$
and $V_{fi}$, the scattering amplitude $\mathcal{M}_{fi}$ can be
related to the corresponding effective potential in the momentum
space $V(\mathbf{q})$ \cite{Berestetsky:1982aq}
\begin{eqnarray}
V_{fi}(\mathbf{q})=-\frac{\mathcal{M}_{fi}}{\sqrt{\Pi_f 2 p_f^0
\Pi_i 2p_i^0}}\approx -\frac{\cal{M}_{fi}}{\sqrt{\Pi_f 2 m_f^0 \Pi_i
2m_i^0}}\, ,
\end{eqnarray}
where $p_{f(i)}$ and $m_{f(i)}$ denote the four-momentum and mass
of the final (initial) state, respectively.

Generally, one also needs to introduce the form factor in each
interaction vertex, which reflects the off-shell effect of the
exchanged meson and the structure effect, because the components of
the molecular state and exchanged mesons are not elementary
particles. Although various form factors were adopted in dealing
with different systems
\cite{Tornqvist:1993ng,Close:2010wq,Machleidt:2000ge,Machleidt:1987hj},
we take the simple monopole form factor as an example
\begin{eqnarray}
F(q)=\frac{\Lambda^2-m_E^2}{\Lambda^2-q^2}\, ,
\end{eqnarray}
where $m_E$ and $q$ denote the mass and four-momentum of the exchanged
meson, respectively, and $\Lambda$ is a cutoff.

As $q^2\to 0$ and $\Lambda\gg m_E$, the form factor approaches to the
unity. As $q^2\to \infty$, the form factor approaches to zero. Within
the framework of the meson exchange model, a constituent hadron
is treated as a whole. Its inner structure should not be explored
by the exchanged meson. The large momentum contribution from the
meson exchange should be suppressed. Otherwise, such a formalism is
not self-consistent. In other words, the form factor is introduced
to cut off the ultraviolet contribution \cite{Liu:2008xz}.

By performing the Fourier transformation to $V_{fi}(\mathbf{q})$,
one obtains the effective potential $V_{fi}(\bf r)$ in the coordinate
space, which can be applied to search for the bound state
solution by solving the Schr\"odinger equation.

\subsubsection{Predictions for the hidden-charm pentaquarks before LHCb's discovery}
\label{sect:3.1.2}

Before LHCb's discovery of the $P_c(4380)$ and
$P_c(4450)$~\cite{Aaij:2015tga}, the possible hidden-charm molecular baryons composed of an S-wave anti-charmed meson
and an S-wave charmed baryon were studied extensively in the framework of one boson exchange (OBE)
model in 2011, where the existence of hidden-charm pentaquarks were predicted \cite{Yang:2011wz}.
We need to specify that the interaction between various charmed mesons and
charmed baryons was first studied within the framework of the coupled channel unitary approach 
with the local hidden gauge formalism~\cite{Wu:2010jy,Wu:2010vk}, and
several meson-baryon dynamically generated narrow $N^*$ and $\Lambda^*$ resonances were predicted
with mass above 4 GeV and width smaller than 100 MeV.
In this subsection, we first introduce the prediction of the OPE model, and
we shall detailly review the prediction of the channel unitary approach in Sec.~\ref{Sect.3.2}.

\begin{figure}[htb]
\begin{center}
\scalebox{0.53}{\includegraphics{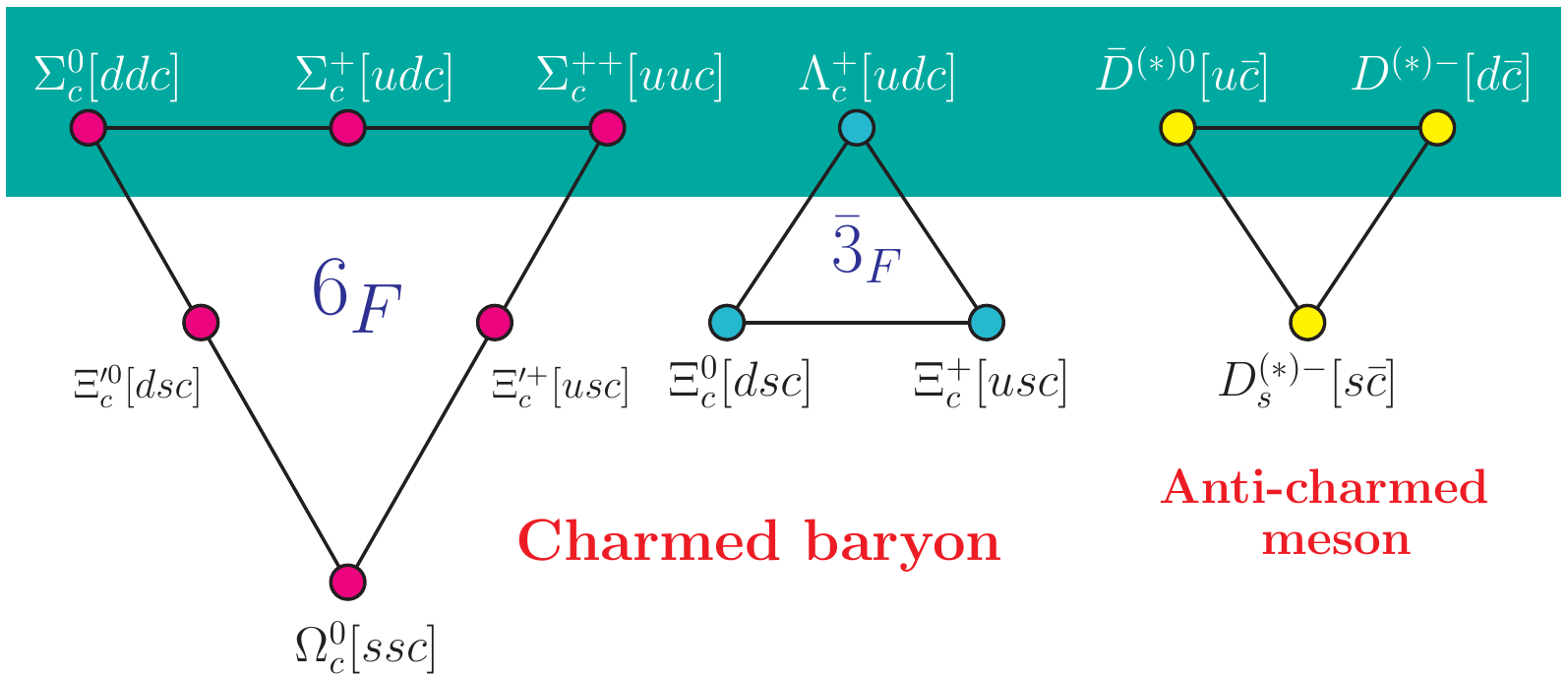}}
\caption{(Color online) The S-wave charmed baryons with $J^P=1/2^+$ and the S-wave
anti-charmed pseudoscalar/vector mesons contributing to the
hidden-charm molecular baryons. Taken from Ref. \cite{Yang:2011wz}.
\label{Table.3.1.mb} }
\end{center}
\end{figure}

As shown in Fig. \ref{Table.3.1.mb}, the S-wave charmed baryons belong to
either the symmetric $6_F$ or antisymmetric $\bar{3}_F$ flavor
representation with $J^P=1/2^+$ or $3/2^+$ for $6_F$ and $J^P=1/2^+$
for $\bar{3}_F$. Additionally, the pseudoscalar and vector
anti-charmed mesons form an S-wave anti-charmed meson family. In
Ref. \cite{Yang:2011wz}, the authors mainly focused on the
hidden-charm molecular states composed of the charmed baryons and
anti-charmed mesons in the green range of Fig. \ref{Table.3.1.mb}.

The effective meson exchange potentials of the hidden-charm systems
$\Lambda_c\bar{D}$ with $I(J^P)=\frac{1}{2}(\frac{1}{2}^-)$, $\Lambda_c
\bar{D}^*$ with
$\frac{1}{2}(\frac{1}{2}^-),\,\frac{1}{2}(\frac{3}{2}^-)$,
$\Sigma_c\bar{D}$ with
$\frac{1}{2}(\frac{1}{2}^-),\,\frac{3}{2}(\frac{1}{2}^-)$,
$\Sigma_c\bar{D}^*$ with
$\frac{1}{2}(\frac{1}{2}^-),\,\frac{3}{2}(\frac{3}{2}^-),\,\frac{1}{2}(\frac{1}{2}^-),\,\frac{3}{2}(\frac{3}{2}^-)$
were extracted in the OBE model. After solving the coupled-channel
Schr\"odinger equation, the numerical results indicated that there do
not exist the $\Lambda_c \bar{D}$ and $\Lambda_c \bar{D}^*$
molecular states.

There do exist molecular bound state solutions for five channels:
the $\Sigma_c\bar{D}^*$ system with
$I(J^P)=\frac{1}{2}(\frac{1}{2}^-)$, $\frac{1}{2}(\frac{3}{2}^-)$,
$\frac{3}{2}(\frac{1}{2}^-)$, $\frac{3}{2}(\frac{3}{2}^-)$ and the
$\Sigma_c\bar{D}$ system with $\frac{3}{2}(\frac{1}{2}^-)$
\cite{Yang:2011wz}. Especially, the predicted $\Sigma_c\bar{D}^*$
molecular state with $I(J^P)=\frac{1}{2}(\frac{3}{2}^-)$
\cite{Yang:2011wz} seems to match one of the two pentaquark states
observed by the LHCb Collaboration in 2015~\cite{Aaij:2015tga}.

The LHCb Collaboration reported two enhancement structures
$P_c(4380)$ and $P_c(4450)$ in the $J/\psi p$ invariant mass
spectrum of $\Lambda_b\to J/\psi pK$~\cite{Aaij:2015tga}. The
experimental information was reviewed in Sec. \ref{Sect.2.6}. Since
their final states are $J/\psi p$, each of these $P_c$ states with
isospin $I=1/2$ contains a pair of $c\bar c$.

The $P_c(4380)$ lies near the $\Sigma_c(2455)\bar{D}^*(2010)$ and
$\Sigma_c^*(2520)\bar{D}(1870)$ thresholds while the $P_c(4450)$
is very close to the $\Sigma_c(2520)\bar{D}^*$ and
$\Sigma_c^*(2520)\bar{D}^*$ thresholds. It is interesting to note that
the mass gap between the $P_c(4450)$ and $P_c(4380)$ is almost
the same as the mass difference between the
$\Sigma_c^*(2520)$ and $\Sigma_c(2455)$.

\subsubsection{Molecular assignments after LHCb's discovery}

After LHCb's discovery, several groups explored the molecular
assignment of these two $P_c$ states \cite{Chen:2015loa,He:2015cea}. In
Ref. \cite{Chen:2015loa}, the assignments of the $P_c(4380)$ and
$P_c(4450)$ as the hidden-charm $\Sigma_c(2455)\bar{D}^*$ and
$\Sigma_c^*(2520)\bar{D}^*$ molecular pentaquarks and their partner
states were investigated carefully with the one pion exchange (OPE)
model. Their flavor wave functions $|I,I_3\rangle$ were constructed
as
\begin{eqnarray*}\left\{\begin{array}{c}
\left|\frac{1}{2},\frac{1}{2}\right\rangle =
     \sqrt{\frac{2}{3}}\left|\Sigma_c^{(*)++}{D}^{*-}\right\rangle
     -\frac{1}{\sqrt{3}}\left|\Sigma_c^{(*)+}\bar{D}^{*0}\right\rangle\\
\left|\frac{1}{2},-\frac{1}{2}\right\rangle =
     \frac{1}{\sqrt{3}}\left|\Sigma_c^{(*)+}{D}^{*-}\right\rangle
     -\sqrt{\frac{2}{3}}\left|\Sigma_c^{(*)0}\bar{D}^{*0}\right\rangle
     \end{array}\right.,\quad
\left\{\begin{array}{l}
\left|\frac{3}{2},\frac{3}{2}\right\rangle = \left|\Sigma_c^{(*)++}\bar{D}^{*0}\right\rangle\\
\left|\frac{3}{2},\frac{1}{2}\right\rangle =
     \frac{1}{\sqrt{3}}\left|\Sigma_c^{(*)++}{D}^{*-}\right\rangle+\sqrt{\frac{2}{3}}\left|\Sigma_c^{(*)+}\bar{D}^{*0}\right\rangle\\
\left|\frac{3}{2},-\frac{1}{2}\right\rangle =\sqrt{\frac{2}{3}}
    \left|\Sigma_c^{(*)+}{D}^{*-}\right\rangle+ \frac{1}{\sqrt{3}}\left|\Sigma_c^{(*)0}\bar{D}^{*0}\right\rangle\\
\left|\frac{3}{2},-\frac{3}{2}\right\rangle =
     \left|\Sigma_c^{(*)0}{D}^{*-}\right\rangle
     \end{array}\right. .
\end{eqnarray*}

The effective Lagrangians, which were constructed with the heavy
quark symmetry and chiral symmetry
\cite{Yan:1992gz,Burdman:1992gh,Wise:1992hn,Casalbuoni:1996pg,Falk:1992cx,Liu:2011xc},
were adopted to obtain the OPE effective potentials of the
$\Sigma_c(2455)\bar{D}^*$ and $\Sigma_c^*(2520)\bar{D}^*$ systems,
\begin{eqnarray}
\mathcal{L}&=&
     ig\text{Tr}\left[\bar{H_a}^{(\bar{Q})}\gamma^{\mu}A^{\mu}_{ab}\gamma_5H_b^{(\bar{Q})}\right]\, ,
     \label{Eq.3.1.lag01}\\
\mathcal{L}&=&
-\frac{3}{2}g_1\varepsilon^{\mu\nu\lambda\kappa}v_{\kappa}\text{Tr}
      \left[\bar{\mathcal{S}}_{\mu}A_{\nu}\mathcal{S}_{\lambda}\right]\, . \label{Eq.3.1.lag02}
\end{eqnarray}
The multiplet field composed of the pseudoscalar $P$ and vector
$P^{*(\bar{Q})}$ with $P^{*(\bar{Q})}=(\bar{D}^{*0}, D^{*-})^T$ is
defined as $H_a^{(\bar{Q})}=
[P_a^{*(\bar{Q})\mu}\gamma_{\mu}-P_a^{(\bar{Q})}\gamma_5]\frac{1-\rlap\slash
v}{2}$ with $v=(1,\vec{0})$. The superfield
$\mathcal{S}_{\mu}$ is composed of spinor operators as
$\mathcal{S}_{\mu} =
-\sqrt{\frac{1}{3}}(\gamma_{\mu}+v_{\mu})\gamma^5\mathcal{B}_6
       +\mathcal{B}_{6\mu}^*$,
where $\mathcal{B}_6$ and $\mathcal{B}_{6}^*$ are the multiplets
corresponding to $J^P=1/2^+$ and $J^P=3/2^+$ in the $6_F$ flavor
representation, respectively. The axial current $A_{\mu} =
\frac{1}{2}(\xi^{\dag}\partial_{\mu}\xi-\xi\partial_{\mu}\xi^{\dag})$
with $\xi=\exp(i\mathbb{P}/f_{\pi})$ and $f_{\pi}=132$ MeV. The
expressions of $\mathbb{P}$, $\mathcal{B}_6$, and
$\mathcal{B}_{6}^*$ are
\begin{eqnarray}
\mathbb{P} = \left(\begin{array}{cc}
\frac{\pi^0}{\sqrt{2}} &\pi^+\\
\pi^- &-\frac{\pi^0}{\sqrt{2}}
\end{array}\right) \, , \,
\mathcal{B}_6 = \left(\begin{array}{cc}
         \Sigma_c^{++}              &\frac{\Sigma_c^{+}}{\sqrt{2}}\\
         \frac{\Sigma_c^{+}}{\sqrt{2}}      &\Sigma_c^{0}
\end{array}\right) \, , \,
\mathcal{B}_6^* = \left(\begin{array}{cc}
         \Sigma_c^{*++}              &\frac{\Sigma_c^{*+}}{\sqrt{2}}\\
         \frac{\Sigma_c^{*+}}{\sqrt{2}}      &\Sigma_c^{*0}
\end{array}\right).
\end{eqnarray}

Eqs. (\ref{Eq.3.1.lag01}) and (\ref{Eq.3.1.lag02}) can be further expanded as
\begin{eqnarray}
\mathcal{L}_{\bar{D}^*\bar{D}^*\mathbb{P}} &=&
           i\frac{2g}{f_{\pi}}v^{\alpha}\varepsilon_{\alpha\mu\nu\lambda}
           \bar{D}_{a}^{*\mu\dag}\bar{D}_{b}^{*\lambda}\partial^{\nu}\mathbb{P}_{ab},\\
\mathcal{L}_{\mathcal{B}_6\mathcal{B}_6\mathbb{P}} &=&
      i\frac{g_1}{2f_{\pi}}\varepsilon^{\mu\nu\lambda\kappa}v_{\kappa}
      \text{Tr}\left[\bar{\mathcal{B}_6}\gamma_{\mu}\gamma_{\lambda}
      \partial_{\nu}\mathbb{P}\mathcal{B}_6\right],\\
\mathcal{L}_{\mathcal{B}_6^*\mathcal{B}_6^*\mathbb{P}} &=&
      -i\frac{3g_1}{2f_{\pi}}\varepsilon^{\mu\nu\lambda\kappa}v_{\kappa}
      \text{Tr}\left[\bar{\mathcal{B}}_{6\mu}^{*}\partial_{\nu}\mathbb{P}
      \mathcal{B}_{6\nu}^*\right],
\end{eqnarray}
where the coupling constant $g=0.59\pm 0.07\pm 0.01$ was extracted
from the $D^*$ decay width \cite{Isola:2003fh} and $g_1= 0.94$ fixed
in Refs. \cite{Liu:2011xc,Yang:2011wz}.

With the standard procedure of the meson exchange model, one gets
the general expressions of the effective potentials for the
$\Sigma_c\bar{D}^*$ and $\Sigma_c^*\bar{D}^*$ systems,
\begin{eqnarray}
V_{\Sigma_c\bar{D}^*}({r}) &=&
            \frac{1}{3}\frac{gg_1}{f_{\pi}^2}
            \nabla^2Y(\Lambda,m_{\pi},{r})\,\mathcal{J}_0\,\mathcal{G}_0,\label{Eq.3.1.v1}\\
V_{\Sigma_c^*\bar{D}^*}(r) &=&
            \frac{1}{2}\frac{gg_1}{f_{\pi}^2}
            \nabla^2Y(\Lambda,m_{\pi},{r})\,\mathcal{J}_1\,\mathcal{G}_1,\label{Eq.3.1.v2}
\end{eqnarray}
where the $Y(\Lambda,m,{r}) $ function reads
\begin{eqnarray}
Y(\Lambda,m,{r}) &=&\frac{1}{4\pi r}\left(e^{-mr}-e^{-\Lambda
r}\right)-\frac{\Lambda^2-m^2}{8\pi \Lambda}e^{-\Lambda
r} \, .
\end{eqnarray}
In Eqs. (\ref{Eq.3.1.v1}) and (\ref{Eq.3.1.v2}), the coefficients $\mathcal{J}_i$
and $\mathcal{G}_i$ ($i=0,1$) for different isospin and $^{2S+1}L_J$
quantum numbers are collected in Table \ref{Table.3.1.factor}.
\renewcommand{\arraystretch}{1.5}
\begin{table}[htbp]
\caption{The values of the $\mathcal{J}_i$ and $\mathcal{G}_i$
coefficients for the S-wave $\Sigma_c(2455)\bar{D}^*$ and
$\Sigma_c^*(2520)\bar{D}^*$ systems. Here, $S$, $L$, and $J$ denote
the spin, orbital, and total angular quantum numbers, respectively.
Taken from Ref. \cite{Chen:2015loa}.}\label{Table.3.1.factor}
\begin{center}
  \begin{tabular}{cccccc}
\toprule[1pt] $I$\quad  &\quad$\mathcal{G}_0$\quad
&\quad$\mathcal{G}_1$\quad\quad
&\quad$\left|{}^{2S+1}L_{J}\right\rangle$\quad
&\quad$\mathcal{J}_0$\quad  &\quad$\mathcal{J}_1$\\\midrule[1pt]
1/2      &1         &-1       &$\left|{}^2\mathbb{S}_{\frac{1}{2}}\right\rangle$    &-2        &5/3\\
3/2      &-1/2      &1/2      &$\left|{}^4\mathbb{S}_{\frac{3}{2}}\right\rangle$    &1         &2/3\\
\ldots   &\ldots    &\ldots
&$\left|{}^6\mathbb{S}_{\frac{5}{2}}\right\rangle$ & \ldots & -1
\\\bottomrule[1pt]
\end{tabular}
\end{center}
\end{table}

\begin{figure}[htbp]
  \centering
  \includegraphics[width=5in]{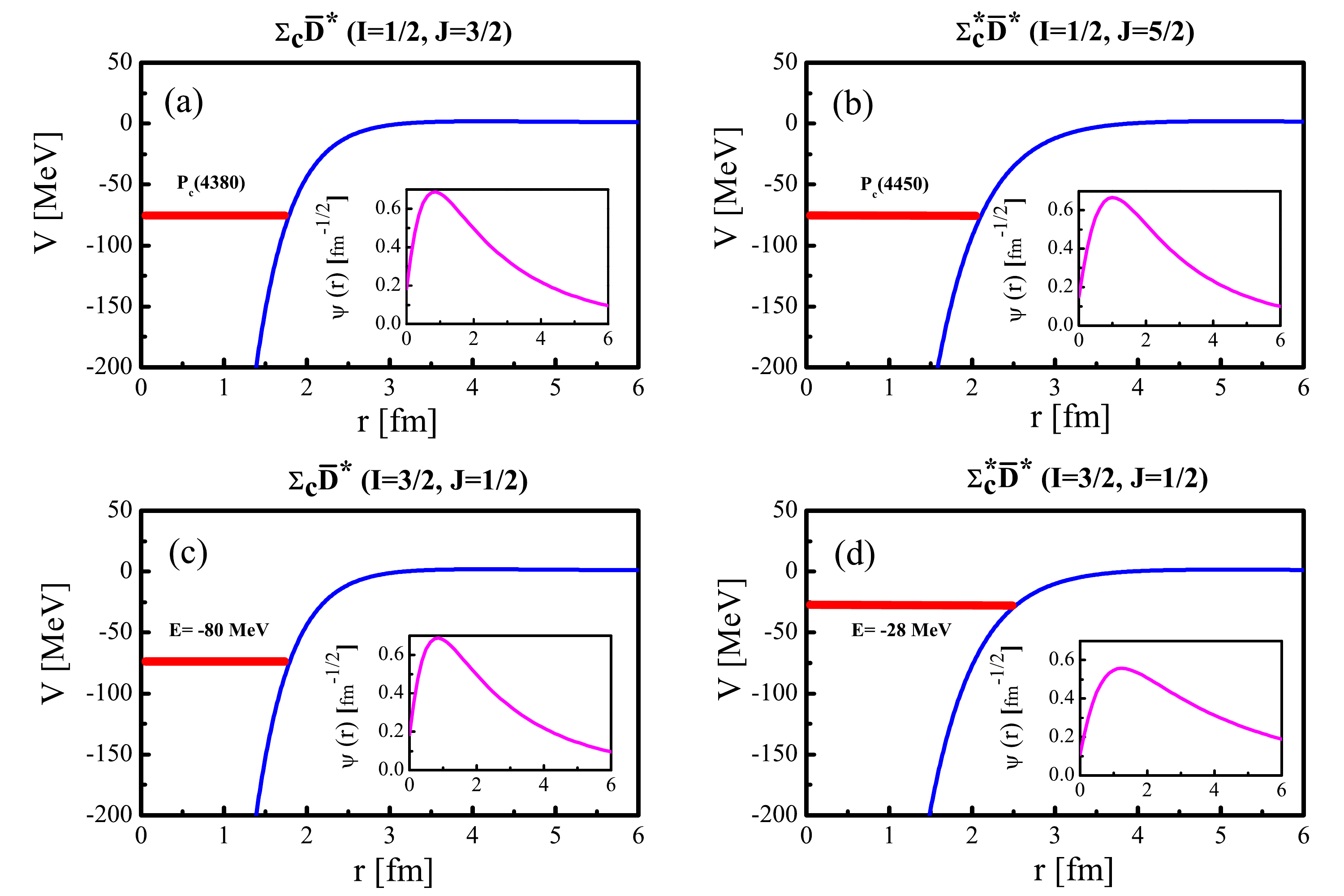}\\
\caption{(Color online) The variation of the OPE potential and the radial wave
function with $r$ for the $\Sigma_c^{(*)}\bar{D}^*$ systems.
$\Lambda=2.35$ GeV and $\Lambda=1.77$ GeV for the
$\Sigma_c\bar{D}^*$ and $\Sigma_c^*\bar{D}^*$ systems, respectively.
The blue curve is the effective potentials, and the red line is the
energy level. Taken from Ref. \cite{Chen:2015loa}.}\label{Table.3.1.energy}
\end{figure}

We summarize some interesting observations in Ref.
\cite{Chen:2015loa}:
\begin{itemize}
\item Under the molecular scheme, the masses of the $P_c(4380)$ and $P_c(4450)$
are reproduced very well as shown in Figs. \ref{Table.3.1.energy} (a) and
\ref{Table.3.1.energy} (b). Moreover their spatial extension is quite large,
around 1-2 fm, which is a characteristic feature of a hadronic
molecule.

\item If the $P_c$ states are S-wave molecular states, both of them
carry negative parity. They can transit into $J/\psi p$ via the
exchange of an $S$-wave charmed meson. The $\Sigma_c\bar{D}^*$ state
with $(I=1/2,J=3/2)$ decays into $J/\psi p$ via S-wave while the
$\Sigma_c^*\bar{D}^*$ state with $(I=1/2,J=5/2)$ decays into $J/\psi
p$ via D-wave. The $D$-wave decay is strongly suppressed by small
phase space. In other words, the $P_c(4450)$ is quite narrow while
the $P_c(4380)$ is broad \cite{Aaij:2015tga}. These two $P_c$ states
also decay into $\eta_c p$.

\item There exist several isospin partners of the $P_c$ states as
shown in Fig. \ref{Table.3.1.energy} (c) and (d). With the same set of
parameters, the binding energy of the $\Sigma_c\bar{D}^*$ system
with $(I=3/2,J=1/2)$ is the same as that of the $\Sigma_c\bar{D}^*$
system with $(I=1/2,J=3/2)$. With the same set of parameters, the
binding energy of the $\Sigma_c^*\bar{D}^*$ system with
$(I=3/2,J=1/2)$ is $28$ MeV, which is smaller than that of the
$\Sigma_c^*\bar{D}^*$ system with $(I=1/2,J=5/2)$. The allowed decay
modes of these two $I=3/2$ states include $\Delta(1232)J/\psi$ and
$\Delta(1232)\eta_c$.

\end{itemize}

\subsubsection{Configuration mixing}

Besides the assignment of the $P_c(4380)$ and $P_c(4450)$ as the
hidden-charm $\Sigma_c(2455)\bar{D}^*$ and
$\Sigma_c^*(2520)\bar{D}^*$ molecular pentaquarks in Ref.
\cite{Chen:2015loa}, He also studied the molecular baryons
\cite{He:2015cea}. He derived the OBE effective potentials and
solved the Bethe-Saltpeter equation with a spectator quasipotential
approximation, where the $P_c(4380)$ and $P_c(4450)$ were explained
as the $\bar{D}\Sigma_c^*$ molecular state with $J^P=3/2^-$ and the
$\bar{D}^*\Sigma_c$ molecular state with $J^P=5/2^+$, respectively.

In the heavy quark symmetry limit, the $(D, D^*)$ pair forms a
degenerate doublet. The $(\Sigma_c, \Sigma_c^*)$ pair is also
degenerate. Except for the isospin and spin factors, the
interactions between the $(D, D^*)$ and $(\Sigma_c, \Sigma_c^*)$
pairs are essentially the same and governed by the same coupling
constants. For example, the $\bar{D}\Sigma_c^*$ molecular state with
$J^P=3/2^-$ discussed in Ref. \cite{He:2015cea} is essentially the
same as the $\Sigma_c(2455)\bar{D}^*$ molecular state with
$J^P=3/2^-$ in Ref. \cite{Chen:2015loa} in the heavy quark symmetry
limit.

Generally speaking, several degenerate flavor configurations
contribute to the same hidden-charm molecular baryons with fixed
$I(J^P)$ in the heavy quark symmetry limit. There exists strong
configuration mixing. In the real world, the charm mass is around
1.5 GeV. The mass degeneracy of the heavy hadron pair is removed by
the $1/M_c$ correction with a mass splitting around 100 MeV. The
mass gap between different mass thresholds is around several tens
MeV, which is comparable with (or even larger than) the binding
energy for the hidden-charm molecular baryons. The coupled-channel
effects due to the flavor configuration mixing may turn out to be
important. Such an investigation is still missing at present.

\subsubsection{Orbital excitations and the $P_c$ parity}

If both the $P_c(4380)$ and $P_c(4450)$ are the S-wave hidden-charm
molecular states, their parities are negative which seems in
conflict with LHCb's measurement that the $P_c(4380)$ and
$P_c(4450)$ have opposite parities \cite{Aaij:2015tga}.
Recall that the D-wave contribution is only a few percent in the
case of the deuteron, where the binding energy is around 2 MeV.
However, the D-wave component contributes significantly to the
formation of the shallow bound state through the tensor force and
S-D wave mixing. The lesson is that the orbital excitation is important!

In the case of the hidden-charm molecular baryons, the P-wave,
D-wave or even higher orbital excitations may accompany the lowest
S-wave state if the binding energy of the hadronic molecule ground
state reaches several MeV to several tens MeV. Especially, the P-wave
state may lie very close to the S-wave ground state with an
excitation energy around several to tens MeV. These two levels are
almost degenerate but carry opposite parities. They may completely
overlap with each other.

For the P-wave orbitally excited molecular baryons, they decay into
the $J/\psi P, \eta_c P$ modes via P-wave while their S-wave decay
modes $\chi_{cJ} p$ are either kinematically forbidden or strongly
suppressed by phase space.

Compared with the S-wave decay, the P-wave decay width is suppressed
by the factor $(k/M)^2$ because of the centrifugal barrier, where
$k$ is decay momentum and $M$ is the pentaquark mass. For the
$J/\psi p$ mode of the $P_c(4450)$ state, the suppression factor is
around 30. In other words, the P-wave hidden-charm molecular
pentaquarks are expected to be quite narrow.

As pointed out in Ref. \cite{Wang:2015qlf}, there may exist two or
more resonant signals around 4380 MeV which are close to each other
but may carry different parity. If the P-wave or higher excitation
is very broad with a width around 500 MeV, such a state may easily
be mistaken as the background. On the other hand, if an excitation
lies several MeV within 4380 MeV but with a width as narrow as
several MeV, then it may probably be buried by the $P_c(4380)$
resonance with a width around 205 MeV! The same situation may also
occur around 4450 MeV.

The above speculation may partly explain why the different
assignments of the spins and parities for these two $P_c$ states yielded
roughly the same good fit \cite{Aaij:2015tga}. The identification of
the nearly degenerate resonances with different parities and widths
may require huge amount experimental data.

We want to emphasize that the possible existence of the P-wave
excitation together with the S-wave ground state is the first
intrinsic property of the hadronic molecular scheme.

The second intrinsic feature of the hidden-charm molecular states is
that the open-charm decay modes should dominate the hidden-charm
decay modes. This observation is supported by the current
experimental measurements of the decay modes of the charged $Z_c$
(or $Z_b$) states, where the open-charm decay width is much larger
than the $J/\psi \pi$ partial width.

In other words, the $J/\psi p$ is not necessarily the dominant decay
mode of the $P_c(4380)$ state although it was observed in the very
clean $J/\psi p$ final state. Instead, the broad Breit-Wigner
distribution of the $P_c(4380)$ ensures that it could decay into the
open-charm modes such as
 $\bar{D}\Sigma_c^*$, $\Sigma_c\bar{D}^*$, $\bar{D}\Sigma_c\pi $, $\bar{D}\Lambda_c^*$,
$\bar{D}^*\Lambda_c$ etc.

\subsection{Dynamically generated resonance}
\label{Sect.3.2}
%=====================================================================================
%=====================================================================================
%

In Refs.~\cite{Wu:2010jy,Wu:2010vk}, Wu, Molina, Oset and Zou
studied the interaction between various charmed mesons and charmed
baryons within the framework of the coupled channel unitary approach
with the local hidden gauge formalism. The hidden-charm baryons are
generated dynamically~\cite{Wu:2010jy,Wu:2010vk}. The same/similar
method was also used to study the hidden-charm baryons in a series
of papers~\cite{Wu:2012md,Molina:2012mv,Garcia-Recio:2013gaa,Xiao:2013yca,Uchino:2015uha,Garzon:2015zva,Xiao:2015fia},
all of which are done before the LHCb's discovery of the $P_c(4380)$ and
$P_c(4450)$~\cite{Aaij:2015tga}. The same/similar approach was
applied to study hidden-bottom baryons~\cite{Wu:2010rv,Xiao:2013jla}
and their
productions~\cite{Roca:2015dva,Feijoo:2015cca,Miyahara:2015cja,Chen:2015sxa},
some of which will be discussed in Sec.~\ref{sect:3.7} and Sec.~\ref{sec:3.8}, respectively. In
this review we introduce this method and review their results
briefly. However, we shall not discuss its application for open-charm
baryons~\cite{GarciaRecio:2012db,Liang:2014eba,Liang:2014kra,Lu:2014ina,Garcia-Recio:2015jsa}.

\begin{figure}[htbp]
\centering
\includegraphics[width=4in]{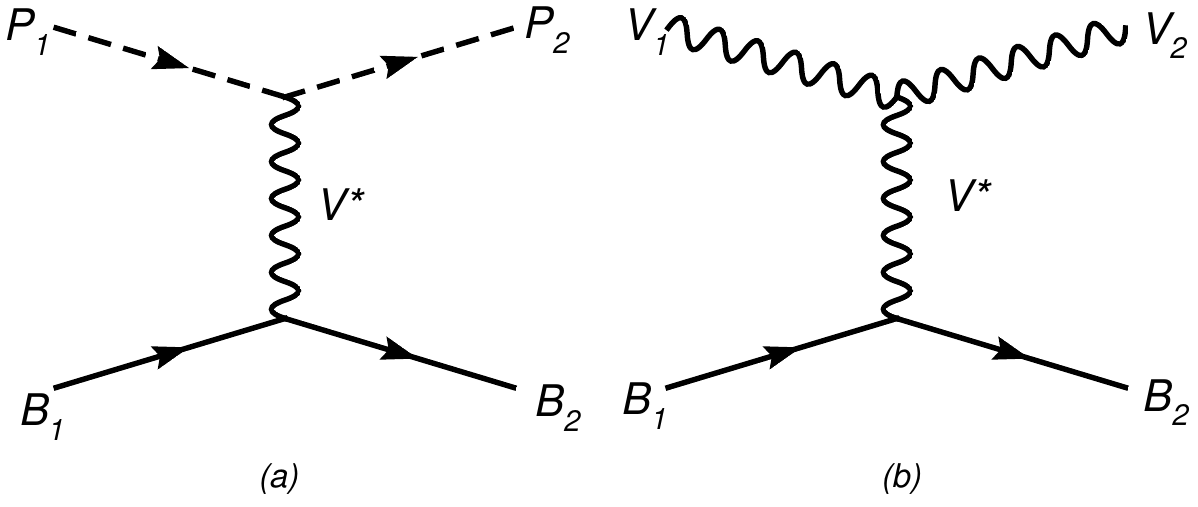}\\
\caption{Feynman diagrams for the pseudoscalar-baryon (a) and
vector-baryon (b) interactions via the exchange of a vector meson.
Taken from Ref. \cite{Wu:2010vk}.} \label{Fig.3.2.Feynman}
\end{figure}

In Refs.~\cite{Wu:2010jy,Wu:2010vk}, Wu {\it et al.} considered the $PB
\to PB$ and $VB \to VB$ interactions by exchanging a vector meson
$V^*$, where $P$ denotes the pseudoscalar charmed mesons $D$ and $D_s$,
$V$ denotes the vector charmed mesons $D^*$ and $D_s^*$, and $B$ denotes
the charmed baryons $\Sigma_c$, $\Lambda_c$, $\Xi_c$, $\Xi^\prime_c$ and
$\Omega_c$. The corresponding Feynman diagrams are shown in
Fig.~\ref{Fig.3.2.Feynman}, where the exchanged vector meson $V^*$ can be
the $\rho$, $\omega$, $K^*$ and $\phi$. The vertices for exchanging
vector mesons come from the following three Lagrangians
\begin{eqnarray}
{\cal L}_{VVV}&=&ig\langle V^\mu[V^{\nu},\partial_\mu V_{\nu}]\rangle \, ,\nonumber\\
{\cal L}_{PPV}&=&-ig\langle V^\mu[P,\partial_\mu P]\rangle \, ,\\
{\cal L}_{BBV}&=&g (\langle\bar{B}\gamma_\mu
[V^\mu,B]\rangle+\langle\bar{B}\gamma_\mu B\rangle\langle
V^\mu\rangle)\, .\nonumber
\end{eqnarray}
Here, the first Lagrangian was constructed based on the hidden gauge
interaction for vector
mesons~\cite{Bando:1984ej,Bando:1987br,Harada:2003jx,Nagahiro:2008cv};
the third Lagrangian was introduced in
Refs.~\cite{Oset:2009vf,Klingl:1997kf,Ecker:1994gg,Bernard:1995dp,Jenkins:1991es}
for the case of three flavors, and was extended to the case of four
flavors in Refs.~\cite{Wu:2010jy,Wu:2010vk}, where
\begin{eqnarray}
P &=& \left(
\begin{array}{cccc}
\frac{\pi^0}{\sqrt{2}}+\frac{\eta_8}{\sqrt{6}} +\frac{\tilde{\eta}_c}{\sqrt{12}}+\frac{\tilde{\eta}_c'}{\sqrt{4}}  &\pi^+     & K^{+}    &\bar{D}^{0}\\
\pi^-      & -\frac{\pi^0}{\sqrt{2}}+\frac{\eta_8}{\sqrt{6}}+\frac{\tilde{\eta}_c}{\sqrt{12}}+\frac{\tilde{\eta}_c'}{\sqrt{4}}& K^{0}    &D^{-}\\
K^{-}      & \bar{K}^{0}       &\frac{-2\eta_8}{\sqrt{6}}+\frac{\tilde{\eta}_c}{\sqrt{12}}+\frac{\tilde{\eta}_c'}{\sqrt{4}}                            &D^{-}_s\\
D^{0}&D^{+}&D^{+}_s&-\frac{3\tilde{\eta}_c}{\sqrt{12}}+\frac{\tilde{\eta}_c'}{\sqrt{4}}\\
\end{array}
\right)\, ,
\end{eqnarray}
\begin{eqnarray}
V_\mu &=& \left(
\begin{array}{cccc}
\frac{\rho^0}{\sqrt{2}}+\frac{\omega_8}{\sqrt{6}} +\frac{\tilde{\omega}_c}{\sqrt{12}}+\frac{\tilde{\omega}_c'}{\sqrt{4}}& \rho^+ & K^{*+}&\bar{D}^{*0}\\
\rho^- &-\frac{\rho^0}{\sqrt{2}}+\frac{\omega_8}{\sqrt{6}}+\frac{\tilde{\omega}_c}{\sqrt{12}}+\frac{\tilde{\omega}_c'}{\sqrt{4}} & K^{*0}&D^{*-}\\
K^{*-} & \bar{K}^{*0} &\frac{-2\omega_8}{\sqrt{6}}+\frac{\tilde{\omega}_c}{\sqrt{12}}+\frac{\tilde{\omega}_c'}{\sqrt{4}}&D^{*-}_s\\
D^{*0}&D^{*+}&D^{*+}_s&-\frac{3\tilde{\omega}_c}{\sqrt{12}}+\frac{\tilde{\omega}_c'}{\sqrt{4}}\\
\end{array}
\right)_\mu\ .
\end{eqnarray}
However, the $BBV$ vertex in the case of four flavors does not have
a simple representation as in the case of three flavors. Wu {\it et
al.} evaluated the matrix elements using SU(4) symmetry in terms of
the Clebsch-Gordan coefficients and reduced matrix elements.

Recall that the SU(3) flavor symmetry is broken at the level of
$20\%\sim 30\%$. In general, the flavor SU(4) symmetry is badly
broken. One should be very cautious about the uncertainty of the
$J/\psi DD$ and $J/\psi BB$ couplings derived from the equations
under the SU(4) symmetry. However, one should note that
with the exchange of light vectors, the heavy quarks are spectators,
and one is using effectively SU(3) symmetry in the dominant terms.

The transition potential corresponding to the diagrams of
Fig.~\ref{Fig.3.2.Feynman} was given by~\cite{Wu:2010jy,Wu:2010vk}
\begin{eqnarray}
V_{ab(P_{1}B_{1}\rightarrow
P_{2}B_{2})}&=&\frac{C_{ab}}{4f^{2}}(q^0_1+q^0_2) \, ,
\\
V_{ab(V_{1}B_{1}\rightarrow
V_{2}B_{2})}&=&\frac{C_{ab}}{4f^{2}}(q^0_1+q^0_2)\vec{\epsilon}_1\cdot\vec{\epsilon}_2\, ,
\end{eqnarray}
where $q^0_1$ and $q^0_2$ are the energies of the initial and final
mesons, respectively. The $C_{ab}$ coefficients can be found in
Refs.~\cite{Wu:2010jy,Wu:2010vk} for six cases with different
isospin and strangeness $(I, S) = (3/2, 0)$, $(1/2, 0)$, $(1/2, -2)$, $(1,
-1)$, $(0, -1)$, $(0, -3)$.

In the derivation of the above transition potential, the authors
assumed that the three-momentum were much smaller than its mass and
kept $\gamma^0$ and the time component only. In this way they
derived the contact interaction for the charmed meson and baryon.
Such an approximation may not work very well when the
states found are very bound.

The scattering matrix $T$ was evaluated by solving the coupled
channels Bethe-Salpeter equation in the on-shell factorization
approach \cite{Oller:2000fj,Oller:1998zr,Nieves:1999bx}
\begin{equation}
T = [1 - V \, G]^{-1}\, V \, ,
\end{equation}
where $G$ is the loop function of a meson and a baryon, and was
evaluated in the dimensional regularization as \cite{Oller:2000fj}
\begin{eqnarray}
G(s) &=&i \int\frac{d^{4}q}{(2\pi)^{4}}\frac{2M_{B}}{(P-q)^{2}-M^{2}_{B}+i\varepsilon}\,\frac{1}{q^{2}-M^{2}_{P}+i\varepsilon},\\
&=&\frac{2M_{B}}{16\pi^2}\big\{a_{\mu}+\textmd{ln}\frac{M^{2}_{B}}{\mu^{2}}+\frac{M^{2}_{P}-M^{2}_{B}+s}{2s}\textmd{ln}\frac{M^{2}_{P}}{M^{2}_{B}}\nonumber\\
&&+\frac{q_{cm}}{\sqrt{s}}\big[\textmd{ln}(s-(M^{2}_{B}-M^{2}_{P})+2q_{cm}\sqrt{s})+\textmd{ln}(s+(M^{2}_{B}-M^{2}_{P})+2q_{cm}\sqrt{s})\nonumber\\
&&-\textmd{ln}(-s-(M^{2}_{B}-M^{2}_{P})+2q_{cm}\sqrt{s})-\textmd{ln}(-s+(M^{2}_{B}-M^{2}_{P})+2q_{cm}\sqrt{s})\big]\big\}\, ,\nonumber
\end{eqnarray}
where the regularization scale $\mu=1$ GeV and the parameter $a_\mu$
is fixed around $-2.3$.

Wu {\it et al.} also took into account some decay mechanisms by
considering the decay of the states to a light baryon plus either a
light meson or a charmonium through heavy charmed meson
exchanges, as done in Refs.~\cite{Molina:2008jw,Geng:2008gx}.

\begin{table}[hbt]
\setlength{\tabcolsep}{0.15cm}
\caption{Mass ($M$), total width
($\Gamma$), and the partial decay width ($\Gamma_i$) for the states from $PB\to PB$, with units in MeV, taken from Refs.~\cite{Wu:2010jy,Wu:2010vk}.}
\label{Table.3.2.pbwidth}
\begin{center}
\begin{tabular}{ccccccccc}\toprule[1pt]
$(I, S)$       & $M$ & $\Gamma$ & \multicolumn{6}{c}{$\Gamma_i$ }\\
\midrule[1pt]
$(1/2, 0)$     &      &             & $\pi N$ & $\eta N$ & $\eta' N$ & $K \Sigma$ &  & $\eta_cN$\\
            & $4261$ & $56.9$        & $3.8$     & $8.1 $     & $3.9$       & $17.0$  & & 23.4\\
\midrule[1pt]
$(0, -1)$  &      &       & $\bar{K} N$  & $\pi\Sigma$ & $\eta\Lambda$ & $\eta'\Lambda$ & $K\Xi$ & $\eta_c\Lambda$\\
           & $4209$ & $32.4$   & $15.8$  & $2.9$       & $3.2 $        & $1.7$          & $2.4$    & 5.8 \\
           & $4394$ & $43.3$     & $0 $  & $10.6$      & $7.1 $        & $3.3 $         & $5.8 $   &16.3  \\
\bottomrule[1pt]\end{tabular}
\end{center}
%\end{table}
%\begin{table}[ht]
%       \renewcommand{\arraystretch}{1.1}
\setlength{\tabcolsep}{0.13cm}
\caption{Mass ($M$), total width ($\Gamma$), and the
partial decay width ($\Gamma_i$) for the states from $VB\to VB$ with
units in MeV, taken from Refs.~\cite{Wu:2010jy,Wu:2010vk}.}
 \label{Table.3.2.vbwidth}
 \begin{center}
\begin{tabular}{ccccccccc}\toprule[1pt]
$(I, S)$  & $M$ & $\Gamma$  & \multicolumn{6}{c}{$\Gamma_i$  }\\
\midrule[1pt]
$(1/2, 0)$  &      &        & $\rho N$ & $\omega N$ & $K^{*} \Sigma$  &  &  &  $J/\psi N$\\
           & $4412$ & $47.3$   & $3.2$      & $10.4  $      &  $13.7$ &  &  &  19.2    \\
\midrule[1pt]$(0, -1)$ &      &       & $\bar K^*N$ & $\rho\Sigma$ & $\omega\Lambda$ & $\phi\Lambda$ & $K^*\Xi$ & $J/\psi\Lambda$\\
          & 4368 & 28.0  & 13.9        & 3.1          & 0.3             & 4.0           & 1.8      & 5.4 \\
          & 4544 & 36.6  & 0           & 8.8          & 9.1             & 0             & 5.0      & 13.8 \\
\bottomrule[1pt]
\end{tabular}
\end{center}
\end{table}

Their results of the pole positions and coupling constants are
listed in Tables~\ref{Table.3.2.pbwidth} and \ref{Table.3.2.vbwidth}, where two
$N_{c\bar c}^*$ states and four $\Lambda_{c\bar c}^*$ states were
found in the $PB$ and $VB$ scattering
channels~\cite{Wu:2010jy,Wu:2010vk}. All these states have masses
larger than 4200 MeV due to their $c\bar c$ components. Their decay
properties were discussed and cross sections for their production
were estimated, suggesting that the $\eta_{c}N$ and $\eta_{c}\Lambda$
are possible decay modes for the $PB$ channels, and the $J/\psi N$ and
$J/\psi \Lambda$ are possible modes for the $VB$ channels. These
results were used by Molina, Xiao, and Oset to study the interaction
of the $J/\psi$ with nuclei in
Ref.~\cite{Molina:2012mv}. They evaluated the total inelastic cross
section of the $J/\psi N$ and found a maximum around $\sqrt{s} =
4415$ MeV, where the $J/\psi N$ couples to a resonance predicted in
Refs.~\cite{Wu:2010jy,Wu:2010vk} (see 4412 MeV in Table \ref{Table.3.2.vbwidth}).

Later, Wu, Lee, and Zou considered several coupled-channel models
derived from relativistic quantum field theory~\cite{Wu:2012md}.
They used both a unitary transformation
method~\cite{Hyodo:2002pk,Sato:1996gk}, and the three-dimensional
reductions of the Bethe-Salpeter Equation~\cite{Hung:2001pz}, and
found that all models gave very narrow molecular-like nucleon
resonances with hidden-charm in the mass range $4.3-4.5$ GeV. Their
results are consistent with the previous
predictions~\cite{Wu:2010jy,Wu:2010vk}.

In Ref.~\cite{Xiao:2013yca}, Xiao, Nieves, and Oset improved these
results by including the leading order constraints of heavy quark
spin symmetry~\cite{Isgur:1989vq,Neubert:1993mb,Manohar:2000dt}, and
developed a series of relationships for the transition potentials
between the different meson-baryon channels in different
combinations of spin and isospin. They found seven states with
different energies or different spin-isospin quantum numbers, all of
which have $I = 1/2$:
\begin{eqnarray*}
\nonumber && {\rm the}~J = 1/2~{\rm sector:}~(4261.87 + i17.84)~{\rm
MeV} \, , \, (4410.13 + i29.44)~{\rm MeV} \, , \, (4481.35 +
i28.91)~{\rm MeV} \, ,
\\ \nonumber && {\rm the}~J = 3/2~{\rm sector:}~(4334.45 + i19.41)~{\rm MeV} \, , \, (4417.04 + i4.11)~{\rm MeV} \, , \, (4481.04 + i17.38)~{\rm MeV} \, ,
\\ && {\rm the}~J = 5/2~{\rm sector:}~(4487.10 + i0)~{\rm MeV} \, .
\end{eqnarray*}
These poles can be easily classified as four basic states: a) the
first pole $(4261.87 + i17.84)$ MeV, corresponding to a $\bar D
\Sigma_c$ state; b) the fourth pole $(4334.45 + i19.41)$ MeV,
corresponding to a $\bar D \Sigma_c^*$ state; c) the second pole
$(4410.13 + i29.44)$ MeV and the fifth pole $(4417.04 + i4.11)$ MeV,
both corresponding to a $\bar D^* \Sigma_c$ state; d) the third pole
$(4481.35 + i28.91)$ MeV, the sixth pole $(4481.04 + i17.38)$ MeV,
and the seventh pole $(4487.10 + i0)$ MeV, all corresponding to a
$\bar D^* \Sigma_c^*$ state. All these states are bound with about
50 MeV with respect to the corresponding $\bar D B$ thresholds.

In 2013, Garcia-Recio, Nieves, Romanets, Salcedo and Tolos studied
the hidden charm $N$ and $\Delta$ resonances using a similar method
constrained by the extended SU(8) spin-flavour symmetry, but their
obtained masses were substantially smaller than the former
values~\cite{Garcia-Recio:2013gaa}. For the $J^P={1/2}^-$,
${3/2}^-$, and ${5/2}^-$ hidden-charm resonances, their masses were
predicted to be $3918\sim 3974$ MeV, $3946\sim 4006$ MeV, $4027$ MeV,
respectively. The total decay widths of the above three resonances were
less than 10 MeV.

The results of Refs.~\cite{Wu:2010jy,Wu:2010vk} were used by Xiao
and Meissner to investigate the elastic and inelastic cross sections
of the $J/\psi N$, $\eta_c N$, $\Upsilon N$ and $\eta_b N$ channels,
where the predicted neutral partners $P^0_c$ of the $P^+_c$ states
may be found in Ref.~\cite{Xiao:2015fia}.

In a recent work~\cite{Uchino:2015uha}, Uchino, Liang, and Oset once
more improved their results by considering two types of additional
interactions as a box diagram correction. They found six states
whose masses, widths, dominant components, and main decay channel
are collected in Table VIII of Ref.~\cite{Uchino:2015uha}.

In short summary, there exist extensive theoretical investigations
of the hidden-charm baryons within the coupled channel unitary
approach before and after the LHCb's discovery of the $P_c(4380)$ and
$P_c(4450)$~\cite{Aaij:2015tga}. We summarize several general
features of the unitary approach.
\begin{itemize}

\item Only the hidden-charm baryons with negative parity are
generated dynamically through the S-wave charmed meson and baryon
scattering.

\item Nearly all the model calculations indicated that the total
width of the hidden-charm baryons were less than 60 MeV. In other
words, these resonances are very narrow.

\item The $J/\psi N$ (or $\eta_c N$) is one of the decay modes of
the non-strange hidden-charm baryons, but channels with open charm or
channels in the light sector still dominate.

\item The charm-less decay modes contribute significantly to the
total decay width of the non-strange hidden-charm baryons.

\item In general, several hidden-charm baryons are generated
dynamically for one set of $J^P$.

\item Within the same unitary approach, slightly different
models lead to diverse predictions for the mass of the hidden-charm
baryon~\cite{Wu:2010jy,Wu:2010vk,Wu:2012md,Garcia-Recio:2013gaa,Xiao:2013yca,Uchino:2015uha},
which indicate inherent uncertainties of this
framework~\cite{Uchino:2015uha}.

\end{itemize}

Sometimes spurious resonances are dynamically generated within the
same approach, which are excluded experimentally. A recent
investigation indicated that the inclusion of the higher order
corrections may change the mass prediction significantly,
and ``the LECs (note by the authors: low energy constants) should be determined in a more reliable way in
order to study the effects of higher-order potentials''~\cite{Lu:2014ina}.

%================================================================================
\subsection{QCD sum rules}
\label{sect:3.3}
%================================================================================

In the previous two subsections, we reviewed the meson exchange
model and the chiral unitary approach. These two methods used the
interactions between the charmed mesons and charmed baryons at the
hadronic level to interpret the hidden-charm pentaquarks discovered
by the LHCb Collaboration~\cite{Aaij:2015tga}. It is also
interesting to study the internal structures of these pentaquarks at
the quark gluon level. There have been several investigations of the
hidden-charm pentaquarks with the method of QCD sum rule.

\subsubsection{A short introduction to the method of QCD sum rule}

QCD sum rule techniques have proven to be a powerful and successful
non-perturbative method over the past few decades. Various aspects
of this formalism have been reviewed in
Refs.~\cite{Shifman:1978bx,Reinders:1984sr,Narison:2002pw,Colangelo:2000dp,Nielsen:2009uh}.
It can also be applied in the framework of heavy quark effective theory~\cite{Grinstein:1990mj,Eichten:1989zv,Falk:1990yz} to study heavy mesons~\cite{Bagan:1991sg,Neubert:1991sp,Broadhurst:1991fc,Ball:1993xv,Huang:1994zj,Dai:1996yw,Dai:1993kt,Dai:1996qx,Colangelo:1998ga,Dai:2003yg,Zhou:2014ytp,Zhou:2015ywa} and heavy baryons~\cite{Shuryak:1981fza,Grozin:1992td,Dai:1995bc,Dai:1996xv,Groote:1996em,Zhu:2000py,Lee:2000tb,Huang:2000tn,Wang:2003zp,Liu:2007fg,Chen:2015kpa,Mao:2015gya}.

In addition to the operator product expansion (OPE), a key idea of
the QCD sum rule is the quark-hadron duality, i.e., the equivalence
of the (integrated) correlation functions at both the hadronic level
and the quark-gluon level. One considers the correlation function
%
%%%%%%%%%%%%%%%%%%%%%%%%%%%%%%%%%%%%%%%%%%%%%%%%%%%%%%%%%%%%%%%%%%%%%%%%%%%%%%
\begin{eqnarray}
\Pi(q^2) &=& i \int d^4x e^{iqx} \langle 0 | T J(x) J^\dagger (0) |
0 \rangle \, , \label{Eq.3.3.pi}
\end{eqnarray}
%%%%%%%%%%%%%%%%%%%%%%%%%%%%%%%%%%%%%%%%%%%%%%%%%%%%%%%%%%%%%%%%%%%%%%%%%%%%%%
%
where $J(x)$ is an interpolating current which has the same quantum
numbers as the hadron $H$ we want to study. The strength of $J(x)$
coupling to $H$ is defined as $f_H$:
\begin{eqnarray}
\langle 0 | J(0) | H \rangle \equiv f_H \, .
\end{eqnarray}
At the hadronic level, one expresses the correlation function in the
form of the dispersion relation:
%
%%%%%%%%%%%%%%%%%%%%%%%%%%%%%%%%%%%%%%%%%%%%%%%%%%%%%%%%%%%%%%%%%%%%%%%%%%%%%%
\begin{equation}
\Pi(q^2)={\frac{1}{\pi}}\int^\infty_{s_<}\frac{{\rm Im}
\Pi(s)}{s-q^2-i\varepsilon}ds \, ,
\end{equation}
%%%%%%%%%%%%%%%%%%%%%%%%%%%%%%%%%%%%%%%%%%%%%%%%%%%%%%%%%%%%%%%%%%%%%%%%%%%%%%
%
where the integration starts from the physical threshold. The
imaginary part of the two-point correlation function is the spectral
density
%
%%%%%%%%%%%%%%%%%%%%%%%%%%%%%%%%%%%%%%%%%%%%%%%%%%%%%%%%%%%%%%%%%%%%%%%%%%%%%%
\begin{eqnarray}
\rho_{\rm phen}(s) \equiv {1 \over \pi}{\rm Im} \Pi(s) =
\sum_n\delta(s-M^2_n)\langle 0| J |n\rangle\langle n| J^\dagger
|0\rangle \, .
\end{eqnarray}
%%%%%%%%%%%%%%%%%%%%%%%%%%%%%%%%%%%%%%%%%%%%%%%%%%%%%%%%%%%%%%%%%%%%%%%%%%%%%%
%
One usually adopts a parametrization of one pole dominance for the
ground state $H$ and a continuum contribution:
%
%%%%%%%%%%%%%%%%%%%%%%%%%%%%%%%%%%%%%%%%%%%%%%%%%%%%%%%%%%%%%%%%%%%%%%%%%%%%%%
\begin{eqnarray}
\rho_{\rm phen}(s) =  f^2_H \delta(s-M^2_H) +
\rm{higher\,\,states}\, .
\end{eqnarray}
%%%%%%%%%%%%%%%%%%%%%%%%%%%%%%%%%%%%%%%%%%%%%%%%%%%%%%%%%%%%%%%%%%%%%%%%%%%%%%
%

At the quark-gluonic level, one computes $\Pi(q^2)$ in the operator
product expansion and evaluates $\rho(s) = \rho_{\rm OPE}(s)$ up to
certain order in the expansion. The sum rule analysis is then
performed by using the Borel transformation
%
%%%%%%%%%%%%%%%%%%%%%%%%%%%%%%%%%%%%%%%%%%%%%%%%%%%%%%%%%%%%%%%%%%%%%%%%%%%%%%
\begin{equation}
\Pi^{(all)}(M_B^2)\equiv\mathcal{B}_{M_B^2}\Pi(p^2) =
\int^\infty_{s_<} e^{-s/M_B^2} \rho(s) ds \, .
\end{equation}
%%%%%%%%%%%%%%%%%%%%%%%%%%%%%%%%%%%%%%%%%%%%%%%%%%%%%%%%%%%%%%%%%%%%%%%%%%%%%%
%
Assuming the contribution from the continuum states can be
approximated well by the OPE spectral density above a threshold
value $s_0$, one arrives at the sum rule relation
%
%%%%%%%%%%%%%%%%%%%%%%%%%%%%%%%%%%%%%%%%%%%%%%%%%%%%%%%%%%%%%%%%%%%%%%%%%%%%%%
\begin{equation}
f^2_H e^{-M_H^2/M_B^2} = \int^{s_0}_0 e^{-s/M_B^2}\rho(s)ds
\label{Eq.3.3.fin} \, .
\end{equation}
%%%%%%%%%%%%%%%%%%%%%%%%%%%%%%%%%%%%%%%%%%%%%%%%%%%%%%%%%%%%%%%%%%%%%%%%%%%%%%
%
Differentiating Log[Eq.~(\ref{Eq.3.3.fin})] with respect to
${1}/{M_B^2}$, finally one obtains
%
%%%%%%%%%%%%%%%%%%%%%%%%%%%%%%%%%%%%%%%%%%%%%%%%%%%%%%%%%%%%%%%%%%%%%%%%%%%%%%
\begin{equation}
M^2_H = \frac{\int^{s_0}_0 e^{-s/M_B^2} s \rho(s) ds} {\int^{s_0}_0
e^{-s/M_B^2} \rho(s) ds}\, .
\end{equation}
%%%%%%%%%%%%%%%%%%%%%%%%%%%%%%%%%%%%%%%%%%%%%%%%%%%%%%%%%%%%%%%%%%%%%%%%%%%%%%
%

\subsubsection{Pentaquark currents}

The hidden-charm local pentaquark interpolating currents of spins $J
= {1\over2}/{3\over2}/{5\over2}$ were systematically constructed in
Refs.~\cite{Chen:2015moa,prepare} based on the results of
Refs.~\cite{Chen:2008qv,Chen:2009sf,Chen:2010ba,Chen:2011rh,Dmitrasinovic:2011yf,Chen:2013efa,Chen:2012vs,Chen:2012ex,Chen:2006hy,Chen:2006zh,Chen:2007xr,Chen:2012ut,Chen:2013gnu,Chen:2013jra,Chen:2014vha} which studied light baryon, tetraquark and dibaryon currents. We pick
out two of them and discuss their internal color and flavor
structure:
\begin{eqnarray}
\eta_{1\mu}^{c \bar c uud}(x) &=& [\bar c_d(x) \gamma_\mu c_d(x)]
[\epsilon_{abc} \left(u^T_a(x) \mathbb{C} d_b(x)\right) \gamma_5
u_c(x)] \, ,
\\ J^{\bar D^*\Sigma_c}_{\mu}(x) &=& [\bar c_d(x) \gamma_\mu d_d(x)]  [\epsilon_{abc} \left(u^T_a(x) \mathbb{C} \gamma_\nu u_b(x)\right) \gamma^\nu \gamma_5 c_c(x)]  \, .
\end{eqnarray}
Here, the sum over repeated indices ($\mu$, $\nu, \cdots$ for Dirac
spinor indices, and $a, b, \cdots$ for color indices) is taken; the
superscript $T$ represents the transpose of the Dirac indices only;
$\mathbb{C}$ is the charge-conjugation operator; $u(x)$, $d(x)$,
$c(x)$, and $\bar c(x)$ are the $up$, $down$, $charm$, and
$anti$-$charm$ quark fields at location $x$, respectively.

These two currents contain the same quark content $c \bar c uud$ and
have the same quantum number $J^P = 3/2^-$. However, their internal
structures are totally different. The first one $\eta_{1\mu}^{c \bar
c uud}$ consists of two color-singlet components: $\bar c_d
\gamma_\mu c_d$ and $\epsilon_{abc} (u^T_a C d_b) \gamma_5 u_c$,
which have quantum numbers $J^P = 1^-$ and $1/2^+$, and couple to
the $J/\psi$ and proton, respectively. The second one $J^{\bar
D^*\Sigma_c}_{\mu}$ consists of two color-singlet components: $\bar
c_d \gamma_\mu d_d$ and $\epsilon_{abc} (u^T_a C \gamma_\nu u_b)
\gamma^\nu \gamma_5 c_c$, which also have quantum numbers $J^P =
1^-$ and $1/2^+$, but couple to the $\bar D^*$ and $\Sigma_c$,
respectively. Hence, the $\eta_{1\mu}^{c \bar c uud}$ couples well to
the combination of the $J/\psi$ and $p$, and the $J^{\bar
D^*\Sigma_c}_{\mu}$ couples well to the combination of $\bar D^*$
and $\Sigma_c$.

These different internal structures suggest that the choice of
currents could be important when applying the method of QCD sum rule
to study multiquark states, in which cases there always exist many
currents of different color, flavor, orbit and spin
structures~\cite{prepare}. For example, if one wants to study a
physical pentaquark state of the molecular type, it would be best to
choose an interpolating current also of the molecular type to ensure
a large overlap with that state. Moreover, if one obtains a good sum
rule result by using such a current, the relevant state might be a
molecular state. But we note that even in this case it is difficult
to fully distinguish between a tightly-bound pentaquark structure
and a weakly-bound molecular structure, because all the quark and
anti-quark fields inside the current are at the same space-time
point.

Furthermore, the $\eta_{1\mu}^{c \bar c uud}$ and $J^{\bar
D^*\Sigma_c}_{\mu}$ still have some overlap and can be partly
related to each other through the color rearrangement and the Fierz
transformation (see the example given in Ref.~\cite{prepare}),
despite of their different internal structures. The color
rearrangement for pentaquarks was given in Ref.~\cite{Chen:2015moa}
\begin{eqnarray}
\delta^{de} \epsilon^{abc} &=& \delta^{da} \epsilon^{ebc} +
\delta^{db} \epsilon^{aec} + \delta^{dc} \epsilon^{abe} \, ,
\end{eqnarray}
which can be used to relate the two color configurations, $[\bar c_d
c_d][\epsilon^{abc}q_a q_b q_c]$ and $[\bar c_d
q_d][\epsilon^{abc}c_a q_b q_c]$. Two similar ones for tetraquarks
are given in Ref.~\cite{Chen:2006hy}
\begin{eqnarray}
\delta^{ad} \delta^{bc} &=& {1\over3} \delta^{ab} \delta^{cd} +
{1\over2}\lambda_n^{ab} \lambda_n^{cd} \, ,
\\ \lambda_n^{ad} \lambda_n^{bc} &=& {16\over9} \delta^{ab} \delta^{cd} - {1\over3}\lambda_n^{ab} \lambda_n^{cd} \, .\nonumber
\end{eqnarray}
Based on these formulae, the color structure used in
Ref.~\cite{Wang:2015epa,Wang:2015ava} can be simplified to
\begin{eqnarray}
\epsilon^{abf} \epsilon^{ceg} \epsilon^{dfg} &=& - \delta^{da}
\epsilon^{ebc} - \delta^{db} \epsilon^{aec} \, .
\end{eqnarray}
There are several formulae related to the Fierz transformation,
which were given and discussed in
Ref.~\cite{Chen:2006hy,Chen:2008qv,prepare}:
\begin{enumerate}

\item
Products of two Dirac matrices without Lorentz indices:
\begin{eqnarray}
\begin{tiny}
\left (
\begin{array}{l}
\mathbf{1} \otimes \gamma_5
\\ \gamma_\mu \otimes \gamma^\mu  \gamma_5
\\ \sigma_{\mu\nu} \otimes \sigma^{\mu\nu} \gamma_5
\\ \gamma_{\mu} \gamma_5 \otimes \gamma^{\mu}
\\ \gamma_5 \otimes \mathbf{1}
\end{array} \right )_{a b, c d}
= \left (
\begin{array}{lllll}
{1\over4} & - {1\over4} & {1\over8} & {1\over4} & {1\over4}
\\ - 1 & - {1\over2} & 0 & - {1\over2} & 1
\\ 3 & 0 & -{1\over2} & 0 & 3
\\ 1 & -{1\over2} & 0 & - {1\over2} & - 1
\\ {1\over4} & {1\over4} & {1\over8} & - {1\over4} & {1\over4}
\end{array} \right )
\left (
\begin{array}{l}
\mathbf{1} \otimes \gamma_5
\\ \gamma_\mu \otimes \gamma^\mu  \gamma_5
\\ \sigma_{\mu\nu} \otimes \sigma^{\mu\nu} \gamma_5
\\ \gamma_{\mu} \gamma_5 \otimes \gamma^{\mu}
\\ \gamma_5 \otimes \mathbf{1}
\end{array} \right )_{a d, b c}.
\end{tiny}
\end{eqnarray}

\item
Products of two Dirac matrices with one Lorentz index:
\begin{eqnarray}
\begin{tiny}\left (
\begin{array}{l}
\mathbf{1} \otimes \gamma^\mu
\\ \gamma^\mu \otimes \mathbf{1}
\\ \gamma_5 \otimes \gamma_\mu \gamma_5
\\ \gamma_\mu \gamma_5 \otimes \gamma_5
\\ \gamma^{\nu} \otimes \sigma_{\mu\nu}
\\ \sigma_{\mu\nu} \otimes \gamma^{\nu}
\\ \gamma^{\nu} \gamma_5 \otimes \sigma_{\mu\nu} \gamma_5
\\ \sigma_{\mu\nu} \gamma_5 \otimes \gamma^{\nu} \gamma_5
\end{array} \right )_{a b, c d}
= \left (
\begin{array}{llllllll}
{1\over4} & {1\over4} & {1\over4} & - {1\over4} & -{i\over4} &
{i\over4} & {i\over4} & {i\over4}
\\ {1\over4} & {1\over4} & -{1\over4} & {1\over4} & {i\over4} &
-{i\over4} & {i\over4} & {i\over4}
\\ {1\over4} & - {1\over4} & {1\over4} & {1\over4} & {i\over4} &
{i\over4} & -{i\over4} & {i\over4}
\\ -{1\over4} & {1\over4} & {1\over4} & {1\over4} & {i\over4} &
{i\over4} & {i\over4} & -{i\over4}
\\ {3i\over4} & -{3i\over4} & -{3i\over4} & -{3i\over4} & -{1\over4} &
-{1\over4} & -{1\over4} & {1\over4}
\\ -{3i\over4} & {3i\over4} & -{3i\over4} & -{3i\over4} & -{1\over4} &
-{1\over4} & {1\over4} & -{1\over4}
\\ -{3i\over4} & -{3i\over4} & {3i\over4} & -{3i\over4} & -{1\over4} &
{1\over4} & -{1\over4} & -{1\over4}
\\ -{3i\over4} & -{3i\over4} & -{3i\over4} & {3i\over4} & {1\over4} &
-{1\over4} & -{1\over4} & -{1\over4}
\end{array} \right )
\left (
\begin{array}{l}
\mathbf{1} \otimes \gamma^\mu
\\ \gamma^\mu \otimes \mathbf{1}
\\ \gamma_5 \otimes \gamma_\mu \gamma_5
\\ \gamma_\mu \gamma_5 \otimes \gamma_5
\\ \gamma^{\nu} \otimes \sigma_{\mu\nu}
\\ \sigma_{\mu\nu} \otimes \gamma^{\nu}
\\ \gamma^{\nu} \gamma_5 \otimes \sigma_{\mu\nu} \gamma_5
\\ \sigma_{\mu\nu} \gamma_5 \otimes \gamma^{\nu} \gamma_5
\end{array} \right )_{a d, b c}.
\end{tiny}
\end{eqnarray}

\item
Products of two Dirac matrices with two anti-symmetric Lorentz
indices:
\begin{eqnarray}
\begin{tiny}\left (
\begin{array}{l}
\mathbf{1} \otimes \sigma_{\mu\nu} \gamma_5
\\ \gamma_5 \otimes \sigma_{\mu\nu}
\\ \sigma_{\mu\nu} \otimes \gamma_5
\\ \sigma_{\mu\nu} \gamma_5 \otimes \mathbf{1}
\\ \epsilon_{\mu\nu\rho\sigma} \sigma_{\rho l} \otimes \sigma_{\sigma l}
\\ \gamma_\mu \otimes \gamma_\nu \gamma_5 - (\mu \leftrightarrow \nu)
\\ \gamma_\mu \gamma_5 \otimes \gamma_\nu - (\mu \leftrightarrow \nu)
\\ \epsilon_{\mu\nu\rho\sigma} \gamma_\rho \otimes \gamma_\sigma
\\ \epsilon_{\mu\nu\rho\sigma} \gamma_\rho \gamma_5 \otimes \gamma_\sigma \gamma_5
\end{array} \right )_{a b, c d}
= \left (
\begin{array}{lllllllll}
{1\over4} & {1\over4} & {1\over4} & {1\over4} & {1\over4} &
{i\over4} & - {i\over4} & {1\over4} & - {1\over4}
\\ {1\over4} & {1\over4} & {1\over4} & {1\over4} & {1\over4} &
- {i\over4} & {i\over4} & - {1\over4} & {1\over4}
\\ {1\over4} & {1\over4} & {1\over4} & {1\over4} & - {1\over4} &
- {i\over4} & {i\over4} & {1\over4} & - {1\over4}
\\ {1\over4} & {1\over4} & {1\over4} & {1\over4} & - {1\over4} &
{i\over4} & - {i\over4} & - {1\over4} & {1\over4}
\\ 1 & 1 & - 1 & - 1 & 0 & 0 & 0 & 0 & 0
\\ - {i\over2} & {i\over2} & {i\over2} & - {i\over2} & 0 &
0 & 0 & {i\over2} & {i\over2}
\\ {i\over2} & - {i\over2} & - {i\over2} & {i\over2} & 0 &
0 & 0 & {i\over2} & {i\over2}
\\ {1\over2} & - {1\over2} & {1\over2} & - {1\over2} & 0 &
-{i\over2} & - {i\over2} & 0 & 0
\\ - {1\over2} & {1\over2} & - {1\over2} & {1\over2} & 0 &
-{i\over2} & - {i\over2} & 0 & 0
\end{array} \right )
\left (
\begin{array}{l}
\mathbf{1} \otimes \sigma_{\mu\nu} \gamma_5
\\ \gamma_5 \otimes \sigma_{\mu\nu}
\\ \sigma_{\mu\nu} \otimes \gamma_5
\\ \sigma_{\mu\nu} \gamma_5 \otimes \mathbf{1}
\\ \epsilon_{\mu\nu\rho\sigma} \sigma_{\rho l} \otimes \sigma_{\sigma l}
\\ \gamma_\mu \otimes \gamma_\nu \gamma_5 - (\mu \leftrightarrow \nu)
\\ \gamma_\mu \gamma_5 \otimes \gamma_\nu - (\mu \leftrightarrow \nu)
\\ \epsilon_{\mu\nu\rho\sigma} \gamma_\rho \otimes \gamma_\sigma
\\ \epsilon_{\mu\nu\rho\sigma} \gamma_\rho \gamma_5 \otimes \gamma_\sigma \gamma_5
\end{array} \right )_{a d, b c}.
\end{tiny}
\end{eqnarray}

\item
Products of two Dirac matrices with two symmetric Lorentz indices:
\begin{eqnarray}
\begin{tiny}
\left (
\begin{array}{l}
g_{\mu\nu} \mathbf{1} \otimes \mathbf{1}
\\ g_{\mu\nu} \gamma_\rho \otimes \gamma^\rho
\\ g_{\mu\nu} \sigma_{\rho\sigma} \otimes \sigma^{\rho\sigma}
\\ g_{\mu\nu} \gamma_{\rho} \gamma_5 \otimes \gamma^{\rho} \gamma_5
\\ g_{\mu\nu} \gamma_5 \otimes \gamma_5
\\ \gamma_\mu \otimes \gamma_\nu + (\mu \leftrightarrow \nu)
\\ \gamma_\mu \gamma_5 \otimes \gamma_\nu \gamma_5 + (\mu \leftrightarrow \nu)
\\ \sigma_{\mu \rho} \otimes \sigma_{\nu \rho} + (\mu \leftrightarrow \nu)
\end{array} \right )_{a b, c d}
= \left (
\begin{array}{llllllll}
{1\over4} & {1\over4} & {1\over8} & -{1\over4} & {1\over4} & 0 & 0 &
0
\\ 1 & -{1\over2} & 0 & -{1\over2} & -1 & 0 & 0 & 0
\\ 3 & 0 & -{1\over2} & 0 & 3 & 0 & 0 & 0
\\ -1 & -{1\over2} & 0 & -{1\over2} & 1 & 0 & 0 & 0
\\ {1\over4} & -{1\over4} & {1\over8} & {1\over4} & {1\over4} & 0 & 0 & 0
\\ {1\over2} & -{1\over2} & {1\over4} & -{1\over2} & -{1\over2} & {1\over2} & {1\over2} & -{1\over2}
\\ -{1\over2} & -{1\over2} & -{1\over4} & -{1\over2} & {1\over2} & {1\over2} & {1\over2} & {1\over2}
\\ {3\over2} & {1\over2} & -{1\over4} & -{1\over2} & {3\over2} & -1 & 1 & 0
\end{array} \right )
\left (
\begin{array}{l}
g_{\mu\nu} \mathbf{1} \otimes \mathbf{1}
\\ g_{\mu\nu} \gamma_\rho \otimes \gamma^\rho
\\ g_{\mu\nu} \sigma_{\rho\sigma} \otimes \sigma^{\rho\sigma}
\\ g_{\mu\nu} \gamma_{\rho} \gamma_5 \otimes \gamma^{\rho} \gamma_5
\\ g_{\mu\nu} \gamma_5 \otimes \gamma_5
\\ \gamma_\mu \otimes \gamma_\nu + (\mu \leftrightarrow \nu)
\\ \gamma_\mu \gamma_5 \otimes \gamma_\nu \gamma_5 + (\mu \leftrightarrow \nu)
\\ \sigma_{\mu \rho} \otimes \sigma_{\nu \rho} + (\mu \leftrightarrow \nu)
\end{array} \right )_{a d, b c}.
\end{tiny}
\end{eqnarray}

\end{enumerate}
We note that these equations only change the Lorentz structures, and
the minus sign due to the exchange of quark fields is not included
yet.

\subsubsection{Operator Product Expansion}

After the current is fixed, one can calculate the correlation
function, Eq.~(\ref{Eq.3.3.pi}). For example
%
%%%%%%%%%%%%%%%%%%%%%%%%%%%%%%%%%%%%%%%%%%%%%%%%%%%%%%%%%%%%%%%%%%%%%%%%%%%%%%
\begin{eqnarray}
\Pi_{\mu\nu}^{\bar D^*\Sigma_c}(q^2) &=& i \int d^4x e^{iqx} \langle
0 | T J^{\bar D^*\Sigma_c}_{\mu}(x) \bar J^{\bar
D^*\Sigma_c}_{\nu}(0) | 0 \rangle
\\ \nonumber &=& 2 \epsilon_{a_1b_1c_1} \epsilon_{a_2b_2c_2} \times {\rm Tr}[ \mbox{S}^{d_1 d_2}_d(x) \gamma_{\nu} \mbox{S}^{d_2 d_1}_c(-x) \gamma_{\mu} ] \times {\rm Tr}[ \mbox{S}^{b_1 b_2}_u(x) \gamma_{\rho_2} \mathbb{C} \mbox{S}^{a_1 a_2}_u(x) \mathbb{C} \gamma_{\rho_1} ] \times \gamma^{\rho_1} \gamma_5 \mbox{S}^{c_1 c_2}_c(x) \gamma^{\rho_2} \gamma_5 \, .
\label{Eq.3.3.contract}
\end{eqnarray}
%%%%%%%%%%%%%%%%%%%%%%%%%%%%%%%%%%%%%%%%%%%%%%%%%%%%%%%%%%%%%%%%%%%%%%%%%%%%%%
%
Then the light quark propagator can be expanded in the OPE expansion
in the fixed-point gauge as
%
%%%%%%%%%%%%%%%%%%%%%%%%%%%%%%%%%%%%%%%%%%%%%%%%%%%%%%%%%%%%%%%%%%%%%%%%%%%%%%
\begin{eqnarray}\nonumber
\mbox{S}^{ab}_q(x)&\equiv&\langle0|\mbox{T}[q^a(x)\bar{q}^b(0)]|0\rangle
\\ \nonumber &=&\frac{i\delta^{ab}}{2\pi^2x^4} x\!\!\!\slash
+\frac{i}{32\pi^2}\frac{\lambda^n_{ab}}{2}\mbox{g}_c\mbox{G}^n_{\mu\nu}\frac{1}{x^2}(\sigma^{\mu\nu}
x\!\!\!\slash + x\!\!\!\slash \sigma^{\mu\nu})
-\frac{\delta^{ab}}{12}\langle\bar{q}q\rangle
+\frac{\delta^{ab}x^2}{192}\langle g_c\bar{q}\sigma Gq\rangle
\\ &&-\frac{\delta^{ab}m_q}{4\pi^2x^2}
+\frac{i\delta^{ab}m_q }{48}\langle\bar{q}q\rangle x\!\!\!\slash
+\frac{i\delta^{ab}m_q^2}{8\pi^2x^2} x\!\!\!\slash \, ,
\end{eqnarray}
%%%%%%%%%%%%%%%%%%%%%%%%%%%%%%%%%%%%%%%%%%%%%%%%%%%%%%%%%%%%%%%%%%%%%%%%%%%%%%
%
and the heavy quark propagator can be expanded as
%
%%%%%%%%%%%%%%%%%%%%%%%%%%%%%%%%%%%%%%%%%%%%%%%%%%%%%%%%%%%%%%%%%%%%%%%%%%%%%%
\begin{eqnarray}\nonumber
\mbox{S}^{ab}_c(p)&\equiv& \int {d^4x} e^{i p \cdot x}
\mbox{S}^{ab}_c(x) = \int {d^4x} e^{i p \cdot x}
\langle0|\mbox{T}[c^a(x)\bar{c}^b(0)]|0\rangle
\\ &=& i {p\!\!\!\slash + m_c \over p^2 - m_c^2} \delta^{ab}
+
\frac{i}{4}\frac{\lambda^n_{ab}}{2}\mbox{g}_c\mbox{G}^n_{\mu\nu}\frac{\sigma^{\mu\nu}
(p\!\!\!\slash + m_c) + (p\!\!\!\slash + m_c) \sigma^{\mu\nu}}{(p^2
- m_c^2)^2} + \frac{i\delta^{ab}
m_c}{12}\langle\mbox{g}_c^2\mbox{G}^2\rangle { p^2 + m_c
p\!\!\!\slash \over (p^2 - m_c^2)^4} \, .
\end{eqnarray}
%%%%%%%%%%%%%%%%%%%%%%%%%%%%%%%%%%%%%%%%%%%%%%%%%%%%%%%%%%%%%%%%%%%%%%%%%%%%%%
%
After inserting these two equations into Eq.~(\ref{Eq.3.3.contract}) and
performing the Borel transformation
\begin{eqnarray}
\mathcal{B}_{M_B^2} [\Pi(q^2)] &=& \lim_{
\begin{tiny}
\begin{array}{c}
-q^2,n\to\infty
\\ -q^2/n = M_B^2
\end{array}
\end{tiny}}
{(-q^2)^{n+1} \over n\!} \left( { d \over dq^2} \right)^n \Pi(q^2)
\, ,
\end{eqnarray}
we can obtain the spectral density $\rho(s)$. In
Ref.~\cite{Chen:2015moa}, $\rho(s)$ was evaluated up to dimension
eight, including the perturbative term, the quark condensate
$\langle \bar q q \rangle$, the gluon condensate $\langle g_s^2 GG
\rangle$, the quark-gluon mixed condensate $\langle g_s \bar q
\sigma G q \rangle$, and their combinations $\langle \bar q q
\rangle^2$ and $\langle \bar q q \rangle\langle g_s \bar q \sigma G
q \rangle$. For example, when the pentaquark current $J^{\bar
D^*\Sigma_c}_{\mu}(x)$ is used, one can obtain
\begin{eqnarray}
\rho^{\bar D^*\Sigma_c}_{\mu\nu}(s) &=& \mathbf{1} \times g_{\mu\nu}
\times m_c \times \left (
\rho^{pert}_1(s)+\rho^{\qq}_1(s)+\rho^{\GGa}_1(s)+\rho^{\qq^2}_1(s)+\rho^{\qGqa}_1(s)+\rho^{\qq\qGqa}_1(s)
\right )
\\ \nonumber &+& q\!\!\!\slash\times g_{\mu\nu} \times \left ( \rho^{pert}_2(s)+\rho^{\qq}_2(s)+\rho^{\GGa}_2(s)+\rho^{\qq^2}_2(s)+\rho^{\qGqa}_2(s)+\rho^{\qq\qGqa}_2(s) \right )
+ \cdots \, ,
\end{eqnarray}
where $\cdots$ denotes other Lorentz structures, such as $\mathbf{1}
\times \sigma_{\mu\nu}$, etc.. The expressions of the spectral
densities can be found in Ref.~\cite{Chen:2015moa}.

\subsubsection{Parity of Pentaquarks}

In the previous section, two sum rules are obtained when one
pentaquark current $J^{\bar D^*\Sigma_c}_{\mu}(x)$ is used. One is
proportional to $\mathbf{1} \times g_{\mu\nu}$, and the other is
proportional to $q\!\!\!\slash \times g_{\mu\nu}$. They can be used
to calculate the mass of the pentaquark and determine its parity at
the same time. This technique was used in Ref.~\cite{Chen:2015moa},
and can be applied for other baryons and pentaquarks.

Although a pentaquark current has definite parity, it can couple to
states of both positive and negative parities via (see discussions in
Refs.~\cite{Chung:1981cc,Jido:1996ia,Kondo:2005ur,Ohtani:2012ps}):
\begin{eqnarray}
\langle 0 | J | H \rangle &=& f_H u(p) \, , \label{Eq.3.3.gamma0}
\\ \langle 0 | J | H^\prime \rangle &=& f_{H^\prime} \gamma_5 u^\prime(p) \, ,
\label{Eq.3.3.gamma5}
\end{eqnarray}
where $| H \rangle$ has the same parity as $J$, and $| H^\prime
\rangle$ has the opposite parity. Oppositely, the current $J$ and
its partner $\gamma_5 J$ can also couple to the same state $H$.
%Indeed, they just lead to the same sum rule result, which will be discussed in the following.

In Ref.~\cite{Chen:2015moa}, the non-$\gamma_5$ coupling in
Eq.~(\ref{Eq.3.3.gamma0}) was used,
\begin{eqnarray}
\langle 0 |J^{\bar D^*\Sigma_c}_{\mu} | [\bar D^*\Sigma_c] \rangle
&=& f_{\bar D^*\Sigma_c} u_\mu (p) \, .
\end{eqnarray}
Then the two-point correlation functions can be written as:
%%%%%%%%%%%%%%%%%%%%%%%%%%%%%%%%%%%%%%%%%%%%%%%%%%%%%%%%%%%%%%%%%%%%%%%%%%%%%%
\begin{eqnarray}
\label{Eq.3.3.piDsS32} \Pi^{\bar D^*\Sigma_c}_{\mu \nu}\left(q^2\right)
\nonumber &=& i \int d^4x e^{iq\cdot x} \langle 0 | T\left[J^{\bar
D^*\Sigma_c}_{\mu}(x) \bar J_{\nu}^{\bar D^*\Sigma_c} (0)\right] | 0
\rangle
\\ &=& \left(\frac{q_\mu q_\nu}{q^2}-g_{\mu\nu}\right) (q\!\!\!\slash + M_H) \Pi^{\bar D^*\Sigma_c}\left(q^2\right) + \cdots \, ,
\end{eqnarray}
%%%%%%%%%%%%%%%%%%%%%%%%%%%%%%%%%%%%%%%%%%%%%%%%%%%%%%%%%%%%%%%%%%%%%%%%%%%%%%
where the spin $1/2$ components are all contained in $\cdots$, such
as $q_\mu q_\nu (q\!\!\!\slash + m) \Pi_{1/2}^{\bar
D^*\Sigma_c}\left(q^2\right)$, etc.

One can also use the $\gamma_5$ couplings in Eq.~(\ref{Eq.3.3.gamma5}).
The resulting two-point correlation function is similar to
Eq.~(\ref{Eq.3.3.piDsS32}), but with $(q\!\!\!\slash + M_H)$ replaced by
$(- q\!\!\!\slash + M_H)$. This difference would tell us the parity
of $H$. If the two sum rules from these two tensor structures lead to
almost the same numerical results, the current $J^{\bar
D^*\Sigma_c}_{\mu}(x)$ couples to a state having the same parity,
that is $P = -$. We note that the result does not change when using
$\gamma_5 J^{\bar D^*\Sigma_c}_{\mu}$ having the opposite parity.

\subsubsection{Numerical results and discussions}

\begin{figure}[hbtp]
\begin{center}
\includegraphics[width=6cm]{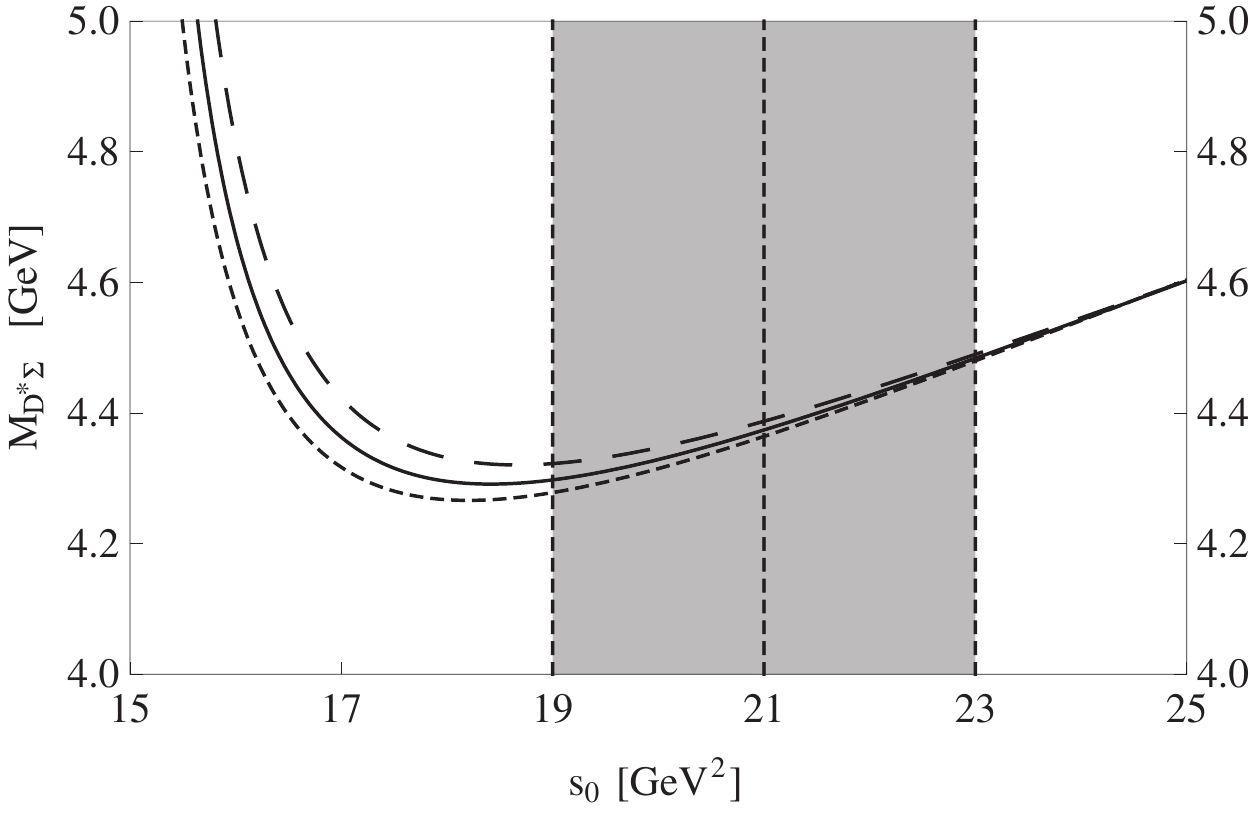}
\includegraphics[width=6cm]{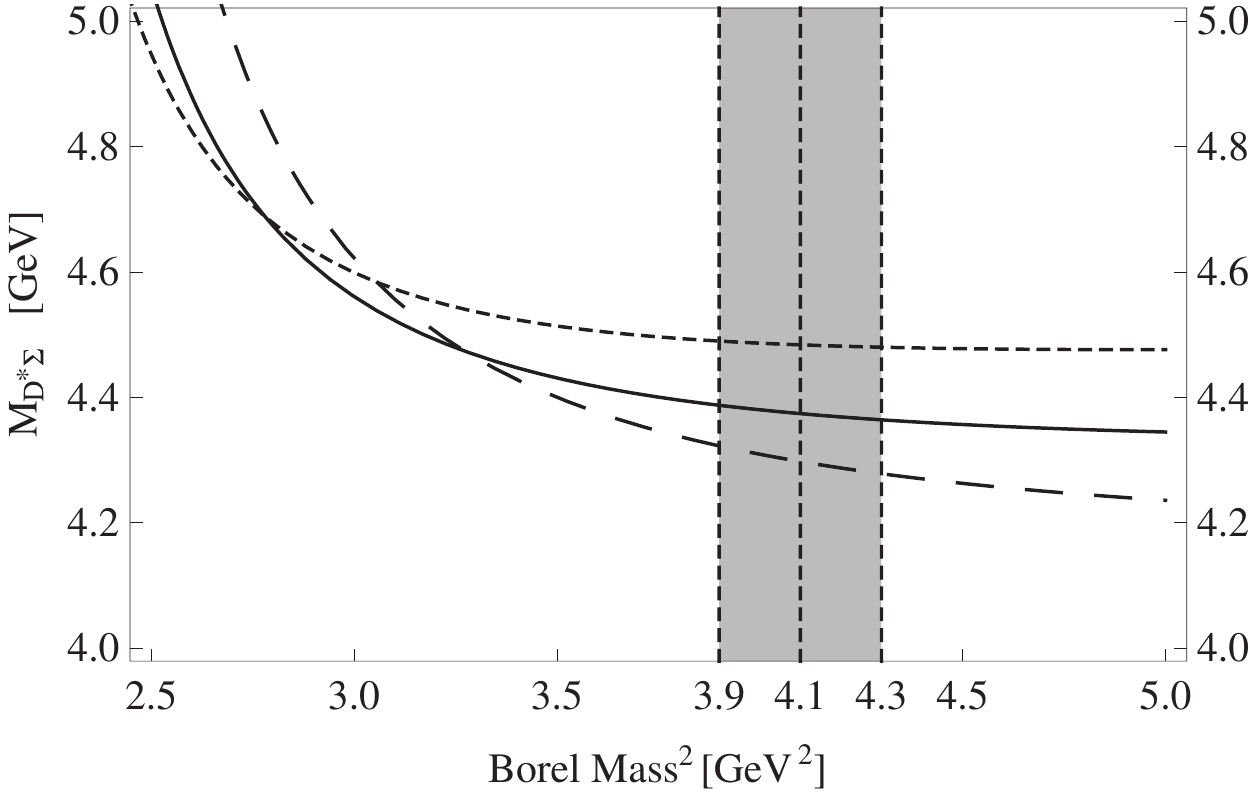}
\caption{The variation of $M_{[\bar D^*\Sigma_c]}$ with respect to
the threshold value $s_0$ (left) and the Borel mass $M_B$ (right),
taken from Ref.~\cite{Chen:2015moa}.} \label{Fig3.3.pentaquarksumrule1}
\end{center}
\end{figure}

Several currents were used in Ref.~\cite{Chen:2015moa} to perform
QCD sum rule analyses. The current
\begin{eqnarray}
J^{\bar D^*\Sigma_c}_{\mu} &=& [\bar c_d \gamma_\mu d_d]
[\epsilon_{abc} (u^T_a C \gamma_\nu u_b) \gamma^\nu \gamma_5 c_c]
\, ,
\end{eqnarray}
was used to obtain:
\begin{eqnarray}
M_{[\bar D^*\Sigma_c],{3/2^-}} = 4.37^{+0.19}_{-0.12} \mbox{ GeV} \,
.
\end{eqnarray}
The results are shown in Fig.~\ref{Fig3.3.pentaquarksumrule1}. 
The mass value is consistent with the
experimental results of the $P_c(4380)$~\cite{Aaij:2015tga},
supporting it to be a $[\bar D^*\Sigma_c]$ hidden-charm pentaquark.
Its quantum numbers are evaluated to be $J^P=3/2^-$ at the same time. 
This $[\bar D^*\Sigma_c]$
structure may be interpreted as a tightly-bound pentaquark structure
or a $[\bar D^*\Sigma_c]$ molecular state. But in both cases, it can
easily decay into $\bar D^*\Sigma_c$ final states if its mass is
above the $\bar D^*\Sigma_c$ threshold. Moreover, the current
$J^{\bar D^*\Sigma_c}_{\mu}$ has some overlap with $\eta_{1\mu}^{c
\bar c uud}(x)$, suggesting that it can also decay into S-wave
$J/\psi p$ final states.

\begin{figure}[hbtp]
\begin{center}
\includegraphics[width=6cm]{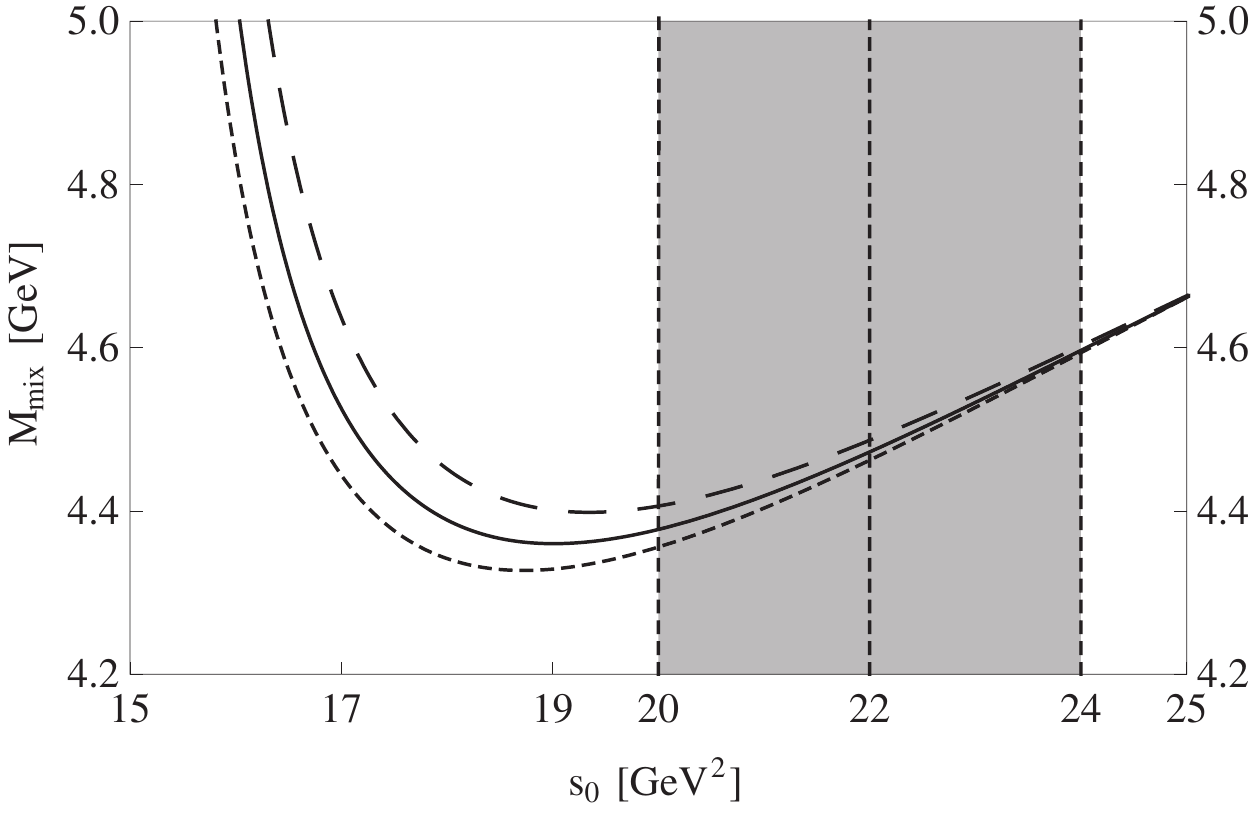}
\includegraphics[width=6cm]{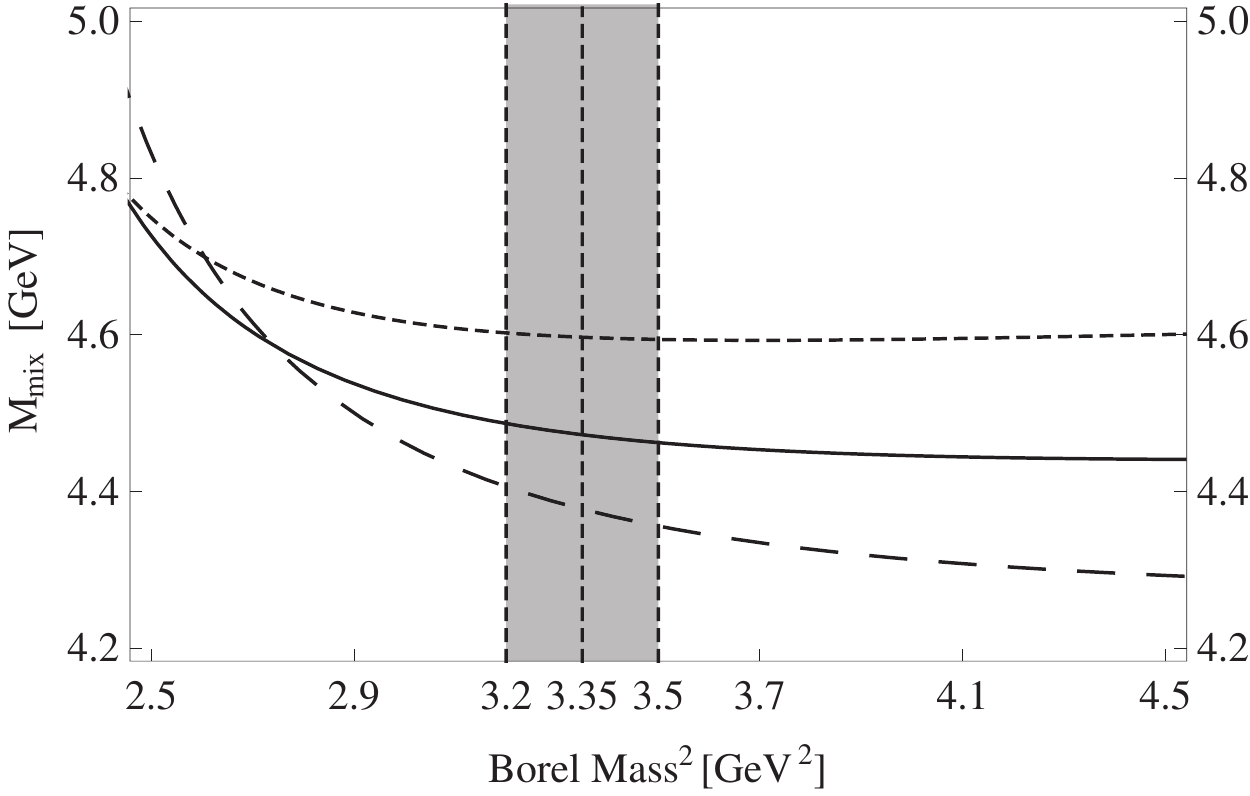}
\caption{The variation of $M_{[\bar D\Sigma_c^*\&\bar
D^*\Lambda_c]}$ with respect to the threshold value $s_0$ (left) and
the Borel mass $M_B$ (right).}
\label{Fig3.3.pentaquarksumrule2}
\end{center}
\end{figure}

A mixed current consisting of $J^{\bar D\Sigma_c^*}_{\{\mu\nu\}}$
and $J^{\bar D^*\Lambda_c}_{\{\mu\nu\}}$ was used in
Ref.~\cite{Chen:2015moa}:
\begin{eqnarray}
J^{\bar D\Sigma_c^*\&\bar D^*\Lambda_c}_{\{\mu\nu\}} = \sin\theta
\times J^{\bar D\Sigma_c^*}_{\{\mu\nu\}} + \cos\theta \times J^{\bar
D^*\Lambda_c}_{\{\mu\nu\}} \, ,
\end{eqnarray}
where
\begin{eqnarray}
J^{\bar D\Sigma_c^*}_{\{\mu\nu\}} &=& [\bar c_d \gamma_\mu \gamma_5
d_d] [\epsilon_{abc} (u^T_a C \gamma_\nu u_b) c_c] + \{ \mu
\leftrightarrow \nu \} \, ,
\\ J^{\bar D^*\Lambda_c}_{\{\mu\nu\}} &=& [\bar c_d \gamma_\mu u_d] [\epsilon_{abc} (u^T_a C \gamma_\nu \gamma_5 d_b) c_c] + \{ \mu \leftrightarrow \nu \} \, .
\end{eqnarray}
The mixing angle $\theta$ was fine-tuned to be $-51\pm5^\circ$, and
the hadron mass was extracted as
\begin{eqnarray}
M_{[\bar D\Sigma_c^*\&\bar D^*\Lambda_c],{5/2^+}} =
4.47^{+0.20}_{-0.13} \mbox{ GeV} \, .
\end{eqnarray}
The results are shown in Fig.~\ref{Fig3.3.pentaquarksumrule2}. 
The mass value is consistent with the
experimental results of the $P_c(4450)$~\cite{Aaij:2015tga},
supporting it to be an admixture of $[\bar D^* \Lambda_c]$ and
$[\bar D \Sigma_c^*]$. Its quantum numbers are evaluated
to be $J^P=5/2^+$ at the same time. According to
its internal structure described by $J^{\bar D\Sigma_c^*\&\bar
D^*\Lambda_c}$, its main decay modes include the P-wave $\bar D^*
\Lambda_c$ and $\bar D \Sigma_c^*$. Moreover, the P-wave $J/\psi p$
decay mode is also possible.

According to Ref.~\cite{Chen:2015moa}, there may also exist a $[\bar
D\Sigma_c^*]$ hidden-charm pentaquark and a $[\bar D^*\Sigma_c^*]$
one, having masses around 4.5 GeV, and two hidden-bottom pentaquarks
as partners of the $P_c(4380)$ and $P_c(4450)$, having masses around
11.6 GeV. Their possible decay modes were also discussed. The same
approach was used  to systematically study hidden-charm pentaquarks
of spins $J = {1\over2}/{3\over2}/{5\over2}$~\cite{prepare}, where
a full classification of the local hidden-charm pentaquark
interpolating currents is given.

Besides Refs.~\cite{Chen:2015moa,prepare}, Wang applied the method
of QCD sum rule to study the hidden-charm pentaquarks $P_c(4380)$
and $P_c(4450)$, but used the diquark-diquark-antiquark type
interpolating currents~\cite{Wang:2015epa,Wang:2015ava}
\begin{eqnarray}
\epsilon^{ila} \bar c_a [ \epsilon^{ijk} u_j d_k][\epsilon^{lmn} u_m
c_n] \, .
\end{eqnarray}
His results also favor assigning the $P_c(4380)$ and $P_c(4450)$ to
be the $3/2^-$ and $5/2^+$ pentaquark states, respectively.

\subsection{Tightly bound pentaquark state in the quark model}
\label{sect:3.4}
%=====================================================================================
%=====================================================================================
%

In the previous subsection we reviewed the application of the QCD sum
rule to the hidden-charm pentaquarks. There are many hidden-charm
pentaquark interpolating currents, suggesting the internal
color-flavor structures of hidden-charm pentaquarks can be
extraordinarily complicated. The uncertainties in the masses are very large, 
of the order of 300 MeV. Among these various internal
structures, the tightly bound pentaquark state is very interesting
where the five quarks are confined within one MIT bag. In this
subsection we review those models which are based on this
interpretation. Even for the tightly bound pentaquark picture, there
exist many variations, such as the antiquark-diquark-diquark picture
and the diquark-triquark picture etc.

Recall that the mass gap between $\Lambda_c$ and $\Sigma_c$ is
around 170 MeV, which indicates that the attraction between the up
and down quarks is strong when the $[ud]$ pair stays in the scalar
isoscalar color anti-triplet configuration. The diquark or triquark
is sometimes used to denote the spin, isospin and color correlation
between the quarks. From the very beginning, we want to emphasize
that there do not exist point-like or extremely compact colored
building blocks such as diquarks or triquarks. Within the nucleons,
the scalar and axial vector diquarks transform into each other
freely. Moreover, all the tetraquark or pentaquark color
configurations can be rigorously decomposed into the sum of a series
of product of two color-singlet hadrons, as shown in the previous
subsection.

\subsubsection{Chiral quark model} \label{Sec:3.3.2}
%================================================================================
%================================================================================

The constituent quark model is very successful in the description of
the static properties of hadrons. On the other hand, chiral symmetry
of QCD and its spontaneous breaking play a pivotal role in the low
energy sector. The chiral quark model includes the interaction
between the constituent quark and chiral fields ($\pi$, $K$, $\eta$)
besides the quark gluon interaction. Replacing the nucleon fields in
the linear $\sigma$ model \cite{GellMann:1960np} with quarks, one
gets the SU(2) $\sigma$ model at the quark level
\cite{Fernandez:1993hx}. The interaction term reads
\begin{eqnarray}
{\cal L}_I=-g\bar{\psi}(\sigma+i\gamma_5\boldsymbol{\tau}\cdot\boldsymbol{\pi})\psi,
\end{eqnarray}
where $\boldsymbol{\tau}$ denotes the Pauli matrix and the coupling
constant $g$ can be related to the nucleon-nucleon-pion coupling
constant $g_{N N \pi}$. Zhang {\it et al.} extended this model to the three
flavor case and named it as the chiral SU(3) quark model
\cite{Zhang:1994pp,Zhang:1997ny}
\begin{eqnarray}
\left(\begin{array}{c}u\\d\end{array}\right)&\to&\left(\begin{array}{c}u\\d\\s\end{array}\right),\\
(\sigma+i\gamma_5\boldsymbol{\tau}\cdot\boldsymbol{\pi}) &\to &
(\sum_{a=0}^8\lambda_a\sigma_a+i\sum_{a=0}^8\lambda_a\pi_a),
\end{eqnarray}
where $\lambda_0$ is the unit matrix and $\lambda_i(i=1,\cdots,8)$
are Gell-Mann matrices.

To figure out whether the one gluon exchange (OGE) or vector meson
exchange is a right mechanism for the short range strong
interaction, they further extended the model to the case involving
vector mesons, which was called the extended chiral SU(3) quark
model \cite{Zhang:2003fb}. This version of the chiral quark model
contains the one-boson-exchange potential, one-gluon-exchange
potential, and confinement potential. For the hadron-hadron
interactions, the short-range quark exchange effects between two
hadrons were also taken into account.

Both of the three-flavor chiral quark models can reasonably describe
the spectrum of the ground state baryons, the binding energy of the
deuteron, the baryon-baryon scattering data, and the kaon-nucleon
scattering phase shifts etc. Within this framework, various
di-hadron bound states were investigated such as $\Delta\Delta$
\cite{Yuan:1999pg}, $\Omega\Omega$ \cite{Dai:2006gs},
$N\bar{\Omega}$ \cite{Zhang:2006dy}, $\Delta K$
\cite{Huang:2004ke,Huang:2005jf}, $N\phi$ \cite{Huang:2005gw},
$\Sigma K$ \cite{Huang:2005cj}, $\Omega\pi$ \cite{Wang:2006jg}, and
$\Omega\omega$ \cite{Wang:2007bf}.

The chiral SU(3) quark model and its extended version were also
employed to study the heavy quark systems
\cite{Liu:2008qb,Liu:2008mi}. In Ref. \cite{Wang:2011rga}, the
authors studied the $\Sigma_c\bar{D}$ and $\Lambda_c\bar{D}$ systems
by solving a resonating group method (RGM) equation, which is a
well-established method for studying interactions between composite
particles or quark clusters \cite{Oka:1981ri}. The extension to the
heavy quark case introduces additional parameters: the charm quark
mass, a coupling constant of OGE, and confinement strengths
involving the charm quark. Except the charm quark mass, these
parameters can be fixed with the masses of heavy quark hadrons
$\Sigma_c$, $\Lambda_c$, $D$, $D^*$, $J/\psi$, and $\eta_c$. The
heavy quark mass is treated as an adjustable parameter.

Before LHCb's discovery of the $P_c(4380)$ and $P_c(4450)$
resonances, there were some explorations of the hidden-charm
pentaquarks in chiral constituent quark model \cite{Wang:2011rga}.
In Ref. \cite{Wang:2011rga}, the authors performed a dynamical
investigation of the $\Sigma_c\bar D$ and $\Lambda_c\bar D$ systems
with spin $S=1/2$ and isospin $I=1/2$ by solving a resonating group
method equation. They calculated the effective Hamiltonian of the
$\Sigma_c\bar D$ and $\Lambda_c\bar D$ systems in three different
models (according to the different coupling between the vector meson
field and the quark field). They found that the interaction between
$\Sigma_c$ and $\bar D$ is attractive and a $\Sigma_c\bar D$ bound
state can be formed by considering either the linear confinement
potential or the quadratic one. The energy of the $\Sigma_c\bar D$
bound state was predicted as $4.279 - 4.312$ GeV in the linear
confinement potential. They found no $\Lambda_c\bar D$ bound state
due to the repulsive interaction between $\Lambda_c$ and $\bar D$.
In a coupled-channel calculation, they found that the
coupled-channel effect of the $\Sigma_c\bar D$ and $\Lambda_c\bar D$
can be negligible due to the large mass difference between the
$\Sigma_c\bar D$ and $\Lambda_c\bar D$ thresholds and the small
off-diagonal matrix elements of the $\Sigma_c\bar D$ and
$\Lambda_c\bar D$. Thus, there was no $\Sigma_c\bar
D$-$\Lambda_c\bar D$ resonance as a result. These predicted
$\Sigma_c\bar{D}$ bound states are potential candidates of
hidden-charm pentaquarks, although they have lower spin than the
observed $P_c(4380)$ and $P_c(4450)$ resonances.

Inspired by the discovery of the $P_c(4380)$ and $P_c(4450)$ states
in LHCb \cite{Aaij:2015tga}, further investigation of the
hidden-charm pentaquarks in the chiral quark model was performed in Ref.
\cite{Huang:2015uda} by considering the quark delocalization color
screening effect to solve the RGM equation.

It is known that electron delocalization contributes to the
formation of chemical bonds in molecular physics. Noticing the
similarity between the force of hadrons and that of atoms, quarks
confined in one nucleon are speculated to delocalize into another
nucleon \cite{Wang:1992wi}, which leads to the screened confinement
potential between quarks belonging to different nucleons. Along this
line, there were many investigations on the various aspects of
hadron properties
\cite{Wu:1996fm,Ping:1998si,Wu:1998wu,Ping:2000cb,Pang:2001xx}. It
was shown that the modified confinement potential gives an
equivalent description for the contribution from the hidden-color
channels \cite{Huang:2011kf}.

The modified chiral quark model used in Ref. \cite{Huang:2015uda}
contains the $\pi$-exchange potential, OGE potential, and a quark
delocalization color screening confinement potential, which also
describes well the $NN$ and $YN$ interactions. For the heavy quark
systems, one has to determine the heavy quark mass and a color
screening parameter. The former is fixed from heavy hadron masses
while the latter is treated as an adjustable parameter. All the
other parameters are determined in the studies of the light quark
systems \cite{Chen:2007qn,Ping:2008tp,Ping:2000dx,Chen:2011zzb}.

The authors of Ref. \cite{Huang:2015uda} introduced a
phenomenological color screening confinement potential in the model
Hamiltonian with the color screening parameter as an adjustable
parameter. They investigated the possible hidden-charm molecular
pentaquarks composed of the $D/D^*$ and
$\Sigma_c/\Sigma_c^*/\Lambda_c$ with quantum numbers $Y=1,
I=\frac{1}{2}, \frac{3}{2}$ and $J^P=\frac{1}{2}^-, \frac{3}{2}^-,
\frac{5}{2}^-$. For the $IJ^P=\frac{1}{2}\frac{1}{2}^-,
\frac{1}{2}\frac{3}{2}^-$ systems. They found that the potentials
are all attractive for the channels $\Sigma_c\bar D$,
$\Sigma_c^*\bar D$ and $\Sigma_c^*\bar D^*$ but repulsive for the
channels $\Lambda_c\bar D$ and $\Lambda_c\bar D^*$. For the
$IJ^P=\frac{1}{2}\frac{5}{2}^-$ systems, only the $\Sigma_c^*\bar
D^*$ channel has an attractive potential. For the isospin
$I=\frac{3}{2}$ systems, all channels are repulsive except the
$\Sigma_c^*\bar D$ and $\Sigma_c^*\bar D^*$ which have a very week
attractive potential. The binding energies of the attractive
channels were calculated by solving the RGM equation and the masses
were obtained. They also considered the channel coupling effects in
the evaluations. They concluded that the $P_c(4380)$ was a mixed
structure of the $\Lambda_c\bar D^*$, $\Sigma_c^*\bar D$,
$\Sigma_c^*\bar D$ and $\Sigma_c^*\bar D^*$ with
$IJ^P=\frac{1}{2}\frac{3}{2}^-$ and the main channel was the
$\Sigma_c^*\bar D$, while the $P_c(4450)$ state was a
$\Sigma_c^*\bar D^*$ resonance with $IJ^P=\frac{1}{2}\frac{5}{2}^-$.
They also predicted another pentaquark state $\Sigma_c\bar D$ with
$IJ^P=\frac{1}{2}\frac{1}{2}^-$. The corresponding hidden-bottom
pentaquark partners were also discussed.

In Ref. \cite{Yang:2015bmv}, the authors performed a dynamical
calculation of five-quark systems with quantum numbers $I={1\over 2}$ and
$J^P={1\over 2}^\pm, {3\over 2}^\pm, {5\over 2}^\pm$ in the
framework of the chiral quark model with the gaussian expansion method.
The authors pointed out that the negative parity states could
be bound states while all the positive parity states are the
scattering states. The $P_c(4380)$ and $P_c(4450)$ were treated as the
bound states of the $\Sigma_c^*\bar D$ and $\Sigma_c {\bar D}^*$
respectively. For the first time, the authors calculated the
average distance between each quark, some of which may reach 2 fm!
The authors concluded that the distances between quarks confirm the
molecular nature of the $P_c$ states.

\subsubsection{The diquark/triquark model}

In Ref.~\cite{Maiani:2015vwa}, Maiani, Polosa, and Riquer discussed
the hidden-charm pentaquarks discovered by the LHCb
Collaboration~\cite{Aaij:2015tga}, and pointed out that they can be
a natural expectation of an extended picture of hadrons where quarks
and diquarks are fundamental units~\cite{Maiani:2004vq}. The diquark
[$qq$] here is a color anti-triplet member, similar to an anti-quark
$\bar q$~\cite{Jaffe:1976ig,Close:2002zu,Jaffe:2003sg}. They used
the combinations of one antiquark [$\bar c$], one heavy diquark
[$cq$], and one light diquark [$qq$]:
\begin{eqnarray}
P_c(4380,3/2^-) &=& {\bar c [cq]_{s=1} [q¡äq¡ä¡ä]_{s=1},L = 0} \, ,
\\
P_c(4450,5/2^+) &=& {\bar c [cq]_{s=1} [q¡äq¡ä¡ä]_{s=0},L = 1} \, ,
\end{eqnarray}
to explain the newly observed $P_c(4380)$ and $P_c(4450)$. The
$S$-wave state $P_c(4380)$ has negative parity, and the $P$-wave
state $P_c(4450)$ has positive parity. The mass difference between
the $P_c(4380)$ and $P_c(4450)$ is about 70 MeV, a) partly due to
the orbital excitation, which is of order 280 MeV, estimated for
$XYZ$ mesons~\cite{Maiani:2014aja}, and b) partly due to the mass
difference between diquarks with spin $s = 1$ and $s = 0$, which is
of order 200 MeV, estimated from charm and beauty baryons
spectra~\cite{DeRujula:1975ge}. They also studied the flavor
structures of the pentaquarks, and proposed several flavor partners
of the $P_c(4380)$ and $P_c(4450)$ to be observed in the $\Xi_b$ and
$\Omega_b$ decays. The idea was extended by the same authors for
charmed dibaryons in Ref.~\cite{Maiani:2015iaa} where they proposed
several possible channels to search for them in the
$\Lambda_b(5620)$ decays.

The same antiquark-diquark-diquark system was applied to study the
hidden-charm pentaquarks in later
studies~\cite{Anisovich:2015cia,Anisovich:2015zqa,Li:2015gta,Ghosh:2015ksa}.
In Ref.~\cite{Anisovich:2015cia,Anisovich:2015zqa}, Anisovich {\it et al.} gave the spin and isospin
structure of these states and estimated their masses, based on their
previous study of the tetraquark states~\cite{Anisovich:2015caa}.
Their results suggested that the $P_c(4450)$ is an
antiquark-diquark-diquark state with spin-parity $5/2^-$, while the
$P_c(4380)$ is the result of rescatterings in the $pJ/\psi$
spectrum, and can be related with possible resonances of the
$K^-J/\psi$ channel located in the mass region 4000-4500 MeV.

In Ref.~\cite{Ghosh:2015ksa}, Ghosh, Bhattacharya, and Chakrabarti
used the quasi particle diquark model, where diquarks are supposed
to behave like a quasi particle in an analogy with an electron in
the crystal lattice which behaves as a quasi particle. They
estimated the masses of the $P_c(4380)$ in both $[ud]_0[uc]_1 \bar c$
and $[ud]_1[uc]_0\bar c$ configurations to be 4403 MeV and 4345 MeV,
respectively, and the mass of the $P_c(4450)$ in the $[ud]_1[uc]_1
\bar c$ configuration to be 4443 MeV, both of which are consistent
with the LHCb experiment~\cite{Aaij:2015tga}.

A similar combination of diquark [$cq$] and triquark [$\bar c(ud)$]
was proposed by Lebed to study the hidden-charm pentaquarks in
Ref.~\cite{Lebed:2015tna}, which was based upon a mechanism proposed
to study tetraquark states~\cite{Brodsky:2014xia}. The $P_c(4380)$
and $P_c(4450)$ were described in terms of a confined but rapidly
separating color-antitriplet diquark $[cu]$ and color-triplet
antitriquark $[\bar c(ud)]$. The separations between diquark and
antitriquark were estimated to be $0.64$ fm for the $P_c(4380)$ and
$0.70$ fm for the $P_c(4450)$. These distances were achieved before
the hadronization, providing a qualitative explanation for the
suppression of the measured widths of the $P_c(4380)$ and
$P_c(4450)$. Later in Ref.~\cite{Lebed:2015dca}, Lebed applied the
same method to investigate the hidden-strangeness pentaquarks $P_s$,
and proposed to observe it in the $\Lambda_c \to P_s^+ \pi^0 \to
\phi p \pi^0$ decay.

This diquark-triquark system was also used by Zhu and Qiao in
Ref.~\cite{Zhu:2015bba}, where both diquark and triquark are not
compact objects, and have nonzero sizes. They analyzed the color
attractive configuration for the triquark and defined the nonlocal
wave functions for the pentaquark state, which are later used to
construct an effective diquark-triquark Hamiltonian based on
spin-orbital interaction. The pentaquark spectrum were obtained using
this Hamiltonian, where the pentaquark state with mass 4.349 GeV can explain the $P_c(4380)$,
and the pentaquark state with mass 4.453 GeV can explain the
$P_c(4450)$. Their mass splitting near 100 MeV is also consistent
with the LHCb experiment~\cite{Aaij:2015tga}. They further suggested
to analyze the $J/\psi \Sigma^+$ and $J/\psi \Lambda$ mass spectrum
near $4.682$ GeV in $\Xi_b^0 \to J/\psi \Sigma^+K^- $ and $\Xi_b^-
\to J/\psi\Lambda K^-$ decays, to look for the charged and neutral
hidden charm pentaquarks with $J^P=5/2^+$ and strange number $S=-1$.

\subsection{Kinematical effect}
\label{Sec:3.5}

There also exist the non-resonant explanations to the LHCb's
observation \cite{Aaij:2015tga}. In Refs. \cite{Guo:2015umn,
Mikhasenko:2015vca, Liu:2015fea}, these authors proposed various
rescattering mechanisms to show that the narrow $P_c(4450)$ state
might arise from the kinematical effect. Before reviewing these
results, we note that the strength of this kinematical effect 
(singularity) can not be calculated due to the lack of knowledge 
on the relevant couplings.

Guo {\it et al.} \cite{Guo:2015umn} studied the possibility of
the ${P^+_c(4450)}$ as the kinematical effect since it is around the
${\chi_{c1}}p$ threshold. The $J^P$ quantum number of the P-wave
${\chi_{c1}} p$ system matches the quantum number of the
${P^+_c(4450)}$ \cite{Aaij:2015tga}. Two typical decay mechanisms
for the $\Lambda_b\to K^-J/\psi p$ were introduced, i.e., (a)
$\Lambda_b\to K^- p \chi_{c1} \to K^- p J/\psi$ via $p\chi_{c1}\to p
J/\psi$ rescattering, (b) $\Lambda_b\to \Lambda^* \chi_{c1} \to K^-
p J/\psi$ via exchanging a proton. In the above two processes, the
$p\chi_{c1}\to p J/\psi$ rescattering occurs via exchanging soft
gluons. When dealing with the first process, they adopted a similar
method to that in Ref. \cite{Guo:2014iya} to get a loop function.
With a phenomenological amplitude which includes a constant
background contribution, the Argand diagram given by LHCb
\cite{Aaij:2015tga} can be reproduced \cite{Guo:2015umn}.

Guo {\it et al.} further considered the second mechanism and
discussed whether that the $P_c(4450)$ signal can arise from the
triangle singularity. By solving the Landau equation, the leading
Landau singularities for the triangle diagram were extracted.
Through investigating the motion of the solutions in the complex
${\sqrt{s}}$ plane, they noticed that the ${\Lambda^*}$ as the
intermediate state introduced in this decay mechanism can only
correspond to the ${\Lambda(1890)}$. The corresponding triangle
singularity is close to the ${\chi_{c1} p}$ threshold, which can
produce a threshold enhancement. Thus, they concluded that the
${P_c^+(4450)}$ can also be the kinematical effects around the
${\chi_{c1}p}$ threshold \cite{Guo:2015umn}.

Later, Liu {\it et al.} also investigated the possibility of the
$P_c(4450)$ as kinematical effects \cite{Liu:2015fea} and considered
three typical decay mechanisms which include (a) $\Lambda_b\to
\Lambda^*  \chi_{cJ} \to K^- p J/\psi$ via exchanging a proton which
is the same as one of the mechanisms proposed in Ref.
\cite{Guo:2015umn}, (b) $\Lambda_b\to D_s^{**}  \Lambda_c^* \to K^-
p J/\psi$ through exchanging $\bar{D}^{(*)}$, where
$D_s^{**}$ denotes a P-wave charmonium state,
(c) $\Lambda_b\to
K\bar{D}^{(*)}\Sigma_c^*\to K J/\psi p$ via
$\bar{D}^{(*)}\Sigma_c^*\to J/\psi p$ rescattering. To deal with the
triangle loop diagrams, the same approach as in Ref.
\cite{Liu:2015taa} was adopted, where the possible $\chi_{cJ}p$ and
$\Lambda_c^* \bar{D}^{(*)}$ threshold enhancements due to the
triangle singularities were considered.

Although both Guo {\it et al.}  \cite{Guo:2015umn} and Liu {\it et
al.} \cite{Liu:2015fea} studied similar issues of $\Lambda_b\to
\Lambda^* \chi_{cJ} \to K^- p J/\psi$, there exist some difference.
In Ref. \cite{Liu:2015fea}, Liu {\it et al.} indicated that the mass
thresholds for the ${p \chi_{cJ}~(J=0,~1,~2)}$ are all close to the
peak masses for the ${P_c^+}(4450)$. Including these thresholds in
their calculation, they found that the corresponding mass of the
${\Lambda^*}$ is larger than 2 GeV, which is in contrast with
$m(\Lambda^*)=1.89\sim 2.11$ GeV in Ref. \cite{Guo:2015umn}.

In Ref. \cite{Guo:2015umn}, the authors pointed out that the
charmonium was produced by the weak current [${\bar{c} \gamma^\mu
(1-\gamma_5) c}$] at the leading order. This current has no
projection onto the ${\chi_{c0}}$ or ${\chi_{c2}}$, although the
rescattering interaction strength for ${\chi_{c0,2} p}\to {J/\psi
p}$ is of the similar size to that for the ${\chi_{c1} p}\to {J/\psi p}$
scattering due to heavy quark spin symmetry. So the production of
these two charmonium states $\chi_{c0,2} $ in the $b$ decays can only come
from higher-order QCD corrections which are suppressed. According to
the above qualitative analysis, there do not exist enhancements
around the $\chi_{c0}p$ and $\chi_{c2}p$ thresholds in the $J/\psi p$
invariant mass distribution of the ${\Lambda_b}\to K^-J/\psi p$ decay.
Thus, Guo {\it et al.} only presented the $\chi_{c1}p$ peak
structure in the mass distribution of the $J/\psi p$, while Liu {\it et
al.} gave two peak structures around the $\chi_{c1}p$ and
$\chi_{c2}p$ thresholds.

Liu {\it et al.} further analyzed the second mechanism
\cite{Liu:2015fea}, and pointed out that there also exist the
conditions required by the triangle singularity. When the
${\Lambda_c^{(*)} \bar{D}^{(*)}_{sJ}}$ threshold is close to the
${\Lambda_b}$ mass, a sizeable enhancement at the ${\Lambda_c^{(*)}
\bar{D}^{(*)}}$ thresholds is allowed, where the
$\Lambda_c(2595)\bar{D}$ threshold results in an enhancement around
4.45 GeV. If the $\Lambda_c(2595)\bar{D}$ is in S-wave, the quantum
number of the obtained enhancement around 4.45 GeV in Ref.
\cite{Liu:2015fea} must be $J^P=1/2^+$, which does match the
experimental measurement of the $P_c(4450)$ quantum number. Under
this situation, the $P_c(4450)$ cannot be explained as the
enhancement due to the S-wave $\Lambda_c(2595)\bar{D}$ threshold.
They also argued that there still exists the possible enhancement
near 4.45 GeV when the $\Lambda_c(2595)\bar{D}$ is in P-wave, which
is suppressed compared with the S-wave case \cite{Liu:2015fea}. At
present, a quantitative study of this point is still absent.

Besides discussing $\Lambda_b\to \Lambda^*  \chi_{cJ} \to K^- p
J/\psi$ via exchanging a proton and $\Lambda_b\to D_s^{**}
\Lambda_c^* \to K^- p J/\psi$ through exchanging the $\bar{D}^{(*)}$,
Liu {\it et al.} also studied the mechanism $\Lambda_b\to
K\bar{D}^{(*)}\Sigma_c^*\to K J/\psi p$ via the
$\bar{D}^{(*)}\Sigma_c^*\to J/\psi p$ rescattering, which generated
the cusp structure around 4.45 GeV in the $J/\psi p$ mass
distribution. As indicated in Ref. \cite{Liu:2015fea}, the cusp
structure can be ignored comparing with the enhancement
from the triangle singularity.

Mikhasenko studied the triangle diagram $\Lambda_b\to
{D_s^{*-}}{\Sigma_c^+}\to K^- J/\psi p$ via a $D^{*0}$ exchange,
where there exists the $\Sigma_c^+ D^{*0}\to J/\psi p$ interaction
\cite{Mikhasenko:2015vca}. To some extent, this triangle diagram is
similar to the third decay mechanism proposed in Ref.
\cite{Liu:2015fea} since this triangle diagram in Ref.
\cite{Mikhasenko:2015vca} can be abbreviated to the third decay
mechanism in Ref. \cite{Liu:2015fea} when this $D_s^*$ exchange is
absorbed into the vertex. Mikhasenko used the ${D_s^{*-}}$,
${\Sigma_c^+}$ and ${D^{*0}}$ as building blocks of the triangle
loop and found an enhancement near the ${\Sigma_c^+ D^{*0}}$
threshold from the triangle singularity, which may correspond to the
observed ${P_c^+(4450)}$ structure \cite{Mikhasenko:2015vca}.

%================================================================================
%================================================================================
\subsection{Other theoretical schemes} \label{Sec:3.6}
%================================================================================
%================================================================================
Besides the theoretical works introduced in previous subsections,
there are other theoretical schemes to explain the nature of the
hidden-charm pentaquarks $P_c(4380)$ and $P_c(4450)$, such as the
topological soliton model \cite{Scoccola:2015nia}, the two-channel
framework \cite{Meissner:2015mza}, doublet-exotic molecule structure
\cite{Mironov:2015ica} and so on.

Based on the existence of a $C=-1$ ($C$ is charm quantum number)
meson bound state \cite{Oh:1994np,Oh:1994ux}, Scoccola, Riska and
Rho employed the soliton-$\bar DD$ system to study hidden-charm
pentaquark states in the topological soliton model in Ref.
\cite{Scoccola:2015nia}. They discussed both the Naive Skyrme Model
(NSM) and its incorporation with heavy quark symmetry (HQS), named
the SMHQS model. Using the single meson spectra obtained from the
NSM and SMHQS formulations, they first estimated the masses of the
pentaquark candidates with quantum numbers $J^P=3/2^-, 5/2^+$ and
isospin $I=1/2$ without the contributions of the non-adiabatic
corrections. Then they considered the first order perturbation
theory of the rotational corrections to find the mass of a system
composed of a soliton and two bound mesons (one $C=-1$ and the other
with $C=+1$) in the NSM model. They did not give predictions in the
SMHQS scheme because of the absence of the hyperfine splittings,
which were needed in the calculation of the rotational corrections.
Their results suggested the existence of a soliton-$\bar DD$
pentaquark-type state with $(I, J^P)=(1/2, 3/2^-)$ and mass
consistent with the $P_c(4380)$ resonance. In the case of the
$5/2^+$ channel, the predicted mass was in the range of $4.57 -
4.71$ GeV, which was too high as compared with the mass of the
$P_c(4450)$. Besides, they also predicted two $1/2^+$ and one
$3/2^+$ pentaquarks.

In Ref. \cite{Meissner:2015mza}, Meissner and Oller analyzed the
composite nature of the $P_c(4450)$ in a two-channel scenario: the
lower mass channel $J/\psi p$ and the heavier one $\chi_{c1}p$.
Using a probabilistic interpretation of the compositeness relation,
they calculated the couplings and partial decay widths of the
$P_c(4450)$ resonance to the two-body channels $J/\psi p$ and
$\chi_{c1}p$. Their result showed that the coupling to the heavier
mass channel $\chi_{c1}p$ is much larger than that to the lower one
$J/\psi p$, which results in a larger partial decay width of the
former decay channel. This is very similar to the scalar meson
$f_0(980)$, where the higher $K\bar K$ channel has a strong coupling
while the coupling to the lower $\pi\pi$ channel is suppressed
although the second one has a larger phase space. They concluded
that the $P_c(4450)$ is almost entirely a $\chi_{c1}p$ resonance.

The authors of Ref. \cite{Mironov:2015ica} discussed the possibility
of the two pentaquarks $P_c(4380)$ and $P_c(4450)$ composed of two
colored constituents with configuration $[q_1\bar
q_2]_{\mathbf{8_c}}[q_3q_4q_5]_{\mathbf{8_c}}$, where the subscript
$\mathbf{8_c}$ denotes the color structure. Such a configuration
results in a doublet of pentaquark states.

In Ref. \cite{Burns:2015dwa}, Burns studied the model-independent
phenomenology of the $P_c(4380)$ and $P_c(4450)$ pentaquark states
based on the meson-baryon molecular configuration. He analyzed
possible spin-parity assignments for these two states and speculated
various decay patterns and production processes. The author argued
that these two $P_c$ states were mixtures of isospins $1/2$ and
$3/2$ among several possible meson-baryon pairs. In the same
configuration, the neutral partners of the $P_c(4380)$ and
$P_c(4450)$ states and other possible states with different quantum
numbers were predicted.

In Ref. \cite{Eides:2015dtr}, the authors interpreted the $P_c(4450)$ as a
bound state of the charmonium $\psi(2S)$ and the nucleon. The binding
potential arises from the charmonium-nucleon interaction in the form
of the product of the charmonium chromoelectric polarizability and
the nucleon energy-momentum distribution. The authors estimated the
quarkonium polarizability and calculated the nucleon properties in
the framework of the mean-field picture of light baryons in the
large $N_c$ limit. They arrived at two almost degenerate states
$J^P=(1/2)^-$ and $J^P=(3/2)^-$ around 4450 MeV with a narrow width.

Motivated by the surface potential in the Hasenfratz-Kuti model, the
authors of Ref. \cite{Kahana:2015tkb} assumed that there exists a surface
potential between the proton and the $J/\psi$ in the form, $V_S(l)=V_s
x^{\alpha-1} (1-x)^{\beta-1}$, where $x=r/r_0$, $\alpha=9$,
$\beta=5$, $r_0=0.85$ fm and $l$ is the orbital angular momentum.
Both the $P_c(4450)$ and $P_c(4380)$ were interpreted as the molecular
resonances of the proton and $J/\psi$. They adjusted the well depth
$V_s=2350$ MeV to obtain the resonance at the observed excitation
energy. They identified the $l=4$ state as the $P_c(4450)$, fixed its
excitation energy at $E=410$ MeV and derived its width to be around
50 MeV. The $l=3$ resonance is identified with the $P_c(4380)$, which
has a width around 142 MeV and excitation energy around 340 MeV if
the well depth $V_s=1410$ MeV.

\subsection{Production and decay patterns}
\label{sect:3.7}

\subsubsection{Production of the $P_c$ via weak decays}
\label{sect:3.7.1}

Considering the LHCb's measurement of the $K^-p$ and $J/\psi p$
invariant mass distributions \cite{Aaij:2015tga}, Roca, Nieves and
Oset studied the $\Lambda_b\to K^- J/\psi p$ process. They
introduced three mechanisms, i.e., (a) a quark-level basic process
to produce the $J/\psi K^-p$ final state through the weak decay of
$\Lambda_b$, (b) a hadronic level description of $\Lambda_b\to K^-
J/\psi p\to K^-J/\psi p$ with the $K^-p\to K^-p$ final state
interaction in coupled channels, (c) a hadronic level description of
$\Lambda_b\to K^- J/\psi p\to K^-J/\psi p $ with the $J/\psi p\to
J/\psi p$ final state interaction in coupled channels
\cite{Roca:2015dva}. The first two mechanisms reflect the
contribution from the $\Lambda(1405)$, while the third mechanism
contains the $J/\psi N$ final sate interaction in coupled channels
including the $\bar{D}^*\Lambda_c$, $\bar{D}^*\Sigma_c$,
$\bar{D}\Sigma_c^*$, $\bar{D}^*\Sigma_c^*$, which may produce poles
at $4334\pm 19 i$, $4417+4i$ and $4481+17i$ MeV \cite{Xiao:2013yca}.
Since these poles couple with the $J/\psi p$, the corresponding
resonance shape appears in the $J/\psi p$ invariant mass spectrum.
After comparing the invariant mass spectra of the $J/\psi p$ and
$K^-p$ with the experimental data, the authors of
Ref.~\cite{Roca:2015dva} noticed that the shape and relative
strength of the $K^-p$ invariant mass spectrum near the threshold
and the structure $P_c(4450)$ in the $J/\psi p$ invariant mass
distribution can be reproduced simultaneously, which supports the
$P_c(4450)$ as a $J^P=3/2^-$ hidden-charm molecular state composed
of the $\bar{D}^*\Sigma_c$ and $\bar{D}^*\Sigma_c^*$
\cite{Roca:2015dva}. Wang {\it et al.} considered the $\Lambda_b\to
\pi^- J/\psi p$ reaction in analogy to the $\Lambda_b\to K^- J/\psi
p$ process and studied the possible manifestation of the
hidden-charm pentaquarks in the sharp structure of the $J/\psi p$
mass distribution around 4450 MeV \cite{Wang:2015pcn}.
Additionally, Lu {\it et al.} studied the $\Lambda_{b} \to J/\psi K^{0} \Lambda$
reaction by taking into account a hidden-charm state with
strangeness that couples to $J/\psi\Lambda$~\cite{Lu:2016roh}.

The authors of Refs. \cite{Li:2015gta,Cheng:2015cca} carried out the
analysis of the production of hidden-charm pentaquarks via weak
decays of bottom baryons based on SU(3) flavor symmetry, which were
extensively applied to study the $B$ decay and $CP$ violation. Under the
assumption of the $P_c(4380)$ and $P_c(4450)$ with configurations
$\{\bar{c}[cq]_{s=1}[q'q'']_{s=1},L=0\}$ and
$\{\bar{c}[cq]_{s=1}[q'q'']_{s=0},L=1\}$ respectively as suggested
in Ref. \cite{Maiani:2015vwa},  Li {\it et al.} \cite{Li:2015gta}
derived the amplitude relations under SU(3) flavor symmetry. There
exist $\mathbf{3}\otimes\mathbf{\bar 3}=\mathbf{1}\oplus\mathbf{8}$ and $\mathbf{3}\otimes\mathbf{6}=\mathbf{8}\oplus\mathbf{10}$ multiplets for
pentaquarks. The invariant weak decay amplitude of bottom baryons
into an octet or a decuplet pentaquark plus a light pseudoscalar
octet meson was decomposed into several terms. For the hidden-charm
pentaquarks with the same $J^{P}$ quantum number, there exist
amplitude relations which can be tested experimentally
\cite{Li:2015gta}.

Cheng and Chua studied the weak decays of the bottom baryons in the
$\bar{\textbf{3}}$ representation and $\Omega_{b}^{-}$ in the
$\textbf{6}$ representation into a pseudoscalar meson and an octet
(a decuplet) hidden-charm pentaquark \cite{Cheng:2015cca}. The decay
amplitudes of $(\Lambda_{b}^{0},\Xi_{b}^{0},\Xi_{b}^{-})\rightarrow
P_{8}+M$, $(\Lambda_{b}^{0},\Xi_{b}^{0},\Xi_{b}^{-})\rightarrow
P_{10}+M$, $\Omega_b^-\rightarrow P_{8}+M$, $\Omega_b^-\rightarrow
P_{10}+M$ were calculated, where $P_8$ and $P_{10}$ stand for the
hidden-charm pentaquarks in octet and decuplet, respectively. The
authors suggested that the channel $\Xi_{b}^{0}\rightarrow
P_{\Sigma^{+}}K^-$, $\Xi_b^{-}\rightarrow P_{\Sigma^{-}}\bar{K}^{0}$,
$\Omega_b^{-}\rightarrow P_{\Xi^{-}}{\bar K^{0}}$,
and $\Omega_b^{-}\rightarrow P_{\Xi^{0}}{K^{-}}$ may have
contributions comparable with that of $\Lambda_{b}^{0}\rightarrow
P_{p}^{+}K^{-}$.

Hsiao {\it et al.} \cite{Hsiao:2015nna} assumed that $\Lambda_b\to
J/\psi p K^-$ occurs via $b\to c\bar{c}s$ and ignored those
contributions via the non-resonant $\Lambda_b\to pK^-$ and resonant
$\Lambda_b\to \Lambda^*\to pK^-$. Thus, the $P_c(4380)$ and $P_c(4450)$
would be produced mainly by the charmless $\Lambda_b$ decays with $b\to
u\bar{u}s$. The two observed $P_c$ pentaquarks are produced from the
intrinsic charms within $\Lambda_b$. According to this mechanism,
the ratio of branching ratios was obtained,
$\mathcal{B}(\Lambda_b\to \pi^-(P_c(4380)/P_c(4450))\to \pi^- J/\psi
p)/\mathcal{B}(\Lambda_b\to K^-(P_c(4380)/P_c(4450))\to K^- J/\psi
p)=0.58\pm0.05$. Additionally, the direct CP violating asymmetries
were also predicted as ${\cal{A}}_{CP}(\Lambda_b\to
\pi^-(P_c(4380)/P_c(4450))\to \pi^- J/\psi p)=(-7.4\pm0.9)\%$ and
${\cal{A}}_{CP}(\Lambda_b\to K^-(P_c(4380)/P_c(4450))\to K^- J/\psi
p)=(+6.3\pm0.2)\%$ \cite{Hsiao:2015nna}.

The authors of Ref.~\cite{Roca:2015tea} discussed the $\Lambda_b\to
J/\psi \Lambda(1405)$ decay. Similar production mechanisms of hidden
charm pentaquarks with strangeness from $\Lambda_b$ and $\Xi_b$
decays were investigated in
Refs.~\cite{Chen:2015sxa,Feijoo:2015kts,Lu:2016roh}.

\subsubsection{Photo-production of the $P_c$}
\label{sect:3.7.2}

The authors of Ref. \cite{Wang:2015jsa} studied the photo-production
of the two hidden-charm pentaquark states $P_c(4380)$ and
$P_c(4450)$ via $\gamma p\to J/\psi p$. The estimated total and
differential cross sections of $\gamma p\to J/\psi p$ depend on the
unknown coupling of the $P_c(4380)/P_c(4450)$ with $J/\psi p$. Under the
assignment of the $P_c(4450)$ as the $\Sigma_c \bar{D}^*$ molecular
state, Karliner and Rosner \cite{Karliner:2015voa} made an estimate
of the $P_c(4450)$ production cross section through the vector
dominance. They pointed out that the events of the $P_c(4450)$ produced
at CLAS12 and forthcoming GlueX are considerable. Voloshin also
studied the $\gamma p\to P_c(4380)/P_c(4450)\to J/\psi +p$ process
and estimated the cross section \cite{Kubarovsky:2015aaa}. These
investigations indicate that there exists good potential to search
for the two $P_c$ states through the photo-production.

In Ref. \cite{Lu:2015fva}, the authors proposed to search for the
production of the neutral hidden-charm pentaquarks $P_c(4380)^0$ and
$P_c(4450)^0$ via the $\pi^-p\to J/\psi n$ reaction. Their
calculation shows that there exist clear signals corresponding to
the $P_c(4380)^0$ and $P_c(4450)^0$ with the cross section around $1$
$\mu$b, which suggests that the $\pi^-p\to J/\psi n$ reaction is
suitable to produce the two neutral $P_c$ states \cite{Lu:2015fva}.
In Ref. \cite{Ouyang:2015rre}, the authors discussed the production of
the hidden-charm baryon $N^*_{c\bar c}(4261)$ with $J^P={1\over 2}$
in the above reaction process.

\subsubsection{Strong decay patterns of the $P_c$ states}
\label{sect:3.7.3}

Assuming the $P_c(4380)$ and $P_c(4450)$ to be molecular states composed
of $\bar{D}^{(*)}$ and $\Sigma_c^{(*)}$, their strong decay patterns
were studied with the spin rearrangement scheme in Ref.
\cite{Wang:2015qlf}. Several typical ratios of the partial decay
widths of the hidden-charm pentaquarks were obtained. Especially,
for the three S-wave $(\bar{D}\Sigma_c^*)$, $(\bar{D}^*\Sigma_c)$
and $(\bar{D}^*\Sigma_c^*)$ molecular pentaquarks with
$J^P={{3}/{2}}^{-}$, the obtained ratio of their $J/\psi N$ decay
widths is $\Gamma\left[(\bar
D{\Sigma^{\ast}_c})\right]:\Gamma\left[({\bar
D}^{\ast}{\Sigma}_c)\right]:\Gamma\left[({\bar
D}^{\ast}{\Sigma^{\ast}_c})\right]=2.7:1.0:5.4$. These ratios are
model independent.

\subsection{The hidden-bottom and doubly heavy pentaquark states}\label{sec:3.8}

If there exist the hidden-charm pentaquarks, their hidden-bottom
partners should also exit. The hidden-bottom and doubly heavy
pentaquark states were predicted with different theoretical models.
In the following, we briefly review these states.

The hidden-bottom molecular pentaquarks composed of a bottom meson
and a bottom baryon were studied extensively within the OBE model
\cite{Yang:2011wz}. There exist hidden-bottom molecular pentaquarks,
which include the $\Sigma_b B$ with $\frac{3}{2}(\frac{1}{2}^-)$,
$\Sigma_b^* B^*$ with $\frac{1}{2}(\frac{1}{2}^-)$,
$\frac{1}{2}(\frac{3}{2}^-)$, $\frac{3}{2}(\frac{1}{2}^-)$,
$\frac{3}{2}(\frac{3}{2}^-)$, while the $\Lambda_b B$ with
$\frac{1}{2}(\frac{1}{2}^-)$, $\Lambda_b B^*$ with
$\frac{1}{2}(\frac{1}{2}^-)$, $\frac{1}{2}(\frac{3}{2}^-)$ do not
exist.

\renewcommand{\arraystretch}{1.4}
\begin{table}[hbtp]
\caption{The typical values of the obtained bound state solutions
$[E (\text{MeV}), \Lambda (\text{GeV}])$ for the hidden-bottom
$\Sigma_b^{(*)}B^*$ and $B_c$-like $\Sigma_c^{(*)}B^*$ and
$\Sigma_b^{(*)}\bar{D}^*$ systems. Based on the experience of the
S-wave $\Sigma_c^{(*)}\bar{D}^*$ systems, the bound state solution
is searched for in the range of $\Lambda<2.35$ GeV for
$\Sigma_{b}B^*$ and the range of $\Lambda<1.77$ GeV for
$\Sigma_{b}^*B^*$. Taken from Ref.
\cite{Chen:2015loa}.}\label{SigmabS}
\begin{center}
\begin{tabular}{ccccccc}
\toprule[1pt] {$(I,J)$}
 &$\Sigma_c{B}^*$    &$\Sigma_b\bar{D}^*$    &$\Sigma_b{B}^*$
 &$\Sigma_c^*{B}^*$    &$\Sigma_b^*\bar{D}^*$    &$\Sigma_b^*{B}^*$
 \\\midrule[1pt]
 (1/2,1/2)   &$\times$     &$\times$     &$\times$
             &$\times$     &$\times$     &$\times$ \\
{(1/2,3/2)}  &[-0.27, 1.22]    &[-0.26, 1.34]      &[-0.27, 0.84]
             &$\times$     &$\times$     &$\times$ \\
             &[-2.58, 1.32]    &[-2.62, 1.44]      &[-2.36, 0.94]
             &$\times$     &$\times$     &$\times$ \\
             &[-7.48, 1.42]    &[-7.63, 1.54]      &[-6.88, 1.04]
             &$\times$     &$\times$     &$\times$ \\
{(1/2,5/2)}  &$\times$     &$\times$     &$\times$
             &[-0.28, 0.88]     &[-0.14, 0.96]    &[-0.30, 0.64]\\
             &$\times$     &$\times$     &$\times$
             &[-3.18, 0.98]     &[-2.78, 1.06]    &[-3.11, 0.74]\\
             &$\times$     &$\times$     &$\times$
             &[-9.67, 1.08]     &[-8.97, 1.16]    &[-9.51, 0.84]\\
{(3/2,1/2)}  &[-0.27, 1.22]    &[-0.26, 1.34]      &[-0.27, 0.84]
             &[-0.42, 1.02]     &[-0.30, 1.12]    &[-0.28, 0.72]\\
             &[-2.58, 1.32]    &[-2.62, 1.44]      &[-2.36, 0.94]
             &[-3.33, 1.12]     &[-3.03, 1.22]    &[-2.74, 0.82]\\
             &[-7.48, 1.42]    &[-7.63, 1.54]      &[-6.88, 1.04]
             &[-9.37, 1.22]     &[-8.91, 1.32]    &[-8.19, 0.92]\\
{(3/2,3/2)}              &$\times$     &$\times$     &$\times$
                                       & $\times$    &$\times$     &[-0.28, 1.44]\\
                                       &$\times$     &$\times$     &$\times$
                                       &$\times$     &$\times$     &[-3.28, 1.60]\\
                                       &$\times$     &$\times$     &$\times$
                                       &$\times$     &  $\times$   &[-9.13, 1.74]
\\
(3/2,5/2)  &$\times$     &$\times$     &$\times$   &$\times$     &$\times$     &$\times$ \\

\bottomrule[1pt]
\end{tabular}
\end{center}
\end{table}

In a subsequent work, Chen {\it et al.} studied the hidden-bottom
molecular pentaquarks and $B_c$-like pentaquarks with the OPE model,
which is an important extension of the hidden-charm molecular
pentaquark states reviewed in Sec. \ref{sect:4.1.1}. The results are
collected in Table \ref{SigmabS}, which indicate that the
$\Sigma_{b}B^*$, $\Sigma_c B^*$, $\Sigma_b \bar{D}^*$ bound states
with either $(I=1/2,J=3/2)$ or $(I=3/2,J=1/2)$ may exist. The decay
modes of the $\Sigma_{b}B^*$, $\Sigma_c B^*$, and $\Sigma_b \bar{D}^*$
states with $(I=1/2,J=3/2)$ include the
{$\Upsilon(1S)N$/$\Upsilon(2S)N$, $B_c(1^-) N$, and
$\bar{B}_c(1^-)N$}, respectively. The $\Sigma_{b}B^*$, $\Sigma_c
B^*$, and $\Sigma_b \bar{D}^*$ states with $(I=3/2,J=1/2)$ can decay
into the {$\Upsilon(1S)\Delta(1232)$, $B_c(1^-) \Delta(1232)$, and
$\bar{B}_c(1^-)\Delta(1232)$}, respectively \cite{Chen:2015loa}.

For the $\Sigma_{b}^*B^*$, $\Sigma_c^* B^*$, $\Sigma_b^* \bar{D}^*$
S-wave systems with $(I=1/2,J=3/2)$ and $(I=3/2,J=1/2)$, there also
exist the bound state solutions. The $\Sigma_b^* B^*$ state can also
carry $(I=3/2,J=3/2)$. The $\Upsilon(1S)N$/$\Upsilon(2S)N$,
$B_c(1^-) N$, and $\bar{B}_c(1^-)N$ are the main decay modes of the
$\Sigma_{b}^*B^*$, $\Sigma_c^* B^*$, $\Sigma_b^* \bar{D}^*$ with
$(I=1/2,J=3/2)$, respectively, while the main decay channels of the
$\Sigma_{b}^*B^*$, $\Sigma_c^* B^*$, $\Sigma_b^* \bar{D}^*$ states
with $(I=3/2,J=1/2)$ are the {$\Upsilon(1S)\Delta(1232)$, $B_c(1^-)
\Delta(1232)$, and $\bar{B}_c(1^-)\Delta(1232)$}, respectively. The
$\Sigma_b^* B^*$ state with $(I=3/2,J=3/2)$ mainly decays into the
$\Upsilon(1S)\Delta(1232)$ \cite{Chen:2015loa}.

In Ref. \cite{Karliner:2015ina}, Karliner and Rosner also suggested
the existence of doubly heavy molecular pentaquarks. The above
hidden-bottom and doubly heavy pentaquark states from the OBE and
OPE models may be accessible at future experiments.
In a recent reference~\cite{Karliner:2016ith}, Karliner and Rosner considered the $\eta$
exchange, which can be important for hadrons without $u$ and $d$
light quark, such as the $D_s$. They suggested to observe the
$\Lambda_c \bar D_s^*$ resonance in the process
$\Lambda_b \to J/\psi \Lambda (\pi^+ \pi^- \mbox{ or } \eta)$.

The coupled channel unitary approach with the local hidden gauge
formalism, reviewed in Sec.~\ref{Sect.3.2}, was also applied to
study the hidden bottom pentaquarks in
Refs.~\cite{Wu:2010rv,Xiao:2013jla}. In Ref.~\cite{Wu:2010rv} Wu,
Zhao, and Zou predicted two $N^*_{b \bar b}$ states with mass and
width $(M,\Gamma) = (11052, 1.38)$ MeV and $(11100, 1.33)$ MeV and
four $\Lambda^*_{b \bar b}$ states with $(M,\Gamma) = (11021, 2.21)$
MeV, $(11191, 1.24)$ MeV, $(11070, 2.17)$ MeV, and $(11239, 1.19)$
MeV. In Ref.~\cite{Xiao:2013jla}, Xiao and Oset found seven hidden
bottom pentaquark states, all of which have $I = 1/2$:
\begin{eqnarray}
\nonumber && {\rm the}~J = 1/2~{\rm sector:}~(10963.04 + i8.59)~{\rm
MeV} \, , \, (11002.81 + i19.97)~{\rm MeV} \, , \,
(11023.55+i22.75)~{\rm MeV} \, ,
\\ \nonumber && {\rm the}~J = 3/2~{\rm sector:}~(10984.43 + i9.19)~{\rm MeV} \, , \, (11007.28 + i3.00)~{\rm MeV} \, , \, (11019.00 + i14.80)~{\rm MeV} \, ,
\\ && {\rm the}~J = 5/2~{\rm sector:}~(11026.10+i0)~{\rm MeV} \, .\nonumber
\end{eqnarray}
These poles were classified as four basic states: a) the first pole
$(10963.04 + i8.59)$ MeV, corresponding to a $B \Sigma_b$ state; b)
the fourth pole $(10984.43+i9.19)$ MeV, corresponding to a $B
\Sigma_b^*$ state; c) the second pole $(11002.81 + i19.97)$ MeV and
the fifth pole $(11007.28 + i3.00)$ MeV, both corresponding to a
$B^* \Sigma_b$ state; d) the third pole $(11023.55 + i22.75)$ MeV,
the sixth pole $(11019.00 + i14.80)$ MeV, and the seventh pole
$(11026.10+i0)$ MeV, all corresponding to a $B^* \Sigma_b^*$ state.
All these states are bound with about 50-130 MeV with respect to the
corresponding $B^{(*)} \Sigma_b^{(*)}$ thresholds.

Two hidden-bottom pentaquarks were predicted in Ref.~\cite{Chen:2015moa}
using the method of QCD sum rule as partners of the $P_c(4380)$ and
$P_c(4450)$. Their masses were extracted as
$11.55^{+0.23}_{-0.14}$ MeV and $11.66^{+0.28}_{-0.27}$ MeV, and
spin-parity quantum numbers $J^P = 3/2^-$ and $5/2^+$, respectively.

Zhu and Qiao used the constituent diquark-triquark model to study
hidden bottom pentaquarks~\cite{Zhu:2015bba}, where they have used
the diquark mass $m_{\delta} = m_{[bq]} = 5.249$ GeV and the
triquark mass $m_\theta = m_{[ud\bar b]} = 5.618$ GeV.

\subsection{Theoretical and experimental challenges}\label{sec:3.9}

The presence of the heavy quarks (or charmed/bottomed hadrons)
lowers the kinetic energy of the system, which favors the formation
of either the ``genuine" pentaquark states or the molecular baryons.
For the molecular scheme, the light quarks are also essential since
the meson exchange force between the light quarks, especially the
long-range pionic interaction binds the hadronic system.

The discovery of the $P_c(4350)$ and $P_c(4450)$ opens a new era in the
exploration of the multiquark states. There remain many theoretical
and experimental challenges, some of which are highlighted below:
\begin{itemize}

\item At present, the $P_c(4380)$ and $P_c(4450)$ were only reported
by the LHCb Collaboration \cite{Aaij:2015tga}. These states should
be confirmed in other processes and analyses.

\item BelleII, CLAS12, GlueX and JPARC may have the potential to search for the
hidden-charm $P_c$ states through the $e^+e^-$ annihilation, photoproduction, or
$\pi^-p\to J/\psi n$ reaction.

\item Since it is very close to the $\chi_{c1} p$ threshold,
could the $P_c(4450)$ arise from kinematical effects through various
rescattering mechanisms such as the triangle singularity?

\item Various theoretical approaches predict many isospin and spin
partner states of the $P_c(4380)$ and $P_c(4450)$. Where and how can these states be observed?

\item The experimental identification of the parity for each $P_c$
state is crucial for the discrimination of various models. For
example, the S-wave molecular pentaquark has negative parity while
its nearby P-wave excitation carries positive parity in the
framework of the molecular scheme. In contrast, only the pentaquarks
with negative parity are dynamically generated through S-wave
rescattering within the unitary approach.

\item The identification of the dominant decay modes of the two $P_c$
states is important. Although they were observed in the clean
$J/\psi p$ final state, the $J/\psi p$ mode is not necessarily their
main decay mode. Recall the similar situations in the case of the charged
hidden-charm/bottom tetraquark states. The $Z_c(3900)$ was first
observed in the hidden-charm mode. Later its
open-charm decay width was measured to be much larger. As molecular
baryons, the open-charm decay modes of the $P_c$ states would be
dominant. As a compact ``genuine" pentaquark state confined within
one MIT bag, the $J/\psi p$ mode may be the dominant mode of the
$P_c$ state.

\item For a dynamically generated hidden-charm baryon, its total
width was quite small and less than 60 MeV according to most of the
model calculations within the unitary approach. However, the sum of
the partial decay widths of the charm-less decay modes are larger
than the decay width of their main hidden-charm modes. This feature
is characteristic of the dynamically generated resonance and the
unitary approach. The observation of significant charm-less decay
modes will support the $P_c$ states as the dynamically generated
resonance within the unitary framework.

\item Is it possible to produce the hidden-bottom pentaquark states
experimentally?

\item Is it feasible to simulate the scattering of the charmed
baryon and anticharmed meson on the lattice? Can the molecular
bound states be isolated from the scattering states? Or can the
``genuine" pentaquark resonances be observed directly on the lattice?

\item Besides hidden-charm pentaquarks, the hidden-charm (or hidden-bottom)
systems with six quarks may also exist
\cite{Lee:2011rka,Li:2012bt,Li:2014gra,Karliner:2015voa}. There may
exist some potential to observe the molecular systems composed of a pair of
charmed baryons and a pair of charmed and anti-charmed baryons at
LHCb \cite{Lee:2011rka,Li:2012bt}.

\item Through the decay products $\Sigma_c^{(*)}$ and $\Lambda_c^{(*)}$
of the $\Lambda_b$ weak decay, one may explore the production
mechanism of the $P_c$ states within the molecular scheme. There
exist three types of Feynman diagrams which contribute to the
$\Lambda_b \to J/\psi p K$ decay, as shown in Fig. \ref{Fig:3.9Pentaquarkproduction}.
Interested readers may also consult reviews in Ref.~\cite{Oset:2016lyh}.

\begin{figure}[hbtp]
\begin{center}
\includegraphics[width=13cm]{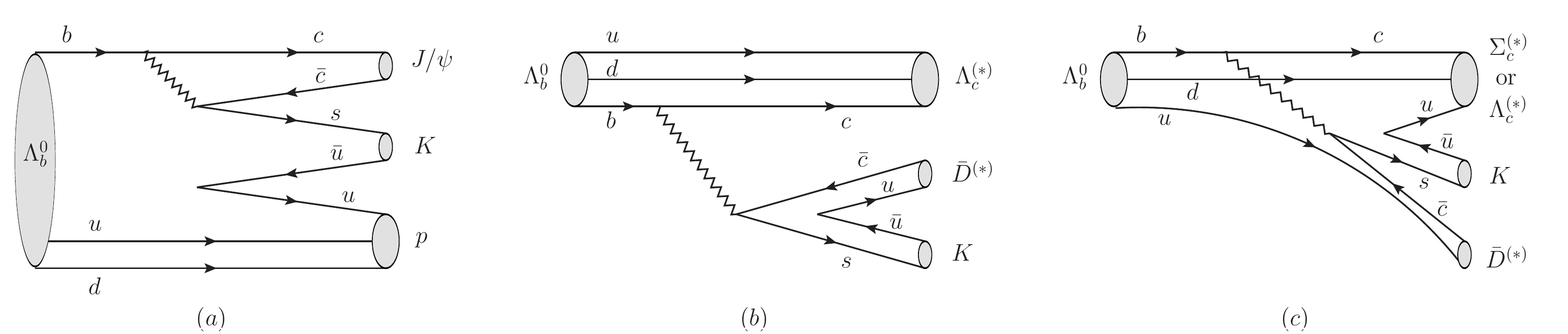}
\caption{The types of production mechanisms of the $P_c$ states
via the $\Lambda_b$ weak decay.} \label{Fig:3.9Pentaquarkproduction}
\end{center}
\end{figure}

\begin{itemize}
\item (a) The $b$ quark decays weakly via $b\to c \bar c s$ and
the $c\bar c$ pair transforms into the $J/\psi$ while the scalar
isoscalar $[ud]$ diquark within the $\Lambda_b$ is not pulled apart
and acts as the spectator throughout the whole process. This
mechanism contributes to the non-resonant $J/\psi p $ production and
explains naturally why only $\Lambda^*$ and no $\Sigma^*$
intermediate states were observed by the LHCb Collaboration.

\item (b) The $b$ quark decays weakly via $b\to c \bar c s$. The $c$
quark combines with the scalar isoscalar $[ud]$ diquark to form the
$\Lambda_c^{(*)}$. At the same time, a pair of light quarks $q\bar
q$ is produced from the vacuum. The $\bar c$ quark picks up the
quark $q$ and becomes a ${\bar D}^{(*)}$ meson while the remaining
$s\bar q$ forms the kaon. Some of the $\Lambda_c^{(*)}{\bar
D}^{(*)}$ rescatter into the $J/\psi p$ final states. However, the
interaction between the $\Lambda_c^{(*)}$ and ${\bar D}^{(*)}$ is rather weak.
Therefore, most of the $\Lambda_c^{(*)}{\bar D}^{(*)}$ events are
non-resonant. In other words, one would expect many
$\Lambda_c^{(*)}$ in the final states.

\item (c) The $b$ quark decays weakly via $b\to c \bar c s$. A pair of
light quarks $q\bar q$ are produced from the vacuum while the scalar
isoscalar $[ud]$ diquark within $\Lambda_b$ is pulled apart into
$q_1$ and $q_2$ by non-perturbative QCD interaction. The $\bar c$
quark picks up the quark $q_1$ from the original diquark and becomes
a ${\bar D}^{(*)}$ meson. The $s \bar q$ forms the kaon. The charm
quark and remaining $q_2 q$ transforms into a charmed baryon. Since
the total isospin of the $q_2 q$ pair is either 0 or 1, the charmed
baryon could be $\Lambda_c^{(*)}$ or $\Sigma_c^{(*)}$, the
probability of which is 50\% respectively. According to the
molecular scheme, the hidden-charm pentaquark states are produced
through the rescattering process $\Sigma_c^{(*)} {\bar D}^{(*)}\to
P_c \to J/\psi p$. Besides this resonant process, there exist many
$\Sigma_c^{(*)} {\bar D}^{(*)}$ and $\Lambda_c^{(*)}{\bar D}^{(*)}$
events. One would expect many $\Sigma_c^{(*)}$ and more
$\Lambda_c^{(*)}$ in the final states.

\end{itemize}

\end{itemize}

%%-------------Section 4-----------------------------------
\section{Theoretical interpretations of the $XYZ$ states}
\label{sect:4}

\subsection{$Z_b(10610)$ and $Z_b(10650)$}
\label{Sect:4.1}

\subsubsection{Molecular scheme}\label{sect:4.1.1}

\paragraph{Meson exchange model}\label{sect:4.1.1.1}

In 2008, the authors of Ref. \cite{Liu:2008fh} discussed the
possible S-wave molecular states composed of $D\bar{D}^*$ and
$B\bar{B}^*$ and pointed out that ``{\it there probably exists a
loosely bound S-wave $B\bar{B}^*$ molecular state}". Within the
meson exchange model, the possible attraction between the heavy
mesons arises from the exchange of the light mesons. Especially the
chiral interaction between the light quarks and pions plays a
central role in the formation of the shallow heavy meson bound
states, which does not depend on the heavy quark mass. As the heavy
quark mass increases, the heavy meson mass increases and the kinetic
energy of the dimeson system decreases, while the potential between
the heavy meson pair remains roughly the same. The formation of the
bound states relies on the competition between the kinetic energy
and attractive potential of the system. In other words, the
$B\bar{B}^*$ meson pairs are easier and more likely to form a
loosely bound molecular state than the $D\bar{D}^*$
\cite{Liu:2008fh}.

In 2011, the charged bottomonium-like states $Z_b(10610)$ and
$Z_b(10650)$ were reported by the Belle Collaboration in five
hidden-bottom dipion decays of $\Upsilon(5S)$ \cite{Belle:2011aa}.
Their experimental information was summarized in Sec.~\ref{Sec:2.7}.
The discovery of these states stimulated further theoretical studies
of the $Z_b(10610)$ and $Z_b(10650)$ as the
$B\bar{B}^*$ and $B^*\bar{B}^*$ molecular states \cite{Sun:2011uh},
respectively. Their flavor wave functions read \cite{Sun:2011uh}
\begin{eqnarray}
|{Z_b(10610)}^\pm\rangle&=&\frac{1}{\sqrt{2}}\big(|B^{*\pm}\bar{B}^0\rangle+|B^\pm\bar{B}^{*0}\rangle\big),\\
|{Z_b(10610)}^0\rangle&=&\frac{1}{2}\Big[\big(|B^{*+}B^-\rangle-|B^{*0}\bar{B}^0\rangle\big)
+\big(|B^+B^{*-}\rangle-|B^0\bar{B}^{*0}\rangle\big)\Big], \\
|{Z_b(10650)}^\pm\rangle&=&|B^{*\pm}\bar{B}^{*0}\rangle,\\
|{Z_b(10650)}^0\rangle&=&\frac{1}{\sqrt{2}}\big(|B^{*+}B^{*-}\rangle-|B^{*0}\bar{B}^{*0}\rangle\big).
\end{eqnarray}
The total effective potential of the $Z_b(10610)$ system is
\begin{eqnarray}
\mathcal{V}^{Z_b(10610)}&=&
V^{\mathrm{Direct}}_\sigma-\frac{1}{2}V^{\mathrm{Direct}}_\rho
+\frac{1}{2}V^{\mathrm{Direct}}_\omega+\frac{1}{4}\bigg(-2V^{\mathrm{Cross}}_\pi
+ \frac{2}{3}V^{\mathrm{Cross}}_ \eta-2V^{\mathrm{Corss}}_\rho+2
V^{\mathrm{Cross}}_\omega\bigg)
\end{eqnarray}
with these subpotentials from the $\pi$, $\eta$, $\sigma$, $\rho$
and $\omega$ meson exchanges, i.e.,
\begin{eqnarray}
\nonumber V^{\mathrm{Cross}}_\pi&=&
-\frac{g^2}{f_\pi^2}\bigg[\frac{1}{3}(\mbox{\boldmath$\epsilon$}_2\cdot
\mbox{\boldmath$\epsilon$}^\dag_3)Z(\Lambda_2,m_2,r)+\frac{1}{3}S(\hat{\mbox{\boldmath$r$}},\mbox{\boldmath$\epsilon$}_2,
\mbox{\boldmath$\epsilon$}^\dag_3)T(\Lambda_2,m_2,r)\bigg],
\\
\nonumber V^{\mathrm{Cross}}_ \eta&=& -\frac{g^2}{f_\pi^2}
\bigg[\frac{1}{3}(\mbox{\boldmath$\epsilon$}_2\cdot
\mbox{\boldmath$\epsilon$}^\dag_3)Z(\Lambda_3,m_3,r)
+\frac{1}{3}S(\hat{\mbox{\boldmath$r$}},\mbox{\boldmath$\epsilon$}_2,
\mbox{\boldmath$\epsilon$}^\dag_3)T(\Lambda_3,m_3,r)],\\
\nonumber V^{\mathrm{Direct}}_\sigma&=& -g_s^2(\mbox{\boldmath$\epsilon$}_2\cdot \mbox{\boldmath$\epsilon$}^\dag_4 ) Y(\Lambda,m_\sigma,r),\\
V^{\mathrm{Direct}}_{\rho}&=& -\frac{1}{2}\beta^2g_V^2 (\mbox{\boldmath$\epsilon$}_2\cdot \mbox{\boldmath$\epsilon$}^\dag_4)Y(\Lambda,m_\rho,r),\\
\nonumber V^{\mathrm{Cross}}_{\rho}&=&2\lambda^2g_V^2\bigg[\frac{2}{3}(\mbox{\boldmath$\epsilon$}_2\cdot
\mbox{\boldmath$\epsilon$}^\dag_3)Z(\Lambda_0,m_0,r)-\frac{1}{3}S(\hat{\mbox{\boldmath$r$}},\mbox{\boldmath$\epsilon$}_2,
\mbox{\boldmath$\epsilon$}^\dag_3)T(\Lambda_0,m_0,r)\bigg],\\
\nonumber V^{\mathrm{Direct}}_{\omega} &=& -\frac{1}{2}\beta^2g_V^2 (\mbox{\boldmath$\epsilon$}_2\cdot \mbox{\boldmath$\epsilon$}^\dag_4)Y(\Lambda,m_\omega,r),\\
\nonumber V^{\mathrm{Cross}}_{\omega}&=& 2 \lambda^2g_V^2\bigg[\frac{2}{3}
(\mbox{\boldmath$\epsilon$}_2\cdot
\mbox{\boldmath$\epsilon$}^\dag_3)Z(\Lambda_1,m_1,r)-\frac{1}{3}
S(\hat{\mbox{\boldmath$r$}},\mbox{\boldmath$\epsilon$}_2,\mbox{\boldmath$\epsilon$}^\dag_3)
T(\Lambda_1,m_1,r)\bigg],
\end{eqnarray}
where $\Lambda_{0,1,2,3}^2=\Lambda^2-(m_{B^*}-m_B)^2$,
$m_{0,1,2,3}^2=m^2_{\rho,\omega,\pi,\eta}-(m_{B^*}-m_B)^2$ and
$S(\hat{\mbox{\boldmath$r$}},\mathbf{a},\mathbf{b})=3(\hat{\mbox{\boldmath$r$}}\cdot\mathbf{a})
(\hat{\mbox{\boldmath$r$}}\cdot\mathbf{b})-\mathbf{a}\cdot\mathbf{b}$.
The functions $Y(\Lambda,m,r)$, $Z(\Lambda,m,r)$ and $T(\Lambda,m,r)$
are defined as:
\begin{eqnarray}
\nonumber Y(\Lambda,m_E,r) &=& \frac{1}{4\pi r}(e^{-m_E\,r}-e^{-\Lambda r})-\frac{\Lambda^2-m_E^2}{8\pi \Lambda }e^{-\Lambda r}\, ,\\
Z(\Lambda,m_E,r) &=& \bigtriangledown^2Y(\Lambda,m_E,r) = \frac{1}{r^2} \frac{\partial}{\partial r}r^2 \frac{\partial}{\partial r}Y(\Lambda,m_E,r) \, ,\\
\nonumber T(\Lambda,m_E,r) &=&  r\frac{\partial}{\partial
r}\frac{1}{r}\frac{\partial}{\partial r}Y(\Lambda,m_E,r) \, .
\end{eqnarray}
One notes that $m_2$ is much smaller than $m_\pi$, which implies the
OPE potential decreases very slowly. In Ref. \cite{Sun:2011uh}, both
the S-wave and D-wave interactions between the $B$ and $\bar{B}^*$
mesons were considered. In the derivation of the final effective
potential in the form of the $2\times 2$ matrix, one makes the
following replacement in the subpotentials,
\begin{eqnarray}
\left. \begin{array}{c}
(\mbox{\boldmath$\epsilon$}_2\cdot \mbox{\boldmath$\epsilon$}^\dag_3) \\
(\mbox{\boldmath$\epsilon$}_2\cdot
\mbox{\boldmath$\epsilon$}^\dag_4)
\end{array}\right\}
\rightarrowtail \left(\begin{array}{cc}
1&0\\
0&1\\
\end{array}\right)\, , \,
S(\hat{\mbox{\boldmath$r$}},\mbox{\boldmath$\epsilon$}_2,\mbox{\boldmath$\epsilon$}^\dag_3)
\rightarrowtail \left(\begin{array}{cc}
0&-\sqrt{2}\\
-\sqrt{2}&1\\
\end{array}\right).
\end{eqnarray}
Similarly, the total effective potential of the $Z_b(10650)$ system
is
\begin{eqnarray}
\mathcal{V}^{Z_b(10650)}&=& W_\sigma-\frac{1}{2}W_\rho +\frac{1}{2}W_\omega-\frac{1}{2}W_\pi+\frac{1}{6}W_\eta,
\end{eqnarray}
where the corresponding subpotentials from the $\pi$, $\eta$,
$\sigma$, $\rho$ and $\omega$ meson exchanges are
\begin{eqnarray}
\nonumber W_\pi&=&
-\frac{g^2}{f_\pi^2}\bigg[\frac{1}{3}(\mbox{\boldmath$\epsilon$}_1\times
\mbox{\boldmath$\epsilon$}^\dag_3)\cdot(\mbox{\boldmath$\epsilon$}_2\times
\mbox{\boldmath$\epsilon$}^\dag_4)Z(\Lambda,m_\pi,r)+\frac{1}{3}S(\hat{\mbox{\boldmath$r$}},\mbox{\boldmath$\epsilon$}_1\times
\mbox{\boldmath$\epsilon$}^\dag_3,
\mbox{\boldmath$\epsilon$}_2\times
\mbox{\boldmath$\epsilon$}^\dag_4)T(\Lambda,m_\pi,r)\bigg],
\\
\nonumber W_\eta&=& -\frac{g^2}{f_\pi^2}
\bigg[\frac{1}{3}(\mbox{\boldmath$\epsilon$}_1\times
\mbox{\boldmath$\epsilon$}^\dag_3)\cdot(\mbox{\boldmath$\epsilon$}_2\times
\mbox{\boldmath$\epsilon$}^\dag_4)Z(\Lambda,m_\eta,r)
+\frac{1}{3}S(\hat{\mbox{\boldmath$r$}},\mbox{\boldmath$\epsilon$}_1\times
\mbox{\boldmath$\epsilon$}^\dag_3,
\mbox{\boldmath$\epsilon$}_2\times \mbox{\boldmath$\epsilon$}^\dag_4)T(\Lambda,m_\eta,r)\bigg],\\
W_\sigma&=& -g_s^2(\mbox{\boldmath$\epsilon$}_1\cdot \mbox{\boldmath$\epsilon$}^\dag_3 ) (\mbox{\boldmath$\epsilon$}_2\cdot \mbox{\boldmath$\epsilon$}^\dag_4 ) Y(\Lambda,m_\sigma,r),\\
\nonumber W_{\rho}&=& -\frac{1}{4} \bigg\{2\beta^2g_V^2(\mbox{\boldmath$\epsilon$}_1\cdot \mbox{\boldmath$\epsilon$}^\dag_3)(\mbox{\boldmath$\epsilon$}_2\cdot \mbox{\boldmath$\epsilon$}^\dag_4)Y(\Lambda,m_\rho,r)-8\lambda^2g_V^2\bigg[\frac{2}{3}(\mbox{\boldmath$\epsilon$}_1\times \mbox{\boldmath$\epsilon$}^\dag_3)\cdot(\mbox{\boldmath$\epsilon$}_2\times \mbox{\boldmath$\epsilon$}^\dag_4)Z(\Lambda,m_\rho,r)\nonumber\\
\nonumber &&-\frac{1}{3}S(\hat{\mbox{\boldmath$r$}},\mbox{\boldmath$\epsilon$}_1\times \mbox{\boldmath$\epsilon$}^\dag_3,\mbox{\boldmath$\epsilon$}_2\times \mbox{\boldmath$\epsilon$}^\dag_4)T(\Lambda,m_\rho,r)\bigg]\bigg\},\\
\nonumber W_{\omega} &=& -\frac{1}{4} \bigg\{2\beta^2g_V^2(\mbox{\boldmath$\epsilon$}_1\cdot \mbox{\boldmath$\epsilon$}^\dag_3)(\mbox{\boldmath$\epsilon$}_2\cdot \mbox{\boldmath$\epsilon$}^\dag_4)Y(\Lambda,m_\omega,r)-8\lambda^2g_V^2\bigg[\frac{2}{3}(\mbox{\boldmath$\epsilon$}_1\times \mbox{\boldmath$\epsilon$}^\dag_3)\cdot(\mbox{\boldmath$\epsilon$}_2\times \mbox{\boldmath$\epsilon$}^\dag_4)Z(\Lambda,m_\omega,r)\nonumber\\
\nonumber &&-\frac{1}{3}S(\hat{\mbox{\boldmath$r$}},\mbox{\boldmath$\epsilon$}_1\times
\mbox{\boldmath$\epsilon$}^\dag_3,\mbox{\boldmath$\epsilon$}_2\times
\mbox{\boldmath$\epsilon$}^\dag_4)T(\Lambda,m_\omega,r)\bigg]\bigg\}.
\end{eqnarray}
There also exists the S-wave and D-wave mixing for the
$B^*\bar{B}^*$ system. The total effective potential of the $Z_b(10650)$
with $J=1$ is a $3\times 3$ matrix, where the polarization related
terms in the subpotentials should be replaced by the following
expressions
{\small\begin{eqnarray*}
&&(\mbox{\boldmath$\epsilon$}_1\cdot
\mbox{\boldmath$\epsilon$}^\dag_3)(\mbox{\boldmath$\epsilon$}_2\cdot
\mbox{\boldmath$\epsilon$}^\dag_4)
 \rightarrowtail \left(\begin{array}{ccc}
1&0&0\\
0&1&0\\
0&0&1\\
\end{array}\right),\,
(\mbox{\boldmath$\epsilon$}_1\times
\mbox{\boldmath$\epsilon$}^\dag_3)\cdot(\mbox{\boldmath$\epsilon$}_2\times
\mbox{\boldmath$\epsilon$}^\dag_4) \rightarrowtail
\left(\begin{array}{ccc}
1&0&0\\
0&1&0\\
0&0&-1\\
\end{array}\right),\,
S(\hat{\mbox{\boldmath$r$}},\mbox{\boldmath$\epsilon$}_1\times
\mbox{\boldmath$\epsilon$}^\dag_3,
\mbox{\boldmath$\epsilon$}_2\times
\mbox{\boldmath$\epsilon$}^\dag_4) \rightarrowtail
\left(\begin{array}{ccc}
0&-\sqrt{2}&0\\
-\sqrt{2}&1&0\\
0&0&1\\
\end{array}\right).
\end{eqnarray*}}

\renewcommand{\arraystretch}{1.4}
\begin{table}[hbtp]
\caption{The bound state solutions (binding energy $E$ and
root-mean-square radius $r_{\mathrm{RMS}}$) for the $Z_b(10610)$ and
$Z_b(10650)$ systems. Here the results of Ref.~\cite{Sun:2011uh} for two situations are listed, i.e.,
including all OBE contribution and only considering the OPE potential. \label{Table.4.1.BBS1}}
\begin{center}
\begin{tabular}{c|ccc|cccc}\toprule[1pt]
&\multicolumn{3}{c|}{OBE}&\multicolumn{3}{c}{OPE}\\\midrule[1pt]
 State& $\Lambda$ & $E$ (MeV)& $r_{\mathrm{RMS}}$
(fm) & $\Lambda$ & $E$ (MeV)& $r_{\mathrm{RMS}}$ (fm)
\\\midrule[1pt]
&2.1&-0.22&3.05&2.2&-8.69&0.62\\
                          $Z_{b}(10610)$ &2.3&-1.64&1.31&2.4&-20.29&0.47\\
                          &2.5&-4.74&0.84&2.6&-38.54&0.36\\\midrule[1pt]

       \multirow{4}*{$Z_b(10650)$}& 2.2&-0.81&1.38&2&-2.17&1.15\\
                &2.4&-3.31&0.95&2.2&-8.01&0.68\\
                &
        2.6&-7.80&0.68&2.4&-19.00&0.48\\
                &
        2.8&-14.94&0.52&2.6&-36.36&0.38\\
\bottomrule[1pt]
\end{tabular}
\end{center}
\end{table}

The bound state solutions for the $Z_{b}(10610)$ and $Z_{b}(10650)$
systems are listed in Table \ref{Table.4.1.BBS1}. Let's summarize
\cite{Sun:2011uh}:
\begin{itemize}
\item
The long-range one-pion-exchange (OPE) force alone is strong enough
to form the loosely bound isovector molecular states composed of the
$B\bar{B}^*$ and $B^*\bar{B}^*$.

\item
The short-range vector meson exchange force provides some effective
repulsion in these channels and prevents the heavy meson from moving
very close to the anti-meson. As can be seen in the case of the
$Z_{b}(10650)$ with $\Lambda=2.2$ GeV, the binding energy from the
OPE potential is 8 MeV and the root-mean-square radius is 0.68 fm.
In contrast, the binding energy from the OBE potential is 0.81 MeV
and its radius is 1.38 fm.

\item
When the binding energies of the $B\bar{B}^*$ and $B^*\bar{B}^*$
systems are less than 1 MeV, their root-mean-square radius may reach
1.4-3 fm, which is characteristic of the molecular states.

\item
With the molecular scheme, the mass splittings $M(Z_b(10650)) -
M(Z_b(10610))=M(B^*)-M(B)$ and $M(Z_c(4020)) -
M(Z_c(3900))=M(D^*)-M(D)$ are governed by the spin-spin interaction
and scale with the heavy quark masses as expected in QCD if
$Z_c(4020)$ and $Z_c(3900)$ are molecular resonances (see also discussions
in Sec.~\ref{sect:4.2}).

\end{itemize}

\paragraph{Other models}

In Ref.~\cite{Yang:2011rp}, Yang, Ping, Deng, and Zong applied the
chiral quark model to study the possible molecular states composed
of a pair of heavy mesons, $B\bar B$, $B \bar B^*$, $B^* \bar B^*$,
and $B_s \bar B$, in the $S$-wave sector. They found the $B \bar
B^*$ and $B^* \bar B^*$ bound states with quantum numbers $I(J^{PC})
= 1(1^{+-})$, which are good candidates of the $Z_b(10610)$ and
$Z_b(10650)$, respectively. They also predicted three bound states:
$B \bar B^*$ with $I(J^{P C}) = 0(1^{++})$, $B^* \bar B^*$ with
$I(J^{P C}) = 1(0^{++})$ and $0(2^{++})$.

The coupled channel unitary approach with the local hidden gauge
formalism, reviewed in Sec.~\ref{Sect.3.2}, was also applied to
study the $B \bar B^*$ and $B^* \bar B^*$ interactions. In
Ref.~\cite{Ozpineci:2013qza}, Ozpineci, Xiao, and Oset investigated
the meson-meson interaction with hidden beauty in both $I = 0$ and
$I = 1$ sectors. They found the interactions are too weak in the $I
= 1$ sector to create any bound state.

In Ref.~\cite{Dias:2014pva}, Dias, Aceti, and Oset used the local
hidden gauge approach to study the $B \bar B^*$ and $B^* \bar B^*$
interactions in the $I = 1$ sector. They considered the
contributions due to the exchange of two pions and heavy vector
mesons. They found a loosely bound state with mass in the range
10587-10601 MeV for the $B \bar B^*$ interaction, very close to the
experimental value of the $Z_b(10610)$ and a cusp at 10650 MeV for
the $B^* \bar B^*$ interaction for $J = 0, 1, 2$.

The method of QCD sum rules was also used to study the two $Z_b$
states
\cite{Zhang:2011jja,Cui:2011fj,Wang:2013daa,Wang:2014gwa,Chen:2015ata,Chen:2013omd,Wang:2010rt,Wang:2013zra}.
Using the $B\bar B^*$ and $B^*\bar B^*$ molecule-type interpolating
currents with $I^GJ^P=1^+1^+$, their extracted masses were
roughly consistent with the masses of the $Z_b(10610)$ and
$Z_b(10650)$ mesons, respectively.

\paragraph{Symmetry analysis}

Within the molecular picture, Bondarn {\it et al.} discussed the
heavy quark spin structure and the $Z_b$ states. Especially, they
noticed that the $b \bar b$ pair within the $Z_b(10610)$ and
$Z_b(10650)$ resonances is a mixture of a spin-triplet and a
spin-singlet of equal amplitude \cite{Bondar:2011ev}. Voloshin
investigated their isoscalar analogs and proposed to observe them in
the $I^G(J^P) = 0^-(1^+)$ channel~\cite{Voloshin:2012yq}. He
pointed out that the ratio of the yield for the pairs of the charged
and neutral $B^{(*)}$ mesons in the processes $\Upsilon(5S) \to
\pi^0 B \bar B^*$ and $\Upsilon(5S) \to \pi^0 B^* \bar B^*$ is very
sensitive to the interaction between the mesons due to significant
isospin breaking by the Coulomb force.

The heavy quark flavor symmetry was applied to study the heavy meson
hadronic molecules in Refs.~\cite{Nieves:2011vw,Guo:2013sya}. In
Ref.~\cite{Nieves:2011vw}, Nieves and Valderrama discussed the
possible $B \bar B^*$ bound states using the analogue to the weakly
bound $X(3872)$ state and under certain assumptions about the short
range dynamics. In Ref.~\cite{Guo:2013sya}, Guo, Hidalgo-Duque,
Nieves, and Valderrama investigated the consequences of the heavy quark
flavor symmetry and predicted many new hadronic molecules.

In Ref.~\cite{Cleven:2011gp}, Cleven, Guo, Hanhart, and Meissner used
the measured invariant mass distributions for the transitions of the
$\Upsilon(5S)$ to the final states $h_b \pi^+ \pi^-$ and $h_b(2P)
\pi^+ \pi^-$ to test the molecular picture. They made use of the
power counting for the bottom meson loops in the framework of a
nonrelativistic effective field theory. Their results showed the
data~\cite{Collaboration:2011gja} is consistent with the assumption
that the main components of $Z_b$ states are $S$-wave $B
\bar B^*$ and $B^* \bar B^*$ bound states, although a small compact
tetraquark component can not be excluded. Possible power counting
schemes were also discussed in Ref.~\cite{Valderrama:2012jv}.

Assuming the binding mechanism between two heavy mesons is
mostly molecular-like isospin-exchange attraction, Karliner and
Rosner studied the bottomonium-like and charmonium-like multiquark
states in Ref.~\cite{Karliner:2015ina}. They also interpreted the
$Z_b(10610)$ and $Z_b(10650)$ as the weakly bound molecular states
composed of the $\bar B B^*$ and $\bar B^* B^*$, respectively.

\subsubsection{The tetraquark assignment}
\label{sect:4.1.2}

\paragraph{QCD sum rules}

The color configurations of the tetraquarks include
{\small$\left(\mathbf{\bar 3}_{[q Q]}\otimes\mathbf{3}_{[\bar Q \bar
q]}\right)$}, {\small$\left(\mathbf{6}_{[q Q]}\otimes\mathbf{\bar 6}_{[\bar
Q \bar q]}\right)$} and {\small$\left(\mathbf{8}_{[q\bar
Q]}\otimes\mathbf{8}_{[Q\bar q]}\right)$}. The first two types
correspond to the diquark configuration.

In 2011, Chen and Zhu had performed an extensive investigation of
the hidden-bottom tetraquark systems with quantum numbers
$J^{PC}=1^{+-}$ in QCD sum rules in Ref. \cite{Chen:2010ze}. They
considered all six kinds of diquark fields $q^T_a CQ_b$, $q^T_a
C\gamma_5Q_b$, $q^T_aC\gamma_\mu Q_b$,
$q^T_aC\gamma_\mu\gamma_5Q_b$, $q^T_a C\sigma_{\mu\nu}Q_b$, and
$q^T_aC\sigma_{\mu\nu}\gamma_5Q_b$ to construct the eight tetraquark
interpolating currents
\begin{eqnarray}
\begin{split}
J_{1\mu}&=q^T_aCQ_b(\bar{q}_a\gamma_{\mu}\gamma_5C\bar{Q}^T_b+\bar{q}_b\gamma_{\mu}\gamma_5C\bar{Q}^T_a)
-
q^T_aC\gamma_{\mu}\gamma_5Q_b(\bar{q}_aC\bar{Q}^T_b+\bar{q}_bC\bar{Q}^T_a)\,
, \\
J_{2\mu}&=q^T_aCQ_b(\bar{q}_a\gamma_{\mu}\gamma_5C\bar{Q}^T_b-\bar{q}_b\gamma_{\mu}\gamma_5C\bar{Q}^T_a)
-
q^T_aC\gamma_{\mu}\gamma_5Q_b(\bar{q}_aC\bar{Q}^T_b-\bar{q}_bC\bar{Q}^T_a)\,
, \\
J_{3\mu}&=q^T_aC\gamma_5Q_b(\bar{q}_a\gamma_{\mu}C\bar{Q}^T_b+\bar{q}_b\gamma_{\mu}C\bar{Q}^T_a)
-
q^T_aC\gamma_{\mu}Q_b(\bar{q}_a\gamma_5C\bar{Q}^T_b+\bar{q}_b\gamma_5C\bar{Q}^T_a)\,
, \\
J_{4\mu}&=q^T_aC\gamma_5Q_b(\bar{q}_a\gamma_{\mu}C\bar{Q}^T_b-\bar{q}_b\gamma_{\mu}C\bar{Q}^T_a)
-
q^T_aC\gamma_{\mu}Q_b(\bar{q}_a\gamma_5C\bar{Q}^T_b-\bar{q}_b\gamma_5C\bar{Q}^T_a)\,
, \\
J_{5\mu}&=q^T_aC\gamma^{\nu}Q_b(\bar{q}_a\sigma_{\mu\nu}\gamma_5C\bar{Q}^T_b+\bar{q}_b\sigma_{\mu\nu}\gamma_5C\bar{Q}^T_a)
-
q^T_aC\sigma_{\mu\nu}\gamma_5Q_b(\bar{q}_a\gamma^{\nu}C\bar{Q}^T_b+\bar{q}_b\gamma^{\nu}C\bar{Q}^T_a)\,
, \\
J_{6\mu}&=q^T_aC\gamma^{\nu}Q_b(\bar{q}_a\sigma_{\mu\nu}\gamma_5C\bar{Q}^T_b-\bar{q}_b\sigma_{\mu\nu}\gamma_5C\bar{Q}^T_a)
-
q^T_aC\sigma_{\mu\nu}\gamma_5Q_b(\bar{q}_a\gamma^{\nu}C\bar{Q}^T_b-\bar{q}_b\gamma^{\nu}C\bar{Q}^T_a)\,
, \\
J_{7\mu}&=q^T_aC\gamma^{\nu}\gamma_5Q_b(\bar{q}_a\sigma_{\mu\nu}C\bar{Q}^T_b+\bar{q}_b\sigma_{\mu\nu}C\bar{Q}^T_a)
-
q^T_aC\sigma_{\mu\nu}Q_b(\bar{q}_a\gamma^{\nu}\gamma_5C\bar{Q}^T_b+\bar{q}_b\gamma^{\nu}\gamma_5C\bar{Q}^T_a)\, , \\
J_{8\mu}&=q^T_aC\gamma^{\nu}\gamma_5Q_b(\bar{q}_a\sigma_{\mu\nu}C\bar{Q}^T_b-\bar{q}_b\sigma_{\mu\nu}C\bar{Q}^T_a)
-
q^T_aC\sigma_{\mu\nu}Q_b(\bar{q}_a\gamma^{\nu}\gamma_5C\bar{Q}^T_b-\bar{q}_b
\gamma^{\nu}\gamma_5C\bar{Q}^T_a)\, ,
\label{Sec:4.1.currentsZb}
\end{split}
\end{eqnarray}
where $q$ represents the up or down quark and $Q$ the bottom quark. The
color structures are symmetric $\mathbf 6 \otimes \mathbf{\bar 6}$ for
the currents $J_1, J_3, J_5, J_7$, and antisymmetric $\mathbf{\bar
3} \otimes \mathbf 3$ for the currents $J_2, J_4, J_6, J_8$. All
these tetraquark currents in Eq.~(\ref{Sec:4.1.currentsZb})
can couple to both isotriplet and isosinglet hadron states
\begin{equation}
\begin{split}
J_\mu\sim qQ\bar q\bar Q\sim \left\{\begin{array}{l} \left.
\begin{array}{l}
Z^+: dQ\bar u\bar Q\\
\, Z^0: uQ\bar u\bar Q +dQ\bar d\bar Q \\
Z^-: uQ\bar d\bar Q\\
\end{array}\right\},~~\mbox{Isovector with}~I=1\, , \\
\left. \, \, \, Z^0: uQ\bar u\bar Q - dQ\bar d\bar Q\, ,
~~\mbox{Isoscalar with}~I=0\, . \right.
\end{array}\right. \label{Eq.4.1.Tetraquarkquarkcontent}
\end{split}
\end{equation}
The extracted masses of the hidden-bottom tetraquarks are collected
in Table \ref{Table.4.1.ZbTetraquarkQSR}.

Besides the diquark configuration, the color octet-octet
$\left(\mathbf{8}_{[q\bar Q]}\otimes\mathbf{8}_{[Q\bar q]}\right)$
types of interpolating currents were also considered for the
hidden-bottom tetraquark systems in Refs.
\cite{Chen:2015ata,Chen:2013omd}
\begin{eqnarray}
\begin{split}
J_{1\mu}^{(\mathbf 8)}&=(\bar q_a\gamma_5\lambda^n_{ab} Q_b)(\bar{Q}_c\gamma_{\mu}\lambda^n_{cd}q_d)+(\bar q_a\gamma_{\mu}\lambda^n_{ab}Q_b)(\bar{Q}_c\gamma_5\lambda^n_{cd}q_d)\, ,
\\  J_{2\mu}^{(\mathbf 8)}&=(\bar q_a\lambda^n_{ab}Q_b)(\bar{Q}_c\gamma_{\mu}\gamma_5\lambda^n_{cd}q_d)-(\bar q_a\gamma_{\mu}\gamma_5\lambda^n_{ab}Q_b)(\bar{Q}_c\lambda^n_{cd}q_d)\, ,
\\
J_{3\mu}^{(\mathbf 8)}&=(\bar q_a\gamma^{\alpha}\lambda^n_{ab}Q_b)(\bar{Q}_c\sigma_{\alpha\mu}\gamma_5\lambda^n_{cd}q_d)-(\bar q_a\sigma_{\alpha\mu}\gamma_5\lambda^n_{ab}Q_b)(\bar{Q}_c\gamma^{\alpha}\lambda^n_{cd}q_d)\, ,
\\
J_{4\mu}^{(\mathbf 8)}&=(\bar q_a\gamma^{\alpha}\gamma_5\lambda^n_{ab}Q_b)(\bar{Q}_c\sigma_{\alpha\mu}\lambda^n_{cd}q_d)+(\bar q_a\sigma_{\alpha\mu}\lambda^n_{ab}Q_b)(\bar{Q}_c\gamma^{\alpha}\gamma_5\lambda^n_{cd}q_d)\, . \label{Eq.4.1.currents1}
\end{split}
\end{eqnarray}
The extracted masses are collected in Table
\ref{Table.4.1.ZbQSR}.

From Tables \ref{Table.4.1.ZbQSR} and
\ref{Table.4.1.ZbTetraquarkQSR}, the extracted masses of the
tetraquark states using both the diquark-diquark and color
octet-octet types of interpolating currents are roughly the same.
The bottomonium-like $qb\bar q\bar b$ and $sb\bar s\bar b$
tetraquark states are nearly degenerate around $9.9-10.2$ GeV. These
values are much lower than the masses of the $Z_b(10610)$ and
$Z_b(10650)$ states.

There exist colored forces between the $[q\bar b]_{\mathbf{8}}$ and
$[b\bar q]_{\mathbf{8}}$ clusters or the diquark antidiquark pair.
The tetraquarks may be more compact than the color singlet-singlet
hadron molecules. In other words, the numerical results from the QCD
sum rule approach do not support the tetraquark interpretation of
these two charged $Z_b$ mesons \cite{Chen:2010ze}.

%%%%%%%%%%%%%%%%%%%%%%%%%%%%%%%%%%%%%%%
\renewcommand{\arraystretch}{1.6}
\begin{table}[htb]
\caption{Numerical results for the $J^{PC}=1^{+-}$ bottomonium-like
tetraquark states using the diquark-diquark interpolating currents \cite{Chen:2010ze}.
\label{Table.4.1.ZbTetraquarkQSR}}
\begin{center}
\begin{tabular}{cccccc} \toprule[1pt]
& Current ~~&~~ $s_0\,(\mbox{GeV}^2)$ ~~&~~ \mbox{Borel window} $(\mbox{GeV}^2)$ ~~&~~ $m_X$ \mbox{(GeV)} ~~&~~ \mbox{PC\,(\%)} \\
\midrule[1pt]
{\multirow{4}{*}{$qb\bar q\bar b$}}        & $J_{3\mu}$         &  $10.6^2$           & $7.5-8.5$           & $10.08\pm0.10$    & 45.9  \\
                        & $J_{4\mu}$         &  $10.6^2$           & $7.5-8.5$           & $10.07\pm0.10$    & 46.2  \\
                        & $J_{5\mu}$         &  $10.6^2$           & $7.5-8.4$           & $10.05\pm0.10$    & 45.3  \\
                        & $J_{6\mu}$         &  $10.7^2$           & $7.5-8.7$           & $10.15\pm0.10$    & 47.6
\vspace{5pt} \\
{\multirow{4}{*}{$sb\bar s\bar b$}}        & $J_{3\mu}$         &  $10.6^2$           & $7.5-8.3$           & $10.11\pm0.10$    & 43.8  \\
                        & $J_{4\mu}$         &  $10.6^2$           & $7.5-8.4$           & $10.10\pm0.10$    & 44.1  \\
                        & $J_{5\mu}$         &  $10.6^2$           & $7.5-8.3$           & $10.08\pm0.10$    & 43.7  \\
                        & $J_{6\mu}$         &  $10.7^2$           & $7.5-8.5$           & $10.18\pm0.10$    & 46.5  \\
\bottomrule[1pt]
\end{tabular}
\end{center}
\end{table}
%%%%%%%%%%%%%%%%%%%%%%%%%%%%%%%%%%

%%%%%%%%%%%%%%%%%%%%%%%%%%%%%%%%%%%%%%%
\renewcommand{\arraystretch}{1.6}
\begin{table}[htb]
\caption{Numerical results for the $J^{PC}=1^{+-}$ bottomonium-like
tetraquark states using the color octet-octet interpolating
currents \cite{Chen:2015ata,Chen:2013omd}.
\label{Table.4.1.ZbQSR}}
\begin{center}
\begin{tabular}{cccccc} \toprule[1pt]
& Current ~~&~~ $s_0\,(\mbox{GeV}^2)$ ~~&~~ \mbox{Borel window} $(\mbox{GeV}^2)$ ~~&~~ $m_X$ \mbox{(GeV)} ~~&~~ $f_X$ $(10^{-2}\,\mbox{GeV}^5)$ \\
\midrule[1pt]
& $J_{1\mu}^{(\mathbf 8)}(B\bar B^*)$   & 108   & $7.5 - 8.8 $    & $ 9.93\pm0.15$   & $1.02\pm0.30$ \\
& $J_{3\mu}^{(\mathbf 8)}(B^*\bar B^*)$   & 108   & $7.8 - 8.7 $ & $
9.92\pm0.15$   & $2.17\pm0.62$ \\\bottomrule[1pt]
\end{tabular}
\end{center}
\end{table}
%%%%%%%%%%%%%%%%%%%%%%%%%%%%%%%%%%

\paragraph{Diquark model}

Motivated by the $Y_b(10890)$ resonance observed by the Belle
Collaboration~\cite{Abe:2007tk}, Ali {\it et al.} studied the
spectroscopy and decays of the bottomonium-like tetraquarks in
Ref.~\cite{Ali:2009pi}. Assuming the existence of the tightly
bound diquarks ($bq$) and antidiquarks ($\bar b \bar
q$)~\cite{Ali:2009pi,Ali:2009es}, they adopted the effective
Hamiltonian
\begin{eqnarray}
H = 2 m_\mathcal{Q} + H^{(\mathcal{Q}\mathcal{Q})}_{SS} +
H^{(\mathcal{Q}\mathcal{\bar Q})}_{SS} + H_{SL} + H_{LL} \, ,
\end{eqnarray}
which includes the constituent diquark mass $m_\mathcal{Q}$,
spin-spin interaction inside the single diquark
$H^{(\mathcal{Q}\mathcal{Q})}_{SS}$, spin-spin interaction between
quark and antiquark belonging to two diquarks
$H^{(\mathcal{Q}\mathcal{\bar Q})}_{SS}$, spin-orbit term $H_{SL}$,
and purely orbital term $H_{LL}$. One notes that there does not
exist any confinement dynamics in the above Hamiltonian. The
dominant contribution arises from the constituent diquark mass.
Various hyperfine interactions have to be determined through fitting
to data under the assumption that some observed states are
tetraquark states. In contrast, the hyperfine interactions in the
quark model were derived rigorously with the help of the one gluon
exchange potential and linear confinement potential.

Later in Ref.~\cite{Ali:2011ug}, Ali, Hambrock, and Wang reproduced
the observed masses of the $Z_b(10610)$ and $Z_b(10650)$ in terms of
the decay widths for the $h_b(2P)\pi^\pm$. They obtained a ratio for the
relative decay amplitudes in the decays $Z_b(10610)/Z_b(10650) \to
h_b(mP) \pi^\pm$, which agrees with the experimental
data~\cite{Belle:2011aa}. Under the tetraquark hypothesis, the
$Z_b(10610)$ and $Z_b(10650)$ were further
investigated with the non-relativistic QCD factorization scheme~\cite{Ali:2013xba}. Ali {\it et al.} identified the $Z_b(10610)$ and
$Z_b(10650)$ as the $S$-wave $J^{PG} = 1^{++}$ states with the
diquark spin distribution~\cite{Ali:2014dva}:
\begin{eqnarray}
Z_b(10610) &=&\frac{1}{\sqrt{2}} \left[{|1_{[bq]}, 0_{[\bar b\bar q]}\rangle - |0_{[bq]}, 1_{[\bar b\bar q]}\rangle}\right] \, , \\
Z_b(10650) &=& |1_{[bq]}, 1_{[\bar b\bar q]}\rangle_{J=1} \, ,
\end{eqnarray}
where $s_{[bq]}$ and $s_{[\bar b\bar q]}$ are the diquark and
antidiquark spins, respectively. Moreover, there also exists the
partner of the $X(3872)$ in the bottom sector $X_b$ with $J^{PC} =
1^{++}$
\begin{eqnarray}
X_b &=&\frac{1}{\sqrt{2}} \left[{|1_{[bq]}, 0_{[\bar b\bar q]}\rangle + |0_{[bq]}, 1_{[\bar
b\bar q]}\rangle}\right] \, .
\end{eqnarray}
%Within the tetraquark scheme, the mass difference $M(Z_b(10650)) -
%M(Z_b(10610))$ and $M(Z_c(4020)) - M(Z_c(3900))$ also scales with
%the heavy quark mass. {\color{red} I am not sure what is the mean of
%this sentence.}

\subsubsection{Kinematical effect}
\label{Sect:4.1.3}

Besides the above resonant interpretations of the ${Z_b(10610)}$ and
${Z_b(10650)}$, Bugg proposed the ${Z_b(10610)}$ and ${Z_b(10650)}$ as
the cusp effect around the ${\bar{B} B^*}$ and ${\bar{B}^* B^*}$
thresholds \cite{Bugg:2011jr}. Szczepaniak analyzed the properties
of the partial waves %which were derived from projection of a 4-legged amplitude with crossed-channel exchanges
in the kinematic
region of the direct channel which corresponds to the ${Z_b(10610)}$. He
pointed out that the triangle singularities would also give
considerable contributions to the ${Z_b(10610)}$ peak
\cite{Szczepaniak:2015eza}.

\begin{figure}[hbtp]
\begin{center}
\scalebox{0.5}{\includegraphics{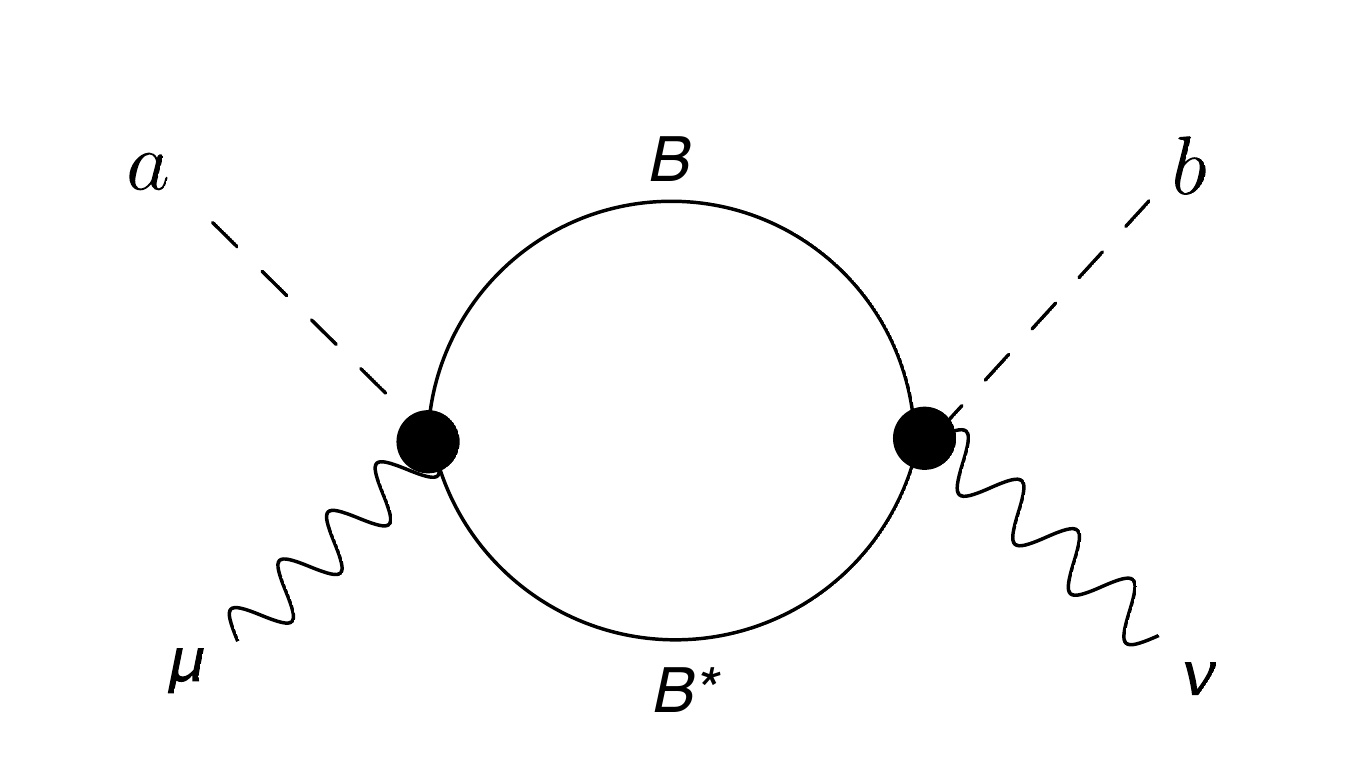}}
\end{center}
\caption{The $\Upsilon\pi$ scattering by including the
coupled-channel effect. Taken from Ref. \cite{Swanson:2014tra}.
\label{Fig.4.1.figcusp}}
\end{figure}

Swanson also discussed the cusp hypothesis of the $Z_b(10610)$,
$Z_b(10650)$, $Z_c(3900)$ and $Z_c(4025)$ \cite{Swanson:2014tra}.
The decay amplitude $\Upsilon\to \pi\pi \Upsilon$ can be related to
the scattering amplitude of $\pi\Upsilon\to \pi \Upsilon$ in Fig.~\ref{Fig.4.1.figcusp}.
He considered the angular momentum barrier factors
and reproduced the experimental data of $\Upsilon(5S)\to
\Upsilon(mS)\pi\pi$ $(m=1,2,3)$ and $\Upsilon(5S)\to h_b(nP)\pi\pi$
($n=1,2$) (see Figures 3-5 of Ref. \cite{Swanson:2014tra} for more
details). He concluded that the $Z_b(10610)$ and $Z_b(10650)$ may arise
from the kinematical effect \cite{Swanson:2014tra}.

\begin{figure*}[htb]
\centering
\includegraphics[width=0.23\textwidth]{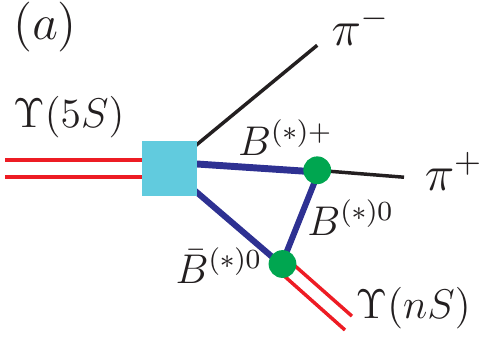}\qquad\qquad\includegraphics[width=0.23\textwidth]{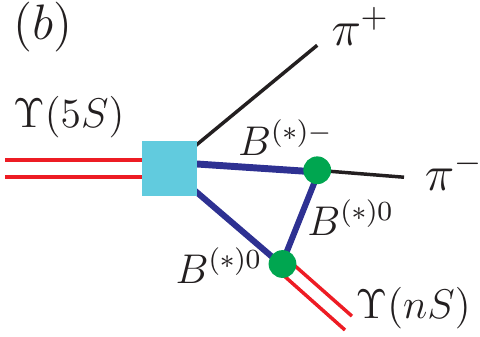}
\caption{(Color online) The schematic diagrams describing the ISPE
mechanism of the $\Upsilon(5S)$. Here, we use $\Upsilon(5S)\to
\Upsilon(nS)\pi^+\pi^-$ as an example. Taken from Ref.
\cite{Chen:2011pv}.}
\label{Fig.4.1.ISPE1}       % Give a unique label
\end{figure*}

\begin{figure*}[htbp]
\centering
\includegraphics[width=0.9\textwidth]{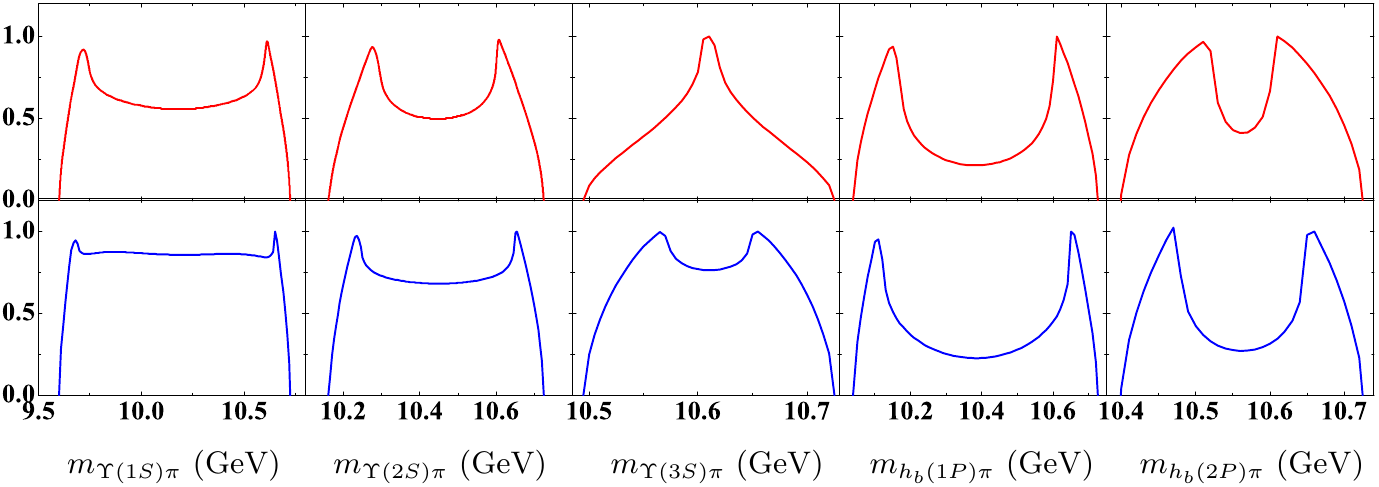}
\caption{(Color online) The invariant mass spectra of the
$\Upsilon(nS)\pi^\pm$ ($n=1,2,3$) and $h_b(mP)\pi^\pm$ ($m=1,2$) in
$\Upsilon(5S)\to \Upsilon(nS)\pi^+\pi^-$ and $\Upsilon(5S)\to
h_b(mP)\pi^+\pi^-$ decays. Here, the maximum of the theoretical line
shape is normalized to 1. Taken from Ref. \cite{Chen:2011pv}.}
\label{Fig.4.1.ISPE2}       % Give a unique label
\end{figure*}

In Ref. \cite{Chen:2011pv}, Chen and Liu proposed the {\it initial
single pion emission} (ISPE) mechanism in the hidden-bottom dipion
decays of the $\Upsilon(5S)$. The direct emission of the single pion
from the $\Upsilon(5S)$ ensures that the intermediate
$B^{(*)}\bar{B}^{(*)}$ pairs carry low momenta. Then meson pairs
interact with each other to transit into final states through the
exchange of one $B^{(*)}$ meson (see Fig. \ref{Fig.4.1.ISPE1}). There exist sharp structures around
10610 MeV and 10650 MeV in the line shapes of the
${d\Gamma(\Upsilon(5S\to
\Upsilon(nS)\pi^+\pi^-))}/{dm_{\Upsilon(nS)\pi^+}}$ and
${d\Gamma(\Upsilon(5S\to h_b(mP)\pi^+\pi^-))}/{dm_{h_b(mP)\pi^+}}$
distributions in Fig. \ref{Fig.4.1.ISPE2}, which could correspond to the
$Z_b(10610)$ and $Z_b(10650)$ structures.

\subsubsection{Production and decay patterns}
\label{Sect:4.1.4}

Assuming the $Z_b(10610)$ and $Z_b(10650)$ to be hadronic molecules
composed of the $\bar BB^*$ and $\bar B^* B^*$, their radiative
productions from the $\Upsilon(5S)$ were studied in the heavy quark spin
symmetry limit in Ref.~\cite{Voloshin:2011qa}. Their two-body strong
decays $Z^+_b(10610) \to \Upsilon(nS) \pi^+$ and $Z^+_b(10650) \to
\Upsilon(nS) \pi^+$ were evaluated in a phenomenological Lagrangian
approach in Ref.~\cite{Dong:2012hc}. Their transitions to the
bottomonium with the emission of a pion were investigated with the
application of the leading (dipole) term in the QCD multipole
expansion in Ref.~\cite{Li:2012uc}. These transitions were also
studied via the intermediate $B \bar B$ meson loops in
Ref.~\cite{Li:2012as} and via the triangle diagrams in
Ref.~\cite{Ohkoda:2013cea}.

In Ref. \cite{Ma:2014ofa}, the authors adopted the spin
rearrangement scheme in the heavy quark limit and extensively
investigated three classes of the radiative decays: $\mathfrak{M}\to
(b\bar{b})+\gamma$, $(b\bar{b})\to \mathfrak{M}+\gamma$, $
\mathfrak{M} \to \mathfrak{M}^\prime+\gamma$, corresponding to the
electromagnetic transitions between one molecular state and
bottomonium, one bottomonium and molecular state, and two molecular
states respectively. Some model independent ratios were derived when
the initial or final states belong to the same spin flavor
multiplet.

With the spin rearrangement scheme in the heavy quark limit, the
authors of Ref. \cite{Ma:2014zva} performed a comprehensive
investigation of the decay pattern and production mechanism of the
hidden bottom di-meson states, which are either composed of a P-wave
bottom meson and an S-wave bottom meson or two S-wave bottom mesons.
The model-independent ratios can be measured by future experiments
like BESIII, Belle, LHCb and the forthcoming BelleII, which will
provide important clues to the inner structures of the exotic
states.

Line shapes in the vicinity of the $B^{(*)} \bar B^{(*)}$ thresholds as
well as two-body decay rates of the $Z_b(10610)$ and $Z_b(10650)$
were studied using the heavy quark symmetry in
Ref.~\cite{Mehen:2011yh}. The mixing of the S-D partial waves
for the heavy meson pairs in the decays $\Upsilon(5S) \to B^* \bar B
\pi$ and $\Upsilon(5S) \to B^* \bar B^* \pi$ was studied in
Ref.~\cite{Voloshin:2013ez}, where the $Z_b(10610)$ and $Z_b(10650)$
were taken into account. Decays and productions of the $Z_b$ and
other $B\bar B$ molecules resonances via the bottomonium were
studied under the heavy quark symmetry in Ref.~\cite{Ohkoda:2012rj}.
The contribution of the $Z_b$ resonances to $\Upsilon(5S) \to
\pi\pi\pi\chi_b$ was studied in Ref.~\cite{Li:2014cia}.

The decays of the $Z_b(10610)$ and $Z_b(10650)$ to $\Upsilon(nS) \pi$,
$h_b(mP)\pi$ and $\chi_{bJ} (mP)\gamma$ ($n = 1, 2, 3$, $m = 1, 2$
and $J = 0, 1, 2$) were investigated within a nonrelativistic
effective field theory in Ref.~\cite{Cleven:2013sq}. The authors
argued that the decays to $\Upsilon(nS) \pi$ suffer from potentially
large higher order corrections. However, the P-wave transitions of
the $Z_b$ states are dominated by a single one-loop diagram and
therefore offer the best possibility to confirm the nature of the
$Z_b$ states as molecular states and to further study their
properties. In Ref. \cite{Chen:2015jgl}, the contribution of the
charged $Z_b$ states to the $\Upsilon(3S) \to \Upsilon(1S) \pi \pi$ decays was
discussed.

\subsubsection{A short summary}
\label{Sect:4.1.5}

\begin{itemize}
\item
The $Z_b(10610)$ and $Z_b(10650)$ are charged hidden-bottom states with
narrow widths, which are close to the thresholds of the $B\bar{B}^*$ and
$B^*\bar{B}^{*}$, respectively. The $B{\bar B}^*$ branching ratio is
$(86.0\pm 3.6)\%$ for the $Z_b(10610)$. For the $Z_b(10650)$, the $B^{(*)}{\bar
B}^*$ branching ratio is $(73.4\pm 7.0)\%$. Both the $Z_b(10610)$ and
$Z_b(10650)$ are very probably either the $B^*\bar{B}^{*}$ or
$B^*\bar{B}^{*}$ molecular states respectively.

\item
There exists attraction in the isovector axial vector channel. In
fact the one-pion-exchange force alone is strong enough to form the
shallow bound states composed of the $B\bar{B}^*$ and $B^*\bar{B}^*$.
There also exist their isoscalar molecular partners.

\item
The $\Upsilon(5S)$ [or $\Upsilon (6S)$] is the ideal factory of the
heavy molecular states, which shall be produced abundantly at
BelleII in the near future! The masses of the $B{\bar B}^*\pi$ and
$B^*{\bar B}^*\pi$ are 10.744 GeV and 10.790 GeV respectively, which
are very close to the $\Upsilon (5S)$ mass 10.860 GeV. Because of
the tiny decay phase space, the relative motion between the
$B^{(*)}\bar{B}^*$ pair is very slow, which is favorable to the
formation of the $B^{(*)}\bar B^*$ molecular states.

\item
If the four quarks $q\bar qb\bar b$ are confined within the MIT bag
to form an isovector axial vector tetraquark state, such a system
will decay into the $\Upsilon (1S)\pi$ via S-wave easily and has a
very large decay width around several hundred MeV. The $\Upsilon
(1S)\pi$ should be its dominant decay mode.

\item
To explain the narrow total width and the dominant open-bottom decay
modes, the diquark within the tetraquark was assumed to be tightly bound
and compact, which awaits future experimental confirmation.

\end{itemize}

\subsection{$Z_c(3900)$, $Z_c(4020)$ and $Z_c(4025)$}\label{sect:4.2}

\subsubsection{Molecular scheme}\label{Sect:4.2.1}

\paragraph{Meson exchange model}

In 2008, the possible $D^{(*)}\bar{D}^{(*)}$ molecular states within
the OBE model were discussed in Ref. \cite{Liu:2008tn}. Later, Sun
{\it et al.} considered the S-D wave mixing effect and performed an
intensive study of the $D\bar{D}^*$ and $D^*\bar{D}^*$ molecular
state systems in Ref. \cite{Sun:2012zzd}.

\renewcommand{\arraystretch}{1.5}
\begin{table}[htbp]
\caption{The obtained bound state solutions (binding energy $E$ and
root-mean-square radius $r_{\mathrm{RMS}}$) for the $D\bar{D}^*$
system. Taken from Ref. \cite{Sun:2012zzd}.
\label{Table.4.2.BBS2}}
\begin{center}
\begin{tabular*}{150mm}{@{\extracolsep{\fill}}cccccccccccc}
\toprule
&&\multicolumn{3}{c}{OBE}&\multicolumn{3}{c}{OPE}\\
$I^G(J^{PC})$ & State& $\Lambda$ & $E$ (MeV)& $r_{\mathrm{RMS}}$ (fm) & $\Lambda$ & $E$ (MeV)& $r_{\mathrm{RMS}}$ (fm) \\\hline%\midrule[1pt]
\multirow{4}*{$1^+(1^{+-})$}&\multirow{4}*{$\hat\Phi^{*}$}&\multirow{4}*{-}&\multirow{4}*{-}&\multirow{4}*{-}
                          &4.6&-0.85&1.46\\
                          &&&&&4.7&-3.42&1.17\\
                          &&&&&4.8&-7.18&0.93\\
                          &&&&&4.9&-12.40&0.75\\
                          %\hline
\multirow{1}*{$1^-(1^{++})$}&\multirow{1}*{$\Phi^{*}$}&-&-&-&-&-&-\\
                  %\hline
\multirow{4}*{$0^-(1^{+-})$}&\multirow{4}*{$\hat\Phi^{*}_8$}&1.3&-&-&3.4&-0.11&1.74\\
                &
        &1.4&-1.56&1.61&3.5&-2.03&1.50\\
                &
        &1.5&-12.95&0.98&3.6&-4.79&1.26\\
                &
        &1.6&-35.73&0.69&3.7&-9.62&1.06\\
%\hline
\multirow{4}*{$0^+(1^{++})$}&\multirow{4}*{$\Phi^{*}_8$}&1.1&-0.61&&1.7&-3.01&1.37\\
                                   &&1.2&-4.42&1.38&1.8&-7.41&1.06\\
                                   &&1.3&-11.78&1.05&1.9&-14.15&0.84\\
                                   &&1.4&-21.88&0.86&2&-23.82&0.68\\
\bottomrule
\end{tabular*}%
\end{center}
\end{table}

Except the one-pion-exchange potential, the effective potential of
the $D\bar{D}^*$ system is similar to that of the $B\bar{B}^*$ system
listed in Sec.~\ref{sect:4.1.1.1}. The mass gap between $m_{D^*}$ and
$m_D$ is larger than the pion mass, which is different from the case
of the $B\bar{B}^*$ system. In the derivation of the OPE potential
in the coordinate space, one generally keeps the principal value
only when making the Fourier transformation to the scattering
amplitude in the momentum space. The OPE potential of the
$D\bar{D}^*$ system not only oscillates but also decreases very
slowly \cite{Liu:2008fh,Liu:2008tn}, which is an inherent
uncertainty of the OBE model when the mass gap of the final and
initial states in the crossed diagram is larger than the exchanged
meson mass. The potential from the $\pi$ meson exchange is
\begin{eqnarray}
V^{\mathrm{Cross}}_\pi&=&
-\frac{g^2}{f_\pi^2}\bigg[\frac{1}{3}(\mbox{\boldmath$\epsilon$}_2\cdot
\mbox{\boldmath$\epsilon$}^\dag_3)Z^{DD^*}_\pi(\Lambda_4,m_4,r)+\frac{1}{3}S(\hat{\mbox{\boldmath$r$}},\mbox{\boldmath$\epsilon$}_2,
\mbox{\boldmath$\epsilon$}^\dag_3)T^{DD^*}_\pi(\Lambda_4,m_4,r)\bigg],
\end{eqnarray}
where
\begin{eqnarray}
\nonumber Y^{DD^*}_\pi(\Lambda_4,m_4,r) &=& \frac{1}{4\pi r}\bigg(-e^{-\Lambda_4 r}-\frac{r(\Lambda_4^2+m_4^2)}{2\Lambda_4}e^{-\Lambda_4 r}+\cos({m_4r})\bigg),\\
Z^{DD^*}_\pi(\Lambda_4,m_4,r) &=& \bigtriangledown^2Y^{DD^*}_\pi(\Lambda_4,m_4,r) = \frac{1}{r^2} \frac{\partial}{\partial r}r^2 \frac{\partial}{\partial r} Y^{DD^*}_\pi(\Lambda_4,m_4,r),\\
\nonumber T^{DD^*}_\pi(\Lambda_4,m_4,r) &=&  r\frac{\partial}{\partial
r}\frac{1}{r}\frac{\partial}{\partial
r}Y^{DD^*}_\pi(\Lambda_4,m_4,r).
\end{eqnarray}
The parameters $\Lambda_4$ and $m_4$ are defined as $\Lambda_4 =
\sqrt{\Lambda^2-(m_{D^*}-m_D)^2}$ and $m_4 =
\sqrt{(m_{D^*}-m_D)^2-m_\pi^2}$. The potentials of the
$D^*\bar{D}^*$ system and $B^*\bar{B}^*$ system have the same form
(see Sec.~\ref{sect:4.1.1.1}).

\begin{table}[htbp]
\caption{The obtained bound state solutions (binding energy $E$ and
root-mean-square radius $r_{\mathrm{RMS}}$) for the $D^*\bar{D}^*$
system. Taken from Ref. \cite{Sun:2012zzd}.
\label{Table.4.2.DSDS1}}
\begin{center}
\begin{tabular*}{150mm}{@{\extracolsep{\fill}}cccccccccccc}
\toprule
&&\multicolumn{3}{c}{OBE}&\multicolumn{3}{c}{OPE}\\%\midrule[1pt]
$I^G(J^{PC})$ & State& $\Lambda$ & $E$ (MeV)& $r_{\mathrm{RMS}}$
(fm)& $\Lambda$ & $E$ (MeV)& $r_{\mathrm{RMS}}$ (fm)\\\hline
\multirow{4}*{$1^{+}(0^+)$}&\multirow{4}*{$\Phi^{**}[J=0]$}
                        &3.6&-0.94&1.74&2.8&-2.03&1.47\\
                        &&3.8&-6.16&1.00&2.9&-6.10&1.00\\
                        &&4&-16.44&0.66&3&-12.51&0.74\\
                        &&4.2&-33.23&0.49&3.1&-21.56&0.59\\
\multirow{4}*{$0^{-}(0^{+-})$}&\multirow{4}*{$\Phi_8^{**}[J=0]$}&1.4&-1.72&1.62&3&-5.70&1.24\\
            &&
        1.5&-17.98&0.88&3.1&-12.15&0.96\\
            &&
        1.6&-54.60&0.47&3.2&-21.83&0.78\\
\multirow{4}*{$1^{+}(1^+)$}&\multirow{4}*{$\Phi^{**}[J=1]$}&\multirow{4}*{-}&\multirow{4}*{-}&\multirow{4}*{-}&4.7&-6.96&0.94\\
                &&           &&&4.8&-12.29&0.73\\
                &&           &&&4.9&-19.36&0.60\\
                &&           &&&5&-28.31&0.51\\
\multirow{4}*{$0^{-}(1^{+-})$}&\multirow{4}*{$\Phi_8^{**}[J=1]$}&1.3&-&-&3.6&-9.91&1.01\\
                                   &&1.4&-3.44&1.44&3.7&-15.25&0.87\\
                                   &&1.5&-16.57&0.90&3.8&-22.07&0.76\\
                                   &&1.6&-41.25&0.66&3.9&-30.53&0.68\\
\multirow{1}*{$1^{+}(2^+)$}&\multirow{1}*{$\Phi^{**}[J=2]$}&-&-&-&-&-&\\

\multirow{4}*{$0^{-}(2^{+-})$}&\multirow{4}*{$\Phi_8^{**}[J=2]$}&1.1&-0.61&1.72&1.6&-3.89&1.28\\
                                   &&1.2&-7.50&1.19&1.7&-9.64&0.98\\
                                   &&1.3&-19.22&0.89&1.8&-18.38&0.77\\
                                   &&1.4&-35.93&0.73&1.9&-30.71&0.64\\
\bottomrule
\end{tabular*}%
\end{center}
\end{table}

In Tables \ref{Table.4.2.BBS2} and \ref{Table.4.2.DSDS1}, the
obtained bound state solutions for the $D\bar{D}^*$ and
$D^*\bar{D}^*$ systems are correlated with the cutoff parameter
$\Lambda$ in the monopole form factor
$F(q)=(\Lambda^2-m_E^2)/(\Lambda^2-q^2)$, which is introduced to
suppress the contribution from the large momentum exchange. When
$\Lambda$ is very large, $F(q)\approx 1$ for the soft momentum
exchange. As we emphasized, the deuteron is the only well
established di-hadron molecule. Various meson and nucleon coupling
constants are known well. After fitting to the experimental data,
the value of $\Lambda$ turns out to be 1-2 GeV in the deuteron case.
Such a value is regarded as ``reasonable" and used in the discussion
of the hidden-charm molecular states and other molecular systems.
The readers should be cautious about the uncertainty of this
criteria.

From the numerical results in Tables \ref{Table.4.2.BBS2} and \ref{Table.4.2.DSDS1},
we list some interesting observations from the OPE and OBE models
below.
\begin{itemize}
\item There exists the long-range attraction due to the pion exchange
in the $I^GJ^P=1^+1^+$ $D\bar{D}^*$ and $D^*\bar{D}^*$ systems.

\item There exist loosely bound $D\bar{D}^*$ and $D^*\bar{D}^*$
molecular states in the isoscalar channel for a reasonable cutoff
around 1-2 GeV~\cite{Navarra:2001ju}, where the short-range vector meson exchange provides
additional attraction.

\item With a large cutoff around 4.7 GeV which corresponds to the form
factor $F(q)\approx 1$, the one-pion-exchange force is
strong enough to form the isovector $D\bar{D}^*$ and $D^*\bar{D}^*$
molecular bound states. However, the short-range interaction from
the vector meson exchange in the OBE model tends to dissolve these
loosely bound systems.

\item Increasing the pionic coupling constant $g$ is helpful to
form the bound states~\cite{He:2013nwa}.

\item There do not exist the isovector $D\bar{D}^*$ and $D^*\bar{D}^*$
molecular bound states if one uses the value of the pionic coupling
extracted from the $D^*$ decay width and insists a cutoff around 1-2
GeV. However, the cutoff requirement arises from the experience with
the deuteron only. One needs to keep this point in mind.

\item The $Z_c(3900)$ and $Z_c(4020)$ may be the molecular resonances
generated by the long-range one-pion-exchange force, although the
OPE force might not be strong enough to form the molecular bound
states below the threshold with a cutoff around 1-2 GeV.

\end{itemize}

\paragraph{Other molecular models on the $Z_c(3900)$ and $Z_c(4025)$}

The observation of the $Z_c(3900)$ has trigged intensive discussions. In
Ref.~\cite{Zhao:2014gqa}, Zhao, Ma, and Zhu considered the
spin-orbit force and recoil corrections in the $D\bar{D}^*$
interaction in the OBE model. They found two isoscalar $D \bar D^*$
molecular states with $J^{PC} = 1^{++}$ and $1^{+-}$, the first of
which corresponds to the $X(3872)$. However, they found it not easy to
accommodate the $Z_c(3900)$ as the candidate of the isovector
molecular bound state of the $D \bar D^*$. Later, He investigated the $D
\bar D^*$ systems in the framework of the Bethe-Salpeter approach
with the quasipotential approximation \cite{He:2014nya}, where both
direct and cross diagrams in the one-boson-exchange potential were
considered. His results indicated the existence of an isoscalar bound
state $D \bar D^*$ with $J^{PC} = 1^{++}$, which may be related to
the $X(3872)$. But no bound state was produced from the $D \bar D^*$
interaction in the isovector sector \cite{He:2014nya}.

The authors of Refs. \cite{Sun:2012zzd,Zhao:2014gqa,He:2014nya}
noticed that it is not so easy to accommodate $Z_c(3900)$ as an
isovector $D \bar D^*$ molecular bound state, which is supported by
several Lattice
studies~\cite{Prelovsek:2013xba,Prelovsek:2014swa,Chen:2014afa,Chen:2015jwa}.
Prelovsek and Leskovec searched for the $Z_c(3900)$ on the lattice
in the channel with $J^{PC} = 1^{+-}$ and $I = 1$ without
success~\cite{Prelovsek:2013xba}. Instead, they found discrete scattering
states $D \bar D^*$ and $J/\psi \pi$ only. Later in
Ref.~\cite{Prelovsek:2014swa}, a search for the $Z_c^+$ with mass
below 4.2 GeV was performed for the $\bar cc \bar du$ channel with
$I^G(J^{PC}) = 1^+(1^{+-})$. The authors of
Ref.~\cite{Prelovsek:2014swa} were able to find all the expected
signals. But again they found no convincing signal for an
extra $Z^+_c$ state. In Ref.~\cite{Chen:2014afa}, Chen {\it et al.}
analyzed the low-energy scattering of the $D \bar D^*$ meson system
using Lattice QCD with $N_f = 2$ twisted mass fermion configurations
with three pion mass values. Their results indicated a weak repulsive
interaction between the two mesons $D$ and $\bar D^*$, and did not
support a bound state in this channel corresponding to the $Z_c(3900)$.
In Ref.~\cite{Chen:2015jwa}, the low-energy scattering of the $D^*
\bar D^*$ meson system was studied by Chen {\it et al.} by Lattice
QCD calculation. Their results indicated a weak repulsive interaction
between the two vector charmed mesons, and did not support a bound
state in this channel corresponding to the $Z_c(4020)/Z_c(4025)$.

In Ref.~\cite{Aceti:2014uea}, Aceti {\it et al.} studied the $D \bar
D^*$ interaction in the isovector channel in the local hidden gauge
approach with heavy quark spin symmetry. They found a state with a
mass of 3869-3875 MeV and a width around 40 MeV with $I = 1$ and
positive $G$-parity, which is interpreted as the isospin partner of
the $X(3872)$. They reanalyzed the $e^+e^- \to \pi^\pm (D \bar
D^*)^\mp$ reaction~\cite{Ablikim:2013xfr} and found a solution with
a mass of 3875 MeV and a width around 30 MeV. But they did not
firmly interpret it as the $Z_c(3900)$. In
Ref.~\cite{Aceti:2014kja}, Aceti {\it et al.} studied the isovector
$D^* \bar D^*$ interaction in the local hidden gauge approach, and
interpreted the $Z_c(4020)/Z_c(4025)$ as a possible $2^+$ bound
state of the $D^* \bar D^*$ with $I = 1$.

Besides the light meson exchange, He also considered the additional
very short-range attraction from the $J/\psi$ exchange in
Ref.~\cite{He:2015mja} and found that the $Z_c(3900)$ can be
interpreted as a resonance above the threshold from the $D \bar D^*$
interaction. In Ref.~\cite{Karliner:2015ina}, Karliner and Rosner
interpreted the $Z_c(3900)$, together with the $X(3872)$, as weakly bound
molecular states with the $D\bar D^*$ component and the $Z_c(4020)/Z_c(4025)$ as a $D^*\bar D^*$ molecular state.

The molecular type of interpolating currents were employed to
investigate the $Z_c(3900)$ and $Z_c(4020)$ states in QCD sum rules
in Refs
\cite{Chen:2015ata,Chen:2013omd,Zhang:2013aoa,Cui:2013yva,Wang:2013daa,Wang:2014gwa,Cui:2013vfa,Khemchandani:2013iwa}.
The extracted mass agrees with the experimental values within
errors. However, one should be cautious in the interpretation of the
extracted resonances as the molecular states even if the
color singlet-singlet $\left(\mathbf{1}_{[q\bar c]}\otimes\mathbf{1}_{[c\bar
q]}\right)$ molecular type of interpolating currents
were used. The four quarks in the currents have the same space-time
position. Moreover, the interpolating current ``sees" only the
quantum numbers of the resonance. Different interpolating currents
with the same quantum numbers can generally couple to the same physical
state.

\subsubsection{Tetraquark state assignment}\label{Sect:4.2.2}

\paragraph{QCD sum rules}\label{Sect:4.2.2.1}

In 2010, Chen and Zhu studied the mass spectrum of the hidden-charm
tetraquark states with $J^{PC}=1^{+-}$ in QCD sum rules in Ref.
\cite{Chen:2010ze}. They used the interpolating currents listed in
Eq.~(\ref{Sec:4.1.currentsZb}). The numerical results are
collected in Table \ref{Table.4.2.ZcTetraquarkQSR}. Later in
Refs. \cite{Chen:2015ata,Chen:2013omd}, the interpolating currents
listed in Eq.~(\ref{Eq.4.1.currents1}) with the color
octet-octet $\left(\mathbf{8}_{[q\bar c]}\otimes\mathbf{8}_{[c\bar
q]}\right)$ configuration were also used. The results are collected
in Table \ref{Table.4.2.ZcMoleculeQSR}. The extracted masses of
the $J^{PC}=1^{+-}$ hidden-charm tetraquark states are roughly
consistent with the masses of the $Z_c(3900)$ and $Z_c(4020)/Z_c(4025)$.
There were some other QCD sum rule investigations of
the charged $Z_c$ mesons as charmonium-like tetraquarks in Refs
\cite{Wang:2013llv,Qiao:2013dda,Wang:2015nwa}.

%%%%%%%%%%%%%%%%%%%%%%%%%%%%%%%%%%%%%%%%%%%%%%%%%%%%%%%%%%%%%%%%%%%%%%%%%%%%%%%%%%%%%%%%%%%%%%%%%%%%%%%%%%%%%%%%%%%%%%%%%%%%%%%%%%%%%%%
\renewcommand{\arraystretch}{1.6}
\begin{table}[htb]
\caption{Numerical results for the $J^{PC}=1^{+-}$ hidden-charm
tetraquark states with the diquark-antidiquark interpolating
currents \cite{Chen:2010ze}.} \label{Table.4.2.ZcTetraquarkQSR}
\begin{center}
\begin{tabular}{cccccc} \toprule[1pt]
& Current ~~&~~ $s_0(\mbox{GeV}^2)$ ~~&~~ \mbox{Borel window} $(\mbox{GeV}^2)$ ~~&~~ $m_X$ \mbox{(GeV)} ~~&~~ \mbox{PC(\%)} \\
\midrule[1pt]
{\multirow{4}{*}{$qc\bar q\bar c$}}     & $J_{3\mu}$         &  $4.6^2$            & $3.0-3.4$           & $4.16\pm0.10$     & 46.2  \\
                     & $J_{4\mu}$         &  $4.5^2$            & $3.0-3.3$           & $4.02\pm0.09$     & 44.6  \\
                     & $J_{5\mu}$         &  $4.5^2$            & $3.0-3.4$           & $4.00\pm0.11$     & 46.0  \\
                     & $J_{6\mu}$         &  $4.6^2$            & $3.0-3.4$           & $4.14\pm0.09$     & 47.0
\vspace{5pt} \\
{\multirow{4}{*}{$sc\bar s\bar c$ }}    & $J_{3\mu}$         &  $4.7^2$            & $3.0-3.6$           & $4.24\pm0.10$     & 49.6  \\
                     & $J_{4\mu}$         &  $4.6^2$            & $3.0-3.5$           & $4.12\pm0.11$     & 47.3  \\
                     & $J_{5\mu}$         &  $4.5^2$            & $3.0-3.3$           & $4.03\pm0.11$     & 44.2  \\
                     & $J_{6\mu}$         &  $4.6^2$            & $3.0-3.4$           & $4.16\pm0.11$     & 46.0  \\
\bottomrule[1pt]
\end{tabular}
\end{center}
\end{table}
%%%%%%%%%%%%%%%%%%%%%%%%%%%%%%%%%%%%%%%%%%%%%%%%%%%%%%%%%%%%%%%%%%%%%%%%%%%%%%%%%%%%%%%%%%%%%%%%%%%%%%%%%%%%%%%%%%%%%%%%%%%%%%%

%%%%%%%%%%%%%%%%%%%%%%%%%%%%%%%%%%%%%%%%%%%%%%%%%%%%%%%%%%%%%%%
\renewcommand{\arraystretch}{1.6}
\begin{table}[htb]
\caption{Numerical results for the $J^{PC}=1^{+-}$ hidden-charm
tetraquark states with the color octet-octet interpolating currents
\cite{Chen:2015ata,Chen:2013omd}.
\label{Table.4.2.ZcMoleculeQSR}}
\begin{center}
\begin{tabular}{cccccc} \toprule[1pt]
& Current ~~&~~ $s_0(\mbox{GeV}^2)$ ~~&~~ \mbox{Borel window} $(\mbox{GeV}^2)$ ~~&~~ $m_X$ \mbox{(GeV)} ~~&~~ $f_X$ $(10^{-2}\mbox{GeV}^5)$ \\
\midrule[1pt]
& $J_{1\mu}^{(\mathbf 8)}(D\bar D^*)$     & 18   & $2.8 - 3.7 $ & $3.90\pm0.12$ & $0.69\pm0.21$ \\
& $J_{3\mu}^{(\mathbf 8)}(D^*\bar D^*)$   & 18   & $3.1 - 3.9 $ & $3.85\pm0.11$ & $1.51\pm0.46$ \\
& $J_{4\mu}^{(\mathbf 8)}(D_1\bar D_1)$   & 20   & $2.8 - 3.1$ & $4.03\pm0.18$ & $0.59\pm0.23$ \\
\bottomrule[1pt]
\end{tabular}
\end{center}
\end{table}
%%%%%%%%%%%%%%%%%%%%%%%%%%%%%%%%%%%%%%%%%%%%%%%%%%%%%%%%%%%%%%%%%%%

\paragraph{Diquark model}

The $Z_c(3900)$ and $Z(4020)/Z_c(4025)$ were interpreted as tightly
bound tetraquark states composed of diquarks. In
Refs.~\cite{Maiani:2007wz,Faccini:2013lda}, Maiani {\it et al.}
interpreted the $Z_c(3900)$ as a diquark-antidiquark charmonium-like
tetraquark state
and investigated its decay modes. This idea was further developed in
Ref.~\cite{Maiani:2014aja}, where Maiani, Piccinini, Polosa and
Riquer proposed a ``type-II'' diquark-antidiquark model. This is the
extension of their ``type-I''  diquark-antidiquark
model~\cite{Maiani:2004vq}, which will be discussed in
Sec.~\ref{sect:4.5.2}. In this model, the S-wave tetraquarks can
be written in the spin basis as $|s, \bar s\rangle_J$, where $s =
s_{qc}$ and $\bar s = s_{\bar q \bar c}$ are the diquark and
antidiquark spins, respectively. The authors identified the $X(3872)$ to
be $X_1 = {1\over\sqrt2} \left( | 1,0 \rangle_1 + | 0,1 \rangle_1
\right)$, and the $Z_c(3900)$ and $Z(4020)$ to be the linear
combinations of $Z = {1\over\sqrt2} \left( | 1,0 \rangle_1 - | 0,1
\rangle_1 \right)$ and $Z^\prime = | 1,1 \rangle_1$\, .
They also identified the three other $S$-wave tetraquarks to be
$X_0 = | 0,0 \rangle_0$, $X_0^\prime = | 1,1 \rangle_0$ and $X_2 = | 1,1 \rangle_2$.
They further
used the approximation that the dominant spin interactions are
within each diquark:
\begin{equation}
H \approx 2\kappa_{qc} \; \left(\bm s_q\cdot \bm s_c +  \bm s_{\bar
q}\cdot \bm s_{\bar c}\right) = \kappa_{qc}\left[ s(s+1)+\bar s
(\bar s +1)-3\right] \, ,
\end{equation}
which leads to
\begin{equation}
X(3872) = X_1 \, , \, Z(3900) \approx Z \, , \, Z(4020) \approx
Z^\prime \,
\end{equation}
with the mass ordering
\begin{equation}
M(X_1) \approx M(Z) \, , \, M(Z^\prime) - M(Z) \approx 2 \kappa_{qc}
\, .
\end{equation}
The parameter $\kappa_{qc}$ is fixed to be $67$ MeV. The other three
$S$-wave tetraquarks have the masses $M(X_0)
\approx 3770$ MeV and $M(X_0^\prime) = M(X_2) \approx 4000$ MeV. The
last two states, $X_0^\prime$ and $X_2$, are identified with the
$X(3940)$ and $X(3916)$, while the first one $X_0$ has a smaller
mass than that of the $X(3872)$ and not yet identified. In this scheme, the
authors identified the $Z^+(4430)$ to be the first radial excitation
of the $Z(3900)$, which will be reviewed in Sec.~\ref{Sect:4.3.2}.

The same diquark-antidiquark picture is used in
Ref.~\cite{Patel:2014vua} to study the $Z_c(3900)/Z_c(3885)$ and
$Z_c(4020)/Z_c(4025)$ states. In Ref.~\cite{Patel:2014vua}, Patel,
Shah, and Vinodkumar used the non-relativistic interaction potential
\begin{eqnarray}
V(r) = V_V + V_S = k_s {\alpha_s \over r} + \sigma r \, ,
\end{eqnarray}
which is just the Cornell potential consisting of the Coulomb
potential and the linear confining potential. They assigned the
$Z_c(3900)$ and $Z_c(4025)$ as $Q \bar q - \bar Qq$ molecular-like
four quark states, and the $Z_c(3885)$ as a diquark-antidiquark
tetraquark state.

In Ref.~\cite{Deng:2015lca}, Deng, Ping, Huang, and Wang
systematically investigated the charged tetraquark states $[cu][\bar
c \bar d]$ using the color flux-tube model with a four-body
confinement potential. Their Hamiltonian was given as follows:
\begin{eqnarray}
H_4 = \sum_{i=1}^4 \left( m_i + {{\bf p}^2_i \over 2 m_i} \right) -
T_C + \sum_{i>j}^4 V_{ij} + V^C_{min} + V^{C,SL}_{min} \, ,
\end{eqnarray}
where $T_C$ is the center-of-mass kinetic energy of the state, and
$V^C_{min}$ ($V^C_{min}$ and $V^{C,SL}_{min}$) is the quadratic
confinement potential. $V_{ij}$ contains the one-boson-exchange
potential $V^B_{ij}$, the $\sigma$-meson exchange potential
$V^\sigma_{ij}$, the one-gluon-exchange potential $V^G_{ij}$. They
identified the $Z_c(3900)/Z_c(3885)$ as the tetraquark state
$[cu][\bar c \bar d]$ with the quantum numbers $1^3S_1$ and $J^P=1^+$,
and the $Z_c(4020)/Z_c(4025)$ as the tetraquark state $[cu][\bar c \bar
d]$ with $1^5S_2$ and $J^P=2^+$. This is an extension of their previous
study~\cite{Deng:2014gqa}, where Deng, Ping, and Wang interpreted
the $Z_c(3900)$ and $Z_c(4025)/Z_c(4020)$ as the S-wave tetraquark
states $[cu][\bar c \bar d]$ with quantum numbers $I = 1$ and $J =
1$ and $2$, respectively.

Within the framework of the color-magnetic interaction, the mass
spectra of the hidden-charm and hidden-bottom tetraquark states
were studied systematically by Zhao, Deng and Zhu in
Ref.~\cite{Zhao:2014qva}. They considered the chromomagnetic
interaction which was derived from one gluon exchange
\begin{eqnarray}
H_{CM} &=& - \sum_{i>j} v_{ij} \vec \lambda_i \cdot \vec \lambda_j
\vec \sigma_i \cdot \vec \sigma_j \, ,
\end{eqnarray}
where $\vec \lambda_i$ is the quark color operator and $\vec
\sigma_i$ the spin operator. For the tetraquark system $q_1 q_2
\bar q_3 \bar q_4$ with four different flavors, the interaction
matrix element between two $SU(6)_{cs}$ eigenstates $|k\rangle$ and
$|l\rangle$ is
\begin{eqnarray}
V_{CM}(q_1 q_2 \bar q_3 \bar q_4) = \langle k | H_{CM} |l \rangle =
V_{12}(q_1 q_2) + V_{13}(q_1 \bar q_3) + V_{14}(q_1 \bar q_4) +
V_{23}(q_2 \bar q_3) + V_{24}(q_2 \bar q_4) + V_{34}(\bar q_3 \bar
q_4) \, .
\end{eqnarray}
They considered the configurations $qc \bar q\bar c$, $qc \bar s\bar
c$, $sc \bar s\bar c$, $qb\bar q\bar b$, $qb \bar s\bar b$, $sb \bar
s\bar b$ with $J^P = 1^+$, $0^+$, $2^+$. They used two schemes and
found that it was impossible to accommodate all the three charged
states $Z_c(3900)$, $Z_c(4025)$ and $Z_c(4200)$ as tightly bound
tetraquark states, and at least one or two of these states is a
molecular state or has some other structures. Furthermore, they tended to
conclude that both the $Z_c(3900)$ and $Z_c(4025)$ are good molecular
candidates, while the $Z_c(4200)$ is a very promising candidate of the
lowest axial-vector hidden-charm tetraquark state.

The internal structure of the $Z_c(3900)$ was discussed in
Ref.~\cite{Voloshin:2013dpa} by Voloshin. To differentiate
the molecular model as well as the hadro-charmonium and tetraquark
schemes, he urged the measurements of the quantum numbers of the
resonance and its decay rates into yet unseen channels
$\pi\psi^\prime$, $\pi h_c$, $\rho \eta_c$ and into pairs of heavy
mesons $D^* \bar D$ and $D \bar D^*$.

\subsubsection{Kinematical effect}\label{Sect:4.2.3}

\paragraph{ISPE mechanism}

\begin{figure}[hbtp]
\begin{center}
\scalebox{1}{\includegraphics{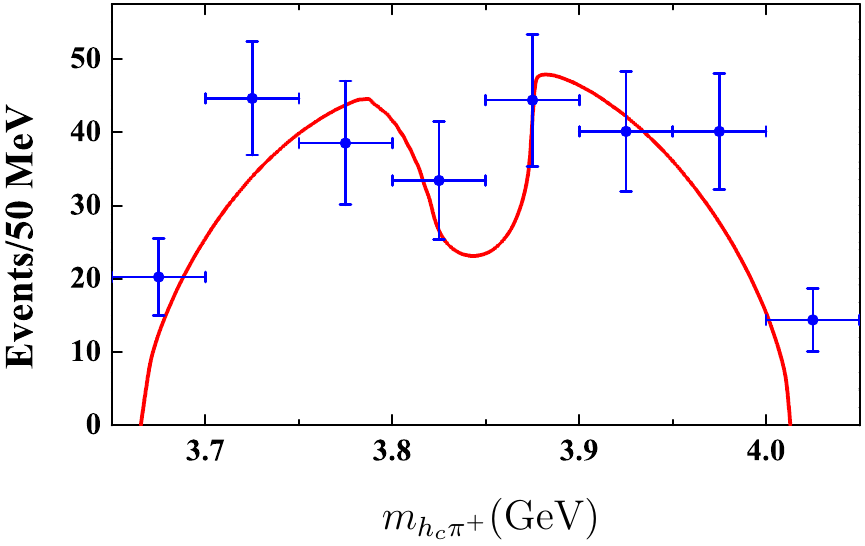}}
\end{center}
\caption{(Color online) A comparison of the $h_c\pi^\pm$ mass
distribution of $\psi(4160)\to h_c(1P) \pi^+\pi^-$ (red solid line)
from the ISPE mechanism and measurement by CLEO-c (blue points with
errors) \cite{CLEO:2011aa}. Taken from Ref. \cite{Chen:2011xk}.
\label{Fig.4.2.com}}
\end{figure}

\begin{figure*}[htb]
\centering%
\begin{tabular}{ccc}
\scalebox{0.75}{\includegraphics{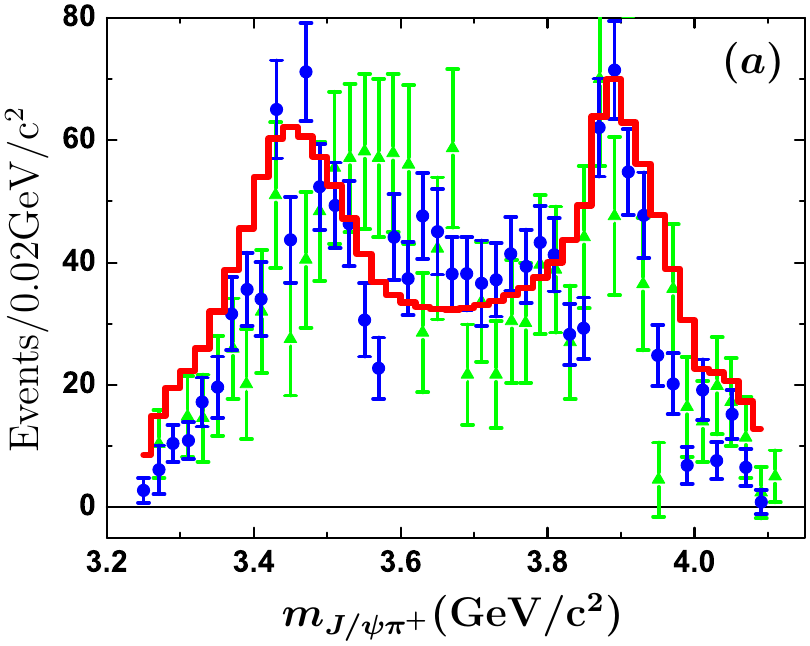}} & \hspace{5mm} &%
\scalebox{0.75}{\includegraphics{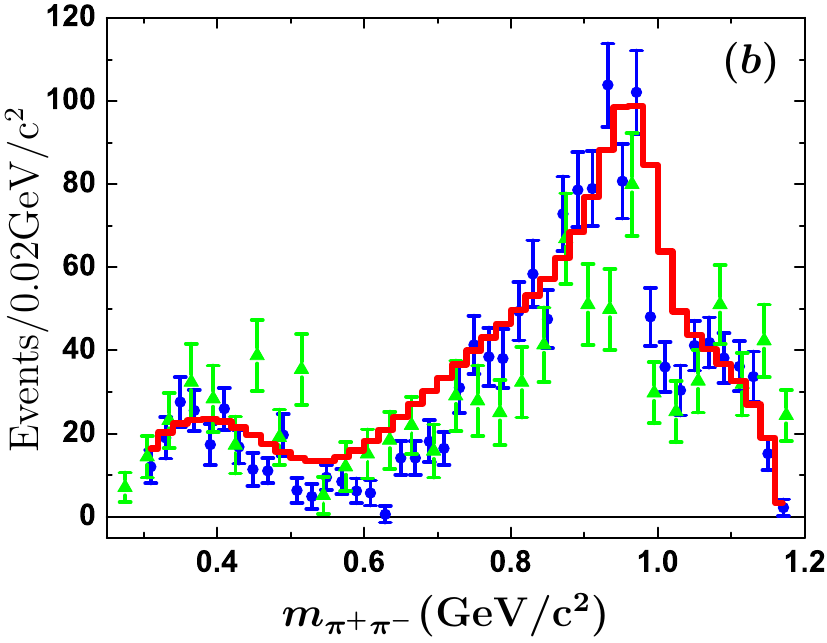}}%
%(a)
\end{tabular}
\caption{(Color online) The distributions of the $J/\psi\pi^+$ and
$\pi^+\pi^-$ invariant mass spectra of $Y(4260) \to \pi^+ \pi^-
J/\psi$. The blue dots and green triangles with error bars are the
experimental data given by BESIII \cite{Ablikim:2013mio} and Belle
\cite{Liu:2013dau}, respectively. The red histograms are our results
considering contributions of the ISPE mechanism to the $Y(4260) \to
\pi^+ \pi^- J/\psi$ decay. Taken from Ref. \cite{Chen:2013coa}.
\label{Fig.4.2.Fit}}
\end{figure*}

The ISPE mechanism was also used to study the hidden-charm dipion
decays of higher charmonia \cite{Chen:2011xk}. The line shapes of
$d\Gamma/d m_{J/\psi\pi^+}$, $d\Gamma/d m_{\psi(2S)\pi^+}$ and
$d\Gamma/d m_{h_c(1P)\pi^+}$ of the $\psi(4040)$, $\psi(4160)$,
$\psi(4415)$, $Y(4260)$ decays into $J/\psi\pi^+\pi^-$,
$\psi(2S)\pi^+\pi^-$, $h_c(1P)\pi^+\pi^-$ showed the existence of the
charged charmonium-like structures near the $D\bar{D}^*$ and
$D^*\bar{D}^*$ thresholds \cite{Chen:2011xk}. The $h_c\pi^\pm$ mass
distribution of $\psi(4160)\to h_c(1P) \pi^+\pi^-$ (red solid line)
from the ISPE mechanism \cite{Chen:2011xk} agrees roughly with the
measurement by CLEO-c \cite{CLEO:2011aa} (see Fig. \ref{Fig.4.2.com}). The $Z_c(3900)$ structure
and its reflection observed by the BESIII and Belle collaborations in
$Y(4260)\to \pi^+\pi^- J/\psi$ \cite{Ablikim:2013mio,Liu:2013dau}
can also be reproduced with the ISPE mechanism \cite{Chen:2013coa}.
The results are shown in Fig. \ref{Fig.4.2.Fit}.

\paragraph{Coupled channel cusp}

In Ref. \cite{Swanson:2014tra}, Swanson proposed the charged
$Z_c(3900)$ and $Z_c(4025)$ states as the coupled channel cusp
\cite{Swanson:2014tra}. He pointed out that similar $Z_c$ structures
may exist in $\bar{B}^0\to J/\psi \pi^0\pi^0$ and $B^\pm\to
J/\psi\pi^\pm \pi^0$. Neutral charmonium-like structures may exist
near the $D_s\bar{D}_s^*$ and $D_s^*\bar{D}_s^*$ thresholds in
$\bar{B}_s\to J/\psi\phi\phi$ and $\bar{B}^0\to J/\psi\phi K$
decays, and near the $D\bar{D}^*$,
$D^*\bar{D}^*$, $D_s\bar{D}_s^*$ and $D_s^*\bar{D}_s^*$ thresholds
in $\bar{B}_0\to J/\psi \eta K$. Later, Szczepaniak also indicated
that triangle singularities would give considerable contributions
to the ${Z_c(3900)}$ \cite{Szczepaniak:2015eza}. However,
Guo {\it et. al.} insisted that these $XYZ$ states cannot be purely
kinematic effects, and the genuine S-matrix poles corresponding to
states should be introduced \cite{Guo:2014iya}. In Ref.
\cite{Swanson:2015bsa}, Swanson used a causal and analytic model of
final state rescattering to describe the current experimental data
on the $Z_c(3900)$ and $Z_c(4025)$ without poles in the scattering
matrix. In Ref. \cite{Albaladejo:2015lob}, the authors analyzed the
$Z^\pm_c(3900)/Z^\pm_c(3885)$ structure using both energy dependent
and independent $D\bar D^*$ S-wave interaction. The authors
noticed that the $Z_c$ peak is either due to a resonance with a mass
around the $D\bar D^*$ threshold or arises from a virtual state of
the molecular nature. They concluded that a $\bar D^* D$ bound state
solution is not allowed.

\subsubsection{Production and decay patterns}\label{Sect:4.2.4}

Under the $D^*\bar{D}^*$ molecular state assumption, the pionic,
dipionic, and radiative decays of the $Z_c(3900)$ and $Z_c(4025)$
and their productions via excited charmonia decays were studied
extensively with the spin rearrangement scheme in
Refs.~\cite{He:2013nwa,Ma:2014ofa,Ma:2014zva}. The electromagnetic
structure of the $Z_c(3900)$ as an axial vector molecule was
discussed using an effective theory with contact interactions in
Ref.~\cite{Wilbring:2013cha}. Hidden-charm decays of the $Z_c(3900)$
and $Z_c(4025)$ were investigated via the intermediate $D^{(*)} \bar
D^{(*)}$ meson loop in Ref.~\cite{Li:2013xia}. Strong decays of the
$Z_c(3900)$ were studied using a phenomenological Lagrangian
approach in Ref.~\cite{Dong:2013iqa}. The decay rates of the
$Z_c(3900)$ to $J/\psi \pi$, $\psi(2S)\pi$, $\eta_c \rho$ and $D^*
\bar D^*$ were studied within the light front model in
Ref.~\cite{Ke:2013gia}. The $Z_c(3900)/Z_c(4020) \to \eta_c \rho$
decays were studied within both tetraquark and molecular pictures in
Ref.~\cite{Esposito:2014hsa}. The counterparts of $Z_c$ and their
decay modes were studied in Ref.~\cite{Li:2014pfa} using the
effective Lagrangian based on the heavy quark symmetry.

Radiative and dilepton decays of the
$Z_c(3900)$ were studied using a phenomenological Lagrangian
approach in Ref.~\cite{Gutsche:2014zda}. Radiative decays of the
neutral $Z_c(3900)$ were studied in a hadronic molecule scenario in
Ref.~\cite{Chen:2015igx}.
The photoproduction of the $Z_c(3900)$ in the $\gamma p \to Z_c(3900)^+
n$ process was proposed in Ref.~\cite{Lin:2013mka}. The possible
contribution of the $Z_c(4025)$ resonance in the $e^+ e^- \to  (D^*
\bar D^*)^\pm \pi^\mp$ reaction was reanalyzed in
Ref.~\cite{Torres:2013lka}.

\subsubsection{A short summary}
\label{Sect:4.2.5}

\begin{itemize}

\item The $Z_c(3900)$ and $Z_c(4025)$ structures may arise from
some kinematical effects, such as the ISPE mechanism, triangle
singularities, and coupled channel cusp effects.

\item There exists the long-range attraction due to the pion exchange
in the isovector $J^P=1^+$ $D\bar{D}^*$ and $D^*\bar{D}^*$ systems.
However, the attraction may not be strong enough to form molecular
bound states. The $Z_c(3900)$ and $Z_c(4025)$ states may be
molecular resonances.

\item In the diquark-antidiquark model, the $Z_c(3900)$ and
$Z(4020)/Z_c(4025)$ were interpreted as the S-wave tetraquark
states. To explain their narrow decay widths and the
dominance of the open-charm decay modes, very special dynamics has
to be introduced such as the existence of the tightly bound
diquarks.

\end{itemize}

\subsection{$Z^+(4430)$}
\label{Sect:4.3}
\subsubsection{Molecular state scheme}\label{Sect:4.3.1}

\paragraph{$\bar D_1 D^*$ molecule}

After its discovery, the subsequent experimental progress on the
$Z^+(4430)$ was accompanied by surprises. There are valuable lessons
for us to learn from the history of this interesting particle.

After the observation of the $Z^+(4430)$ \cite{Choi:2007wga}, Meng and
Chao suggested the $Z^+(4430)$ as the S-wave $\bar D_1 D^*$ (or $\bar
D_1^\prime D^*$) molecular state in Ref. \cite{Meng:2007fu} since the
$Z^+(4430)$ is close to the $D^*(2010)D_1(2420)$ and
$D^*(2010)D_1^\prime(2430)$ thresholds. Under this assignment, the
$Z^+(4430)$ can be produced through the $B^+$ decay. The branching
ratio of the Cabibbo-Kobayashi-Maskawa (CKM) favored decay mode $B^+\to \bar{D}^{*0}(2010)
D_s^{*+}$ is large, where the $D_s^{*+}$ denotes the ground or excited
state, which can decay into $D^+K^0$ and $D^0 K^+$ or
$D_1^+(2420)K^0$ and $D_1^0(2420)K^+$. The $Z^+(4430)$ can be
formed in
the $\bar{D}^{*0}D_1^+(2420)$ scattering process \cite{Meng:2007fu}.
This scheme was also proposed by Rosner in Ref.
\cite{Rosner:2007mu}.

To answer whether the $\bar D_1(\bar D_1^\prime)$ and $D^*$
interaction is strong enough to form a bound state corresponding to the
$Z^+(4430)$, the dynamical calculation of the S-wave $\bar D_1 D^*$
(or $\bar D_1^\prime D^*$) system with $J^P=0^-$, $1^-$, $2^-$ was
carried out with the OPE model \cite{Liu:2007bf}, where $D_1^\prime$ and
$D_1$ belong to $(0^+,1^+)$ and $(1^+,2^+)$ doublets, respectively.
The potential from the crossed diagram is much larger than that from
the diagonal scattering diagram. With various trial wave functions,
the numerical results indicated that the attraction from the OPE
potential alone is not strong enough to form a bound state with
realistic pionic coupling constants deduced from the decay widths of
$D_1$ and $D_1^\prime$ \cite{Liu:2007bf}.

Later, the authors of Ref. \cite{Liu:2008xz} reexamined this issue
by considering both the pion and $\sigma$ meson exchange potentials,
where the form factor was introduced to take into account the
structure effect of the interaction vertex. Different from the
treatment in Ref. \cite{Liu:2007bf}, the authors solved the
Schr\"odinger equation with the obtained effective potentials
\cite{Liu:2008xz}. They found the OPE potential from the crossed
diagram plays a dominant role in the formation of the S-wave $\bar
D_1^\prime D^*$ or $\bar D_1 D^*$ molecular system. There exists the
bound state solution for the S-wave $\bar{D}_1^\prime D^*$ system
with $J^P=0^-$, $1^-$, $2^-$, where the contribution from the $\sigma$
meson exchange is small. Whether the broad width of $D_1^\prime$
disfavors the formation of a molecular state should be carefully
studied in the future \cite{Liu:2008xz}. The S-wave $\bar{D}_1D^*$
molecular state with $J^P=0^-$ may exist if taking appropriate
parameters as input. Different from the case of the S-wave
$\bar{D}_1^\prime D^*$ system, the contribution from the $\sigma$
meson exchange is significant for the S-wave $\bar D_1D^*$ molecular
state.

Ding {\it et al.} \cite{Ding:2008mp} also studied the $D^*\bar{D}_1$
interaction in the quark model. The interaction between a quark and
an antiquark includes the short distance one-gluon exchange potential
and the long distance confinement interaction \cite{Ding:2008mp}.
After solving the multichannel Schr\"odinger equation, they found
that the $Z^+(4430)$ can be explained as a loosely $\bar D_1D^*$
molecular state with $J^P=0^-$ \cite{Ding:2008mp}, which is
consistent with the conclusion from the OPE model \cite{Liu:2008xz}. In
Refs. \cite{Lee:2007gs,Zhang:2009vs}, the $Z^+(4430)$ was proposed as a
$D^*\bar D_1$ molecule with $J^P=0^-$ using the molecular type of
interpolating currents in QCD sum rules.

Liu and Zhang applied a chiral quark model to study the S-wave
$D_1\bar{D}^*$ ($D_1^\prime \bar D^*$) system by solving the
resonating group method equation \cite{Liu:2009wb}. Their results
disfavored the assignment of the $Z^+(4430)$ as the S-wave $D_1\bar{D}^*$
($D_1^\prime \bar D^*$) molecular state \cite{Liu:2009wb}. Later in
Ref.~\cite{Li:2013bca}, Li {\it et al.} used the SU(3) chiral quark
model to study the interaction potentials between one S-wave and
one P-wave heavy mesons systematically. Their results also
disfavored the assumption that the $Z^+(4430)$ is an isovector $D_1
\bar D^*$ charged molecule.

The quenched lattice QCD using L\"uscher formalism was adopted to
study the interaction between the $D^*$ and $\bar{D}_1$ in Refs.
\cite{Meng:2009qt,CLQCD:2010sna}. The authors concluded that the interaction of
$D^*$ and $\bar D_1$ is attractive in the $J^P=0^-$ channel but
not strong enough to form a bound state \cite{Meng:2009qt,CLQCD:2010sna}.
Both the phenomenological models
\cite{Liu:2007bf,Liu:2008xz,Ding:2008mp,Liu:2009wb} and the quenched
lattice QCD calculations \cite{Meng:2009qt,CLQCD:2010sna} indicated that the S-wave
$D_1\bar{D}^*$ interaction is attractive. The authors of
Ref.~\cite{Barnes:2014csa} proposed that the $Z^+(4430)$ is either a
$D^*\bar D_1$ state dominated by the long-range pion exchange, or a
$D \bar D^*(1S, 2S)$ state with important short-range components.

\paragraph{$Z^+(4430)$ as a $J^P=1^+$ molecule}

After Belle and LHCb established its spin-parity $J^P=1^+$, it is
obvious that the S-wave $\bar{D}_1D^*$ (or $\bar{D}_1^\prime D^*$)
molecular state assignment of the $Z^+(4430)$ does not hold. The
authors of Ref. \cite{Ma:2014zua} proposed three possible molecular
configurations: (1) the $Z^+(4430)$ as the P-wave excitation of the
S-wave $D_1\bar{D}^*$ or $D_2\bar{D}^*$ molecule; (2) the
$Z^+(4430)$ as the S-wave molecule composed of a $D$ or $D^*$ meson
and a D-wave vector $D$ meson; (3) the $Z^+(4430)$ as the cousin
molecular state of the $Z_c(3900)$ and $Z_c(4020)$ composed of a $D$
or $D^*$ meson and their radial excitations. In the heavy quark
symmetry, they further investigated the radiative and strong decay
patterns of the $Z^+(4430)$, and found \cite{Ma:2014zua}
\begin{itemize}
\item If the charm quark mass goes to infinity and heavy quark
symmetry is exact, the S-wave molecule composed of a $D$ or $D^*$
meson and a D-wave vector $D$ meson does not decay into the
$\psi'\pi$ final states, which is the discovery mode of the $Z^+(4430)$.
Unfortunately, the charm quark mass is only 1.5 GeV.

\item The $Z^+(4430)$ could be the P-wave excitation of the $D_1 {\bar
D}^\ast$ or $D_2 {\bar D}^\ast$ molecule, their radiative and strong
decay patterns can be investigated with the spin rearrangement
scheme in the heavy quark symmetry limit, together with their S-wave
molecular ground states. Their radiative decays are presented in
Ref. \cite{Ma:2014ofa}. Within this scheme, the non-observation of the
$Z^+(4430)$ in the $J/\psi\pi$ mode is always a serious challenge.
There exists no manifest symmetry forbidding this mode. The same
challenge holds for the tetraquark interpretation. Moreover, if the
$Z^+(4430)$ is the P-wave molecule, where is the ground state?

\item If the $Z^+(4430)$ happens to be the  molecular cousin of the $Z_c(3900)$
and $Z_c(4020)$ composed of a $D$ or $D^*$ meson and their radial
excitations, it would decay into the $J/\psi\pi$ and $\psi'\pi$ easily.
However, it would not decay into the $\psi(1^3D_1)\pi$ in the heavy quark
symmetry limit. The neutral component will also decay into the
$\chi_{cJ}$ through the $M1$ transition. The resulting decay width
ratio is 1:3:5. Since the $Z^+(4430)$ contains one radial excitation as
its molecular component, one may expect that the $Z^+(4430)$ may decay
into the final state containing a radial excitation more easily.

\end{itemize}

\subsubsection{The tetraquark assignment}\label{Sect:4.3.2}

In Ref.~\cite{Maiani:2007wz}, Maiani, Polosa and Riquer suggested
the $Z^+(4430)$ as the first radial excitation of the tetraquark basic
supermultiplet to which the $X(3872)$ belong. Later in their
``type-II'' diquark-antidiquark model~\cite{Maiani:2014aja},
reviewed in Sec.~\ref{Sect:4.2.2}, Maiani {\it et al.} identified the
$Z^+(4430)$ as the first radial excitation of the $Z(3900)$. The mass
difference between the $Z_c(4430)$ and $Z_c(3900)$ is 593 MeV,
which is very close to the mass difference between the $\psi(2S)$
and $\psi(1S)$ with 589 MeV.

A similar but relativistic diquark-antidiquark picture was proposed
in Ref.~\cite{Ebert:2008kb}, where Ebert, Faustov, and Galkin
calculated the masses of the excited heavy tetraquarks with hidden-charm
in the diquark-antidiquark picture. They used the dynamical
approach based on the relativistic quark model, and took into
account the diquark structure by calculating the diquark-gluon form
factor. They used $S$ and $A$ to denote scalar and axial vector
diquarks which are flavour antisymmetric and symmetric, respectively.
Their studies suggested that the $Z^+(4430)$ can be either the $1^+$
$2S$ $[cu][\bar c \bar d]$ tetraquark state consisting of $(S\bar A
- \bar S A)/\sqrt2$, or the $0^+$ $2S$ $[cu][\bar c \bar d]$
consisting of $A\bar A$.

In Ref.~\cite{Patel:2014vua}, Patel, Shah, and Vinodkumar once more
identified the $Z^+(4430)$ as the first radial excitation of the
$Z_c(3885)$ using a non-relativistic quark model. Later in
Ref.~\cite{Hadizadeh:2015cvx}, Hadizadeh and Khaledi-Nasab studied
heavy tetraquarks with hidden-charm and hidden-bottom by solving the
homogeneous Lippmann-Schwinger integral equation in momentum space.
The $Z^+(4430)$ was again explained as the $2S$ $cq \bar c \bar q$
tetraquark state consisting of $A\bar A$, with $A$ the axial vector
diquark. Its mass was evaluated to be 4535 MeV and 4469 MeV from
nonrelativistic and relativistic Lippmann-Schwinger equations,
respectively.

Hence, the idea that the $Z^+(4430)$ is the first radial excitation
of the tetraquark basic supermultiplet to which the $X(3872)$ and
$Z_c(3900)$ belong is accepted in many models which interpreted the
charged charmonium-like states as the tightly bound
diquark-antidiquark states. However, in Ref.~\cite{Deng:2015lca},
Deng {\it et al.} systematically investigated the charged tetraquark
states $[cu][\bar c \bar d]$ using the color flux-tube model with a
four-body confinement potential. They can not describe the
$Z^+(4430)$ as such a tetraquark state.

Especially, the Belle experiment measured the ratio
$\mathcal{B}(Z^+(4430) \to \psi(2S) \pi^+)/\mathcal{B}(Z^+(4430) \to
J/\psi \pi^+)$ to be about 10~\cite{Mohanty:2014hfa}. As discussed
in Ref.~\cite{Liu:2014eka}, it would be extremely challenging to
accommodate this ratio if the $Z_c(4430)$ is an S-wave tetraquark
ground state or its radial excitation. Hence, the tetraquark
interpretations of the $Z_c(3900)$ (as the ground state of the
$Z^+(4430)$) and its partner $X(3872)$, would be crucial for the
diquark-antidiquark picture, which has been/will be reviewed in
Sec.~\ref{Sect:4.2.2} and Sec.~\ref{sect:4.5.2}, respectively.

Assuming the $Z^+(4430)$ to be a tetraquark bound state made up of
$(cu)(\bar c \bar d)$, its bottom partners were investigated in
Ref.~\cite{Cheung:2007wf}. First, Cheung, Keung, and Yuan replaced
one of the charm quarks by a bottom quark, and obtained the mass of
$Z_{bc}$ to be around 7.6 GeV. Then, they replaced both the charm quark
and antiquark by the bottom quark and antiquark, and obtained the mass
of $Z_{bb}$ to be about 10.7 GeV. They also proposed two channels to
observe them, i.e., $Z^{++}_{bc} \to B^+_c (2 ^3S_1) \pi^+$ and
$Z^+_{bb} \to \Upsilon(2S) \pi^+$ .

To distinguish whether the tetraquarks are segregated into di-meson
molecules, diquark-antidiquark pairs, or more democratically
arranged four-quark states, Brodsky and Lebed proposed a number of
experimentally straightforward and feasible tests in
Ref.~\cite{Brodsky:2015wza}, which can be applied to tetraquark
candidates such as the $X(3872)$ and $Z^+(4430)$. A new dynamical
picture was introduced in Ref.~\cite{Lebed:2015sxa}, where Lebed
proposed that some subset of charmonium-like states are bound
(not molecular) states of color $\mathbf{3}-\mathbf{\bar3}$ compact diquarks,
which have achieved
substantial separation due to the large energy release of the
process in which they are formed. This mechanism relies on the
existence of compact diquarks and gives a qualitative picture of the
strong preference of the $Z^+(4430)$ to decay to the $\psi(2S)$ rather
than the $J/\psi$. However, a dynamical quantitative calculation is
still missing.

\subsubsection{Cusp effect}\label{Sect:4.3.3}

Besides the above exotic assignments, the $Z^+(4430)$ was proposed as
the ${D^*(2010) \bar D_1(2420)}$ threshold cusp effect in Ref.
\cite{Bugg:2008wu}. The author adopted the unitarized quark model
\cite{Tornqvist:1995kr} with the T-matrix method. The ${Z(4430)}$
structure in the Belle data could be reproduced by either a
resonance or a bare cusp. For the later case, although there was no
second or third-sheet pole in the vicinity of the cusp, the Argand
diagram and the peak were all reproduced well.

\subsubsection{Production, decay patterns, and other theoretical schemes}\label{Sect:4.3.4}

The $Z^+(4430)$ was also studied using various theoretical
frameworks, including the utilization of the SU(3) flavor
symmetry~\cite{Li:2007bh}, the large N scalar QCD in two
dimensions~\cite{Grinstein:2008wm}, the semirelativistic quark
potential model~\cite{Matsuki:2008gz}, and the $\pi \psi'$
interaction~\cite{Danilkin:2009ak}.

The authors of Ref. \cite{Du:2012pn} studied the charmonium-like
tetraquark states with $J^{PC}=0^{+-}$ by using the following
interpolating currents in QCD sum rule,
\begin{equation}
\begin{split}
&\eta_1(x)=u_{a}^TC\gamma^\mu c_{b}(\bar{d}_{a}\stackrel{\leftrightarrow}{D}_\mu C\bar{c}_{b}^T+\bar{d}_{b}\stackrel{\leftrightarrow}{D}_\mu C\bar{c}^T_{a})-u_{a}^TC\stackrel{\leftrightarrow}{D}_\mu c_{b}(\bar{d}_{a}\gamma^\mu C\bar{c}_{b}^T+\bar{d}_{b} \gamma^\mu C\bar{c}^T_{a}),\\
&\eta_2(x)=u_{a}^TC\gamma^\mu c_{b}(\bar{d}_{a}\stackrel{\leftrightarrow}{D}_\mu C\bar{c}_{b}^T-\bar{d}_{b}\stackrel{\leftrightarrow}{D}_\mu C\bar{c}^T_{a})-u_{a}^TC\stackrel{\leftrightarrow}{D}_\mu c_{b}(\bar{d}_{a}\gamma^\mu C\bar{c}_{b}^T-\bar{d}_{b} \gamma^\mu C\bar{c}^T_{a}),\\
&\eta_3(x)=u_{a}^TC\gamma^\mu \gamma_5c_{b}(\bar{d}_{a}\stackrel{\leftrightarrow}{D}_\mu \gamma_5C\bar{c}_{b}^T+\bar{d}_{b}\stackrel{\leftrightarrow}{D}_\mu \gamma_5C\bar{c}^T_{a})-u_{a}^TC\stackrel{\leftrightarrow}{D}_\mu\gamma_5 c_{b}(\bar{d}_{a}\gamma^\mu \gamma_5C\bar{c}_{b}^T+\bar{d}_{b} \gamma^\mu \gamma_5C\bar{c}^T_{a}),\\
&\eta_4(x)=u_{a}^TC\gamma^\mu
\gamma_5c_{b}(\bar{d}_{a}\stackrel{\leftrightarrow}{D}_\mu
\gamma_5C\bar{c}_{b}^T-\bar{d}_{b}\stackrel{\leftrightarrow}{D}_\mu
\gamma_5C\bar{c}^T_{a})-u_{a}^TC\stackrel{\leftrightarrow}{D}_\mu\gamma_5
c_{b}(\bar{d}_{a}\gamma^\mu \gamma_5C\bar{c}_{b}^T-\bar{d}_{b}
\gamma^\mu \gamma_5C\bar{c}^T_{a}), \label{equ:current0+-}
\end{split}
\end{equation}
where
$\stackrel{\leftrightarrow}{D}_\mu=\stackrel{\rightarrow}{D}_\mu-\stackrel{\leftarrow}{D}_\mu$
is the covariant derivative operator. The masses of these scalar
charmonium-like tetraquark states were extracted as $4.8-5.0$ GeV,
which are much higher than that of the $Z^+(4430)$. As a byproduct,
the masses of the bottomonium-like tetraquark states with
$J^{PC}=0^{+-}$ were predicted to be around $11.2-11.3$ GeV
\cite{Du:2012pn} .

In Ref.~\cite{Liu:2008qx}, Liu, Zhao, and Close studied the
photoproduction of the $Z^+(4430)$ in the $\gamma p \to Z^+(4430) n \to
\psi^\prime \pi^+ n$ process. This process was further studied in
Ref.~\cite{Galata:2011bi}, where Galata presented a model for high
energy and forward angle $Z^+(4430)$ photoproduction in an effective
Lagrangian approach.

In Ref.~\cite{Ke:2008kf}, Ke and Liu studied the $Z^+(4430)$ in the
nucleon-antinucleon scattering, and discussed the production of the
$Z^+(4430)$ in the PANDA experiment. Production of the neutral
$Z^0(4430)$ in the $p\bar p \to \psi^\prime \pi^0$ reaction was
discussed using an effective Lagrangian in Ref.~\cite{Wang:2015pfa}.
The $(D \bar D^*)^+ \to \psi(2S)\pi^+$ rescattering process in the
decay chain $B \to D_s(2S)^-D \, , \, D_s(2S)^- \to \bar D^* K$ was
investigated to explain the peak structure in the $\psi(2S)\pi^+$
mass spectrum around 4.43 GeV in
Refs.~\cite{Pakhlov:2011xj,Pakhlov:2014qva}.

The decay properties of the $Z^+(4430)$ were generally discussed
under the QCD-string based explanation in
Ref.~\cite{Gershtein:2007vi}. The hidden-charm and radiative decays
of the $Z^+(4430)$ were studied in Ref.~\cite{Branz:2010sh}, and its
open-charm decays were studied in Ref.~\cite{Liu:2008yy}, assuming
it as a $D_1 \bar D^*$ molecular state. Line shapes of
the $Z^+(4430)$ in the $\psi^\prime \pi^+$ decay channel and in $D^*
\bar D^* \pi$ decay channels were investigated in
Ref.~\cite{Braaten:2007xw}. The typical radiative and hidden-charm
and open-charm strong decay patterns of the $Z^+(4430)$ were
investigated with the help of the heavy quark symmetry in
Ref.~\cite{Ma:2014zua}. The phase motion in the $Z^-(4430)$
amplitude in the $B^0 \to \psi^\prime \pi^- K^+$ decay was studied using
the isobar-based amplitude difference method in
Ref.~\cite{Bediaga:2015tga}.

\subsubsection{A short summary}
\label{Sect:4.3.5}

\begin{itemize}

\item Some model calculations and one quenched lattice QCD calculation
indicated that the S-wave $D_1 \bar{D}^*$ interaction is
attractive. There may exist bound states in the S-wave and P-wave
$\bar D_1 D^*$ system. The $Z^+(4430)$ may be the P-wave
$D_1\bar{D}^*$ or $D_2\bar{D}^*$ molecular state.

\item The $Z^+(4430)$ may be the  molecular cousin of the $Z_c(3900)$
and $Z_c(4020)$ composed of a $D$ or $D^*$ meson and their radial
excitations, which decays into the $J/\psi\pi$ and $\psi'\pi$ easily.
However, it will not decay into the $\psi(1^3D_1)\pi$ in the heavy quark
symmetry limit.

\item In many diquark-antidiquark models, the $Z^+(4430)$ was
described as the first radial excitation of the basic tetraquark
supermultiplet containing the $X(3872)$ and $Z_c(3900)$.

\item
The presence of one radial excitation within the $Z^+(4430)$ may
help to explain the measured ratio $\mathcal{B}(Z^+(4430) \to
\psi(2S) \pi^+)/\mathcal{B}(Z^+(4430) \to J/\psi \pi^+)$.

\end{itemize}

\subsection{Other charged states: $Z^+(4051)$, $Z^+(4248)$ and $Z^+(4200)$}
\label{Sect:4.4}
\subsubsection{Molecular state scheme}\label{Sect:4.4.1}

Like the $Z^+(4430)$, the $Z^+(4051)$ and $Z^+(4248)$ must contain at least
four quarks if they are resonances, which also inspired discussions
whether they can be hadronic molecular states.

The possible molecular states composed of S-wave charmed and
anti-charm mesons were systematically studied in Ref.
\cite{Liu:2008tn}, which can be categorized into a flavor octet and
a singlet. Through the OBE model, the effective potentials were
obtained to check whether the corresponding bound state solution can
be found. The total effective potentials of the $D^*\bar{D}^*$
systems with $J=0,1$ and $J=2$ are attractive and repulsive in the
range $r<1$ fm respectively (see Fig. 5 of Ref. \cite{Liu:2008tn}).
They found that there does not exist the $D^*\bar{D}^*$ molecular
bound state with a reasonable cutoff, so the $Z^+(4051)$ is probably
not a molecular bound state \cite{Liu:2008tn}. However, the
possibility of interpreting it as the $D^*\bar{D}^*$ molecular
resonance was still not excluded.

In Ref. \cite{Liu:2008mi}, the authors systematically studied the
bound state problem of the S-wave heavy meson-antimeson systems in a
chiral SU(3) quark model by solving the resonating group method
equation. There does not exist the isovector (charm-anticharm)
molecular state. The assignment of the $Z^+(4051)$ as an S-wave
$D^*\bar{D}^*$ molecule was disfavored \cite{Liu:2008mi}. Moreover,
the $Z^+(4248)$ can not be explained as the $D^* \bar D^*_0$ molecule
according to the analysis in the chiral quark model~\cite{Li:2013bca}.
The $Z^+(4248)$ was suggested as a $D_1\bar D$ molecular state using
QCD sum rule approach \cite{Lee:2008gn}. Ding applied the OBE model
to study the interaction between the $D_1$ and $\bar D$, and he concluded
that the $Z^+(4248)$ as a $D_1\bar D$ molecular state was disfavored
due to the large cutoff \cite{Ding:2008gr}.

Various theoretical investigations do not support the molecular
assignment of the $Z^+(4051)$ and $Z^+(4248)$
\cite{Liu:2008tn,Liu:2008mi,Ding:2008gr}.

\subsubsection{Tetraquark state assignment}\label{Sect:4.4.2}

The $Z^+(4051)$ and $Z^+(4248)$ were discussed as the
diquark-antidiquark states in
Refs.~\cite{Ebert:2008kb,Patel:2014vua,Deng:2015lca}. In
Ref.~\cite{Ebert:2008kb}, Ebert, Faustov, and Galkin found no
tetraquark candidates for the $Z^+(4051)$ structure, but they found
that the $Z^+(4248)$ can be interpreted as the charged partner of
the $1^-$ $1P$ $[cu][\bar c \bar d]$ tetraquark state consisting of
$S \bar S$, or the $0^-$ $1P$ $[cu][\bar c \bar d]$ tetraquark state
consisting of $(S \bar A \pm \bar S A)/\sqrt2$, where $S$ and $A$
are the scalar and axial vector diquarks respectively. In
Ref.~\cite{Patel:2014vua}, Patel, Shah, and Vinodkumar assigned the
$Z^+(4050)$ as a $Q \bar q - \bar Qq$ molecular-like tetraquark
state. In Ref.~\cite{Deng:2015lca}, Deng, Ping, Huang, and Wang
identified the $Z^+(4051)$ as the tetraquark state $[cu][\bar c \bar
d]$ with the quantum numbers $1^3P_1$ and $1^-$, the $Z^+(4248)$ as
the tetraquark state $[cu][\bar c \bar d]$ with $1^5D_1$ and $1^+$,
and the $Z^+(4200)$ as the tetraquark state $[cu][\bar c \bar d]$
with $1^3D_1$ and $1^+$.

\subsubsection{Production and decay patterns}\label{Sect:4.4.3}

As a tetraquark state, the hadronic decays $Z_c(4200)^+\to
J/\psi\pi^+, Z_c(4200)^+\to\eta_c\rho^+$ and $Z_c(4200)^+\to D^+\bar
D^{\ast 0}$ were calculated with the three-point functions in the
framework of the QCD sum rules in Ref. \cite{Chen:2015fsa}. The decay
widths of the dominant decay modes $\eta_c\rho$ and $J/\psi\pi$ are
253 MeV and 87 MeV, respectively. Because of the suppression of the
phase space, the decay width of its open-charm mode is around
several MeV. Including all these channels, the full decay width of
the $Z_c(4200)^+$ state is consistent with the experimental value
reported by the Belle Collaboration, supporting the tetraquark
interpretation of the $Z^+(4200)$ \cite{Chen:2015fsa}.

The photoproduction of the $Z^+(4200)$ was investigated using an
effective Lagrangian approach and the Regge trajectories model in
Ref.~\cite{Wang:2015lwa}. Productions of the neutral $Z^0(4200)$ in
$p\bar p \to J/\psi \pi^0$ reaction was discussed in
Ref.~\cite{Wang:2015uua}. The rescattering effects in the $e^+e^-
\to D^{(*)}\bar D^{(*)}$ process were investigated to understand the
$Z^+(4051)$, $X(3872)$ and the relevant bound state problem in a
meson exchange model in Ref.~\cite{Liu:2010xh}.

\subsubsection{A short summary}
\label{Sect:4.4.4}

\begin{itemize}

\item Further experimental confirmation of the charged states
$Z^+(4051)$, $Z^+(4248)$ and $Z^+(4200)$ will be helpful.

\item Various theoretical investigations do not support molecular
assignments of the $Z^+(4051)$ and $Z^+(4248)$.

\item The tetraquark assignments for the $Z^+(4051)$, $Z^+(4248)$
and $Z_c(4200)$ were discussed using several theoretical models.

\end{itemize}

\subsection{$X(3872)$}\label{Sect:4.5}

As the first observed state in the $XYZ$ family, the $X(3872)$ has
attracted extensive attentions from both theoretical and
experimental groups all over the world. As shown in
Sec.~\ref{Sect:2.1.1}, the experimental information of the $X(3872)$
is the most abundant among all the observed $XYZ$ states. However,
we still do not fully understand its nature, although more than ten
years passed since its observation in 2003. During these years,
various pictures/interpretations have been proposed to explain the
nature of the $X(3872)$. In the following, we mainly focus on several popular
theoretical schemes of the $X(3872)$, i.e., the $D\bar{D}^*$
molecular state, the axial vector tetraquark assignment, and the
radial excitation of the axial vector charmonium state, etc.

\subsubsection{Molecular scheme}\label{Sect:4.5.1}

Before reviewing the theoretical progress on the molecular
assignment of the $X(3872)$, we note that the mass of the
$\chi_{c1}^\prime(2P)$ charmonium state was estimated to be $3.95$
GeV~\cite{Godfrey:1985xj}, which is significantly higher (around 80
MeV) than the observed mass of the $X(3872)$~\cite{pdg}.

\paragraph{Swanson's model}

Swanson proposed to interpret the $X(3872)$ as a $J^{PC} = 1^{++}$
$D^0 \bar D^{*0}$ hadronic resonance stabilized by the admixture of
$\omega J/\psi$ and $\rho J/\psi$~\cite{Swanson:2003tb}, due to the
proximity of this state to the $D\bar{D}^*$ threshold. He analysed
the $X(3872)$ based on a microscopic model, which incorporates both
the quark exchange induced effective interaction
\begin{equation}
\sum_{i<j}{\bm{\lambda}(i) \over 2}\cdot {\bm{\lambda}(j) \over 2}
\left \{ {\alpha_s \over r_{ij}} - {3\over 4} br_{ij} - {8 \pi
\alpha_s \over 3 m_i m_j } \bm{S}_i \cdot \bm{S}_j \left ( {\sigma^3
\over \pi^{3/2} } \right ) e^{-\sigma^2 r_{ij}^2} \right \},
\end{equation}
as well as the pion exchange induced effective
interaction~\cite{Tornqvist:1993vu,Tornqvist:1993ng,Tornqvist:2004qy}
\begin{equation}
V_{\pi} = - \gamma V_0 \left[ \left(\begin{array}{cc} 1 & 0
\\
0 & 1
\end{array}\right)
C(r) + \left(\begin{array}{cc} 0 & -\sqrt2
\\
-\sqrt2 & 1
\end{array}\right)
T(r) \right] \, . \label{Vpi}
\end{equation}
He found that the quark exchange effects can cause binding in the
coupled $D \bar D^*$, $\omega J/\psi$ or $\rho J/\psi$ systems, but
the potential depth is not sufficient to form a resonance. On the
other hand, the pion exchange effects can not bind the $D \bar D^*$
system with canonical parameters. However, the combined pion and
quark induced effective interactions are sufficient to cause
binding, and he found there is only one binging $D \bar D^*$ state
of $1^{++}$, which can be used to explain the $X(3872)$. He found no
other $J^{PC}$, no charged modes, and no $D \bar D$ molecules exist
in this model.

Under this molecule picture, Swanson calculated the ratio of
$\mathcal{B}(X(3872)\to \psi^\prime \gamma)$ with respect to
$\mathcal{B}(X(3872)\to J/\psi \gamma)$ to be $4\times 10^{-3}$
\cite{Swanson:2003tb,Swanson:2004pp}, which largely deviated from
the experimental data $3.4 \pm 1.4$~\cite{Aubert:2008ae} and $2.46
\pm 0.64 \pm 0.29$~\cite{Aaij:2014ala}. He derived the ratio of
${\cal{B}}(X(3872)\to \gamma J/\psi)$ with respect to ${\cal{B}}(X(3872)\to
J/\psi\pi^+\pi^-)$ to be around $10^{-2}$~\cite{Swanson:2003tb,Swanson:2004pp}, while the experimental value
is $0.14 \pm 0.05$~\cite{Abe:2005ix} and $0.33 \pm
0.12$~\cite{Aubert:2008ae}. Within this model, the ratio of
${\cal{B}}(X(3872)\to D^0\bar{D}^{0}\pi^0)$ with respect to ${\cal{B}}(X(3872)\to
J/\psi\pi^+\pi^-)$ was inconsistent with experimental data.

\paragraph{$X(3872)$ as the $D\bar D^{*}$ molecule}

Wong \cite{Wong:2003xk} applied a quark-based model to study the
molecular states composed of two heavy mesons, in terms of a
four-body non-relativistic Hamiltonian with pairwise effective
interactions. The calculated masses of the $D^0 \bar D^{*0}$ and
$D^+D^{*-}$ molecular systems are 3863.67 MeV and 3871.77 MeV,
respectively. He suggested that the molecular states $D^0 \bar
D^{*0}$, $D^+ \bar D^{*-}$, and $D^-\bar D^{*+}$ are mixed to form
components of $I = 0$ and $I = 1$ states, and the $I = 0$ state can
be interpreted as the $X(3872)$.

Such a molecule assignment was used in many theoretical studies to
investigate the $X(3872)$. The authors of Ref. \cite{AlFiky:2005jd}
used an effective Lagrangian to describe the $X(3872)$, which is
consistent with the heavy-quark and chiral symmetries needed. They
modified the Weinberg's approach to describe bound states
\cite{Weinberg:1991um,Weinberg:1990rz}, and found that the $X(3872)$
can be a molecular bound state of the $D^{*0}$ and $\bar D^0$
mesons. They also proposed the molecular bound state $X_b$ of the
$B^{*0}$ and $\bar B^0$ with the mass of $10604$ MeV. Fleming {\it
et al.} \cite{Fleming:2007rp} also developed an effective field
theory of non-relativistic pions and $D$ mesons, and applied it to
describe the $X(3872)$ as a bound state of the $D^0\bar{D}^{*0}$ and
$\bar D^0 {D}^{*0}$. They calculated the next-to-leading-order
correction to the partial decay width $X(3872)\to
D^0\bar{D}^0\pi^0$.

However, Suzuki pointed out that some of the observed properties of
the $X(3872)$ are incompatible with the molecule interpretation
\cite{Suzuki:2005ha}. Especially, there is no long-range force to
bind the $D$ and $\bar D^*$ into a deuteron-like state, and the
observed production rates of the $X(3872)$ in $B$ decay and $pp$
collision are too large for a very loosely bound state. Alternately,
he interpreted the $X(3872)$ as the excited $^3P_1$ charmonium state
mixing with $D$ and $\bar D^*$ mesons. Detailed reviews about this
charmonium picture can be found in Sec.~\ref{sect:4.5.3}.

At the same time, many experiments were devoted to study the
$X(3872)$, and more and more experimental information was available.
Some of them are very interesting, see reviews in
Sec.~\ref{Sect:2.1.1}. More research groups jointed the debate on
whether the interaction between the $D$ and $\bar{D}^*$ mesons is
large enough to form a bound state.

The one pion exchange potential alone does not bind the proton and
neutron pair into the deuteron, and the strong attractive force in
the intermediate range has to be introduced in order to form the
deuteron, which is modeled as the sigma meson exchange potential.
This is well-known in nuclear physics. Liu {\it et al.}
\cite{Liu:2008fh} performed a dynamical calculation of the $D^0 \bar
D^{*0}$ system taking into account both the pion and sigma meson
exchange potential. Their analysis disfavored the interpretation of
the $X(3872)$ as a loosely bound molecular state with the
experimental $D^*D \pi$ coupling constant $g = 0.59$ and a
reasonable cutoff around 1 GeV. In contrast, they proposed that
there exists a loosely bound $S$-wave $B \bar B^*$ molecular state.

In Ref.~\cite{Thomas:2008ja}, Thomas and Close studied the pion
exchange between charm and bottom mesons. They found that the
$X(3872)$ can be a bound state, but the results are very sensitive
to a poorly constrained parameter, such as a cutoff around 1500 MeV
and the relatively large $L = 2$ components, etc. They confirmed the
results obtained in Ref. \cite{Liu:2008fh}, as well as confirmed
that bound states in the $B \bar B$ sector are possible.

Later in Ref.~\cite{Liu:2008tn}, Liu {\it et al.} further considered
the vector meson exchange, besides the pseudoscalar and scalar meson
exchanges~\cite{Liu:2008fh}. They applied this OBE model to
systematically study possible molecular states composed of S-wave
charmed and anti-charm mesons, such as the $D \bar D$, $D\bar{D}^*$,
and $D^* \bar D^*$. They found that the vector meson exchange
provides strong attraction in the $D^* \bar D$ system together with
the pion exchange, and the $X(3872)$ may be accommodated as a
molecular state.

Lee {\it et al.} \cite{Lee:2009hy} also discussed this issue in the
framework of a potential model generated by the exchange of
pseudoscalar, scalar and vector mesons. They considered both charged
and neutral $D\bar{D}^*$ components, and both $S$-wave and $D$-wave
contributions. Additionally, the isospin symmetry breaking effects
were fully taken into account. Their result showed that there exists
a bound state in the $D\bar{D}^*$ system with $J^{PC}=1^{++}$ for a
reasonable value of the meson-exchange regularization parameter,
$\Lambda \sim 1.2$ GeV. They also suggested that the $B\bar B^*$ bound
states can be bounded in the isoscalar limit for $J^{PC} = 1^{++}$
and $1^{+-}$.

A systematic analysis of four-quark hidden-charm states as both
compact four-quark states and meson-meson molecules was performed in
Ref.~\cite{FernandezCarames:2009zz} by Fernandez-Carames, Valcarce,
and Vijande. The authors found a $D \bar D^*$ bound state slightly
below the threshold with quantum numbers $I~J^{P C} = 0~ 1^{++}$,
which could correspond to the $X(3872)$. Further studies can be
found in Refs.~\cite{Carames:2010zz,Vijande:2014cfa}.

After the observation of the two charged $Z_b$
states~\cite{Belle:2011aa}, the authors of Ref. \cite{Sun:2012zzd}
performed an extensive study of the possible $B^*\bar{B}$,
$B^*\bar{B}^*$, $D^*\bar{D}$, $D^*\bar{D}^*$ molecule states in the
framework of the OBE model. They considered neutral and charged
$D\bar{D}^*$ modes and S-wave and D-wave mixing. Their results
indicated that there exists a bound state solution in the
$D\bar{D}^*$ system. See Table~\ref{Table.4.2.BBS2} in Sec.~\ref{Sect:4.2.1},
where $\Phi_8^*$ corresponds to the $X(3872)$.

The unitarized heavy meson chiral perturbation theory was applied in
Ref.~\cite{Wang:2013kva} by Wang and Wang to study the $D \bar D^*$
scattering with the pion exchange and a contact interaction. They
found a loosely bound state $X(3872)$, with the pole position being
$3871.70 - i 0.39$ MeV, which is not sensitive to the strength of
the contact interaction. Hence, their calculation provides a
theoretical confirmation of the existence of the $1^{++}$ state
$X(3872)$, and the light quark mass dependence of the pole position
indicates that the $X(3872)$ has a predominately $D \bar D^*$
molecular nature. Their analysis is reexamined in
Ref.~\cite{Baru:2015nea} by Baru {\it et al.} However, assuming the
$X(3872)$ to be a $D \bar D^*$ molecular state, they concluded that
the pion mass dependence of its pole position is expected to depend
strongly on the pion mass dependence of the $D \bar D^*$ interaction
at short range. They also argued that a more deeply bound $X(3872)$
for an increased pion mass as found in Ref.~\cite{Prelovsek:2013cra}
does not contradict its molecular nature.

\paragraph{Isospin violation, S-D wave mixing and coupled channel
effects}

The authors of \cite{Li:2012cs} further investigated the $X(3872)$ as a
$J^{PC} = 1^{++}$ $D\bar D^*$ molecular state in the OPE model and
the OBE model. They not only took into account the $S$-$D$ wave
mixing effect, but also considered the isospin breaking and the
coupled-channel effect. In order to find out the specific role of
the charged $D\bar{D}^*$ mode, the isospin breaking and the channel
coupling of the $X(3872)$ to the $D^*\bar{D}^*$ in forming the shallow
bound state, the authors first considered the neutral component
$D^0\bar{D}^{*0}$ only and included the S-D wave mixing, which
corresponds to Case I. Then the charged $D^+D^{*-}$ component was
added to form the exact $D\bar{D}^*$ isospin singlet with the S-D
mixing, which is Case II. Since the $1^{++}$ $D^*\bar{D}^*$ channel
lies only 140 MeV above and couples strongly to the $D\bar{D}^*$
channel, the authors further introduced the coupling of the $D\bar{D}^*$
to $D^*\bar{D}^*$ in Case III. Finally, they took into account the
explicit mass splitting between the charged and neutral $D(D^\ast)$
mesons, which is the physical Case IV. The authors considered six
channels of these four cases in Table~\ref{Channel}.

\begin{table}
\renewcommand{\arraystretch}{1.0} \caption{The different channels for Cases
I, II, III and IV of the $X(3872)$ with $J^{PC}=1^{++}$.
%For simplicity,
%we adopt the following short-hand notations,
%$\left[D^{0}\bar{D}^{*0}\right]\equiv {1\over
%\sqrt{2}}\left(D^{0}\bar{D}^{*0}-D^{*0}\bar{D}^{0}\right)$,
%$\left[D^+D^{*-}\right]\equiv{1\over
%\sqrt{2}}\left(D^+D^{*-}-D^{*+}D^-\right)$,
%$\left\{D^*\bar{D}^*\right\}\equiv {1\over
%\sqrt{2}}\left(D^{*0}\bar{D}^{*0}+D^{*+}D^{*-}\right)$ and
%$\left(D\bar{D}^*\right) \equiv {1\over
%2}\left[\left(D^{0}\bar{D}^{*0}-D^{*0}\bar{D}^{0}\right)
%+\left(D^{+}D^{*-}-D^{*+}D^{-}\right)\right]$.
``$ -$ " means the corresponding channel does not exist.
Taken from Ref. \cite{Li:2012cs}. }\label{Channel}
\begin{center}
\begin{tabular*}{16cm}{@{\extracolsep{\fill}}ccccccc}
%\toprule[0.8pt]
\toprule[0.8pt] \addlinespace[2pt]
      ~       & \multicolumn{6}{c}{Channels}\\
      Cases   &   1   &   2   &   3   &    4   &   5   &   6  \\
\specialrule{0.6pt}{1pt}{3pt}
       I      &$\left[D^{0}\bar{D}^{*0}\right]|^3S_1\rangle$&
              $\left[D^{0}\bar{D}^{*0}\right]|^3D_1\rangle$ &
               $ - $  & $-$   &  $-$  &  $-$  \\ [3pt]
       II     &$\left(D\bar{D}^{*}\right)|^3S_1\rangle$&
               $\left(D\bar{D}^{*}\right)|^3D_1\rangle$&
               $ - $  & $-$   &  $-$  &  $-$  \\ [3pt]
       III    & $\left(D\bar{D}^{*}\right)|^3S_1\rangle$&
                $\left(D\bar{D}^{*}\right)|^3D_1\rangle$ &
                $-$& $-$&
               $\left\{D^*\bar{D}^*\right\}|^3S_1\rangle$&
               $\left\{D^*\bar{D}^*\right\}|^3D_1\rangle$\\ [3pt]
       IV(Phy)&$\left[D^0\bar{D}^{*0}\right]|^3S_1\rangle$&
              $\left[D^0\bar{D}^{*0}\right]|^3D_1\rangle$ &
              $\left[D^+D^{*-}\right]|^3S_1\rangle$&
              $\left[D^+D^{*-}\right]|^3D_1\rangle$&
              $\left\{D^*\bar{D}^*\right\}|^3S_1\rangle$&  $\left\{D^*\bar{D}^*\right\}|^3D_1\rangle$\\ [3pt]
\bottomrule[0.8pt] %\bottomrule[0.8pt]
\end{tabular*}\end{center}
\end{table}

From Tables \ref{Numerical:OPE} and \ref{Numerical:OBE}, one notes
that the mass difference between the charged and neutral charmed
mesons leads to large isospin violation in the probability of the
$\left[D^{0}\bar{D}^{*0}\right]$ and $\left[D^+D^{*-}\right]$ components
in the flavor wave functions of the $X(3872)$. Moreover, the isospin
breaking effect is amplified by the tiny binding energy. As
an example, when the binding energy is 0.3 MeV, they found that the
isovector component inside the $X(3872)$ is 26\%, and the ratio of
the two hidden-charm decay modes $\mathcal{B}(X(3872) \to \pi^+
\pi^- \pi^0 J/\psi)/\mathcal{B}(X(3872) \to \pi^+ \pi^- J/\psi)$ was
estimated to be 0.42, consistent with the experimental
value~\cite{Abe:2005ix,delAmoSanchez:2010jr}.

\begin{table}[htp]
\renewcommand{\arraystretch}{1.3}
\caption{The molecular solutions of the $X(3872)$ with the OPE
potential.
%$\Lambda$ is the cutoff parameter. ``B.E." is the binding
%energy while ``Mass" is the calculated mass of the $X(3872)$. $r_{rms}$
%and ``$P_i$" are the root-mean-square radius and the probability of
%the ith channel, respectively.
``$\times$" means no binding solutions, and ``$-$" denotes that the corresponding component does
not exist. Taken from Ref. \cite{Li:2012cs}.}\label{Numerical:OPE}
{\small
\begin{center}
  \begin{tabular*}{16cm}{@{\extracolsep{\fill}}cccccccccccl}
  %\toprule[0.8pt]
  \toprule[0.8pt]\addlinespace[3pt]
    Cases         &$\Lambda$ (GeV)& B.E. (MeV)& Mass (MeV)&$r_{rms}$ (fm)& $P_1\,(\%)$&
    $P_2\,(\%)$     & $P_3\,(\%)$    & $P_4\,(\%)$ &$P_5\,(\%)$  & $P_6\,(\%)$ \\
    \specialrule{0.6pt}{3pt}{3pt}
    \multirow{3}*{I}&            &           &          &           &         &
        ~         &    $-$       & $-$       & $-$      &  $-$     \\
        ~         &$0.80\sim2.0$ &\multicolumn{5}{c}{$\times$}
                  &    $-$       &  $-$      & $-$      &   $-$    \\
        ~         &              &           &          &           &         &
        ~         &    $-$       & $ -$      & $ - $    &   $-$  \\ [5pt]
\multirow{5}*{II} &    1.55      & $0.32$    & 3871.49  &   4.97 &
98.81   &
        1.19      & $ -$         & $ - $     &  $-$     &   $-$  \\
        ~         &    1.60      & $0.92$    & 3870.89  &   3.51   & 98.39   &
        1.61      &    $ -$      & $ - $     &  $-$     &  $-$   \\
        ~         &    1.65      & $1.90$    & 3869.91  &   2.56   & 98.01   &
        1.99      &    $ -$      & $ - $     &  $-$     &  $-$   \\
        ~         &    1.70      & $3.31$    & 3868.50  &   1.99   & 97.69   &
        2.31      &   $ -$       & $ - $     &  $-$     &   $-$  \\
        ~         &    1.80      & $7.70$    & 3864.11  &   1.36   & 97.18   &
        2.82      &    $ -$      & $ - $     & $-$      &  $-$   \\  [5pt]
\multirow{6}*{III}&    1.10      & $0.76$    & 3871.05  &   3.79 &
97.82    &
        0.73      & $-$          & $-$       &    1.24  &   0.20 \\
         ~        &    1.15      & $2.72$    & 3869.09  &   2.17   & 96.15   &
        0.82      &  $-$         & $-$       &  2.64    &   0.40 \\
        ~         &    1.20      & $6.25$    & 3865.56  &   1.49   & 94.26   &
        0.77      &    $-$       &  $-$      &    4.37  &   0.60 \\
        ~         &    1.25      & $11.66$   & 3860.15  &   1.13   & 92.20   &
        0.67      &   $-$        &  $-$      &  6.32    &   0.81 \\
        ~         &    1.30      & $19.21$   & 3852.60  &   0.91   & 90.05   &
        0.55      &   $-$        &  $-$      & 8.38     &   1.02 \\
                  &    1.55      &  95.79    &  3776.02 &   0.47   &  80.68  &
        0.16      &    $-$       &   $-$     &  17.37   &   1.80 \\   [5.0pt]
\multirow{5}*{IV(Phy)} &    1.15      & $0.26$    &  3871.55 & 4.79
&  85.68  &
   0.22           &    12.29     &  0.24     &   0.36   &  0.21  \\
       ~          &    1.17      & $1.03$    &  3870.78 &   2.99   &  76.37  &
  0.30            &    20.27     &  0.33     &   2.39   &  0.35  \\
       ~          &    1.20      & $2.93$    &  3868.88 &   1.84   &  66.18  &
  0.34            &   28.74      &  0.36     &   3.84   &  0.54  \\
       ~          &    1.25      & $7.99$    &  3863.82 &   1.20   &  56.72  &
  0.32            &    35.76     &  0.34     &   6.08   &  0.79  \\
       ~          &     1.30     & $15.36$   &  3856.45 &   0.93   &  51.59  &
  0.27            &    38.61     &  0.28     &   8.25   &  1.01  \\ [3pt]
    \bottomrule[0.8pt]
   % \bottomrule[0.8pt]
\end{tabular*}
\end{center}}
\end{table}

\begin{table}[htp]
\caption{The molecular solutions of the $X(3872)$ with the OBE
potential.
%$\Lambda$ is the cutoff parameter. ``B.E." is the binding
%energy while ``Mass" is the calculated mass of the $X(3872)$. $r_{rms}$
%and ``$P_i$" are the root-mean-square radius and the probability of
%the ith channel, respectively.
``$\times$" means no binding
solutions, and ``$-$" denotes that the corresponding component does
not exist. Taken from Ref. \cite{Li:2012cs}.}\label{Numerical:OBE}
{\small
\begin{center}
  \begin{tabular*}{16cm}{@{\extracolsep{\fill}}ccccccccccc}
  %\toprule[0.8pt]
  \toprule[0.8pt]\addlinespace[3pt]
    Cases       &$\Lambda$ (GeV)& B.E. (MeV)& Mass (MeV)&$r_{rms}$ (fm)&$P_1\,(\%)$&
    $P_2\,(\%)$   & $P_3\,(\%)$& $P_4\,(\%)$ & $P_5\,(\%)$ & $P_6\,(\%)$   \\
    \specialrule{0.6pt}{3pt}{3pt}
    \multirow{4}*{I}&1.85  & $0.21$    & 3871.60   &   5.36       &99.54    &
        0.46    & $-$      & $ - $     &  $-$      &  $-$        \\
        ~       &    1.90  & $0.53$    & 3871.28   &   4.32       &99.27    &
        0.63    & $-$      & $ -$      &  $-$      &  $-$        \\
        ~       &    1.95  & $0.96$    & 3870.85   &   3.48       &99.18    &
        0.82    & $-$      & $ -$      &  $-$      & $-$         \\
        ~       &    2.00  & $1.51$    & 3870.30   &   2.88       & 98.99   &
        1.01    & $ -$     & $ -$      &  $-$      &  $-$        \\ [5pt]
\multirow{5}*{II}&    1.10 & $0.61$    & 3871.20   &   4.21 &98.82
&
          1.18  & $ -$     & $ -$      &  $-$      &  $-$        \\
        ~       &    1.15  & $2.15$    & 3869.66   &   2.54       &98.27    &
          1.73  & $ -$     & $-$       &  $-$      &  $-$        \\
        ~       &    1.20  & $4.58$    & 3867.23   &   1.84       &97.28   &
          2.18  & $ -$     & $ -$      &  $-$      &  $-$        \\
        ~       &    1.25  & $7.84$    & 3863.97   &   1.48       &97.40   &
          2.60  & $ -$     & $ -$      &  $-$      &  $-$        \\
        ~       &    1.30  & $11.87$   & 3859.94   &   1.26       &97.01   &
          2.99  & $ -$     & $ -$      &  $-$      &  $-$        \\  [5pt]
\multirow{6}*{III}&  1.00  & $0.74$    & 3871.07   &   3.92 &98.38
&
          0.79  &  $-$     &  $-$      &   0.66    &  0.18      \\
         ~      &    1.10  & $5.69$    & 3866.12   &   1.66       &96.39   &
          1.07  &   $-$    &  $-$      &    1.91   &  0.62       \\
        ~       &    1.15  & $9.67$    & 3862.14   &   1.34       &95.51   &
          1.12  &   $-$    &  $-$      &   2.46    &   0.92      \\
        ~       &    1.20  & $14.51$   & 3857.30   &   1.15       &94.65   &
          1.15  &   $-$    &  $-$      &  2.94     &   1.26      \\
        ~       &    1.25  & $20.18$   & 3851.63   &   1.02       &93.82   &
          1.17  &   $-$    &  $-$      &  3.35     &   1.67     \\
        ~       &    1.30  & $26.68$   & 3845.13   &   0.92       &92.98   &
          1.18  &   $-$    &  $-$      &   3.71    &  2.14       \\ [5.0pt]
\multirow{6}*{IV(Phy)}&   1.05  &  $0.30$   &  3871.51  &   4.76  &
86.80 &
 0.27           &  11.77  &   0.28    &   0.67    &  0.20      \\
       ~        &   1.06   &  $0.60$   &  3871.21  &   3.85       &  82.83 &
  0.33          &  15.35   &  0.34     &   0.88    &  0.27      \\
       ~        &  1.08    &  $1.43$   &  3870.38  &   2.69       &  75.80 &
  0.41          & 21.68    &  0.42     &   1.28    &  0.41      \\
      ~         &  1.10    &  $2.53$   &  3869.28  &   2.09       &  70.44 &
  0.46          & 26.46    &  0.47     &   1.62    &  0.54      \\
      ~         & 1.12     &  $3.84$   &  3867.97  &   1.75       &  66.40 &
  0.50          & 30.00    &  0.51     &   1.92    &  0.67      \\
      ~         & 1.15     &  $6.16$   &  3865.65  &   1.46       &  62.03 &
  0.53          & 33.72    &  0.54     &   2.31    &  0.87      \\
     ~          & 1.20     &  $10.83$  &  3860.98  &   1.19       &  57.38 &
  0.56          & 37.42    &  0.56     &   2.85    &  1.23      \\ [3pt]
 %\bottomrule[0.8pt]
 \bottomrule[0.8pt]

\end{tabular*} \end{center}}
\end{table}

The authors emphasized that the existence of the shallow bound state
$X(3872)$ and very large isospin violation in its hidden-charm decay
arise from the very delicate combined efforts of the several
driving forces including the long-range one-pion exchange, the S-D
wave mixing, the mass splitting between the charged and neutral
$D(D^\ast)$ mesons, and the coupled-channel effects
\cite{Li:2012cs}.

In a recent work \cite{Zhao:2014gqa}, Zhao, Ma, and Zhu further
studied the $D\bar{D}^*$ system in the framework of the OBE model. They
examined the spin-orbit force correction up to $\mathcal{O}(1/M)$
and the recoil correction up to $\mathcal{O}(1/M^2)$, and found that
the former one is important for the very loosely bound state. Their
result suggested that there exists an isoscalar molecular state in
the $D\bar{D}^*$ system with $J^{PC}=1^{++}$.  Their result also
suggested that the recoil correction may be larger than the binding
energy of the $X(3872)$, which may partly force the $X(3872)$ to
become a very shallow bound state~\cite{Zhao:2014gqa}.

\begin{figure}[htbp]
\begin{center}
\begin{tabular}{c}
\scalebox{0.43}{\includegraphics{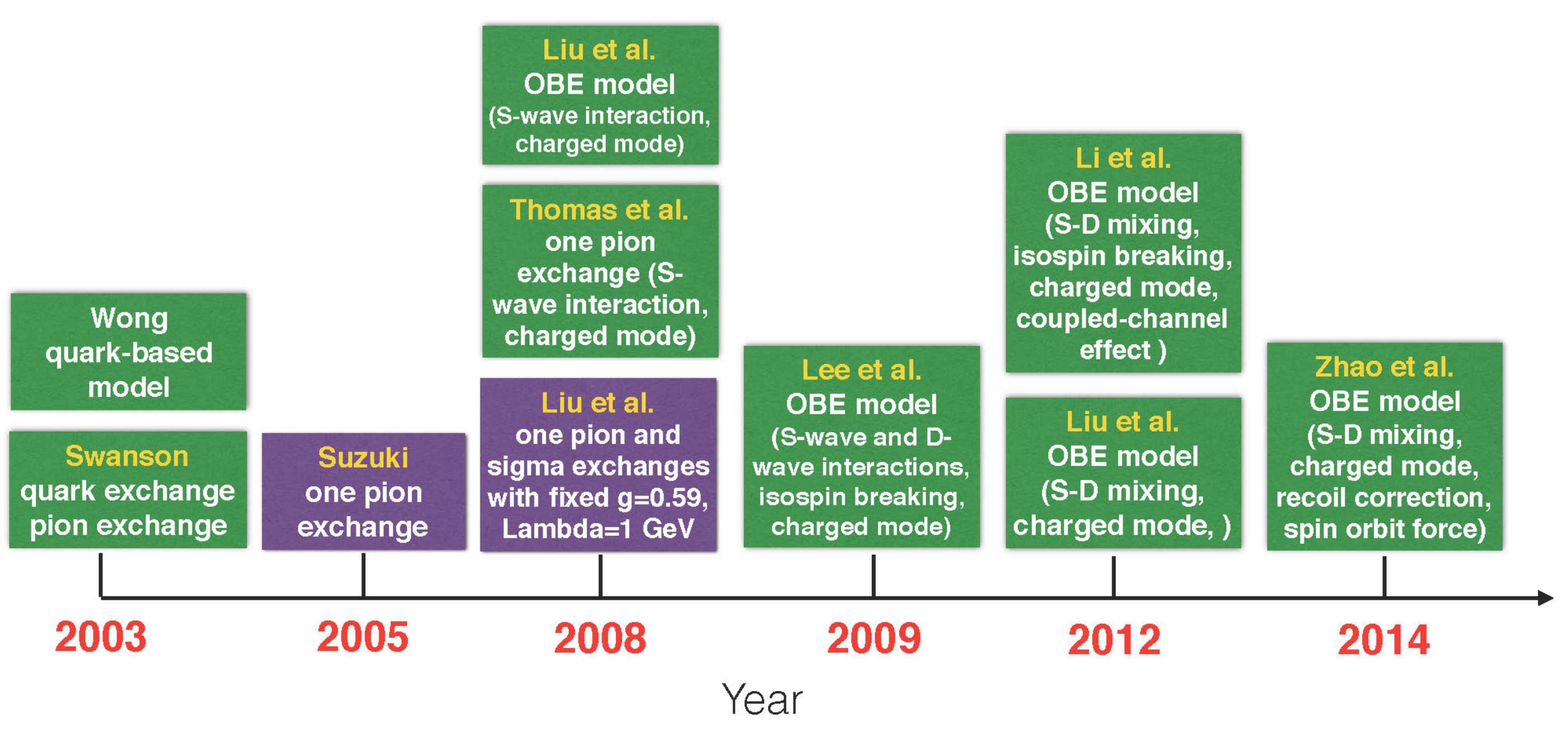}}
\end{tabular} \caption{(Color online) A summary of the theoretical progresses on the dynamical studies of the $D\bar{D}^*$ molecular state of the $X(3872)$. Here, these studies are marked by green and purple backgrounds when the corresponding conclusion of whether the $X(3872)$ is a $D\bar{D}^*$ molecular state  is positive and negative, respectively.
\label{Fig.4.5.3872summary}}
\end{center}
\end{figure}

In Fig. \ref{Fig.4.5.3872summary}, we summarized the above
investigations, which attempted to answer whether the $X(3872)$ is a
$D\bar{D}^*$ molecular state or not. Until now, many theoretical
studies suggested that the interaction in the $D\bar{D}^*$ system
with $I^G(J^{PC})=0^+(1^{++})$ is attractive, which can result in a
shallow bound state. However, further experimental and theoretical
studies are needed to test this hadronic molecular state assignment
to the $X(3872)$. Especially, studies on its decay behaviors are
important, which will be reviewed in Sec.~\ref{sect:4.5.4}.

\paragraph{QCD sum rule}

In QCD sum rule, Chen {\it et al.} described the $X(3872)$ as a
mixed state of charmonium hybrid and $\bar DD^*$ molecular state
using the following current with $J^{PC}=1^{++}$ \cite{Chen:2013pya}
\begin{equation}
J_\nu^\xi=\sqrt{1-\xi^2}J_\nu^m+\xi \sigma J_\nu^h\, , \label{mix_hmcurrent}
\end{equation}
where the charmonium hybrid current $J_\nu^h$ and molecular current $J_\nu^m$ are
\begin{align}
 J_\mu^h &= \frac{1}{2} g \bar{c} \gamma^\nu \lambda^a \tilde{G}_{\mu\nu}^a c \, , \quad \tilde{G}_{\mu\nu}^a = \frac{1}{2} \epsilon_{\mu\nu\alpha\beta} \tilde{G}^{\alpha\beta}_a \,,
 \\
J_\nu^m &= \frac{1}{\sqrt{2}}\left(\bar{q}_a\gamma_5 c_a \bar{c}_b \gamma_\nu q_b - \bar{q}_a\gamma_\nu c_a \bar{c}_b \gamma_5 q_b \right) \, .
\end{align}
The parameter $\sigma$ in Eq. \eqref{mix_hmcurrent} is a mass scale
that accounts for the different mass dimensions of the hybrid and
molecular currents. $\xi$ is a dimensionless mixing parameter. The
predicted mass increased with the mixing parameter until it reached
a maximum value in agreement with the mass of $X(3872)$.

Lee, Nielsen and Wiedner studied a $D^*\bar D_s$
molecule with $J^P=1^+$ in Ref. \cite{Lee:2008uy}. They proposed
such molecule as a natural generalized state to the strangeness
sector of the $X(3872)$. They calculated the two-point correlation
function by keeping the $m_s$ proportional term in the OPE series.
As an extension, they studied the $D^0\bar D^{*0}-D^{*0}\bar D^{0}$
molecule to assign the $X(3872)$ meson by taking $m_s=0$. They
obtained $m_{D^*\bar D}=(3.88\pm0.06)$ GeV, in agreement with the
experimental value of the $X(3872)$ meson \cite{Lee:2008uy}. This
result was consistent with the calculations of the $D^*\bar D$ state
in Refs. \cite{Zhang:2009vs,Wang:2013daa,Chen:2015ata}. In Ref.
\cite{Lee:2008tz}, Lee, Morita and Nielsen extended their discussion
to include the total width by employing the Breit-Wigner function to
the pole term, using the same molecular current with that in Ref.
\cite{Lee:2008uy}. They found that introducing the width slightly
modified the predicted mass and resulted in a better reproduction of
the $X(3872)$.

However, the narrow decay width of the $X(3872)$ can not be
explained in QCD sum rule if it is a pure four-quark state
\cite{Navarra:2006nd}. To reproduce the small width of the
$X(3872)$, it was considered as a mixture between charmonium and
molecular state with $J^{PC}=1^{++}$ in Ref. \cite{Matheus:2009vq}.
They found a small mixing angle $5^\circ\leq\theta\leq13^\circ$ to
reproduce the parameters $m_X=(3.77\pm0.18)$ GeV and $\Gamma(X\to
J/\psi\pi^+\pi^-)=(9.3\pm6.9)$ MeV. They concluded that the
$X(3872)$ is approximately $97\%$ a charmonium state with $3\%$
molecule (admixture of $88\%$ $D^0 \bar D^{*0}$ and $12\%$ $D^+\bar
D^{*-}$). Later, this configuration was used to study the radiative
decay of the $X(3872)$ in Ref. \cite{Nielsen:2010ij}. The authors
calculated the three-point functions for the vertex
$X(3872)J/\psi\gamma$ to study the partial decay width of
$X(3872)\to J/\psi\gamma$. They obtained the branching ration
$\Gamma(X(3872)\to J/\psi\gamma)/\Gamma(X(3872)\to
J/\psi\pi^+\pi^-)=0.19\pm0.13$, which is consistent with the
experiment result for the radiative decay of the $X(3872)$
\cite{Abe:2005ix,Aubert:2006aj}.

Besides the above references, the $X(3872)$ was also studied as a
molecular state in
Refs.~\cite{Wang:2009aw,Ortega:2012rs,Guo:2014hqa,Tomaradze:2015cza,Karliner:2015ina,Baru:2015tfa}.

\subsubsection{The axial vector tetraquark state}
\label{sect:4.5.2}

\paragraph{Diquark model}

A diquark-antidiquark model was proposed in Ref.~\cite{Maiani:2004vq}
by Maiani, Piccinini, Polosa, and Riquer to explain the $X(3872)$,
based on their previous studies on the lightest scalar
mesons~\cite{Maiani:2004uc}. We note that this is the ``type-I''
diquark-antidiquark model, and the ``type-II'' diquark-antidiquark
model~\cite{Maiani:2014aja} has been reviewed in
Sec.~\ref{Sect:4.2.2}.

\begin{figure}[hbtp]
\begin{center}
\scalebox{0.5}{\includegraphics{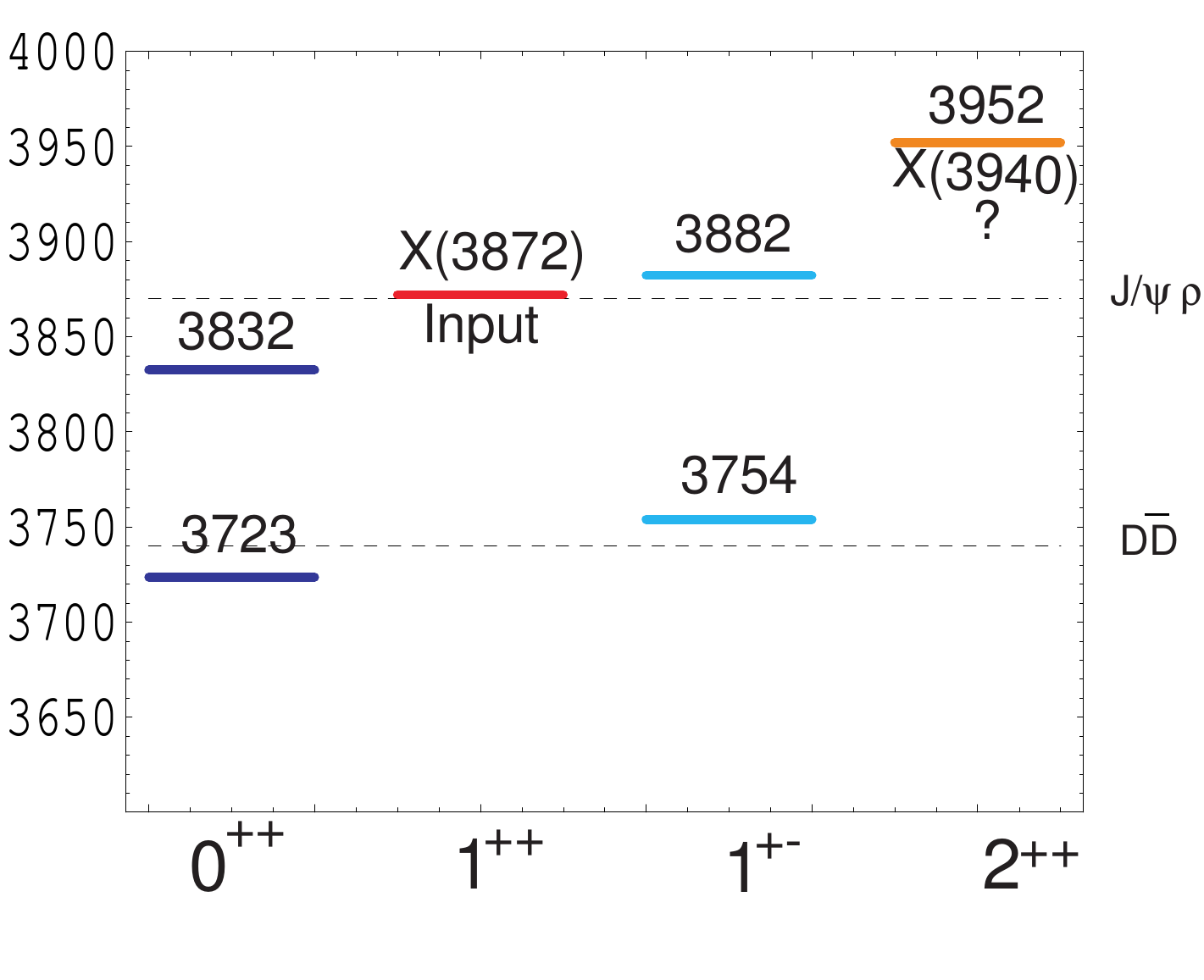}}
%\scalebox{0.5}{\includegraphics{figs/selexspecf.pdf}}
\end{center}
\caption{(Color online) The mass spectrum of the $X$ particles.
Taken from Ref. \cite{Maiani:2004vq}. \label{Fig.4.5.MaianiX3872}}
\end{figure}

In this model the hadron masses depend on three ingredients: quark
composition, constituent quark masses and spin-spin interactions.
The Hamiltonian is
\begin{equation}
H = \sum_i m_i + \sum_{i<j} 2 \kappa_{ij} (S_i \cdot S_j) \, ,
\end{equation}
where the coefficients $\kappa_{ij}$ depend on the flavor of the
constituents $i$, $j$ and the particular color state of the pair.
All these parameters can be derived from the quark-antiquark mesons
and three-quark baryons. Especially, they considered the ``good''
diquark having $S=0$ and the ``bad'' diquark having $S=1$, whose
masses are derived from the light scalar mesons as well as the
$X(3872)$. Using these components, they constructed six $[cq][\bar c
\bar q^\prime]$ states, including two states with $J^{PC} = 0^{++}$,
one state with $J^P = 1^{++}$, two states with $J^P = 1^{+-}$, and
one state with $J^{PC} = 2^{++}$:
\begin{eqnarray}
\nonumber |0^{++}\rangle &=& |0_{cq}, 0_{\bar c \bar q^\prime}; J=0\rangle \,
,
\\ \nonumber |0^{++\prime}\rangle &=& |1_{cq}, 1_{\bar c \bar q^\prime}; J=0\rangle \, ,
\\ |1^{++}\rangle &=& {1\over\sqrt2} \Big(|0_{cq}, 1_{\bar c \bar q^\prime}; J=1\rangle + |1_{cq}, 0_{\bar c \bar q^\prime}; J=1\rangle\Big) \, ,
\\ \nonumber |1^{+-}\rangle &=& {1\over\sqrt2} \Big(|0_{cq}, 1_{\bar c \bar q^\prime}; J=1\rangle - |1_{cq}, 0_{\bar c \bar q^\prime}; J=1\rangle\Big) \, ,
\\ \nonumber |1^{+-\prime}\rangle &=& |1_{cq}, 1_{\bar c \bar q^\prime}; J=1\rangle \, ,
\\ \nonumber |2^{+-\prime}\rangle &=& |1_{cq}, 1_{\bar c \bar q^\prime}; J=2\rangle \, .
\end{eqnarray}
The authors used the $1^{++}$ state to fit the $X(3872)$, and
calculated masses of the other five states as well as masses of the
six $[cq][\bar s \bar q^\prime]$ states, as shown in
Fig.~\ref{Fig.4.5.MaianiX3872}. They also studied the isospin
breaking effects, and predicted that the $X(3872)$ was made of two
components with a mass difference related to $m_u - m_d$ and
discussed the production of the $X(3872)$ and of its charged partner
$X^\pm$ in the weak decays of $B^{+,0}$~\cite{Bigi:2005fr}. This
idea was further developed in Ref.~\cite{Maiani:2007vr} by Maiani,
Polosa, and Riquer, where they proposed four states: $X_u =
[cu][\bar c \bar u]$, $X_d = [cd][\bar c \bar d]$, $X^+ = [cu][\bar
c \bar d]$, and $X^- = [cd][\bar c \bar u]$, and used $X_u$ and
$X_d$ to explain the mass difference between the $X(3872)$ state
decaying into
$J/\psi\pi^+\pi^-$~\cite{Choi:2003ue,delAmoSanchez:2010jr} and the
one decaying into $D^0 \bar D^0 \pi^0$~\cite{Abazov:2004kp}. The
quantum numbers of the $X(3872)$, both $J^{PC}=2^{-+}$ and $1^{++}$,
were discussed by Burns, Piccinini, Polosa, and Sabelli in
Ref.~\cite{Burns:2010qq}. According to the prediction for the
charged partner of the $X(3872)$ in Ref.~\cite{Maiani:2007vr}, BaBar
carried out a careful search for them in the $B\to X^-K$, $X^-\to
J/\psi \pi^- \pi^0$ \cite{Aubert:2004zr}. However, the charged
partners of the $X(3872)$ have not been observed in experiment.

The diquark-antidiquark picture was later used to calculate the
masses of heavy tetraquarks by Ebert, Faustov and Galkin in the
relativistic quark model in Ref.~\cite{Ebert:2005nc}. They also
found that the $X(3872)$ can be the neutral charm tetraquark state,
and concluded that one more neutral and two charged tetraquark
states must exist with close masses. Similar conclusions were
obtained in Ref.~\cite{Terasaki:2007uv} by Terasaki that the
$X(3872)$ consists of two iso-singlet tetra-quark mesons, $X^\pm
\sim {[cn](\bar c \bar n) \pm (cn)[\bar c \bar n]}_{I=0}$, with
opposite $G$ parities. Here the parentheses and the square brackets
denote symmetry and anti-symmetry, respectively, of the wave
function under the exchange of flavors between them.

\paragraph{QCD sum rule}

Using the diquark-antidiquark configuration, Chen and Zhu
studied the hidden-charm tetraquark systems with $J^{PC}=1^{++}$ in
the QCD sum rules in a systematical way \cite{Chen:2010ze}. The
constructed all charmonium-like tetraquark interpolating currents
with quantum numbers $J^{PC}=1^{++}$ without derivative operators \cite{Chen:2010ze,Chen:2013jra,Chen:2015fwa}
\begin{eqnarray}
\begin{split}
J_{1\mu}&=q^T_aCQ_b(\bar{q}_a\gamma_{\mu}\gamma_5C\bar{Q}^T_b+\bar{q}_b\gamma_{\mu}\gamma_5C\bar{Q}^T_a)
+
q^T_aC\gamma_{\mu}\gamma_5Q_b(\bar{q}_aC\bar{Q}^T_b+\bar{q}_bC\bar{Q}^T_a)\,
, \\
J_{2\mu}&=q^T_aCQ_b(\bar{q}_a\gamma_{\mu}\gamma_5C\bar{Q}^T_b-\bar{q}_b\gamma_{\mu}\gamma_5C\bar{Q}^T_a)
+
q^T_aC\gamma_{\mu}\gamma_5Q_b(\bar{q}_aC\bar{Q}^T_b-\bar{q}_bC\bar{Q}^T_a)\,
, \\
J_{3\mu}&=q^T_aC\gamma_5Q_b(\bar{q}_a\gamma_{\mu}C\bar{Q}^T_b+\bar{q}_b\gamma_{\mu}C\bar{Q}^T_a)
+
q^T_aC\gamma_{\mu}Q_b(\bar{q}_a\gamma_5C\bar{Q}^T_b+\bar{q}_b\gamma_5C\bar{Q}^T_a)\,
, \\
J_{4\mu}&=q^T_aC\gamma_5Q_b(\bar{q}_a\gamma_{\mu}C\bar{Q}^T_b-\bar{q}_b\gamma_{\mu}C\bar{Q}^T_a)
+
q^T_aC\gamma_{\mu}Q_b(\bar{q}_a\gamma_5C\bar{Q}^T_b-\bar{q}_b\gamma_5C\bar{Q}^T_a)\,
, \\
J_{5\mu}&=q^T_aC\gamma^{\nu}Q_b(\bar{q}_a\sigma_{\mu\nu}\gamma_5C\bar{Q}^T_b+\bar{q}_b\sigma_{\mu\nu}\gamma_5C\bar{Q}^T_a)
+
q^T_aC\sigma_{\mu\nu}\gamma_5Q_b(\bar{q}_a\gamma^{\nu}C\bar{Q}^T_b+\bar{q}_b\gamma^{\nu}C\bar{Q}^T_a)\,
, \\
J_{6\mu}&=q^T_aC\gamma^{\nu}Q_b(\bar{q}_a\sigma_{\mu\nu}\gamma_5C\bar{Q}^T_b-\bar{q}_b\sigma_{\mu\nu}\gamma_5C\bar{Q}^T_a)
+
q^T_aC\sigma_{\mu\nu}\gamma_5Q_b(\bar{q}_a\gamma^{\nu}C\bar{Q}^T_b-\bar{q}_b\gamma^{\nu}C\bar{Q}^T_a)\,
, \\
J_{7\mu}&=q^T_aC\gamma^{\nu}\gamma_5Q_b(\bar{q}_a\sigma_{\mu\nu}C\bar{Q}^T_b+\bar{q}_b\sigma_{\mu\nu}C\bar{Q}^T_a)
+
q^T_aC\sigma_{\mu\nu}Q_b(\bar{q}_a\gamma^{\nu}\gamma_5C\bar{Q}^T_b+\bar{q}_b\gamma^{\nu}\gamma_5C\bar{Q}^T_a)\, , \\
J_{8\mu}&=q^T_aC\gamma^{\nu}\gamma_5Q_b(\bar{q}_a\sigma_{\mu\nu}C\bar{Q}^T_b-\bar{q}_b\sigma_{\mu\nu}C\bar{Q}^T_a)
+
q^T_aC\sigma_{\mu\nu}Q_b(\bar{q}_a\gamma^{\nu}\gamma_5C\bar{Q}^T_b-\bar{q}_b
\gamma^{\nu}\gamma_5C\bar{Q}^T_a)\, ,
\label{Eq.4.5.currents1pp}
\end{split}
\end{eqnarray}
where $q$ the represents up or down quark and $Q$ the charm quark. The
color structures are symmetric
$\mathbf 6 \otimes \mathbf{\bar 6}$
for the currents $J_1, J_3, J_5, J_7$ and antisymmetric $\mathbf{\bar
3} \otimes \mathbf 3$ for the currents $J_2, J_4, J_6, J_8$. As
shown in Eq.~(\ref{Eq.4.1.Tetraquarkquarkcontent}), all these
interpolating currents in Eq.~(\ref{Eq.4.5.currents1pp}) can
couple to both isotriplet and isosinglet hadron states.

Using these interpolating currents, the authors obtained the mass
spectra of the charmonium-like and bottomonium-like tetraquark
states with $J^{PC}=1^{++}$ in Table \ref{Table.4.5.X3872QSR}.
The hidden-strange tetraquark states were also studied. Using the
current $J_{4\mu}$ in Eq.~(\ref{Eq.4.5.currents1pp}), the mass
of the $qc\bar q\bar c$ tetraquark state was extracted as
$m_X=(4.03\pm0.11)$ GeV, slightly above the mass of the $X(3872)$.
The interpolating current $J_{4\mu}$ was also used to study the
$X(3872)$ as a tetraquark state in Refs.
\cite{Matheus:2006xi,Wang:2013vex}, where the extracted hadron
masses were consistent with the result in Table
\ref{Table.4.5.X3872QSR}.

%%%%%%%%%%%%%%%%%%%%%%%%%%%%%%%%%%%%%%%%%%%%%%%%%%%%%%%%%%%%%%%%%%%%%%%%%%%%%%%%%%%%%%%%%%%%%%%%%%%%%%%%%%%%%%%%%%%%%%%%%%%%%%%%%%%%%%%
\renewcommand{\arraystretch}{1.6}
\begin{table}[htb]
\caption{Mass spectra for the $J^{PC}=1^{++}$ charmonium-like and
bottomonium-like tetraquark states in diquark-antidiquark
configuration \cite{Chen:2010ze}. \label{Table.4.5.X3872QSR}}
\begin{center}
\begin{tabular}{cccccc} \toprule[1pt]
& Current ~~&~~ $s_0\,(\mbox{GeV}^2)$ ~~&~~ \mbox{Borel window} $\,(\mbox{GeV}^2)$ ~~&~~ $m_X$\, \mbox{(GeV)} ~~&~~ \mbox{PC\,(\%)} \\
\midrule[1pt]
\multirow{2}{*}{$qc\bar q\bar c$}   & $J_{3\mu}$         &  $4.6^2$         & $3.0-3.4$           & $4.19\pm0.10$     & 47.3 \\
                   & $J_{4\mu}$         &  $4.5^2$         & $3.0-3.3$           & $4.03\pm0.11$     & 46.8
\vspace{5pt} \\
\multirow{4}{*}{$qb\bar q\bar b$ }  & $J_{3\mu}$         &  $10.9^2$        & $8.5-9.5$           & $10.32\pm0.09$    & 47.0 \\
                   & $J_{4\mu}$         &  $10.8^2$        & $8.5-9.2$           & $10.22\pm0.11$    & 44.6 \\
                   & $J_{7\mu}$         &  $10.7^2$        & $7.8-8.4$           & $10.14\pm0.10$    & 44.8  \\
                   & $J_{8\mu}$         &  $10.7^2$        & $7.8-8.4$           & $10.14\pm0.09$    & 44.8  \\
\bottomrule[1pt]
\end{tabular}
\end{center}
\end{table}
%%%%%%%%%%%%%%%%%%%%%%%%%%%%%%%%%%%%%%%%%%%%%%%%%%%%%%%%%%%%%%%%%%%%%%%%%%%%%%%%%%%%%%%%%%%%%%%%%%%%%%%%%%%%%%%%%%%%%%%%%%%%%%%

\paragraph{Chromomagnetic interaction}

The chromomagnetic interaction
\begin{eqnarray}
H = \sum_i m_i  + H_{CM} = \sum_i m_i - \sum_{i>j} v_{ij} \vec
\lambda_i \cdot \vec \lambda_j  \vec \sigma_i \cdot \vec \sigma_j \,
,
\end{eqnarray}
was also applied to study the $X(3872)$ in
Refs.~\cite{Hogaasen:2005jv,Cui:2006mp,Buccella:2006fn}. In
Ref.~\cite{Hogaasen:2005jv}, Hogaasen, Richard, and Sorba found that
the chromomagnetic interaction, with proper account for
flavour-symmetry breaking, can be used to explain the mass and
coupling properties of the $X(3872)$ resonance as a $J^{PC} =
1^{++}$ state consisting of a heavy quark-antiquark pair and a light
one. This study was extended to study the S-wave configurations
containing two quarks and two antiquarks in
Ref.~\cite{Buccella:2006fn} by Buccella, Hogaasen, Richard, and
Sorba, where they investigated light, charmed, charmed and strange,
hidden-charm and double-charm mesons, as well as their analogues
with bottom quarks. In Ref.~\cite{Cui:2006mp}, Cui, Chen, Deng, and Zhu
performed a schematic study of the masses of possible heavy
tetraquarks using the chromomagnetic interaction with the flavor
symmetry breaking corrections, and they found that the
chromomagnetic interaction is repulsive for the $2^+$ heavy tetraquarks,
while the $0^+$ $qc\bar q\bar c$ states will also exist if the
$X(3872)$ is a $1^+$ tetraquark.

\paragraph{Constituent quark model}

However, in Ref.~\cite{Vijande:2007fc}, Vijande, Weissman, Barnea,
and Valcarce studied the four-quark system $c \bar cq \bar q$ in the
framework of the constituent quark model. They solved the four-body
Schr\"odinger equation by means of the hyperspherical harmonic
formalism using different types of quark-quark potentials, and ruled
out the possibility that the $X(3872)$ is a compact tetraquark
system, unless additional correlations, either in the form of
diquarks or at the level of the interacting potential, not
considered in simple quark models do contribute.

\subsubsection{Radial excitation of the axial vector charmonium}\label{sect:4.5.3}

\paragraph{Quark model}

\begin{figure}[htbp]
\begin{center}
\begin{tabular}{c}
\scalebox{0.73}{\includegraphics{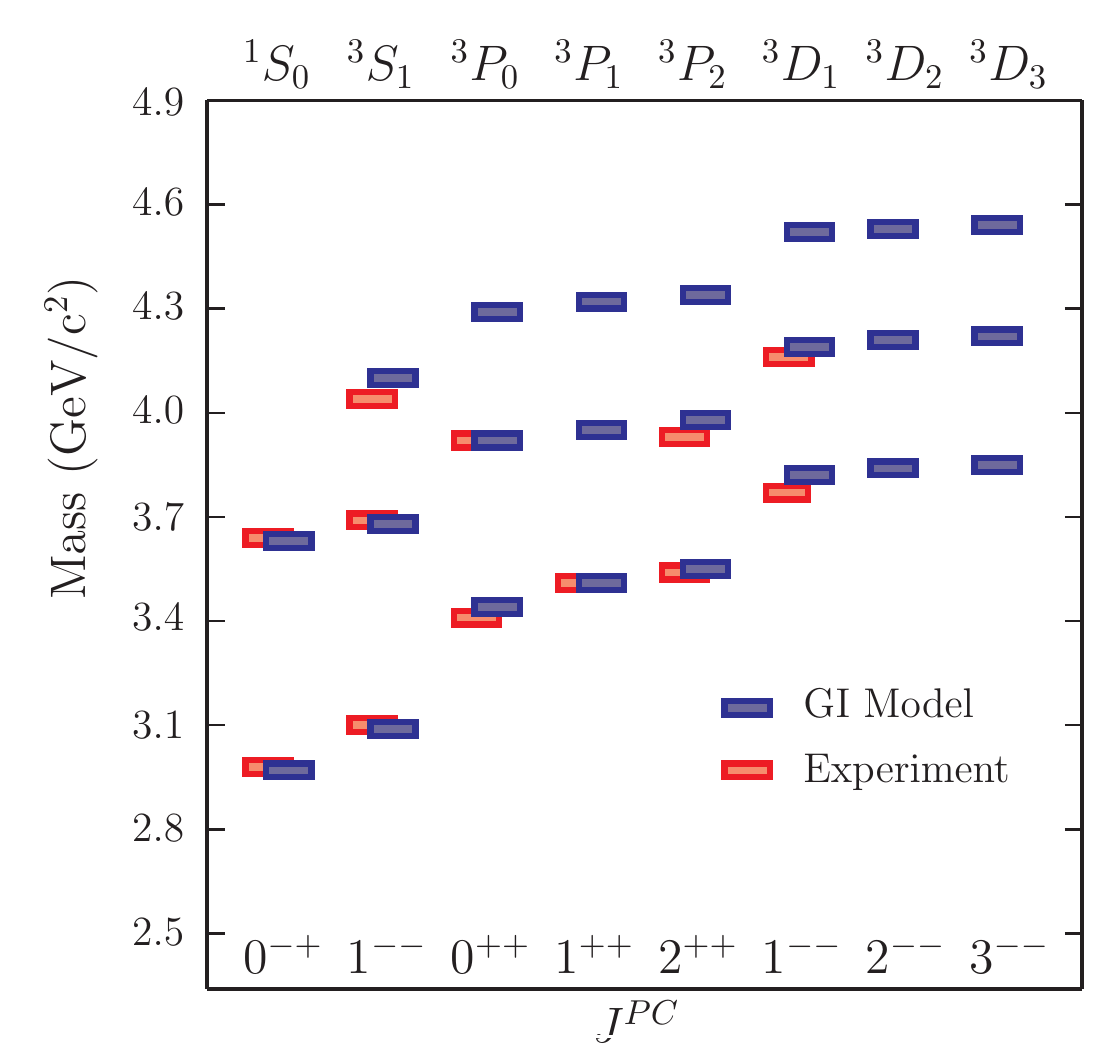}}
\end{tabular} \caption{(Color online) The comparison between the result from the GI model  \cite{Godfrey:1985xj} and the experimental data \cite{pdg} for charmonium family.
\label{Fig.4.5.CF}}
\end{center}
\end{figure}

There are lots of discussions on the interpretation of the $X(3872)$
as a P-wave charmonium
state~\cite{Close:2003sg,Barnes:2003vb,Kong:2006ni,Ortega:2010qq,Kalashnikova:2010hv,Yang:2010am,Takizawa:2012hy,Ferretti:2013faa,Butenschoen:2013pxa,Ferretti:2014xqa,Takeuchi:2014rsa,Aliev:2014doa,Aliev:2014uda,Meng:2014ota,Yang:2015tsa}.
But there exist two major difficulties: (1) the mass of the
$X(3872)$ is significantly lower than the predictions of quark
models, for example, see Fig.~\ref{Fig.4.5.CF} obtained using the GI
model~\cite{Godfrey:1985xj}; (2) the large isospin violation in the
$X(3872)\to J/\psi\rho$ decay was observed in
Refs.~\cite{Abe:2005ix,delAmoSanchez:2010jr}, as shown in
Eqs.~(\ref{Eq.2.1.3872iso1}) and (\ref{Eq.2.1.3872iso2}) in
Sec.~\ref{Sect:2.1.1}. One natural speculation is that the $X(3872)$
may not be a pure $\chi_{c1}^\prime(2P)$ charmonium state. Instead,
the $X(3872)$ may be a mixture of the bare $c\bar c$ charmonium
state and the $D\bar{D}^*$ molecule component. In other words, the
coupled channel effect may play a very important role in the case of
the $X(3872)$.

Comparing the mass spectrum of charmonium states calculated
by the GI model~\cite{Godfrey:1985xj} with the current
experimental data~\cite{pdg} (see Fig.~\ref{Fig.4.5.CF}), one quickly
notices that the charmonium states below 3.9 GeV (or the thresholds
of the charmed meson pair) can be produced reasonably well by the GI
model. However, many $XYZ$ states above 3.9 GeV can not be simply
categorized into the charmonium family, especially when they are
close to the thresholds of the charmed meson pair, where the
coupled-channel effects become important.

\paragraph{Coupled channel effects}

Kalashnikova employed the simplest version of the coupled-channel
model to study the charmonium mass spectrum
\cite{Kalashnikova:2005ui}. First he calculated the mass of the bare
$2^3P_1$ state to be 4180 MeV, using a simple potential model with
Hamiltonian $H_0=p^2/m_c+V(r)+C$ with $V(r)=\delta
r-4\alpha_s/(3r)$. Then he considered the coupling of this state to
the $D \bar D^*$ channel. Together with the $1^{++}$ resonance with
the mass of 3990 MeV, a near-threshold virtual state with the width
about 0.3 MeV was generated, and may be identified with the
$X(3872)$.

In Ref. \cite{Zhang:2009bv}, Zhang, Meng, Zheng used the
coupled-channel Flatt\'e formula to perform an analysis of the Belle
data on the $X(3872)$~\cite{Adachi:2008sua}. They found the co-existence
of two poles: one is a sheet II (or sheet IV) pole very close to the
$D^{*0} \bar D^0$ threshold, and the other one is a sheet III pole below
the $D^{*0} \bar D^0$ threshold. They pointed out that the $X(3872)$ can
be a conventional $2^3P_1$ $c\bar c$ state strongly polluted by
coupled channel effects since there exist two poles around the
$D^0\bar{D}^{*0}$ threshold. They also analyzed the data from
BaBar~\cite{Aubert:2007rva}. But as suggested in
Ref.~\cite{Kalashnikova:2009gt} by Kalashnikova and Nefediev, the
description was not very satisfactory, reflecting incompatibility of
the Belle and BaBar data.

In Ref.~\cite{Kalashnikova:2009gt}, Kalashnikova and Nefediev
analyzed the same data from Belle~\cite{Adachi:2008sua} and
BaBar~\cite{Aubert:2007rva}. They found that the BaBar
data~\cite{Aubert:2007rva} is more compatible with the assumption
of the $X(3872)$ being a virtual state of a dynamical nature in the
$D \bar D^*$ system, and the charmonium admixture is small. In
contrast, they found that the Belle data~\cite{Adachi:2008sua}
clearly indicated a sizeable $c\bar c$ $2^3P_1$ component in the
$X(3872)$ wave function, which conclusion is similar to that in
Ref.~\cite{Zhang:2009bv}. The conclusion in Ref. \cite{Zhang:2009bv}
was partly supported by the coupled-channel analysis of the $X(3872)$ in
Ref. \cite{Danilkin:2010cc}, where Danilkin and Simonov adopted a
coupled-channel model developed in Ref. \cite{Danilkin:2009hr} to
carry out a pole analysis and the calculation of the $D\bar{D}^*$
production cross section. They obtained a sharp peak structure at
the $D^0D^{*0}$ threshold, where the original position of 3954 MeV
for the $2^3P_1$ $c\bar{c}$ state was shifted by the coupled-channel
effect from the $D\bar{D}^*$ channel \cite{Danilkin:2010cc}.

In Ref. \cite{Li:2009zu}, Li and Chao studied the higher charmonia
with the screened potential. One notes that the screened potential
and the coupled-channel model roughly play the same role in
lowering the mass of the $\chi_{c1}^\prime(2P)$ charmonium state,
which can reach 3901 MeV and is close to the mass of the
$X(3872)$~\cite{Li:2009ad}.

The nature of the $X(3872)$ enhancement was analyzed in the
framework of the resonance-spectrum expansion in
Ref.~\cite{Coito:2010if} by Coito, Rupp, and Beveren. They studied
the $X(3872)$ as a regular $J^{PC} = 1^{++}$ charmonium state,
though strongly influenced and shifted by the open-charm decay
channels, and found a very delicate interplay among the $D^0\bar
D^{*0}$, $\rho^0J/\psi$, and $\omega J/\psi$ channels. Their results
suggested that the $X(3872$) is a very narrow axial-vector $c \bar
c$ resonance, with a pole at or slightly below the $D^0\bar D^{*0}$
threshold. Later in Ref.~\cite{Coito:2012vf}, Coito, Rupp, and
Beveren studied the $X(3872)$ as a confined $^3P_1$ $c\bar c$ state
coupling to the almost unbound $S$-wave $D^0 \bar D^{*0}$ channel
via the $^3P_0$ mechanism. They calculated the two-component wave
function for different values of the binding energy and the
transition radius $a$, and found a significant $c \bar c$ component.
In the case of a small binding energy of 0.16 MeV and $a$ between 2
and 3 GeV$^{-1}$, the $c\bar c$ probability can be strongly limited
to be roughly around 7-11\%. Then the X(3872) r.m.s. radius and the
$S$-wave $D^0\bar D^{*0}$ scattering length are 7.8 fm and 11.6 fm,
respectively. Hence, they concluded that the $X(3872)$ is not a
genuine meson-meson molecule, nor actually any other mesonic system
with non-exotic quantum numbers, due to the inevitable mixing with
the corresponding quark-antiquark states.

\subsubsection{Lattice QCD}

In Ref.~\cite{Chiu:2006hd}, the TWQCD Collaboration used a molecular
type operator composed of the $D$ and $\bar{D}^*$, $(\bar q \gamma_i
c)(\bar c \gamma_5 q) - (\bar c \gamma_i q)(\bar q \gamma_5 c)$, and
detected a $1^{++}$ resonance with a mass around $3890 \pm 30$ MeV
in quenched lattice QCD simulation with exact chiral symmetry, which
was identified as the $X(3872)$. They also used a
diquark-antidiquark operator, $(q^T C \gamma_i c)(\bar q C \gamma_5
\bar c^T) - (\bar q^T C \gamma_i \bar c)(q C \gamma_5 c^T)$, and
detected the same resonance. Later in Ref.~\cite{Chiu:2006us}, this
study was extended to the mass spectrum of the $1^+$ exotic mesons
with quark content $(cs\bar c\bar q)/(cq\bar c\bar s)$, and they
detected a $1^+$ resonance with mass around $4010 \pm 50$ MeV.

The lattice QCD simulation was also applied to study the $X(3872)$
as a charmonium state. In Ref.~\cite{Liu:2012ze} the Hadron Spectrum
Collaboration studied the highly excited charmonium mesons up to
around 4.5 GeV using dynamical QCD configurations. They found that
the $D$-wave $2^{-+}$ charmonium state is around 30 MeV below the
$X(3872)$, while the first radial excitation of the P-wave $1^{++}$
state is around 110 MeV above the $X(3872)$. In
Ref.~\cite{Yang:2012mya} the CLQCD Collaboration also studied the
$2^{-+}$ charmonium in quenched lattice QCD, and its mass was
determined to be $3.80\pm0.03$ GeV, which is close to the mass of
the $D$-wave charmonium $\psi(3770)$ and in agreement with quark
model predictions, but again significantly smaller than the mass of
the $X(3872)$.

Many dynamical studies suggested that the $D$ and $D^*$ interaction
is strongly attractive
\cite{Swanson:2003tb,Wong:2003xk,Suzuki:2005ha,Liu:2008fh,Thomas:2008ja,Liu:2008tn,Lee:2009hy,Sun:2012zzd,Li:2012cs,Zhao:2014gqa}.
Hence, a pure $\chi_{c1}^\prime(2P)$ with $J^{PC}=1^{++}$ can easily
couple to the $S$-wave $D\bar{D}^*$ scattering state if their masses
are similar. Moreover, their mixture can result in a small mass
\cite{Suzuki:2005ha,Meng:2005er}, which value can be significantly
lower than the mass of the pure $\chi_{c1}^\prime(2P)$ state
predicted in the GI model~\cite{Godfrey:1985xj}, but close to the
mass of the $X(3872)$.

This picture was supported by the lattice QCD
calculations~\cite{Prelovsek:2013cra,Padmanath:2015era}. In
Ref.~\cite{Prelovsek:2013cra}, Prelovsek and Leskovec found a
candidate for the $X(3872)$ using dynamical $N_f = 2$ lattice
simulation with $J^{PC} = 1^{++}$ and $I = 0$, in addition to the
nearby $D \bar D^*$ and $J/\psi \omega$ discrete scattering states.
In their simulation, they chose the interpolating fields that couple
to $\bar cc$ as well as the scattering states, i.e.,
$\mathcal{O}^{\bar c c}$, $\mathcal{O}^{D D^*}$,
$\mathcal{O}^{J/\psi \omega}$ (for $I=0$), and $\mathcal{O}^{J/\psi
\rho}$ (for $I=1$). They extracted large and negative $D \bar D^*$
scattering length, $a^{DD^*}_0 = -1.7 \pm 0.4$ fm, and the effective
range, $r^{DD^*}_0 = 0.5\pm0.1$ fm. They did not find a candidate
for the $X(3872)$ in the $I =1$ channel, which may be due to the
exact isospin symmetry in their simulation.

\begin{figure}[htbp]
\begin{center}
\begin{tabular}{c}
\scalebox{0.73}{\includegraphics{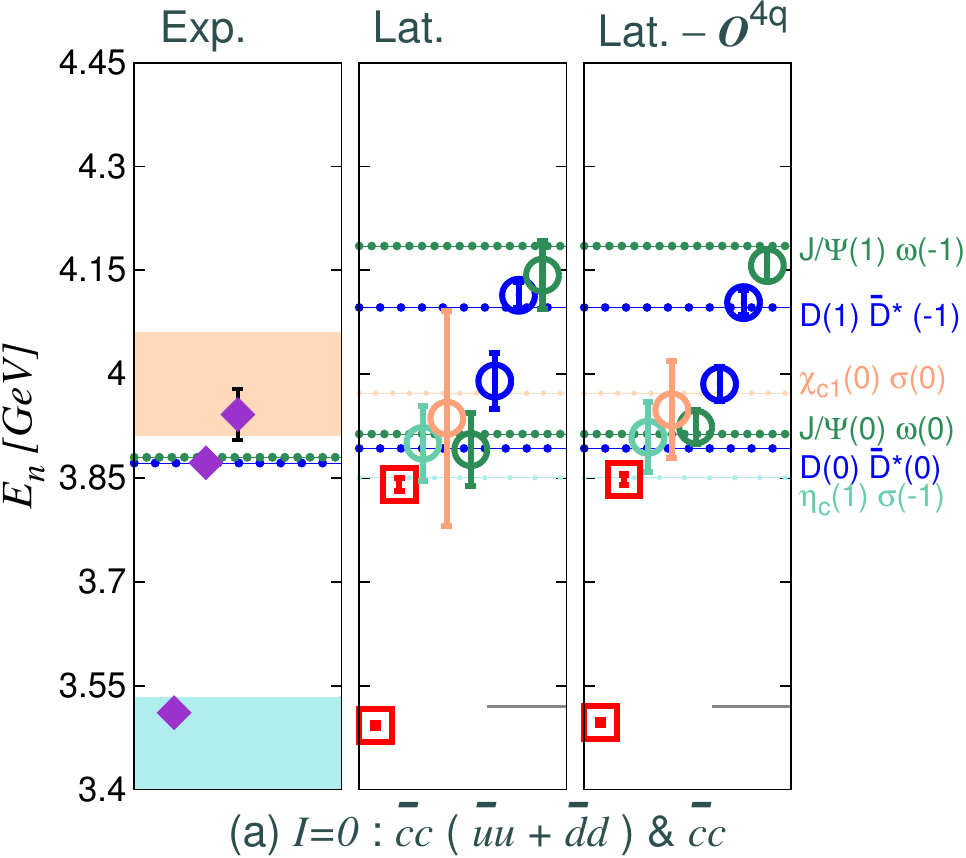}}
\scalebox{0.73}{\includegraphics{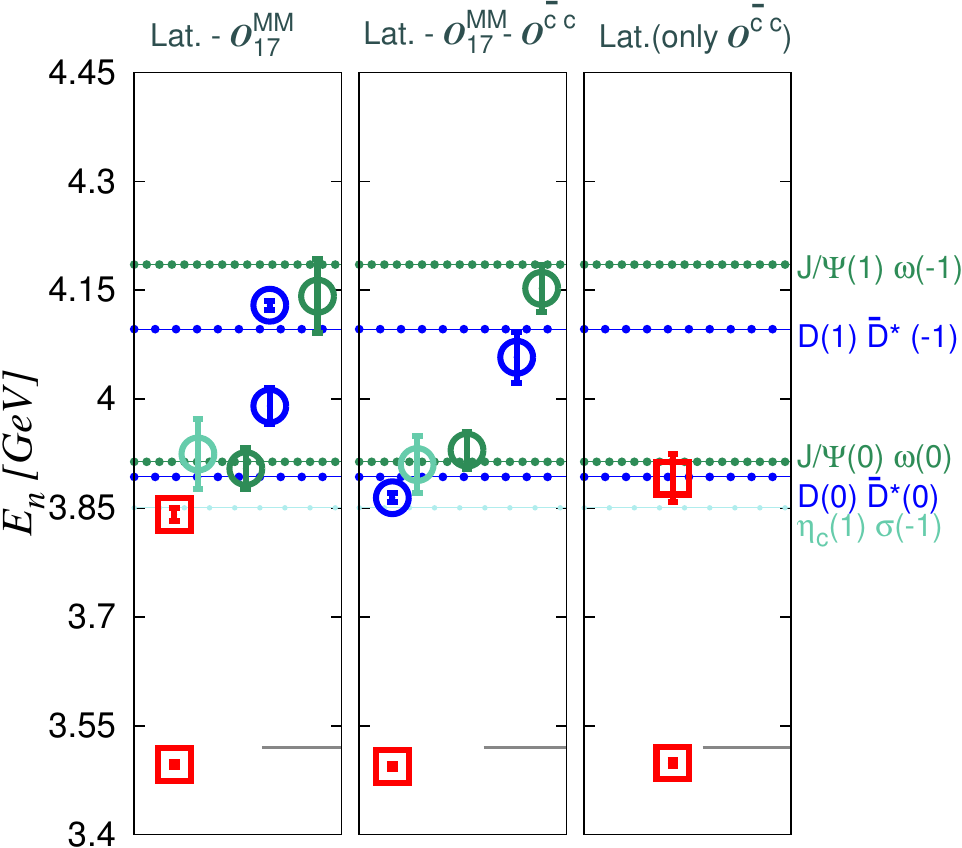}}
\end{tabular} \caption{(Color online) The spectra of states with $J^{PC}=1^{++}$ for the cases with $u/d$ valence quarks, taken from Ref.~\cite{Padmanath:2015era}.
From the left to the right: a) physical thresholds and possible experimental candidates, including the $\chi_{c1}$, $X(3872)$, and $X(3940)$;
b) the discrete spectrum determined from the optimized basis of Ref.~\cite{Padmanath:2015era}, including the
$\bar cc$, two-meson and diquark-antidiquark operators;
c) the spectrum obtained from the optimized basis, without the two-meson operators;
d) the spectrum obtained from the optimized basis, without one of the two-meson operators, $O^{MM}_{17}$;
e) the spectrum obtained from the optimized basis, without $\bar cc$ and $O^{MM}_{17}$ operators;
f) the spectrum obtained with only $\bar cc$ operators.
\label{Fig.4.5.lattice}}
\end{center}
\end{figure}

This study was extended in Ref.~\cite{Padmanath:2015era} by
Padmanath, Lang, and Prelovsek, where a large basis of interpolating
fields was utilized, including $\bar cc$, two-meson and
diquark-antidiquark ones. The obtained discrete spectrum are shown
in Fig.~\ref{Fig.4.5.lattice}. Again they found a lattice candidate
for the $X(3872)$ with $J^{PC} = 1^{++}$ and $I=0$, but only if both
$\bar cc$ and $D \bar D^*$ interpolators are included. However, this
candidate can not be found if the diquark-antidiquark and $D \bar
D^*$ are used in the absence of $\bar cc$. Moreover, no candidate
for the neutral or charged $X(3872)$, or any other exotic candidates
are found in the $I = 1$ channel, and no signatures of the exotic
$\bar cc\bar ss$ candidates are found below 4.2 GeV. In other words,
the most recent dynamical lattice QCD simulation strongly disfavors
either the diquark-antidiquark or various four-quark
interpretations of the $X(3872)$.

This dynamical lattice QCD simulation was performed with $N_f=2$ and
$m_\pi=266$ MeV. The pion mass on the lattice is still much larger
than its physical mass 140 MeV. Within the molecular scheme, the
long range one-pion-exchange force plays a dominant role in the
formation of the loosely bound molecular state
\cite{Li:2012cs,Zhao:2014gqa}, which decays exponentially as the
pion mass increases. The present lattice simulation with the pion
mass $m_\pi=266$ MeV is still unable to explore and judge whether
the $X(3872)$ is a molecular state or not.

\subsubsection{Other theoretical schemes, production and decay patterns}
\label{sect:4.5.4}

\paragraph{Other schemes}

There exist some other theoretical schemes. The $X(3872)$ was
interpreted as a $c \bar c g$ hybrid state~\cite{Li:2004sta}, a
vector glueball mixed with the neighboring vector
charmonium~\cite{Seth:2004zb}, and a dynamically generated mixed
state of a $DD^*$ molecule and $\chi_{c1}(2P)$~\cite{Ortega:2009hj}
etc. Whether the $X(3872)$ is due to the cusp effect and threshold
effect was discussed in
Refs.~\cite{Bugg:2004rk,Bugg:2004sh,Rosner:2006vc,Bugg:2008wu,Guo:2014iya,Blitz:2015nra}.
Since the $X(3872)$ is very close to the $D^{*0}\bar D^0$ threshold,
the rescattering effects of the $D$ and $D^*$ mesons were studied in
Refs.~\cite{Voloshin:2007hh,Dunwoodie:2007be,Canham:2009zq,Guo:2010ak,Zhou:2013ada},
and the influence of thresholds was studied in
Ref.~\cite{Carames:2012th}. Constraints from precision measurements
on the hadron-molecule interpretation of $XYZ$ resonances were
discussed in Ref.~\cite{Polosa:2015tra}.

The chiral unitary approach in coupled channels was applied to study
the $X(3872)$ in
Refs.~\cite{Gamermann:2007fi,Gamermann:2007mu,Gamermann:2009fv,Gamermann:2009uq,Gamermann:2010ga},
where they found that the charged components of the $D\bar D^*$ play an important role in
describing the ratio of its decay to $J/\psi \rho$ and $J/\psi \omega$.
The $X(3872)$ was also investigated using the AdS/QCD in
Ref.~\cite{Carlucci:2007um}, and a simple string model in
Ref.~\cite{Vijande:2007ix}. An effective field theory, called XEFT,
was proposed in
Refs.~\cite{Braaten:2010mg,Fleming:2011xa,Jansen:2013cba,Braaten:2015tga,Jansen:2015lha}
to study the $X(3872)$ as a loosely-bound charm-meson molecule. A
dynamical picture to explain the nature of the exotic $XYZ$ states
was proposed in Ref.~\cite{Brodsky:2014xia} based on a
diquark-antidiquark open-string configuration, while the three-body
$D \bar D \pi$ dynamics for the $X(3872)$ was investigated in
Refs.~\cite{Braaten:2007ct,Baru:2011rs}. The selection rules for
hadronic transitions between $Q\bar Q$ mesons were investigated in
Refs.~\cite{Braaten:2014ita,Braaten:2014qka} based on
Born-Oppenheimer potentials, and the Born-Oppenheimer approximation
was also used in Ref.~\cite{Liu:2014dla} to study the four-quark
bound states.

Heavy quark spin selection rule and power counting schemes were
investigated in
Refs.~\cite{Voloshin:2004mh,Valderrama:2012jv,Liu:2013rxa,Alhakami:2015uea}.
Line shapes of the $X(3872)$ were studied in
Refs.~\cite{Braaten:2005jj,Braaten:2007dw,Braaten:2007ft,Artoisenet:2010va,Chen:2013upa}.
The lattice QCD simulation and related studies can be found in
Refs.~\cite{Albaladejo:2013aka,Baru:2013rta,Garzon:2013uwa,Kalinowski:2015bwa}.
The analogous states of the $X(3872)$, such as $X_b$ involving $b$
quarks, were investigated in
Refs.~\cite{Hou:2006it,Gerasyuta:2009pc,Nieves:2012tt,Hogaasen:2013nca,Hidalgo-Duque:2013pva,Guo:2014sca,Li:2014uia,Guo:2014ura,Karliner:2014lta,Li:2015uwa,Albaladejo:2015dsa}.

\paragraph{Decay}

Voloshin pointed out in Ref.~\cite{Voloshin:2003nt} that the
internal structure of the $X(3872)$ can be studied by measuring the
rate and the spectra in the decays $X(3872) \to D^0 \bar D^0 \pi^0$
and $X(3872) \to D^0 \bar D^0 \gamma$. The hadronic transitions from
the $X(3872)$ to $\chi_{cJ}$ were investigated in
Refs.~\cite{Dubynskiy:2007tj,Fleming:2008yn,Mehen:2015efa,Braaten:2013poa},
which can also be used to test different theoretical proposals
related to the $X(3872)$. Later, various methods/models were applied to
study the radiative transitions of the $X(3872)$ such as $\psi(4160) \to
\gamma X(3872)$, $X(3872) \to \gamma J/\psi(\psi^\prime)$, $X(3872)
\to D^0 \bar D^0 \gamma$~\cite{AbdEl-Hady:2006nbt,Colangelo:2007ph,Dong:2008gb,DeFazio:2008xq,Wang:2010ej,Jia:2010jn,Harada:2010bs,Mehen:2011ds,Dubnicka:2011mm},
which may play a fundamental role in the determination of the nature
of the $X(3872)$.

The isospin-violating branching fraction observed by
Belle~\cite{Abe:2005ix} and BaBar~\cite{delAmoSanchez:2010jr}
experiments (see Eqs. (\ref{Eq.2.1.3872iso1})-(\ref{Eq.2.1.3872iso2})) are very
interesting. The related two-pion and three-pion decays, $X(3872)
\to J/\psi \rho (\to \pi^+ \pi^-)$ and $X(3872) \to J/\psi \omega
(\to \pi^+ \pi^- \pi^0)$, were studied in
Refs.~\cite{Pakvasa:2003ea,Kim:2004cz,Rosner:2004ac,Braaten:2005ai,Melikhov:2006ec,Terasaki:2009in,Hanhart:2011tn}.
The $D^0 \bar D^0 \pi^0$ mode was studied in Refs.~\cite{Hanhart:2007yq,Stapleton:2009ey,Dubnicka:2010kz,Fleming:2007rp}.
The $D\bar{D}^*$ molecular state assignment to the $X(3872)$ not only
answers why the $X(3872)$ is close to the $D^0D^{*0}$ threshold, but
also explains its isospin violating in the $J/\psi \rho$ decay
mode~\cite{Swanson:2003tb,Swanson:2004pp,Li:2012cs}.

The ratio of $\mathcal{B}(X(3872)\to \psi^\prime \gamma)$ to
$\mathcal{B}(X(3872)\to J/\psi \gamma)$ was measured to be $3.4 \pm
1.4$~\cite{Aubert:2008ae} and $2.46 \pm 0.64 \pm
0.29$~\cite{Aaij:2014ala}. Under the molecule picture, several
groups studied the above ratio~\cite{Dong:2009uf,Guo:2014taa}.
In Refs.~\cite{Dong:2009yp,Dong:2009uf}, Dong {\it et al.} studied
the $\gamma J/\psi$ and $\gamma \psi^\prime$ decay modes of the
$X(3872)$ using a phenomenological Lagrangian approach. They noticed
that a nontrivial interplay between a possible charmonium and the
molecular components in the $X(3872)$ can explain the ratio of
$\mathcal{B}(X(3872) \to \psi^\prime \gamma)$ with respect to
$\mathcal{B}(X(3872) \to J/\psi \gamma)$. In
Ref.~\cite{Guo:2014taa}, Guo {\it et al.} studied the radiative
decays of the $X(3872)$ into $\gamma J/\psi$ and $\gamma
\psi^\prime$ using an effective field theory. Their results also
suggested that their experimental
ratio~\cite{Aubert:2008ae,Aaij:2014ala} is not in conflict with the
hadronic molecular picture that the $X(3872)$ is dominated by the $D\bar
D^*$ component.

The ratio of ${\cal{B}}(X(3872)\to \gamma J/\psi)$ with respect to
${\cal{B}}(X(3872)\to J/\psi\pi^+\pi^-)$ was measured to be $0.14
\pm 0.05$~\cite{Abe:2005ix} and $0.33 \pm
0.12$~\cite{Aubert:2008ae}. This ratio also encodes important
information on the underlying structure of the $X(3872)$, which was
reexamined using a phenomenological Lagrangian approach in
Refs.~\cite{Dong:2009yp,Dong:2009uf}. This ratio was explained using
the molecular components in the $X(3872)$. Moreover, the authors
noticed that this ratio suggests that the $c\bar c$ component plays
a subleading role only~\cite{Dong:2009yp,Dong:2009uf}.
The $D\bar D^*$ picture was also used in Ref.~\cite{Aceti:2012cb} to evaluate
this radiative decay together with the $J/\psi \rho$ and $J/\psi \omega$
decay channels. The ratios were found to be compatible with experiment
and the relevance of the charged $D\bar D^*$ components was assessed
to obtain the correct ratios.

Li and Chao considered the $X(3872)$ as a $\chi_{c1}^\prime(2P)$
charmonium with the mixture of the $D\bar{D}^{*}$
channel~\cite{Li:2009zu}. The isospin violating decay process,
$X(3872)\to J/\psi \rho (\to \pi^+\pi^-)$, can happen through final
state interactions with the intermediate $D\bar{D}^{*}$ loop
\cite{Liu:2006df}. With this mechanism, the ratio $R_{\rho/\omega}
\equiv {\Gamma(X(3872)\to J/\psi \rho) \over \Gamma(X(3872)\to
J/\psi \omega)}\simeq 1$ \cite{Meng:2007cx}. With the screening
potential, the ratio of $\mathcal{B}(X(3872)\to \psi^\prime \gamma)$
to $\mathcal{B}(X(3872)\to J/\psi \gamma)$ was
$1.3-6.0$~\cite{Li:2009zu}, which is consistent with the present
experimental
data \cite{Aubert:2008ae,Aaij:2014ala}.

\paragraph{Production}

The production of the $X(3872)$ in $B$ meson decays was also studied in
Refs.~\cite{Braaten:2004jg,Braaten:2004rw,Braaten:2004fk,Wang:2007fs,Wang:2007sxa}.
Its production in the charmonia radiative decays was studied in
Ref.~\cite{Guo:2013nza}. Its production in $e^+ e^-$ annihilations
was studied in
Refs.~\cite{Dubynskiy:2006cj,Li:2013nna,Denig:2014fha}. Its
production in high energy heavy ion collisions was studied in
Ref.~\cite{Torres:2014fxa}. Its hadronic effects in heavy ion
collisions were studied in Ref.~\cite{Cho:2013rpa}. Its production
at PANDA was studied in Refs.~\cite{Chen:2008cg,Larionov:2015nea}.
Its production at the Tevatron and LHC was studied in
Refs.~\cite{Artoisenet:2009wk,Bignamini:2009sk,Guerrieri:2014gfa}. Its production at
CDF was studied in Ref.~\cite{Bignamini:2009fn}.

Braaten and Kusunoki \cite{Braaten:2004ai} found that the branching
ratio of $B^0\to X(3872)K^0$ is one order of magnitude smaller than
that of $B^+\to X(3872)K^+$ assuming the $X(3872)$ as a $D\bar{D}^*$
molecular state. However, the ratio of ${\cal{B}}(B^0 \to K^0
X(3872))$ to ${\cal{B}}(B^+ \to K^+ X(3872))$ was measured to be
$0.82 \pm 0.22 \pm 0.05$~\cite{Adachi:2008te}, $1.26 \pm 0.65 \pm
0.06$~\cite{Adachi:2008sua}, and $0.50 \pm 0.14 \pm
0.04$~\cite{Choi:2011fc}. This difference was reexamined in their
later work~\cite{Braaten:2007ft}, which investigated line shapes of
the $X(3872)$. They pointed out that the prediction of
Ref.~\cite{Braaten:2004ai} was based on the current-current
approximation and heavy quark symmetry, and a conceptual error was
identified as the implicit assumption that the scattering parameters
$\gamma_0$ and $\gamma_1$ are small compared to $\kappa_1(0)$.
Actually, they suggested that $\gamma_0$ and $\gamma_1$ could be
determined phenomenologically from ratios of rates for $B^0 \to K^0
X(3872)$ and $B^+ \to K^+ X(3872)$.

%%%%%%%%%%%%%%%%%
\begin{table}[htb]
\caption{Integrated cross sections for $pp/\bar p\to X(3872)$, in units of $nb$.
The results of Ref.~\cite{Guo:2014sca} were obtained using Herwig and Pythia,
which are written outside and inside brackets, respectively.}
\label{Tab.4.5.X3872compare}
\begin{center}
\begin{tabular}{cccccc}
\toprule[1pt]
\multirow{2}{*}{$\sigma(pp/p\bar p\to X(3872))$} & \multirow{2}{*}{Experiment} & \multirow{2}{*}{Ref.~\cite{Bignamini:2009sk}} & \multirow{2}{*}{Ref.~\cite{Artoisenet:2009wk}} & Ref.~\cite{Guo:2014sca}  & Ref.~\cite{Guo:2014sca}
\\ & & & & with {$\Lambda=0.5$ GeV} & with {$\Lambda=1$ GeV} \\\hline
%  & &  &  &$\Lambda=1$ GeV & \\\hline
Tevatron & 37--115~\cite{Bauer:2004bc,Guo:2014sca}
& $<0.085$& $1.5$--$23$ &10(7) & 47(33) \\
LHC with $\sqrt{s} = 7$ TeV & 13--39~\cite{Chatrchyan:2013cld,Guo:2014sca}
& -- & 45--100 & 16(7) & 72(32) \\
\bottomrule[1pt]
\end{tabular}
\end{center}
\end{table}

In Ref.~\cite{Guo:2014sca}, Guo, Meissner, Wang, and Yang used the
Monte Carlo event generator tools Pythia and Herwig to simulate the
production of bottom/charm meson and antimeson pairs at hadron
colliders in proton-proton/antiproton collisions, and then derived
an order-of-magnitude estimate for the production cross sections of
the $X(3872)$ as a $D \bar D^*$ molecular state at the LHC and
Tevatron experiments. Their results are consistent with the
experimental measurement by the CDF~\cite{Bauer:2004bc}
and CMS~\cite{Chatrchyan:2013cld} collaborations, which are
shown in Table~\ref{Tab.4.5.X3872compare} together with the
predictions of Refs.~\cite{Bignamini:2009sk,Artoisenet:2009wk}.
They also simulated the production of
the bottom analogues and the spin partner of the $X(3872)$,
including $X_b$ of $1^{++}$, and $X_{b2}$ and $X_{c2}$ of $2^{++}$.
They found that the cross sections are at the $nb$ level for the
$X_b$ and $X_{b2}$, which are two orders of magnitude larger than
that for the $X_{c2}$. They also proposed a search for these states
at the Tevatron and LHC.

In Ref.~\cite{Meng:2005er}, Meng, Gao, and Chao treated charmonia as
nonrelativistic bound states in QCD factorization and obtained
${\cal{B}}(B^0 \to K^0 X(3872)) = {\cal{B}}(B^+ \to K^+ X(3872))
\approx 2 \times 10^{-4}$, which might imply that the $X(3872)$
contains a dominant $J^{PC} = 1^{++}(2P)$ $c\bar c$ component and a
substantial $D^0 \bar D^{*0}$ continuum component.

\begin{figure}[htbp]
\begin{center}
\begin{tabular}{c}
\scalebox{0.73}{\includegraphics{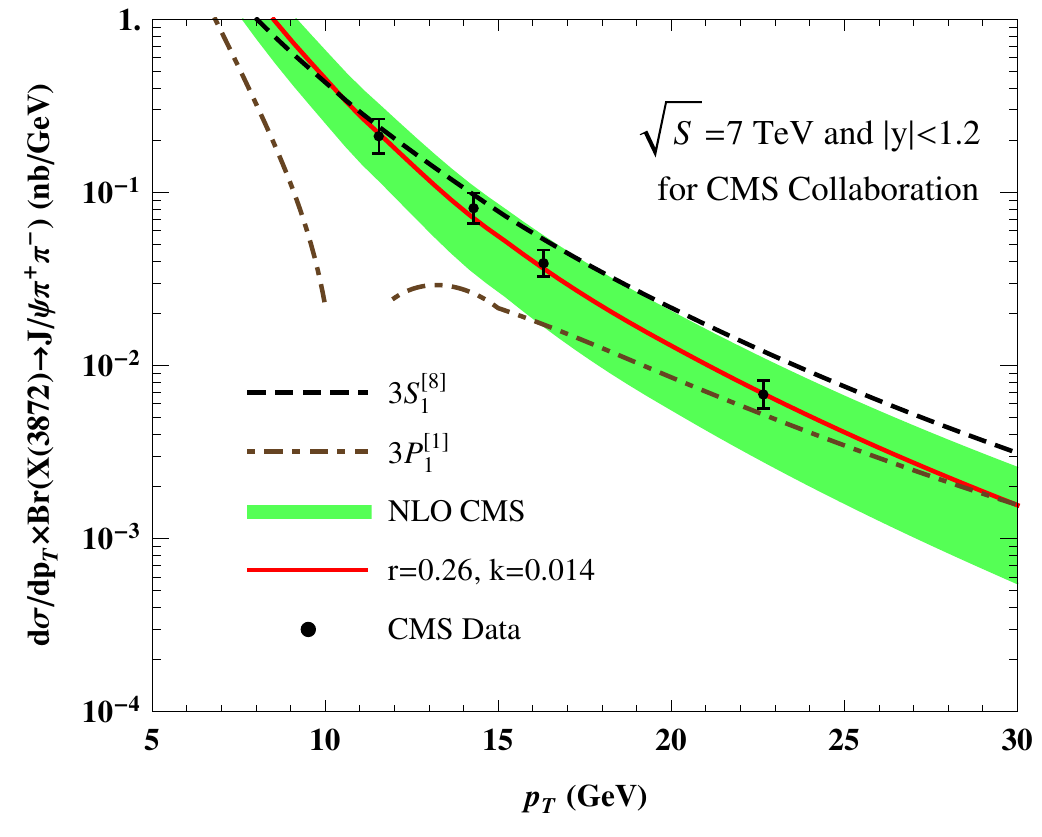}}
\end{tabular} \caption{(Color online) The fit of the CMS $p_T$ distribution data~\cite{Chatrchyan:2013cld}. Taken from Ref.~\cite{Meng:2013gga}.
\label{Fig.4.5.diagrams}}
\end{center}
\end{figure}

Especially, the production rate of the $X(3872)$ is comparable to that
of $\psi'$ at hadron
colliders~\cite{Acosta:2003zx,Bignamini:2009sk,Chatrchyan:2013cld},
which may
also imply a $c\bar c$ core within the $X(3872)$. In
Ref.~\cite{Meng:2013gga}, Meng, Han, and Chao assumed that the short
distance production of the $X(3872)$ proceeds dominantly through its
$\chi_{c1}^\prime$ component, and evaluated its production cross
sections at Tevatron and LHC at NLO in $\alpha_s$ within the
framework of NRQCD factorization. The results are shown in
Fig.~\ref{Fig.4.5.diagrams}. They fit the CMS experimental
data~\cite{Chatrchyan:2013cld} and obtained the ratio $r = m_c^2 \langle
\mathcal{O}^{\chi_{c1}^\prime}(^3S_1^{[8]}) \rangle / \langle
\mathcal{O}^{\chi_{c1}^\prime}(^3P_1^{[1]}) \rangle = 0.26 \pm
0.07$, which is almost the same with that for
$\chi_{c1}$~\cite{Ma:2010vd}. With this input, the authors were able
to account for the CDF data~\cite{Acosta:2003zx,Bignamini:2009sk}.
The fit of the production cross section of the $X(3872)$ at hadron
colliders leads to the same value of $k = Z_{c\bar c} \cdot
\mathcal{B}(X(3872) \to J/\psi \pi^+\pi^-)$ constrained by the $B$
meson decay data~\cite{pdg}.

In Ref.~\cite{Butenschoen:2013pxa}, Butenschoen, He, and Kniehl
studied the inclusive hadroproduction of the $\chi_{c1}(2P)$ within
the factorization formalism of nonrelativistic QCD at the
next-to-leading order. They tested the hypothesis that the $X(3872)$
is a pure $\chi_{c1}(2P)$ charmonium state using the data from the
CDF~\cite{Acosta:2003zx,Bauer:2004bc},
CMS~\cite{Chatrchyan:2013cld}, and LHCb~\cite{Aaij:2011sn}
collaborations. The authors concluded that NLO NRQCD is inconsistent
with the hypothesis $X(3872) \equiv \chi_{c1}(2P)$, because they
either obtained an unacceptably high value of $\chi^2$, a value of
$|R^\prime_{2P}(0)|$ incompatible with well-established potential
models, or an intolerable violation of the NRQCD velocity rules.

\subsubsection{A short summary}
\label{Sect:4.5.5}

The $X(3872)$ is the first observed charmonium-like state in the
$XYZ$ family. It's very interesting to quote the prediction which
Swanson wrote in 2003~\cite{Swanson:2003tb}: ``Thus the discovery
of the $X(3872)$ may be the entr\'ee into a new regime of hadronic
physics which will offer important insight into the workings of
strong QCD and should help clarify many open issues in light quark
spectroscopy.''

\begin{itemize}
\item
The dynamical lattice QCD simulation with hidden-charm tetraquark
operators and $m_\pi=266$ MeV is unable to reproduce the $X(3872)$
signal on the lattice~\cite{Padmanath:2015era}. Moreover, if it is
a tetraquark state, the $X(3872)$ will always be accompanied with
several charged partners having similar masses, which have not been
observed in the B meson decays and other experiments.

\item
The $X(3872)$ is extremely close to the $D^0D^{*0}$ threshold. Its
$J/\psi\rho$ decay mode is isospin violating. All these features can
be naturally explained by the molecular assignment of the $X(3872)$
as an S-wave $D \bar D^*$ bound state. In fact, there exists enough
attraction in the isoscalar $D\bar D^*$ system to form a shallow
bound state. Within the molecular scheme, both the decay and
production behaviors can be accounted for naturally.
The charged components of the $D \bar D^*$ play an
important role in describing the ratio of decay to $J/\psi \rho$ and 
$J/\psi \omega$~\cite{Gamermann:2007fi,Gamermann:2007mu,Gamermann:2009fv,Gamermann:2009uq,Gamermann:2010ga}.

\item
The existence of the $X(3872)$ as a shallow bound state is extremely
sensitive to the one-pion-exchange force, which decreases exponentially
as the pion mass increases. It will be very desirable to perform a
dynamical lattice QCD simulation with (1) the $D \bar D^*$
interpolators only, (2) explicit isospin violation and (3) the pion
mass around 140 MeV.

\item
The assignment of the $X(3872)$ as $\chi_{c1}'$ is also feasible if
there exist strong coupled channel effects between the bare $c\bar
c$ state in the quark model and the $\bar DD^{*}$ scattering state,
which helps to lower the mass of the bare $c\bar c$ state and
explain the isospin violating decay mode. In fact, a lattice
candidate for the $X(3872)$ with $J^{PC} = 1^{++}$ and $I=0$ was
found only if both $\bar cc$ and $D \bar D^*$ interpolators are
included~\cite{Padmanath:2015era}. As a mixture of $c\bar c$ and
$\bar DD^{*}$, the large $\psi' \gamma$ decay ratio of the $X(3872)$
and its large production rate at hadron colliders can be understood
easily.

\item
The extreme proximity of the $X(3872)$ to the $D \bar D^*$ threshold
requires a large $D \bar D^*$ component in its wave function.
Otherwise, such proximity seems too accidental to be convincing.

\item
If the $X(3872)$ turns out to be a mixture of the $J^{PC} =
1^{++}(2P)$ $c\bar c$ component and a substantial $D^0 \bar D^{*0}$
continuum component, one may expect the existence of another
$X(3872)$-like state. The flavor wave function of this state is
orthogonal to that of the $X(3872)$. Its mass may be higher than
3872 MeV and has the same quantum numbers as the $\chi_{c1}'$. This
state may be quite broad due to the existence of the open-charm
decay modes. In other words, the $X(3872)$ may be more molecule-like
while the other state is more $c\bar c$-like.

\item
The experimental identification of the $\chi_{c1}'$ state is
extremely important, which shall shed light not only on the
$X(3872)$ but also on the $c\bar c (2P)$ states.

\end{itemize}

\subsection{$Y(4260)$}\label{Sect:4.6}

The $Y(4260)$, which has been reviewed in Sec.~\ref{Sect:2.2.1},
also attracted great attentions from both experimentalists and
theorists. However, its nature is still controversial. In the
following, we mainly focus on several major theoretical aspects of
the $Y(4260)$, i.e., the hybrid charmonium, the vector tetraquark
state, the molecular state and non-resonant explanation.

\subsubsection{Is $Y(4260)$ a higher charmonium?}

Since the $Y(4260)$ is directly produced from the $e^+e^-$
annihilation process, its spin-parity quantum number must be
$J^{PC}= 1^{--}$, which is consistent with that of a vector
charmonium state. Thus, theorists tried to categorize it into the
vector charmonium family. Different pictures were proposed, such as
the $3D$ ($3^3D_1$) and the $4S$  ($4^3S_1$) charmonium states, and so on.

In Ref.~\cite{LlanesEstrada:2005hz}, Llanes-Estrada endorsed the
$Y(4260)$ as the $\psi(4260)$, corresponding to the $4S$ vector
charmonium state, where the S-D wave interference was used to
explain the lack of a signal in $e^+e^-$ annihilations. They also
suggested some avenues that can exclude exotic meson assignments. In
Ref.~\cite{Zhang:2006td}, Zhang studied the charmonium spectrum by
combining the linearity and parallelism of the Regge trajectories
with a hyperfine splitting relation in multiplet, and interpreted
the $Y(4260)$ as the $3^3D_1$ charmonium state. In
Ref.~\cite{Li:2009zu}, Li and Chao calculated the masses,
electromagnetic decays, and E1 transitions of charmonium states in
the screened potential model. In their model, the mass of the
$\psi(4S)$ was predicted to be 4273 MeV, which is roughly compatible
with the observed masses of the $Y(4260)$. In
Ref.~\cite{Shah:2012js}, Shah, Parmar, and Vinodkumar studied the
masses of the S-wave quarkonia based on the Martin-like potential.
They also found that the $Y(4260)$ can be interpreted as the $4S$
charmonium state.

However, in Ref.~\cite{Eichten:2005ga}, Eichten, Lane, and Quigg
refined the Cornell coupled-channel model for the coupling of the
$c\bar c$ levels to two-meson states in light of new experimental
information. Especially, they calculated the decay behavior of the
$2^3D_1$ charmonium state and excluded this assignment of the
$Y(4260)$. In Ref.~\cite{Segovia:2008zz}, Segovia, Yasser, Entem,
and Fernandez studied the energy spectrum, electromagnetic, and
strong decays of the $J^{PC}=1^{--}$ hidden charm resonances in a
constituent quark model,  in order to assert if they are $c \bar c$
states or more complicated structures. They found that the new
$Y(4360)$ state can be identified as the $4S$ state and the
$\psi(4415)$ as the $3D$ state. However, they found that the
$Y(4260)$ cannot be categorized into the charmonium family.

In Ref.~\cite{Dai:2012pb}, Dai, Shi, Tang, and Zheng studied the
property of the $Y(4260)$ resonance by re-analyzing the experimental
data till March 2015. They took into account the final state
interactions of the $\pi\pi$ and $K \bar K$ couple channels, and
found a sizable coupling between the $Y(4260)$ and the
$\omega\chi_{c0}$. They found two nearby poles in the $Y(4260)$
propagator, indicating that the $Y(4260)$ is most likely a confining
state. They argued that the small value of $\Gamma_{e^+e^-}$ is
consistent with the hybrid scenario, and also consistent with the
explanation that the $Y(4260)$ is the $3D$ charmonium state.
However, the difficulty of the $3D$ explanation comes from the role
of the $X(4160)$, which is considered as a good candidate of the
$3D$ charmonium state in quark model.

\begin{figure}[htbp]
\begin{center}
\scalebox{0.46}{\includegraphics{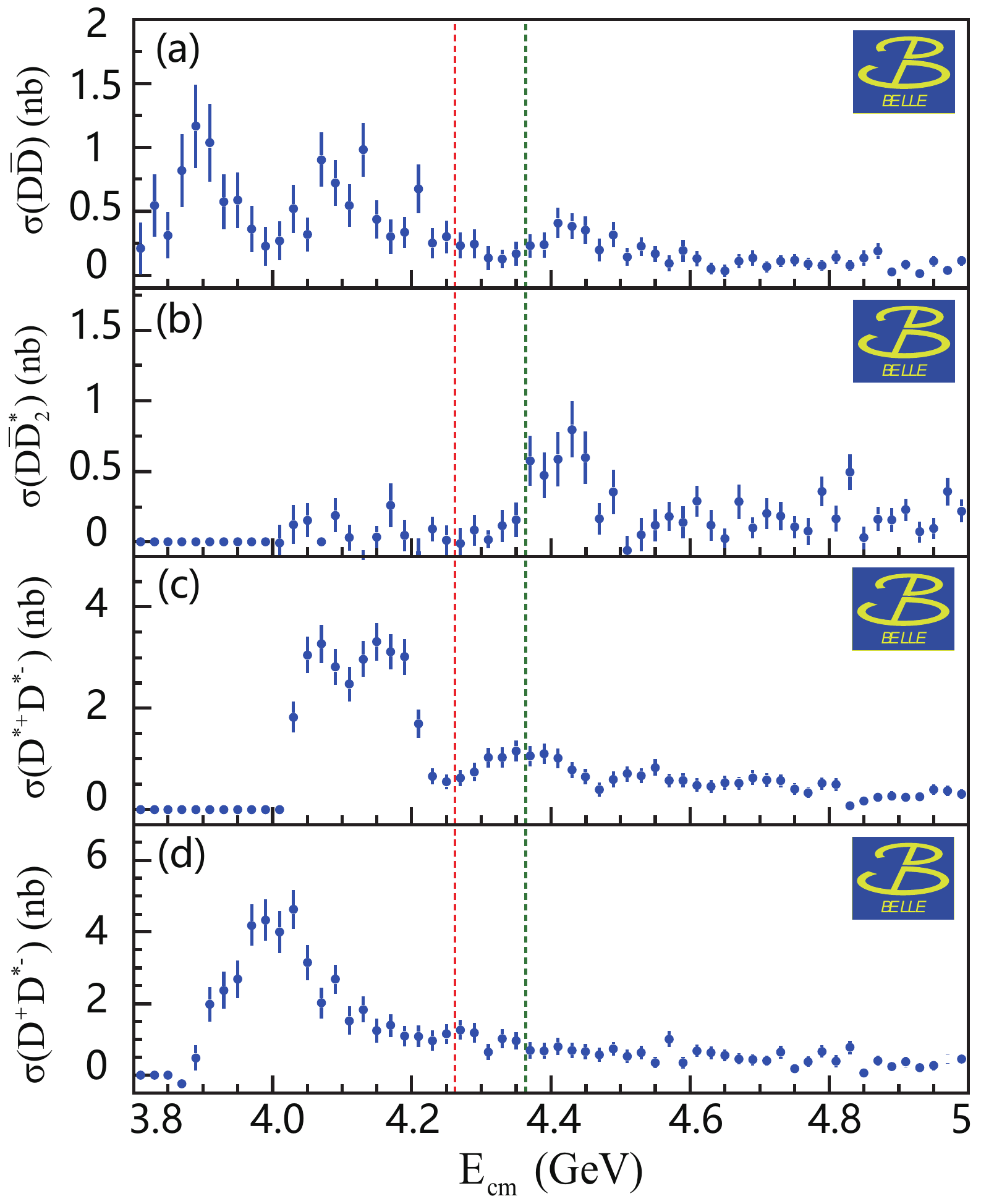}}
\caption{(Color online) The experimental data of the open-charm decay channels from
$e^+e^-$ annihilation. Here, (a) $e^+e^-\to D\bar{D}$
\cite{Pakhlova:2008zza}; (b) $e^+e^-\to D^0 D^- \pi^+$
\cite{Pakhlova:2007fq}; (c) $e^+e^-\to D^{*+}{D}^{*-}$
\cite{Abe:2006fj}; (d) $e^+e^-\to D^+ {D}^{*-}$ \cite{Abe:2006fj}.
The red and green dashed vertical lines correspond to the central
values of the masses of $Y(4260)$ and $Y(4360)$ \cite{pdg},
respectively. \label{Fig.4.6.open-charm-Y4260}}
\end{center}
\end{figure}

The resonant parameters of the $Y(4260)$ listed in Table~\ref{Table:2.2.Y4260} show that it has a large width. Under the
higher charmonium assignment, the open-charm decays of the $Y(4260)$ are
probably dominant. However, the $Y(4260)$ was only observed in its
hidden-charm decay mode $J/\psi\pi^+\pi^-$, but missing in any open-charm
decay mode (see Fig.~\ref{Fig.4.6.open-charm-Y4260}). In fact, the
non-observation of the $Y(4260)$ in the open-charm modes is challenging
for all theoretical interpretations.

\begin{figure}[htbp]
\begin{center}
\scalebox{0.58}{\includegraphics{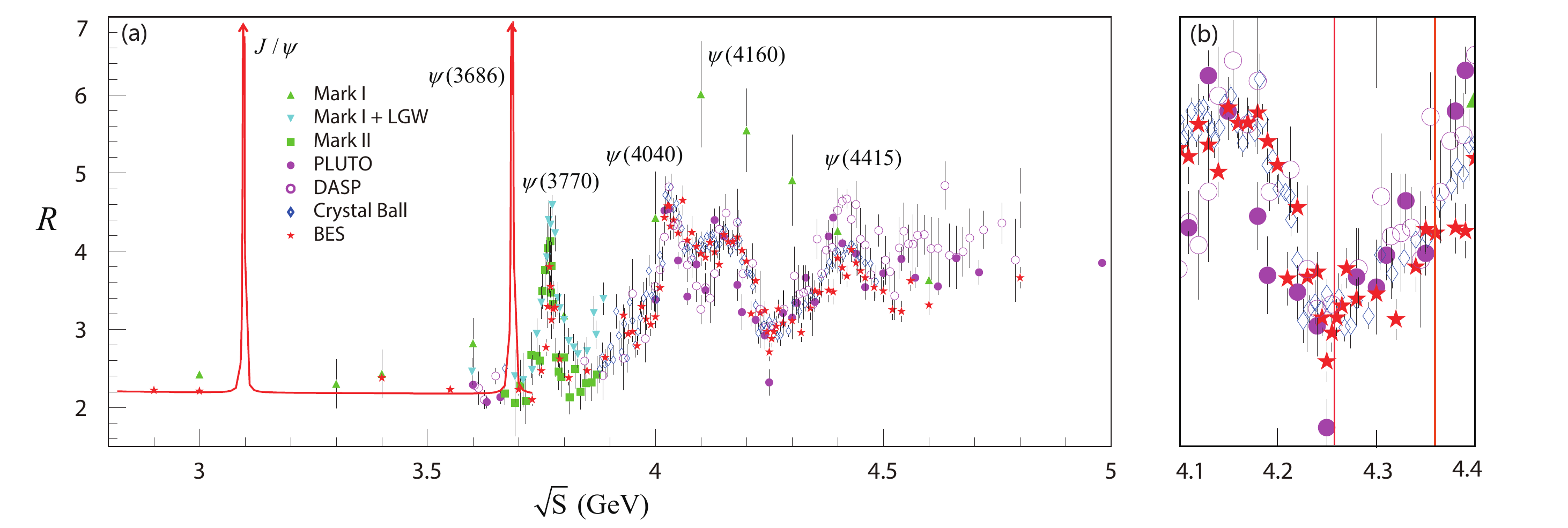}}
\caption{(Color online) (a) The
experimental data of $R$ value from PDG \cite{pdg}. (b) The detailed
data corresponding to the range $\sqrt{s}=4.1\sim 4.4$ GeV from PDG \cite{pdg}. Here,
the vertical lines correspond to the central values of the masses of
the $Y(4260)$ and $Y(4360)$ \cite{pdg}, respectively.
All the data here are taken from http://pdg.lbl.gov/current/xsect/.
\label{Fig.4.6.R-4260}}
\end{center}
\end{figure}

In addition, the $R$ value scan (the ratio of $\sigma(e^+e^-\to
{\mathrm{hadrons}})$ and $\sigma(e^+e^-\to \mu^+\mu^-)$) is applied
to identify vector resonances like $\rho$, $\omega$, $\phi$, and
$J/\psi$. However, one cannot find an enhancement structure
corresponding to the $Y(4260)$ from the $R$ value scan (see
Fig.~\ref{Fig.4.6.R-4260}).

\subsubsection{The hybrid charmonium}\label{Sect:4.6.1}

Among various assignments, the hybrid charmonium configuration is
particularly interesting. After its observation in 2005, the author
of Ref. \cite{Zhu:2005hp} proposed the hybrid charmonium
interpretation of the $Y(4260)$. He discussed several possible
structures of the $Y(4260)$ meson with $J^{PC}=1^{--}$ as a
conventional $c\bar c$ state, couple-channel effect, hadron
molecule, tetraquark state, glueball and charmonium hybrid. The main
points are summarized below \cite{Zhu:2005hp}
\begin{itemize}
\item The $Y(4260)$ does not look like a conventional $c\bar c$
state with $J^{PC}=1^{--}$. In PDG \cite{pdg}, the $1^{--}$ radially
excited S-wave states $\psi(2S)$, $\psi(3S)$, $\psi(4S)$ and
D-wave state $\psi(2D)$ were established to be $\psi^\prime$,
$\psi(4040)$, $\psi(4415)$ and $\psi(4160)$ respectively. The
$\psi(3D)$ and $\psi(5S)$ were predicted above 4.5 GeV in quark
model. It is nearly impossible to accommodate $Y(4260)$ as a
conventional charmonium state.

\item It is very difficult to shift the mass of the $\psi(3D)$
from above 4.5 GeV down to 4.26 GeV by the couple-channel effects.
As indicated in Ref. \cite{Eichten:2004uh}, the couple-channel
effects of the open-charm thresholds can only cause around tens MeV
mass shift of the $c\bar c$ state, which is too small compared with
the mass difference between the $\psi(3D)$ and the $Y(4260)$ meson.

\item The $Y(4260)$ seems not a hadronic molecule.
Its quantum number $J^{PC}=1^{--}$ excludes the possibility of the
$\bar D_sD_{s0}(2317)$ molecule, which is only 26 MeV above the
$Y(4260)$ meson. The total decay width of the $Y(4260)$ disfavors
the assignments of the $\bar DD_1$, $\bar DD_1^\prime$, $\bar
D_0D^*$ and $\bar D^*D_1^\prime$ molecules.

\item The $J^{PC}=1^{--}$ glueball is disfavored by its distinct
decay patterns. If it is a glueball, the $Y(4260)$ meson should
mainly decay into multiple light mesons due to the large phase
space. However, it was observed in the decay mode
$J/\psi\pi^+\pi^-$.

\item The tetraquark hypothesis is also not favored by the
not-so-large total width of the $Y(4260)$ meson and the absence of
the open-charm $D\bar D$ decay mode. The charmonium-like tetraquark
states can easily decay into $D\bar D$ final state via the so called
``fall-apart'' mechanism. The decay width of such modes would be
expected to be more than several hundreds MeV due to the big phase
space. This is in conflict with total width $80-130$ MeV of the
$Y(4260)$ meson, as shown in Table~\ref{Table:2.2.Y4260}. To
completely exclude the tetraquark possibility, Zhu suggested to
search for the isovector partner of $Y(4260)$.

\item The charmonium hybrid interpretation of the $Y(4260)$ is
strongly favored by the experimental data. As a hybrid meson, its
two S-wave meson decay modes are suppressed according to QCD sum
rule calculations \cite{Zhu:1998sv,Zhu:1999wg} and flux tube model
analysis \cite{Close:1994hc,Close:2003mb}. Instead the final states
with one S-wave meson and one P-wave meson are potentially
important. Such a decay pattern is consistent with the experimental
data of the $Y(4260)$ meson, in which the open-charm $D\bar D$ decay
mode was not observed. The decay mode $Y(4260)\to
\omega+\chi_{c0,1,2}\to 3\pi+\chi_{c0,1,2}$ was suggested to be
important in the charmonium hybrid configuration \cite{Zhu:2005hp}.

\end{itemize}
As a result, the author of Ref. \cite{Zhu:2005hp} excluded the
possibility of the $Y(4260)$ being a conventional $c\bar c$ state, a
molecule, a glueball and a tetraquark state and concluded that the
$Y(4260)$ meson was a good candidate of the charmonium hybrid state.

Kou and Pene supported this charmonium hybrid interpretation with
several important dynamical arguments in Ref. \cite{Kou:2005gt}.
They proposed that the $Y(4260)$ was a $1^{--}$ charmonium hybrid
state ($H_B$) containing a pseudoscalar colour-octet $0^{-+}$ $\bar
cc$ and a magnetic constituent gluon in P-wave. They proved a
selection rule that the symmetries of the wave function forbid the
decay into two S-wave charmed mesons $D^{(*)}\bar D^{(*)}$ in any
potential model. For the decay processes $H_B\to D^{(*)}\bar
D^{(*)}$, the spatial overlap was described as
\begin{equation}
I=\int \int \frac{d{\vec{p}_{c\bar c}}\
d\vec{k}}{\sqrt{2\omega}(2\pi)^6} \ \Psi_{l_{H_B}}^{m_{H_B}}
(\vec{p}_{c\bar c}, \ \vec{k}) \Psi_{l_B}^{m_B\ *}(\vec{p}_B) \
\Psi_{l_C}^{m_C\ *}(\vec{p}_C) d\Omega_f Y_l^{m\ *}(\Omega_f)\, ,
\end{equation}
where $\Psi_{l_{H_B}}^{m_{H_B}}$, $\Psi_{l_B}^{m_B\ *}(\vec{p}_B)$,
and $\Psi_{l_C}^{m_C\ *}(\vec{p}_C)$ were the spacial wave functions
for the initial hybrid state and the final $ D^{(*)}$ and
$\bar{D}^{(*)}$ states, respectively. The authors of Ref.
\cite{Kou:2005gt} proved that this overlap integral vanishes in the
case of two S-wave mesons final states. Thus, the decay $H_B\to
D^{(*)}\bar D^{(*)}$ was forbidden in any potential model.
Therefore, the hybrid charmonium $H_B$ had a relatively narrow
width, which matches the experimental observation for the $Y(4260)$.
They also suggested the decay $Y(4260)\to D^{**}\bar D^{(*)}\to
D^{(*)}\bar D^{(*)}\pi^\prime$ ($D^{**}$ denotes a P-wave charmonium
state) to be dominant.

Later, Close and Page also proposed the charmonium hybrid assignment
to the $Y(4260)$ \cite{Close:2005iz}. They assessed the experimental
information of the $Y(4260)$, including the near $D_1(2420)\bar D$
threshold mass, the dominant $J/\psi\sigma$, $J/\psi f_0(980)$, $J/\psi a_0(980)$
decay modes and the small partial decay width $\Gamma(Y(4260)\to
e^+e^-)$. All these properties were inconsistent with those of
conventional $c\bar c$ states. They discussed the previous studies of
hybrid meson decays \cite{Zhu:1999wg,Kou:2005gt} and the mass
predictions for charmonium hybrids in the flux-tube model
\cite{Barnes:1995hc,Merlin:1986tz}, the UKQCD quenched lattice QCD
\cite{Lacock:1996ny,Manke:1998yg,Juge:1999aw} and quenched lattice
NRQCD \cite{Drummond:1999db,Manke:1999ru}. Accordingly, they
proposed the charmonium hybrid interpretation of the $Y(4260)$ and
suggested experimental searches for the
$Y(4260)$ in the $Y(4260)\to \{\sigma, \eta\} h_c$ processes
\cite{Close:2005iz}.

In Ref. \cite{Liu:2012ze}, Liu \textit{et al.} presented a mass
spectrum of the highly excited charmonium mesons and the charmonium
hybrid mesons using dynamical lattice QCD simulation. They used the
operator of the general form
$\bar{\psi}\Gamma\stackrel{\leftrightarrow}{D}_i\stackrel{\leftrightarrow}{D}_j\cdots\psi$
to evaluate the two-point correlation functions, where the
$\stackrel{\leftrightarrow}{D}=\stackrel{\rightarrow}{D}-\stackrel{\leftarrow}{D}$
is the covariant derivative operator. Using distillation and the variational
method with a large basis of operators, they successfully
computed the dynamical spectrum of charmonium hybrids. They
identiÞed the lightest hybrid supermultiplet consisting of states
with quantum numbers $J^{PC}=(0, 1, 2)^{-+}, 1^{--}$, as well as an
excited hybrid supermultiplet. The mass of the $1^{--}$ charmonium
hybrid in their mass spectrum was around 4.2 GeV, which allows an
interpretation of the $Y(4260)$ as a vector hybrid meson.

The mass spectra of heavy quarkonium hybrids were also studied in
Coulomb gauge QCD with gluon degrees of freedom in the mean field
approximation in Ref. \cite{Guo:2008yz}. Their predictions of the
hybrid masses were systematically higher compared to lattice. For
example, they found that the $1^{-+}$ and $1^{--}$ hybrid states lie
at 4.47 GeV. This value was heavier than the mass of the $Y(4260)$
meson \cite{Guo:2008yz}. The mass of the charmonium hybrid meson was
also studied in QCD sum rules in Ref.
\cite{Qiao:2010zh,Chen:2013zia,Chen:2013eha,Govaerts:1986pp,Govaerts:1984hc,Govaerts:1984bk,Harnett:2012gs,Berg:2012gd}.
%%%%%%%%%%%%%%%%%%%%%%%%%%%%%%%%%%%%%%%%%%%%%%%%%%%%%%%%%%%%%%%%%%%%

\subsubsection{The vector tetraquark state}
\label{sect:4.6.2}

In Ref.~\cite{Maiani:2005pe}, Maiani, Riquer, Piccinini, and Polosa
proposed that the $Y(4260)$ is the first orbital excitation of a
diquark-antidiquark state $[cs][\bar c\bar s]$, and predicted that
the $Y(4260)$ should decay predominantly in $D_s \bar D_s$, as well
as be seen in $B$ non-leptonic decays in association with one
kaon~\cite{Bigi:2005fr}. Later in Ref.~\cite{Drenska:2009cd},
Drenska, Faccini, and Polosa studied the $[cs][\bar c\bar s]$
diquark-antidiquark particles with different $J^{PC}$ quantum
numbers. They computed their spectrum and decay modes within a
constituent diquark-antidiquark model, and predicted $m_Y = 4330 \pm
70$ MeV for the $Y(4260)$.

This idea was updated in their ``type-II'' diquark-antidiquark
model, reviewed in Sec.~\ref{Sect:4.2.2}, by taking into account the
orbital angular momentum between diquark and antidiquark. For this
case, they used the notation $|s, \bar s; S, L \rangle _{J}$ to
denote the excited tetraquark states with the total spin $S = s +
\bar s$ and total angular momentum $J$, where $s = s_{qc}$ and $\bar
s = s_{\bar q \bar c}$ are the diquark and antidiquark spins,
respectively. For tetraquark states of $J^{PC} = 1^{--}$, there are
four states having $L=1$ and one state having $L=3$:
\begin{eqnarray}
\nonumber Y_1 &=& |0, 0; 0, 1\rangle_1 \, ,
\\ \nonumber Y_2 &=& {1\over\sqrt2} \Big( |1, 0; 1, 1\rangle_1 + |0, 1; 1, 1\rangle_1 \Big) \, ,
\\ Y_3 &=& |1, 1; 0, 1\rangle_1 \, ,
\\ \nonumber Y_4 &=& |1, 1; 2, 1\rangle_1 \, ,
\\ \nonumber Y_5 &=& |1, 1; 2, 3\rangle_1 \, .
\end{eqnarray}
They used a Hamiltonian containing both spin-orbit and spin-spin
interactions
\begin{eqnarray}
H = B_c {\bm L^2 \over 2} - 2a \bm L \cdot \bm S +
2\kappa^\prime_{qc} \; \left(\bm s_q\cdot \bm s_c +  \bm s_{\bar
q}\cdot \bm s_{\bar c}\right),
\end{eqnarray}
and discussed possible interpretations of $Y(4008)$, $Y(4260)$,
$Y(4630)$, etc.

Especially, they fixed the assignment that $Y (4260) = Y_2 =
{1\over\sqrt2} \Big( |1, 0; 1, 1\rangle_1 + |0, 1; 1, 1\rangle_1
\Big)$. Hence, in their model the $Y(4260)$ is just the first
orbital excitation of the $X(3872) = {1\over\sqrt2} \Big( |1, 0; 1,
0\rangle_1 + |0, 1; 1, 0\rangle_1 \Big)$. This idea was later used
in Ref.~\cite{Chen:2015dig} by Chen, Maiani, Polosa, and Riquer to
calculate the radiative transition $Y(4260) \to \gamma X(3872)$,
$\Gamma_{\rm rad} \equiv \Gamma(Y(4260) \to \gamma X(3872))$, using
a non-relativistic calculation of the electric dipole term of a
diquarkonium bound state. Specializing to $I = 0$ for the $X(3872)$,
they found $\Gamma_{\rm rad} = 496$ keV for the $Y(4260)$ with $I =
0$ and $\Gamma_{\rm rad} = 179$ keV for $I = 1$. They derived upper
bounds to $\mathcal{B}(Y(4260) \to J/\psi + \pi + \pi)$ and to
$\Gamma(Y(4260) \to \mu^ + \mu^-)$, which can be confronted with
future data from electron-positron and hadron colliders.

The diquark-antidiquark picture for the $Y(4260)$ was also studied
in Refs.~\cite{Ebert:2005nc,Ebert:2008kb} by Ebert, Faustov, and
Galkin in the framework of the relativistic quark model. They
treated the dynamics of the light quark in a heavy-light diquark
completely relativistically, and investigated the internal structure
of the diquark by calculating the diquark-gluon form factor in terms
of the diquark wave functions. They found that the $Y(4260)$ cannot
be interpreted as the $1^{--}$ $1P$ state of the charm-strange
diquark-antidiquark tetraquark, i.e., ($[cs]_{S=0}[\bar c \bar
s]_{S=0}$). Instead, they found that a more natural tetraquark
interpretation for the $Y(4260)$ is the $1^{--}$ $1P$ state of
$([cq]_{S=0}[\bar c\bar q]_{S=0})_{P-wave}$. The other two
possibilities are the $1^{--}$ $1P$ states of
${1\over\sqrt2}([cq]_{S=0}[\bar c\bar q]_{S=1})_{P-wave} -
[cq]_{S=1}[\bar c\bar q]_{S=0})_{P-wave})$ and $([cq]_{S=1}[\bar
c\bar q]_{S=1})_{P-wave}$.

The authors of Ref. \cite{Chen:2010ze} had studied the
charmonium-like tetraquark states with $J^{PC}=1^{--}$ in QCD sum
rules. They constructed the diquark-antidiquark tetraquark
interpolating currents
\begin{eqnarray}
\begin{split}
J_{1\mu}&=q_{a}^TC\gamma_5Q_{b}(\bar{q}_{a}\gamma_{\mu}\gamma_5C\bar{Q}^T_{b}+\bar{q}_{b}\gamma_{\mu}\gamma_5C\bar{Q}^T_{a})
-
q_{a}^TC\gamma_{\mu}\gamma_5Q_{b}(\bar{q}_{a}\gamma_5C\bar{Q}^T_{b}+\bar{q}_{b}\gamma_5C\bar{Q}^T_{a})\,
, \\
J_{2\mu}&=q_{a}^TC\gamma^{\nu}Q_{b}(\bar{q}_{a}\sigma_{\mu\nu}C\bar{Q}^T_{b}-\bar{q}_{b}\sigma_{\mu\nu}C\bar{Q}^T_{a})
-
q_{a}^TC\sigma_{\mu\nu}Q_{b}(\bar{q}_{a}\gamma^{\nu}C\bar{Q}^T_{b}-\bar{q}_{b}\gamma^{\nu}C\bar{Q}^T_{a})\,
, \\
J_{3\mu}&=q_{a}^TC\gamma_5Q_{b}(\bar{q}_{a}\gamma_{\mu}\gamma_5C\bar{Q}^T_{b}-\bar{q}_{b}\gamma_{\mu}\gamma_5C\bar{Q}^T_{a})
-
q_{a}^TC\gamma_{\mu}\gamma_5Q_{b}(\bar{q}_{a}\gamma_5C\bar{Q}^T_{b}-\bar{q}_{b}\gamma_5C\bar{Q}^T_{a})\,
, \\
J_{4\mu}&=q_{a}^TC\gamma^{\nu}Q_{b}(\bar{q}_{a}\sigma_{\mu\nu}C\bar{Q}^T_{b}+\bar{q}_{b}\sigma_{\mu\nu}C\bar{Q}^T_{a})
-
q_{a}^TC\sigma_{\mu\nu}Q_{b}(\bar{q}_{a}\gamma^{\nu}C\bar{Q}^T_{b}+\bar{q}_{b}\gamma^{\nu}C\bar{Q}^T_{a})\,
, \\
J_{5\mu}&=q_{a}^TCQ_{b}(\bar{q}_{a}\gamma_{\mu}C\bar{Q}^T_{b}+\bar{q}_{b}\gamma_{\mu}C\bar{Q}^T_{a})
-
q_{a}^TC\gamma_{\mu}Q_{b}(\bar{q}_{a}C\bar{Q}^T_{b}+\bar{q}_{b}C\bar{Q}^T_{a})\,
, \\
J_{6\mu}&=q_{a}^TC\gamma^{\nu}\gamma_5Q_{b}(\bar{q}_{a}\sigma_{\mu\nu}\gamma_5C\bar{Q}^T_{b}+\bar{q}_{b}\sigma_{\mu\nu}\gamma_5C\bar{Q}^T_{a})
-
q_{a}^TC\sigma_{\mu\nu}\gamma_5Q_{b}(\bar{q}_{a}\gamma^{\nu}\gamma_5C\bar{Q}^T_{b}+\bar{q}_{b}\gamma^{\nu}
\gamma_5C\bar{Q}^T_{a})\, , \\
J_{7\mu}&=q_{a}^TCQ_{b}(\bar{q}_{a}\gamma_{\mu}C\bar{Q}^T_{b}-\bar{q}_{b}\gamma_{\mu}C\bar{Q}^T_{a})
-
q_{a}^TC\gamma_{\mu}Q_{b}(\bar{q}_{a}C\bar{Q}^T_{b}-\bar{q}_{b}C\bar{Q}^T_{a})\,
, \\
J_{8\mu}&=q_{a}^TC\gamma^{\nu}\gamma_5Q_{b}(\bar{q}_{a}\sigma_{\mu\nu}\gamma_5C\bar{Q}^T_{b}-\bar{q}_{b}\sigma_{\mu\nu}\gamma_5C\bar{Q}^T_{a})
-
q_{a}^TC\sigma_{\mu\nu}\gamma_5Q_{b}(\bar{q}_{a}\gamma^{\nu}\gamma_5C\bar{Q}^T_{b}-\bar{q}_{b}\gamma^{\nu}
\gamma_5C\bar{Q}^T_{a})\, , \label{Eq.4.6.tetraquarkcurrents1}
\end{split}
\end{eqnarray}
where Q is the charm quark for charmonium-like tetraquark systems
and bottom quark for bottomonium-like systems. Using these
interpolating currents, the two-point correlation functions and
spectral densities were calculated up to the dimension eight
condensates in the OPE series. The mass spectrum of the $1^{--}$
charmonium-like tetraquark states was collected in Table~\ref{Table:4.6.tetraquark}. The mass of the vector charmonium-like
$qc\bar q\bar c$ tetraquark state was extracted as $4.5-4.8$ GeV
within the uncertainties. These values were much higher than the
mass of the $Y(4260)$ meson but consistent with that of the
$Y(4660)$. The numerical results didn't support the tetraquark
interpretation of the $Y(4260)$ state, which was consistent with the
discussion in Ref. \cite{Zhu:2005hp}. The vector hidden-charm and
hidden-bottom tetraquark states were also studied in QCD sum rules
in Refs. \cite{Wang:2009gx,Zhang:2010mw}.

%%%%%%%%%%%%%%%%%%%%%%%%%%%%%%%%%%%%%%%%%%%%%%%%%%%%%%%%%%%%%%%
\renewcommand{\arraystretch}{1.6}
\begin{table}[htb]
\caption{Mass spectrum of the charmonium-like $qc\bar q\bar c$ and
$sc\bar s\bar c$ tetraquark states with $J^{PC}=1^{--}$ \cite{Chen:2010ze}.
\label{Table:4.6.tetraquark}}
\begin{center}
\begin{tabular}{cccccc} \toprule[1pt]
& Current ~~&~~ $s_0(\mbox{GeV}^2)$ ~~&~~ \mbox{Borel window} $(\mbox{GeV}^2)$ ~~&~~ $m_X$ \mbox{(GeV)} ~~&~~ \mbox{PC(\%)} \\
\midrule[1pt]
\multirow{3}{*}{$qc\bar q\bar c$   }     & $J_{1\mu}$      &  $5.0^2$         & $2.9-3.6$           & $4.64\pm0.09$     & 44.1  \\
                        & $J_{4\mu}$      &  $5.0^2$         & $2.9-3.6$           & $4.61\pm0.10$     & 46.4  \\
                        & $J_{7\mu}$      &  $5.2^2$         & $2.9-4.1$           & $4.74\pm0.10$     & 47.3
\vspace{5pt} \\
\multirow{6}{*}{$sc\bar s\bar c$ }       & $J_{1\mu}$      &  $5.4^2$         & $2.8-4.5$           & $4.92\pm0.10$     & 50.3  \\
                        & $J_{2\mu}$      &  $5.0^2$         & $2.8-3.5$           & $4.64\pm0.09$     & 48.6  \\
                        & $J_{3\mu}$      &  $4.9^2$         & $2.8-3.4$           & $4.52\pm0.10$     & 45.6  \\
                        & $J_{4\mu}$      &  $5.4^2$         & $2.8-4.5$           & $4.88\pm0.10$     & 51.7  \\
                        & $J_{7\mu}$      &  $5.3^2$         & $2.8-4.3$           & $4.86\pm0.10$     & 46.0  \\
                        & $J_{8\mu}$      &  $4.8^2$         & $2.8-3.1$           & $4.48\pm0.10$     & 43.2   \\
\bottomrule[1pt]
\end{tabular}
\end{center}
\end{table}
%%%%%%%%%%%%%%%%%%%%%%%%%%%%%%%%%%%%%%%%%%%%%%%%%%%%%%%%%%%%%%%%%%%%%%%%%%%%%%%%%%%%%%%%%%%%%%%%%%%%%%%%%%%%%%%%%%%%%%%%%%%%%%%%%%%

\subsubsection{The molecular state}
\label{sect:4.6.3}

There are several molecular interpretations for the $Y(4260)$ state.
In Ref.~\cite{Yuan:2005dr}, Yuan, Wang, and Mo interpreted the
$Y(4260)$ as an $\omega \chi_{c1}$ molecular state and discussed
both its production and decay properties~\cite{Yuan:2005dr}.

Later in Ref.~\cite{Ding:2008gr}, Ding performed a dynamical study
of the $Y(4260)$ and $Z^+_2(4250)$ simultaneously in the framework
of the meson exchange model to see whether they could be the
$D_1\bar D$ or $D_0\bar D^*$ hadronic molecule. He employed the
heavy meson chiral Lagrangian, which combines the heavy quark
symmetry and the chiral symmetry. He found that the off-diagonal
interaction induced by the $\pi$ exchange plays a dominant role. The
diagonal interactions contain the $\sigma$ exchange and the light
vector meson exchange. The contribution of the $\sigma$ exchange
does not favor the formation of the molecular state with
$I^G(J^{PC}) = 0^-(1^{--})$, but favors the binding of the molecule
with $I^G(J^P) = 1^-(1^-)$. He suggested that the $Y(4260)$ could be
accommodated as a $D_1D$ and $D_0D^*$ molecule. He also studied the
bottom analog of the $Y(4260)$ and proposed to observe it in the
$\pi^+\pi^-\Upsilon$ channel.

In Refs.~\cite{Close:2009ag,Close:2010wq}, Close and Downum, and
Thomas studied the strong $S$-wave pion exchange effects, and
suggested that a spectroscopy of quasi-molecular states may arise in
the case of charmed mesons $D$, $D^*$, $D_0$, $D_1$, which are
consistent with enigmatic charmonium states observed above 4 GeV in
$e^+e^-$ annihilations. They discussed the possible interpretations
of the $Y(4260)$ being $D\bar D_1$ and $D^* \bar D_1$ bound states,
and proposed to observe the $D\bar D\pi\pi\pi$ channel to compare
with the $D\bar D\pi\pi$ channel, which can be used to reveal the
mixing between $D^* \bar D_1$ and $D\bar D_1/D^* \bar D_0$ molecular
systems.

The interaction potentials between one S-wave and one P-wave
heavy mesons as well as the potentials between two $P$-wave heavy
mesons were deduced based on a chiral quark model by Li, Wang, Dong,
and Zhang in Ref.~\cite{Li:2013bca}. They concluded that the
$Y(4260)$ can not be explained as the $D^* \bar D^{*0}$ molecule,
but might be explained as a $0^-(1^{--})$ $D \bar D_1$ molecule.

The interpretation of the $Y(4260)$ as a $D_1 \bar D$ molecule was
also discussed in Ref.~\cite{Cleven:2013mka} by Cleven {\it et al.}. They
demonstrated that the nontrivial cross section line shapes of
$e^+e^- \to J/\psi \pi\pi$ and $h_c \pi \pi$ can be naturally
explained by the molecular scenario, and found a significantly
smaller mass for the $Y(4260)$. They also predicted an unusual line
shape of the $Y(4260)$ in the $D \bar D^*$ channel, which could be a
smoking gun for a predominantly molecular nature of the $Y(4260)$.

The lattice QCD calculation of $D$ and $\bar{D}_1$ interaction can
be found in Ref.~\cite{Chiu:2005ey}. In this reference the TWQCD
Collaboration used a molecular type operator composed of $D$ and
$\bar{D}_1$ mesons, $(\bar q \gamma_5 \gamma_i c)(\bar c \gamma_5 q)
- (\bar c \gamma_5 \gamma_i q)(\bar q \gamma_5 c)$, and detected a
$1^{--}$ signal with a mass around $4238 \pm 31$ MeV in quenched
lattice QCD simulations with exact chiral symmetry, which was
identified with the $Y(4260)$.

\subsubsection{Non-resonant explanations}

In Ref.~\cite{vanBeveren:2009jk}, van Beveren and Rupp argued that
the puzzling branching ratios of open-charm decays in $e^+e^-$
annihilations can be reasonably described with a simple form factor,
which strongly suppresses open channels far above the threshold.
They applied this idea to study the $e^+e^- \to J/\psi \pi\pi$ data
on the $Y(4260)$ enhancement, and obtained a good fit with a simple
nonresonant cusp structure around the $D^*_s \bar D^*_s$ threshold.
Moreover, they found the data shows an oscillatory pattern between a
fast (OZI-allowed) and a slow (OZI-forbidden) $J/\psi f_0(980)$
mode.

In Ref.~\cite{vanBeveren:2010mg}, van Beveren, Rupp, and Segovia
reconstructed the shape of the $Y(4260)$ observed in $e^+e^- \to
J/\psi \pi^+ \pi^-$ by a stepwise study, where they considered the
contributions from the open-charm thresholds like $D\bar{D}$,
$D\bar{D}^*$, $D^*\bar{D}^*$, $D_s\bar{D}_s$, $D_s\bar{D}^*_s$,
$D_s^*\bar{D}_s^*$, $\Lambda_c\bar{\Lambda}_c$, and all well-known
vector charmonia ($\psi(4040)$, $\psi(4160)$, and $\psi(4415)$) with
mass above 4 GeV. Additionally, they concluded that the $\psi(3D)$
charmonium state has been observed in a range of 4.53-4.58 GeV with
a width around 40-70 MeV.

\begin{figure}[htb]
\centering \scalebox{0.5}{\includegraphics{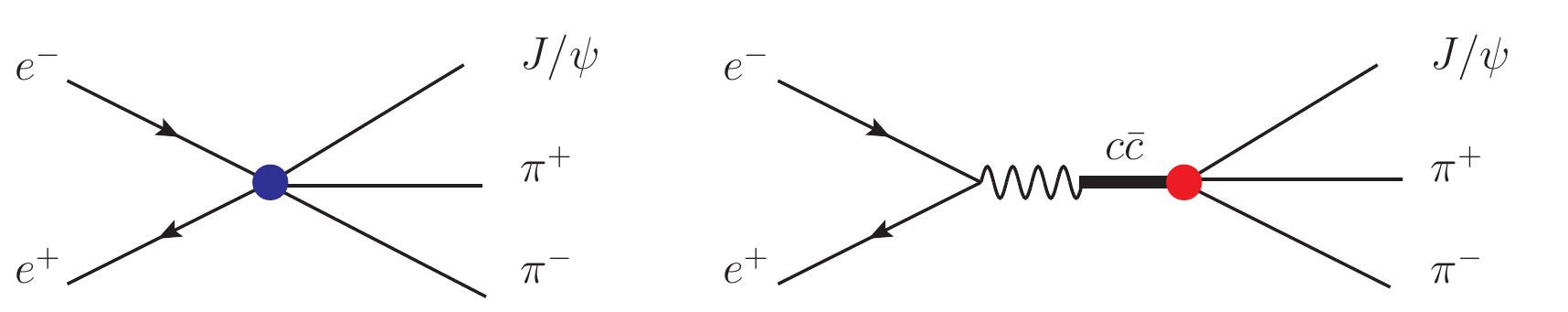}}
\put(-210,-20){$(a)$}%
\put(-80,-20){$(b)$}%
\caption{(Color online) The diagrams relevant to $e^{+} e^{-} \to
J/\psi \pi^+ \pi^-$. Here, (a) corresponds to the direct
$e^+e^-$ annihilation into $J/\psi\pi^+\pi^-$. (b) is from
the contributions of the intermediate charmonia. Taken from Ref.
\cite{Chen:2010nv}. \label{Fig.4.6.Fig-Feyn1}}
\end{figure}

\begin{figure}[htb]
\begin{center}
\includegraphics[bb=25 340 560 780,scale=0.45]{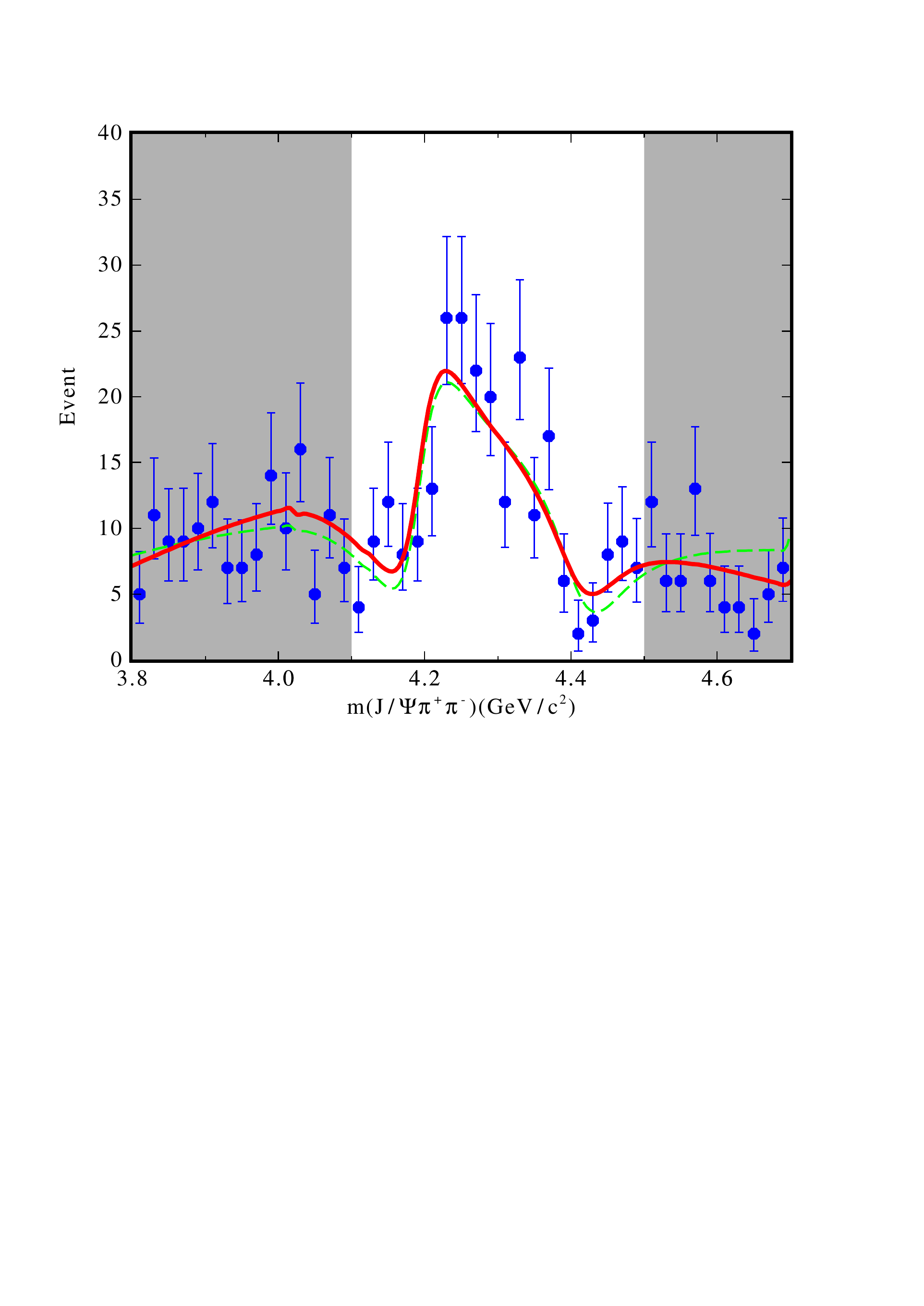}
\caption{(Color online) The comparison of the obtained fitting
result (solid red line) with the experimental data (blue dots with
error bar) measured by BaBar \cite{Aubert:2005rm}. The result is normalized to the experimental
data. Taken from Ref. \cite{Chen:2010nv}. \label{Fig.4.6.resulttotal}}
\end{center}
\end{figure}

Chen, He and Liu also proposed a non-resonant explanation for the
$Y(4260)$ structure observed in the $e^+e^-\to J/\psi\pi^+\pi^-$
process \cite{Chen:2010nv}, where they considered the interference
of the production amplitudes of the $e^+e^-\to J/\psi\pi^+\pi^-$
process via the direct $e^+e^-$ annihilation and through
intermediate charmonia $\psi(4160)/\psi(4415)$ (see Fig.~\ref{Fig.4.6.Fig-Feyn1}). The $Y(4260)$ structure was reproduced well, which
is shown in Fig.~\ref{Fig.4.6.resulttotal}. Since the $Y(4260)$ is not a genuine
resonance within this scheme \cite{Chen:2010nv}, it naturally
answers why there is no evidence of the $Y(4260)$ in the exclusive
open-charm decay channels
\cite{Pakhlova:2008zza,Pakhlova:2007fq,Abe:2006fj} and $R$-value
scan \cite{pdg}.

Very recently, Chen, Liu, Li and Ke \cite{Chen:2015bft} further
pointed out that this nonresonant explanation to the $Y(4260)$ suggested
in Ref. \cite{Chen:2010nv} is similar to the Fano interference
effect, which extensively exists in atomic physics, condensed matter
physics and even nuclear physics, where the asymmetric line shape of
the $Y(4260)$ can be reflected by the Fano-like interference picture
\cite{Chen:2015bft}.

\subsubsection{Other theoretical schemes, production and decay patterns}
\label{sect:4.6.4}

The $Y(4260)$ was interpreted as a molecular state composed of two
colored baryons, i.e., a baryonium state in
Refs.~\cite{Qiao:2005av,Qiao:2007ce}. The dynamical generation of the
$Y(4260)$ in the $J/\psi\pi\pi$ and $J/\psi K \bar K$ systems was
studied in Ref.~\cite{MartinezTorres:2009xb}, showing that the $K\bar K$
system clustered around the $f_0(980)$
resonance, hence providing a natural explanation for the important $J/\psi
f_0(980)$ decay mode. The $Y(4260)$ was
studied in Ref.~\cite{Li:2013ssa} within the picture of
hadrocharmonium, where a compact charmonium was embedded in a light
quark mesonic excitation. The coupled-channel effects and
nonresonant explanation for the $Y(4260)$ structure were studied in
Refs.~\cite{vanBeveren:2006ih,Chen:2010nv,Chen:2011xk}, while its
related threshold effects were studied in
Refs.~\cite{Rosner:2006vc,Liu:2013vfa}. The bottom counterparts of
the $Y(4260)$, $Y_b$, were studied in Ref.~\cite{Hou:2006it}.

Weak productions of the $Y(4260)$ in semi-leptonic $B_c$ decays were
studied in Ref.~\cite{Wang:2007fs}, where the $Y(4260)$ was treated as an
excited charmonium state. Assuming the $Y(4260)$ to be a tetraquark
state with a hidden $c\bar c$ quark pair, its production at the LHC
and Tevatron was studied via the Drell-Yan Mechanism in
Ref.~\cite{Ali:2011qi}. Assuming the $Y(4260)$ to be a mixture of
the charmonium and exotic tetraquark state, its production in $B$
meson decay was studied  using QCD sum rules in
Ref.~\cite{Albuquerque:2015nwa}.

Assuming the $Y(4260)$ as a hybrid state, its decays into $J/\psi
\pi \pi$ and open charm mesons were studied in
Ref.~\cite{Kou:2005gt}, and its relative decay rates into various $S$
and $P$ wave charm meson pairs were calculated using QCD string
model in Ref.~\cite{Kalashnikova:2008qr}. Assuming the $Y(4260)$ to
be a $D_1(2420) \bar D$ molecular state, its hidden-charm, charmed
pair, and charmless decay channels were studied via the intermediate
$D_1 \bar D$ meson loops with an effective Lagrangian approach in
Refs.~\cite{Li:2013yla,Li:2014gxa}, while, its strong decay modes
$Z_c(3900)^\pm \pi^\mp$, $J/\psi \pi^+ \pi^-$, and
$\psi(nS)\pi^+\pi^-$, were studied in Ref.~\cite{Dong:2013kta}. The
authors of Ref.~\cite{Burns:2007hk} generalized results of lattice
QCD to determine the spin-dependent symmetries and factorization
properties of the meson production in OZI allowed processes, which
were applied to establish the structure of the $Y(4260)$ from its S-wave
decays. The upper limit of the electron width of the $Y(4260)$,
$\Gamma(Y(4260) \to e^+ e^-)$, was determined to be 580 eV at 90\%
C.L. in Ref.~\cite{Mo:2006ss} by Mo, {\it et al.}.

With the spin rearrangement, the authors of Ref. \cite{Ma:2015nmy}
performed a comprehensive investigation of the decay patterns of the
$Y(4260)$ with different inner structures such as the conventional
charmonium, the molecule, the P-wave tetraquark and the hybrid
charmonium. The $J/\psi \left(\pi\pi\right)_{\mathrm{S-wave}}$ mode
is suppressed in the heavy quark symmetry limit if the $Y(4260)$ is
a molecular state and $\left(\pi\pi\right)_{\mathrm{S-wave}}$ arises
from either $\sigma$ or $f_0(980)$. Moreover the hybrid charmonium
and hidden-charm tetraquark have very similar decay patterns. Both
of them decay into the $J/\psi \pi\pi$ and open charm modes easily.

\subsubsection{A short summary}
\label{sect:4.6.5}

\begin{itemize}

\item There are also some non-resonant explanations for the $Y(4260)$,
such as the coupled-channel effects, the threshold effects, and the
Fano interference effect, etc.

\item The $Y(4260)$ may be the $\psi(4S)$ or $\psi(3D)$
charmonium state in some quark models. However, there does not exist
an enhancement structure corresponding to the $Y(4260)$ from the $R$
value scan. The absence of the open-charm decay channels is also
hard to explain.

\item Both the $Y(4260)$ and the $X(3872)$ were proposed as
tetraquark states composed of a pair of diquark and antidiquark. The
$Y(4260)$ was interpreted as the $P$-wave excitation of the
$X(3872)$ state. If the $Z_c(3885)$ and $Z_c(4025)$ also turn out to
be tetraquark states, the decay process $e^+ e^- \to Y(4260) \to
\pi^- Z_c(3885)^+ (\to (D \bar D^*)^+)$ and $e^+ e^- \to Y(4260) \to
\pi^- Z_c(4025)^+ (\to (D^* \bar D^*)^+)$ can be understood
naturally. However, some QCD sum rule calculations indicate the
vector tetraquark state may lie around 4.6 GeV.

\item The $Y(4260)$ was also suggested as a $D_1\bar D$ molecular
state. If the $Z_c(3885)$ and $Z_c(4025)$ are also molecular states,
the decay process $e^+ e^- \to Y(4260) \to \pi^- Z_c(3885)^+ (\to (D
\bar D^*)^+)$ and $e^+ e^- \to Y(4260) \to \pi^- Z_c(4025)^+ (\to
(D^* \bar D^*)^+)$ can also be understood easily. However, the
discovery mode $J/\psi \left(\pi\pi\right)_{\mathrm{S-wave}}$ is
strongly suppressed in the heavy quark symmetry limit if the $Y(4260)$
is a molecular state and $\left(\pi\pi\right)_{\mathrm{S-wave}}$ arises
from either $\sigma$ or $f_0(980)$.

\item The $Y(4260)$ is a good candidate of the hybrid charmonium
state with $J^{PC}=1^{--}$. This interpretation explains the current
experimental information, and was supported by the lattice QCD
simulations.

\end{itemize}

\subsection{$Y(3940)$, $Y(4140)$ and $Y(4274)$}
\label{Sect:4.7}

The $Y(3940)$ and $Y(4140)$, which were reviewed in
Sec.~\ref{Sect:2.1.2} and \ref{Sect:2.1.3}, were observed in the
mass spectrum of $J/\psi+{light\, vector\, meson}$ in the $B$ meson
decay
\begin{eqnarray*}
B\to K+\Bigg\{\begin{array}{cc} Y(3940) &\Longrightarrow {J/\psi \omega}\\
Y(4140) & \Longrightarrow {J/\psi \phi}\end{array} \, .
\end{eqnarray*}
Besides the $Y(4140)$, the $Y(4274)$ reviewed in Sec.~\ref{Sect:2.1.3}
was also observed in the $J/\psi \phi$ mass spectrum in the $B^+\to
J/\psi \phi K^+$ decay process.

\subsubsection{Molecular state scheme}
\label{Sect:4.7.1}

The $Y(3940)$ and $Y(4140)$ are close to the thresholds of
$D^*\bar{D}^*$ and $D_s^*\bar{D}_s^*$, respectively, and satisfy an
almost exact mass relation
\begin{eqnarray}
M_{Y(4140)} - 2M_{D_s^*} \approx M_{Y(3940)}-2M_{D^*} \, .
\end{eqnarray}
Hence, a uniform molecular picture of the $Y(4140)$ and $Y(3940)$ was proposed in
Refs.~\cite{Liu:2009ei,Liu:2008tn}, where the flavor wave functions
of the $Y(3940)$ and $Y(4140)$ are:
\begin{eqnarray}
|Y(4140)\rangle&=&|D_s^{*+}D_s^{*-}\rangle,\\
|Y(3940)\rangle&=&\frac{1}{\sqrt{2}}\Big[|D^{*0}\bar{D}^{*0}\rangle+|D^{*+}D^{*-}\rangle\Big].
\end{eqnarray}
Moreover, the authors observed a selection rule for the quantum
numbers of the $Y(3940)$ and $Y(4140)$ under the $D^*\bar{D}^*$ and
$D_{s}^*\bar{D}_{s}^*$ molecular state assignments. They argued that
their widths are narrow naturally, because both the hidden-charm and
open charm two-body decays occur through the rescattering of the
vector components within the molecular states while the three- and
four-body open charm decay modes are forbidden kinematically. The
possible quantum numbers of the S-wave vector-vector system are
$J^{P}=0^+, 1^+, 2^+$. However, they can only have $J^{P}=0^+$ and
$2^+$, for the neutral $D^\ast {\bar D}^\ast$ system with $C=+$, due
to $C=(-1)^{L+S}$ and $J=S$ with $L=0$. This provides an important
criterion to test the molecular explanation for the $Y(3940)$ and
$Y(4140)$.

The $Y(4274)$ was interpreted as the $S$-wave $D_s \bar
D_{s0}(2317)$ molecular state with $J^P = 0^-$ in
Ref.~\cite{Liu:2010hf}. This interpretation
was supported by dynamical study of the system composed of the
pseudoscalar and scalar charmed mesons. They also investigated the
$S$-wave $D\bar D_0(2400)$ molecular charmonium as the molecular
partner of the $Y(4274)$, which is in accord with the enhancement
structure appearing at 4.2 GeV in the $J/\psi \omega$ invariant mass
spectrum from $B$ decays~\cite{Abe:2004zs,Aubert:2007vj}. There
might also exist structures around the thresholds of the
$D_s\bar{D}_{s1}^\prime(2460)$, $D_s^*\bar{D}_{s0}(2317)$,
$D_{s}\bar{D}_{s1}(2536)$, $D_{s}\bar{D}_{s2}(2573)$,
$D_{s}^*\bar{D}_{s1}^\prime(2460)$ and $D_{s}^*\bar{D}_{s1}(2536)$.

\begin{figure}[htb]
\begin{center}
\scalebox{0.47}{\includegraphics{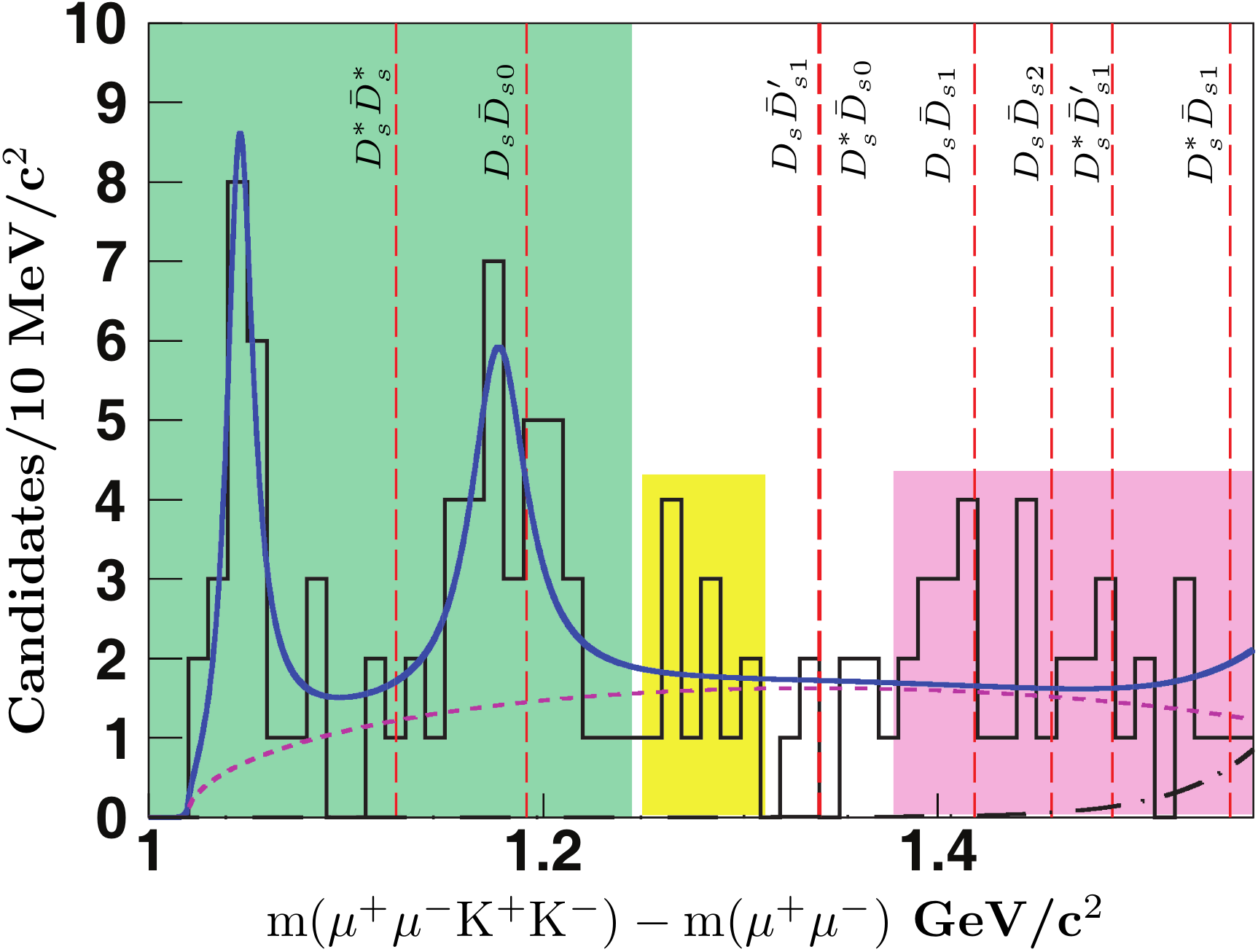}} \caption{(Color online) The mass difference $\Delta
M=m(\mu^+\mu^-K^+K^-)-m(\mu^+\mu^-)$ distribution (histogram) for
events in the $B^+$ mass window \cite{Aaltonen:2011at}. Besides the
$Y(4140)$, one explicit enhancement appears around $4274$ MeV. Here,
the purple dashed line is the background from the three-body phase
space. The blue solid line is the fitting result with resonance
parameters of the $Y(4140)$ and $Y(4270)$ resonances in Ref.
\cite{Aaltonen:2011at}. The vertical red dashed lines denote the
thresholds of the $D_s^*\bar{D}_s^*$, $D_{s}\bar{D}_{s0}(2317)$,
$D_{s}\bar{D}_{s1}^\prime(2460)$, $D_{s}^*\bar{D}_{s0}(2317)$,
$D_{s}\bar{D}_{s1}(2536)$, $D_{s}\bar{D}_{s2}(2573)$,
$D_{s}^*\bar{D}_{s1}^\prime(2460)$ and $D_{s}^*\bar{D}_{s1}(2536)$.
Taken from Ref. \cite{Liu:2010hf}. \label{Fig.4.7.BW}}
\end{center}
\end{figure}

The molecular explanation for the $Y(3940)$ and $Y(4140)$ was also
studied in
Refs.~\cite{Liu:2008tn,Mahajan:2009pj,Branz:2009yt,Ding:2009vd,Chen:2014fda}.
In Ref.~\cite{Liu:2008tn}, the authors performed a dynamical
calculation to study whether $D^*\bar{D}^*$ or $D_s^*\bar{D}_s^*$
system can be bound. They used the effective Lagrangian approach.
The exchanged mesons between the $D^*$ and $\bar{D}^*$ ($D_{s}^*$ and $\bar{D}^*_s$)
include the pseudoscalar, vector and $\sigma$ mesons. The S-wave
molecular solution was found for the $Y(4140)$ and $Y(3940)$ with
$J^P=0^+,\,2^+$.

In Ref.~\cite{Mahajan:2009pj}, Mahajan argued that the $Y(4140)$ is
more likely to be a $D^*_s\bar{D}^*_s$ molecular state or an exotic
$J^{PC} = 1^{-+}$ hybrid charmonium. He also discussed decay modes
which would allow unambiguous the identification of the hybrid
charmonium option.

In Ref.~\cite{Branz:2009yt}, Branz, Gutsche, and Lyubovitskij
suggested that the $Y(3940)$ and $Y(4140)$ are heavy hadron
molecular states with quantum number $J^{PC} = 0^{++}$. They
evaluated widths of the strong decays $Y(3940) \to J/\psi \omega$,
$Y(4140) \to J/\psi\phi$ and radiative decay $Y(3940)/Y(4140) \to
\gamma\gamma$ in a phenomenological Lagrangian approach, supporting
the molecular interpretation of the $Y(3940)$ state as a
superposition of the $D^{*+}D^{*-}$ and $D^{*0}\bar D^{*0}$, and the
$Y(4140)$ as a bound state of the $D^{*+}_s$ and $D^{*-}_s$ mesons. They
also investigated the alternative assignment of $J^{PC} = 2^{++}$,
and obtained similar results for the strong decay widths.

In Ref.~\cite{Ding:2009vd}, Ding dynamically studied the
interpretation of the $Y(4140)$ as a $D_{s}^*\bar{D}_{s}^*$ molecule in
the one boson exchange approach, where $\sigma$, $\eta$ and $\phi$
exchanges were included. He suggested the most favorable quantum
number for the $Y(4140)$ is $J^{PC} = 0^{++}$. However, $0^{-+}$
and $2^{++}$ can not be excluded. He also proposed to search for the
$1^{+-}$ and $1^{--}$ partners in the $J/\psi\eta$ and
$J/\psi\eta^\prime$ final states, which is an important test of the
molecular hypothesis of the $Y(4140)$.

In Ref.~\cite{Chen:2014fda}, Chen and Lu studied the general form of
the Bethe-Salpeter wave functions for the bound states composed of
two vector fields of arbitrary spin and definite parity, and applied
this framework to study the $Y(3940)$ as a molecule state consisting
of the $D^{*0}$ and $\bar D^{*0}$. They considered the attractive potential
between the $D^{*0}$ and $\bar D^{*0}$ including one light meson
($\sigma$, $\pi$, $\omega$, and $\rho$) exchange, and found the
obtained mass of the $Y(3940)$ is consistent with the experimental
value.

In QCD sum rules, the $Y(4140)$ was studied as a scalar $D_s^*\bar
D_s^*$ molecular state in Refs.
\cite{Zhang:2009st,Albuquerque:2009ak,Wang:2014gwa}. All these
calculations obtained the masses around $4.0-4.2$ GeV and thus
supported the molecular interpretation of the $Y(4140)$. However, a
negative result was obtained in Refs. \cite{Wang:2009ue,Wang:2009ry},
in which the mass of the scalar $D_s^*\bar D_s^*$ molecule was
extracted at $4.3-4.6$ GeV. The $Y(3940)$ was studied as a $D^*\bar
D^*$ molecular state in Refs.
\cite{Albuquerque:2009ak,Wang:2014gwa}. These calculations
disfavored the molecular interpretation of this state.

A mixed charmonium-molecule scenario was employed to explain the
$Y(3940)$ in Ref. \cite{Albuquerque:2013owa}, in which the authors
used a $\chi_{c0}-D^*\bar D^*$ current with $J^{PC}=0^{++}$ to
compute the correlation function. For the mixing angle
$\theta=(76.0\pm5.0)^\circ$, the mass was extracted as
$M=(3.95\pm0.11)$ GeV and the partial decay width $\Gamma_{Y\to
J/\psi\omega}=(1.7\pm0.6)$ MeV.

In Refs. \cite{Finazzo:2011he,Wang:2011uk}, the $Y(4274)$ was
studied as a $\bar D_sD_{s0}$ hidden-strange charmonium-like
molecular state with $J^{PC}=0^{-+}$. The extracted masses were
heavier than the mass of the $Y(4274)$. The mass of the $\bar
DD_{0}$ was extracted to be $(4.55\pm0.49)$ GeV.

\subsubsection{Other theoretical schemes, production and decay patterns}
\label{Sect:4.7.2}

Besides the molecular picture, the $Y(3940)$, $Y(4140)$ and
$Y(4274)$ were investigated in some other theoretical
frameworks~\cite{Gershtein:2006ng,Zhang:2006td,Maiani:2004vq,Ebert:2005nc}.
In Ref.~\cite{Gershtein:2006ng}, Gershtein, Likhoded, and Luchinsky
systematically studied heavy quarkonia from Regge trajectories on
$(n,M^2)$ and $(M^2, J)$ planes, and interpreted the $Y(3940)$ as
the $\chi_{c0}(2P)$ charmonium state. This was further discussed
combining the linearity and parallelism of Regge trajectories with a
hyperfine splitting relation in multiplet in
Ref.~\cite{Zhang:2006td}.

In the ``type-I'' diquark-antidiquark model proposed by Maiani,
Piccinini, Polosa, and Riquer~\cite{Maiani:2004vq}, the $Y(3940)$
was interpreted as the $2^{++}$ $S$-wave state, i.e., $Y(3940) =
|2^{++}\rangle = |1_{cq}, 1_{\bar c\bar q^\prime}; J = 2\rangle$. In
Ref.~\cite{Ebert:2005nc}, Ebert, Faustov, and Galkin studied masses
of heavy tetraquarks with hidden charm and bottom in the framework
of the relativistic quark model, and their results supported
the $Y(3940)$ as a $2^{++}$
diquark-antidiquark tetraquark, i.e., $([cq]_{S=1}[\bar c \bar
q]_{S=1})_{J=2}$.

In Ref.~\cite{Stancu:2009ka}, Stancu studied the spectrum of
tetraquarks of type $c \bar cs\bar s$ within a simple quark model
with chromomagnetic interaction, and suggested that the $Y(4140)$
could possibly be the strange partner of the $X(3872)$ in a tetraquark
interpretation. Later in Ref.~\cite{Patel:2014vua}, Patel, Shah, and
Vinodkumar calculated masses of the low-lying four-quark states in
the hidden charm sector within the framework of a non-relativistic
quark model, and they found that the $Y(4140)$ can be interpreted as
the diquark-antidiquark ($c q - \bar c\bar q$) state, while it can
also be interpreted as a $c\bar q - \bar cq$ molecular-like state
only if its parity is positive.

The coupled-channel effects, threshold effects and nonresonant
explanation for the $Y(3940)$, $Y(4140)$ and $Y(4274)$ were
investigated in
Refs.~\cite{Bugg:2004sh,Danilkin:2009hr,vanBeveren:2009dc}. They
were also proposed as dynamically generated resonances from the
vector-vector interaction within the framework of the hidden gauge
formalism in Refs.~\cite{Molina:2009ct,Branz:2010rj}. Their
productions in the $B$ meson decays were studied using the same
approach in Ref.~\cite{Liang:2015twa}. In
Ref.~\cite{HidalgoDuque:2012pq}, Hidalgo-Duque, Nieves, and Pavon
Valderrama proposed an effective field theory incorporating light
SU(3)-flavour and heavy quark spin symmetry to describe charmed
meson-antimeson bound states. Assuming that the $X(3915)$ and
$Y(4140)$ are $D^* \bar D^*$ and $D^*_s \bar D^*_s$ molecular
states, they determined the full spectrum of molecular states with
isospin $I = 0$, $1/2$ and $1$.

The weak productions of the $Y(3940)$, $Y(4140)$, and $Y(4274)$ in the
semi-leptonic $B_c$ decays were studied using light-cone QCD sum
rules in Ref.~\cite{Wang:2007fs}. The inclusive production of
the $\chi_{cJ}$ in the $\eta_b$ decays was studied in
Ref.~\cite{He:2009sm}, which may help resolve whether some of these
states are excited charmonia. In Ref.~\cite{He:2009yda}, He and Liu
investigated the discovery potential of the $Y(3940)$ via the
photoproduction process $\gamma p \to Y(3940)p$.

Many authors investigated the decay behavior of the $Y(3940)$,
$Y(4140)$, and $Y(4274)$, including their hidden-charm decay,
open-charm decay, radiative decay and double-photon decay. The
hidden charm decay of the $Y(3940)$ was investigated in
Ref.~\cite{Liu:2006df}, considering the final state interaction
effect. As indicated in Ref.~\cite{Liu:2009ei}, the line shapes of
the photon spectrum of $Y(4140)\to {D}_s^{\ast+} D_s^- \gamma$ and
$Y(3940)\to{D}^{\ast+} D^-\gamma$ are crucial to test the molecule
assignment of the $Y(4140)$ and $Y(3940)$. The
radiative decay of the $Y(3940)$ and $Y(4140)$ was later calculated in
Ref.~\cite{Liu:2009pu}.

In Ref.~\cite{Liu:2009iw}, Liu studied the hidden-charm decay of
the $Y(4140)$ assuming it as the second radial excitation of the
P-wave charmonium $\chi_{cJ}^{\prime\prime}$ ($J = 0, 1$).
The upper limit of the branching ratio of the
hidden charm decay $Y(4140)\to J/\psi\phi$ is of the order of
$10^{-4}\sim 10^{-3}$ for both charmonium assumptions for
the $Y(4140)$, which disagrees with the large hidden charm decay pattern
indicated by the CDF experiment~\cite{Aaltonen:2009tz}.
The assumption of the $Y(4140)$ as the second radial excitation of the
P-wave charmonium $\chi_{cJ}^{\prime\prime}$ ($J=0,\,1$) is problematic.

In Ref.~\cite{He:2011ed}, He and Liu investigate decay widths
and line shapes of the open-charm radiative and pionic decays of
the $Y(4274)$ with the $D_s \bar D_{s0}(2317)$ molecular charmonium
assignment. Their calculation indicated that the decay widths of
$Y(4274) \to D^+_s D^{*-}_s \gamma$ and $Y(4274) \to D^+_s D^-_s
\pi^0$ can reach up to 0.05 keV and 0.75 keV, respectively. The
authors suggested future experiments to search for
the open-charm radiative and pionic decays of the $Y(4274)$.

\subsubsection{A short summary}
\label{Sect:4.7.3}

\begin{itemize}

\item The $Y(3940)$, $Y(4140)$, and $Y(4274)$ can be interpreted as the
$D^*\bar{D}^*$, $D_s^*\bar{D}_s^*$, and $D_s \bar D_{s0}(2317)$
molecular states, respectively.

\item Some authors interpreted these states as the charmonium states,
diquark-antidiquark tetraquark states, etc. There are also
non-resonant explanations.

\end{itemize}

\subsection{Other charmonium-like states}\label{Sect:4.8}

\subsubsection{$Y(4008)$ and $Y(4360)$}
\label{Sect:4.8.1}

There are not so many theoretical studies on the $Y(4008)$ and
$Y(4360)$, whose experimental information has been reviewed in
Sec.~\ref{Sect:2.2.1} and \ref{Sect:2.2.2}, respectively. In the
following, we separately introduce their theoretical research
status.

\paragraph{$Y(4008)$}

In Ref.~\cite{Liu:2007ez}, Liu discussed some possible assignments
for the $Y(4008)$, including both the $\psi(3S)$ charmonium states
and the $D^*\bar{D}^*$ molecular state. Within both pictures, he
found that the branching ratio of $Y(4008) \to J/\psi \pi^0 \pi^0$
is comparable with that of $Y(4008) \to J/\psi \pi^+ \pi^-$. He also
studied other hidden-charm and open-charm decays, and proposed
further experiments to search for missing channels $D \bar D$, $D
\bar D^*$, $\chi_{cJ} \pi^+ \pi^- \pi^0$, and $\eta_c \pi^+ \pi^-
\pi^0$.

These two assignments were also investigated in later studies. In
Ref.~\cite{Li:2009zu}, Li and Chao studied the higher charmonium
states in the non-relativistic screened potential model, and
interpreted the $Y(4008)$ as the $\psi(3S)$ charmonium state. In
Ref.~\cite{Chen:2010nv}, Chen, Ye, and Zhang studied strong decays
of the radially excited $\psi(3^3S_1)$ state within the $^3P_0$
model. They found that the $Y(4008)$ is hard to be identified with a
$\psi(3^3S_1)$ charmonium if it is confirmed to be below the $D^*
\bar D^*$ threshold by experiment. However, it is probably a
$\psi(3^3S_1)$ charmonium once it is above the $D^* \bar D^*$
threshold.

In Ref.~\cite{Ding:2009zq}, Ding studied the $D^*\bar{D}^*$ system
dynamically in the one boson exchange model, and found the
interpretation of the $Y(4008)$ as a $D^* \bar D^*$ molecule is not
favored by its huge width, although it is close to the $D^* \bar
D^*$ threshold. However, in Ref.~\cite{Xie:2013uha}, Xie, Mo, Wang,
and Cotanch studied tetraquark states with hidden charm within an
effective Coulomb gauge Hamiltonian approach, and found that the $Y
(4008)$ can be interpreted as the lightest $1^{--}$ molecule with
the $\eta h_c$ type structure.

The $Y(4008)$ was studied in Ref.~\cite{Maiani:2014aja} by Maiani, Piccinini, Polosa, and Riquer
in their ``type-II'' diquark-antidiquark model, and interpreted as
$Y(4008) = Y_1 = |0_{cq}, 0_{\bar c \bar q}; 0, 1\rangle_1$. In Ref.~\cite{Zhou:2015frp}, Zhou,
Deng, and Ping also interpreted the $Y(4008)$ as a tetraquark state $[cq][\bar c \bar q]$ with
$I~J^{PC} = 0~1^{--}$. They used a color flux-tube model with a
four-body confinement potential, and interpreted the $Y(4008)$ as a
tetraquark state $[cq][\bar c\bar q]$ with $n^{2S+1}L_J$ of
$1^1P_1$.

In Ref.~\cite{Chen:2015bft}, Chen {\it et al.} proposed that the $Y(4008)$ is
not a genuine resonance, where the broad structure corresponding to
the $Y(4008)$ in $e^+e^-\to J/\psi\pi^+\pi^-$ can be reproduced when
introducing the interference between the continuum and background
contributions.

\paragraph{$Y(4360)$}

The $Y(4360)$ was
interpreted as the $\psi(3D)$ charmonium state in
Ref.~\cite{Li:2009zu} by Li and Chao using the nonrelativistic
screened potential model. In Ref.~\cite{Ding:2007rg}, Ding, Zhu, and
Yan also interpreted the $Y(4360)$ as a $3^3D_1$ $c \bar c$ state,
and applied the flux tube model to evaluate its $e^+e^-$ leptonic
widths, E1 transitions, M1 transitions and the open flavor strong
decays. It was interpreted as the $2S$ bound state in the $D_1 \bar
D^*$ system~\cite{Close:2009ag,Close:2010wq}, a hadrocharmonium
state~\cite{Li:2013ssa}, a tetraquark
state~\cite{Ebert:2008kb,Zhou:2015frp}, a baryonium
state~\cite{Qiao:2007ce,Chen:2011cta}, and a charmonium hybrid
state~\cite{Kalashnikova:2008qr,Qiao:2010zh}.

The $Y(4360)$ was interpreted as the $2S$ bound state in the $D_1
\bar D^*$ system in Refs.~\cite{Close:2009ag,Close:2010wq}, while the
$Y(4260)$ was assumed to be the $1S$ state. In
Ref.~\cite{Li:2013ssa}, Li and Voloshin studied the $Y(4360)$ within
the hadrocharmonium picture, where a (relatively) compact charmonium
was embedded in a light quark mesonic excitation. They suggested
that the $Y(4260)$ and $Y(4360)$ are a mixture of two hadrocharmonium
states, one containing a spin-triplet $c \bar c$ pair and the other
containing a spin-singlet heavy quark pair. Based on this picture,
they found a distinctive pattern of interference between the
resonances.

In Ref.~\cite{Maiani:2014aja}, Maiani, Piccinini, Polosa, and Riquer
studied the $Y(4360)$ in their ``type-II'' diquark-antidiquark model, and interpreted it as
as the radial excitation of the $Y(4008)$, which has been reviewed in the previous subsection, i.e. $Y(4008) = Y_1 = |0_{cq}, 0_{\bar c \bar q}; 0, 1\rangle_1$.
In Ref.~\cite{Ebert:2008kb}, Ebert, Faustov, and Galkin calculated
the masses of the excited heavy tetraquarks with hidden charm within
the relativistic diquark-antidiquark picture, and found that the
$Y(4360)$ can be interpreted as the excited $1^{--}$ $1P$ $[cq][\bar
c \bar q]$ tetraquark state consisting of $A\bar A$, where $A$ is an
axial vector diquark. In contrast, the $Y(4360)$ was interpreted as a
tetraquark state $[cq][\bar c\bar q]$ with $n^{2S+1}L_J$ of $1^5F_1$
using a color flux-tube model by Zhou, Deng, and Ping in
Ref.~\cite{Zhou:2015frp}.

In Ref.~\cite{Qiao:2007ce}, Qiao proposed that the $Y(4360)$,
together with the $Y(4260)$, $Y(4660)$ and $Z^+(4430)$, can be
systematically embedded into an extended baryonium picture. Later in
Ref.~\cite{Chen:2011cta}, Chen and Qiao derived the two-pion
exchange interaction potential between heavy baryon and heavy
anti-baryon to see whether they can form a bound state. They used
the obtained potential to calculate heavy baryonium masses by
solving the Schr\"odinger equation, and found that the $Y(4360)$ could
be interpreted as a $\Lambda_c$-$\bar \Lambda_c$ bound state.

In Ref.~\cite{Kalashnikova:2008qr}, Kalashnikova and Nefediev
employed the QCD string model to calculate the masses and spin
splittings of the lowest charmonium hybrid states with a magnetic
gluon. The mass of the vector charmonium hybrid state is 4.397 GeV.
They argued that strong coupling of the vector hybrid to the $D\bar
D_1$ and $D^* \bar D_0$ modes can cause considerable threshold
attraction, leading to the formation of the $Y(4360)$. Later in
Ref.~\cite{Qiao:2010zh}, Qiao {\it et al.} evaluated the masses of the
$1^{--}$ charmonium and bottomonium hybrids in terms of QCD sum
rules. They found that the hybrid ground state in charm sector lies
in 4.12--4.79 GeV, whose mass resides between the $Y(4360)$ and
$Y(4660)$. Hence, they suggested that the $Y(4360)$, as well as the
$Y(4660)$, might be charmonium hybrid candidates.

Moreover, a non-resonant description of the charmonium-like
structure $Y(4360)$ was proposed in Ref.~\cite{Chen:2011kc}. The
authors found that the $Y(4360)$ structure can be depicted well by
the interference effect of the production amplitudes of $e^+e^- \to
\psi(2S) \pi^+ \pi^-$ via the intermediate charmonia
$\psi(4160)/\psi(4415)$ and direct $e^+e^-$ annihilation into
$\psi(2S)\pi^+\pi^-$ (similar to that in Fig. \ref{Fig.4.6.Fig-Feyn1}). They
argued that the $Y(4360)$ is not a genuine resonance, which explains
why the $Y(4360)$ was only observed in the hidden-charm decay
channel $\psi(2S)\pi^+\pi^-$ and not observed in the exclusive
open-charm decay channel, nor the $R$-value scan (see Figures
\ref{Fig.4.6.open-charm-Y4260} and \ref{Fig.4.6.R-4260} for more details). In
Ref.~\cite{Chen:2015bft}, Chen {\it et al.} further indicated that
the $Y(4360)$, $Y(4260)$ and $Y(4008)$ can be due to the Fano-like
interference~\cite{Chen:2011kc}.

The Initial Single Pion Emission mechanism was also used to study
the hidden-charm dipion decays of the $Y(4360)$ in
Ref.~\cite{Chen:2013bha}. The authors found that there exist charged
charmoniumlike structures near the $D\bar D^*$ and $D^* \bar D^*$
thresholds in the $J/\psi \pi^+$, $\psi(2S)\pi^+$ and $h_c(1P)\pi^+$
invariant mass spectra of the hidden-charm dipion decays of the
$Y(4360)$.

\subsubsection{$Y(4660)$ and $Y(4630)$}
\label{Sect:4.8.2}

\paragraph{$Y(4660)$}

The $Y(4660)$, which was reviewed in Sec.~\ref{Sect:2.2.2}, was
observed in the initial-state radiation process $e^+e^- \to
\gamma_{\mathrm{ISR}}Y(4660) (\to \pi^+\pi^-
\psi(3686))$~\cite{Wang:2007ea}. In Ref.~\cite{Ding:2007rg}, Ding,
Zhu, and Yan suggested that the $Y(4660)$ is a good candidate of the
$5^3S_1$ $c\bar c$ state, and evaluated its $e^+e^-$ leptonic
widths, E1 transitions, M1 transitions and the open flavor strong
decays in the flux tube model. In contrast, the $Y(4660)$ was
assigned as the $\psi(6S)$ charmonium state in the screened
potential model by Li and Chao in Ref.~\cite{Li:2009zu}. However, in
Ref.~\cite{vanBeveren:2010jz}, van Beveren and Rupp analyzed the
shape of the threshold signals in the production cross sections of
the reaction $e^+e^- \to D^* \bar D^*$~\cite{Aubert:2009aq}, and
argued that the $Y(4660)$ should not be associated with the
resonance poles of the $c\bar c$ propagator.

Besides the charmonium state, the $Y(4660)$ was interpreted as a
$f_0(980) \psi^\prime$ bound state in
Refs.~\cite{Guo:2008zg,Guo:2009id} and a tetraquark state in
Refs.~\cite{Ebert:2008kb,Maiani:2014aja}, etc. The hadro-charmonium
picture was also proposed to explain the $Y(4660)$ as a compact
charmonium resonance bound inside an excited state of light hadronic
matter by Dubynskiy and Voloshin in Ref.~\cite{Dubynskiy:2008mq}.

In Ref.~\cite{Guo:2008zg}, Guo, Hanhart, and Meissner assumed that
the $Y(4660)$ is a $f_0(980) \psi^\prime$ bound state, and
calculated the invariant mass spectrum of $\psi^\prime \pi^+ \pi^-$
as well as the corresponding $\pi\pi$ and $\bar K K$ spectra in its
mass range. They obtained a good description of both spectra,
which suggests that the $Y(4660)$ may be generated dynamically
in the $f_0(980) \psi^\prime$ channel. They further proposed to
measure the $\psi^\prime \bar K K$ channel as a nontrivial test of
this hypothesis. Guo, Hanhart, and Meissner also used the heavy
quark spin symmetry to study heavy meson hadronic molecules, and
predicted an $f_0(980) \eta_c^\prime$ bound state as the
spin-doublet partner of the $Y(4660)$~\cite{Guo:2009id}. Its mass
was evaluated to be $4616^{+5}_{-6}$ MeV, and was suggested to
mainly decay into $\eta^\prime \pi\pi$ with a width of $60\pm30$
MeV. They also predicted its decays into $\eta^\prime_c K^+K^-$,
$\eta^\prime_c \gamma \gamma$ and $\Lambda^+_c \Lambda^-_c$, and
proposed to search for this state in the $B^\pm \to \eta^\prime_c K^\pm
\pi^+ \pi^-$ decay.

In Ref.~\cite{Maiani:2014aja}, Maiani, Piccinini, Polosa, and Riquer
studied the $Y(4660)$ in their ``type-II'' diquark-antidiquark
model, and interpreted it as the radial excitation of the $Y(4260)$,
i.e., $Y (4260) = Y_2 = {1\over\sqrt2} \Big( |1, 0; 1, 1\rangle_1 +
|0, 1; 1, 1\rangle_1 \Big)$, which has been reviewed in
Sec.~\ref{Sect:4.6}. In Ref.~\cite{Ebert:2008kb}, Ebert, Faustov,
and Galkin calculated the masses of the excited heavy tetraquarks
with hidden charm within the relativistic diquark-antidiquark
picture, and found that the $Y(4660)$ can be interpreted as the
excited $1^{--}$ $2P$ $[cq][\bar c \bar q]$ tetraquark state
consisting of $S\bar S$, where $S$ is a scalar diquark. The authors
had used the $1^{--}$ $1P$ $[cq][\bar c \bar q]$ state to explain
the $Y(4260)$ in their model.

The authors of Ref. \cite{Chen:2010ze} studied the
charmonium-like tetraquark states with $J^{PC}=1^{--}$ in QCD sum
rules using the interpolating currents listed in Eq.
\eqref{Eq.4.6.tetraquarkcurrents1}. The obtained masses were
collected in Table \ref{Table:4.6.tetraquark}. The masses of the
$qc\bar q\bar c$ and $sc\bar s\bar c$ tetraquark states were
extracted around $4.5-4.8$ GeV and $4.4-5.0$ GeV, respectively.
These masses were consistent with the mass of the $Y(4660)$. The
$Y(4660)$ was also proposed as the $\psi(2S)f_0(980)$ molecular
state in Ref. \cite{Albuquerque:2011ix}. The authors used a $\bar
cc\bar ss$ molecular current in QCD sum rules and obtained the mass
$m=(4.67\pm0.09)$ GeV.

The initial single chiral particle emission mechanism was proposed
to study the hidden-charm di-kaon decays of the $Y(4660)$ in
Ref.~\cite{Chen:2013wca}. The authors calculated the distributions
of differential decay width, and obtained the line shape of the
$J/\psi K^+$ invariant mass spectrum of $Y(4660) \to J/\psi K^+
K^-$. Their results suggested that there may exist enhancement
structures with both hidden-charm and open-strange decays near the
$D \bar D^*_s/D^* \bar D_s$ and $D^* \bar D^*_s/\bar D^* D^*_s$
thresholds. The hidden-charm di-eta decays of the $Y(4660)$ was
studied with the same approach in Ref.~\cite{Chen:2013axa}. The
$Y(4660)$ was used to search for the missing $\psi(4S)$ state in
Ref.~\cite{Chen:2015bma}.

\paragraph{$Y(4630)$}

The $Y(4630)$, reviewed in Sec.~\ref{Sect:2.2.3}, was observed in
the exclusive $e^+e^- \to \Lambda_c \bar \Lambda_c$ cross
section~\cite{Pakhlova:2008vn}, which might be interpreted as a
baryonium state~\cite{Lee:2011rka}. The authors of
Ref.~\cite{Lee:2011rka} performed a systematic study of the possible
loosely bound states composed of two charmed baryons or a charmed
baryon and an anti-charmed baryon within the framework of the OBE
model, where the exchanged bosons include pseudoscalar mesons $\pi$
and $\eta$, vector mesons $\rho$, $\omega$, and $\phi$, and the
scalar meson $\sigma$. They also considered the $S$-$D$ mixing
effects for the spin-triplets. Especially, their investigation
indicated that there does exist strong attraction through the
$\sigma$ and $\omega$ exchanges in the $\Lambda_c \bar \Lambda_c$
channel, which suggested that the $Y(4630)$ may be interpreted as a
$\Lambda_c \bar \Lambda_c$ bound state.

In Ref.~\cite{Simonov:2011jc}, Simonov investigated the
nonperturbative baryon-antibaryon production due to the double quark
pair $(q\bar q)(q\bar q)$ generation inside a hadron. They applied
this mechanism to study the electroproduction of $\Lambda_c \bar
\Lambda_c$, and found an enhancement near 4.61 GeV. This structure
was in agreement with experimental data~\cite{Pakhlova:2008vn}, and
was used to explain the $Y(4630)$.

In Ref.~\cite{Qiao:2007ce}, Qiao proposed that the $Y(4660)$,
together with the $Y(4260)$, $Y(4360)$ and $Z^+(4430)$, can be
systematically embedded into an extended baryonium picture. In
Ref.~\cite{Chen:2011cta}, Chen and Qiao argued that it is not the
$Y(4660)/Y(4630)$, but the $Y(4260)$ and $Y(4360)$, which could be
interpreted as $\Lambda_c \bar \Lambda_c$ bound states.

In Ref. \cite{Liu:2016sip}, the authors studied the
open-charm decay $Y(4630)\to \Lambda_c\bar{\Lambda}_c$ by assuming
that the $Y(4630)$ is a charmonium-like tetraquark made of a
diquark and an anti-diquark. Their results shows that the
$Y(4630)$ could be a radially excited state of the
diquark-antidiquark bound state.

\paragraph{Are $Y(4630)$ and $Y(4660)$ the same state?}

The $Y(4660)$ and $Y(4630)$ were observed in different processes,
i.e., $e^+e^- \to \gamma_{\mathrm{ISR}} \pi^+\pi^-
\psi(3686)$~\cite{Wang:2007ea} and exclusive $e^+e^- \to \Lambda_c
\bar \Lambda_c$~\cite{Pakhlova:2008vn}, respectively. However, their
masses and widths are consistent with each other within
errors~\cite{Pakhlova:2008vn}. Hence, they can be the same
state/structure as pointed out in
Refs.~\cite{Bugg:2008sk,Cotugno:2009ys,Guo:2010tk}.

In Ref.~\cite{Bugg:2008sk}, Bugg used the $Y(4660)$ to explain the
peak observed in $\Lambda_c \bar \Lambda_c$ channel at 4.63
GeV~\cite{Pakhlova:2008vn}, and suggested that a form factor with a
reasonable radius of the interaction can provide an explanation of
the shift of mass between the $Y(4630)$ and $Y(4660)$. In
Ref.~\cite{Cotugno:2009ys}, Cotugno, Faccini, Polosa, and Sabelli
analyzed the data on the $Y(4630) \to \Lambda_c \bar
\Lambda_c$~\cite{Pakhlova:2008vn} and the $Y(4660) \to \psi(3686)
\pi \pi$~\cite{Wang:2007ea}. They suggested that the $Y(4630)$ and
$Y(4660)$ correspond to a single state, called $Y_B(4660)$ with
$M_{Y_B} = 4660.7 \pm 8.7$ MeV and $\Gamma_{Y_B} = 61 \pm 23$ MeV.
They further argued that the $Y_B(4660)$ is an excellent candidate for a
$[cd][\bar c \bar d]$ diquark-antidiquark bound state.

In Ref.~\cite{Guo:2010tk}, Guo {\it et al.} considered the $\Lambda_c \bar
\Lambda_c$ final state interaction and proposed that the $Y(4630)$
may be described as the same state as the $Y(4660)$, which is
assumed as a $f_0(980) \psi^\prime$ bound state~\cite{Guo:2008zg}.
Moreover, the $Y(4660)$ was suggested to have a spin partner, which
is the $f_0(980) \eta_c^\prime$ bound state~\cite{Guo:2009id}. They
discussed this state, and proposed to measure the $B$ decays to
$K \Lambda_c \bar \Lambda_c$ and $K \eta_c^\prime \pi^+ \pi^-$
to test the hypothesis that the $Y(4630)$ and $Y(4660)$ are the same
molecular state.

\subsubsection{$X(3915)$, $X(4350)$ and $Z(3930)$}
\label{Sect:4.8.3}

\paragraph{$Z(3930)$}

Until now, three charmonium-like states have been reported via the
$\gamma\gamma$ fusion process by Belle (see review in Sec.
\ref{Sec:2.4}). Among them, the $Z(3930)$ is a very good candidate of
the charmonium $\chi_{c2}^\prime$ with $n^{2s+1}J_L=2^3P_2$
\cite{Uehara:2005qd,Aubert:2010ab}. According to PDG \cite{pdg},
the $Z(3930)$ is the $\chi^{\prime}_{c2}(2P)$ charmonium state and the $X(3915)$ is
the $\chi_{c0}(2P)$ charmonium state. The hyperfine splitting between the $Z(3930)$ and
$X(3915)$ is only $6\%$ of that between the $\chi_{c2}(1P)$ and
$\chi_{c0}(1P)$ \cite{Esposito:2014rxa}, which is unexpectedly
smaller than the potential model prediction \cite{Barnes:2005pb}.
Such a splitting was also much smaller than the corresponding
splitting of $m_{\chi_{b2}^\prime}-m_{\chi_{b0}^\prime}$ \cite{pdg}.
These puzzles still challenge the P-wave charmonium assignments of
the $Z(3930)$ and $X(3915)$
\cite{Guo:2010ak,Olsen:2014maa,Olsen:2014qna}.

\paragraph{$X(3915)$}

\begin{figure}[htb]
\begin{center}
\scalebox{0.96}{\includegraphics{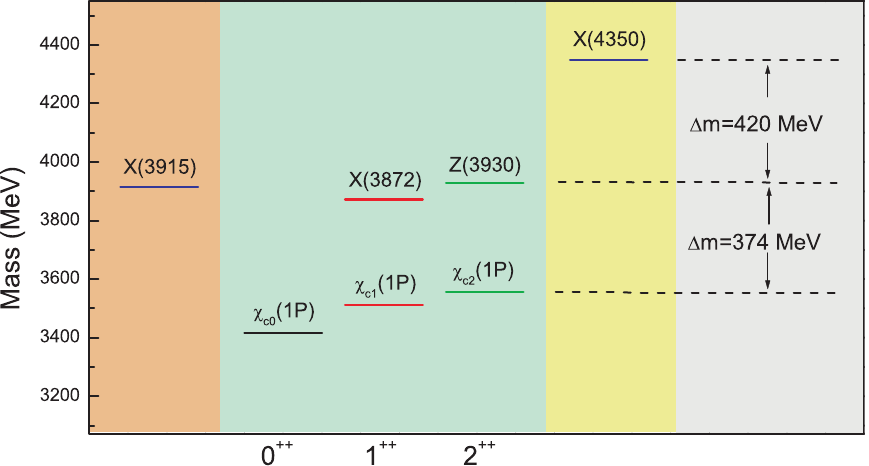}}
\caption{(Color online) The P-wave charmonium states \cite{pdg} and
the candidates for their first radial excitations. Taken from  Ref.
\cite{Liu:2009fe}. \label{Fig.4.8.review}}
\end{center}
\end{figure}

The mass of the $\chi_{c0}^{\prime}$ charmonium state was predicted to be around 3916 MeV
in the GI model \cite{Godfrey:1985xj,Barnes:2005pb}, as shown in
Fig. \ref{Fig.4.5.CF}. In Ref. \cite{Liu:2009fe}, the charmonium-like state
$X(3915)$ was proposed as the first radial excitation of
the $\chi_{c0}(3415)$. The authors argued that the $Z(3930)$, $X(3872)$ and
$X(3915)$ may fill in the spin-triplet 2P charmonium states as shown
in Fig. \ref{Fig.4.8.review} if one considers the strong coupled channel
effects in the $J^{PC}=1^{++}$ channel which may lower the mass of
the $X(3872)$ \cite{Kalashnikova:2005ui,Li:2009zu}.

Assuming the $X(3915)$ is the $\chi_{c0}(2P)$ charmonium state, the authors of Ref.
\cite{Jiang:2013epa} calculated the OZI-allowed open-charm decay
$X(3915)\to D\bar D$ in the $^3P_0$ model with the Bethe-Salpeter
method. They found that the node structure in the $\chi_{c0}(2P)$
wave function leads to the narrow decay width of the $X(3915)$. They
suggested a partial decay width ratio $\Gamma(X(3915)\to
D^+D^-)/\Gamma(X(3915)\to D^0\bar D^0)=2.3$ to test the
$\chi_{c0}(2P)$ assignment of the $X(3915)$ \cite{Jiang:2013epa}.
However, Yang, Xia and Ping calculated the strong decay widths of
the $X(3915)$ as the $\chi_{c0}(2P)$ charmonium state and obtained a much larger strong
decay width than the experiment data \cite{Yang:2009fj}. A similar
conclusion was obtained in Ref. \cite{Zhao:2011sd}.

In Ref. \cite{Guo:2010ak}, Guo {\it et al.} doubted the
$\chi_{c0}^\prime$ assignment of the $X(3915)$. The OZI-allowed
open-charm decay $X(3915)\to D\bar D$ generally has a larger width
than that of the $X(3915)$. The mass splitting
$m_{Z(3930)}-m_{X(3915)}=14\pm6$ MeV was too small compared with
$m_{\chi_{c2}}-m_{\chi_{c0}}=141$ MeV. The mass splitting
$m_{Z(3930)}-m_{X(3915)}$ was even smaller than
$m_{\chi_{b2}^\prime}-m_{\chi_{b0}^\prime}=36.2\pm0.8$ MeV
\cite{pdg}. The authors suggested a signal for the $\chi_{c0}(2P)$
with a mass around 3840 MeV and width about 200 MeV in the Belle
\cite{Uehara:2005qd} and BaBar \cite{Aubert:2010ab} data for
$\gamma\gamma\to D\bar D$ in Ref. \cite{Guo:2012tv}.

\begin{figure}[h!]
\centering \scalebox{0.6}{\includegraphics{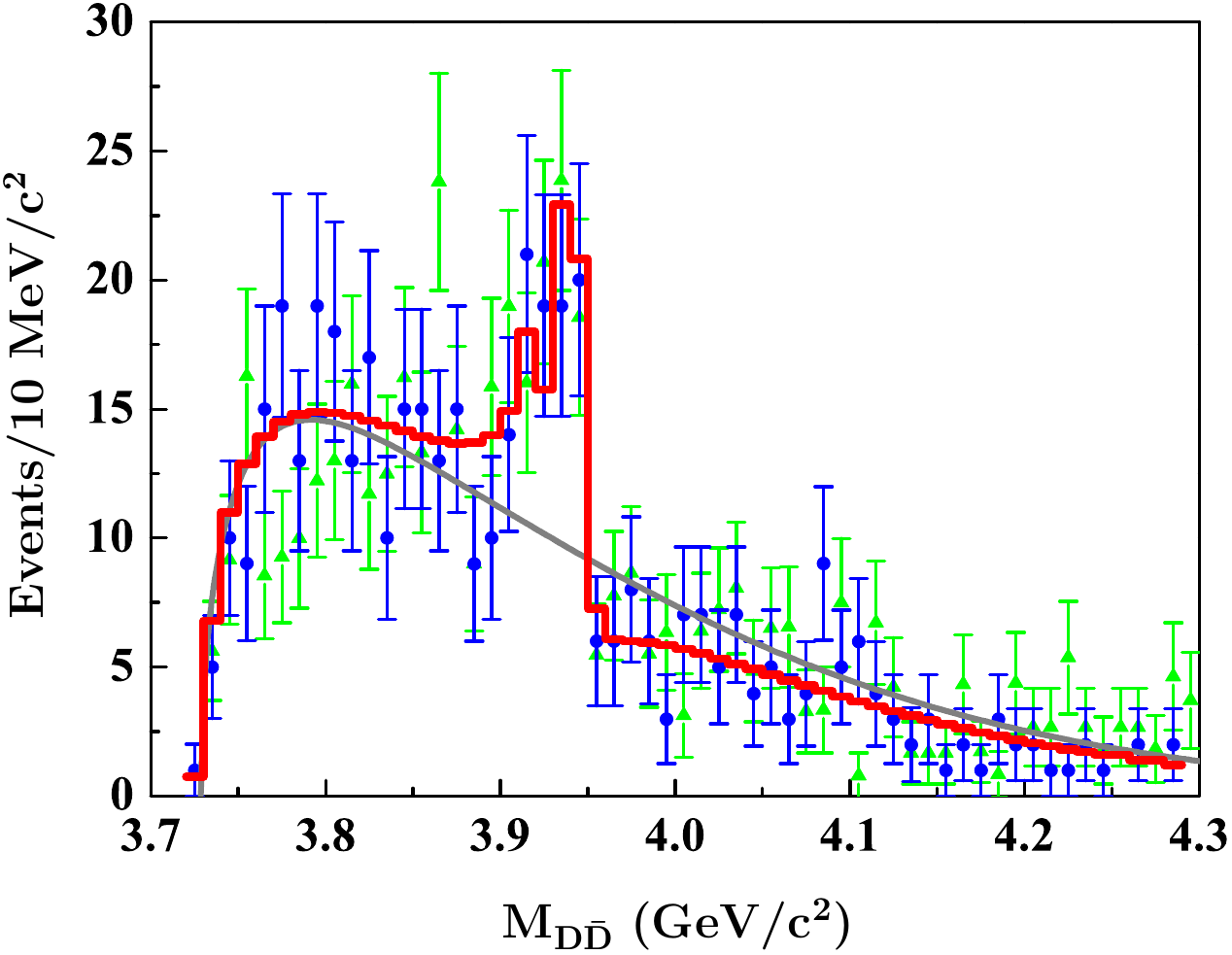}} \centering
\scalebox{0.6}{\includegraphics{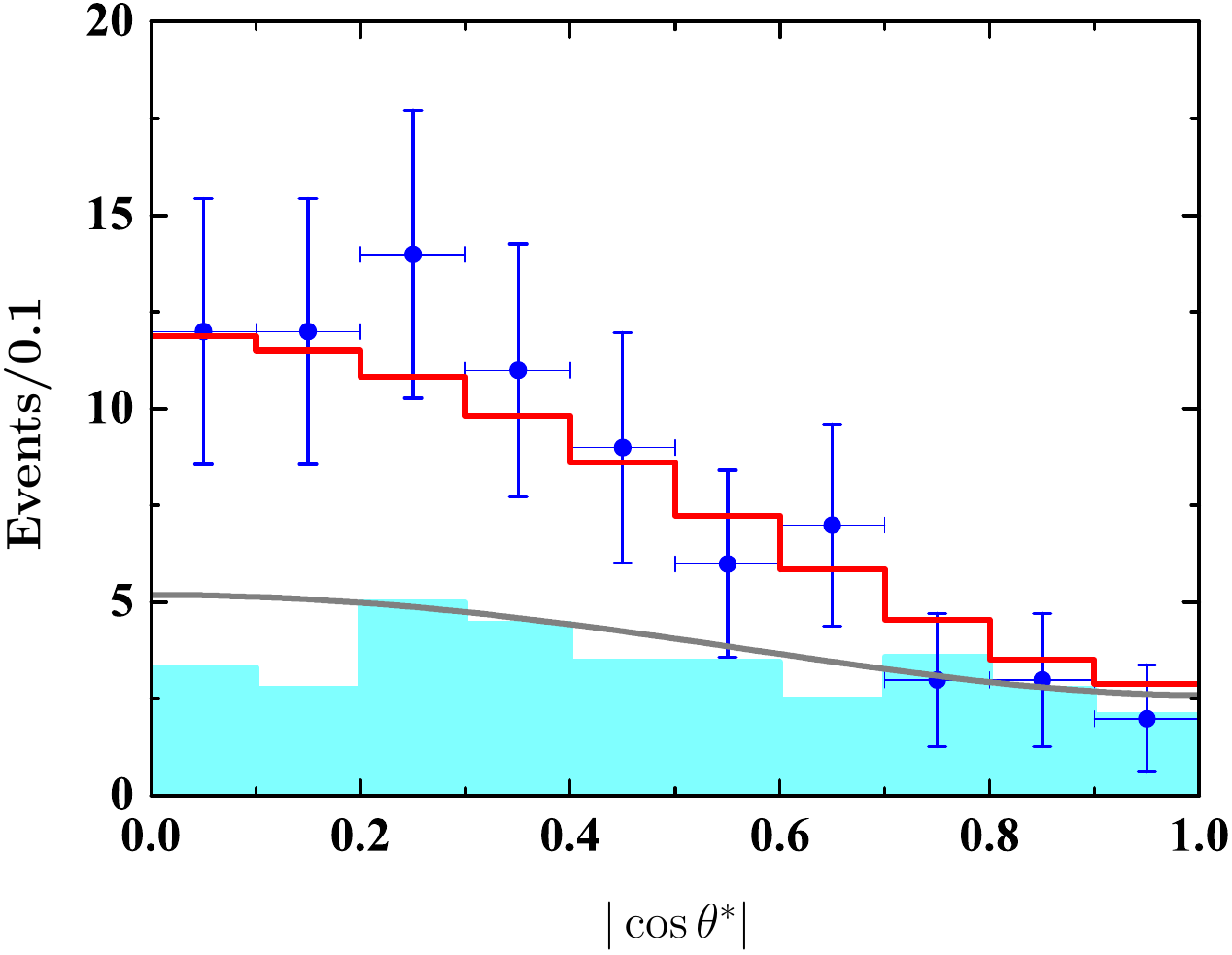}}
\put(-330,-20){$(a)$}%
\put(-100,-20){$(b)$}%
\caption{(Color online) (a) The best fit (red histogram) to the
experimental data of the $D\bar{D}$ invariant mass distributions
given by Belle \cite{Uehara:2005qd} (blue dots with error bar) and
BaBar \cite{Aubert:2010ab} (green triangles with error bar). (b) The
best fit (red histogram) to the $\cos \theta^\ast$ distribution of
$\gamma\gamma\to D\bar{D}$. Taken from Ref. \cite{Chen:2012wy}.
}\label{Fig.4.8.mDD}
\end{figure}

The absence of the $X(3915)$ in $\gamma\gamma\to D\bar{D}$ also
challenges the $\chi_{c0}^\prime(2P)$ assignment of the $X(3915)$ since
the $D\bar{D}$ was argued to be the dominant decay mode of
the $\chi_{c0}^\prime(2P)$
\cite{Liu:2009fe}. In Ref. \cite{Chen:2012wy}, Chen {\it et al.}
pointed out that the $Z(3930)$ enhancement structure may contain
both the $\chi_{c0}(2P)$ and $\chi_{c2}(2P)$ signals, according to their
analysis of the $D\bar{D}$ invariant mass spectrum and
$\cos\theta^\ast$ distribution of $\gamma\gamma\to D\bar{D}$ (see
Fig. \ref{Fig.4.8.mDD} ).

In Refs. \cite{Olsen:2014maa,Olsen:2014qna}, Olsen pointed out that
the $\chi_{c0}(2P)$ assignment of the $X(3915)$ was in conflict with
the experiment data. If the $X(3915)$ was the $\chi_{c0}(2P)$, the
branching fraction was estimated to be ${\cal
B}(\chi_{c0}(2P)\to\omega J/\psi)<7.8\%$, which was smaller than the
$14.3\%$ lower limit derived for the same quantity from the $B\to
KX(3915)$ decay rate.

The authors of Ref. \cite{Zhou:2015uva} combined the analysis for
the amplitude and the angular distribution of the $\gamma\gamma\to
D\bar D$ and $\gamma\gamma\to J/\psi\omega$ data from BaBar
\cite{Aubert:2010ab,Uehara:2009tx}. They found that the assignment
of $2^{++}$ to the $X(3915)$ was more consistent with the data. The
authors argues that the $X(3915)$ and $Z(3930)$ are the same tensor
state.

\paragraph{$X(4350)$}

The mass of the $\chi_{c2}(3P)$ was predicted to be $4337$ MeV in the GI
model \cite{Barnes:2005pb}. Hence, the $X(4350)$ was interpreted as
the $\chi_{c2}(3P)$ charmonium state in Ref. \cite{Liu:2009fe}. The authors studied its
two-body open-charm decay behaviors within the quark pair creation
model. The calculations from QCD sum rules in Refs.
\cite{Chen:2010ze,Albuquerque:2010fm,Mo:2014nua,Wang:2015pea}
disfavored the assignment of the $X(4350)$ as the exotic
charmonium-like tetraquark or
molecular state. The $X(3915)$ and $X(4350)$ states were also
studied with other methods in Refs.
\cite{Wang:2009wk,Yang:2010sf,Abud:2009rk,Albuquerque:2010fm,Ma:2010xx,Chen:2012wy,Li:2015iga,Lin:2013ppa,Chen:2013yxa}

\subsubsection{$X(3940)$ and $X(4160)$}
\label{Sect:4.8.4}

The $X(3940)$ and $X(4160)$ states were discovered in the double
charmonium production process $e^+e^-\to J/\psi X$
\cite{Abe:2007jna,Shen:2009vs}. They have positive $C$-parity. Since
the $X(3940)$ decays into $D\bar D^*$, its spin-parity quantum
numbers can be $J^P=1^+, 0^-, 1^-, 2^-$ etc. The $J^{PC}=1^{-+}$
combination was disfavored unless it's a hybrid charmonium. As a
candidate of the charmonium state, its allowed quantum numbers are
$J^{PC}=1^{++}$, $0^{-+}$ and $2^{-+}$ etc.

\paragraph{$X(3940)$}

The $1^{++}$ assignment of the $X(3940)$ was disfavored  by
analyzing the recoil mass distribution of the $J/\psi$
\cite{Swanson:2006st}. In Fig. \ref{Fig:2.3.X3940}, there were four
visible enhancements corresponding to the $\eta_c$ at 2980 MeV, the
$\chi_{c0}$ at 3415 MeV, the $\eta_c(2S)$ at 3638 MeV and the
$X(3940)$ state. There was no evidence of the $\chi_{c1}$ in Fig.
\ref{Fig:2.3.X3940}. There was no reason to expect
the $\chi_{c1}^\prime$ to be a stronger signal than the $\chi_{c1}$
\cite{Swanson:2006st}.

In Ref. \cite{Rosner:2005gf}, Rosner proposed the $X(3940)$ as the
$\eta_c(3S)$ charmonium state. The Regge trajectory also indicated
that the $X(3940)$ could be the $0^{-+}$ charmonium state
\cite{Gershtein:2006ng}. In Ref. \cite{Braguta:2006py}, the authors
studied the $e^+e^-\to J/\psi X(3940)$ process in the framework of
the light cone formalism. They found that the production cross
section was in agreement with the experiment if the $X(3940)$ is
the $\eta_c(3S)$ \cite{Braguta:2006py}. In Ref. \cite{He:2014xna}, He
{\it et al.} calculated the two-body open-charm decay widths of
the $X(3940)$ as the $\eta_c(3S)$ charmonium state. However, a problem of the $\eta_c(3S)$
interpretation of the $X(3940)$ is that its mass is a bit lower than
theoretical predictions \cite{Barnes:2005pb,Li:2009zu}.

\paragraph{$X(4160)$}

The $X(4160)$ was observed only in the $D^*\bar D^*$ final states
\cite{Abe:2007sya}. In Ref. \cite{Chao:2007it}, Chao discussed
possible interpretations for the $X(4160)$ based on NRQCD
calculations. The author first proposed the $X(4160)$ as the D-wave
charmonium state $2 ^1D_2$ with $J^{PC}=2^{-+}$ and calculated its
production rate, which was only $5\%$ of that for $e^+e^-\to
J/\psi+\eta_c(1S)$, incompatible with the experimental observation of
the $X(4160)$. Then, Chao considered the $X(4160)$ as the $\psi(4160)$.
However, such a possibility was completely ruled out since the
production rate was much smaller than that for $e^+e^-\to
J/\psi+J/\psi$. Chao found that the production rate of $e^+e^-\to
J/\psi+\eta_c(4S)$ was not too small in NRQCD, which was consistent
with the experiment. Therefore, the $X(4160)$ was a good candidate
of the $\eta_c(4S)$ charmonium state \cite{Chao:2007it}. If so, the absence of the $D\bar
D$ decay channel can be understood easily since such a decay is
forbidden. However, the $\eta_c(4S)$ was predicted to lie higher
than the $X(4160)$ \cite{Barnes:2005pb,Li:2009zu}. He {\it et al.}
indicated that the $\eta_c(4S)$ may have a very narrow width according
to their calculation \cite{He:2014xna}.

The mass of the $\chi_0(3P)$ was predicted to be 4131 MeV in Ref.
\cite{Li:2009zu}. Chao also considered the $\chi_0(3P)$
interpretation \cite{Chao:2007it}. Such an assignment was particularly
interesting if the observed broad peak around $3.8-3.9$ GeV in the
recoil mass of the $D\bar D$ against $J/\psi$ in $e^+e^-\to J/\psi D
\bar D$ was due to the $\chi_0(2P)$.

Molina and Oset proposed the $X(4160)$ as an isoscalar $D_s^*\bar
D_s^*$ molecular state with $J^{PC}=2^{++}$ in Ref.
\cite{Molina:2009ct}. They studied the vector-vector interaction
within the framework of the hidden gauge formalism and found there
exists strong interaction to bind the $D_s^*\bar D_s^*$ system. A
bound state was found around 4157 MeV with $I^GJ^{PC}=0^+2^{++}$,
which was identified as the $X(4160)$ \cite{Molina:2009ct}.

Using the tetraquark interpolating currents composed in Ref.
\cite{Jiao:2009ra}, Chen and Zhu studied the charmonium-like
tetraquark states with $J^{PC}=0^{-+}$ and extracted the mass around
$4.55$ GeV \cite{Chen:2010jd}. This value is much higher than the
masses of the $X(3940)$ and $X(4160)$ and does not support them to
be charmonium-like tetraquark states. The charmonium-like tetraquark
states with the exotic quantum numbers $J^{PC}=0^{--}$ were also
studied in Ref. \cite{Chen:2010jd}.

\subsubsection{Narrow enhancement structures around 4.2 GeV in the
hidden-charm channels}
\label{Sect:4.8.5}

\begin{figure}[htbp]
\begin{center}
\includegraphics[scale=0.4]{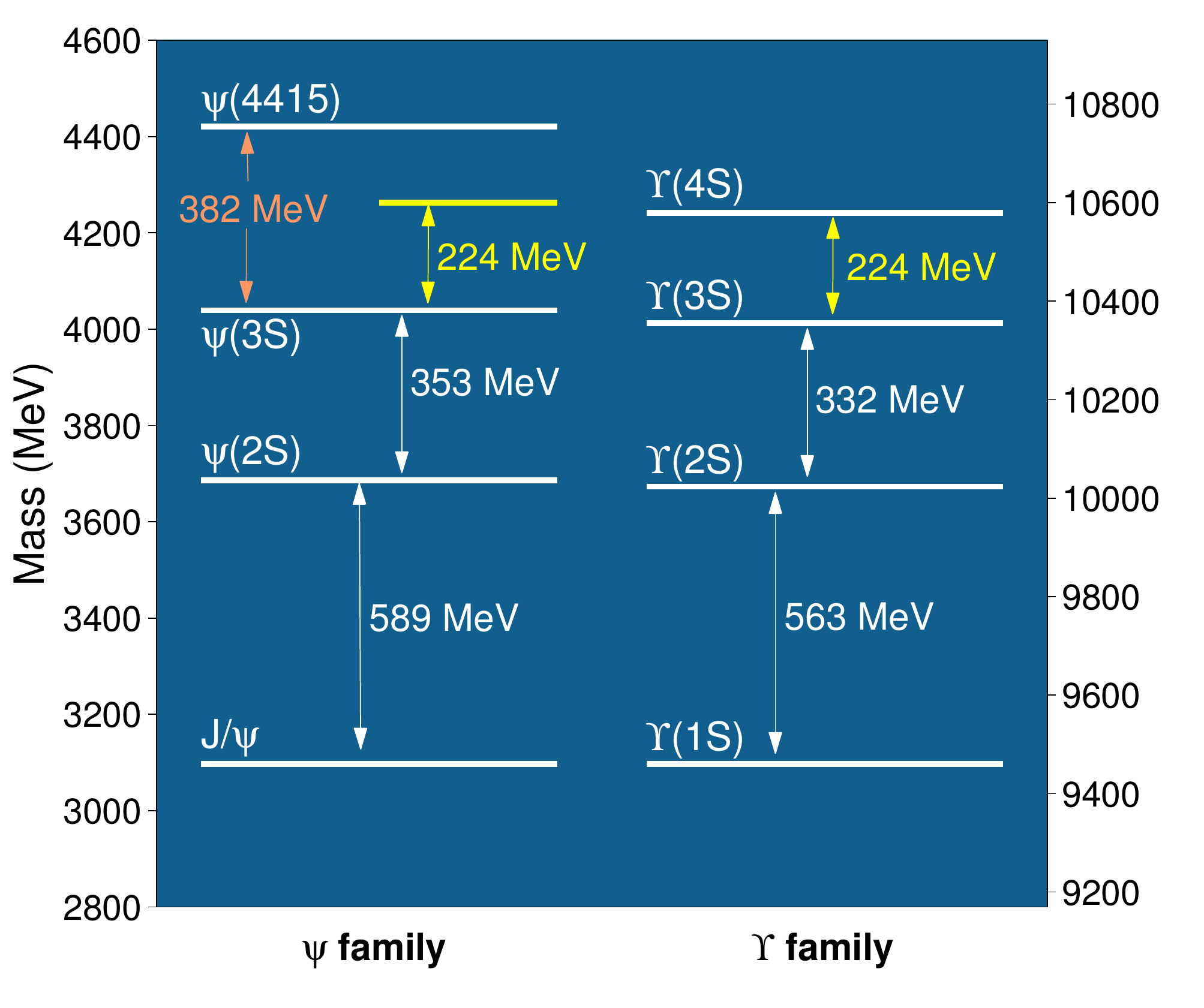}
\caption{(Color online) A comparison between the $J/\psi$ and
$\Upsilon$ families. Taken from Ref. \cite{He:2014xna}.
\label{Fig.4.8.mass}}
\end{center}
\end{figure}

In Ref. \cite{He:2014xna}, He {\it et al.} noticed the similarity between
vector charmonium and bottomonium families. The mass gap between
the $\psi(2S)$ and $J/\psi$ is almost the same as that between
the $\Upsilon(2S)$ and $\Upsilon(1S)$ and
$M_{\psi(3S)}-M_{\psi(2S)}\approx
M_{\Upsilon(3S)}-M_{\Upsilon(2S)}$, where the $\psi(3686)$ and
$\psi(4040)$ are treated as the $\psi(2S)$ and $\psi(3S)$ charmonium states, respectively.
Until now the bottomonia with the radial quantum numbers $n=1,2,3,4$
have been established \cite{pdg}. If this mass gap relation
continues to hold for higher states with $n=3,4$ in $J/\psi$ and
$\Upsilon$ families, the mass of the $\psi(4S)$ should be located at
$4263$ MeV \cite{He:2014xna}. Within this framework, the $\psi(4415)$
cannot be treated as the $\psi(4S)$. Fig. \ref{Fig.4.8.mass} shows the details
of the mass gaps for the $J/\psi$ and $\Upsilon$ families. This simple
estimate of the mass of the $\psi(4S)$ \cite{He:2014xna} is consistent
with the results obtained with the screened potential in
Refs.~\cite{Dong:1994zj,Li:2009zu}, where the mass of the $\psi(4S)$ was
predicted to be 4247 MeV in Ref. \cite{Dong:1994zj} and 4273 MeV in
Ref. \cite{Li:2009zu}.

\begin{figure}[htbp]
\begin{center}
\includegraphics[scale=0.38]{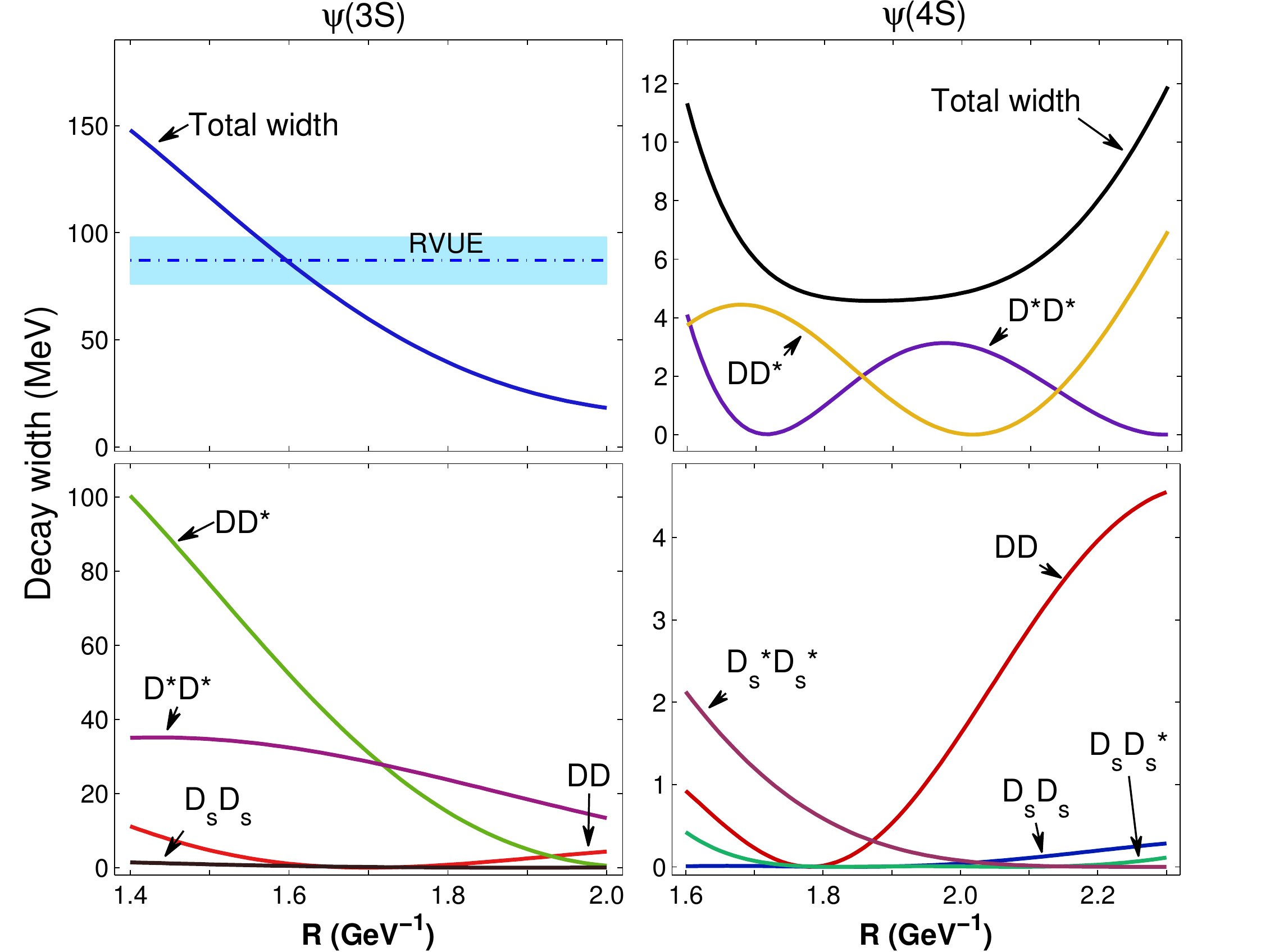}
\caption{(Color online) The total and partial decay widths of
the $\psi(3S)$ (left) and the $\psi(4S)$ (right). Here, the dashed line with
a band (left) is the experimental data from Ref. \cite{Mo:2010bw}.
Taken from Ref. \cite{He:2014xna}. \label{Fig.4.8.3S}}
\end{center}
\end{figure}

He {\it et al.} \cite{He:2014xna} further calculated the partial and total
decay widths of the $\psi(4S)$ through the quark pair creation (QPC)
model and found a very interesting result of the decay behavior of
the $\psi(4S)$ (see Fig. \ref{Fig.4.8.3S}). The total decay width of the $\psi(4S)$
is stable over the corresponding $R$ range adopted, while its
partial decay widths strongly depend on the $R$ value. This
phenomenon is due to the node effects. The predicted charmonium
$\psi(4S)$ has a very narrow width around 6 MeV within the QPC model.
Thus, the charmonium-like states $Y(4260)$ and $Y(4360)$ cannot
correspond to this predicted $\psi(4S)$ state due to their large
decay widths \cite{He:2014xna}.

According to the cross sections of $e^+e^-\to h_c(1P)\pi^+\pi^-$ at
center-of-mass energies $3.90-4.42$ GeV \cite{Ablikim:2013wzq}, Yuan
fitted the corresponding line shape with two Breit-Wigner functions
\cite{Chang-Zheng:2014haa}. He found a narrow structure with mass
$(4216\pm18)$ MeV and width $(39\pm22)$ MeV and another broad structure
with mass $(4293\pm9)$ MeV and width $(222\pm67)$ MeV
\cite{Chang-Zheng:2014haa}. Both structures have $J^{PC}=1^{--}$.
The narrow structure was proposed as a good candidate of the
$\psi(4S)$ \cite{He:2014xna}.

Later, the BESIII Collaboration reported an enhancement structure in
$e^+e^-\to \omega\chi_{c0}$ \cite{Ablikim:2014qwy}, which has a mass
$M=(4230\pm 8)$ MeV and width $\Gamma=(38\pm12)$ MeV. BESIII
indicated that this resonance structure is different from the $Y(4260)$
reported in the analysis of $e^+e^-\to J/\psi\pi^+\pi^-$
\cite{Aubert:2005rm}. Chen, Liu and Matsuki \cite{Chen:2014sra}
introduced the $\psi(4S)$ to explain the enhancement in $e^+e^-\to
\omega\chi_{c0}$. They calculated the branching ratio of
$\psi(4S)\to \omega\chi_{c0}$ from the meson loop contribution and
estimated the upper limit of the branching ratio of $\psi(4S)\to \eta
J/\psi$ to be $1.9\times 10^{-3}$, which is consistent with the
experimental data \cite{Ablikim:2014qwy,Xu-yang:2015aya}.

\begin{figure}[htbp]
\centering%
\scalebox{0.8}{\includegraphics{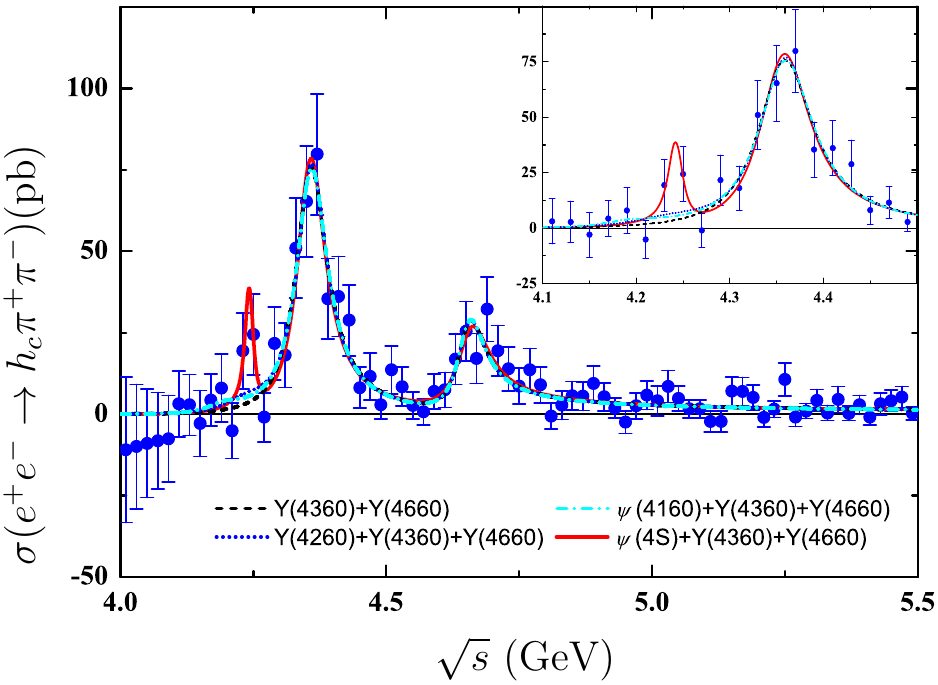}} \caption{(Color online) A comparison of the fits to the cross section for $e^+ e^-
\to \pi^+ \pi^- \psi(2S)$ with different schemes. Taken from Ref.
\cite{Chen:2015bma}. \label{Fig.4.8.compare}}
\end{figure}

In Ref. \cite{Chen:2015bma}, Chen, Liu, and Matsuki further checked
the cross section for $e^+ e^-\to \psi(2S) \pi^+ \pi^-$
\cite{Wang:2007ea} and found a number of events near 4.2 GeV, other
than the structures of the $Y(4360)$ and $Y(4660)$ (see Fig.~\ref{Fig.4.8.compare}). If setting this structure as the $\psi(4S)$, the
upper limit of the branching ratio of $\psi(4S)\to
\psi(2S)\pi^+\pi^-$ can be extracted
\cite{Wang:2007ea}, i.e., ${\mathcal{B}}(\psi(4S)\to
\psi(2S)\pi^+\pi^-)<3\times 10^{-3}$, which can be understood by
hadronic loop contributions \cite{Chen:2015bma}.

Since there exist evidences of narrow structures around 4.2 GeV in
$e^+e^-\to h_c(1P)\pi^+\pi^-$
\cite{Ablikim:2013wzq,Chang-Zheng:2014haa}, $e^+e^-\to
\omega\chi_{c0}$ \cite{Ablikim:2014qwy}, $e^+ e^-\to \psi(2S) \pi^+
\pi^-$ \cite{Wang:2007ea}, Chen, Liu, and Matsuki performed a
combined fit to these hidden-charm decay channels under two schemes
\cite{Chen:2015bma}. The mass and width of the $\psi(4S)$ were extracted
to be $m_{\psi(4S)}=(4234 \pm 5)$ MeV, $\Gamma_{\psi(4S)} =(29 \pm 14)$
MeV and $m_{\psi(4S)}=(4220 \pm 8)$ MeV, $\Gamma_{\psi(4S)} =(43 \pm 9)$
MeV for Scheme I and Scheme II, respectively. The narrow structure
around 4.2 GeV in $e^+e^-\to h_c(1P)\pi^+\pi^-$
\cite{Ablikim:2013wzq,Chang-Zheng:2014haa}, $e^+e^-\to
\omega\chi_{c0}$ \cite{Ablikim:2014qwy}, $e^+ e^-\to \psi(2S) \pi^+
\pi^-$ \cite{Wang:2007ea} may be due to the same state $\psi(4S)$.

\begin{figure}[htbp]
\centering%
\scalebox{0.6}{\includegraphics{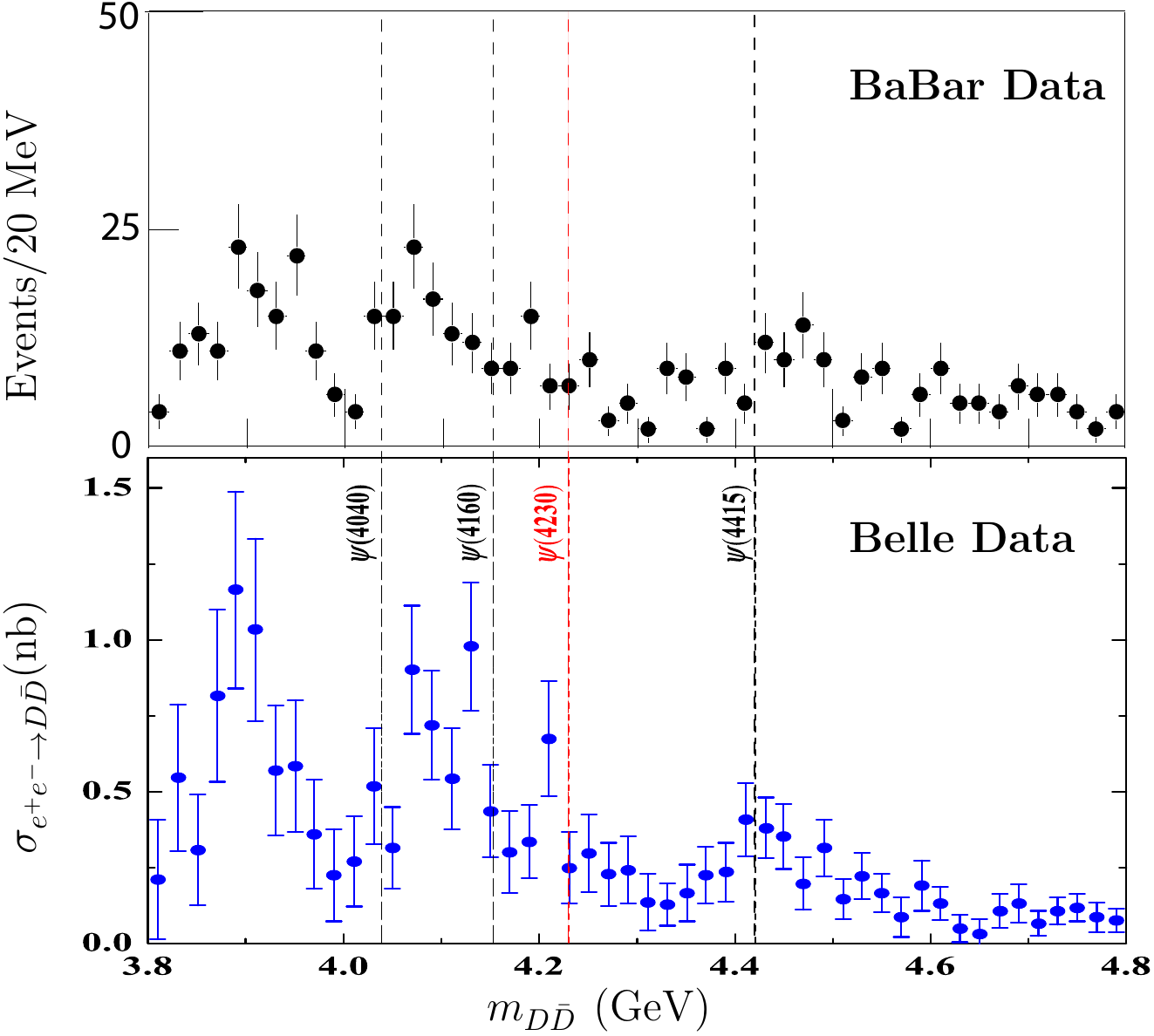}} \caption{(Color online) The experimental data of $e^+e^- \to D\bar{D}$ from the
Belle \cite{Pakhlova:2008zza} and BaBar \cite{Aubert:2009aq}
collaborations and the comparison with the central masses of
$\psi(4040)$, $\psi(4160)$, $\psi(4415)$ and the predicted
$\psi(4S)$. Here, the predicted mass is taken from Ref.
\cite{Aubert:2005rm}. Taken from Ref. \cite{Chen:2014sra}.
\label{Fig.4.8.DD}}
\end{figure}

In Ref. \cite{Chen:2014sra}, Chen, Liu, and Matsuki discussed the
possible evidence of a narrow structure around 4.2 GeV in the Belle
\cite{Pakhlova:2008zza} and BaBar \cite{Aubert:2009aq} data of
$e^+e^- \to D\bar{D}$, which may correspond to the $\psi(4S)$ (see
Fig.~\ref{Fig.4.8.DD} for more details).

\subsubsection{$X(3823)$}
\label{Sect:4.8.6}

As indicated in Sec. \ref{Sect:2.1.7}, the $X(3823)$ agrees with the
theoretical predictions for the $1^3D_2$ state
\cite{Eichten:1978tg,Eichten:1979ms,Buchmuller:1980su,Godfrey:1985xj,Ebert:2002pp,Eichten:2004uh}.
The $1 ^3D_2$ charmonium state with $J^{PC}=2^{--}$ is expected to
be very narrow. Its open-charm decay mode $D\bar D$ is forbidden by
the spin-parity symmetry. The upper limit of $B(X(3823)\to
\chi_{c2}\gamma)/B(X(3823)\to \chi_{c1}\gamma)$ was measured to be
$<0.41$ by Belle \cite{Bhardwaj:2013rmw}, and $<0.42$ by BESIII
\cite{Ablikim:2015dlj}, consistent with the theoretical calculation
for the $1 ^3D_2$ state in Refs.
\cite{Eichten:2002qv,Ebert:2002pp,Ko:1997rn,Qiao:1996ve}. In
addition, the partial decay width for $\psi(1^3D_2)\to
J/\psi\pi^+\pi^-$ obtained in these works was also in agreement with
the observation by E705 \cite{Antoniazzi:1993jz}. Therefore, the
$X(3823)$ is a very good candidate of the $1 ^3D_2$ charmonium
state.

The charmonium state with $J^{PC}=2^{--}$ was investigated in QCD
sum rule using the interpolating current \cite{Chen:2011qu}
\begin{eqnarray}
J_{\mu\nu}=\bar
c(x)(\gamma_{\mu}\gamma_5\stackrel{\leftrightarrow}{D}_{\nu}
  +\gamma_{\nu}\gamma_5\stackrel{\leftrightarrow}{D}_{\mu}-\frac{2}{3}\eta_{\mu\nu}\gamma_5
  \stackrel{\leftrightarrow}{D\!\!\!\slash})c(x)\, . \label{Eq.4.7.current2nn}
\end{eqnarray}
The mass of the $1 ^3D_2$ charmonium state as $m=3.97\pm0.25$ GeV.

In Ref. \cite{Segovia:2013wma}, the authors studied the heavy meson
properties within a non-relativistic constituent quark model,
including the spectroscopy and the electromagnetic, strong and weak
decay processes. They obtained the mass of the $1 ^3D_2$ $c\bar c$
state at 3812 MeV and the ratio $B(X(3823)\to
\chi_{c2}\gamma)/B(X(3823)\to \chi_{c1}\gamma)=0.24$.

Voloshin discussed the $e^+e^-\to \pi\pi X(3823)$ process in the
soft pion limit in Ref. \cite{Voloshin:2015jua}. The amplitude of
this production process was studied up to the second order in the
pion momenta at different energies. The amplitude of $e^+e^-\to
\pi\pi X(3823)$ rapidly grows with the momenta of the pions.

In Ref. \cite{Wang:2015xsa}, Wang \textit{et al.} studied the
$X(3823)\to J/\psi\pi^+\pi^-$ process to identify the significance
of the coupled-channel effects. They computed the partial decay
width distribution with the dipion invariant mass with both the QCD
multipole expansion method and the effective Lagrangian approach.
Neglecting the coupled-channel effects, they found the disagreement
with E705 experiment in Ref. \cite{Antoniazzi:1993jz}. Including the
coupled-channel effect of $D\bar D^*$, they calculated the same
process and found the interference between the direct and
the indirect processes \cite{Wang:2015xsa}.

In Ref. \cite{Deng:2015bva}, the authors studied the
electromagnetic transitions of the charmonium states within a
constituent quark mode. Considering the $X(3823)$ as the
$\psi_2(1D)$ state, they calculated the radiative decay widths
\begin{eqnarray}
\nonumber \Gamma[X(3823)\to \chi_{c0}(1P)\gamma]& \simeq & 1.42 \ \mathrm{keV}\, ,\\
\Gamma[X(3823)\to \chi_{c1}(1P)\gamma]& \simeq & 227 \ \mathrm{keV}\, ,\\
\nonumber \Gamma[X(3823)\to \chi_{c2}(1P)\gamma]& \simeq & 42 \ \mathrm{keV}\,
,
\end{eqnarray}
in which the $\chi_{c1}(1P)\gamma$ was a dominant decay mode. The
ratio $\Gamma[X(3823)\to \chi_{c2}(1P)\gamma]/\Gamma[X(3823)\to
\chi_{c1}(1P)\gamma]\simeq 19\%$ was in agreement with the
observations $< 42\%$~\cite{Ablikim:2015dlj}.

\subsubsection{A short summary}
\label{Sect:4.8.7}

\begin{itemize}

\item The $Y(4008)$ was explained as the $\psi(3S)$ charmonium state,
the $D^* \bar D^*$ or $\eta h_c$ molecular state, and the
diquark-antidiquark tetraquark state. There are also non-resonant
explanations.

\item The $Y(4360)$ was explained as the $\psi(3D)$ charmonium state,
the $2S$ $D_1 \bar D^*$ molecule, a hadrocharmonium state, a
diquark-antidiquark tetraquark state, a baryonium state, and a
charmonium hybrid state. Non-resonant explanation was also proposed,
such as the Fano-like interference.

\item The $Y(4660)$ was assigned as the $5^3S_1$ or $\psi(6S)$
charmonium state, the $f_0(980) \psi^\prime$ bound state, and a
diquark-antidiquark tetraquark state. The $Y(4630)$ was assigned as
the $\Lambda_c \bar \Lambda_c$ baryonium bound state. The $Y(4660)$
and $Y(4630)$ may be the same state.

\item In PDG \cite{pdg}, the $X(3915)$ and $Z(3930)$ were assigned
as the $\chi^{\prime}_{c0}(2P)$ and $\chi^{\prime}_{c2}(2P)$ charmonium
states, respectively. This assignment is challenged by the small
hyperfine splitting and the absence of the $X(3915)$ in
$\gamma\gamma\to D\bar{D}$.

\item The $X(3940)$ and $X(4160)$ were suggested as the
$\eta_c(3S)$ and $\eta_c(4S)$ charmonium state, respectively.

\item The $X(4350)$ was interpreted as the $\chi_{c2}(3P)$ charmonium state.

\item The $X(3823)$ is the $1^3D_2$ charmonium state with $J^{PC}=2^{--}$.

\end{itemize}

%\input{section4.X.tex}
%-------------Section 5-----------------------------------
%
%=====================================================================================
%=====================================================================================
\section{Outlook and summary}
\label{sect:5}

Since 2003, many charmonium-like/bottomonium-like $XYZ$ states have
been observed. Recently two hidden-charm pentaquarks $P_c$ were
observed by LHCb. Some of these states do not fit into the quark
model spectrum easily. They are good candidates of the hidden-charm
tetraquark and pentaquark states and enable us to carry out
intensive studies of the exotic hadronic matter, which is one of the
most important issues in hadron physics. These states provide us an
ideal platform to deepen our understanding of the non-perturbative
QCD.

In the past 13 years, the observations of the $XYZ$ and $P_c$ states
have inspired theorist's extensive interests in revealing their
underlying structures. There have accumulated a huge number of
theoretical papers on these states, which cover their mass
spectroscopy, other static properties, reaction, decay and
production behaviors. Various schemes were proposed such as the
conventional charmonium states, molecular states, tetraquark states,
hybrid charmonium states and so on.

In this report, we have tried our best to summarize the experimental and
theoretical progresses on the hidden-charm multiquark states in
order to gain some valuable lessons from the extensive research in
the past 13 years.

\subsection{Current status and future confirmation of the hidden-charm multiquark states} \label{Sect.5.1}
%=====================================================================================
%=====================================================================================
%

We collect all the hidden-charm tetraquark and pentaquark states in
Tables~\ref{Tab.5.manylabA}, \ref{Tab.5.manylabB} and \ref{Tab.5.onelab}. We award an
overall status $***~*$ to those states which have been firmly
established by two or more collaborations in two or more decay
modes, and $***$ to states which have been well established in a
single decay mode by two or more collaborations. The states with
$***~*$ and $***$ are listed in Tables~\ref{Tab.5.manylabA} and \ref{Tab.5.manylabB}. 
We award $**$ to those states which have been well established in several
decay modes by one collaboration, and $*$ to those states which was
only observed in a single decay mode by one collaboration. The
states with $**$ and $*$ are listed in Table~\ref{Tab.5.onelab}.

The $Y(4660)$ and $Y(4630)$ may be the same state. The $Z_c(3900)$
and $Z_c(3885)$ are probably the same state. The $Z_c(4020)$ and
$Z_c(4025)$ are probably the same state. We award $***~*$ to them.
The Belle and LHCb experiments observed the $Z^+(4430)$, while the
BaBar experiments did not. Moreover, it has only been observed in
the $B \to K Z^+(4430) (\to \psi(3686) \pi^+)$ decay process. So we
awarded $***$ to the $Z^+(4430)$.

Future experimental confirmation of those states which were observed
in a single decay mode by one collaboration is crucial. Some states
were observed in the hidden-charm decay modes. It is also important
to search for their open-charm and charm-less decay modes. For
example, the $Y(4260)$ was observed in $J/\psi \pi\pi$, $\pi
Z_c(3900)$ and $\pi Z_c(4020)$. Its open-charm modes are still
missing.

The two $P_c$ states were observed in the $J/\psi p$ mode. If they
are molecular states, their open-charm decay modes may be dominant.
The compact ``genuine'' pentaquark states confined within one MIT bag
may mainly decay into the $J/\psi p$ mode. The dynamically generated
hidden-charm baryons are generally narrow. The sum of their
charm-less partial decay widths is larger than their
hidden-charm widths. This feature is characteristic of the
dynamically generated resonance and the unitary approach. The
observation of significant charm-less decay modes will support the
$P_c$ states as the dynamically generated resonance within the
unitary framework.

\begin{sidewaystable}[htb]
\small
\centering%
\caption{Hidden-charm multiquark states that have been observed by
more than one collaborations: $X(3872)$ and $Y(4260)$. \label{Tab.5.manylabA}}
\begin{tabular}{c|c|c|c|c|c|c}
\toprule[1pt] States & Status & Mass [MeV] & Width [MeV] &
$I^GJ^{PC}/IJ^{P}$ & Observation & Note
\\ \midrule[1pt]
$X(3872)$ & $***~*$ & $3871.69 \pm 0.17$~\cite{pdg} &
$<1.2$~\cite{pdg} & $0^+1^{++}$ & $\begin{array}{l} B \to K X(3872)
\left\{\begin{array}{l} \to J/\psi \rho^0, J/\psi \pi^+ \pi^-
\\
\to J/\psi \omega (\to \pi^+ \pi^- \pi^0)
\\
\to D^0 \bar D^{*0}, D^0 \bar D^0 \pi^0
\\
\to \gamma J/\psi, \gamma \psi(3686)
\end{array}\right.
\\
p\bar p \to \cdots + X(3872) ( \to J/\psi \pi^+ \pi^- )
\\
pp \to \cdots + X(3872) \left\{\begin{array}{l}
 \to J/\psi \pi^+ \pi^- \\
 \to \gamma J/\psi, \gamma\psi(3686)
\end{array}\right.
\\
e^+ e^- [\to Y(4260)] \to \gamma X(3872) ( \to J/\psi \pi^+ \pi^- )
\end{array}
$ & $\begin{array}{l}
\mbox{Belle~\cite{Choi:2003ue}, BaBar~\cite{Aubert:2004ns}} \\
\mbox{Belle~\cite{Abe:2005ix}, BaBar~\cite{delAmoSanchez:2010jr}} \\
\mbox{Belle~\cite{Gokhroo:2006bt}, BaBar~\cite{Aubert:2007rva}} \\
\mbox{Belle~\cite{Abe:2005ix}, BaBar~\cite{Aubert:2006aj}} \\
\mbox{CDF~\cite{Acosta:2003zx}}, \mbox{D0~\cite{Abazov:2004kp}}\\
\mbox{LHCb~\cite{Aaij:2011sn}}, \mbox{CMS~\cite{Chatrchyan:2013cld}}\\
\mbox{LHCb~\cite{Aaij:2014ala}}\\
\mbox{BESIII~\cite{Ablikim:2013dyn}}
\end{array}
$
\\ \midrule[1pt]
$Y(4260)$ & $***~*$ & $4251 \pm 9$~\cite{pdg} & $120 \pm
12$~\cite{pdg} & $0^-1^{--}$ & $\begin{array}{l} e^+e^- \to
\gamma_{\rm ISR} Y(4260) \left\{\begin{array}{l} \to J/\psi \pi^+
\pi^-
\\
\to J/\psi f_0(980)
\\
\to J/\psi \pi^0 \pi^0
\end{array}\right.
\\
e^+ e^- \to Y(4260) \left\{\begin{array}{l} \to \pi^- Z_c(3900)^+
(\to J/\psi \pi^+)
\\
\to \pi^- Z_c(3885)^+ (\to (D \bar D^*)^+)
\\
\to \pi^- Z_c(4020)^+ (\to h_c \pi^+)
\\
\to \pi^- Z_c(4025)^+ (\to (D^* \bar D^*)^+)
\end{array}\right.
\\
e^+ e^- [\to Y(4260)] \to \gamma X(3872) ( \to J/\psi \pi^+ \pi^- )
\end{array}
$ & $\begin{array}{l}
\mbox{BaBar~\cite{Aubert:2005rm}, CLEO~\cite{He:2006kg}, Belle~\cite{Yuan:2007sj}}\\
\mbox{BaBar~\cite{Lees:2012cn}}\\
\mbox{CLEO~\cite{Coan:2006rv}}\\
\mbox{BESIII~\cite{Ablikim:2013mio}, Belle~\cite{Liu:2013dau}}\\
\mbox{BESIII~\cite{Ablikim:2013xfr}}\\
\mbox{BESIII~\cite{Ablikim:2013wzq}}\\
\mbox{BESIII~\cite{Ablikim:2013emm}}\\
\mbox{BESIII~\cite{Ablikim:2013dyn}}
\end{array}
$
\\ \bottomrule[1pt]
\end{tabular}
\end{sidewaystable}
%%%%%%%%%%%%%%%%%%%%%%%%%%%%%%%%%%%%%%%%%%%%%%%%%%%%%%%%%%%%%%%%%%%%%%%%%%%%%%%%%%%%%%%%%%%%%%%%%%%%%%%%%%%%%%%%%%%%%%%%%%%%%%%%%%%%%%%%%%%%%%
\begin{sidewaystable}[htb]
\small
\centering%
\caption{Hidden-charm multiquark states that have been observed by
more than one collaborations: others. \label{Tab.5.manylabB}}
\begin{tabular}{c|c|c|c|c|c|c}
\toprule[1pt] States & Status & Mass [MeV] & Width [MeV] &
$I^GJ^{PC}/IJ^{P}$ & Observation & Note
\\ \midrule[1pt]
$Y(3940)$ & $***$
&$3919.1{^{+3.8}_{-3.5}}\pm2.0$~\cite{delAmoSanchez:2010jr} &
$31{^{+10}_{-8}}\pm5$~\cite{delAmoSanchez:2010jr} & $0^+?^{?+}$ &
$\begin{array}{l} B \to K Y(3940) (\to J/\psi \omega)
\end{array}
$ & Belle~\cite{Abe:2004zs},
BaBar~\cite{Aubert:2007vj}
\\ \midrule[1pt]
$Y(4140)$ & $***$ & $4148.0 \pm 2.4 \pm
6.3$~\cite{Chatrchyan:2013dma} & $28{^{+15}_{-11}} \pm
19$~\cite{Chatrchyan:2013dma} & $0^+?^{?+}$ & $\begin{array}{l} B
\to K Y(4140) (\to J/\psi \phi)
\end{array}
$ & CDF~\cite{Aaltonen:2009tz}, D0~\cite{Abazov:2013xda},
CMS~\cite{Chatrchyan:2013dma}
\\ \midrule[1pt]
$Y(4274)$ & $***$ & $4274.4{^{+8.4}_{-6.7}} \pm
1.9$~\cite{Aaltonen:2011at} & $32.3{^{+21.9}_{-15.3}} \pm
7.6$~\cite{Aaltonen:2011at} & $0^+?^{?+}$ & $\begin{array}{l} B \to
K Y(4274) (\to J/\psi \phi)
\end{array}
$ & CDF~\cite{Aaltonen:2011at}, CMS~\cite{Chatrchyan:2013dma}
\\ \midrule[1pt]
$X(3823)$ & $***~*$ & $3821.7\pm 1.3\pm 0.7$~\cite{Ablikim:2015dlj}
& $< 16$~\cite{Ablikim:2015dlj} & $0^-2^{--}$ & $\begin{array}{l}
\psi^\prime\to J/\psi\pi^+\pi^- \\
B \to K X(3823) (\to \gamma \chi_{c1})
\\
e^+ e^- \to \pi^+ \pi^- X(3823) (\to \gamma \chi_{c1})
\end{array}
$ & $\begin{array}{l}
\mbox{E705~\cite{Antoniazzi:1993jz}}, \\
 \mbox{Belle~\cite{Bhardwaj:2013rmw}}, \\
\mbox{BESIII~\cite{Ablikim:2015dlj}}
\end{array}$
\\ \midrule[1pt]
$Y(4360)$ & $***$ & $4354 \pm 10$~\cite{pdg} & $78 \pm
16$~\cite{pdg} & $0^-1^{--}$ & $\begin{array}{l} e^+e^- \to
\gamma_{\rm ISR} Y(4360) (\to \psi(3686) \pi^+ \pi^-)
\end{array}
$ & BaBar~\cite{Aubert:2007zz},Belle~\cite{Wang:2007ea}
\\ \midrule[1pt]
$Y(4660)$ &  \multirow{2}{*}{$***~*$} & $4665 \pm 10$~\cite{pdg} &
$53 \pm 16$~\cite{pdg} & \multirow{2}{*}{$0^-1^{--}$} & $\begin{array}{l} e^+e^- \to
\gamma_{\rm ISR} Y(4660) (\to \psi(3686) \pi^+ \pi^-)
\end{array}
$ & Belle~\cite{Wang:2007ea}, BaBar~\cite{Lees:2012pv}
\\ \cdashline{1-1}[1pt/1pt] \cdashline{3-4}[1pt/1pt] \cdashline{6-7}[1pt/1pt]
$Y(4630)$ & & $4634{^{+8}_{-7}}{^{+5}_{-8}}$~\cite{Pakhlova:2008vn}
& $92{^{+40}_{-24}}{^{+10}_{-21}}$~\cite{Pakhlova:2008vn} &
& $\begin{array}{l} e^+e^- \to \gamma_{\rm ISR} Y(4630)
(\to \Lambda_c \bar \Lambda_c)
\end{array}
$ & Belle~\cite{Pakhlova:2008vn}
\\ \midrule[1pt]
$X(3915)$ & $***$ & $3915\pm 3\pm 2$~\cite{Uehara:2009tx} & $17\pm
10\pm 3$~\cite{Uehara:2009tx} & $0^+0^{++}$ & $\begin{array}{l}
\gamma\gamma\to X(3915) (\to J/\psi \omega)
\end{array}
$ & Belle~\cite{Uehara:2009tx}, BaBar~\cite{Lees:2012xs}
\\ \midrule[1pt]
$Z(3930)$ & $***$ & $3929\pm 5\pm 2$~\cite{Uehara:2005qd} & $29\pm
10\pm 2$~\cite{Uehara:2005qd} & $0^+2^{++}$ & $\begin{array}{l}
\gamma\gamma\to Z(3930) (\to D\bar D)
\end{array}
$ & Belle~\cite{Uehara:2005qd}, BaBar~\cite{Aubert:2010ab}
\\ \midrule[1pt]
$Z^+(4430)$ & $***$ & $4478 ^{+15}_{-18}$~\cite{pdg} & $181 \pm
31$~\cite{pdg} & $1^{+}1^{+-}$ & $\begin{array}{l} B \to K Z^+(4430) (\to
\psi(3686) \pi^+)
\end{array}
$ & Belle~\cite{Choi:2007wga},LHCb~\cite{Aaij:2014jqa}
\\ \midrule[1pt]
$Z_c(3900)$ & \multirow{2}{*}{$***~*$} & $3888.7\pm3.4$~\cite{pdg} &
$35\pm7$~\cite{pdg} & \multirow{2}{*}{$1^+1^{+-}$} & $\begin{array}{l} e^+ e^- \to
Y(4260)
\to \pi^- Z_c(3900)^+ (\to J/\psi \pi^+) \\
e^+ e^- \to \psi(4160) \to \pi^- Z_c(3900)^+ (\to J/\psi \pi^+)
\end{array}
$ & $\begin{array}{c}
\mbox{BESIII~\cite{Ablikim:2013mio}, Belle~\cite{Liu:2013dau},}\\
\mbox{Xiao \textit{et al.}~\cite{Xiao:2013iha}}
\end{array}
$
\\ \cdashline{1-1}[1pt/1pt] \cdashline{3-4}[1pt/1pt] \cdashline{6-7}[1pt/1pt]
$Z_c(3885)$ & & $3883.9\pm1.5\pm4.2$~\cite{Ablikim:2013xfr} &
$24.8\pm3.3\pm11.0$~\cite{Ablikim:2013xfr} & &
$\begin{array}{l} e^+ e^- \to Y(4260) \to \pi^- Z_c(3885)^+ (\to (D
\bar D^*)^+)
\end{array}
$ & $\begin{array}{c} \mbox{BESIII~\cite{Ablikim:2013xfr}}
\end{array}
$
\\ \midrule[1pt]
$Z_c(4020)$ & \multirow{2}{*}{$***~*$} &
$4022.9\pm0.8\pm2.7$~\cite{Ablikim:2013wzq} &
$7.9\pm2.7\pm2.6$~\cite{Ablikim:2013wzq} & \multirow{2}{*}{$1^+1^{+-}$} &
$\begin{array}{l} e^+ e^- \to Y(4260) \to \pi^- Z_c(4020)^+ (\to h_c
\pi^+)
\end{array}
$ & $\begin{array}{c} \mbox{BESIII~\cite{Ablikim:2013wzq}}
\end{array}
$
\\ \cdashline{1-1}[1pt/1pt] \cdashline{3-4}[1pt/1pt] \cdashline{6-7}[1pt/1pt]
$Z_c(4025)$ & & $4026.3\pm2.6\pm3.7$~\cite{Ablikim:2013emm} &
$24.8\pm5.6\pm7.7$~\cite{Ablikim:2013emm} & &
$\begin{array}{l} e^+ e^- \to Y(4260) \to \pi^- Z_c(4025)^+ (\to
(D^* \bar D^*)^+)
\end{array}
$ & $\begin{array}{c} \mbox{BESIII~\cite{Ablikim:2013emm}}
\end{array}
$
\\ \bottomrule[1pt]
\end{tabular}
\end{sidewaystable}
\begin{sidewaystable}[htb]
\small
\centering%
\caption{Hidden-charm multiquark states that have been observed by
only one collaboration. \label{Tab.5.onelab}}
\begin{tabular}{c|c|c|c|c|c|c}
\toprule[1pt]
States & Status & Mass [MeV] & Width [MeV] & $I^GJ^{PC}/IJ^{P}$ & Observation & Note\\
\\ \midrule[1pt]
$Y(4008)$ & $*$ & $4008\pm40^{+114}_{-28}$~\cite{Yuan:2007sj} &
$226\pm44\pm87$~\cite{Yuan:2007sj} & $0^-1^{--}$ & $\begin{array}{l}
e^+e^- \to \gamma_{\rm ISR} Y(4008) (\to J/\psi \pi^+ \pi^-)
\end{array}
$ & Belle~\cite{Yuan:2007sj}
\\ \midrule[1pt]
$X(3940)$ & $*$ & $3942^{+7}_{-6}\pm6$~\cite{Abe:2007sya} &
$37^{+26}_{-15}\pm8$~\cite{Abe:2007sya} & $?^??^{?+}$ &
$\begin{array}{l} e^+e^- \to J/\psi X(3940) (\to \bar D D^*)
\end{array}
$ & Belle~\cite{Abe:2007sya}
\\ \midrule[1pt]
$X(4160)$ & $*$ & $4156{^{+25}_{-20}}\pm15$~\cite{Abe:2007sya} &
$139{^{+111}_{-61}}\pm21$~\cite{Abe:2007sya} & $?^??^{?+}$ &
$\begin{array}{l} e^+e^- \to J/\psi X(4160) (\to \bar D^{*} D^{*})
\end{array}
$ & Belle~\cite{Abe:2007sya}
\\ \midrule[1pt]
$X(4350)$ & $*$ & $4350.6^{+4.6}_{-5.1}\pm0.7$~\cite{Shen:2009vs} &
$13^{+18}_{-9}\pm4$~\cite{Shen:2009vs} & $?^?0^{?+}/2^{?+}$ &
$\begin{array}{l} \gamma\gamma\to X(4350) (\to J/\psi \phi)
\end{array}
$ & Belle~\cite{Shen:2009vs}
\\ \midrule[1pt]
$Z^+(4051)$ & $*$ & $4051 \pm 14 {^{+20}_{-41}}$~\cite{Mizuk:2008me}
& $82 {^{+21}_{-17}} {^{+47}_{-22}}$~\cite{Mizuk:2008me} & $??^{?}$
& $\begin{array}{l} B \to K Z^+(4051) (\to \chi_{c1} \pi^+)
\end{array}
$ & Belle~\cite{Mizuk:2008me}
\\ \midrule[1pt]
$Z^+(4248)$ & $*$ & $4248 {^{+44}_{-29}}
{^{+180}_{-35}}$~\cite{Mizuk:2008me} & $177 {^{+54}_{-39}}
{^{+316}_{-61}}$~\cite{Mizuk:2008me} & $??^{?}$ & $\begin{array}{l}
B \to K Z^+(4248) (\to \chi_{c1} \pi^+)
\end{array}
$ & Belle~\cite{Mizuk:2008me}
\\ \midrule[1pt]
$Z^+(4200)$ & $*$ &
$4196{^{+31}_{-29}}{^{+17}_{-13}}$~\cite{Chilikin:2014bkk} &
$370{^{+70}_{-70}}{^{+70}_{-132}}$~\cite{Chilikin:2014bkk} &
$1^+1^{+-}$ & $\begin{array}{l} B \to K Z^+(4200) (\to J/\psi \pi^+)
\end{array}
$ & Belle~\cite{Chilikin:2014bkk}
\\ \midrule[1pt]
$Z^+(4240)$ & $*$ & $4239 \pm 18 {^{+45}_{-10}}$~\cite{Aaij:2014jqa}
& $220 \pm 47 {^{+108}_{-74}}$~\cite{Aaij:2014jqa} & $?0^{-}/?1^+$ &
$\begin{array}{l} B \to K Z^+(4240) (\to \psi(3686) \pi^+)
\end{array}
$ & LHCb~\cite{Aaij:2014jqa}
\\ \midrule[1pt]
$Z_b(10610)$ & $**$ & $10607.2 \pm 2.0$~\cite{Belle:2011aa} & $18.4
\pm 2.4$~\cite{Belle:2011aa} & $1^+1^{+-}$ & $\begin{array}{l}
\Upsilon(5S)\to \pi^\mp Z_b^\pm(10610) \left\{\begin{array}{l} \to
\pi^\pm\Upsilon(nS) (n=1, 2, 3)
\\
\to \pi^\pm h_b(mP) (m=1, 2)
\end{array}\right. \\
\Upsilon(10860)\to \pi^\mp Z^{\pm}_b(10610)(\to [B\bar{B}^*+{\rm
c.c.}]^{\pm})
\end{array}
$ & $\begin{array}{c}
\mbox{Belle~\cite{Belle:2011aa},}\\
\mbox{Belle~\cite{Adachi:2012cx}}
\end{array}
$
\\ \midrule[1pt]
$Z_b(10650)$ & $**$ & $10652.2 \pm 1.5$~\cite{Belle:2011aa} & $11.5
\pm 2.2$~\cite{Belle:2011aa} & $1^+1^{+-}$ & $\begin{array}{l}
\Upsilon(5S)\to \pi^\mp Z_b^\pm(10610) \left\{\begin{array}{l} \to
\pi^\pm\Upsilon(nS) (n=1, 2, 3)
\\
\to \pi^\pm h_b(mP) (m=1, 2)
\end{array}\right. \\
\Upsilon(10860)\to \pi^\mp Z^{\pm}_b(10650)(\to [B^*\bar{B}^*]^\pm)
\end{array}
$ & $\begin{array}{c}
\mbox{Belle~\cite{Belle:2011aa},}\\
\mbox{Belle~\cite{Adachi:2012cx}}
\end{array}
$
\\ \midrule[1pt]
$P_c(4380)^+$ & $*$ & $4380 \pm 8 \pm 29$~\cite{Aaij:2015tga} & $205
\pm 18 \pm 86$~\cite{Aaij:2015tga} & ${1\over2}?^?$ & $\begin{array}{l}
\Lambda_b^0 \to K^- P_c(4380)^+ ( \to J/\psi  p )
\end{array}
$ & $\begin{array}{c} \mbox{LHCb~\cite{Aaij:2015tga}}
\end{array}
$
\\ \midrule[1pt]
$P_c(4450)^+$ & $*$ & $4449.8 \pm 1.7 \pm 2.5$~\cite{Aaij:2015tga} &
$39 \pm 5 \pm 19$~\cite{Aaij:2015tga} & ${1\over2}?^?$ & $\begin{array}{l}
\Lambda_b^0 \to K^- P_c(4450)^+ (\to J/\psi p )
\end{array}
$ & $\begin{array}{c} \mbox{LHCb~\cite{Aaij:2015tga}}
\end{array}
$
\\ \bottomrule[1pt]
\end{tabular}
\end{sidewaystable}

\subsection{Non-resonant schemes} \label{Sect.5.2}
%=====================================================================================
%=====================================================================================
%

The $P_c$ and some charmonium-like and bottomonium-like states are
very close to the two open-charm/bottom hadron thresholds. Various
final state interactions may generate enhancement structures. For
example, the cusp effect was proposed to explain the
$Z_c^+(4430)$~\cite{Bugg:2008wu}, $Z_b(10610)$ and
$Z_b(10650)$~\cite{Bugg:2011jr,Swanson:2014tra}. The initial single pion
emission mechanism was applied to interpret the charged $Z_b$ and
$Z_c$ structures~\cite{Chen:2011pv,Chen:2013coa}. The triangle
singularities were employed to understand the $Z_b$, $Z_c$ and $P_c$
structures~\cite{Szczepaniak:2015eza,Guo:2015umn,Liu:2015fea,Liu:2015taa,Mikhasenko:2015vca}.

There are different versions of the non-resonant schemes and their
conclusions are not totally the same. Generally it is not easy to
distinguish the exotic state assignment from the non-resonant
schemes. With more precise experimental data, the partial wave
analysis and the establishment of the phase motion of the signal
will be helpful to test different scenarios.

\subsection{Partner states} \label{Sect.5.3}
%=====================================================================================
%=====================================================================================
%

The hidden-charm pentaquark states $P_c(4380)$ and $P_c(4450)$ shall
be accompanied by their hidden-bottom and doubly heavy pentaquark
states. The hidden-charm tetraquark states can also have
hidden-bottom partner states. For example, the $X(3872)$ was
interpreted as either a diquark-antidiquark state or a $D \bar
D^{*}$ molecular state. Within both schemes its bottom partner $X_b$
was suggested to exist: the mass of the $1^{++}$ $[bq][\bar b \bar
q]$ tetraquark was predicted to be 10504 MeV in
Ref.~\cite{Ali:2009pi}, and the mass of the $1^{++}$ $B \bar B^*$
molecular state was predicted to be 10580 MeV in
Ref.~\cite{Guo:2013sya}. Hence, the search of the $X_b$ state is of
particular importance to all the hidden-charm and hidden-bottom
tetraquark states. Indeed, the observations of these predicted
partner states would be the most powerful support of the relevant
methods/models.

Within the tetraquark scheme, lots of $XYZ$ partner states were
predicted. Especially, in the ``type-I'' diquark-antidiquark
model~\cite{Maiani:2004vq}, six $[cq][\bar c \bar q^\prime]$ states
were constructed, including two states with $J^{PC} = 0^{++}$, one
state with $J^P = 1^{++}$, two states with $J^P = 1^{+-}$, and one
state with $J^{PC} = 2^{++}$. Among them, the $1^{++}$ state was
used to fit the $X(3872)$. This idea was developed in
Ref.~\cite{Maiani:2007vr} by Maiani, Polosa, and Riquer, where they
further proposed four states with $J^P = 1^{+}$: $X_u = [cu][\bar c
\bar u]$, $X_d = [cd][\bar c \bar d]$, $X^+ = [cu][\bar c \bar d]$,
and $X^- = [cd][\bar c \bar u]$. However, the two charged states, as
partners of the $X(3872)$, were not observed in B meson
decays~\cite{Aubert:2004zr}. Now, this ``type-I'' diquark-antidiquark
model has been updated to the ``type-II'' diquark-antidiquark
model, which can explain the quantum numbers of many $XYZ$ states as
well as their decay patterns, but still can not explain why the
charged partners of the $X(3872)$ have not been observed in B meson
decays and other experiments~\cite{Maiani:2014aja}.

There are also many $XYZ$ partner states within the molecular
scheme. We take the $Z_b(10610)$ and $Z_b(10650)$ as an example,
which were interpreted as the isovector $B\bar{B}^*$ and
$B^*\bar{B}^*$ molecular states of
$I^G J^P = 1^+ 1^+$~\cite{Sun:2011uh}. Several other molecular
states were predicted as
their partners, including isoscalar $B \bar B^*$ molecular states of
$I^G J^{PC} = 0^-1^{+-}$ and $0^+1^{++}$, and isoscalar $B^* \bar
B^*$ molecular states of $0^+0^{++}$, $0^-1^{+-}$, and $0^+2^{++}$,
etc. Generally speaking, the isoscalar molecular states are bound
more tightly than their isovector partners when their other quantum
numbers are the same.

\subsection{Connections between different $XYZ$ states} \label{Sect.5.4}
%=====================================================================================
%=====================================================================================
%

\begin{figure}[hbtp]
\begin{center}
\includegraphics[width=14cm]{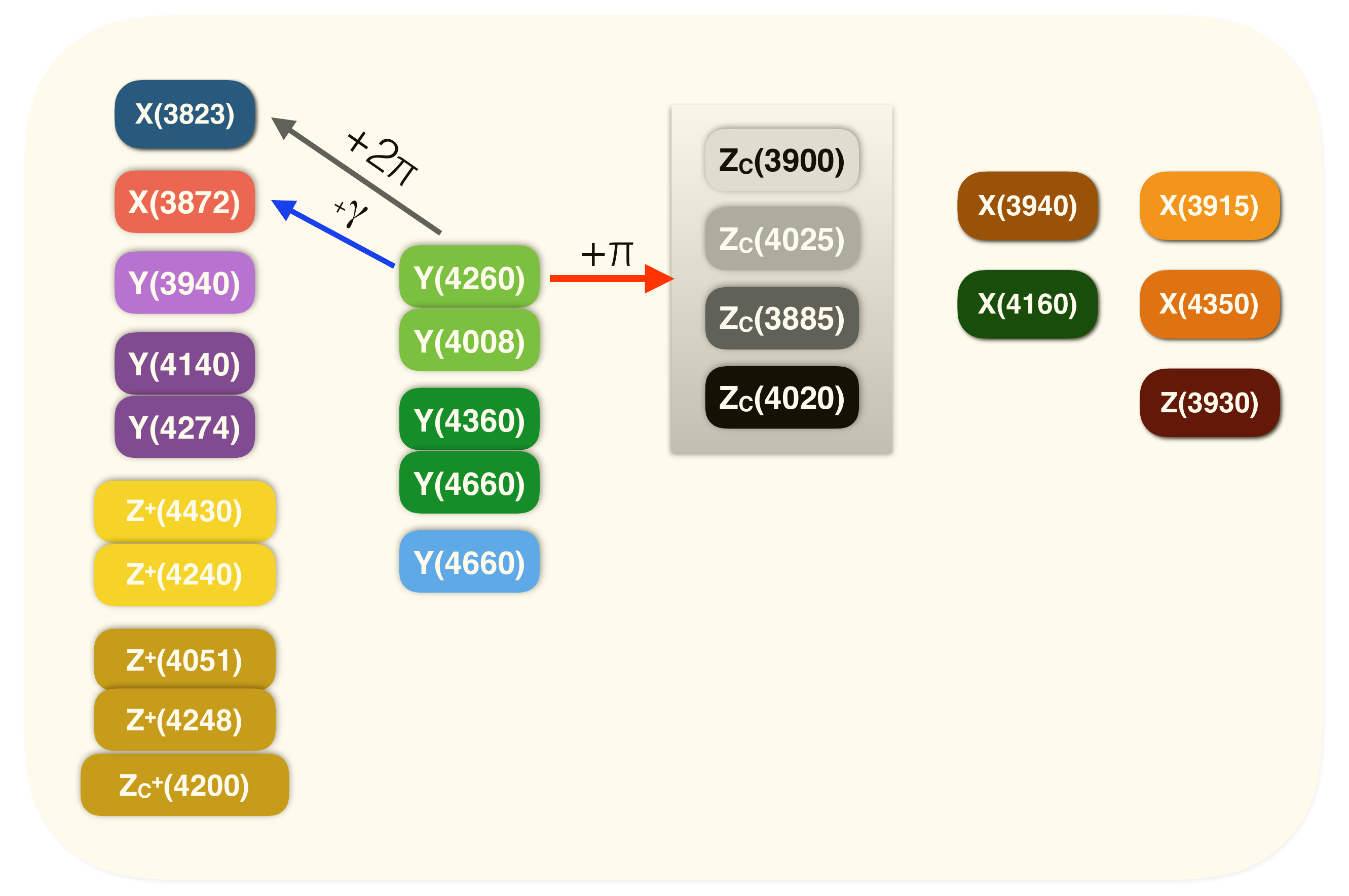}
\caption{(Color online) The connections between different $XYZ$
states. Here, the states are marked by the same color background if
they have the same production and decay modes, while the states
listed in the same column are of similar color background if they
have similar decay mode. The red, blue and grey arrows show that
there exist pionic, dipion and radiative transitions between $XYZ$
states, respectively. Additionally, the states listed in the same
column have the same production mode (see the categorization shown
in Table \ref{Fig:2.0.production}).} \label{Fig:5.2}
\end{center}
\end{figure}

The observed $XYZ$ states are not isolated since there may exist
connections between different $XYZ$ states, which are shown in Fig.
\ref{Fig:5.2}. We need to emphasize the similarities existing in some
of the observed $XYZ$ states, which may provide us some important
clues to their inner structures. For example,

\begin{itemize}
\item The similarity between the $Y(3940)$ and $Y(4140)$ inspired the
explanation of the $D^*\bar{D}^*$ and $D_s^*\bar{D}_s^*$ molecular
states to the $Y(3940)$ and $Y(4140)$, respectively \cite{Liu:2009ei}
(see review in Sec. \ref{Sect:4.7}).

\item A unified Fano-like interference picture was proposed to
explain the $Y(4260)$, $Y(4360)$ and $Y(4008)$ due to their similarity
\cite{Chen:2015bft} (see review in Sections \ref{Sect:4.6} and
\ref{Sect:4.8.1}).

\end{itemize}

The radiative transition $Y(4260)\to \gamma X(3872)$
encodes important information on their inner structure. If $X(3872)=
\chi_{c1}(2P)$, the $Y(4260)$ may favor the higher charmonium
assignment, since $Y(4260)\to \gamma X(3872)$ is a typical E1
transition.

We want to emphasize the importance of finding the connections
between different $XYZ$ states. As shown in Fig. \ref{Fig:5.2}, the
present observed connections between different $XYZ$ states are not
abundant. Future experimental efforts on the transitions between
different $XYZ$ states will shed light on their inner structures.

\subsection{Open-charm, pionic and radiative decays} \label{Sect.5.5}
%=====================================================================================
%=====================================================================================
%

The hidden-charm multiquark states have many different inner
structures, which lead to very different decay patterns. Especially,
the ratio of their open-charm and hidden-charm decay widths is very
sensitive to their underlying structures.

In certain cases, either the $XYZ$ states or the final states belong
to the same multiplet which have the same spatial wave function. In
the heavy quark symmetry limit, there generally exist
model-independent ratios of the decay widths of various decay
processes. These ratios are sensitive to different theoretical
models and encode important information of their inner structures.
Let's take the molecular assignment as an example. For the
charmonium-like molecular states, we collect some ratios of their
strong and radiative decays derived through the spin rearrangement
scheme in the heavy quark symmetry limit~\cite{Ma:2014ofa,Ma:2014zva}:
\begin{enumerate}
\item Assuming the $Y(4260)$ and $Y(4360)$ to be the isoscalar $D_1 \bar D$
and $D_1 \bar D^*$ molecular states, respectively, the ratios of
their pionic decays were evaluated to be
\begin{eqnarray}
\nonumber \Gamma(Y(4260) \to \chi_{c0} \pi^+ \pi^- \pi^0) :
\Gamma(Y(4260) \to \chi_{c1} \pi^+ \pi^- \pi^0) : \Gamma(Y(4260) \to
\chi_{c2} \pi^+ \pi^- \pi^0) = 4 : 3 : 5\, (2.11 : 1 : 1.28) \, ,
\\ \nonumber \Gamma(Y(4360) \to \chi_{c0} \pi^+ \pi^- \pi^0) : \Gamma(Y(4360) \to \chi_{c1} \pi^+ \pi^- \pi^0) : \Gamma(Y(4360) \to \chi_{c2} \pi^+ \pi^- \pi^0)
= 4 : 3 : 5\, (1.94 : 1 : 1.36) \, ,
\end{eqnarray}
where the ratio in the bracket is the result considering the phase
space factors. The ratios of their radiative decays were evaluated
to be
\begin{eqnarray}
\nonumber \Gamma(Y(4260) \to \chi_{c0} \gamma({E1})) :
\Gamma(Y(4260) \to \chi_{c1} \gamma({E1}) : \Gamma(Y(4260) \to
\chi_{c2} \gamma({E1}) = 4 : 3 : 5 \,(1.8 : 1 : 1.4) \, ,
\\ \nonumber \Gamma(Y(4360) \to \chi_{c0} \gamma({E1}) : \Gamma(Y(4360) \to \chi_{c1} \gamma({E1}) : \Gamma(Y(4360) \to \chi_{c2} \gamma({E1})
= 4 : 3 : 5\, (1.8 : 1 : 1.4) \, .
\end{eqnarray}
%The ratios of the decays of $Y(4260)$ and $Y(4260)$ were also evaluated and listed in Refs.~\cite{Ma:2014ofa,Ma:2014zva}

\item Assuming the $X(3872)$ to be the isoscalar $D \bar D^*$ molecular state,
the ratio of their radiative decays was evaluated to be
\begin{eqnarray}
\nonumber \Gamma(X(3872) \to \psi(1^3D_1) \gamma({E1})) :
\Gamma(X(3872) \to \psi(1^3D_2) \gamma({E1}) = 1 : 3\,(1 : 2.9) \, .
\end{eqnarray}

\item Assuming the $Z_c(3900)$ and $Z_c(4020)$ to be the
charged isovector $D\bar D^*$ and $D^*\bar D^*$ molecular states,
respectively, the ratios between their decay widths were evaluated
to be
\begin{eqnarray}
\nonumber {\Gamma(Z_c(3900) \to J/\psi \pi^0) \over \Gamma(Z_c(4020)
\to J/\psi \pi^0)} &=& 1 : 1 \,(1 : 1.07) \, ,
\\ \nonumber {\Gamma(Z_c(3900) \to \eta_c \rho^0) \over \Gamma(Z_c(4020) \to \eta_c \rho^0)} &=& 1 : 1 \,(1 : 2.47) \, ,
\\ \nonumber {\Gamma(Z_c(3900) \to \eta_c \gamma({E1}) \over \Gamma(Z_c(4020) \to \eta_c \gamma({E1})} &=& 1 : 1\, (1 : 1.30) \, .
\end{eqnarray}

\item Assuming the $Y(3940)$ and $Y(4140)$ to be the $D^* \bar D^*$
and $D_s^* \bar D_s^*$ molecular states of $J^{PC} = 0^{++}$,
respectively, the ratios between their decay widths were evaluated
to be
\begin{eqnarray}
\nonumber {\Gamma(Y(3940) \to J/\psi \gamma({E1}) \over
\Gamma(Y(4140) \to J/\psi \gamma({E1})} &=& 1 : 1\, (1 : 1.6) \, ,
\\ \nonumber {\Gamma(Y(3940) \to h_c \gamma({M1}) \over \Gamma(Y(4140) \to h_c \gamma({M1})} &=& 1 : 1 \,(1 : 2.8) \, ,
\\ \nonumber {\Gamma(Y(3940) \to \psi(1^3D_2) \gamma({E1}) \over \Gamma(Y(4140) \to \psi(1^3D_2) \gamma({E1})} &=& 1 : 1 \, .
\end{eqnarray}
While, assuming them to be the $D^* \bar D^*$ and $D_s^* \bar D_s^*$
molecular states of $J^{PC} = 2^{++}$, respectively, their $E1$
transition ratios were evaluated to be
\begin{eqnarray}
\nonumber \Gamma(Y(3940) \to \psi(1^3D_1) \gamma({E1}) :
\Gamma(Y(3940) \to \psi(1^3D_2) \gamma({E1}) : \Gamma(Y(3940) \to
\psi(1^3D_3) \gamma({E1}) = 1 : 15 : 84 \, ,
\\ \nonumber \Gamma(Y(4140) \to \psi(1^3D_1) \gamma({E1}) : \Gamma(Y(4140) \to \psi(1^3D_2) \gamma({E1}) : \Gamma(Y(4140) \to \psi(1^3D_3) \gamma({E1})
= 1 : 15 : 84 \, .
\end{eqnarray}

\end{enumerate}

\subsection{Hidden-charm baryonium or dibaryons with two charm quarks} \label{Sect.5.6}
%=====================================================================================
%=====================================================================================
%

The observation of the $XYZ$ and two $P_c$ states opened a window to
explore the hidden-charm four-quark and five-quark matter. One has
good reasons to expect the existence of the hidden-charm baryonium
\cite{Li:2014gra,Karliner:2015ina}, which is the six-quark matter.

The possible S-wave hidden-charm baryonia with configurations
$\Lambda_c\bar\Lambda_c$, $\Sigma_c\bar{\Sigma}_c$, and
$\Lambda_c\bar{\Sigma}_c$ may lie slightly below the thresholds of
$\Lambda_c\bar\Lambda_c$, $\Sigma_c\bar{\Sigma}_c$, and
$\Lambda_c\bar{\Sigma}_c$, which are 4573 MeV, 4740 MeV, and 4908
MeV, respectively. These states may be searched for through the
hidden-charm decay modes such as $J/\psi$ or $\eta_c$ plus $1\pi,
\eta, 2\pi, 3\pi$ or other light hadrons in the $B$ and $B_s$ decays
at LHCb. Their open-charm decays include a pair of charmed and
anti-charmed mesons, or a pair of charmed baryon and anti-baryon
etc. The hidden-bottom baryonia may also exist.

The existence of the $XYZ$ and $P_c$ states also implies the
possible existence of the doubly charmed/bottomed molecular systems
and doubly charmed/bottomed tetraquark or pentaquark systems.
Especially the dimeson, dibaryon and pentaquark molecular systems
with two charm/bottom quarks are particularly interesting, which
have have been studied using various
models~\cite{Frishman:2013qaa,Karliner:2013dqa,Yang:2009zzp,Vijande:2009kj,Park:2013fda,Bicudo:2015bra},
QCD sum rules~\cite{Du:2012wp,Chen:2013aba} as well as the lattice
QCD~\cite{Bicudo:2012qt,Bicudo:2015vta,Bicudo:2015kna}. These
hadrons may be produced either at LHC or through heavy ion
collisions or through higher bottomonium decays.

\subsection{Future facilities}
\label{Sect.5.7}
%=====================================================================================
%=====================================================================================
%

As shown in Tables \ref{Tab.5.manylabB} and \ref{Tab.5.onelab}, many
$XYZ$ states were only observed in one decay channel. Finding more
decay modes is important to reveal their structures and distinguish
different theoretical models. Although more than 20 charmonium-like
$XYZ$ states and two hidden-charm pentaquark states were observed,
we still do not know much about them.

The hadron spectroscopy is a data-driven field. It is inspiring to
take a look at the current experiments (BESIII, LHCb, and CMS) and
the forthcoming experiments (BelleII and PANDA), all of which have
contributed or will contribute to the study of the charmonium-like
multiquark states. Other collaborations such as COMPASS may also
have the potential to contribute to this field. See also discussions
in Ref.~\cite{Briceno:2015rlt}.
\begin{enumerate}

\item {\bf BESIII}:
The whitepaper of BESIII was finished in 2008~\cite{Asner:2008nq}.
The ``Charmonium Physics'' was one of its important subjects. With
the upgrade of BEPCII's LINAC in 2012, its center-of-mass energy
reaches 4.6 GeV. One year later, the $Z_c(3900)$~\cite{Ablikim:2013mio}
and the $Z_c(4020)$ were observed~\cite{Ablikim:2013emm}. BESIII
has collected lots of data
samples ranging from 3.8 to 4.6 GeV, which is an ideal platform to
study the charmonium-like physics.

\item {\bf LHCb}: The whitepaper of LHCb was also written
in 2008~\cite{Alves:2008zz}. Its primary goal was to study the CP
violation and rare decays of beauty and charm hadrons. The decay
mode $\Lambda_b^0 \to J/\psi p K^-$ was first observed in
2013~\cite{Aaij:2013oha}. Two years later, the two hidden-charm
pentaquark states $P_c(4380)$ and $P_c(4450)$ were
discovered~\cite{Aaij:2015tga}. Its current performance can be found
in Ref.~\cite{Aaij:2014jba}. LHCb has the world's largest sample of
exclusively reconstructed charm and beauty decays, and is also an
ideal platform to study the charmonium-like physics. Moreover, with
the update of LHC, LHCb would probably be able to study the
charm-bottom and bottomonium-like tetraquarks and pentaquarks.

\item {\bf CMS}: The whitepaper of CMS was again written
in 2008~\cite{Chatrchyan:2008aa}, where its primary goal was to
study the Higgs mechanism. To meet this goal, its detector is at the
energy frontier. CMS may also contribute to the charmonium-like
states.

\item {\bf BelleII}: As the update of Belle, BelleII was
designed at the rare/precision frontier to observe signatures of new
particles or processes~\cite{Abe:2010gxa}. The ``Charmonia and new
particles'' is one of its goals~\cite{Aushev:2010bq}. BelleII is
another ideal platform to study the charmonium-like, charm-bottom
and bottomonium-like tetraquarks and pentaquarks.

\item {\bf PANDA}: PANDA is another forthcoming experiment to
study the hadron spectroscopy up to the region of charm
quarks~\cite{Lutz:2009ff}. It is also an ideal platform to study the
charmonium-like tetraquark and pentaquark states. Moreover, it is
designed to study other exotic hadrons, such as the hybrid
charmonium states, glueballs, etc.

\end{enumerate}

\subsection{Outlook}
\label{Sect.5.8}
%=====================================================================================
%=====================================================================================
%

A large basis of interpolating fields are used to study the
$X(3872)$ in a recent lattice QCD simulation, including
$\mathcal{O}^{\bar c c}$, $\mathcal{O}^{D D^*}$,
$\mathcal{O}^{J/\psi \omega}$ (for $I=0$), $\mathcal{O}^{J/\psi
\rho}$ (for $I=1$) and the diquark-antidiquark
interpolators~\cite{Padmanath:2015era}. This dynamical lattice QCD
simulation was
performed with $N_f=2$ and $m_\pi=266$ MeV. The discrete scattering
states were obtained. The contributions of different operators were
isolated~\cite{Padmanath:2015era}, which helps us understand the
structure of the $X(3872)$ greatly. Hopefully future dynamical
simulations on the other important states such as the $Y(4260)$, $Z_b$
and $Z_c$ will be achieved soon. Especially the dynamical
simulations with $m_\pi$ very close to 140 MeV shall play a pivotal
role in the confirmation of the molecular scheme.

On the other hand, BESIII has been continuing collecting data. More
interesting results on the XYZ states are expected. With the
discovery of the two $P_c$ states, more partner states may be
observed at LHCb. One also anticipates more experimental measurements
on the $XYZ$ states from CMS. In the near future, BelleII will
start collecting data, which is an ideal factory of the hidden-charm
multiquark states as emphasized in Sec. \ref{Sect:4.1.5}.

Since the discovery of the $J/\psi$, the past 13 years may be the most
important period in the development of hadron spectroscopy. More
excitement, puzzles and surprises are waiting for us in the coming
golden decade. Let's cherish all the expected and embrace all the
unexpected.

%-------------Acknowledgements-----------------------------
\section*{Acknowledgements}

We would like to express our gratitude to all the collaborators and
colleagues who contributed to the investigations presented here, in
particular to Dian-Yong Chen, Rui Chen, Xiao-Lin Chen, Er-Liang Cui,
Wei-Zhen Deng, Ning Li, Xiao-Hai Liu, Yan-Rui Liu, Zhi-Gang Luo, Li Ma,
T. G. Steele, Zhi-Feng Sun, Takayuki Matsuki, Guan-Juan Wang, Lu Zhao.
We appreciate Eulogio Oset for the careful reading of the manuscript and the valuable suggestions.
We also thank Bo Wang for drawing some of
the figures in the paper, and thank Dan Zhou for helping prepare
some relevant documents. This work was supported in part by the
National Natural Science Foundation of China under Grants No.
11205011, No. 11475015, No. 11375024, No. 11222547, No. 11175073,
No. 11575008, and No. 11261130311, the Ministry of Education of
China (SRFDP under Grant No. 20120211110002 and the Fundamental
Research Funds for the Central Universities), the National Youth
Top-notch Talent Support Program ("Thousands-of-Talents Scheme"), and the Natural
Sciences and Engineering Research Council of Canada (NSERC).


%--- appendix
\appendix
%\input{section0.tex} %---for notations






\section*{References}
\bibliographystyle{elsarticle-num}
\bibliography{ref}

\begin{thebibliography}{100}
\bibitem{Zweig:1981pd}
G.~Zweig, {An SU(3) model for strong interaction symmetry and its breaking.
  Version 1}.
\end{thebibliography}

\end{document}